# SOUTH POLE-AITKEN

## SAMPLE RETURN AND EXPLORATION

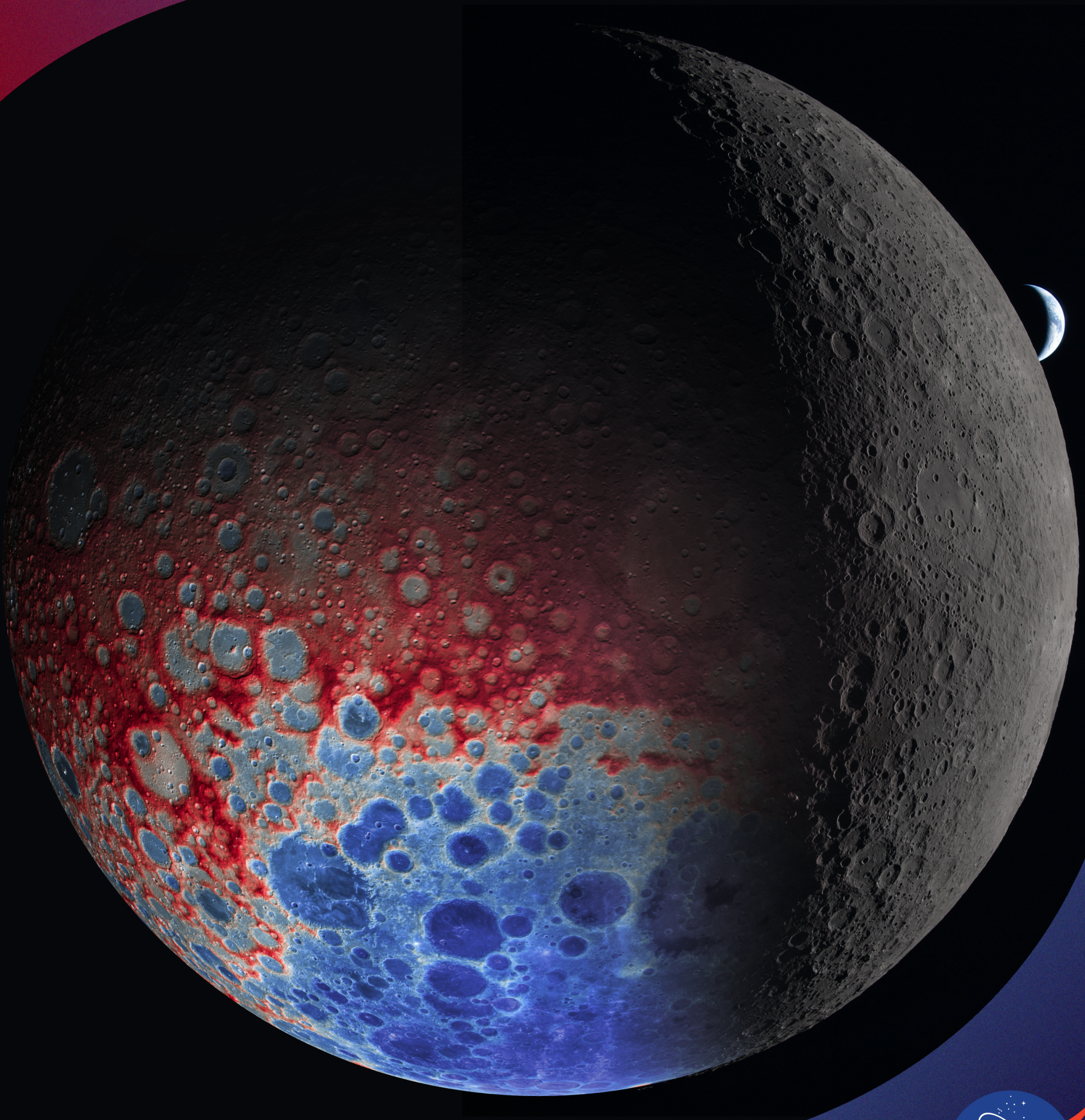

SCIENCE DEFINITION TEAM REPORT

# South Pole-Aitken Basin Sample Return and Exploration (SPARX)
## Science Definition Team Report

**CHAIR:**

Lauren M. Jozwiak, *Johns Hopkins University Applied Physics Laboratory*

**TEAM:**

Robert Citron, *Southwest Research Institute*

Natalie Curran, *Catholic University, NASA Goddard Space Flight Center*

Steve Elardo, *University of Florida*

Caleb Fassett, *Johns Hopkins University Applied Physics Laboratory*

Abigail Fraeman, *Jet Propulsion Laboratory, California Institute of Technology*

Saira Hamid, *Arizona State University*

Lena Heffern, *Metropolitan State University of Denver*

Peter James, *Baylor University*

Daniel Moriarty, *University of Maryland, NASA Goddard Space Flight Center*

Kevin Righter, *University of Rochester*

Jennifer Whitten, *Smithsonian Institute*

**DOCUMENTARIANS:**

Hunter Vannier, *Stoneybrook University*

Zoe Wilbur, *Smithsonian Institute*

**TECHNICAL TEAM LIAISON:**

James T. Keane, *Jet Propulsion Laboratory, California Institute of Technology*

**INTERNATIONAL COLLABORATORS:**

Tomokatsu Morota, *University of Tokyo, Japanese Space Agency*

Hiroshi Nagaoka, *Ritsumeikan University, Japanese Space Agency*

Caroline-Emmanuelle Morisset, *Canadian Space Agency*

Patrick Hill, *Canadian Space Agency*

James Carpenter, *European Space Agency*

Francesca MacDonald, *European Space Agency*

## DISCLAIMERS:

## RECOMMENDED CITATION:

## POINTS OF CONTACT:

SDT Chair: Lauren Jozwiak (lauren.jozwiak@jhuapl.edu)

JPL Technical Team: James Tuttle Keane (james.t.keane@jpl.nasa.gov)

NASA: Ryan Watkins (ryan.n.watkins@nasa.gov)

## FRONT COVER IMAGE:

Composite view of the farside of the Moon and the enormous South Pole-Aitken (SPA) basin as seen by both the crew of Artemis II, and NASA's Lunar Reconnaissance Orbiter (LRO) Lunar Orbiter Laser Altimeter (LOLA). The righthand side of the image is a real image of the lunar farside (and Earth) taken by the crew of Artemis II. The lefthand side of the image is a rendered view of the unilluminated hemisphere showing topography, with blues corresponding to lowlands and reds corresponding to highlands. SPA can be seen as the large, deep (blue) basin on the southern hemisphere of the lunar farside. Image credit: NASA / JPL-Caltech / James Tuttle Keane.

## BACK COVER IMAGE:

Artist concept of the SPARX rover design reference mission, as it approaches Artemis South Pole Moon Base. Image credit: NASA / JPL-Caltech / Daniel Demirdjian.

# TABLE OF CONTENTS:

# SCIENCE TRACEABILITY MATRICES:

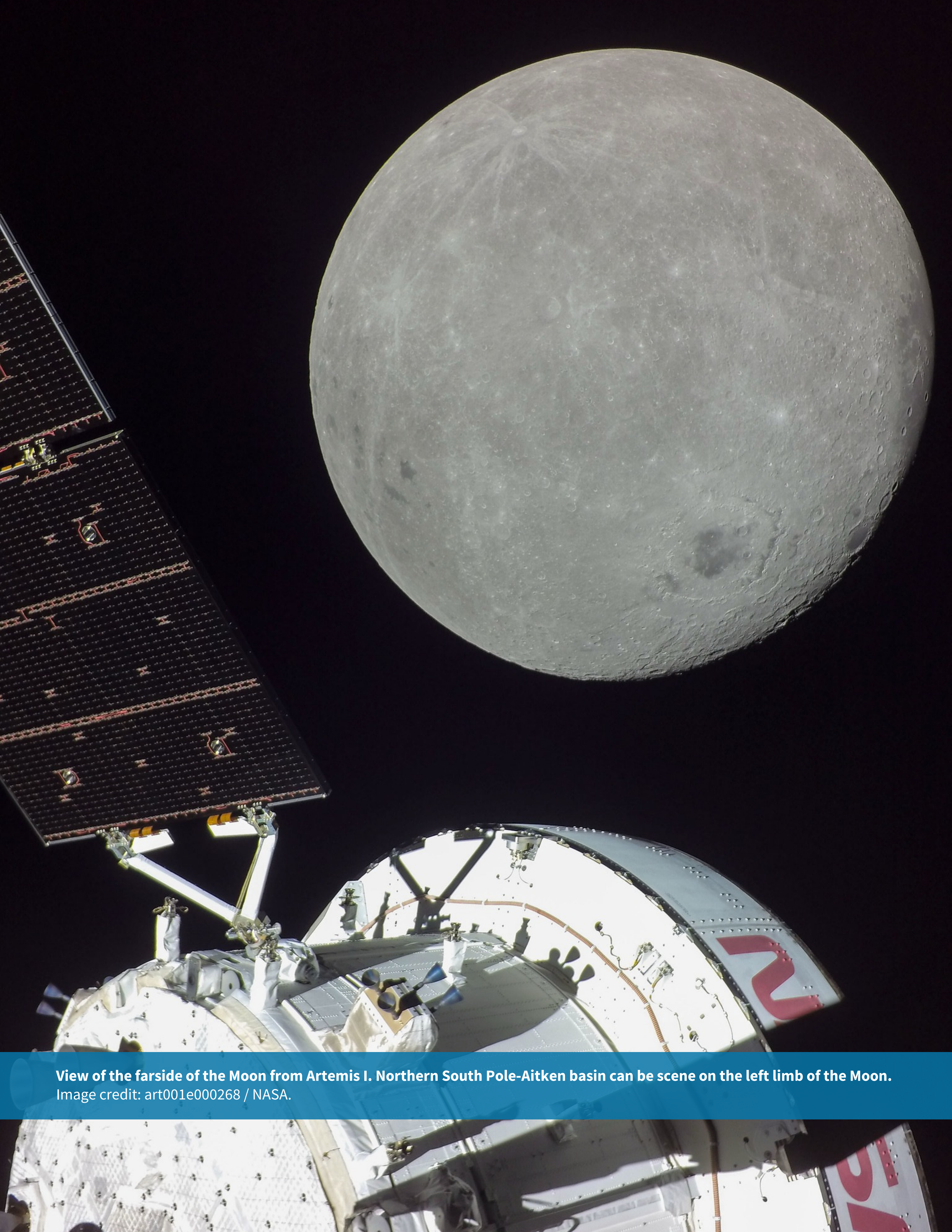

**View of the farside of the Moon from Artemis I. Northern South Pole-Aitken basin can be scene on the left limb of the Moon.**
Image credit: art001e000268 / NASA.

# 1 EXECUTIVE SUMMARY

Sample return from the Moon's South Pole-Aitken Basin (SPA) has been recognized as the highest priority lunar science mission for well over two decades, appearing as a recommended New Frontiers class mission in the two previous Planetary Science Decadal Survey documents (2003–2012 New Frontiers, 2013–2022 Visions & Voyages), and as a prioritized mission for the Lunar Discovery and Exploration Program (LDEP) in the 2023–2032 Origins, Worlds, and Life Decadal Survey (OWL). The South Pole-Aitken basin presents the ideal target destination to test nearly 60 years of lunar science hypotheses owing to a unique combination of its size, antiquity, and location on the lunar farside. Despite the recognized importance of SPA science, past sample return mission designs have been challenged by a combination of cost and technology.

This report defines science goals and objectives for a South Pole-Aitken Basin Sample Return and eXploration (SPARX) mission, the *in situ* and terrestrial laboratory measurements required for the completion of these objectives, and the lithologies, sample properties, and masses of samples that must be returned for the required measurements. This report also explores a range of potential mission implementation and programmatic architectures for achieving various levels of the SPARX science objectives, including a full range of mobility options, commercial capabilities, the role of astronauts, and the potential for multi-landing campaign-style science. From these investigations, **the SPARX SDT has developed a baseline mission concept for a long-distance rover that would collect samples and conduct science at 21 locations across SPA before rendezvousing with Artemis astronauts for sample return**. The proposed baseline mission includes a traverse design that can accomplish the SPARX science goals, optimizes the programmatic understanding of the implementation space, and fits within the cost guidance provided by NASA.

The return of samples from the Moon by the Apollo missions revolutionized not only our understanding of the Moon but also established the foundational lens through which we view the formation and evolution of the Solar System. And yet, as the field of planetary science has continued to evolve, with missions across the Solar System, comparative studies between planetary surfaces, and the re-examination of Apollo samples for a new generation, gaps in these initial models have begun to appear. The time is now, for a new generation of sample return, to address critical questions of Solar System and planetary formation, once again using the Moon as an archetype with scientific implications that will ripple across the rest of the Solar System.

## 1.1 SCIENCE GOALS AND OBJECTIVES

The SPARX science goals and objectives are designed to interrogate Decadal-level science questions, encompassing the earliest stages of the Earth-Moon system, planetary formation and crustal evolution, and the physical consequences of large-scale impacts on planetary surfaces. The SPARX SDT defined three overarching science goals and nine constituent objectives, presented below:

**SG1: What was the chronology of key processes and events during the first half billion years of solar system history?**

- O1.1: Quantify the age of the Moon to distinguish between early- and late-Moon formation models

- O1.2: Distinguish between fast- and slow-cooling models for lunar magma ocean crystallization
- **O1.3: Determine if and when a giant planet instability drove a sharp change in the inner Solar System impactor flux**
- O1.4: Constrain the evolution of the lunar dynamo in the pre-4.0 Ga epoch of lunar history

**SG2: Does the lunar farside preserve evidence of asymmetric magma ocean evolution and planetary crustal formation processes?**

- O2.1: Determine if the farside highlands crust is similar to nearside crust in age and/or composition
- **O2.2: Determine if there was a farside KREEP reservoir, or other late-stage magma ocean products, prior to the SPA impact**
- O2.3: Determine if and how the temporal and compositional evolution of the farside mantle differs from the nearside

**SG3: What are the effects of mega-basin impacts on terrestrial bodies?**

- O3.1: Determine whether there are discontinuities in the structure (crustal thickness, density, magnetic field) across SPA to constrain models for the formation of a mega-basin
- O3.2: Distinguish the relative contributions of impact melt differentiation, crustal backfilling, and volcanic infilling to the interior composition and stratigraphy of SPA

These goals and objectives represent the highest-priority lunar science questions that should be addressed by a mission to SPA, and they trace to multiple OWL Decadal science main question and sub-questions, as well as numerous lunar community documents and other priority NASA documents, including the Moon to Mars objectives. Notably, the SPARX mission achieves Decadal-level (vs Decadal-relevant) science, which is defined at the "significant, unambiguous progress in addressing at least one of the survey's 12 priority science questions." (OWL, footnote on Summary page 5).

At the request of NASA, the SDT has prioritized both the goals and objectives for a SPARX mission (Section 3.4). The goals, as presented, are in priority order, with SG1 being the highest priority goal. The priority of this goal is in recognition of the broad implications of this goal for understanding the early history of the Earth-Moon system, and for establishing the chronology and impact crater flux that define the surface dating systems used for the rest of the Solar System. Therefore, the scientific results of these investigations will have immediate utility for research across the Solar System

The prioritized objective order is as follows: O1.3, O1.1, O2.2, O2.1, O2.3, O3.2, O3.1, O1.2, O1.4. **From this prioritized list, the SDT highlights objectives 1.3, 1.1, and 2.2 as being the highest-priority objectives that should form the science threshold for any mission to SPA**.

## 1.2 SCIENCE MEASUREMENT REQUIREMENTS

The SPARX SDT further defined the *in situ* measurements, terrestrial laboratory measurements, and sample masses and properties required for the completion of the SPARX science objectives.

The SPARX required *in situ* measurements include (1) context imaging, (2) context mineralogy, (3) fine-scale mineralogy, (4) bulk elemental chemistry, (5) gravity field measurements, (6) fine-scale imaging, and (7) magnetic field measurements. **The SPARX baseline mission defines a payload of 5 strawman instruments used to address these 7 measurement classes** (Section 4.2).

The SDT further defines a set of five classes of terrestrial laboratory measurements, including petrologic characterization, bulk elemental chemistry, geochronologic measurements, volatile measurements, and remanent magnetization measurements. Each of these laboratory techniques has different sample requirements and preparations, and these have informed the sample requirements for SPARX. **The SPARX SDT recommends a sampling strategy focused on the acquisition of small rocks (diameter ranges ≥40 mm to ≥3 mm) and unsieved bulk regolith**. This strategy will acquire sufficient material to perform the required analysis, while also allowing a sufficient number of analyses to be performed to generate statistically robust results. Using the required sample properties and masses from the laboratory measurements, combined with NASA sample curation requirements, and scaling factors to account for the natural variability of the lunar regolith, SPARX defines both a baseline and threshold sample return mass requirement (Section 4.3 and 4.4):

- **Baseline Return Sample Mass: ≥20.5 kg**
- **Threshold Return Sample Mass: ≥9.8 kg**

The SPARX TOR requirements for return sample mass indicate a NTE (not to exceed) mass of 100 kg (inclusive of sample containers) for mission designs that incorporate astronaut return from the lunar South Pole, and NTE 10 kg mass for commercial robotic return. The SPARX baseline mission architecture profile will be discussed in further detail below; however, the projected baseline and threshold return sample mass allocations are within the scope provided by the TOR.

## 1.3 MISSION IMPLEMENTATION

The SPARX TOR directed the SDT to investigate a range of possible mission implementation architectures which may accomplish some or all of the SPARX science priorities. The SDT was further directed to offer commentary on the viability of these architectures for achieving the SPARX objectives, with emphasis on the SPARX highest-priority objectives, before selecting a single architecture for the SPARX baseline mission concept. In defining these implementation architectures, the SDT considered the axis of mobility, ranging from static lander to long-distance (>100 km) rover, and sample return architecture, encompassing both robotic return and astronaut-assisted return (Section 5).

Accomplishing either the baseline or threshold SPARX objectives requires the ability to traverse to multiple different regions within SPA, distances significantly greater than 100 km. **A long-distance rover is the only implementation architecture capable of meeting the highest-priority SPARX objectives**.

Given the return mass envelope presented in the TOR, robotic sample return is a viable implementation option for the threshold SPARX mission; however, the lack of currently existing capabilities for either NASA-developed or commercial robotic sample return options increases the risk of committing to this return architecture. **The SDT determined that astronaut-assisted sample return is the preferred implementation for SPARX baseline sample return**. Samples from the Artemis exploration zone and the vicinity of the proposed Moon Base do not address any SPARX science objectives, but requiring traversing this region for astronaut rendezvous does result in multiple additional engineering constraints for the SPARX rover. **A non-polar astronaut landing rendezvous with the SPARX rover at the terminal traverse location is the ideal SPARX architecture, although, designs can also accommodate rendezvous at a south pole Moon Base location as shown in the design reference mission**.

## 1.4 BASELINE MISSION

Using the requirements set by the SDT, the JPL technical team created a design reference mission for a long-distance rover. The purpose of this mission design exercise was to enable an initial feasibility and cost assessment for meeting SPARX science goals. (Section 6)

The baseline SPARX traverse from central SPA to the south pole Moon Base region is ~2,800 km and includes 20 science sampling locations and 1 science non-sampling traverse location. The final ~500 km of this traverse is just to reach the south pole Moon Base. The SPARX threshold traverse includes 10 science sampling locations from central SPA to Poincare basin and is ~1,500 km to the terminal sampling location, although an additional ~1,000 km of traverse would be required to deliver the samples to the South Pole for astronaut-assisted return. This highlights that the highest priority sample sites are far from the Artemis exploration zone and Moon Base near the South Pole.

The cost numbers presented here represent the Phases A-E costs for JPL design reference mission, and assuming a class C+ risk posture. Following TOR guidance, instrument costs are treated as pass-through, with a flat-rate cost of $25M for each of the five strawman payload elements. The costs do not include surface delivery costs, sample return costs, curation costs, or other Phase F costs. **The SPARX baseline mission cost is $840M in FY25 dollars, inclusive of reserves, which is well below the TOR recommended cap of $1.5B.**

## 1.5 CONCLUSIONS

Over half a century after the Apollo missions first returned samples from the lunar surface and bolstered by several generations of remote orbital observation of the Moon, we are now at the dawn of a new era of lunar exploration with sustained human presence as the goal. The SPARX mission will be a crowning scientific achievement of this new era, rewriting the lunar science textbooks, and answering science questions that affect our understanding of the entire Solar System. The goals and objectives defined by this report present a scientific outline for a mission that would transform lunar science, would address multiple high-level Decadal science questions at the only place in the Solar System where they can be credibly addressed, would pioneer a new paradigm for planetary surface exploration in the 21st century, would leverage the burgeoning space economy to pursue ground-breaking science on a more cost-efficient basis, and would unite robotic exploration with human exploration in a model for a future sustained lunar enterprise.

# SPARX

South
Pole
Aitken Basin Sample
Return and
eXploration

MISSION CONCEPT
TARGET LAUNCH DATE:
2031

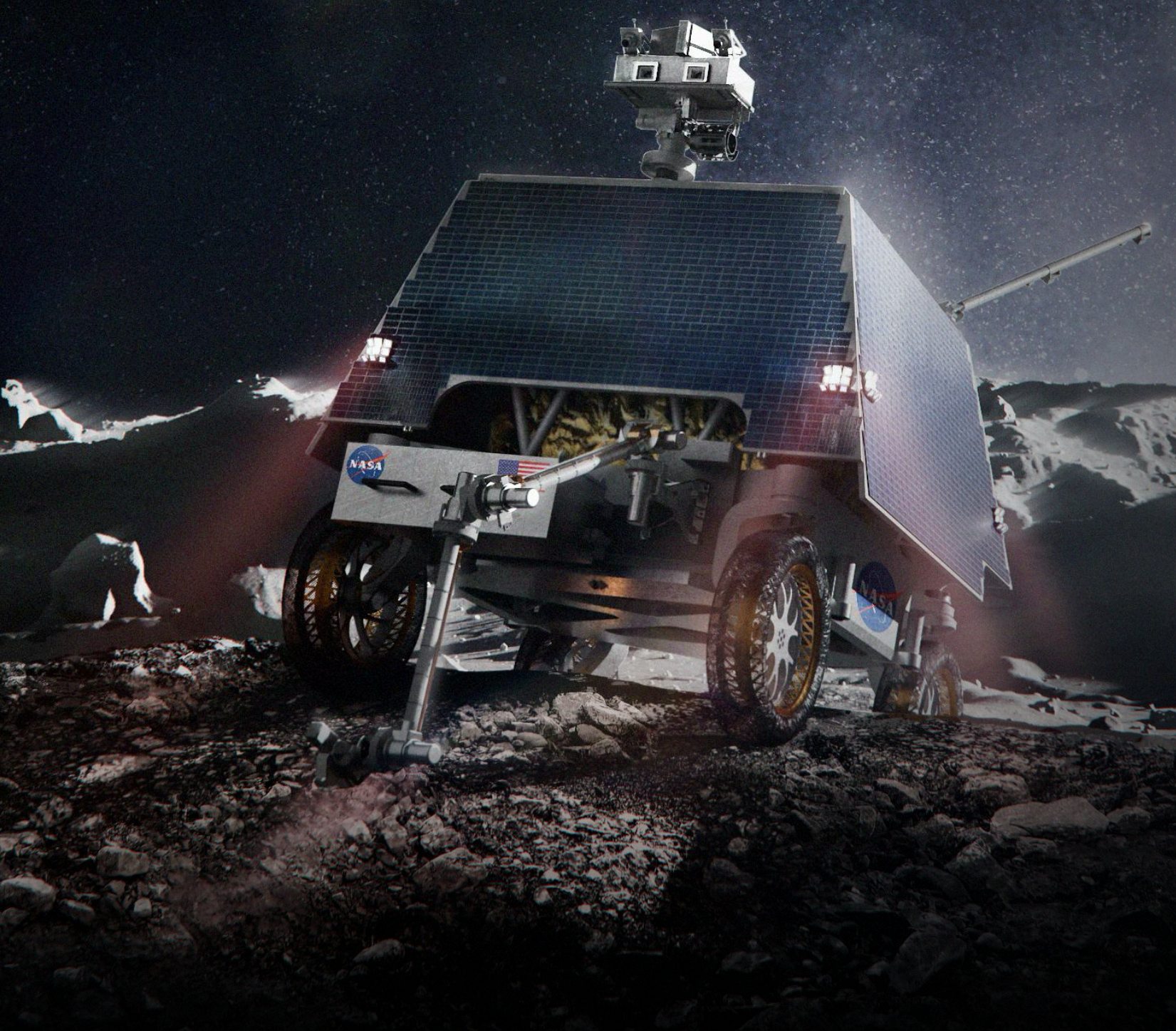

## Bridging robotic discovery with human exploration to unlock the origin of worlds.

**SPARX** is a mission concept for a long-range autonomous rover, capable of traversing over 2,700 km, designed to explore and ultimately collect, cache, and deliver samples for Artemis return from across the **South Pole-Aitken (SPA)** basin on the far side of the Moon. SPA is the largest and oldest impact basin on the Moon and holds critical answers to solar system chronology and planetary evolution.

## Science Goals and Objectives

1. **Determine** the chronology of key processes and events during the first half billion years of solar system history.
2. **Characterize** the lunar far side to investigate evidence of asymmetric magma ocean evolution and planetary crustal formation processes.
3. **Quantify** the effects of mega-basin impacts on terrestrial bodies.

## SPARX Team

SPARX Science Definition Team Chair:
**Dr. Lauren Jozwiak**
Johns Hopkins Applied Physics Laboratory
lauren.jozwiak@jhuapl.edu

JPL Technical Team Liaison:
**Dr. James Tuttle Keane**
Jet Propulsion Laboratory, California Institute of Technology
james.t.keane@jpl.nasa.gov

## Science Payloads

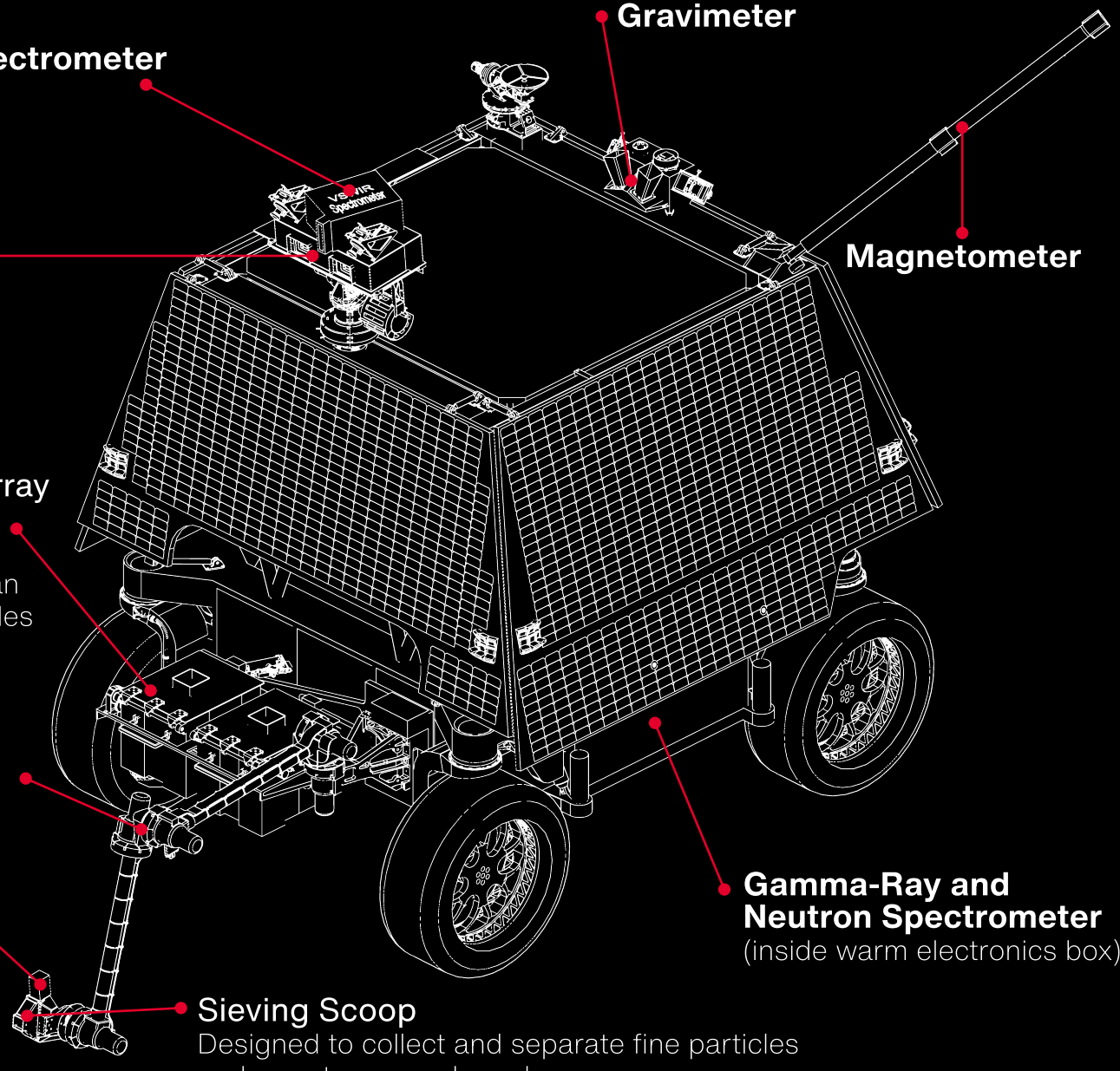

# Mission Facts

- Deliver rover to lunar surface via **Commercial Lunar Payload Services** (CLPS)
- Autonomously traverse over **2,700 kilometers (1,680 miles)** across SPA basin within 5 years
- Collect lunar regolith and rock samples from **20 distinct sites** for a total sample mass of **>20 kg (44 lbs)***
- Carry sufficient energy storage to **survive up to 520 hours** of extreme cold lunar night
- Deliver samples to **Moon Base** for astronaut retrieval and return

* **>20 kg** — approaching the weight limit for a standard airline checked bag

# The vast scale and geologic diversity of SPA **requires long-range mobility**

## Key Terrains

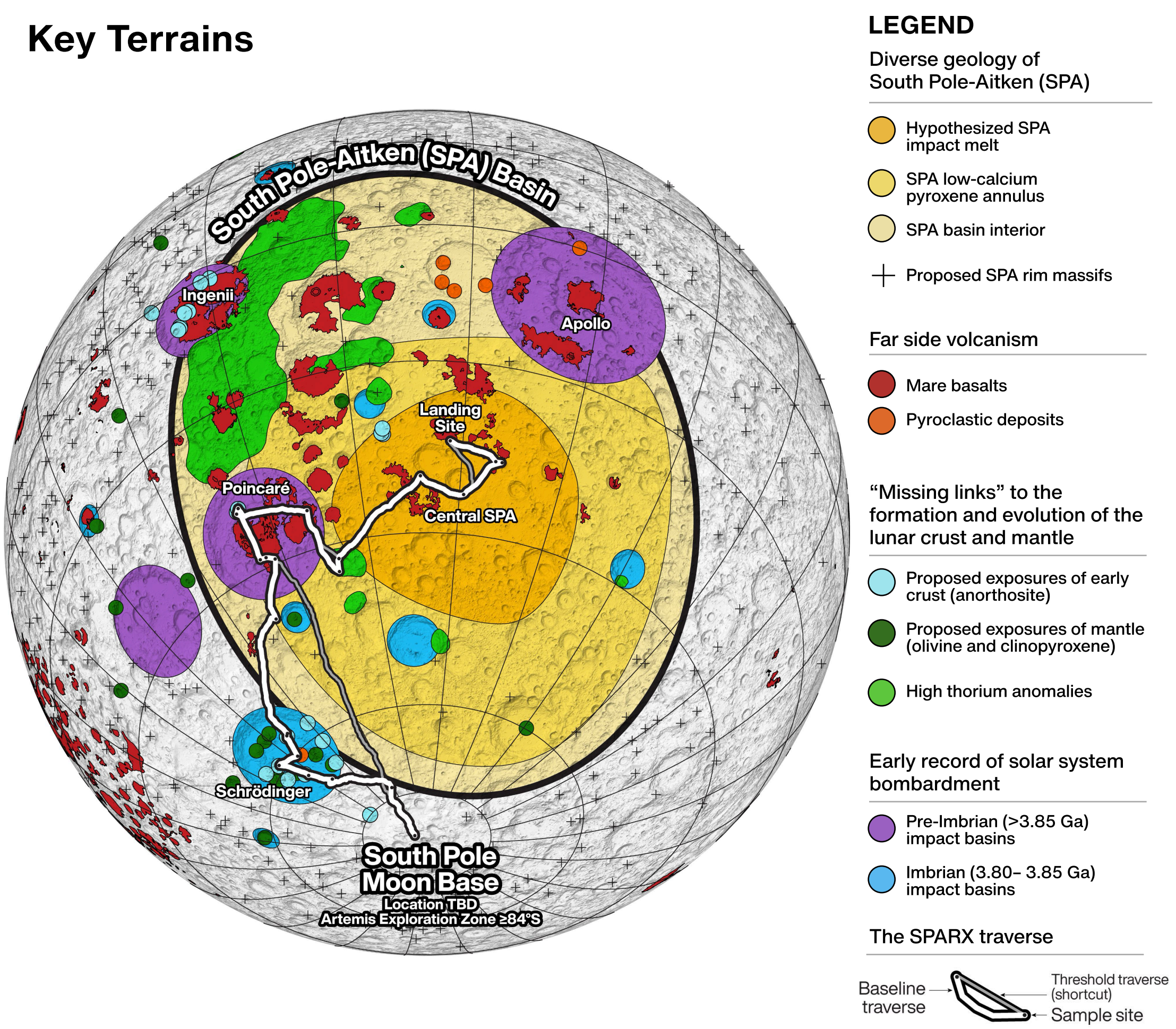

## **SPARX:** Rover Top-Level Schedule

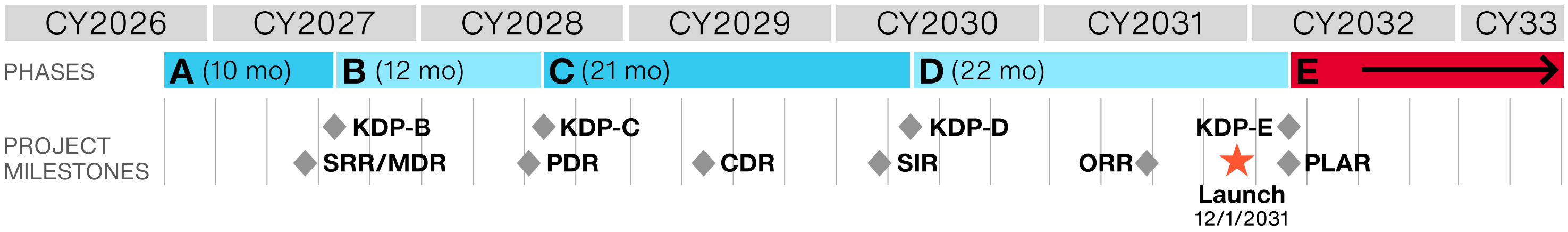

## **Mission Concept of Operations:** Phase E

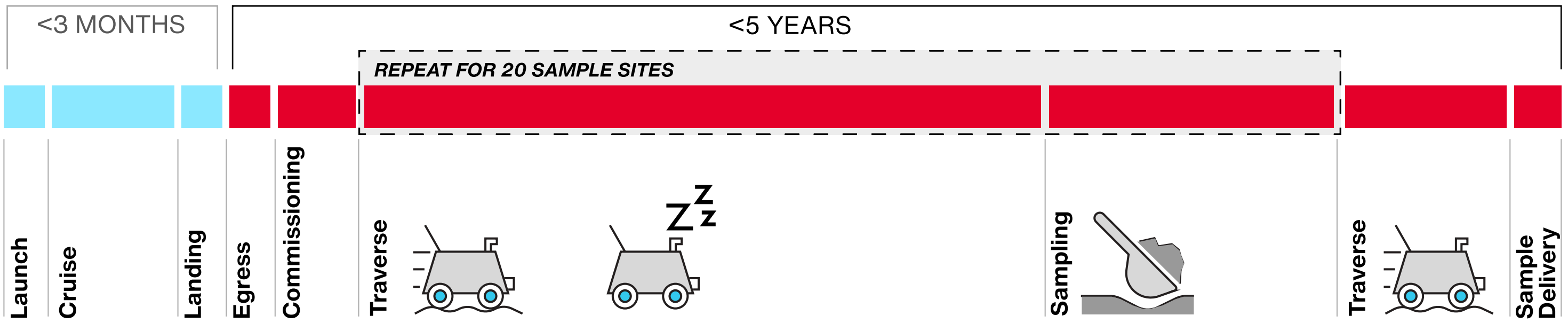

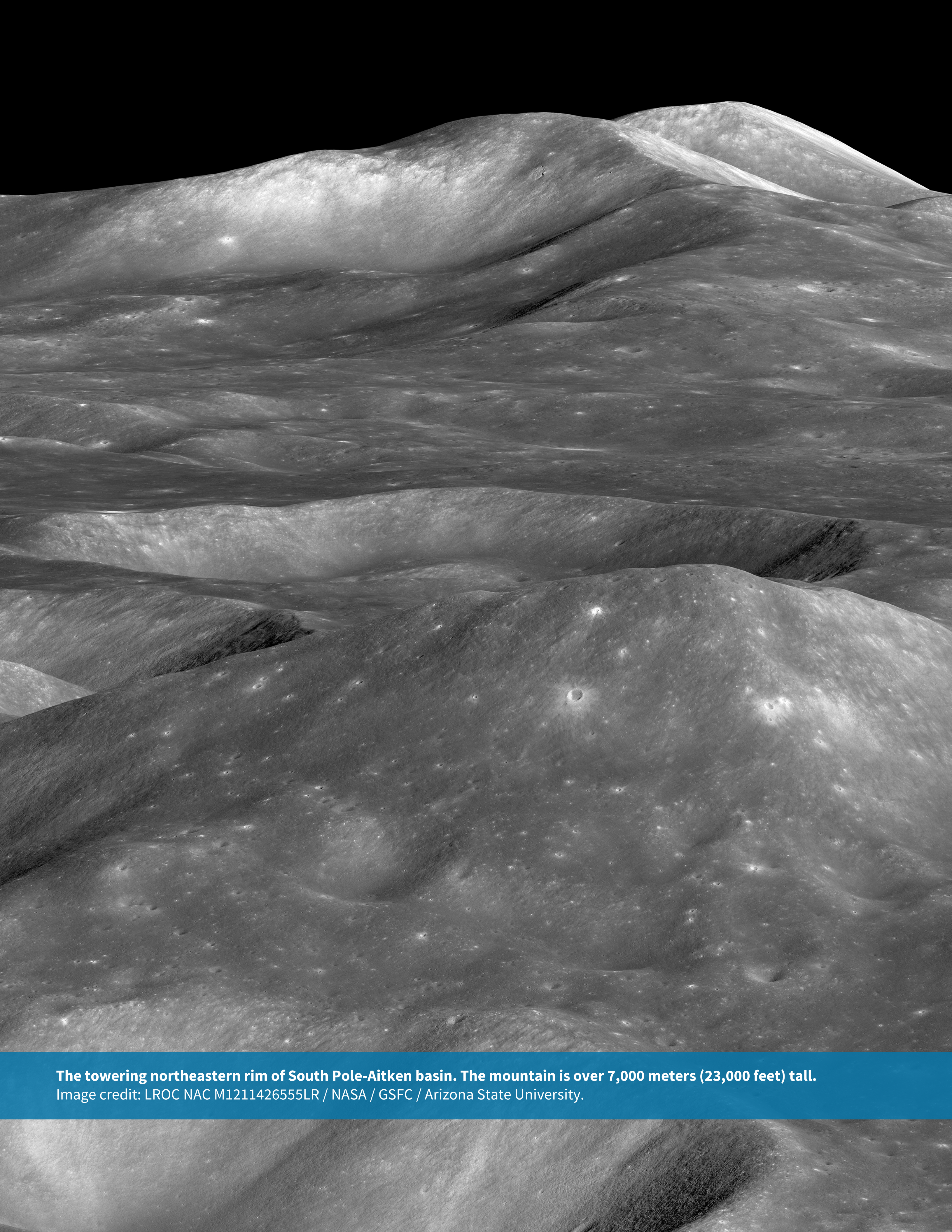

**The towering northeastern rim of South Pole-Aitken basin. The mountain is over 7,000 meters (23,000 feet) tall.**
Image credit: LROC NAC M1211426555LR / NASA / GSFC / Arizona State University.

# 2 OVERVIEW OF SOUTH POLE-AITKEN BASIN SAMPLE RETURN AND EXPLORATION (SPARX)

## 2.1 BACKGROUND TO SPARX

Sample return from the Moon's South Pole-Aitken Basin (SPA) has been recognized as the highest priority lunar science mission for well over two decades, appearing as a recommended New Frontiers class mission in the two previous Planetary Science Decadal Survey documents (2003-2012 New Frontiers, 2013-2022 Visions & Voyages), and as a prioritized mission for the Lunar Discovery and Exploration Program (LDEP) in the 2023-2032 Origins, Worlds, and Life Decadal Survey. The South Pole-Aitken basin presents the ideal target destination to test nearly 60 years of lunar science hypotheses owing to a unique combination of its size, antiquity, and location on the lunar farside. Despite the recognized importance of SPA science, past sample return mission designs have been challenged by a combination of cost and technology.

The 2023-2032 Origins, Worlds and Life (OWL) Decadal Survey presented a new SPA sample return mission concept study named "Endurance" that demonstrated the feasibility of a long-duration, long-traverse mission that could accomplish the majority of defined priority lunar science investigations at a cost cap that was commensurate with New Frontiers scale missions. This mission concept leveraged new developments in rover technology, autonomous systems development, and concepts of operations developed by the Intrepid Pre-decadal Mission Concept Study, in conjunction with technological advances in the commercial exploration marketplace and the progress of the Artemis exploration program. The Endurance mission presented two mission profiles—Endurance-A and Endurance-R—both of which accomplished high-priority lunar sample return science, but which differed in their sample return mechanism. Endurance-A proposed sample return via rendezvous with Artemis astronauts at the lunar South Pole, and Endurance-R proposed fully robotic sample return (similar to Mars Sample Return) from within the South Pole-Aitken basin. The OWL supported development of an SPA Sample Return mission as the highest priority mission for the Lunar Discovery and Exploration Program (LDEP). In response to the OWL recommendation and significant support from the lunar science community, NASA convened the South Pole-Aitken basin sample Return and eXploration (SPARX) Science Definition Team (SDT). NASA elected to convene an SDT to gather broader input from the planetary science community on the science goals of a South Pole-Aitken basin sample return mission, to provide a prioritization of the science objectives, to provide further clarification on the required measurements and sample quantities required by a SPA sample return mission, and to explore a range of alternative implementation approaches not explored in the initial Endurance mission concept study report.

## 2.2 GROUND RULES, APPROACH, AND TERMS OF REFERENCE

Membership in the SPARX SDT was openly competed via a ROSES solicitation, from which the chair, panel members, and early-career documentarians were selected in January 2025. The SDT consisted of 12 team members, 2 documentarians, and 2 non-voting members each from the Japanese Aerospace Exploration Agency (JAXA), the European Space Agency (ESA), and the Canadian Space Agency (CSA). The SDT was assisted in technical analyses by representatives from JPL including a science liaison and a technical liaison, in addition to other science and engineering staff as required to carry out the requested investigations. The SDT operated as a non-consensus body, and the following report and appendices contain records and discussion of non-consensus analyses and voting records for significant SDT findings. The SDT met 6 times over the course of the 2025 calendar year, including 4 hybrid attendance meetings and 2 virtual meetings. Three of the hybrid meetings were supported by non-panel technical staff from the Jet Propulsion Laboratory, who were requested to brief the SDT on specific technical topics relevant to the individual meeting goals. In addition to full team meetings, the team also participated in working group teleconferences, and robust email exchange.

The SPARX Terms of Reference (TOR) document outlined the 11 main tasks for the SDT, and provided a set of guidelines for costing, and ground rules for certain technical and sample return aspects of the prospective mission profiles. One notable change to the TOR occurred in Spring 2025, when the SDT was directed by NASA that all implementation architectures should deprioritize architectures that utilize radioisotope power systems (RPS) following the cancellation of support for these systems in the FY26 President's Budget Request. This requirement was imposed after the first 2 meetings of the SDT and represents a significant departure from the Endurance mission concept, which leveraged a radioisotope thermoelectric generator (RTG) for nighttime operations.

The SPARX SDT Terms of references are summarized below, with the full text presented in Appendix 10.1. Section numbers designate where the relevant materials can be found in the report.

- Define and prioritize science goals for a South Pole-Aitken Basin Sample Return and Exploration mission that is traceable to the science questions and objectives outlined in the OWL and other community documents. (Section 3)
- Define a hierarchy of science objectives that are responsive to the prioritized goals above and identify the criteria that will enable these objectives to be met. (Section 3.4, Section 4.1)
- Discuss the criteria used to prioritize the science goals and objectives. (Section 3.4)
- Determine example location(s) in SPA where the science goals can be met. (Section 4.1, Section 5.2)
- Define the number and type of samples and *in situ* measurements that are needed to address the prioritized science objectives. The committee shall also note the minimum expected number of kilograms of samples required to meet the baseline and threshold missions. (Section 4.1, 4.2, 4.4)
- Define a set of strawman science instrument types that fully address the SDT-defined science objectives, as well as their required measurement capabilities. (Section 4.2, 6.4)
- Develop a Science Traceability Matrix (STM) that flows from the top-level science goals through science objectives and derived measurements, culminating in functional requirements and capabilities for the strawman payload suite, mobility system (if applicable), and sample return system. (Section 3.6)

- Assess the role of astronauts in accomplishing these science goals and objectives. (Section 7.1, Section 7.6.2)
- Determine what technology(ies) need to be matured before the project starts. (Section 6.11, 7.8)
- Define and present implementation approaches to achieving the proposed science objectives, considering a variety of approaches that allow different levels of the hierarchy of science objectives to be met. The committee should also identify key discriminators between the various implementation approaches and outline the rationale for why these approaches are best suited to meeting the science objectives. (Section 5.4–5.8, Section 7.1, Section 7.6)
- From the range of implementation approaches, the SDT shall develop one nominal mission concept that maximizes the overall science return and progress towards NASA's goals within the resources and risk posture constraints provided by NASA. The SDT shall develop both a baseline and threshold mission and describe priorities for scaling the mission concept up or down (in cost and capability) and describe potential trades. (Section 5.8, Section 7.5, Section 8)

The SDT was tasked with investigating and presenting a range of implementation options that also includes a baseline and threshold mission description. For the purposes of the report, the descriptions include all of the designated science objectives, their associated experiments, and example locations where those objectives could be met. The report then presents a baseline and threshold architecture that achieve the outlined set of objectives/ a prioritized subset of the described objectives. The report also includes a discussion of several other traverse implementations and mission architectures that were investigated, as well as an analysis of their ability to achieve the science objectives, and the overall feasibility analysis of the SDT. Per the TOR, costing information is only provided for the baseline architecture investigation, and does not include costs for lander delivery, sample return, or Phase F costs.

## 2.3 REPORT ORGANIZATION

The following report uses an organizational structure that follows that of a Science Traceability Matrix, beginning with the broad science goals of the mission, and progressing through measurements, requirements, and eventually architectures. A brief breakdown of the material in each section is found below:

- **Section 1**: Executive Summary
- **Section 2**: Report goals, organization, and motivation
- **Section 3**: Science goals, objectives, prioritization, and traceability to community priorities and NASA goals
- **Section 4**: Science Requirements (physical observations, measurements, sample requirements, traverse location requirements), *in situ* measurement requirements, terrestrial laboratory measurement requirements, and total sample mass requirements
- **Section 5**: Mission architecture trades, rationale, baseline and threshold architectures
- **Section 6**: Design reference mission for baseline architecture, including top level system requirements, ConOps, traverse design, and cost
- **Section 7**: Programmatic scope, available trades, astronauts, commercial space, evolving lunar exploration program
- **Section 8**: Report conclusions and SDT findings

Throughout the report, we have identified critical findings pertaining to the science and implementation of SPARX. These findings are noted in call-out boxes, and all of the findings can be found in Appendix 10.6 at the end of the report.

## 2.4 CONTEXT FOR THE EXPLORATION OF THE SOUTH POLE-AITKEN BASIN – WHY SPA, WHY NOW?

The return of samples from the Moon by the Apollo missions revolutionized not only our understanding of the Moon but also established the foundational lens through which we view the formation and evolution of the Earth, the Solar System, and planetary systems beyond. Lunar samples, and the resulting lunar magma ocean hypothesis, anchor our understanding of planetary crustal formation, and the Moon remains the archetype for primary crustal formation. In a related manner, the ages of large impact events recorded by Apollo samples led to the hypothesis of a spike in the impact flux rate during the last stages of early Solar System formation. The physical record of this “cataclysm” has been observed in the impact records on other planetary bodies, and the resulting models of impactor populations are directly use to model ages for all surfaces in the Solar System that do not have samples to anchor them (which is to say the overwhelming majority of surfaces). And yet, as the field of planetary science has continued to evolve, with missions across the Solar System, comparative studies between planetary surfaces, and re-examination of Apollo samples for a new generation, gaps in these initial models have begun to appear.

While the initial lunar magma ocean hypothesis seemed to provide a framework for the observed lunar crust, generations of petrologic modeling suggested the initial model was unstable, and a new generation of orbital remote sensing assets indicated compositional variations in crustal units that were once thought to be global. Meanwhile, continued examination of sample ages led to questions of spatial sampling bias, and the true shape and driving mechanisms of early Solar System impactor populations. Resolving these questions has implications, not only for better understanding the Moon, but also for understanding and contextualizing missions to all parts of the Solar System. Despite the seemingly disparate nature of these foundational questions, multiple generations of lunar researchers have persuasively argued that samples from a single region of the Moon could address a preponderance of them—the South Pole-Aitken basin on the lunar farside (Figure 2.1). As the oldest and deepest basin on the Moon, it excavated the earliest generation of the Moon’s crust and likely into the lunar mantle. With a formation age prior to the hypothesized cataclysm, the basin interior was a witness to this critical period of impactor history and is located far enough away from the nearside basins to minimize the potential sampling bias of the existing sample collection. Finally, the incomplete interior modification of the basin retains the geologic fingerprints of successive generations of processes, allowing for both exposure of these most ancient materials, and also investigations of the modification process itself.

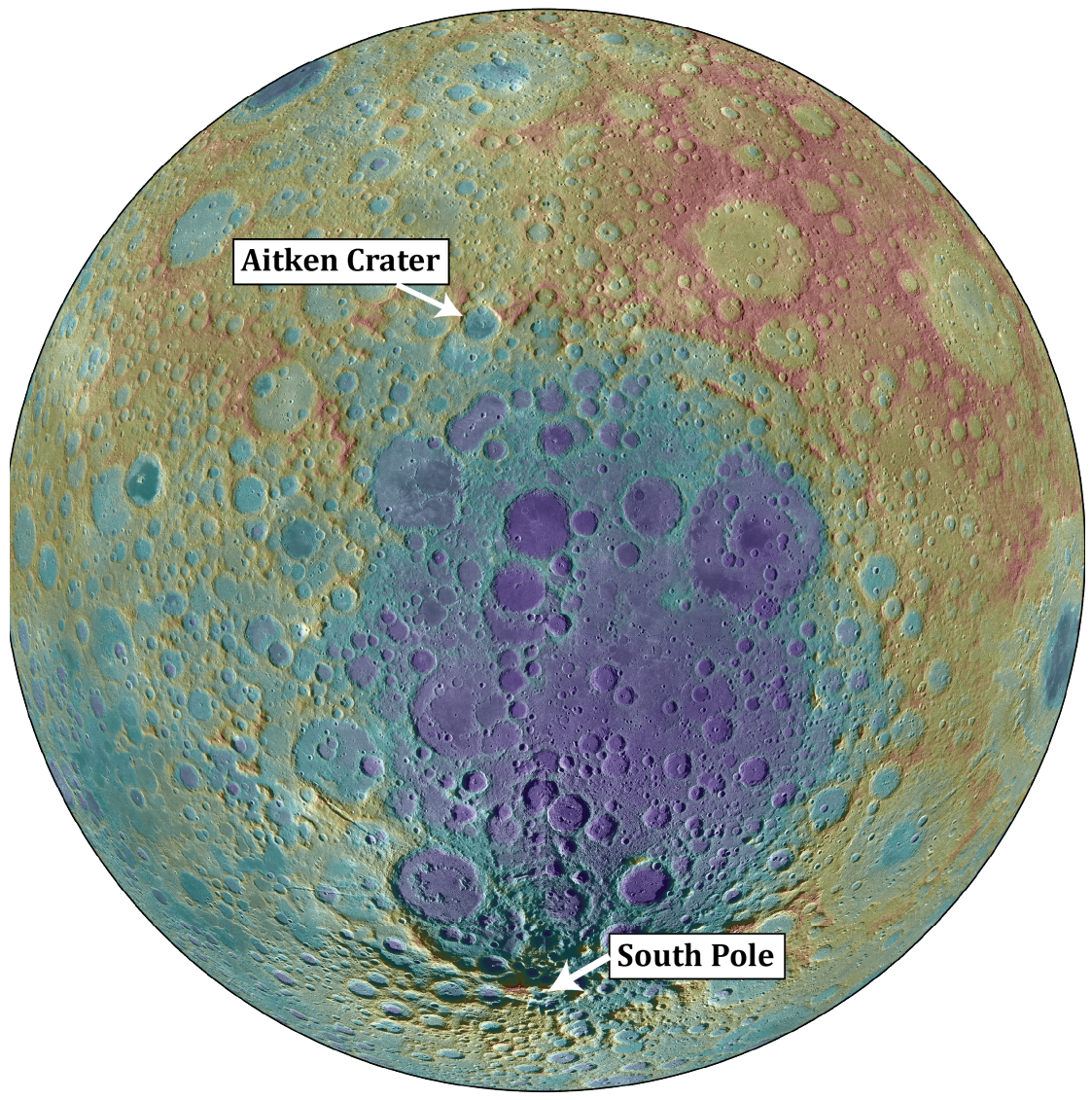

**Figure 2.1.** South Pole-*Aiken basin is the elliptical impact basin spanning from Aitken crater to the lunar* South Pole. *LROC-WAC image mosaic overlain with merged LOLA-Kaguya topography (blue-low; red-high).*

For all these reasons, the South Pole-Aitken basin has, for several decades, been identified as one of the highest priority exploration destinations on the Moon, with sample return from SPA appearing in all National Academy of Sciences Planetary Science and Astrobiology decadal surveys to date. Despite the community consensus about the importance of SPA science, no mission to return samples from SPA has yet been selected. The question then naturally arises, why now?

The re-assessment of a sample return mission to SPA comes from the confluence of several factors: improvements in remote sensing data (e.g., from the Lunar Reconnaissance Orbiter) that enable mission planning, technological advancements in remote surface exploration technology, the rise of a U.S. commercial space industry focused on the lunar surface market, and the programmatic direction of human exploration and sustained presence at the lunar South Pole. The mission profiles presented in this report leverage all of these key advancements to present a scientific outline for a mission that would transform lunar science, would address multiple high-level Decadal science questions at the only place in the Solar System where they can be credibly addressed, would pioneer a new paradigm for planetary surface exploration in the 21$^{st}$ century, would leverage the burgeoning space economy to pursue ground-breaking science on a more cost-efficient basis, and would unite robotic exploration with human exploration in a model for a future sustained lunar enterprise.

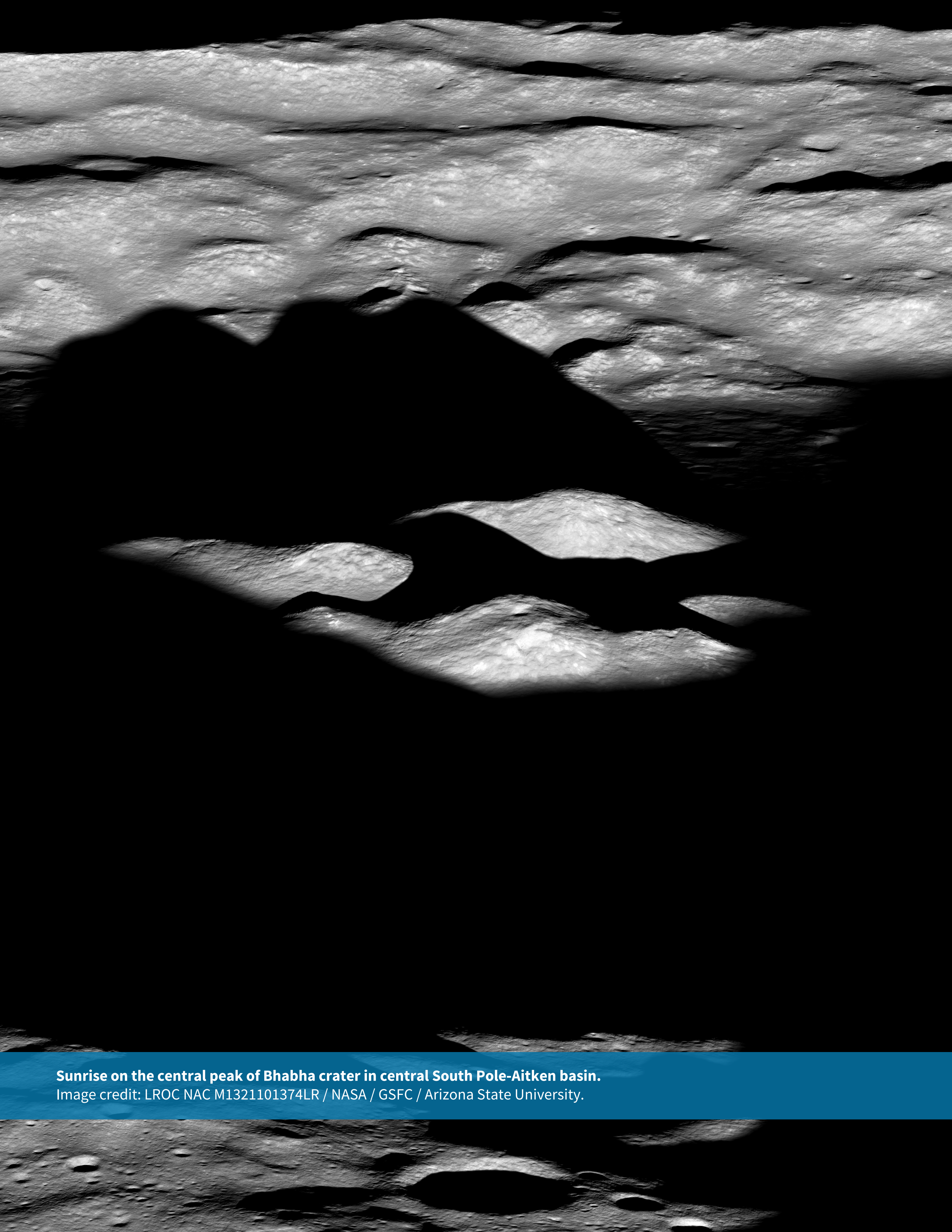

**Sunrise on the central peak of Bhabha crater in central South Pole-Aitken basin.**
Image credit: LROC NAC M1321101374LR / NASA / GSFC / Arizona State University.

# 3 SCIENCE GOALS AND PRIORITIZATION

## 3.1 GOAL 1

**Goal 1: "What Was the Chronology of Key Processes and Events During the First Half Billion Years of Solar System History?"**

From the building of planets to the formation of moons, the first half billion years of Solar System history set the stage for all of the geologic processes that have resulted in the diverse range of bodies observed today (Figure 3.1). For the Earth, in particular, its earliest history saw the formation of the Moon through a giant impact with the proto-Earth, and the subsequent evolution of both bodies following that event (Canup et al., 2023). Following this event, both the Earth and Moon cooled, a magnetic field arose, large impacts bombarded both bodies, and on Earth conditions arose that would set the stage for life to evolve. However, despite the wide array of critical geologic processes thought to occur, there exist precious few markers of this timeframe in the geologic record, with plate tectonics rearranging, recycling, or outright destroying most of Earth's crustal material throughout its history.

Fortunately, unlike the Earth, the Moon preserves abundant material and isotopic fingerprints that provide key information about the first half billion years of solar system history. Returning samples of this material thus gives us an opportunity to learn about fundamental questions in the chronology of the Solar System, including the age (or timing) of specific events and the period (or length of activity) for key processes. **What we seek to determine in Goal 1 includes the timing for the Moon's earliest beginning, the timing and key processes of the Moon's differentiation and crustal formation following the Moon-forming impact, and the shape and nature of the impactor flux during the first half billion years of Solar System history.** While Apollo samples were critical to formulating initial hypotheses and models for all of these events, major questions remain. The importance of these questions is directly reflected in the OWL Question 3 "Origin of the Earth and Inner Solar System Bodies", with the four Goal 1 science objectives directly addressing Main Science Questions 3.2, 3.3, and 3.5, as well as substantial portions of Question 4.2.

The Moon-forming event is canonically hypothesized to be a result of a giant impact of a Mars-sized planetesimal into the proto-Earth (Canup et al., 2023), however, the timing and details of that event remains the subject of significant debate, and interpretation of what follows is limited by the spatial biases of the current sample collection. Directly relevant to understanding the Moon-forming event is understanding the timescale for lunar magma ocean evolution and crustal formation, as the timescale of these processes defines the range of ages used to contextualize the initial formation event. Here too, our understanding of crustal formation is potentially biased by our nearside-focused sample collection, with a scant few samples coming from the lunar farside and highlands. Indeed, the geochemistry of lunar highland meteorites has led to the hypothesis that serial magmatism, rather than just magma ocean differentiation, played a major role in crustal formation (Gross et al., 2014).

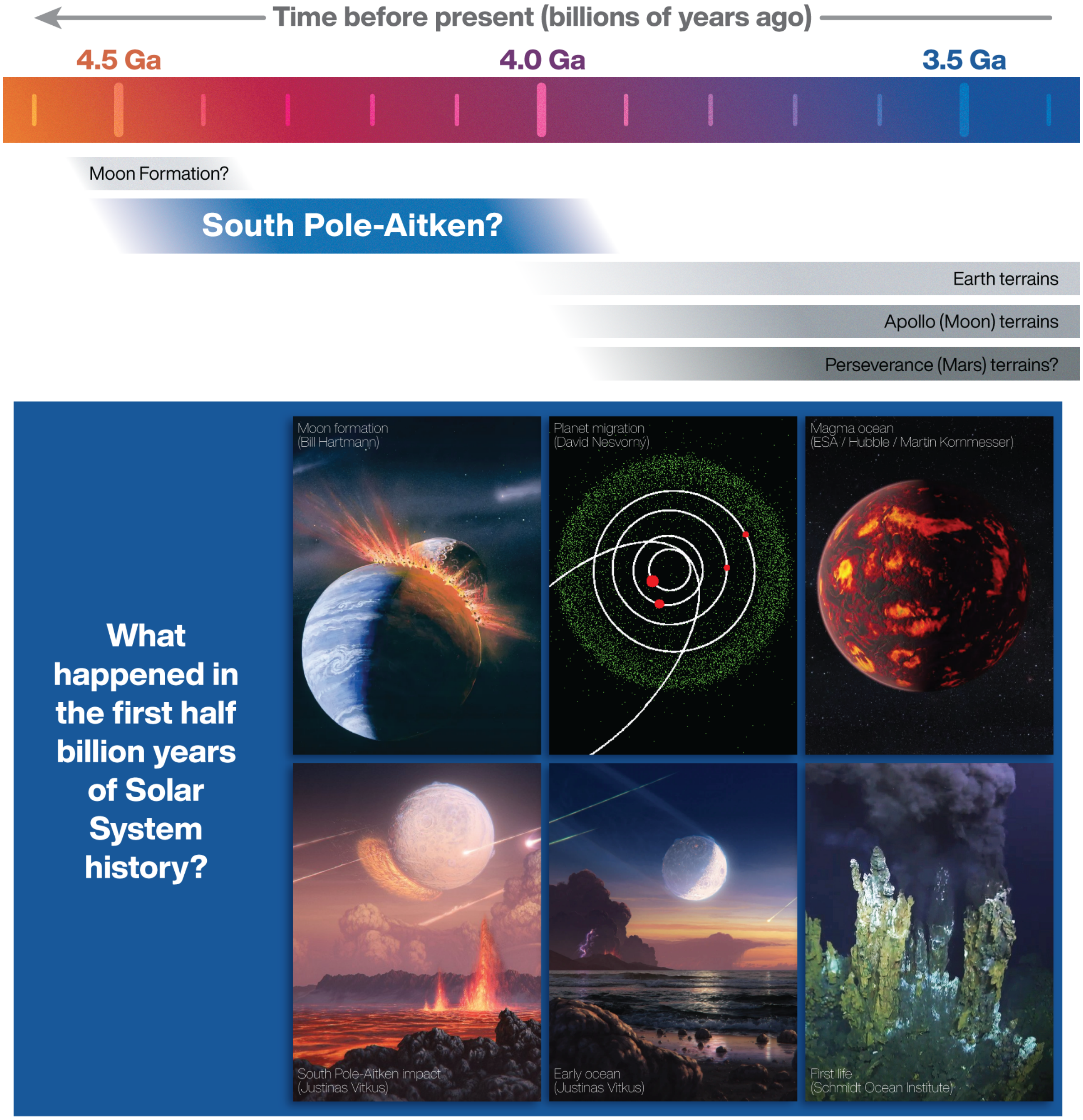

**Figure 3.1.** *Schematic illustration of the first half billion years of Solar System history, and the key Solar System processes that were active in this lost era of Solar System history.*

Both the antiquity and location of the South Pole–Aitken basin work in our favor for investigating processes in the first half billion years. The location on the lunar farside provides access to a region of the Moon well-outside the regions sampled by Apollo and Luna missions, and therefore is not subject to the same potential biases, including the presence of significant amounts of Imbrium basin ejecta, and geochemical signatures of the dregs of the lunar magma ocean, rich in incompatible elements such as Potassium (K), Rare Earth Elements (REE), and Phosphorus (P)—also known as KREEP. This nearside terrain is commonly referred to as the Procellarum KREEP Terrain, or PKT. Meanwhile, the hypothesized age of the basin places it squarely within the critical time period of investigation for Goal 1.

Another important process occurring on the heels of lunar formation and differentiation was the potential evolution of a lunar magnetic field and dynamo. Paleomagnetic signatures measured on Apollo samples have been interpreted to suggest these are the result of a core dynamo that lasted until as late at ~1.0 Ga (Mighani et al., 2020). Unfortunately, the chronology and cause of observed lunar paleomagnetic signatures is disputed. Fundamentally what was gathered from Apollo left us with just enough data to speculate, but a dearth of ancient samples (>3.9 Gyr) has hampered our ability to truly address the process responsible for magnetism. There are good reasons to suspect that the igneous-like processes associated with crystallization of an SPA melt sheet would provide conditions that may have recorded the strength of the early lunar magnetic field. Indeed, this is bolstered by the potentially more iron-rich (mafic) compositions of SPA materials compared to any of the non-mare samples in Apollo. Other ancient material we would expect to encounter during a SPARX traverse may also prove to be valuable targets. The low-resolution crustal magnetic field that can be inferred from orbit would also be greatly improved by traverse magnetometry across SPA, which might be connected to chronology through analysis of the age of returned samples.

Last but not least, the most important geologic force affecting the Moon's surface in the first twenty percent of its history after differentiation was impact cratering. Impacts occur at all scales, but missions from Lunar Reconnaissance Orbiter (LRO)to Gravity Recovery and Interior Laboratory (GRAIL) have given us insight into the existence of some truly gigantic basins on the Moon, including SPA. Presently it is thought that all of the large impact basins with diameters larger than 200 km date to the three earliest lunar epochs (Imbrium, Nectarian, and Pre-Nectarian), indicating the existence of a generally higher impact flux early in solar system history compared to later times.

However, the precise nature of the Moon's impact history has been hotly debated. Was there a short (50–250 Myr) cataclysmic period approximately 4 Gyr ago (a so-called "Late Heavy Bombardment" or "Terminal Lunar Cataclysm"), or did these large basins form more continually from the end of magma ocean crystallization until 3.7 Gya? This question is not just important for understanding the Moon itself. The impactors that struck the Moon have to come from somewhere, and dynamical models have established hypotheses connecting these impactors to reservoirs of small bodies that may have been destabilized by the migration of the giant outer planets (e.g., Tsiganis et al., 2005; Gomes et al., 2005; Morbidelli and Crida, 2007; Nesvorný, 2018). This spike in continent-scale impacts would have affected all bodies in the inner Solar System, with important consequences for Mars, Venus, Mercury, and especially Earth. On Earth, this proposed spike is just at the time that life may have been emerging, but plate tectonics has erased the terrestrial record, if not necessarily the consequences. For every impact on the Moon an order-of-magnitude more would have impacted the Earth, owing to its larger physical and gravitational cross-section. As we describe in further detail in Section 3.4, determining the rate of basin-forming impacts is our highest priority SPARX science goal because of the broad consequences outlined here. SPARX is ideally suited to such a mission as an extended traverse allows access to the required critical terrains, which are widely separated due to the large size of individual impact basins.

### 3.1.1 Objective 1.1

**Objective 1.1: "Quantify the Age of the Moon to Distinguish Between Early Moon and Late Moon Formation Models"**

**The age of formation and differentiation of the Moon is a fundamental anchor point in understanding processes in the early inner solar system and on early Earth** (e.g., Halliday and Canup, 2023). Determining the timing of Moon formation and differentiation has been challenging due to sampling limitations and overlapping ages (e.g., Barboni et al., 2017, 2024; Borg et al., 2019, 2011a, 2011b; Dauphas et al., 2025; Greer et al., 2023; Maurice et al., 202et al., 2009a, 2009b). Measurements of U/Pb, Lu/Hf, Sm/Nd (for example) on lunar materials isotopic systems have yielded a range of ages that span ~170 Ma, from 4.35 to 4.52 Ga. Indeed, there are four different scenarios that are possible given the timing of Moon formation and subsequent differentiation (Figure 3.2). The Moon may have formed early (4.52 Ga) followed by a relatively early series of differentiation events (Figure 3.2A), or followed by a relatively late series of differentiation events (Figure 3.2B; Schneider and Kleine, 2025; Dauphas et al., 2025). On the other hand, the Moon may have formed later (4.45 Ga) followed almost immediately by a series of early differentiation events (Figure 3.2C) or a later series of differentiation events (Figure 3.2D; Maurice et al., 2020; Borg et al., 2015). Any one of these scenarios has some supporting evidence, but evaluation of these scenarios is potentially biased by the predominance of nearside samples upon which they are based. This is a problem because samples sourced from the nearside could be affected by the (unknown) process that resulted in the lunar asymmetry, a young thermal event (delaying cooling below the closure temperature for chronometers) like tidal heating (Nimmo et al., 2024), an impact spike (Hartmann et al., 2000), or a single large impact such as SPA (Joy et al., 2025).

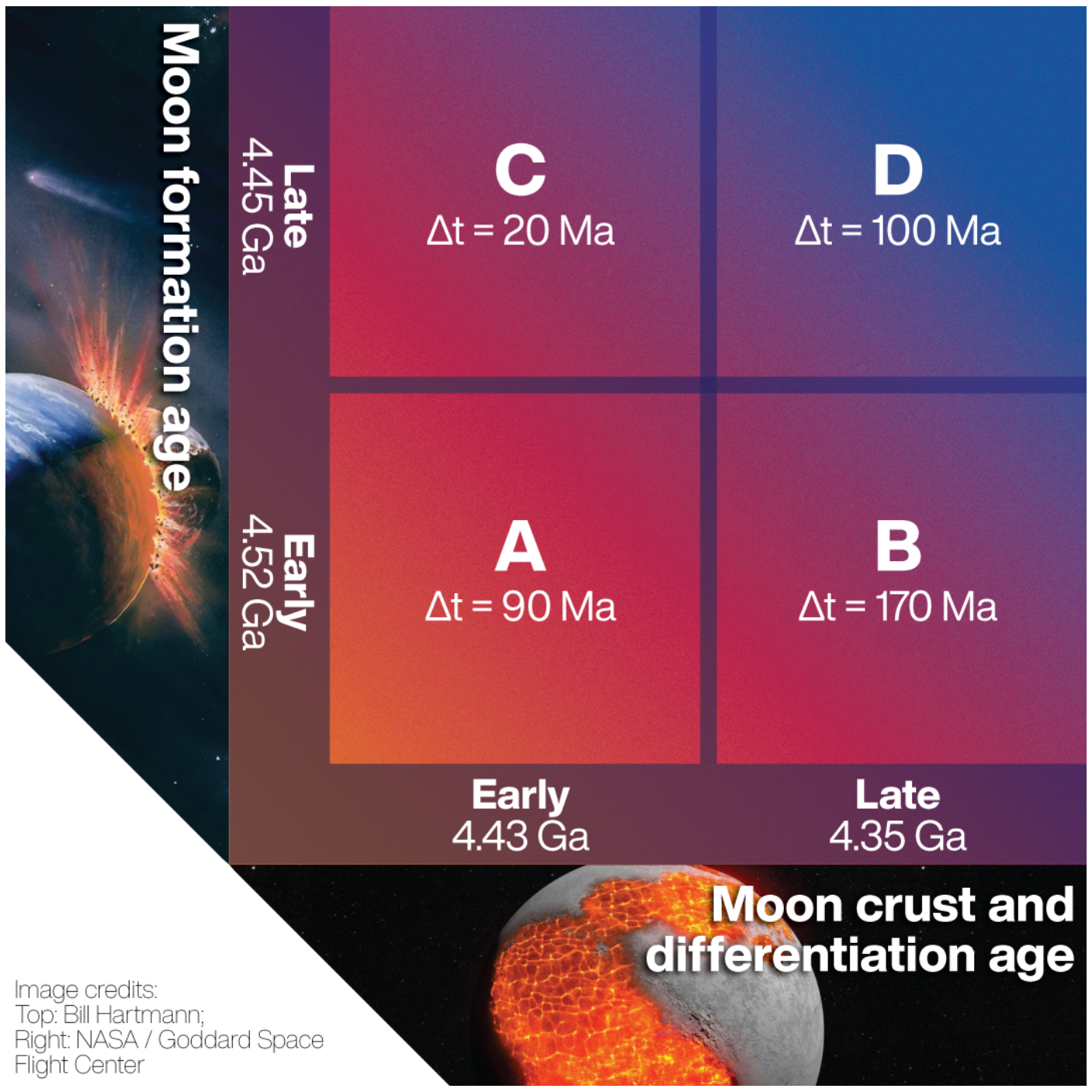

**Figure 3.2.** *The detailed chronology of the early Moon has inherent uncertainty in the possibility of early or late lunar formation age, and in the subsequent crust and differentiation age. The time difference between these events is here called Δt and several scenarios are illustrated: early moon formation followed by early crust formation would lead to a Δt of ~90 million years (A), whereas an early moon formation followed by a late early crust formation would lead to the largest Δt of 170 million years (B). Alternatively, if the moon formation event is late, when combined with an early crust formation event would lead to a Δt of only 20 million years (C), whereas a late moon formation coupled with a late crust formation would lead to Δt = 100 million years (D). Determining which scenario has been limited by sample lithologies and locations and the SPARX samples would provide new leverage on both lithology and location and allow better resolution of this early lunar chronology.*

Constraints on the formation age of the Moon have been provided using several different approaches. For example, an age for the Moon's crust has been determined using Rb-Sr model age for anorthosites (Schneider and Kleine, 2025). Evidence for the earliest crystallization age of lunar materials comes from zircons (e.g., Barboni et al., 2024); while Hf/W ages have been used to determine the timing of lunar core formation (e.g., Thiemens et al., 2019). Similarly, Rb/Sr,

Sm/Nd, and Lu/Hf studies of nearside mare basalts have placed constraints on the ages of mantle source regions from which the basalt units were derived. These ages suggest that the lunar mantle sources are of 4.34 to 4.56 Ga in age (Gaffney et al., 2023; Borg et al., 2019; Elardo et al., 2011; Zhou et al., 2025), but the role of younger thermal event such as impact or tidal heating remain unclear. These different approaches all underline the fundamental challenge in determining the age of the Moon-forming event, that is, there is a gap in time between the impact event we are trying to constrain, and when the earliest materials could start to form and be preserved. It is also devilishly hard to be sure what a particular age represents. For example, subsequent geologic processes on the Moon can modify and reset ages for materials, making the identification and sampling of the most ancient, pristine materials exceptionally challenging. While Apollo, Luna, Chang'e, and lunar meteorite samples have provided materials to have helped constrain the lunar formation age, significant uncertainty remains.

Collection of farside flotation crust, SPA impact melt, and farside mare basalts will provide the opportunity to test whether lunar formation and differentiation are relatively old (4.51 Ga) or young (4.35 Ga) events. Given the multiple possible approaches for determining the age of the Moon, and with the more specific objective of distinguishing between Early Moon and Late Moon formation models, SPARX has outlined a series of 3 Experiments, Ex. 1–3 (Sections 4.1.1-4.1.3) to address this objective. Determination of the composition and age of the anorthositic farside highlands crust, as well as model ages for the magma out of which they formed (Ex. 1, Section 4.1.1), will allow a direct comparison to similar measurements made on nearside rocks (including Pb/Pb in zircons and Rb/Sr, Sm/Nd, Lu/Hf, U/Pb in FANs). Determination of the age of the SPA impact event (Ex. 2, Section 4.1.2) provides the youngest age limit for these events, and will allow comparison to nearside basin ages, as well as preliminary farside impact melt ages (using a range of chronometers such as Pb/Pb, Ar/Ar, and Rb/Sr, Sm/Nd, Lu/Hf, U/Pb isochron systems). Determination of Rb/Sr, Sm/Nd, Lu/Hf, U/Pb model ages of mare basalt source regions from 3 different basalt provinces (Ex. 3, Section 4.1.3) will allow comparison to ages of nearside source regions. Altogether, these three approaches will place age constraints of a more global nature on the formation and differentiation of the Moon, and help to resolve whether these events took place early (4.51 Ga) or late (4.35 Ga).

### 3.1.2 Objective 1.2

**Objective 1.2: "Distinguish Between Fast- and Slow-Cooling Models for Lunar Magma Ocean Crystallization"**

Intimately tied to understanding the formation age of the Moon, and tied to the fundamental challenges described above, is the physical reality that none of these geologic processes occurred instantaneously. There are inherent timescales associated with differentiation, cooling, and crustal formation, the differences in which are model-dependent, and can range from 10s to 100s of millions of years. This is reflected in Figure 3.3. The uncertainties in how the chronology of the early Moon should be interpreted have led to the development of models that attempt to explain how the ages of ancient crustal lithologies and mantle reservoirs fit into the lunar magma ocean paradigm. In any general LMO model (Charlier et al., 2018; Elardo et al., 2011; Smith et al., 1970a, b; Snyder et al., 1992; Warren, 1985; Wood et al., 1970), mare basalt source regions form slightly before or contemporaneously with anorthosites that constitute the primary flotation crust, and the KREEP reservoir, and zircons derived from KREEP-rich lithologies, should then form afterwards. Whether these lithologies and reservoirs would exhibit resolvable differences in crystallization age and/or model age is, to a first order, a product of the cooling duration of the LMO. If the Moon formed and cooled relatively quickly, ≤30 million years for example, then it would be expected that anorthosites would record a narrow range in crystallization ages that would be essentially concordant with isotopic model ages for mare basalt source regions, the KREEP reservoir, and ancient detrital zircons. Conversely, if full crystallization of the LMO was prolonged and lasted on the order of 50–200 million years, there should be resolvable differences in the chronologies of these lithologies and reservoirs with an age progression in line with the predicted crystallization sequence of the LMO. However, thus far the chronology of the early Moon does not fit neatly into either fast-cooling or slow-cooling LMO models and may be complicated by secondary events.

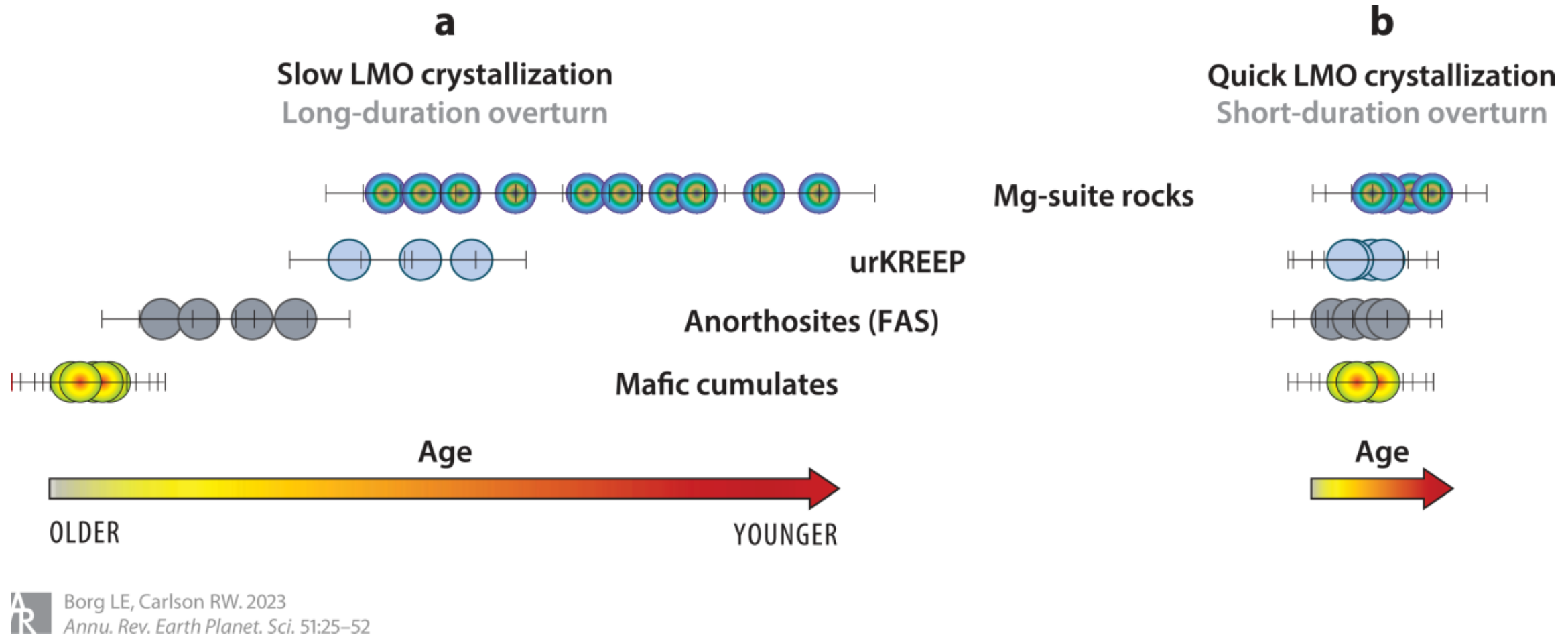

**Figure 3.3.** *Schematic demonstrating the contrasting ranges of ages expected for either slow- or fast-LMO crystallization. Figure 4 from Borg and Carlson, 2023.*

Models for prolonged, slow cooling of the LMO have been driven by several chronologic observations. Ages for ferroan anorthosites that were determined in the 1970s to early 2000s generally indicated a long duration for anorthosite crystallization from >4.50 Ga to as young as ~4.29 Ga (Alibert et al., 1994; Borg et al., 1999; Carlson and Lugmair, 1988; Norman et al., 2003; Nyquist et al., 2006), which has been interpreted as reflecting a long duration for their formation from LMO liquids. Alongside the oldest anorthosite ages, the lunar zircon record contains a small, but important number of analyses with ages from ~4.41–4.46 Ga (Grange et al., 2011, 2009; Greer et al., 2023; Meyer et al., 1996; Nemchin et al., 2009; Taylor et al., 2009), and some interpretations of lunar $^{182}$Hf-$^{182}$W chronometry suggest a Moon-forming giant impact at ≥4.5 Ga (Kleine et al., 2005; Lee et al., 2002, 1997; Touboul et al., 2007; Yin et al., 2002). Those interpretations of an early giant impact and the younger $^{146,147}$Sm-$^{142,143}$Nd, Lu-Hf, and Pb-Pb model ages for mare basalt source regions and the KREEP reservoir between about 4.38 – 4.35 Ga (Borg et al., 2019; Brandon et al., 2009; Carlson and Boyet, 2009; Gaffney and Borg, 2014; McLeod et al., 2014; Nyquist et al., 1995; Rankenburg et al., 2006; Snape et al., 2016) have driven models for prolonged magma ocean crystallization. Since all models agree that magma ocean crystallization will be extremely rapid before establishment of a coherent anorthosite crust, slow-cooling models have focused on the insulating effects of the low-thermal conductivity anorthosite crust, heating and potential crustal remelting due to tidal effects, and the early onset on mantle convection as mechanisms to prolong the final cooling (Elkins-Tanton et al., 2011; Maurice et al., 2020; Michaut and Neufeld, 2022; Zhang et al., 2025). These models suggest that, from a physical standpoint, crystallization of the LMO could have been protracted to the degree required to produce anorthosites over such a large age range.

Conversely, it has been argued that most ages for anorthosites and other lunar crustal lithologies are unreliable and that the apparent large range in ages for these lithologies is artificial. Borg et al., (2015) reviewed the literature data for age determinations of ancient crustal lithologies and applied five quality control criteria to determine the reliability of ages. They argued that most ages determined on anorthosites and Mg-suite samples to that point in time are of low reliability and that there is little evidence for anorthosite ages older than ~4.40 Ga. Additionally, more recent age determinations for anorthosites and Mg-suite samples that rely on the concordance of multiple isotopic systems, consistency with petrologic constraints, and modern standard normalization suggest crystallization ages between ~4.38–4.34 Ga (Borg et al., 2025, 2019, 2017, 2015, 2011; Borg and Carlson, 2023; Carlson et al., 2014; McLeod et al., 2016; Sio et al., 2020), suggesting a late and rapid formation of anorthosites from the LMO followed quickly by crustal intrusions represented by the Mg-suite. Similarly, $^{146,147}$Sm-$^{142,143}$Nd, Lu-Hf, and Pb-Pb model ages for mare basalt source regions and the KREEP reservoir point to the timing of last equilibration with a chondritic reservoir, typically assumed to be the lunar magma ocean, at around 4.38 – 4.35 Ga (Borg et al., 2019; Brandon et al., 2009; Carlson and Boyet, 2009; Gaffney and Borg, 2014; McLeod et al., 2014; Nyquist et al., 1995; Rankenburg et al., 2006; Snape et al., 2016). The overlapping ages of recently dated crustal rocks and multi-system model ages for mantle reservoirs also overlaps with the prominent ~4.34 Ga peak in lunar detrital zircon ages (Barboni et al., 2024; Borg et al., 2015; Grange et al., 2009, 2011, 2013; Meyer et al., 1996; Nemchin et al., 2008; Nemchin et al., 2009a, 2009b; Nemchin et al., 2012; Pidgeon et al., 2007). Taken together, these lines of evidence suggest both the late formation of the Moon and the rapid solidification of the LMO between roughly 4.38 and 4.34 Ga.

Within the confines of current datasets, there does not appear to be a clear resolution to these discrepancies in chronologic interpretations regarding the duration of the LMO, making it a major unanswered and fundamental question regarding the evolution of the early Moon. As is the case with other pressing questions, the bias of the current lunar sample collection to the nearside and the Procellarum KREEP Terrane is extremely limiting to the development of more complete models. The lack of samples and our resulting inability to determine ages for farside rocks and reservoirs is the most significant hindrance to developing models for the chronology of events on the early Moon. SPARX would address this gap in the sample collection, by strategically sampling locations from within SPA. SPARX Experiments 1-3 are designed to provide critical inputs to address the objective of distinguishing between fast- and slow-cooling models for LMO formation. Comparison of the ages of farside anorthosite suites which are interpreted to represent magma ocean flotation cumulates (Ex. 1, Section 4.1.1) would reveal whether there was a delay in the development of the crust on the nearside or whether global primary crust production was simultaneous. The development of model ages for the farside mantle using mare basalts (Ex. 3, Section 4.1.3) and, if present, a farside KREEP reservoir, would reveal the timing of farside mantle reservoir development. Furthermore, dating of the SPA impact (Ex. 2, Section 4.1.2) would allow us to assess whether this event contributed to either any differences in the ages of nearside vs. farside rocks and reservoirs or to their chemical and isotopic re-equilibration.

### 3.1.3 Objective 1.3

**Objective 1.3: "Determine if and when a Giant Planet Instability Drove a Sharp Change in the Inner Solar System Impactor Flux"**

The behavior of the lunar cratering rate during the Moon's earliest history remains an important and unresolved question. Apollo sample analyses show a strong clustering of impact-related ages around ~3.9 Ga (e.g., Tera et al., 1974; Cohen et al., 2000; Cohen et al., 2023), supporting the idea of a high impactor flux at that time. However, both returned samples and dynamical modeling continue to fuel debate about whether the Moon experienced (1) a giant-planet-instability–driven spike in bombardment, (2) a long-term, steady decline in bombardment, or (3) a hybrid of the two (e.g., Hartmann, 1975; Neukum & Ivanov, 1994; Morbidelli et al., 2012; Bottke et al., 2012). These competing hypotheses imply fundamentally different dynamical pathways operating in the early Solar System.

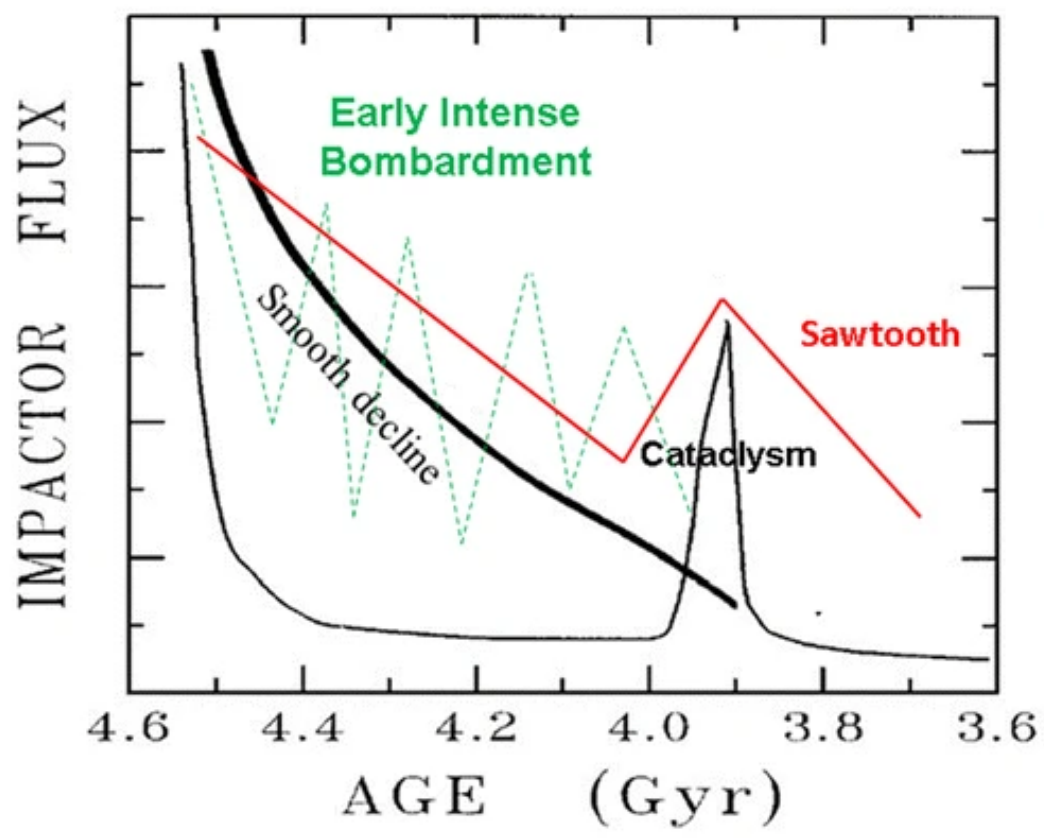

**Figure 3.4.** *Summary of lunar impact scenarios, as presented in Zellner et al., (2017). Graph is by J. W. Delano and modified from Zellner (2001). General flux curves are interpretations of evidence presented in Baldwin (1949, 1964), Hartmann (1965, 1966, 2000), Tera et al. (1974), Neukum et al. (1975, 2001), Neukum and Ivanov (1994), and Morbidelli et al. (2012).*

A steadily declining bombardment could arise from the gradual depletion of leftover planetesimals following terrestrial planet formation. In this scenario, remnant bodies from accretion were slowly delivered into the inner Solar System through processes such as early giant planet migration or the progressive clearing of the primordial asteroid belt (e.g., Bottke et al., 2007; Raymond et al., 2013) ("smooth decline" in Figure 3.4). Such dynamics would produce a monotonic decrease in the impactor population over hundreds of millions of years, yielding a cratering chronology dominated by slow, secular decay rather than abrupt changes.

In contrast, an instability-driven bombardment, triggered by giant planet migration after the dissipation of the solar nebula, would reflect a far more abrupt and violent reorganization of the early Solar System. In the "Nice model" framework, resonance crossing among the giant planets leads to rapid orbital rearrangement, exciting and redistributing large reservoirs of small bodies from the primordial Kuiper belt, trans-Neptunian region, and possibly the outer asteroid belt (e.g., Tsiganis et al., 2005; Gomes et al., 2005; Morbidelli et al., 2007; Nesvorný, 2018). Such an instability could produce a sharp, high-intensity spike in the lunar impactor flux on short timescales while simultaneously sculpting the modern architecture of the Kuiper belt and asteroid belt ("cataclysm" in Figure 3.4). Determining whether the lunar record preserves evidence of such a dynamically violent episode is therefore central to evaluating migration models and reconstructing the early evolution of the inner Solar System.

In addition to altering impact rates, a late giant planet instability would also reshape the architecture of the Solar System. Models predict that this process could truncate and dynamically excite the Kuiper belt, sculpt the present-day distribution of resonant trans-Neptunian objects, and potentially deplete and reorder the main asteroid belt. These wide-ranging consequences make the timing, and even the occurrence, of this instability a central constraint for understanding early Solar System evolution. **Determining whether the lunar cratering record preserves evidence of such a dynamically violent episode is therefore crucial for evaluating models of planetary migration and for reconstructing the conditions under which the terrestrial planets, including Earth, matured.**

Because the South Pole–Aitken (SPA) basin preserves and excavates some of the oldest materials on the Moon, it provides an unparalleled opportunity to probe the early impactor flux and test for any sharp transitions associated with a giant planet instability. SPA's ancient melt sheet, deep-excavated lower-crustal and possibly upper-mantle materials, and limited later resurfacing make it one of the clearest windows into pre-nearside lunar history (e.g., Garrick-Bethell & Zuber, 2009; Vaughan & Head, 2014; Moriarty & Pieters, 2018). These characteristics position SPA as a uniquely powerful site for constraining the absolute lunar impact chronology and determining if and when, a giant planet instability drove a significant change in the impactor flux recorded on the Moon. SPARX has outlined three experiments, Ex. 2, 4, 5, to address this objective. First, determining the age of the SPA impact (Ex. 2, Section 4.1.2) is required to establish a chronological anchor point for the timing of the earliest stratigraphically distinctive event in lunar history. Next, SPARX would acquire sam ples from farside basins that are pre-Imbrian (≥3.9 Ga) and post Imbrian (3–3.9 Ga) (Ex. 4 and 5, Section 4.1.4-4.1.5). This would allow us to establish the form of the basin-forming impact flux, and evaluate if (1) the perceived peak in the impact flux hypothesized based on the lunar nearside samples at 3.9 Ga is observed with samples from the farside, and (2) if the distribution of farside basin ages supports a cataclysmic or smoothly declining impactor flux, or one of the other proposed flux models.

### 3.1.4 Objective 1.4

**Objective 1.4: "Constrain the Evolution of the Lunar Dynamo in the Pre-4.0 Ga Epoch of Lunar History"**

Understanding if, when, and how the Moon generated a magnetic field via a planetary dynamo is important to reconstructing its early thermal and interior evolution. Although the Moon lacks a global magnetic field today, orbital observations reveal strong, localized crustal magnetic anomalies that record ancient magnetic fields and preserve a unique archive of the Moon's earliest history. Many of the most prominent and enigmatic of these anomalies occur in and around SPA (Figure 3.5), a particularly intriguing association given that size and antiquity of the basin. Because the SPA impactor excavated material from depth early in lunar history, exploration of the basin offers a singular opportunity to address major open questions about the origin, intensity, and evolution of the lunar dynamo during the Moon's first billion years—a period that remains poorly sampled by existing lunar materials.

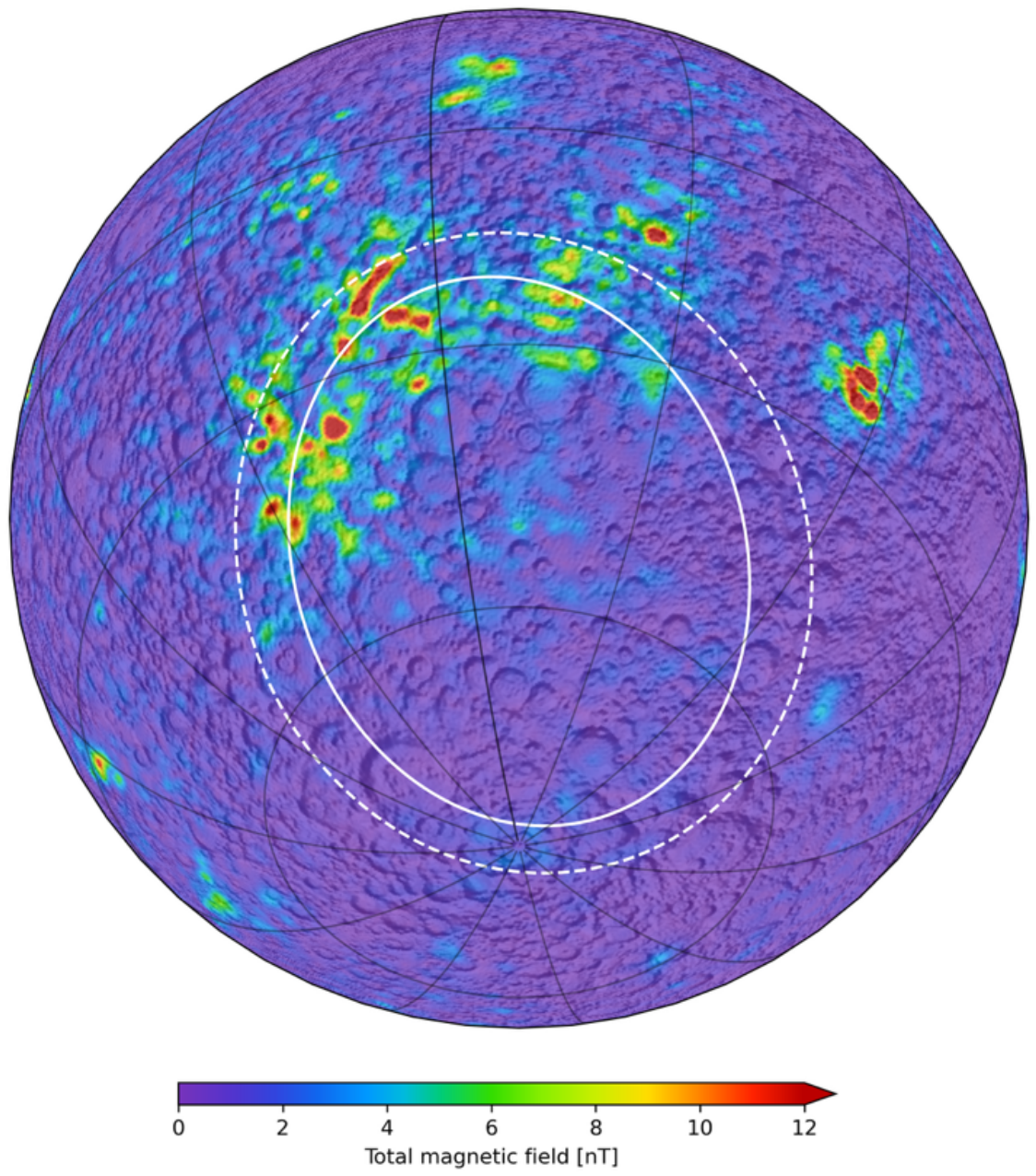

**Figure 3.5.** *Surface magnetic anomalies in and around the* South Pole–*Aitken (SPA) basin. Magnetic field magnitude at 30 km altitude from Kaguya and Lunar Prospector (Tsunakawa et al., 2015) overlaid on Lunar Orbiter Laser Altimeter topography (LOLA; Smith et al., 2010) and shown in an orthographic projection centered at 190°E, 50°S. White lines denote the inner (solid) and outer (dotted) rims of SPA (James et al., 2019).*

A planetary dynamo is generated by the motion of electrically conducting fluid — typically liquid iron — within a planetary core, producing a global magnetic field. Paleomagnetic measurements indicate that the Moon likely hosted such a dynamo and generated a magnetic field peaking at ~100 µT between approximately 4.5 and 3.56 billion years ago (Tikoo et al., 2014, 2017; Mighani et al., 2020; Strauss et al., 2021). These field intensities are comparable to, or even exceed, those of present-day Earth (~20–65 µT; Tauxe, 2010), a surprising result given that the Moon is only a quarter of Earth's diameter and therefore possessed a substantially smaller heat budget. These intensities declined by roughly an order of magnitude by ~3.0 Ga and appear to have shut down entirely by ~1.0 Ga (Mighani et al., 2020). This magnetic history broadly mirrors the evolution of mare volcanism (Figure 3.6), which peaked early and waned as the lunar interior cooled (e.g., Guo et al., 2024), suggesting that both processes may have been governed by common thermal controls tied to the evolution of the Moon's interior and core. In addition, high paleointensities have been linked to high-titanium mare volcanism (Nichols et al., 2026), further pointing to a connection with deep interior processes. However, the physical mechanism that powered the lunar dynamo—whether thermal convection (e.g., Stegman et al., 2003), basal magma ocean convection (e.g., Scheinberg et al., 2018), core crystallization (e.g., Laneuville et al., 2014), impact-driven processes (e.g., Le Bars et al., 2011), downwelling titanium-rich mantle cumulates (e.g., Evans and Tikoo, 2022), tidally-induced mantle precession (e.g., Dwyer et al., 2011), or some combination thereof—remains unresolved. Determining the timing, strength, and longevity of the lunar dynamo, and assessing whether its apparent correlation with volcanism reflects a causal relationship, are central unresolved problems in lunar science.

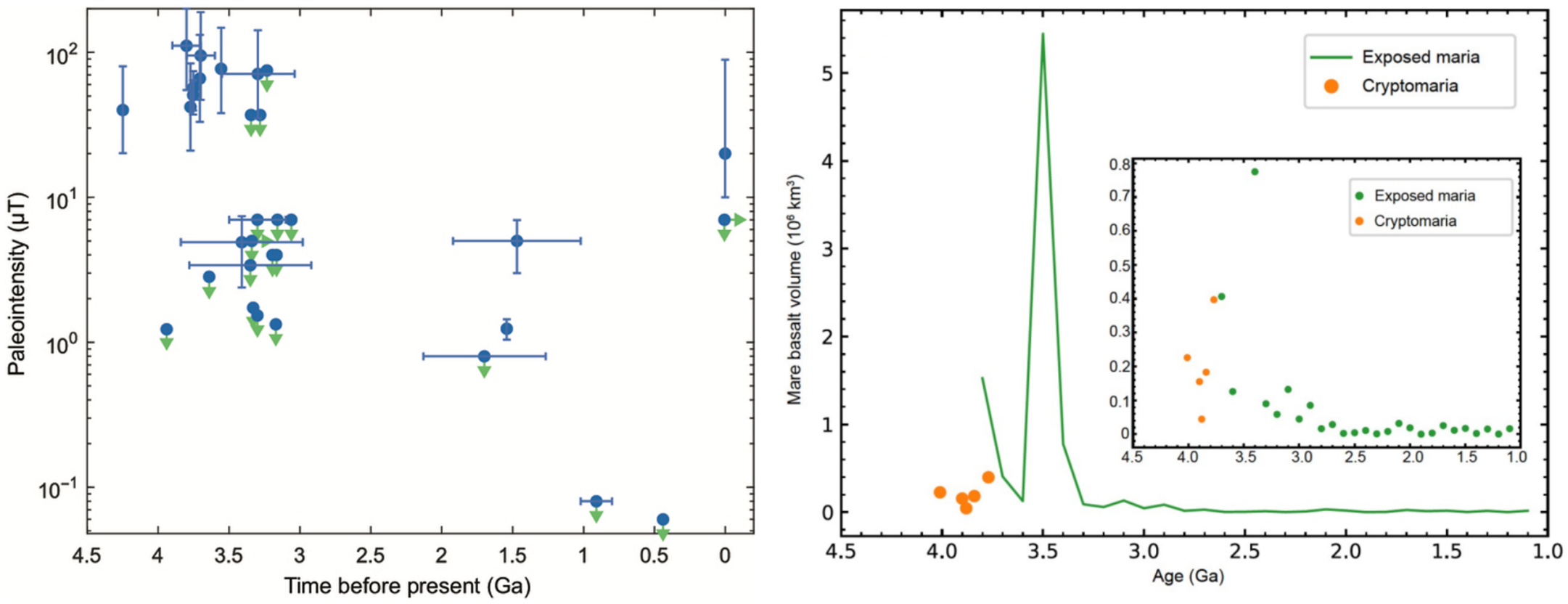

**Figure 3.6.** *Summary of lunar magnetic field paleointensity measurements from Luo et al., (2024) (left) compared with the temporal evolution of mare volcanism from Guo et al., (2024) (right). Blue symbols represent age and paleointensity data from Apollo samples, with horizontal and vertical error bars indicating uncertainties. Green arrows denote upper limits on paleointensity (downward) and age (rightward).*

Progress on these questions is limited by gaps in the existing paleomagnetic record. Nearly all robust lunar paleointensity constraints derive from Apollo samples, which preferentially sample the nearside of the Moon and provide limited coverage of ages older than ~3.9 Ga, with additional gaps between ~3.2 and 1.5 Ga. Apollo troctolite 76535 is potentially the only available Apollo sample that predates 3.9 Ga, with a well-constrained age of ~4.25 Ga. Based on its low peak shock pressure (<6 GPa), inferred depth of origin (45–65 km), and ancient age, this sample has been hypothesized to have been excavated from the South Pole–Aitken basin and may represent the only Apollo paleointensity constraint on the lunar magnetic field prior to ~4.0 Ga (Garrick-Bethell et al., 2020). While Apollo troctolite 76535 provides compelling evidence for a strong early lunar dynamo, its provenance and geological context remain debated (e.g., Bjonnes et al., 2025). As a result, the next phase of lunar magnetic exploration must focus on obtaining more accurate paleointensity measurements tied to well-constrained geological context, particularly from the Moon's oldest terrains in SPA, to resolve the early history of the lunar dynamo.

Addressing this objective requires a different strategy from many of the other SPARX objectives. The critical emphasis of the objective is on acquiring paleomagnetic intensity measurements from samples of a specific age (greater than 4.0 Ga). There is not yet a reliable way to combine precise, sample-specific paleomagnetic measurements with high resolution ages *in situ* on the surface of the Moon, so the samples for this objective instead need to be drawn from a subset of the overall returned sample suite required for other objectives, which meet the age criteria and are suitable for paleomagnetic investigation (Section 4.1.6, Section 4.3.5). The need to return samples for this investigation is also driven by the fact that interpreting crustal magnetization requires accounting for the recording properties of lunar materials. Magnetization depends on both the ambient magnetic field and the thermoremanent susceptibility, which is controlled by magnetic mineralogy, grain size, domain state, and mineral abundance. While lunar materials have sometimes been regarded as poor magnetic recorders, this view likely reflects sampling bias toward coarse-grained lithologies. Recent work suggests that fine-grained phases—particularly in impactites enriched in iron-bearing minerals —may preserve high-fidelity records of ancient magnetic fields (Jung et al., 2025). Such materials are expected to be abundant in SPA and are well suited for modern paleomagnetic techniques. It is important to note that the recommended investigations to address Objective 1.4 as part of SPARX do not include trying to derive information on the vector direction of any detected paleomagnetic fields. Recovering this information requires precise knowledge of the initial sample orientation at the time of magnetization, as well as, preservation of that orientation information throughout the sample return processes. Unlike Mars, the lunar SPA environment provides limited accessible bedrock, and the subsequent requirements that would be levied on drilling, coring, and sample storage systems were deemed to be not commensurate with the probability of scientific success.

Because paleo-field direction cannot be recovered from SPARX acquired samples, additional contextual measurements about the local magnetic environment are desirable. Correctly interpreting the sample measurements requires distinguishing among the primary categories of lunar crustal magnetic anomalies, including those associated with basin melt sheets (e.g., Halekas et al., 2010; Hood, 2011), antipodal regions of major impacts (e.g., Hood and Huang, 1991; Wakita et al., 2021), and more spatially distributed anomalies linked to impactor material, lunar swirls, or intrusive lithologies (e.g., Wieczorek et al., 2012; Wakita et al., 2021; Blewett et al., 2011; Purucker et al., 2012). Each category likely records magnetic fields through distinct physical mechanisms and over different timescales, with fundamentally different implications for the origin and intensity of the magnetizing field. Without separating these source mechanisms, inferred paleofield strengths may conflate signals produced by a long-lived internal dynamo (e.g., Laneuville et al., 2014; Hamid et al., 2023) with those generated or influenced by transient, impact-related fields (e.g., Le Bars et al., 2011; Narrett et al., 2025). A magnetometer survey conducted within SPA directly addresses this challenge by resolving the spatial structure and source depths of crustal remanent magnetization, providing the context needed to link surface magnetization to paleofield intensity, dynamo evolution, and mega-basin formation processes (Section 4.1.11, Section 4.2.7).

By combining *in situ* magnetic field measurements with the return of well-contextualized samples spanning a wide range of ages—including impactites older than 4.0 Ga—an SPA sample return mission provides a unique opportunity to extend the lunar paleointensity record into the Moon's earliest history. These measurements will enable quantitative tests of competing dynamo models and establish how magnetic field generation, interior cooling, and volcanic activity co-evolved. In doing so, SPA exploration represents a critical step toward resolving the origin and evolution of the lunar dynamo and advancing our broader understanding of magnetic field generation in terrestrial planetary bodies.

## 3.2 GOAL 2

### Goal 2: "Does the Lunar Farside Preserve Evidence of Asymmetric Magma Ocean Evolution and Planetary Crustal Formation Processes?"

One of the most remarkable features of the Moon is the asymmetry between its nearside and farside hemispheres. Most visible as the dichotomy between the mare-filled basins of the nearside and the cratered anorthositic highlands of the farside, what is less visible, but even more intriguing, is the persistence of these dichotomies in other properties. These additional hemispheric dichotomies can be seen in geophysical and geochemical datasets (Figure 3.7). Geophysically, the lunar crust transitions from a thinner crust with lower topography on the nearside to a thicker crust with higher topography on the farside (Andrews-Hanna et al. 2023; Smith et al. 2010; Wieczorek et al. 2013). Geochemically, the nearside is dominated by lunar mare (visibly dark-colored remnants of ancient lava flows) concentrated primarily within a region enriched in incompatible elements (KREEP) known as the Procellarum KREEP Terrane (PKT) (Jolliff et al. 2000; Lucey et al. 1994; 1998), whereas the highly-cratered farside is mostly devoid of mare basalts and high concentrations of incompatible elements, indicating hemispheric differences in volcanism and thermal evolution (Elardo et al. 2020; Kamata et al. 2013; Miljkovic et al. 2013). **The nearside-farside asymmetry is one of the Moon's oldest and most fundamental features, and yet, its origin remains unresolved. Understanding its origin has broad implications for the Moon's formation, early history, and subsequent evolution, and also for how we understand planetary formation more broadly.**

Models proposed to explain the origin of the lunar nearside-farside asymmetry generally fall into two categories: (1) either the Moon's crust formed asymmetrically and naturally produced the observed hemispheric nearside-farside differences as the lunar magma ocean (LMO) crystallized, or (2) the Moon's crust/mantle initially formed symmetrically and underwent a subsequent large-scale modification process that redistributed crust and mantle components on a global scale (Figure 3.8).

The lunar crust and mantle are generally agreed to have formed from progressive solidification of an initially global magma ocean; the primordial anorthosite floatation crust would have formed first from plagioclase crystallization, followed by distinct crustal and mantle layers as the LMO continued to cool and crystallize (Smith et al. 1970; Wood et al. 1972), with ilmenite and KREEP-rich materials being the last solids to crystallize from residual melts strongly enriched in incompatible elements. In asymmetric crustal formation hypotheses, the process of LMO solidification proceeded differently between the lunar nearside and farside due to hemispherical differences in LMO dynamics and/or crystallization (Roy et al. 2014; Loper et al. 2002) or tidal heating (Quillen et al. 2019), producing a primordial lunar crust that was initially thicker on the farside and thinner on the nearside. As the LMO continued to solidify beneath an asymmetric crust, late-stage LMO cumulates would have naturally formed and concentrated under the thinner nearside crust (Andrews-Hanna et al. 2025). Asymmetric crustal formation is supported by remote orbital geochemical observations showing hemispheric variation in primordial crustal composition (Arai and Maruyama 2017; Ohtake et al. 2012) and the apparent lack of incompatible element-enriched materials on the farside.

Alternatively, the LMO could have evolved symmetrically, producing an initially uniform primordial lunar crust/mantle that was subsequently altered by a global process such as overturn of a gravitationally unstable late-stage magma ocean cumulate (ilmenite) layer (Hess and Parmentier, 1995; Zhong et al. 2000; Parmentier et al. 2002; Zhang et al. 2017; Maurice et al. 2024) or a nearside mega-impact (Zhu et al. 2019). After-effects of the SPA impact could also contribute to creating asymmetries in an initially symmetric Moon through farside crustal thickening from SPA ejecta (Zuber et al. 1994), antipodal focusing of SPA impact shocks (Schultz et al. 2011; Head et al. 2024), or redistribution of LMO cumulates induced by the SPA impact (Jones et al. 2022, Zhang et al. 2022).

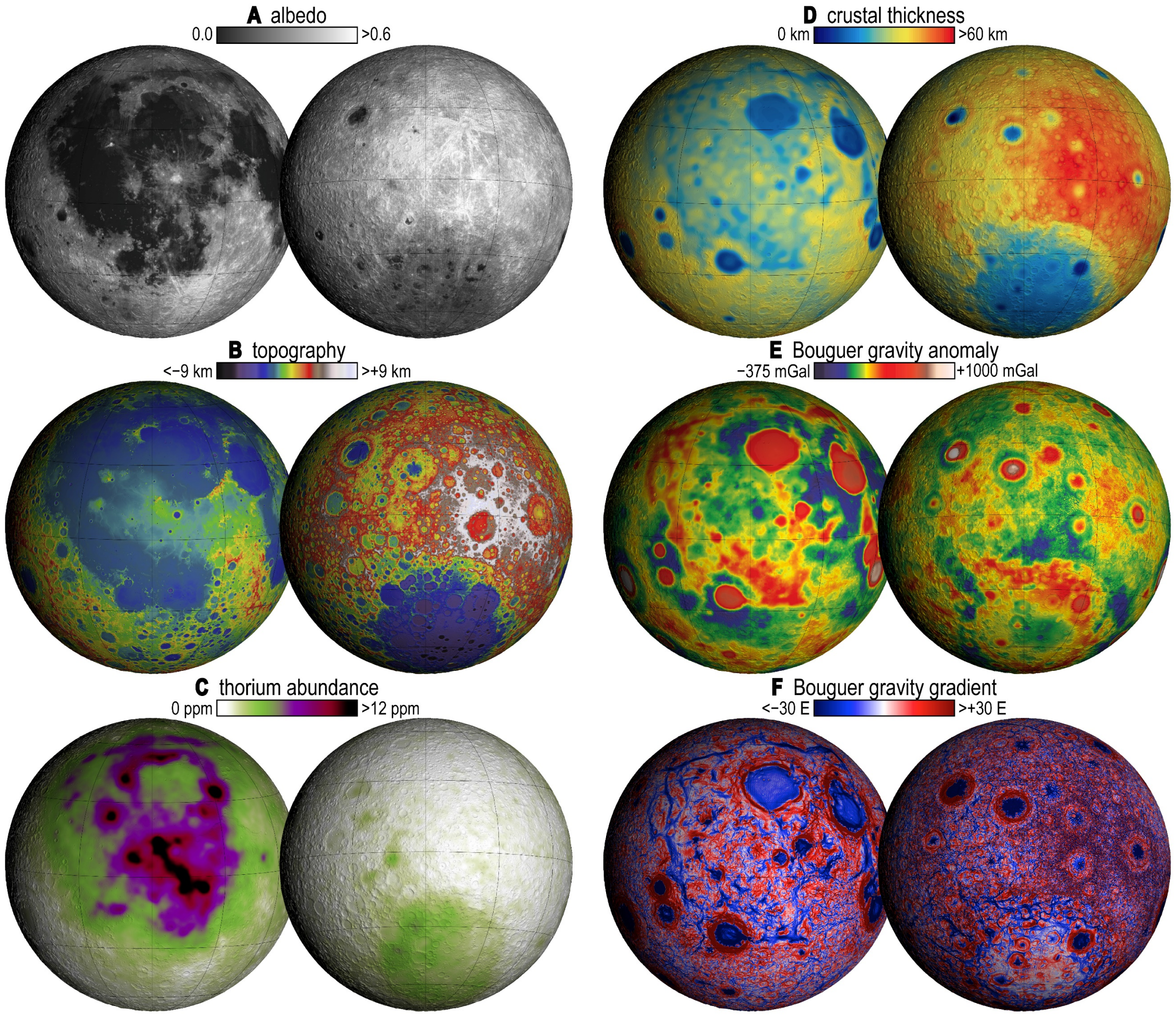

**Figure 3.7.** *Lunar maps showing the nearside/farside asymmetry in (a) visible imagery (b) topography (Smith et al., 2010), (c) thorium (Lawrence et al. 2003) and (d) crustal thickness (Wieczorek et al., 2013) (e) bouguer gravity anomaly, and (f) bouguer gravity gradient (Andrews-Hanna et al., 2013).*

**Sample return from the SPA basin provides a unique opportunity to test hypotheses for forming the lunar asymmetry because the material excavated by the SPA impact allows us access to several distinct layers in the farside lunar stratigraphy** (Figure 3.8). First, SPARX can collect samples from the Moon's primary crust, illuminating differences in original crustal formation and providing constraints for the original undifferentiated mantle composition. Second, the SPA impact would have excavated lower crust and upper mantle material from depths of at least 100 km (Melosh et al. 2017; Citron et al. 2024), providing access to the farside lower crust, farside upper mantle, magnesian-suite intrusions, and any potential farside shallow mantle cumulates, such as KREEP. Samples of this material could reveal if the pre-SPA impact state of the farside contained a layer of late-stage LMO cumulates (e.g., KREEP and ilmenite-rich lithologies) similar to the nearside, which is critical to determining if the LMO crystallized symmetrically or asymmetrically. Finally, deeper regions of the mantle stratigraphy can be sampled through the collection of farside volcanic products, revealing whether they were

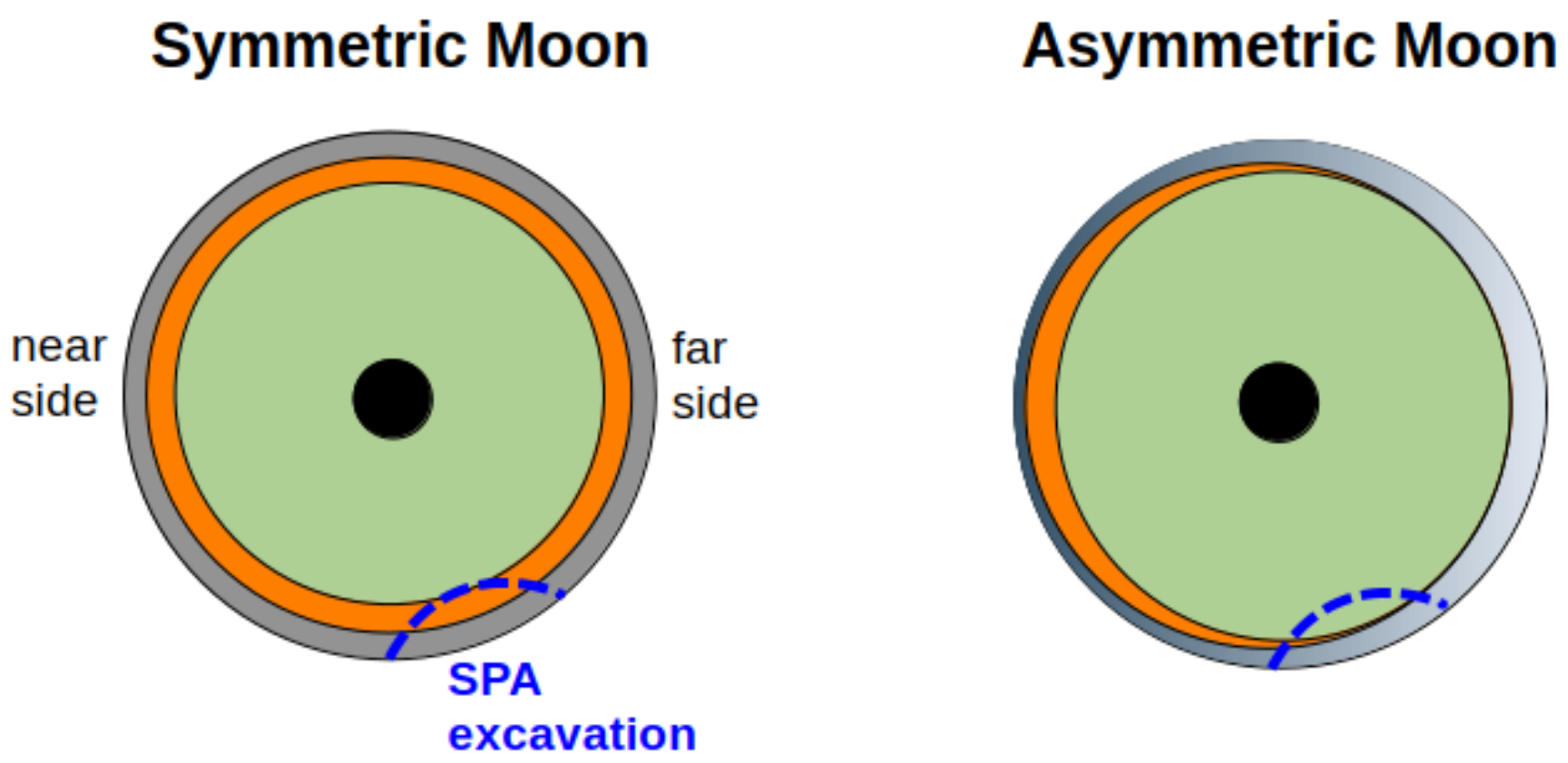

**Figure 3.8.** *Schematic of a Moon that initially formed symmetrically or asymmetrically. In the symmetric case, the LMO crystallizes uniformly producing a global layer of late-stage LMO cumulates (orange) and a uniform crust (gray). If the SPA impact occurred prior to redistribution of the cumulate layer, it would excavate a region near the lunar South Pole (delineated in blue) that would contain significant amounts of LMO cumulates. If the Moon formed asymmetrically, LMO cumulates would preferentially form on the nearside, leaving only a thin, spatially varying layer of LMO cumulates in the SPA excavation zone. Samples of material excavated by SPA could thus differentiate between these two models of the Moon's initial state.*

sourced from a mantle similar in composition to the lunar nearside. Returned samples from SPA could therefore provide a giant leap towards our understanding of planetary differentiation processes and help resolve one of the Moon's most fundamental mysteries in how the nearside-farside asymmetry emerged. In this way, Goal 2 directly addresses OWL question 3.5 ("How and when did the terrestrial planets and moon differentiate"), Question 5.2 ("How have the interiors of solid bodies evolved?"), and specifically Q5.2c ("What are the chemical and physical consequences of cooling and solidification on solid body crusts?").

### 3.2.1 Objective 2.1

**Objective 2.1: "Determine if the Farside Highlands Crust Is Similar to Nearside Crust in Age and/or Composition"**

The prevailing hypothesis for the formation of the lunar crust, the Lunar Magma Ocean (LMO) hypothesis, was proposed after the return of Apollo 11 samples (Wood et al., 1970). The hypothesis posited that the returned anorthositic samples represented the primary lunar crust, and that the crust got more mafic with depth (Ryder and Wood, 1977). However, since its development, this idea of a global lunar crust with a consistent composition and structure has been challenged. Lunar meteorites and granulites in returned samples with more Mg-rich compositions than Apollo ferroan anorthosites challenge the idea that the crust has a globally consistent composition, leading to the idea that the Apollo feldspathic samples are not representative of the global crustal composition (Korotev et al., 2003).

A series of orbital missions (e.g., Kaguya, Changrayaan-1) provided new data about the crustal mineralogy that also challenged the idea of a globally consistent crustal composition. Spectroscopic data indicate there is a nearside-farside dichotomy in crustal composition, with the nearside hosting more iron-rich crustal materials, whereas the farside crust is dominated by more magnesium-rich materials (Arai et al., 2008; Ohtake et al., 2012). Several scenarios have been proposed to explain this difference in crustal mineralogy. One hypothesis suggests that this difference in mineralogy is caused by the asymmetric crystallization of the magma ocean, wherein the more magnesium-rich materials crystallized first and concentrated on the farside, followed by the ferroan crustal rocks that forming later on the nearside (Figure 3.9) (Arai et al., 2008; Ohtake et al., 2012). A second hypothesis explains that the difference in composition is due to remelting of nearside mafic cumulates, generating the ferroan-rich nearside crust. Yet another explanation for the crustal asymmetry is hypothesized to result from the emplacement of basin ejecta, wherein impact events may have excavated more-magnesian lower crustal materials and distributed them over the more ferroan crustal materials at the farside surface (Arai et al., 2008; Korotev et al., 2003). Although mixing between nearside mare and highlands crustal materials has also been suggested to explain the observed mineralogical differences, the observed spatial distributions have led authors to suggest that this is not a favored hypothesis (e.g., Ohtake et al., 2012). Still, other analyses of orbital data have resulted in different conclusions regarding the lateral and vertical structure of the lunar crust. Analyses of crater central peak compositions have suggested that the variations in the lateral composition of the lunar crust can be explained by post-magma ocean processes (e.g., Lemelin et al., 2015), with the initial crustal formation falling into the original symmetric crustal composition scenario.

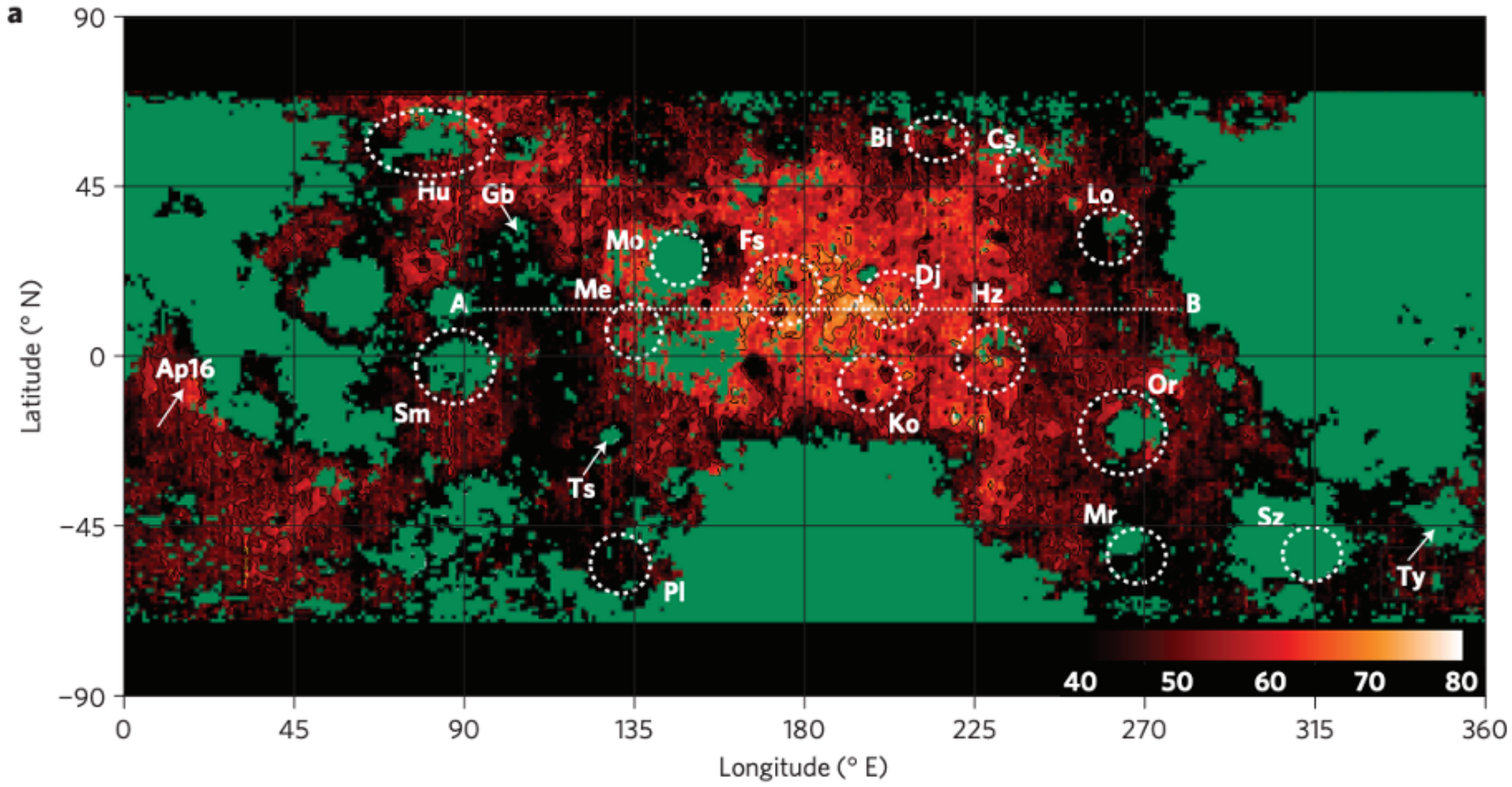

**Figure 3.9.** *Distribution of Mg# (magnesian number) across the lunar highlands. Figure is centered over farside highlands and shows higher Mg# on the farside and low Mg# (more ferroan) on the nearside. Figure 2a from Ohtake et al. (2012).*

Critically, our understanding of lunar evolution and our ability to test all of these different asymmetry formation hypotheses have been hampered by the vast bias of the lunar sample collection to the nearside. Only recently has Chang'e 6 returned a sample from the lunar farside, from within the Apollo impact basin located within SPA. However, Chang'e 6 landed on the mare basalts on the floor of Apollo and, therefore, did not directly sample primary crustal materials. Determining both age and composition of lunar farside crustal materials is critical to refining lunar formation models and assessing whether there is evidence for asymmetric LMO and crustal formation processes. By sampling farside crustal materials from several locations with orbitally-identified anorthosite signatures, SPARX will provide material that can directly assess the lateral continuity of the magma ocean processes recorded in the lunar crustal and mantle stratigraphy, including symmetric vs. asymmetric models for crustal formation. In particular, SPARX will collect samples of anorthositic highlands material from a variety of geologic settings across SPA. Detailed petrologic examination of the composition of mineral components of these samples will reveal whether this portion of the primary crust was formed before a mantle overturn took place (Fe-rich mafic minerals) or after an overturn took place (Mg-rich mafic minerals). The time-sequence of primary crust will be derived by determining ages for each of the samples across a large geographical area and for highlands crustal material with diverse compositions.

The resulting evolution in our understanding of planetary differentiation for the Moon will provide important insight into planetary differentiation in general, as the Moon's LMO model currently serves as the intellectual foundation for models of the differentiation of all terrestrial planets and large asteroids. This objective directly addresses OWL Decadal questions 3.5 ("How and when did the terrestrial planets and Moon differentiate?") and 5.2 ("How have the interiors of solid bodies evolved?").

### 3.2.2 Objective 2.2

**Objective 2.2: "Determine if There Was a Farside KREEP Reservoir, or Other Late-Stage Magma Ocean Products, Prior to the SPA Impact"**

Under the LMO hypothesis, the Moon forms as an initially molten body. As LMO crystallization proceeds, the composition of the residual melt continuously changes, with various minerals forming and either precipitating downwards towards the core, or floating to the surface, in the case of plagioclase. The final liquid dregs of this magma ocean are increasingly enriched in incompatible minor and trace elements including Th, U, K, rare earth elements, and P (collectively referred to as KREEP) (Hubbard et al., 1971). The late- stage LMO cumulates crystallizing from these dregs include Fe-rich pyroxene and ilmenite-bearing cumulates (IBCs), and the final residual liquid (i.e., urKREEP), which end up separated, but sandwiched between the plagioclase flotation crust and mafic mantle in the LMO cumulate stratigraphy (Figure 3.10).

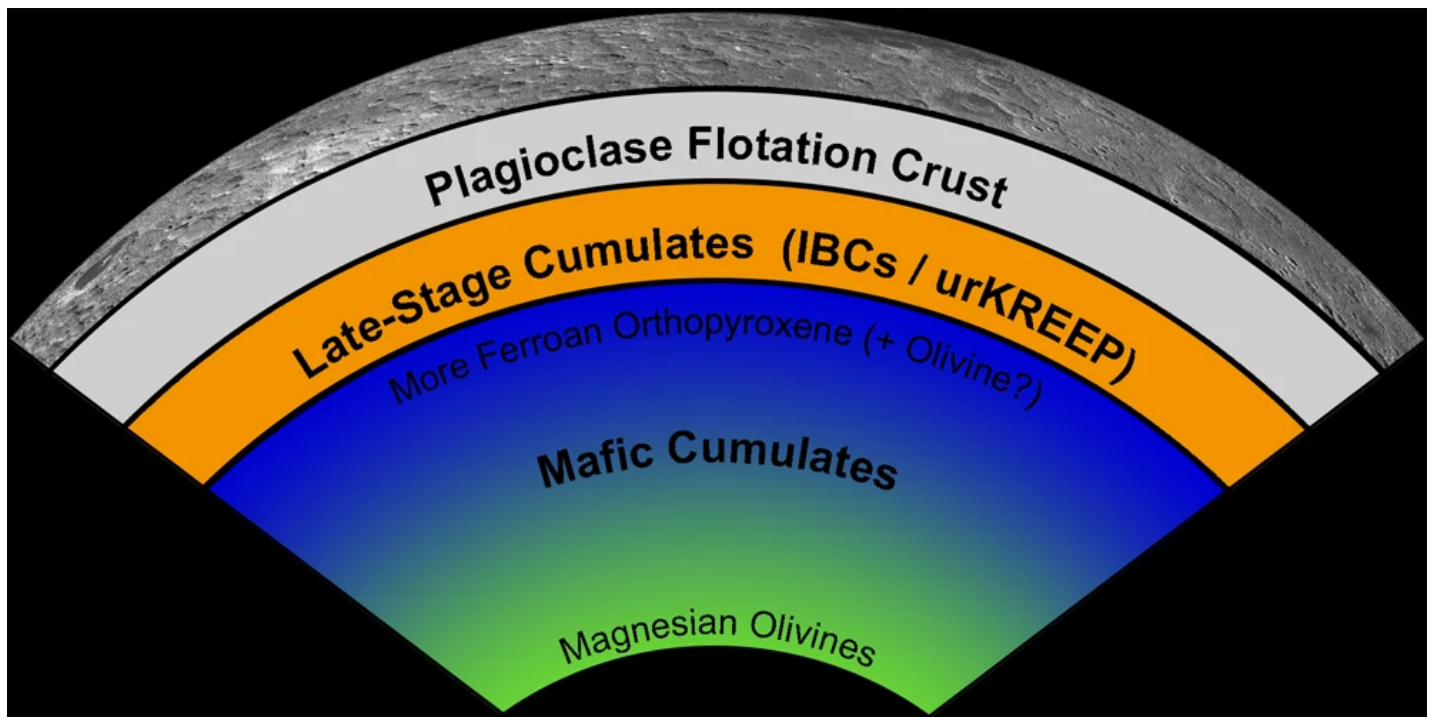

**Figure 3.10.** *A schematic overview of the LMO cumulate stratigraphy (Figure 1 from Moriarty et al., 2021) highlighting the position of late stage cumulates including IBCs and urKREEP.*

However, the initial LMO cumulate stratigraphy is gravitationally unstable, as the late-stage cumulates are cooler and denser than the underlying ultramafic mantle. Accordingly, this cumulate stratigraphy is commonly theorized to undergo gravitational restructuring. The timescales, depth and length scales, dynamics, and ensuring repercussions for the compositional structure of the lunar crust and mantle are complex and not fully understood. For example, restructuring of the early mafic cumulates and downwelling of dense IBCs may initiate prior to complete LMO crystallization (Hess and Parmentier, 1995; Boukaré et al., 2018; Yu et al., 2019; Zhao et al., 2019), and could therefore be one of the sources of the nearside/farside compositional asymmetry. The long-term stability of these late-state cumulates in the shallow mantle remains unresolved (Zhao et al., 2019), as does the potential for mixing (or remixing) of these materials with the lunar mantle during gravitational restructuring (Boukaré et al., 2018).

Our understanding of the formation and distribution, both spatially and with depth, of KREEP and other late stage cumulates comes almost exclusively from orbital datasets anchored by nearside-derived samples. Similar to our understanding of the feldspathic crust, several observations hint that the fate of KREEP differed between the lunar nearside and farside. In particular, the nearside exhibits a thinner overall crust, a higher surface abundance of KREEP-related elements, and a higher degree of volcanic resurfacing (Figure 3.7). The relationship of KREEP to these observations could be the result of straightforward evolution processes—a thinner crust on the lunar nearside would allow large impacts to excavate into the shallow KREEP layer, distributing these materials on the surface and creating the appearance of a nearside concentration of KREEP, and also imposing a thinner barrier to extrusive volcanism. However, as discussed above, this seeming asymmetry in lunar crustal thickness could also be symptomatic of asymmetric magma ocean crystallization. The lack of a shallow KREEP layer on the farside could be the result of the thicker crust essentially squeezing the residual LMO liquid toward the nearside, and the absence of concentrated radiogenic elements results in truncated volcanic resurfacing. In this way, although KREEP represents the last stages of magma ocean crystallization, understanding how this process proceeded on the Moon is vital to understanding the Moon we observe today.

Furthermore, the observed remotely elevated abundances of Th, an element often used as a proxy for KREEP in orbital data sets, in SPA (Figure 3.7) may not be related to KREEP. Elemental and isotopic analyses of basaltic lunar meteorites and the basalt returned by Chang'e 5 show that melts with elevated, KREEP-like abundances of REEs and Th (~2.5 to 5 ppm) were produced without the involvement of KREEP as evidenced the highly depleted Nd and Sr isotopic compositions of these lavas, which more definitively rule out KREEP involvement (e.g., Borg et al., 2009; Elardo et al., 2014; 2025; Tian et al., 2021). Thorium is an incompatible element, and thus low-degree partial melting of the mantle and subsequent fractional crystallization will result in its concentration in the melt. Additionally, fractional crystallization the SPA impact melt sheet would result in the concentration of Th in the last melt (e.g., Hurwitz and Kring, 2014), which would reside close to the surface where it would be incorporated into the SPA terrane by impact gardening. Only laboratory elemental and radiogenic isotopic analyses can definitively reveal whether the materials with enhanced Th in SPA are related to the KREEP reservoir or were produced by other processes.

For this reason, returning samples of Th- and other incompatible element-bearing materials excavated from the lunar interior by the SPA-forming impact, to determine whether a KREEP reservoir existed on the lunar farside, remains a high priority. In particular, understanding the associations of KREEP (e.g., with volcanic materials, with exposed Ur-KREEP, or with primary crustal material) is essential for addressing these fundamental questions related to the initial differentiation and subsequent evolution of the Moon, addressing OWL questions 3.5 ("How and when did terrestrial planets and moons differentiate?"), 5.2a ("What mechanisms have contributed to post-accretion or post-differentiation planetary cooling and heating?"), and 5.2c ("What are the chemical and physical consequences of cooling and solidification on solid body crusts?").

### 3.2.3 Objective 2.3

**Objective 2.3: "Determine if and how the Temporal and Compositional Evolution of the Farside Mantle Differs from the Nearside"**

Although the initial LMO hypothesis does not postulate hemispheric differences in the composition and evolution of the mantle, the observed differences in crustal thickness, composition and volcanic resurfacing (Figure 3.7) could indicate that evolutionary dichotomies extend all the way into the mantle. Samples of nearside basalts, which record mantle processes, show evidence for heterogeneities in mantle composition at both a lateral and vertical scale. This is evidenced by strong variation in the Ti and Fe content (Figure 3.11) (e.g., BVSP, 1981; Taylor et al., 1991; Giguere et al., 2000; Sato et al., 2017), as well as variable levels of "contamination" by KREEP materials suggesting melting and mixing between reservoirs in the shallow mantle (Binder, 1982). Numerous questions remain about the nature of these observed heterogeneities, including whether they represent a globally stratified mantle, or if they are the result of asymmetric gravitational restructuring, in which case the farside mantle could preserve an entirely distinct composition and stratigraphy from the nearside. However, the scarcity of farside samples has made direct comparison of material sourced from nearside and farside mantle reservoirs challenging. It therefore remains a major outstanding question as to if near-surface nearside/farside differences record fundamentally divergent mantle histories or if lunar mantle with similar compositions underlies hemispheric variations in crust and KREEP distributions.

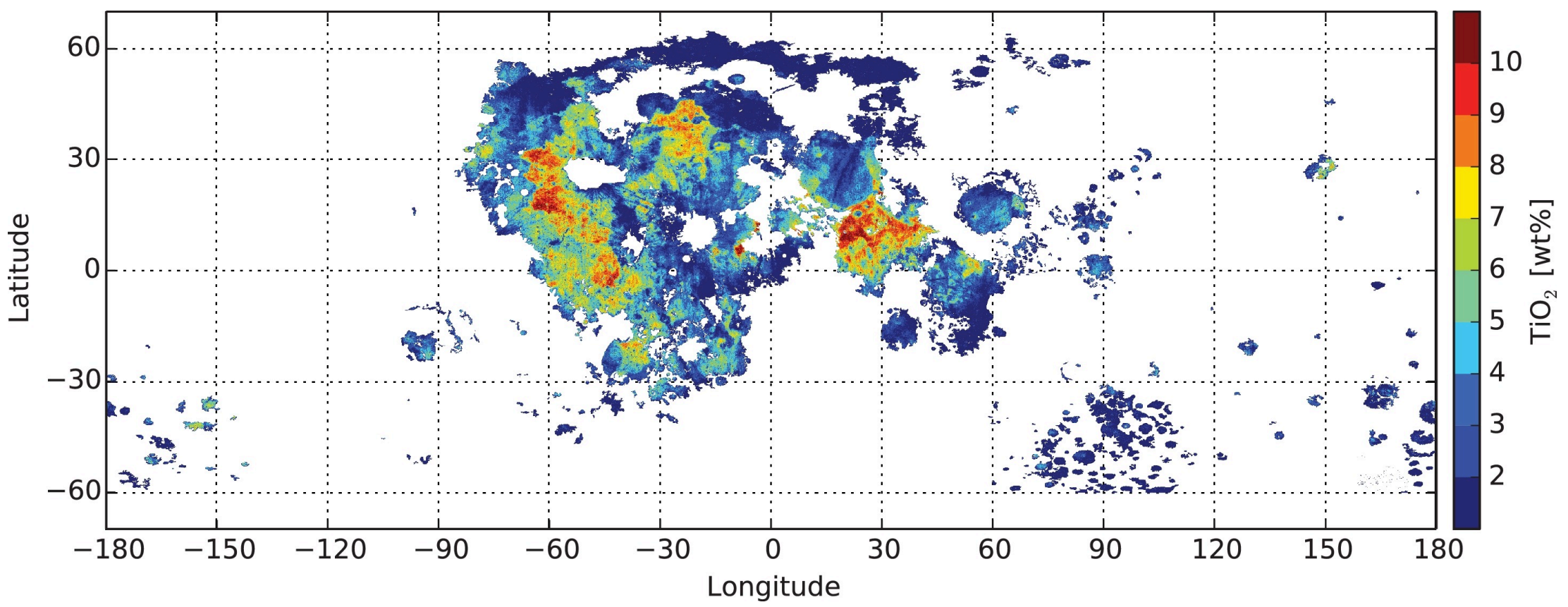

**Figure 3.11.** *TiO2 abundance map for lunar mare derived from LROC WAC measurements. (Figure 4 from Sato et al., 2017).*

Initial analysis of farside basalt samples returned by Chang'e 6 indicate some differences in farside mantle sources potentially resulting from the SPA impact. Chang'e 6 samples indicate that the farside mantle may have a lower oxygen fugacity than the nearside, which could result from either asymmetric crystallization of the LMO or from degassing of the farside mantle driven by the SPA impact (Zhang et al. 2025). Heavier K isotope concentrations in Chang'e 6 basalts also indicate SPA-driven volatile loss (Tian et al. 2026). Whether or not this difference is representative of the farside interior or specific to the Chang'e 6 basalts is an open question. Divergent thermal histories are also suggested by sample analysis and observations of the Moon's tidal response (Park et al. 2025; He et al. 2025). Analysis of additional farside volcanic deposits and possible upper mantle ejecta excavated by SPA will enable a more complete analysis of the differences between nearside and farside mantle reservoirs across space and time.

Farside samples from within SPA will resolve fundamental questions regarding lunar mantle evolution both by sampling lower crust and potentially upper mantle material excavated by SPA, and by sampling farside volcanic products (e.g., basalt and pyroclastic deposits) that constrain deeper mantle melt sources and melting conditions. If mantle sources for farside lavas record similar trace element and isotopic compositions to nearside mantle sources, it would indicate that LMO differentiation and potential overturn produced a broadly similar mantle composition globally, and nearside-farside differences mainly arise from hemispheric variation in crustal thickness/insulation and the localization of near-surface heat producing element reservoirs (PKT/KREEP). Alternatively, farside mantle source compositions that differ significantly from nearside sources (e.g., extreme incompatible element depletion) could indicate that mantle differentiation proceeded asymmetrically between the nearside and farside hemispheres. An initially symmetric mantle could also have undergone later hemisphere-scale modification through processes such as spatially varying mantle overturn or modification/redistribution of mantle reservoirs by the SPA impact; modification of an initially symmetric mantle could result in mantle source compositions that indicate incomplete mixing, localized enrichment/depletion, signatures of impact melt/source mixing, or distinctive reset ages. Addressing this objective, in conjunction with Obj. 2.1 and 2.2, will answer OWL decadal questions 3.5 and 5.2, and this objective in particular will also address OWL question 5.3: “How Have Surface/Near-Surface Characteristics and Compositions of Solid Bodies Been Modified by, and Recorded, Interior Processes?”

## 3.3 GOAL 3

### Goal 3: "What Are the Effects of Mega-Basin Impacts on Terrestrial Bodies?"

The scale of the South Pole–Aitken basin is striking: over 2,500 km in diameter and up to 8 km in depth. As a result, the impact event that formed its structure had consequences for the entire Moon. Impact basins of this scale, where the basin diameter is comparable to the diameter of the host planet or moon, are colloquially termed "mega-basins." Mega-basins seem to have affected almost every planetary body in the Solar System (Figure 3.12). These large impact basins events were clearly a dominant process during early Solar System bombardment and late accretion (Section 3.1.3), and likely influenced subsequent evolution of surface geology and interior dynamics (e.g., Ivanov et al. 2010) (Section 3.1.2). Yet major gaps remain in our understanding of how projectile and target properties influence mega-basin formation, how mega-basin structures evolve from the point of impact to the structures that are observed today, and how mega-basins influence long-term planetary thermal, magmatic, and magnetic evolution. Exploration of SPA, one of the largest known impact basins in the Solar System, will provide unique insight into these processes for mega-basins in general, and more specifically into how the Moon's largest and most ancient basin affected early lunar evolution. Making progress towards understanding the SPA basin's structure fundamentally requires evaluating the three-dimensional stratigraphy of the basin; that is, it requires a way to probe the composition at different depths, the spatial variation of composition, and how variations at the surface and at depth can be linked to structural features and discontinuities across the basin.

Impact basin formation spans stages of projectile compression, target excavation and transient cavity formation, basin modification (including rim collapse and interior uplift), ejecta deposition, and post-impact relaxation and crustal flow (Figure 3.13) (e.g., Solomon et al., 1982; Conrad et al., 2018; Zhu et al., 2025). At planetary scales, these processes involve complex material behavior that remain challenging to capture in numerical models, especially when attempting to link physical models of basin evolution to observable basin structures and composition. The challenges manifest in a range of uncertainties that persist even in our understanding of how we define basin structures. For example, published rim definitions differ substantially across interpretations based on topography, crustal thickness, gravity, and massif distributions, emphasizing the need to better establish what constitutes the basin boundary (James et al. 2019; Bernhardt et al. 2025; Andrews-Hanna et al. 2025). A central unknown in impact basin formation is the mechanism and extent of rim collapse following excavation. One hypothesis for determining the basin rim suggests that excavation may generate a thick circum-basin crustal annulus (e.g., Citron et al. 2024) that subsequently collapses to produce the subdued rim topography observed today; and while a variety of collapse mechanisms have been proposed (Senft and Stewart 2009; Melosh and Ivanov 1999; Crawford and Schultz 2013; Elbeshausen and Melosh 2020), it remains unclear which processes dominate at mega-basin scales.

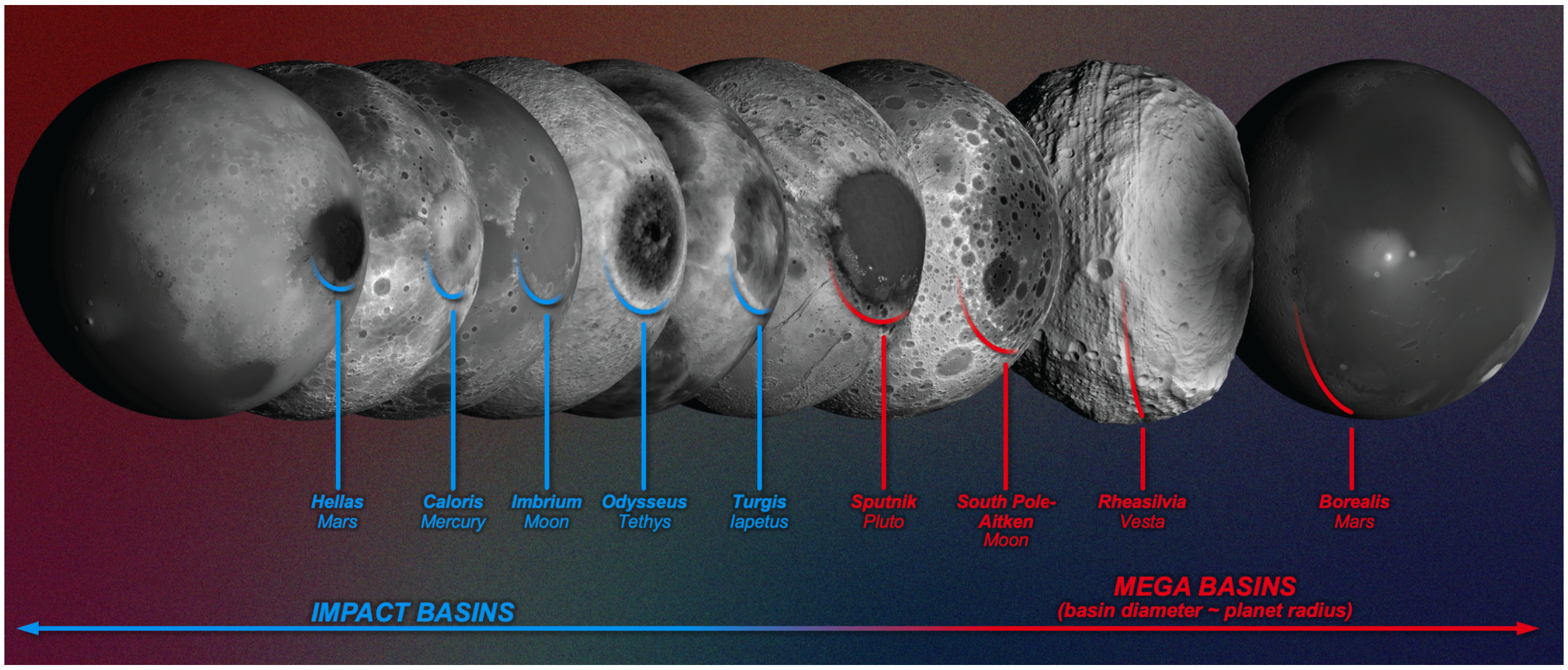

**Figure 3.12.** *Comparison of mega-basin impact structure scales across the Solar System. All worlds are aligned with the impact basin in question in the same orientation, and the size of all worlds are normalized. Grayscale indicates topography (black is low, white is high) and is normalized for each world. Image by James Tuttle Keane.*

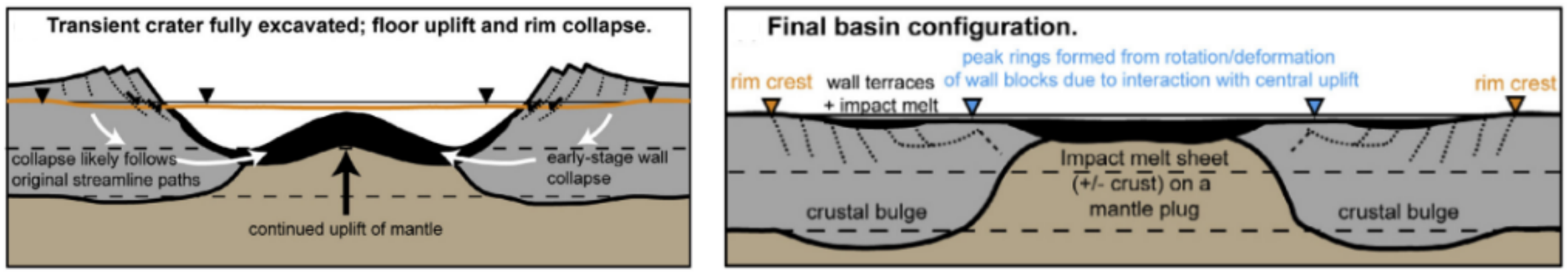

**Figure 3.13.** *Diagram of impact basin collapse, modified from Baker et al. (2016). During impact basin formation, the initial excavation of the transient cavity is followed by collapse of the cavity and basin rim, and uplift of the mantle within the basin floor (left). After the completion of rim collapse the final impact basin is formed (right) where the interior of the basin is a mix of an impact-generated melt sheet and collapsed portions of the basin rim.*

Another related, and critical, unknown is the origin of the geologic material in mega-basin interiors. For SPA in particular, the post-formation interior is hypothesized to be filled by a differentiated impact melt sheet (Hurwitz and Kring, 2014; Vaughan and Head, 2014), or infill/backflow of crustal materials pushed out during the initial rim collapse (Rajšić and Johnson, 2024; Potter et al., 2012), or some combination of both. Further complicating the picture are the subsequent contributions from volcanic infilling, including both mare (e.g., Pasckert et al., 2018) and cryptomare (Izquierdo et al., 2024).

Resolving SPA's formation and structure is also essential for understanding how mega-basin impacts affect planetary evolution. Improved understanding of SPA's geological and geophysical structure will constrain models of the basin's formation and enable more accurate estimates of impactor size, composition, and directionality, as well as inform models of ejecta excavation and emplacement. In turn, this will improve our evolving picture of how the SPA-forming impact influenced lunar surface geology and interior dynamics (including dynamo evolution, crustal magnetic anomalies, and the formation of the lunar hemispheric asymmetry—see Section 3.2), and will help bound the Moon's thermal and crustal state at the time of the impact, including whether a pre-existing crustal thickness gradient existed prior to SPA (e.g., Andrews-Hanna et al. 2025). Answering these questions directly addresses several OWL Decadal questions, including 4.3 (How did collisions affect the geological, geophysical, and geochemical evolution and properties of planetary bodies) and 4.4 (How do the physics and mechanics of impacts produce disruption of and cratering on planetary bodies).

### 3.3.1 Objective 3.1

**Objective 3.1: "Determine Whether There are Discontinuities in the Basin Structure (Crustal Thickness, Density, Magnetic Field) Across SPA to Constrain Models for the Formation of a Mega-Basin"**

Two of the challenges in studying mega-basin structure are (a) the enormous scale of these basins, and (b) that many critical markers, including crustal thickness variation, depth to the mantle, vertical and horizontal changes in composition, and subsurface faults are not directly exposed at the surface. As a result, key measurements for understanding this objective rely on geophysical techniques such as gravity and magnetometry to probe the subsurface. Addressing objective 3.1 requires acquiring high-fidelity geophysical datasets along an extended cross-section of SPA to compare to existing and future geophysical modeling. Specifically, this objective seeks to map discontinuities in basin structure that are seen in various models for SPA formation. Examples of these discontinuities include the extent of the initial excavation cavity; the location and character of SPA's rim; the location of large basin-related circumferential faults (Figure 3.13); the extent of postulated SPA-derived impact melt; and the thickness, extent, and compositional variability of SPA ejecta (Figure 3.14). While a traverse will only map these discontinuities in one section of SPA, the results from the traverse can be coupled to orbital datasets to infer where such discontinuities might occur across the basin. For example, surface traverse mapping of basin rim structures in one section of the basin will constrain what orbital gravity/topography signals indicate rim structures, and can be applied across SPA to constrain the entire basin's rim structure and shape. Similarly, data from the traverse can be used in conjunction with results from the Farside Seismic Suite (FSS) payload. This in turn will constrain impact models that predict different basin structures/shapes based on varied impactor properties and trajectories, enabling an improved understanding of what impactor direction, size, composition, and velocity is most consistent with the structure of SPA.

Relative gravity anomalies arise from variations in bulk density in the lunar subsurface. Thus, surface observations of these anomalies can provide insight into shallow crustal heterogeneity, relief on the crust–mantle interface, and mantle density anomalies. The various lithologies that may be present in the South Pole–Aitken basin are expected to exhibit a wide range of grain densities. Ultramafic rocks originating in the mantle and lower crust have the highest densities, anorthosites are low density, and rocks like basalt and norite are typically intermediate in density between those end members. Bulk density is additionally affected by porosity at all scales due to fragmentation and/or vesicularity. The distribution of these lithologies today is dictated by the basin-forming impact event, the differentiation of its melt sheet, post-basin basaltic volcanism, and subsequent impact bombardment (e.g., Gowman et al. 2026). Ground-based measurements of relative gravity anomalies can thus be used to understand the relative distribution of these materials both laterally and with depth, and these results then compared with different models of basin formation and modification.

The long-wavelength gravity anomaly associated with the South Pole–Aitken basin is well known after the GRAIL mission. One metric of the data quality is the "degree strength", which is the spherical harmonic degree at which the gravity field signal has power equaling the power of the error. The degree strength is at least 600 throughout the basin, corresponding to a spatial scale of 9.1 km (Figure 3.15) (Lemoine et al., 2014). This dataset provides a strong baseline for big-picture understanding of the impact basin structure, however, it lacks the fidelity to address the smaller spatial-scale details of the basin formation process. The spatial resolution of GRAIL gravity is limited by the rapid attenuation of short-wavelength gravity with upward continuation.

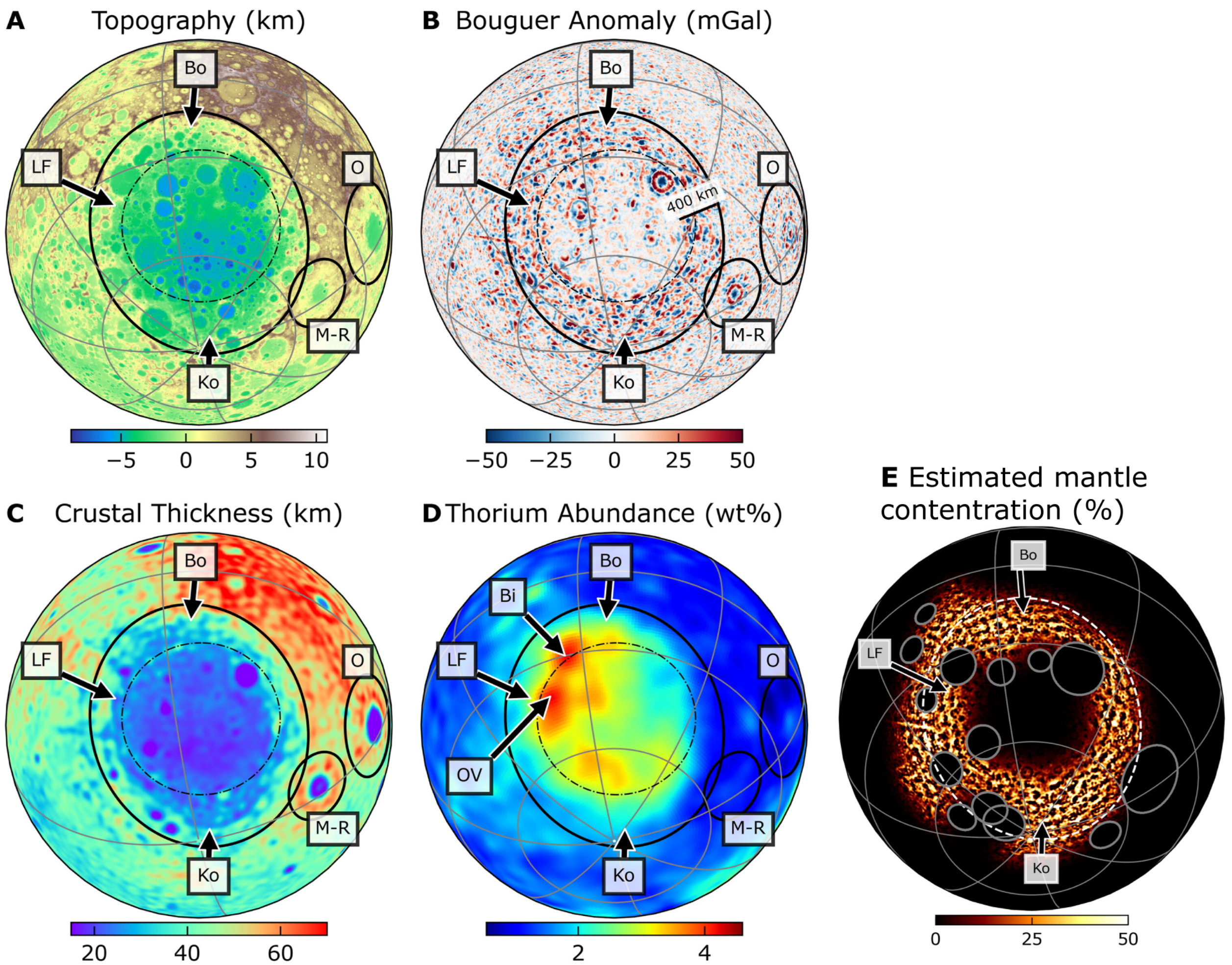

**Figure 3.14.** *Maps of South Pole–Aitken basin from Gowman et al. (2026) showing (a) crustal thickness, (b) Bouguer anomaly, and (c) estimated mantle concentration. The estimated crustal thickness map (a) shows regions of thicker material interior of the basin rim (outlined in black) which may represented sections of collapsed rim. The Bouguer gravity anomaly map (b) shows an annulus of material with positive and negative Bouguer anomaly signals that may represented blocks of dense upper mantle material excavated during the impact. Based on the gravity signatures, Gowman et al. (2026) estimated a potential mantle concentration (c) which shows where excavated upper mantle material is expected to be, within an annulus exterior and interior of the basin rim. The SPA basin ellipse is outlined in black in (a, b) and in white in (c).*

Gravity models for crustal thickness are similarly limited by the non-uniqueness of their solutions, which require a priori assumptions about the distribution of density in the crust (Wieczorek et al., 2013). Ground-based gravity measurements taken at a scale much finer than GRAIL (with effective spatial resolution less than 9 km) can be used to calculate bulk density. These improved calculations of bulk density, taken either across an elevation change (after Nettleton, 1939), or even across flat terrain can be used to characterize the degree of heterogeneity in the subsurface and also to significantly reduce the non-uniqueness of crustal thickness models. Models of crustal thickness can be further improved by results from the Farside Seismic Suite (FSS) (Panning et al., 2022) payload, the results of which can be used to pin the thickness of the crust at a known location. The strongest benefit to this synergy would occur in cases where the SPARX traverse and FSS landing site are within the same general region (e.g. the central portion of Schrödinger basin).

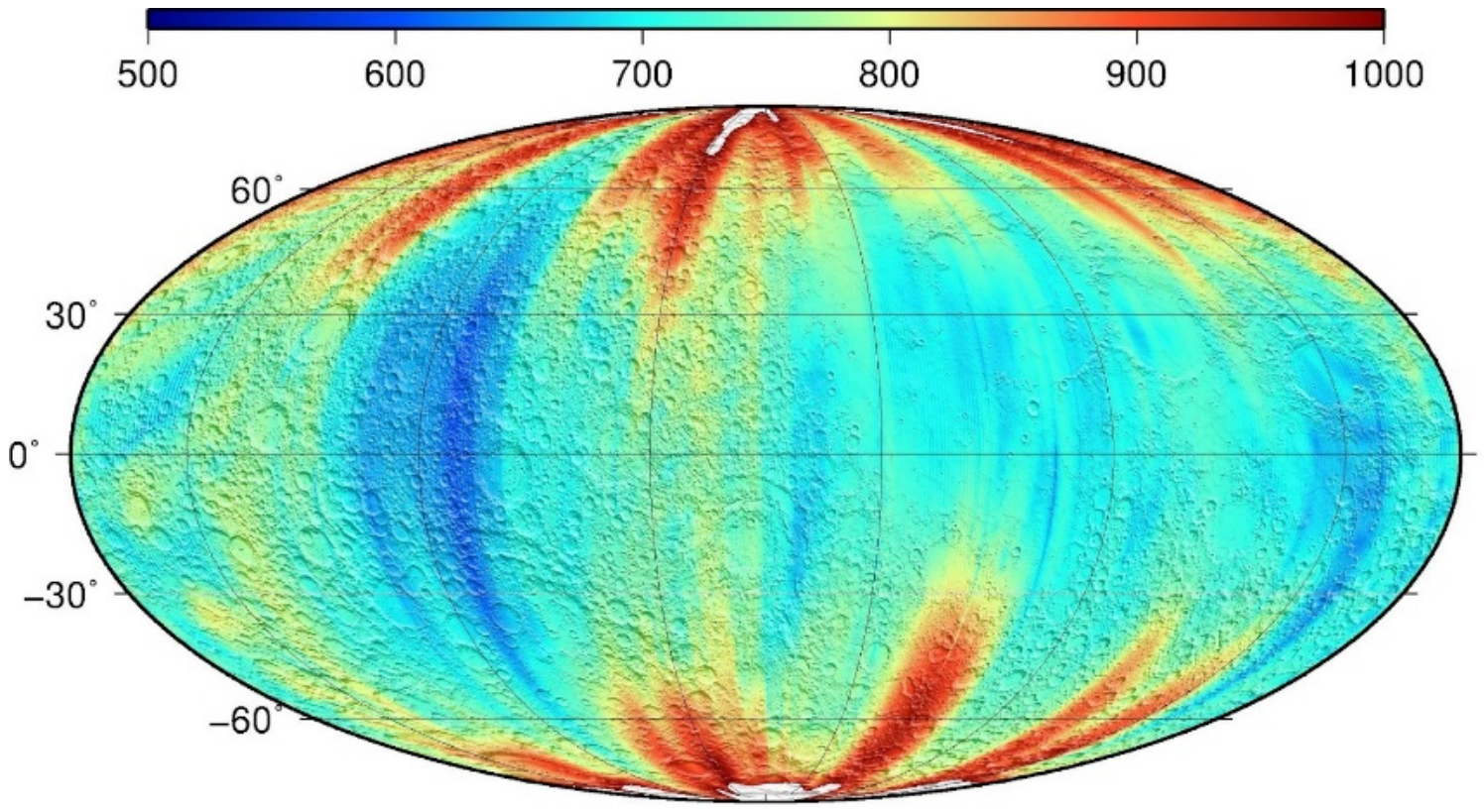

**Figure 3.15.** *GRAIL degree strength map (Lemoine et al., 2014). High degree strength indicates high spatial resolution and signal-to-noise for gravity measurements.*

Another diagnostic set of measurements for constraining discontinuities related to the basin-forming impact event is mapping the distribution of ferromagnetic material, which could correspond to iron-rich impactor materials, basalts, ultramafic mantle, and/or KREEP-bearing materials. The current coarse spatial resolution of orbital datasets reveals a high concentration of remanent crustal magnetism in the northwest region of SPA, which has been hypothesized to be potentially deposited by the SPA-forming projectile and has been used as evidence of a South to North SPA impact trajectory (Wieczorek et al. 2012). However, this interpretation is contested, and other datasets, including topography (Andrews-Hanna et al., 2025) and massif distribution (Bernhardt et al., 2025) are not consistent with this hypothesis. A long traverse with a magnetometer would enable observations of the strength and source depth of remanent crustal magnetism across the SPA basin, constraining various hypotheses for the origin of magnetic crustal material and if magnetic anomalies are related to the formation of SPA. The primary categories of lunar crustal magnetic anomalies include those associated with basin melt sheets (e.g., Halekas et al., 2003; Hood, 2011), antipodal regions of major impacts (e.g., Hood and Huang, 1991; Wakita et al., 2021), and more spatially distributed anomalies linked to impactor material, lunar swirls, or intrusive lithologies (e.g., Wieczorek et al., 2012; Wakita et al., 2021; Blewett et al., 2011; Purucker et al., 2012). Each category likely records magnetic fields through distinct physical mechanisms and over different timescales, with fundamentally different implications for the origin and intensity of the magnetizing field

The spatial resolution required for these gravity, and magnetometry measurements would be impossible to acquire via orbital methods. Therefore, a long-range traverse over the relevant basin-scale features with geophysical instruments provides the only reasonable methods for addressing these questions of basin structure, and ultimately, basin formation. The measurements from Objective 3.1 directly address OWL questions 4.3a (“How did the earliest and/or largest impact events influence the physical evolution of Solar System Worlds”), 4.3b (“How do impacts affect surface and near-surface properties of Solar System worlds?”), and 4.4a (“How does the impact process vary with projectile/target body properties and impact parameters?”).

### 3.3.2 Objective 3.2

**Objective 3.2: "Explore the Relative Role of Impact Melt Differentiation, Crustal Backfilling, and Volcanic Infilling in Emplacing the Materials Found in SPA's Interior"**

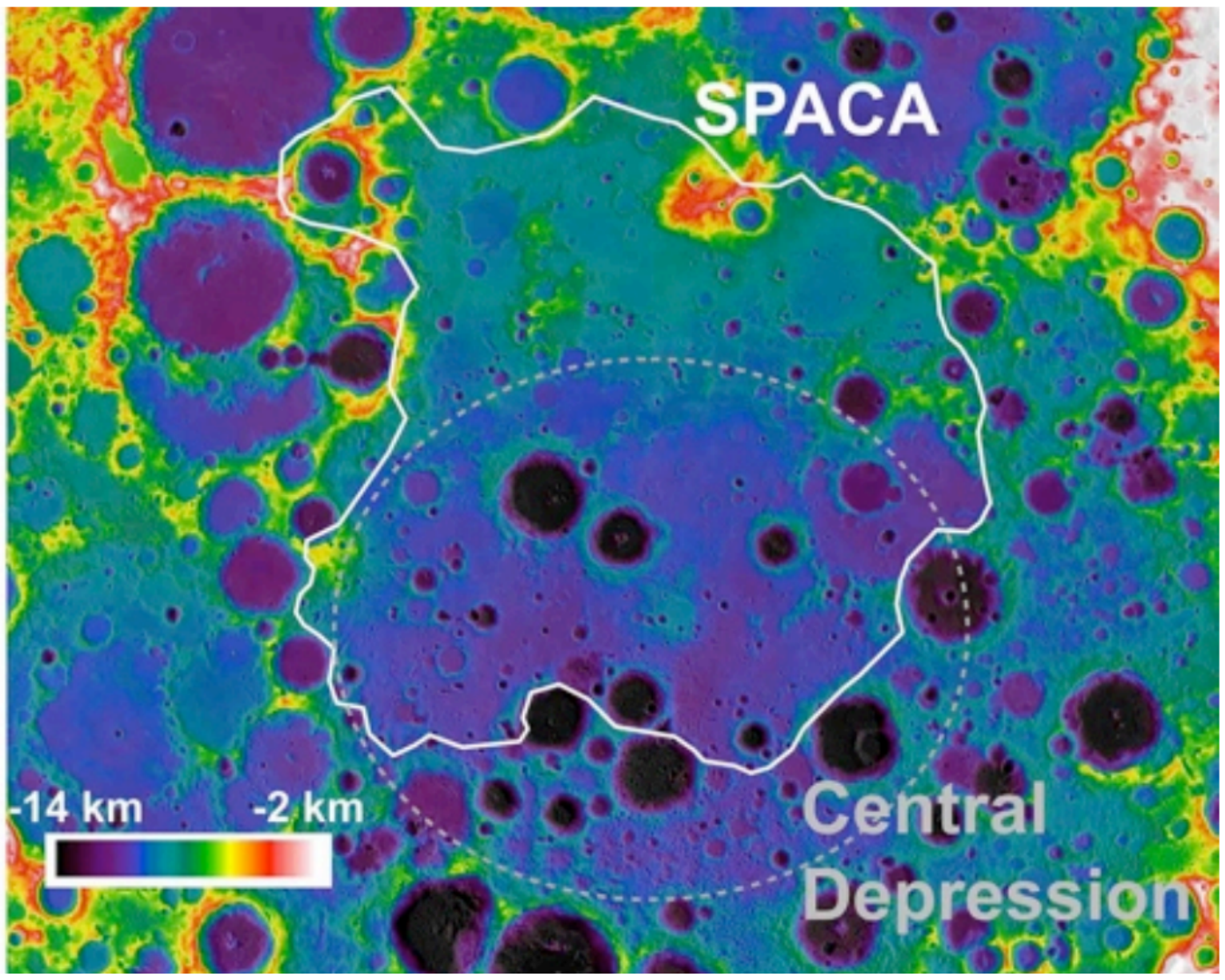

**Figure 3.16.** *A topographic map of central SPA highlighting the presence of a roughly elliptical central depression and widespread smooth plains (From Moriarty and Pieters, 2016).*

As the oldest and largest confirmed impact basin on the Moon, SPA provides an important opportunity to study the formation and evolution of the largest class of impacts. This is aided by the fact that, unlike the nearside basins, SPA has not been fully resurfaced by mare. Instead, it contains a compositional patchwork of units resulting from multiple different processes throughout the basin's history. The central portion of the SPA interior exhibits an extensive central depression, abundant smooth plains, excess mass at depth, unusual volcanic features, and a mafic-bearing surface mineralogy sometimes referred to as the SPA Compositional Anomaly, or SPACA (Figure 3.16) (e.g., Moriarty and Pieters, 2015; 2018; Whitten et al., 2015; Izquierdo et al., 2024; Wang et al., 2024; Chuang et al., 2025; Uemoto et al., 2017; James et al., 2019). Hypotheses rooted in remote sensing observations and impact models suggest some combination of the following sources: basin modification (i.e., collapse of the excavation cavity) (e.g., Rajšić and Johnson 2024; Citron et al., 2024), volcanic resurfacing (e.g., Whitten et al., 2015; Moriarty and Pieters, 2018; Izquierdo et al., 2024; Wang et al., 2024; Chuang et al., 2025), deposition of subsequent basin ejecta (Petro and Pieters, 2008; Meyer et al., 2020), and differentiation of a large SPA impact melt pool (Hurwitz and Kring, 2014; Vaughan and Head, 2014; Uemoto et al., 2017).

Central to understanding the mega-basin formation process is what represents the dominant factors that control the geology of the basin's interior. Impact models demonstrate that the SPA-forming impact should have completely removed the original highlands flotation crust, at least locally (e.g., Melosh et al., 2017; Citron et al., 2024). However, gravity data suggest a ~12–20 km thick crustal layer in the SPA interior (Wieczorek et al., 2013). This implies that the interior is dominated by crust that was reworked or newly generated during or following the SPA impact, most likely from some combination of original crust that infills the basin during transient cavity collapse and a long-lived impact melt sheet that may have undergone differentiation (e.g., Hurwitz and Kring, 2014; Vaughan and Head, 2014). The impact may also have triggered mantle convection via a deep, prolonged thermal pulse (e.g., Jones et al., 2022) that could have added additional crust to the SPA floor via extrusive or intrusive magmatism.

These different processes predict different materials, compositions, and stratigraphies within the basin. Thus, the way to evaluate their relative importance for this objective requires acquiring samples from different depths within the basin. Impact craters act as natural boreholes, allowing sampling of the subsurface from depths that scale with the diameter of the crater (e.g., Ernst et al., 2010). Fortunately, there are multiple impact craters in the central SPA region whose ejecta deposits show evidence of having excavated distinct compositions from depth, acquiring samples from a strategic selection of these craters would allow us to constrain the relative contributions to the bulk of the SPA central depression.

While the processes of impact melt differentiation and crustal backfilling establish the initial state of the basin interior, and are the processes most relevant to understanding wider mega-basin formation processes across the Solar System, they are not the only compositional contributors to the surface of SPA that is observed today. Any attempt to understand the initial post-impact state of the basin, or its subsequent modification needs to also address the contributions from volcanic processes and subsequent impact processes. Volcanic contributions include both surface mare deposits (e.g., Pasckert et al., 2018), and also shallow subsurface deposits in the form of cryptomare (Whitten et al., 2015; Moriarty and Pieters, 2018; Izquierdo et al., 2024; Wang et al., 2024; Chuang et al., 2025) and intrusive magmatic deposits (Jozwiak et al., 2012; 2017). Additionally, light plains deposits formed by deposition of ejecta from post-SPA basins (e.g., Meyer et al., 2020) also contributes locally to SPA (e.g., from Orientale; Apollo). While the morphologic distribution of these units is well-characterized from orbital datasets, the Apollo missions showed that distinct units from orbit often look very different on the surface (Wilhelms 1980). *In situ* measurements could therefore be critical in constraining the composition and origin of geologic units across SPA’s interior, which is essential for understanding basin evolution processes. Through the combination of a carefully selected sample suite, and *in situ* measurements capturing the diversity of geologic units in the basin interior, SPARX will provide critical insight into the history of SPA from the moment of impact to the current day, and address decadal questions 4.3 (“How did collisions affect the geological, geophysical, and geochemical evolution and properties of planetary bodies?”), 5.2 (“How have the interiors of solid bodies evolved?”), and 5.3 (“How have surface/near-surface characteristics and compositions of solid bodies been modified by, and recorded, interior processes?”).

## 3.4 PRIORITIZATION OF SPARX GOALS AND OBJECTIVES

The SPARX TOR requests prioritization of the listed goals and objectives (Sections 3.1–3.3). As the SDT worked to refine the goals and objectives, informal straw-poll prioritization was undertaken at both the March 2025 and April 2025 meetings. The goal of these informal prioritization discussions was to ensure that the questions developed by the SPARX SDT represented the highest-priority questions for a lunar sample return mission to SPA, and to drive consistency in scope across the objectives. However, as the goals and objectives were not yet finalized at this point, the results of those straw polls will not be reported here. Following the finalization of the SPARX goals and objectives during the July 2025 virtual meeting, the SPARX SDT undertook the formal prioritization activity. The results of which are discussed below, and in Appendix 10.2.

### 3.4.1 Prioritization Criteria

The SPARX SDT outlined four classes of criteria on which to evaluate the proposed goals and objectives, outlined below. The purpose of the criteria was to evaluate and prioritize the science goals and objectives within a holistic context of the current lunar exploration environment and any future portfolio.

1. *Traceable Science*
    a. Addressing Explicit Decadal Level Science (OWL, National Academies of Science, 2023)
    b. Transformational Science: degree to which the SPARX defined investigations can make significant progress on achieving the relevant decadal science question(s)
2. *Science Environment*
    a. Science strongly enabled by SPA
    b. Science unique to SPARX given other missions to SPA
    c. Ability to achieve success without returned sample
3. *Programmatic Environment*
    a. Applicability to Moon to Mars Objectives (NASA, 2023)
    b. Applicability to Lunar Community Science Objectives (e.g., SCEM, LER, LEAG Goals)
    c. Science unique to SPARX given the published Artemis first south polar landing science goals (NASA, 2023)
4. *Implementation Considerations*
    a. Confidence that the measurement/sample will address the desired science
    b. Confidence in identifying the correct location to acquire the sample
    c. Difficulty in acquiring required sample relative to other priority samples

The voting criteria above are listed in order of weighted importance, with the Traceable Science and Science Environment carrying the highest weight. The Programmatic Environment and Implementation Considerations criteria were considered second order discriminators between objectives. The SDT notes that while both the Traceable Science and Programmatic Environment factors include community and agency documents, the scope of the documents led the SDT to consider them separately. Given the proposed scope of the SPARX mission, the SDT chose to weight science goals that were both directly traceable to the Decadal survey, and had potential to be transformational for planetary science as a broader field with the highest weight. While the priorities and goals listed in Lunar Community Science Documents and the Moon to Mars Objectives were considered important, the scope of these documents was considered to be narrower. The voting on priority proceeded differently for the Goals and Objectives; goal rankings were assigned based on majority vote (i.e., if the majority voted 4 for a goal, the goal was scored a 4), while objective ranking was assigned based on numerical average of votes. The difference in voting derives from the larger number of objectives, and simplifying voting for the virtual environment. It was decided that for the objectives, it would be more time efficient for members to independently assign values for all objective criteria. These independent values were then averaged together, resulting in the values shown in Figure 3.18. Discussion of the objectives voting followed their average scores. Voting for both categories used the following numerical value scale: 5 = strongly achieves the stated goal (excellent); 4 = achieves the goal well with some caveats (very good); 3 = archives the goal but caveats are present (good); 2 = concerns present to achieve the stated goal (fair); 1 = does not address the stated goal (poor). In the cases of non-consensus votes on prioritization scores, team members with different scores volunteered their rationale for their voting score during discussion, and these discussion points are summarized in Appendix 10.2. In some cases, a revote was undertaken after discussion. The scores post-re-vote are presented below in Figure 3.17 and 3.18.

| | S.G.1 What was the chronology of key processes and events during the first half billion years of Solar System history? | S.G.2 Does the lunar farside preserve evidence of asymmetric magma ocean evolution and planetary crustal formation processes? | S.G.3 What are the effects of mega-basin impacts on terrestrial bodies? |
|---|---|---|---|
| Traceable Science: | | | |
| **Explicit Decadal Level Science** | U | U | U |
| **Transformational Science (degree to which described investigation can make significant progress on proposed objective)** | vote 9-4 | vote 10-2 | vote 8-4 |
| | | | |
| Science Environment: | | | |
| **Science strongly enabled by SPA** | U | U | U |
| Science unique to SPARX given the samples previosuly returned by Chang'E 6 | vote 9-5 | vote 8-7 | U |
| Ability to achieve science without returned sample | U | vote 8-5 | vote 8-4 |
| | | | |
| Programmatic Environment: | | | |
| Applicability to Moon to Mars Objectives | U | U | vote 10-2 |
| Lunar Community Science Objective (e.g. SCEM) | U | U | U |
| *Science unique to SPARX in the context of Artemis science goals* | U | vote 8-7 | vote 7-6 |
| | | | |
| Implementation Considerations: | | | |
| Confidence that the measurement/sample will address the desired science | vote 8-5 | vote 7-5 | vote 7-5 |
| Confidence in identifying the correct location to acquire the sample | vote 10-3 | vote 7-5 | vote 6-5 |
| Ease in acquiring required sample - proximity to other sampling locations | vote 7-5 (one 3) | vote 7-5 | vote 7-5 |

| Categorization Guide |
|---|
| 5 |
| 4 |
| 3 |
| 2 |
| 1 |

**Figure 3.17.** *Prioritization of Goals. The panel votes are recorded. U = unanimous vote.*

### 3.4.2 Goal Prioritization

The prioritization of the goals is reflected in their overall numbering, with Goal 1 being considered the highest priority goal, followed by Goal 2, and then Goal 3. A brief summary of the voting rationale for each goal is presented below, while an expanded discussion of the vote breakdown and rationale is presented in Appendix 10.2.

### Goal 1: What Was the Chronology of Key Processes and Events During the First Half Billion Years of Solar System History?

This goal is considered to be the top priority goal of a SPARX mission by the SDT. Some of the most impactful criteria included the ability of Goal 1 to address several different levels of Decadal Science questions, and the significant applicability of the results to our understanding of processes across the Solar System. The SDT also weighted heavily the criteria that samples for several of the Goal 1 objectives could only be credibly acquired from SPA because of the unique properties of the basin, including age and depth of excavation. **The primacy of Goal 1 stems from (1) the direct applicability of its results to understanding processes across the Solar System, and especially on the early Earth, and (2) the unique environment of SPA being perhaps the only place within the Solar System where samples can be acquired to address these questions.** The balance of these two items significantly influenced the SDT, and team members felt that the magnitude of the potential science return outweighed some of the challenges associated with acquiring and returning these critical samples, reflected in the Implementation Considerations.

### Goal 2: Does the Lunar Farside Preserve Evidence of Asymmetric Magma Ocean Evolution and Planetary Crustal Formation Processes?

This goal is considered to be the second highest priority goal of SPARX, with several team members noting that given the evolving lunar exploration space, this goal has the potential to become the highest SPARX priority. The rationale for this being that Goal 2 science can only be addressed by accessing materials exposed in SPA, while the STM for the first crewed Artemis south polar landing includes several objectives relating to Solar System chronology (Denevi et al., 2025), thereby allowing critical aspects of Goal 1 science to be addressed outside of SPA. Although framed around the Moon-specific nearside/farside asymmetry, **the broader implications of goal 2 reflect important questions about planetary formation processes, including how processes leading to lunar asymmetry deviate from the idealized model for planetary evolution. The applicability of the results of these investigations towards understanding the process of planetary crustal formation throughout the Solar System, and beyond, influenced the SDT's prioritization of this goal.** The SPARX SDT noted that the return of young basalt samples from Chang'e-6 in conjunction with the proposed science goals for the first Artemis landed mission could address portions of this Goal's objectives, although not the highest priority objective within it (Objective 2.2; Figure 3.18).The SDT also specifically notes that the proposed investigations of a SPARX mission would work synergistically with samples returned by either Artemis, Chang'e-6, or analyzed at a future Moon Base by greatly increasing the access to different geologic settings, and by allowing for the analysis of a large number of sample ages from which to construct statistically robust chronology assessments (see Section 4.3.3).

### Goal 3: What are the Effects of Mega-Basin Impacts on Terrestrial Bodies?

This goal is considered to be the lowest priority of SPARX goals; however, the team emphasizes that this ranking does not reflect the scientific importance of the objectives within the goal. The lower ranking relative to Goals 1 and 2 largely reflects that sample return is not required to make significant progress on this goal; therefore, other non-SPARX missions to SPA could feasibly address these objectives, although the degree to which the science could be addressed, and the cost efficiency of such missions is not well constrained. **The science of Goal 3 is tailored to fully explore the unique mega-basin environment of SPA and provide key insights to mega-basin structures that can be applied to other impacts across the Solar System.** The SDT recognized that the nature of the objectives needed to accomplish this Goal is a significant driver in mission architecture and traverse design, to a degree that might outweigh the scientific priority of the objectives, and this will need to be carefully considered in the selection and design of the final mission profile.

### 3.4.3 Objective Prioritization

In addition to prioritizing the overall Science Goals, the SPARX SDT was also tasked with independently prioritizing the Science Objectives of the mission. The results of this prioritization are particularly useful as they highlight the most important investigations for a future SPARX mission. As a result of the long, complex history of the basin, most of the objectives within any specific goal can only be accomplished by traversing to widely-dispersed locations. However, there are locations where objectives from different goals form spatial clusters. By providing a prioritized list of Objectives, the SPARX SDT is providing flexibility for understanding the science option space for a variety of mission designs, not only the nominal SPARX baseline mission design.

SPARX SDT members independently filled out the prioritization matrix for each objective, and the results were then averaged together to provide the following matrix (Figure 3.18). Results are shown as both an evenly weighted average, and also a weighted average. The weighted average assigns 5× weight to the categories of “Explicit Decadal Level Science”, “Transformation Science”, and “Science strongly enabled by SPA”, as these were the categories that the SDT members concurred were of the highest priority for evaluating the relative ranking.  After being compiled, the raw ranking data were then discussed by the SDT to arrive at the final recommended Objective prioritization. The final SPARX Objective prioritization is presented in Table 3.1 below, along with a high-level rationale for the scientific priority of the top three Objectives. The scientific importance and rationale for the remaining objectives is described in detail in the relevant Science Goal sections (Section 3.1–3.3), and all proposed objectives address high-priority decadal science objectives. The top three priority objectives listed here are the foundation that any mission to SPA should address, with the knowledge that any mission designed to address these objectives will, by necessity, also address multiple other critical science objectives outlined by the SPARX STM.

A summary of the original SDT prioritization discussion can be found in Appendix 10.2.

**Table 3.1.** *Prioritization of SPARX science objectives.*

| Objective | Description |
|---|---|
| 1.3 | Determine if and when a giant planet instability drove a sharp change in the inner Solar System impactor flux. |
| 1.1 | Quantify the age of the Moon to distinguish between early- and late-Moon formation models. |
| 2.2 | Determine if there was a farside KREEP reservoir, or other late-stage magma ocean products, prior to the SPA impact. |
| 2.1 | Determine if the farside highlands crust is similar to nearside crust in age and/or composition. |
| 2.3 | Determine if and how the temporal and compositional evolution of the farside mantle differs from the nearside. |
| 3.2 | Distinguish the relative contributions of impact melt differentiation, crustal backfilling, and volcanic infilling to the interior composition and stratigraphy of SPA. |
| 3.1 | Determine whether there are discontinuities in the basin structure (crustal thickness, density, magnetic field) across SPA to constrain models for the formation of a mega-basin. |
| 1.2 | Distinguish between fast- and slow-cooling models for lunar magma ocean crystallization. |
| 1.4 | Constrain the evolution of the lunar dynamo in the pre-4.0Ga epoch of lunar history. |

| | Obj. 1.1 Quantify the age of the Moon to within ± 30 Ma to distinguish between Early Moon and Late Moon formation models | Obj. 1.2 Distinguish between fast-cooling or slow-cooling models for lunar magma ocean crystallization | Obj. 1.3 Determine if and when a giant planet instability drove a sharp change in ISS impactor flux | Obj. 1.4 Constrain the evolution of the lunar dynamo in the pre-4.0 GA epoch | Obj. 2.1 Determine if the farside highlands crust is similar to nearside crust in age and/or composition | Obj. 2.2 Determine if there was a farside KREEP reservoir or other late stage magma ocean product prior to the SPA impact | Obj. 2.3 Determine if and how the temporal and compositional evolution of the farside mantle differs from the nearside | Obj. 3.1 Distinguish between models for the SPA forming impact based on asymmetries in basin structure and materials | Obj. 3.2 Distinguish the relative contributions of impact melt differentiation, crustal backfilling, or volcanic infilling to the interior composition of SPA |
|---|---|---|---|---|---|---|---|---|---|
| Traceable Science: | | | | | | | | | |
| **Explicit Decadal Level Science** | 5.00 | 4.92 | 5.00 | 4.92 | 4.92 | 4.92 | 4.92 | 4.85 | 4.54 |
| **Transformational Science (degree to which described investigation can make significant progress on proposed objective)** | 4.85 | 4.31 | 4.92 | 3.77 | 4.77 | 4.77 | 4.69 | 4.00 | 4.15 |
| Science Environment: | | | | | | | | | |
| **Science strongly enabled by SPA** | 4.69 | 4.31 | 4.69 | 3.31 | 3.85 | 4.85 | 4.54 | 4.85 | 4.92 |
| Science unique to SPARX given the samples previosuly returned by Chang'E 6 | 3.92 | 4.00 | 4.15 | 4.46 | 4.77 | 4.62 | 3.85 | 4.92 | 4.85 |
| Ability to achieve science without returned sample | 1.00 | 1.00 | 1.31 | 1.77 | 1.23 | 1.62 | 1.31 | 3.08 | 2.38 |
| Programmatic Environment: | | | | | | | | | |
| Applicability to Moon to Mars Objectives | 5.00 | 4.92 | 5.00 | 4.54 | 4.85 | 5.00 | 4.92 | 4.77 | 4.54 |
| Lunar Community Science Objective (e.g. SCEM) | 5.00 | 5.00 | 5.00 | 5.00 | 5.00 | 5.00 | 5.00 | 4.92 | 4.85 |
| *Science unique to SPARX in the context of Artemis science goals* | 4.08 | 3.85 | 4.15 | 4.00 | 3.69 | 4.54 | 4.54 | 4.23 | 4.85 |
| Implementation Considerations: | | | | | | | | | |
| measurement/sample will address the | 4.31 | 3.77 | 4.38 | 3.62 | 4.69 | 4.46 | 4.85 | 3.77 | 4.15 |
| Confidence in identifying the correct location to acquire the sample | 4.08 | 4.00 | 4.23 | 3.85 | 4.23 | 4.54 | 4.92 | 3.92 | 4.15 |
| Difficulty in acquiring required sample - proximity to other sampling locations | 4.69 | 4.54 | 4.62 | 3.69 | 4.08 | 3.46 | 4.77 | 2.85 | 3.62 |
| Averages: | **4.24** | 4.06 | **4.31** | 3.90 | 4.19 | **4.34** | 4.39 | 4.20 | 4.27 |
| Weighted Averages | **4.56** | 4.29 | **4.61** | 3.95 | 4.36 | **4.61** | 4.56 | 4.39 | 4.41 |

**Figure 3.18.** *Prioritization of SPARX science objectives. The average scores are presented in the table.*

## SO1.3: Determine if and when a giant planet instability drove a sharp change in the inner Solar System impactor flux.

The highest priority objective of any mission to SPA should be establishing the shape and timing of the early impact flux, including the timing of key events in the formation and evolution of SPA and whether there was a spike in impactor population. This information has applications for understanding processes across the early Solar System, and also forms the basis for interpreting surface ages across the Solar System. Due to the age of these events, and the significant resurfacing that has occurred on most bodies since that time, SPA is perhaps the only place in the Solar System where this objective can be accomplished. The preeminence of this objective is central to the priority of Goal 1.

## SO1.1: Quantify the age of the Moon to distinguish between early- and late-Moon formation models.

The second highest priority objective for any mission to SPA should be acquiring samples to further constrain the age of the Moon. This information is central to understanding models of lunar formation and evolution, as well as, constructing the history of the early Earth. The importance of this scientific objective is also reflected in the science goals and objectives for the first Artemis crewed landing. However, the SDT notes that measurements of the age of the Moon require very specific sample conditions which cannot be verified by *in situ* analysis prior to collection. Therefore, it is strongly in the interest of all missions to ancient lunar regions to prioritize collection of viable samples with the goal of acquiring sufficient measurements to fully constrain the lunar formation age.

#### SO2.2: Determine if there was a farside KREEP reservoir, or other late-stage magma ocean products, prior to the SPA impact.

The highest priority of the Goal 2 objectives, and the highest priority compositional objective for any mission to the lunar farside, is to understand whether or not the farside mantle produced late-stage, incompatible-rich magma ocean products, such as KREEP. This objective is considered to be the most significant because its presence or absence would directly reflect large-scale differences in solidification and crustal formation processes between the lunar nearside and farside. Because of the availability of Apollo samples, the associated LMO hypothesis has become the archetype through which all planetary crustal formation processes are understood. Therefore, it is imperative for any mission to SPA to interrogate the nature and origin of these critical materials.

### 3.4.4 Future Environment for Science Prioritization

The SPARX SDT recognizes that the current lunar exploration environment is particularly dynamic, and this is reflected in several of the prioritization matrix criteria. The SDT has developed the Science Goals and Objectives to best reflect the science priorities laid out by the OWL Decadal survey, various lunar science guiding documents, and the currently understood exploration environment as of early 2026 (Section 3.5). **The unique combination of exploration destination factors presented by SPA, and the incorporation of those factors into the SPARX objectives, makes them relatively robust to increased lunar exploration. That is to say, the SPARX objectives have been particularly crafted to target the highest-priority lunar and planetary science objectives that are only achievable through sample return from regions within SPA.** Nevertheless, the SDT recognizes that given the proposed Artemis landing regions and lunar basecamp locations in the lunar South polar region, certain objectives could be contributed to prior-to, or in-conjunction-with, a future SPARX mission. For these particular cases, we direct future investigators to the Science Objective constituent Experiments (Section 4.1.1–4.1.14), and the requirements described therein for appropriate geologic settings, sample masses, and analyses for addressing the proposed Objectives. Regardless of potential synergies with future Artemis investigations, the goals and objectives outlined by SPARX, with a particular emphasis on the results from returned samples, will contribute significant progress on multiple Decadal-level science questions, and will result in one of the most scientifically wide-reaching investigations achievable in this decade.

The SDT prioritized science investigations that were specific to SPA, and planetary science prioritized by the Planetary Science and Astrobiology Decadal Survey. Consequently, the SDT did not prioritize planetary science investigations that could be accomplished at other locations on the Moon (i.e., site agnostic). For example, the SDT did not prioritize investigations into space weathering, the formation of lunar regolith, or the space environment. Similarly, the SDT did not prioritize science investigations relevant to other NASA science constituents, such as astrophysics, physical sciences, and human/biological sciences. A future SPARX mission could conceivably be relevant to these diverse stakeholders, and the SDT would recommend early discussion of these contributions if that were desired.

**Finding: The SDT finds that the highest priority science goals of a South Pole-Aitken basin mission that includes sample return (e.g., SPARX) are to constrain the chronology of key processes in the earliest period of Solar System formation, to constrain and characterize the lunar crustal asymmetry as a mirror into planetary crustal formation processes, and to characterize the physical and compositional effects of mega-basin impacts as early drivers of planetary evolution.**

**Finding: The SDT reaffirms the findings of the Origins, Worlds, and Life Planetary Science Decadal Survey in establishing sample return from the Moon's South Pole-Aitken basin as the highest priority lunar science mission resulting from the preponderance of Decadal Level Science Questions addressed by the mission, and the Solar System-wide implications of the proposed mission science goals.**

**Finding: The SDT further emphasizes that the Moon's South Pole-Aitken basin is a geologically special region stemming from its location, size, and antiquity. As a result, the critical science questions identified here cannot be addressed by other missions either throughout the Solar System or in other non-SPA regions of the Moon.**

## 3.5 TRACEABILITY OF SPARX SCIENCE TO COMMUNITY DOCUMENTS

The SPARX Science Goals and Objectives are informed by, and trace directly to multiple guiding documents assembled by the United States planetary science, and lunar science communities. Those documents include the National Academies of Sciences, Engineering, and Medicine “Origins, Worlds, and Life: A Decadal Strategy for Planetary Science and Astrobiology 2023–2032” (the OWL Decadal Survey), the National Research Council “Scientific Context for the Exploration of the Moon” (SCEM), NASA Artemis III Science Definition Team Report, NASA Moon to Mars Objectives (2023), Lunar Exploration and Analysis Group Lunar Science Goals “Lunar Exploration Roadmap” (LER), and other documents, such as the annual 2025 NASA Moon to Mars Architecture Definition Document (ADD).

The relevance for each of the goals and objectives to each of these documents is shown below in Foldout 3.1 to NASA/Community Goals.

## 3.6 SPARX SCIENCE TRACEABILITY MATRIX (STM)

The SPARX Science Traceability Matrix (STM) is shown in Foldout 3.2.

| GOALS | | OBJECTIVES | | "Origins, Worlds, and Life" Planetary Science and Astrobiology Decadal Survey | Scientific Context for the Exploration of the Moon (SCEM) | Artemis III Science Definition Team Report | NASA Moon-to-Mars Objectives | LEAG Lunar Science Goals (Lunar Exploration Roadmap) | Other |
|---|---|---|---|---|---|---|---|---|---|
| (G1) | **What was the chronology of key processes and events during the first half billion years of solar system history?** | (O1.1) | **Quantify the age of the Moon to distinguish between Early- and Late-Moon formation models** | **Q3:** How and when did the terrestrial planets, their moons, and the asteroids accrete, what processes determined their initial properties, and to what extent were outer solar system materials incorporated? | **SC2. The structure and composition of the lunar interior provide fundamental information on the evolution of a differentiated planetary body.** | Understand the timing of the collision between the impactor and the proto Earth | **LPS-1(LM):** Uncover the record of solar system origin and early history, by determining how and **when** planetary bodies **formed** | **Investigation-Sci-A-9A:** Determine the bulk composition of the crust.<br>**Investigation-Sci-A-9B:** Determine the bulk composition of the mantle.<br>**Investigation-Sci-A-9C:** Determine the early thermal history of the Moon. | |
| | | (O1.2) | **Distinguish between fast- and slow-cooling models for lunar magma ocean crystallization** | **Q5:** How did the interiors of solid bodies evolve, how is this evolution recorded in a body's physical and chemical properties, and how are solid surface shaped by internal and external processes? | **2b.** Characterize the chemical/physical stratification in the mantle, particularly the nature of the putative 500-km discontinuity and the composition of the lower mantle.<br>**3a.** Determine the extent and composition of the primary feldspathic crust, KREEP layer, and other products of planetary differentiation.<br>**2d.** Characterize the thermal state of the interior and elucidate the workings of the planetary heat engine. | **1a-2.** Constrain the physicochemical conditions and processes that operated at the surface of the lunar magma ocean<br>**1b-1.** Determine the extent and composition of the primary feldspathic crust, KREEP layer, and other products of planetary differentiation<br>**1b-2.** Determine the bulk composition of the crust and mantle<br>Characterize the chemical/physical stratification and lateral heterogeneity in the mantle<br>**1b-3.** Inventory, relationships, and ages of nonmare rocks. | **LPS-1(LM):** Uncover the record of solar system origin and early history, by determining **how** planetary bodies **differentiated** | **Investigation-Sci-A-5A:** Inventory, relationships, and ages of nonmare rocks.<br>**Investigation-Sci-A-5C:** Determine the composition, structure, and variability of the crust.<br>**Investigation-Sci-A-5D:** Determine the composition, structure, and variability of the mantle.<br>**Investigation-Sci-A-9A:** Determine the bulk composition of the crust.<br>**Investigation-Sci-A-9B:** Determine the bulk composition of the mantle.<br>**Investigation-Sci-A-9C:** Determine the early thermal history of the Moon. | |
| | | (O1.3) | **Determine if and when a giant planet instability drove a sharp change in the inner Solar System impactor flux** | **Q4:** How have the population of solar system bodies changed through time, how has bombardment varied across the solar system, and how have collisions affected the evolution of planetary bodies? **Q2:** How and when did the giant planet systems originate... [and] did their orbits migrate? **Q12:** What do the planets, moons, and rings in the solar system tell us about circumstellar disks and exoplanetary systems? | **SC1: The bombardment of history of the inner solar system is uniquely revealed on the Moon.**<br>**1a.** Test the cataclysm hypothesis by determining the spacing in time of the lunar basins.<br>**1b.** Anchor the early Earth-Moon impact flux curve by determining the age of the oldest lunar basin (South Pole-Aitken Basin).<br>**Suggested Implementation:** Sample return from the SPA Basin melt sheet and from floors or ejecta of basins within the SPA Basin for detailed geochemical and isotopic analyses. | Anchor the earliest recorded impact history of the Moon by determining the age of the oldest lunar basin, South Pole-Aitken<br>**3a-1.** Identify impact melt, impact ejecta, and exogenous (impactor) material in lunar samples to address the hypothesized Lunar Cataclysm<br>**3c-1.** Determine the sequence of individual craters and basins that influence local, regional, and global stratigraphy at the Artemis III landing site. | **LPS-1(LM):** Uncover the record of solar system origin and early history, by... characterizing the impact chronology of the inner solar system as recorded on the Moon and Mars, and characterize how impact rates in the inner solar system have changed over time as recorded on the Moon and Mars. | **Investigation-Sci-A-8A:** Understand the impact history of the Moon<br>**Investigation-Sci-B-1A:** Determine the impact flux during the basin-forming epoch. | |
| | | (O1.4) | **Constrain the evolution of the lunar dynamo in the pre-4.0 Ga epoch of lunar history** | **Q5:** How did the interiors of solid bodies evolve, how is this evolution recorded in a body's physical and chemical properties, and how are solid surface shaped by internal and external processes? | **2d.** Characterize the thermal state of the interior and elucidate the workings of the planetary heat engine.<br>"The existence of an early dynamo, which might leave evidence in the remanent magnetization of igneous rocks, would also constrain the thermal history through inferences of the timing of formation and maintenance of a conducting core." | Understand the history of the lunar magnetic field—and the process(es) that generated it—through paleointensity and paleopole determinations | Advance understanding of the geologic processes that affect planetary bodies by **determining the interior structures**, characterizing the magmatic histories, characterizing ancient, modern, and evolution of atmospheres/exospheres, and **investigating how active processes modify the surfaces of the Moon** and Mars. | **Investigation-Sci-C-2B:** Map the Moon's Remanent Crustal Magnetic Fields. | |
| (G2) | **Does the lunar farside preserve evidence of asymmetric magma ocean evolution and planetary crustal formation processes?** | (O2.1) | **Determine if the farside highlands crust is similar to nearside crust in age and/or composition** | **Q5:** How did the interiors of solid bodies evolve, how is this evolution recorded in a body's physical and chemical properties, and how are solid surface shaped by internal and external processes? | **3a.** Determine the extent and composition of the primary feldspathic crust, KREEP layer, and other products of planetary differentiation.<br>**Suggested Implementation:** Sample return from major lunar terranes (e.g., feldspathic highlands, SPA Basin, PKT) for detailed geochemical and isotopic analysis. | **1b-1.** Determine the extent and composition of the primary feldspathic crust, KREEP layer, and other products of planetary differentiation<br>Inventory the variety, age, distribution, and origin of lunar rock types<br>Determine the composition, structure, and lateral variability of the crust on global scales.<br>**1b-2.** Determine the bulk composition of the crust and mantle<br>**1b-3.** Inventory, relationships, and ages of nonmare rocks. | **LPS-1(LM):** Uncover the record of solar system origin and early history, by determining how planetary bodies differentiated | **Investigation-Sci-A-5A:** Inventory, relationships, and ages of nonmare rocks.<br>**Investigation-Sci-A-5C:** Determine the composition, structure, and variability of the crust.<br>**Investigation-Sci-A-9A:** Determine the bulk composition of the crust.<br>**Investigation-Sci-A-9C:** Determine the early thermal history of the Moon. | **Investigation-Sci-A-2A:** Develop a sampling strategy for the Moon.<br>**LER Investigation-Sci-A-2B:** Understand the scientific requirements for sample curation, packaging, and transport to Earth.<br>**Investigation-FF-A-3A:** Test surface mobility systems.<br>**Investigation-FF-A-3C:** Test robotic field assistant technologies.<br>**Investigation-FF-A-8B:** Test technologies for navigating on the lunar surface without a strong magnetic field.<br>**Initiative-Sust-B-5B:** Utilize robotic precursor missions to conduct selected lunar science investigations that focus on characterizing the pristine lunar environment and contribute to identifying lunar resource concentrations.<br>**Initiative-Sust-B-7A:** Provide long-range robotic and pressurized rovers for distances of up to 100 kilometers. |
| | | (O2.2) | **Determine if there was a farside KREEP reservoir, or other late stage magma ocean products, prior to the SPA impact** | **Q5:** How did the interiors of solid bodies evolve, how is this evolution recorded in a body's physical and chemical properties, and how are solid surface shaped by internal and external processes? | **3a.** Determine the extent and composition of the primary feldspathic crust, KREEP layer, and other products of planetary differentiation.<br>**Suggested Implementation:** Sample return from major lunar terranes (e.g., feldspathic highlands, SPA Basin, PKT) for detailed geochemical and isotopic analysis.<br>**3c.** Determine the composition of the lower crust and bulk Moon. | **1b-1.** Determine the extent and composition of the primary feldspathic crust, KREEP layer, and other products of planetary differentiation<br>Inventory the variety, age, distribution, and origin of lunar rock types<br>**1b-2.** Determine the bulk composition of the crust and mantle<br>**1b-3.** Inventory, relationships, and ages of nonmare rocks.<br>Characterize the chemical/physical stratification and lateral heterogeneity in the mantle | **LPS-2(LM):** Advance understanding of the geologic processes that affect planetary bodies by **determining the interior structures, [and] characterizing the magmatic histories.** | **Investigation-Sci-A-5A:** Inventory, relationships, and ages of nonmare rocks.<br>**Investigation-Sci-A-5B:** Inventory, relationships, and ages of mare volcanics and related intrusive rocks.<br>**Investigation-Sci-A-5D:** Determine the composition, structure, and variability of the mantle.<br>**Investigation-Sci-A-9B:** Determine the bulk composition of the mantle.<br>**Investigation-Sci-A-9C:** Determine the early thermal history of the Moon. | |
| | | (O2.3) | **Determine if and how the temporal and compositional evolution of the farside mantle differs from the nearside** | **Q5:** How did the interiors of solid bodies evolve, how is this evolution recorded in a body's physical and chemical properties, and how are solid surface shaped by internal and external processes? | **2b.** Characterize the chemical/physical stratification in the mantle, particularly the nature of the putative 500-km discontinuity and the composition of the lower mantle<br>**Suggested Implementation:** Return of samples representative of the upper mantle (e.g., SPA Basin) provides a geochemical framework for the interpretation of seismic and magnetic sounding data.<br>**3b.** Inventory the variety, age, distribution, and origin of lunar rock types.<br>**Suggested Implementation:** Sample returns from representative and previously unsampled locations (selected from global compositional maps).<br>**3c.** Determine the composition of the lower crust and bulk Moon.<br>**3d.** Quantify the local and regional complexity of the current lunar crust. | **1b-1.** Determine the extent and composition of the primary feldspathic crust, KREEP layer, and other products of planetary differentiation<br>Inventory the variety, age, distribution, and origin of lunar rock types<br>**1b-2.** Determine the bulk composition of the crust and mantle<br>**1b-3.** Inventory, relationships, and ages of nonmare rocks.<br>Characterize the chemical/physical stratification and lateral heterogeneity in the mantle | **LPS-1(LM):** Uncover the record of solar system origin and early history, by determining **how** planetary bodies **differentiated.**<br>**LPS-2(LM):** Advance understanding of the geologic processes that affect planetary bodies by **determining the interior structures [and] characterizing the magmatic histories.** | **Investigation-Sci-A-5A:** Inventory, relationships, and ages of nonmare rocks.<br>**Investigation-Sci-A-5B:** Inventory, relationships, and ages of mare volcanics and related intrusive rocks.<br>**Investigation-Sci-A-5D:** Determine the composition, structure, and variability of the mantle.<br>**Investigation-Sci-A-9B:** Determine the bulk composition of the mantle.<br>**Investigation-Sci-A-9C:** Determine the early thermal history of the Moon. | |
| (G3) | **What are the effects of mega-basin impacts on terrestrial bodies?** | (O3.1) | **Determine whether there are discontinuities in the structure (crustal thickness, density, magnetic field) across SPA to constrain models for the formation of a mega-basin** | **Q4:** How have the population of solar system bodies changed through time, how has bombardment varied across the solar system, and how have collisions affected the evolution of planetary bodies? **Q5:** How did the interiors of solid bodies evolve, how is this evolution recorded in a body's physical and chemical properties, and how are solid surface shaped by internal and external processes? | **2a.** Determine the thickness of the lunar crust (upper and lower) and characterize its lateral variability on regional and global scales.<br>**Suggested Implementation:** Samples from previously unsampled regions to provide constraints on lateral variability of crustal composition and density.<br>**3d.** Quantify the local and regional complexity of the current lunar crust. | Determine the composition, structure, and lateral variability of the crust on global scales.<br>Determine and understand the stages of formation of simple and complex craters, and multi-ring basins.<br>Determine how impacts modify, redistribute, and mix materials.<br>Determine the origin, extent, and differentiation/evolution of basin melt sheets. | **LPS-2(LM):** Advance understanding of the geologic processes that affect planetary bodies by **determining [their] interior structures.** | **Investigation-Sci-A-5C:** Determine the composition, structure, and variability of the crust.<br>**Investigation-Sci-A-7A:** Determine and understand the stages of formation of simple and complex craters, and multi-ring basins.<br>**Investigation-Sci-A-7C:** Determine the origin and evolution of basin melt sheets.<br>**Investigation-Sci-A-7D:** Assess the possibility of impact-triggered magmatism.<br>**Investigation-Sci-A-8A:** Understand the impact history of the Moon<br>**Investigation-Sci-A-8C:** Determine the stratigraphy of the lunar highlands. | |
| | | (O3.2) | **Distinguish the relative contributions of impact melt differentiation, crustal backfilling, and volcanic infilling to the interior composition and stratigraphy of SPA** | **Q5:** How did the interiors of solid bodies evolve, how is this evolution recorded in a body's physical and chemical properties, and how are solid surface shaped by internal and external processes? | **6a.** Characterize the existence and extent of melt sheet differentiation.<br>**6b.** Determine the structure of multi-ring impact basins.<br>**6d.** Measure the extent of lateral and vertical mixing of local and ejecta material. | Inventory the variety, age, distribution, and origin of lunar rock types<br>Determine the origin, extent, and differentiation/evolution of basin melt sheets.<br>Assess the possibility of impact-triggered magmatism. | **LPS-2(LM):** Advance understanding of the geologic processes that affect planetary bodies by determining the interior structures, **characterizing the magmatic histories** and **investigating how active processes modify the surfaces of the Moon** and Mars. | **Investigation-Sci-A-5C:** Determine the composition, structure, and variability of the crust.<br>**Investigation-Sci-A-7A:** Determine and understand the stages of formation of simple and complex craters, and multi-ring basins.<br>**Investigation-Sci-A-7C:** Determine the origin and evolution of basin melt sheets.<br>**Investigation-Sci-A-7D:** Assess the possibility of impact-triggered magmatism.<br>**Investigation-Sci-A-8A:** Understand the impact history of the Moon | |

**Foldout 3.1.** Traceability of SPARX science goals and objectives to community documents.

| SCIENCE GOALS, OBJECTIVES, AND EXPERIMENTS | | | Measurements, Site, and Sample Requirements | | | | |
|---|---|---|---|---|---|---|---|
| **GOALS** | **OBJECTIVES** | **EXPERIMENTS** | **SITE & TRAVERSE REQUIREMENTS*** | **IN SITU MEASUREMENT REQUIREMENTS †^** | **Required Sample Components ‡** | **Sample Lithology Specification ‡** | **Laboratory Analyses of Returned Samples ‡** |
| (G1) **What was the chronology of key processes and events during the first half billion years of solar system history?** | (O1.1) **Quantify the age of the Moon to distinguish between early- and late-Moon formation models** | (E1) **Determine the composition and age of the anorthositic farside highlands crust** | Sample ≥3 (≥[1]) geologically distinct exposures of farside anorthisitic highlands material | **M1:** Context Imaging<br>**M2:** Context Mineralogy<br>**M3:** Fine-Scale Mineralogy<br>**M4:** Bulk Elemental Chemistry<br>*M6: Fine-Scale Imaging* | **Rocks (≥40 mm)**<br>**Rocklets (≥8 mm)**<br>**Bulk regolith** | **Anorthosite**<br>**No requirement** | **Geochronology:** Determine the age and isotopic composition of anorthositic highlands crust utilizing the 147Sm-143Nd isochron system and ≥[2] additional independent geochronometer systems, including Rb-Sr, 146Sm-142Nd, U-Pb, Pb-Pb, Lu-Hf, and Ar-Ar<br>**Petrologic characterization** of returned samples |
| | | (E2) **Determine the age of SPA impact event** | Sample ≥[3] geologically distinct units hypothesized to be excavated or deposited SPA impact melt | **M1:** Context Imaging<br>**M2:** Context Mineralogy<br>**M3:** Fine-Scale Mineralogy<br>**M4:** Bulk Elemental Chemistry<br>*M6: Fine-Scale Imaging* | **Rocklets (≥8 mm)**<br>**Rocklets (≥3 mm)**<br>**Bulk regolith** | **Impact melt**<br>**No requirement** | **Geochronology:** Determine the age of SPA impact melt utilizing the Ar-Ar ± the 147Sm-143Nd isochron system, and ≥[1] additional independent geochronometer system, including Rb-Sr, U-Pb, Pb-Pb, 146Sm-142Nd, Lu-Hf<br>**Petrologic characterization** of returned samples |
| | | (E3) **Determine the model age of the mantle underlying SPA** | Sample ≥3 (≥[2]) spatially and/or temporally distinct mare units as mapped from orbital datasets (e.g., Pasckert et al. 2018) | **M1:** Context Imaging | **Rocklets (≥5 mm)**<br>**Rocklets (≥3 mm)**<br>**Bulk regolith** | **Basalt**<br>**No requirement** | **Geochronology:** Determine the ages and isotopic compositions of mare basalts utilizing the 147Sm-143Nd isochron system and ≥[2] additional independent geochronometer systems, including Rb-Sr, U-Pb, Pb-Pb, Lu-Hf, and Ar-Ar, and calculate the model age of the underlying mantle source utilizing ≥[2] systems including 147Sm-143Nd, 146Sm-142Nd, Lu-Hf, and Pb-Pb<br>**Petrologic characterization** of returned samples |
| | (O1.2) **Distinguish between fast- and slow-cooling models for lunar magma ocean crystallization** | (E1) **Determine the composition and age of the anorthositic farside highlands crust** | Sample ≥3 (≥[1]) geologically distinct exposures of farside anorthisitic highlands material | **M1:** Context Imaging<br>**M2:** Context Mineralogy<br>**M3:** Fine-Scale Mineralogy<br>**M4:** Bulk Elemental Chemistry<br>*M6: Fine-Scale Imaging* | **Rocks (≥40 mm)**<br>**Rocklets (≥8 mm)**<br>**Bulk regolith** | **Anorthosite**<br>**No requirement** | **Geochronology:** Determine the age and isotopic composition of anorthositic highlands crust utilizing the 147Sm-143Nd isochron system and ≥[2] additional independent geochronometer systems, including Rb-Sr, 146Sm-142Nd, U-Pb, Pb-Pb, Lu-Hf, and Ar-Ar<br>**Petrologic characterization** of returned samples |
| | | (E2) **Determine the age of SPA impact event** | Sample ≥[3] geologically distinct units hypothesized to be excavated or deposited SPA impact melt | **M1:** Context Imaging<br>**M2:** Context Mineralogy<br>**M3:** Fine-Scale Mineralogy<br>**M4:** Bulk Elemental Chemistry<br>*M6: Fine-Scale Imaging* | **Rocklets (≥8 mm)**<br>**Rocklets (≥3 mm)**<br>**Bulk regolith** | **Impact melt**<br>**No requirement** | **Geochronology:** Determine the age of SPA impact melt utilizing the Ar-Ar ± the 147Sm-143Nd isochron system, and ≥[1] additional independent geochronometer system, including Rb-Sr, U-Pb, Pb-Pb, 146Sm-142Nd, Lu-Hf<br>**Petrologic characterization** of returned samples |
| | | (E3) **Determine the model age of the mantle underlying SPA** | Sample ≥3 (≥[2]) spatially and/or temporally distinct mare units as mapped from orbital datasets (e.g., Pasckert et al. 2018) | **M1:** Context Imaging | **Rocklets (≥5 mm)**<br>**Rocklets (≥3 mm)**<br>**Bulk regolith** | **Basalt**<br>**No requirement** | **Geochronology:** Determine the ages and isotopic compositions of mare basalts utilizing the 147Sm-143Nd isochron system and ≥[2] additional independent geochronometer systems, including Rb-Sr, U-Pb, Pb-Pb, Lu-Hf, and Ar-Ar, and calculate the model age of the underlying mantle source utilizing ≥[2] systems including 147Sm-143Nd, 146Sm-142Nd, Lu-Hf, and Pb-Pb<br>**Petrologic characterization** of returned samples |
| | (O1.3) **Determine if and when a giant planet instability drove a sharp change in the inner Solar System impactor flux** | (E2) **Determine the age of SPA impact event** | Sample ≥[3] geologically distinct units hypothesized to be excavated or deposited SPA impact melt | **M1:** Context Imaging<br>**M2:** Context Mineralogy<br>**M3:** Fine-Scale Mineralogy<br>**M4:** Bulk Elemental Chemistry<br>*M6: Fine-Scale Imaging* | **Rocklets (≥8 mm)**<br>**Rocklets (≥3 mm)**<br>**Bulk regolith** | **Impact melt**<br>**No requirement** | **Geochronology:** Determine the age of SPA impact melt utilizing the Ar-Ar ± the 147Sm-143Nd isochron system, and ≥[1] additional independent geochronometer system, including Rb-Sr, U-Pb, Pb-Pb, 146Sm-142Nd, Lu-Hf<br>**Petrologic characterization** of returned samples |
| | | (E4) **Determine the age of ≥1 farside Imbrian-aged impact crater (D≥80 km) or basin (D≥300 km) with surface dateable by crater counts** | Sample ≥1 farside imbrian-aged impact crater or basin with a dateable surface (e.g., peak ring or ejecta blanket) | **M1:** Context Imaging<br>**M2:** Context Mineralogy<br>**M4:** Bulk Elemental Chemistry<br>*M3: Fine-Scale Mineralogy*<br>*M6: Fine-Scale Imaging* | **Rocklets (≥8 mm)**<br>**Rocklets (≥3 mm)**<br>**Bulk regolith** | **Impact melt**<br>**No requirement** | **Geochronology:** Determine the age of impact melt utilizing Ar-Ar, and when sample characteristics allow, ≥[2] independent geochronometer systems, including, U-Pb, Pb-Pb, 147Sm-143Nd, Rb-Sr, and Lu-Hf<br>**Petrologic characterization** of returned samples |
| | | (E5) **Determine the age of ≥[1] non-SPA farside pre-Imbrian-aged impact basin (D≥300 km)** | For ≥[1] non-SPA farside pre-Imbrian impact basins (either Poincaré or Ingenii), sample ≥[3] geologically distinct units associated with the crater likely to retain a basin age (e.g., peak ring, rim) | **M1:** Context Imaging<br>**M2:** Context Mineralogy<br>**M4:** Bulk Elemental Chemistry<br>*M3: Fine-Scale Mineralogy*<br>*M6: Fine-Scale Imaging* | **Rocklets (≥8 mm)**<br>**Rocklets (≥3 mm)**<br>**Bulk regolith** | **Impact melt**<br>**No requirement** | **Geochronology:** Determine the age of impact melt utilizing Ar-Ar, and when sample characteristics allow, ≥[2] independent geochronometer systems, including, U-Pb, Pb-Pb, 147Sm-143Nd, Rb-Sr, and Lu-Hf<br>**Petrologic characterization** of returned samples |
| | (O1.4) **Constrain the evolution of the lunar dynamo in the pre-4.0 Ga epoch of lunar history** | (E6) **Determine the age and remanent magnetization signatures from ≥3 returned samples with inferred ages ≥4.0 Ga** | (E6) does not drive sample specific sample site selections; this experiment is completed as part of the coordinated analysis of samples collected for other experiments | *M1: Context Imaging*<br>*M2: Context Mineralogy*<br>*M3: Fine-Scale Mineralogy*<br>*M4: Bulk Elemental Chemistry*<br>*M5: Gravity Field*<br>*M6: Fine-Scale Imaging*<br>*M7: Magnetic Field* | **Does not drive sampling requirements**; this objective is completed as part of the coordinated analysis of samples collected for other objectives | | **Geochronology:** Determine the age of returned samples utilizing Ar-Ar, and, including, U-Pb, Pb-Pb, 147Sm-143Nd, Rb-Sr, and Lu-Hf. when sample characteristics allow, ≥2 independent geochronometer systems<br>**Petrologic characterization** of returned samples<br>**Determine the remanent magnetic field strength of returned samples** |
| (G2) **Does the lunar farside preserve evidence of asymmetric magma ocean evolution and planetary crustal formation processes?** | (O2.1) **Determine if the farside highlands crust is similar to nearside crust in age and/or composition** | (E1) **Determine the composition and age of the anorthositic farside highlands crust** | Sample ≥3 (≥[1]) geologically distinct exposures of farside anorthisitic highlands material | **M1:** Context Imaging<br>**M2:** Context Mineralogy<br>**M3:** Fine-Scale Mineralogy<br>**M4:** Bulk Elemental Chemistry<br>*M6: Fine-Scale Imaging* | **Rocks (≥40 mm)**<br>**Rocklets (≥8 mm)**<br>**Bulk regolith** | **Anorthosite**<br>**No requirement** | **Geochronology:** Determine the age and isotopic composition of anorthositic highlands crust utilizing the 147Sm-143Nd isochron system and ≥[2] additional independent geochronometer systems, including Rb-Sr, 146Sm-142Nd, U-Pb, Pb-Pb, Lu-Hf, and Ar-Ar<br>**Petrologic characterization** of returned samples |
| | (O2.2) **Determine if there was a farside KREEP reservoir, or other late stage magma ocean products, prior to the SPA impact** | (E7) **Determine the composition and age of farside Th-rich materials** | Sample ≥2 (≥[1]) geographically distinct Th-rich regions identified by preexisting orbital datasets to have Th-abundances ≥3 ppm | **M1:** Context Imaging<br>**M2:** Context Mineralogy<br>**M4:** Bulk Elemental Chemistry | **Rocklets (≥8 mm)**<br>**Rocklets (≥3 mm)**<br>**Bulk regolith** | **No pre-defined lithology** (the Th-carrier is unknown from orbital data) | **Geochronology:** Determine the age of returned samples utilizing Ar-Ar, and when sample characteristics allow, ≥2 independent geochronometer systems, including, U-Pb, Pb-Pb, 147Sm-143Nd, 146Sm-142Nd, Rb-Sr, and Lu-Hf<br>**Petrologic characterization** of returned samples<br>**Elemental chemistry** including major elements (Fe, Mg, Ca, K, Si, Ti, Al, Cr, Mn) and trace elements (Th, U, Hf, Zr, Sc, Rb, Sr, Pb, K, P, W, and rare-Earth elements) |
| | (O2.3) **Determine if and how the temporal and compositional evolution of the farside mantle differs from the nearside** | (E8) **Determine the composition and age of farside basaltic units** | Same as (E3); no additional sites are required | **M1:** Context Imaging | **Returned sample requirements are the same as (E3)**; no additional samples are required | | **Geochronology:** Determine the ages and isotopic compositions of mare basalts utilizing the 147Sm-143Nd isochron system and ≥[2] additional independent geochronometer systems, including Rb-Sr, U-Pb, Pb-Pb, Lu-Hf, and Ar-Ar, and calculate the model age of the underlying mantle source utilizing ≥[2] systems including 147Sm-143Nd, 146Sm-142Nd, Lu-Hf, and U-Pb<br>**Petrologic characterization** of returned samples<br>**Elemental chemistry** including major elements (Fe, Mg, Ca, K, Si, Ti, Al, Cr, Mn) and trace elements (Th, U, Hf, Zr, Sc, Rb, Sr, Pb, K, P, and rare-Earth elements) |
| | | (E9) **Sample hypothesized excavated lunar mantle material** | Sample ≥2 geologic unit that is hypothesized to be excavated mantle, including hypothesized low-calcium pyroxene and olivine exposures proposed to trace mantle ejecta | **M1:** Context Imaging<br>**M2:** Context Mineralogy<br>*M4: Bulk Elemental Chemistry* | **Rocks (≥20 mm)**<br>**Rocklets (≥3 mm)**<br>**Bulk regolith** | **Course grained, low-Ca pyroxene and/or olivine**<br>**No requirement** | **Geochronology:** Determine the ages and isotopic compositions of returned samples utilizing Ar-Ar, and when sample characteristics allow, ≥2 independent geochronometer systems, including, U-Pb, Pb-Pb, 147Sm-143Nd, 146Sm-142Nd, Rb-Sr, and Lu-Hf<br>**Petrologic characterization** of returned samples<br>**Elemental chemistry** including major elements (Fe, Mg, Ca, K, Si, Ti, Al, Cr, Mn) and trace elements (Th, U, Hf, Zr, Sc, Rb, Sr, Pb, K, P, and rare-Earth elements) |
| | | (E10) **Determine the volatile content of the farside mantle as recorded by pyroclastic and basaltic materials** | Basaltic materials are collected as part of (E3); no additional sites are required<br>Sample ≥1 pyroclastic deposit | **M1:** Context Imaging<br>*M2: Context Mineralogy*<br>*M4: Bulk Elemental Chemistry* | **Mare basalts are sampled as part of (E3)**; no additional mare basalt samples are required<br>**Bulk regolith** | **Pyroclastic material** | **Geochronology:** Determine the cosmic ray exposure age to ±10 Myr<br>**Petrologic characterization** of returned samples<br>**Elemental chemistry** including major elements (Fe, Mg, Ca, K, Si, Ti, Al, Cr, Mn) and trace elements (Ni, Co, highly-siderophile elements)<br>**Volatile abundance** (H2O, CO, Cl, F, S) |
| (G3) **What are the effects of mega-basin impacts on terrestrial bodies?** | (O3.1) **Determine whether there are discontinuities in the structure (crustal thickness, density, magnetic field) across SPA to constrain models for the formation of a mega-basin** | (E11) **Determine the crustal structure (e.g., thickness, density, remanent magnetic field strength) across the SPA basin (from the center to the putative rim) to constrain mega-basin formation and evolution models** | Perform a traverse of SPA that spans from central SPA out radially ≥[650] km towards the hypothesized rim region | **M1:** Context Imaging<br>**M2:** Context Mineralogy<br>**M5:** Gravity Field<br>*M7: Magnetic Field* | **None**; this experiment does not require returned samples | | **None**; this experiment does not require returned samples |
| | | (E12) **Determine if the massifs in the southwestern portion of the basin constitute the diffuse rim of SPA** | Traverse ≥[1] km in elevation up a hypothesized SPA rim massif | **M5:** Gravity Field | **None**; this experiment does not require returned samples | | **None**; this experiment does not require returned samples |
| | (O3.2) **Distinguish the relative contributions of impact melt differentiation, crustal backfilling, and volcanic infilling to the interior composition and stratigraphy of SPA** | (E13) **Determine the stratigraphy and composition of the unique deposits in central SPA** | Sample the geologic diversity of central SPA including: ≥[1] sample of low-calcium pyroxene material excavated by Bose crate; ≥[1] sample of high-calcium pyroxene material excavated by Bhaba crater; ≥1 sample from the ejecta of a morphologically fresh ≥10-km diameter crater with a well-defined ejecta blanket (plausibly excavating through mare basalts); ≥1 sample of light plains material in central SPA | **M1:** Context Imaging<br>**M2:** Context Mineralogy<br>**M4:** Bulk Elemental Chemistry | **Rocklets (≥8 mm)**<br>**Rocklets (≥3 mm)**<br>**Bulk regolith** | **Site dependent** (including impact melt and high- and low-Ca pyroxene)<br>**No requirement** | **Geochronology:** Determine the ages and isotopic compositions of returned samples utilizing Ar-Ar, and when sample characteristics allow, ≥[2] independent geochronometer systems, including, U-Pb, Pb-Pb, 147Sm-143Nd, 146Sm-142Nd, Rb-Sr, and Lu-Hf<br>**Petrologic characterization** of returned samples<br>**Elemental chemistry** including major elements (Fe, Mg, Ca, K, Si, Ti, Al, Cr, Mn) and trace elements (Th, U, Hf, Zr, Sc, Rb, Sr, Pb, K, P, W, and rare-Earth elements) |
| | | (E14) **Determine the origin and extent of the low-calcium pyroxene annulus enclosing central SPA** | Acquire ≥1 sample from the low-calcium pyroxene annulus enclosing central SPA, but excluding mare basalts and non-SPA impact basin material | **M1:** Context Imaging<br>**M2:** Context Mineralogy<br>**M4:** Bulk Elemental Chemistry | **Rocklets (≥8 mm)**<br>**Rocklets (≥3 mm)**<br>**Bulk regolith** | **Low-Ca pyroxene**<br>**No requirement** | **Geochronology:** Determine the ages and isotopic compositions of returned samples utilizing Ar-Ar, and when sample characteristics allow, ≥[2] independent geochronometer systems, including, U-Pb, Pb-Pb, 147Sm-143Nd, 146Sm-142Nd, Rb-Sr, and Lu-Hf<br>**Petrologic characterization** of returned samples<br>**Elemental chemistry** including major elements (Fe, Mg, Ca, K, Si, Ti, Al, Cr, Mn) and trace elements (Th, U, Hf, Zr, Sc, Rb, Sr, Pb, K, P, W, and rare-Earth elements) |

**Foldout 3.2.** SPARX Science Traceability Matrix (STM). Objectives and experiments shown in grey are considered baseline-only (i.e., not in the threshold mission). Bolded measurements are required, whereas italicized measurements provide additional but non-essential context. *Full traceability of Site and Traverse Requirements can be found in Foldout 5.1. †Full description of In-Situ Measurement Requirements can be found in Foldout 4.1. ‡Full traceability of Sample Requirements can be fud in Foldout 4.2.

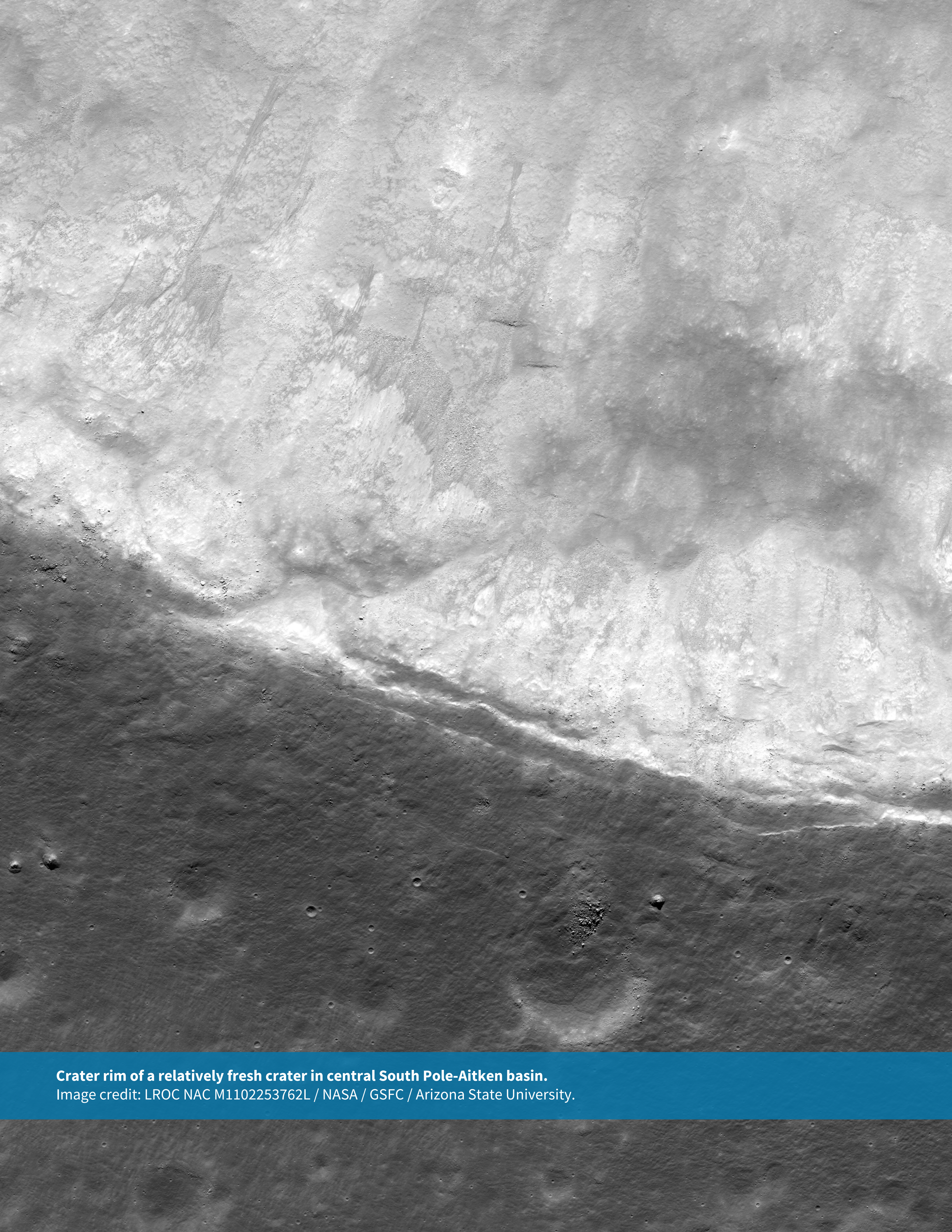

**Crater rim of a relatively fresh crater in central South Pole-Aitken basin.**
Image credit: LROC NAC M1102253762L / NASA / GSFC / Arizona State University.

# 4 SCIENCE REQUIREMENTS

## 4.1 EXPERIMENTS

In the course of the SPARX meeting discussions, there rose a need for an additional scientific narrative link between the standard STM columns of Science Objectives and Physical Measurements. As a result, the SPARX SDT has elected to add a column to the STM titled "Experiments" as a link between the Science Objectives and the Measurements. The rationale for adding this column is two-fold, relating both to the nature of the Science Objectives and the inherent measurement complexities of a sample return mission.

Turning first to the scale of the mission Science Objectives, many objectives require multiple lines of evidence to resolve, which is expected given the size and scope of the mission. For example, O1.1 "Determine the age of the Moon" cannot be achieved by a single measurement or technique, instead it requires knowledge of the timing of the formation of the major lunar constituents (crust and mantle), and also the cooling history that resulted in those ages. From these inputs, the age of lunar formation can then be modeled. Experiments 1–3 are designed to produce the inputs which can then be used to understand the age of the Moon forming event.

The second motivation for the introduction of Experiments is to provide a clear link between the multiple types of measurements (*in situ*, terrestrial laboratory) and requirements (sample type, size, location) in a way that is both clear to future mission science teams, and reflects the unique requirements of each investigation when they might otherwise become opaque. Following the example from above with O1.1 "Determine the age of the Moon", the required terrestrial laboratory measurements for all of the constituent investigations (formation age of crust and mantle, and cooling history) are identical: petrologic characterization of the samples, and determination of ages using high-precision geochronometers. However, the sample lithologies, locations, and masses required for each of these constituents are quite different. By treating each sub-investigation as a discrete Experiment, the STM allows fuller traceability between the practical mission implementation phases and measurements, and the higher-level mission Science Objectives.

The 14 Experiments are detailed in the following sections, including a background description of the Experiment and how it traces to its Science Objective, followed by discussion of the associated physical sample requirements, terrestrial laboratory measurements, sample location and traverse requirements, and requirements for *in situ* measurements. The Experiment descriptions are designed to be self-contained, and therefore necessary rationale or descriptions might be repeated as needed for each individual Experiment. Experiments referenced within the text will use the notation of "E" followed by the experiment number, e.g., E1, E2. Each experiment description includes an assessment of the sample requirements (number and size fraction) required to address the experiment (4.1.x.2), as well as, a total baseline sampling requirement for each sampling location, which includes NASA curation requirements, and should be used to guide mission operations (4.1.x.3). Each experiment also includes a figure of the broad SPA region and locations within the basin where the Experiment might be accomplished/acquire the required samples. The term "context measurements" or "geologic context" refers to a combination of Context Imaging (M1), Context Mineralogy (M2), and M4 (Bulk Elemental Chemistry) acquired for a target region. A full discussion of the required *in situ* measurement capabilities can be found in Section 4.2, and a full discussion of the terrestrial laboratory investigation techniques in Section 4.3

### 4.1.1 Experiment 1: Determine the Composition and Age of the Anorthositic Farside Highlands Crust

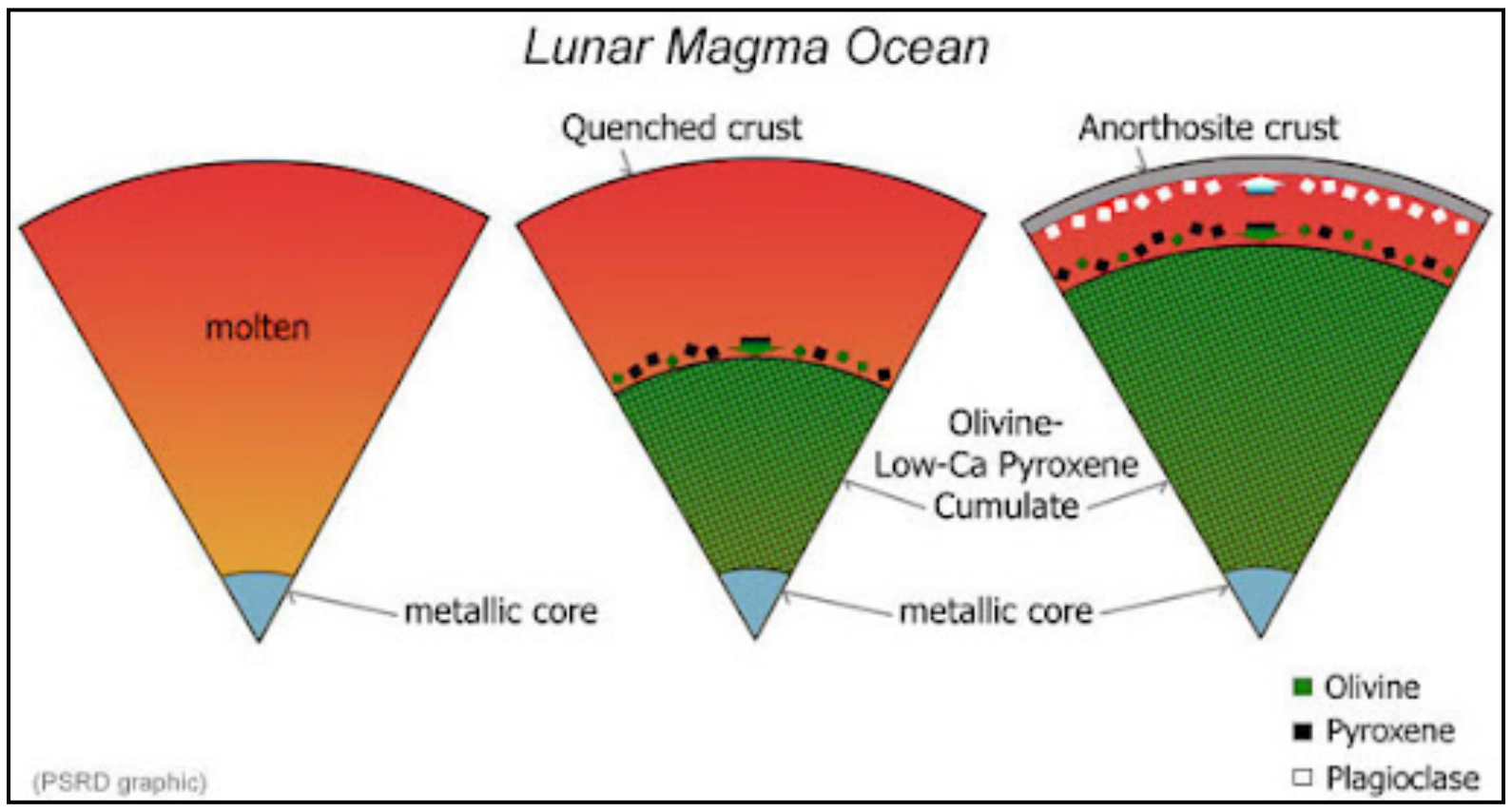

**Figure 4.1.** *Schematic depiction of the canonical lunar magma ocean (LMO) hypothesis. (Figure credit: Taylor, 2011).*

Most theories for the formation of the Moon involve the cooling of the Moon from an initially molten state—a magma ocean. As this molten moon cooled, different materials froze out of the magma ocean and either floated or sank based on their densities (e.g., Wood et al., 1970; Smith et al., 1970) (Figure 4.1) (Section 3.1, 3.2). The Moon's initial crust crystallized from these materials and was dominated by low density plagioclase minerals forming the rock anorthosite. Because of its important role in the Moon's early formation, anorthositic material has long been a high-priority target for lunar sample return; however, this rock type is underrepresented in existing sample collections relative to their global abundance. This underrepresentation owes to a variety of factors, including the locations of past sample return sites (which generally have not landed in anorthosite-rich terranes), and because subsequent geologic processes have broken down, mixed, and altered the initially pristine anorthositic crust. The lunar farside retains some of the most expansive and pristine anorthositic terranes.

#### 4.1.1.1 E1: Science Objective Traceability

Determining the age and composition of the anorthositic farside highlands crust is directly linked to three of the SPARX mission objectives, including Objectives 1.1, 1.2, and 2.1. All three of these objectives can be addressed by studying anorthosite.

**Objective 1.1** seeks to determine whether the Moon formed early or late. The crystallization and floatation of the anorthosite from the magma ocean is a key process in early lunar formation, and the resulting primary floatation crust provides the earliest dateable surface on the Moon.

**Objective 1.2** is to determine whether the Moon cooled slow or fast. Following its formation, the Moon may have experienced a range of potential cooling and differentiation events. These could have included an early series of differentiation events, a younger series including anorthosite formation due to thermal conditions such as a prolonged magma ocean stage, or even a younger thermal event such as tidal heating, higher impact flux, or a single impact (Maurice et al., 2020; Borg et al., 2015; Nimmo et al., 2024; Hartmann et al., 2000; Joy et al., 2025) any of which would produce differences in the ranges of recorded crustal ages and compositions.

**Objective 2.1** seeks to determine the age and composition of the farside crust in comparison to the nearside. Our current understanding of lunar crust formation is based on nearside anorthositic rocks collected by the Apollo, Luna, and Chang'e missions (e.g., Gaffney et al., 2023; Borg and Carlson, 2023). Evaluation of crust formation is thus possibly biased by the nearside sampling, and having farside anorthosite will enable comparisons to be made to the nearside units and allow a more global understanding rather than the potentially localized understanding provided by the nearside units.

#### 4.1.1.2 E1: Sample Requirements

Anorthosites have historically been among the most challenging lunar rocks to date due to several factors (Norman et al., 2003; Gaffney et al., 2011; Borg et al., 2015). Any ancient rock is vulnerable to younger secondary processes that may have reset or affected the radioisotope pairs used for chronology (Nyquist et al., 2006; McCallum et al., 2006). In addition, anorthosite can be a nearly monomineralic (i.e., almost entirely plagioclase feldspar) rock that provides a smaller spread of parent/daughter ratios useful for determining ages (Borg et al., 2011). The concentrations of parent and daughter isotopes utilized for dating can be very small, thus requiring more mass and larger samples than required for many other geochronometers (Borg et al., 2015). As a result of these factors, the *sampling requirements are quite specific and customized to provide the optimal sample to meet the science objectives*, falling into three categories.

First, **this experiment requires sampling larger rocks**. Larger samples, **at least 40 mm in diameter**, are required—as these rocks provide the minimum sufficient mass, mineralogic variation, and sampling options necessary to enable determination of isochrons of highest precision using U-Pb, Rb-Sr, Sm-Nd, and Lu-Hf (e.g., Borg and Carlson, 2023; Borg et al., 2015; Schneider and Kleine, 2025), as well as the geochronometers Pb-Pb and Ar-Ar. From any given sampling location, three such large samples are required to provide a minimum assessment of compositional and chronologic heterogeneity. The requirement for three samples reflects the difficulty in collecting anorthosite samples compatible with the isotopic age dating process. As a point of comparison, from the entire Apollo sample collection, approximately 10 anorthosite samples have ages ascribed to them (Borg, personal communication). The large samples will be used for identifying and isolating mineral separates, including pyroxene (e.g., Borg et al., 2011), that may provide spread of parent/daughter element ratios, thus further reducing the age uncertainty.

Second, **this experiment requires a large number of smaller rocklets sieved (at least 3–20 mm) from the lunar regolith**. Smaller rocks (at least 8 mm in diameter) will be utilized to determine the diversity in composition of anorthosite at these localities. Because the larger samples described above will be collected and utilized primarily for chronology, and 3 large anorthosite samples are unlikely to reflect the compositional diversity at any site, the smaller rocks will be used to determine composition (mineralogy, petrology, and major and trace elements). Coordinated measurements of **at least 4 particles (8–20 mm)** will allow the compositional variation to be documented in anorthositic material at each site. The sieved fraction will also include **approximately ten 5–8 mm particles and ninety 3–5 mm particles** all of which will be used to document and understand the compositional diversity as well as age, utilizing those chronometers that do not require large mass (Rb-Sr, Ar-Ar) (e.g., Schneider and Kleine, 2025; Cohen and Coker, 2010) (Section 4.3). Because these samples will potentially contain a mixture of lithologies (and some non-anorthositic) mass estimates are based on our expectations and understanding of Apollo regolith that ~50% of the material sieved will be anorthositic (e.g., McKay et al., 1991; Carrier, 1973). This estimation is reflected in the total baseline sample mass requirements reported in FO: Sample Requirements (Section 4.3.6) and FO: Site and Traverse Requirements (Section 5.8.3).

Finally, **at least 100 g of bulk regolith** will be scooped at these sites, and bulk regolith is expected to contain a fine-grained fraction that will include zircons and other mineral fragments (e.g., Nemchin et al., 2009; Meyer et al., 1996; Crow et al., 2017; Barboni et al., 2017; 2024; Zhang et al., 2021a, Greer et al., 2023; Dauphas et al., 2025). These zircons will be identified and concentrated to allow chronologic studies, and in particular, to look for ages that might be associated with lunar formation and subsequent crustal differentiation. The number of zircons recovered in Apollo regolith samples, for example, have a very small mass compared to the mass of the volume of material searched and thus most of the material remains unused (e.g., Taylor et al., 2009; Crow et al., 2017) and available for future scientific investigations.

The threshold requirements to accomplish this experiment are the collection of rocks (≥40 mm) and rocklets (≥8 mm). The collection of bulk regolith is baselined.

#### 4.1.1.3 E1: Total Baseline Sample Requirements per Sampling Location

- ≥12 rocks ≥40 mm diameter
- ≥16 rocks ≥8 mm diameter
- ≥400 g bulk regolith

The sampling requirements for this experiment also include sample acquisition sufficient to satisfy NASA sample curation requirements. Total sample numbers and overall sample mass reflect the sum of these two sets of requirements.

#### 4.1.1.4 E1: Site and Traverse Requirements

To address the three associated objectives (1.1, 1.2, 2.1), **anorthosite must be sampled from at least three different and geologically distinct regions or exposures**. Here, *geologically distinct regions* are defined as regions separated by many 10s to 100s of kilometers, with different geologic settings. This geologic diversity is important, as the exact locations for acquiring ideal anorthositic samples are debated. Geologic diversity of this scale is important for spanning a minimum range of competing hypotheses. This approach will ensure the diversity of lithologies to compare to nearside anorthositic units, as well as maximizing the opportunity to sample unique material in age and composition.  Because of uncertainty in the extent of possibly different nearside and farside crust, sampling just one or two locations may lead to incomplete understanding or non-representative sampling of the farside crust (Arai et al., 2008; Gross et al., 2014; Nagaoka et al., 2023).  In addition, the sampling of multiple, distinct localities will maximize opportunity to acquire a range of parent/daughter ratios in rocks and mineral separates, while also avoiding the potential for degradation due to shock or impact that might dominate in a one-site strategy (e.g., Borg et al., 2011; Nyquist et al., 2006; Norman et al., 2003). The sampling of three distinct sites will provide a more global understanding of the age and composition of farside crust rather than risking the potentially local understanding that could result from too limited (1 or 2 site; Nagaoka et al., 2023) sampling.

Identification of locations for anorthosite sampling should utilize a combination of compositional orbital remote sensing data (e.g., $M^3$, Kaguya), and higher-resolution orbital image data (e.g., LROC-NAC) (e. g. Figure 4.2). Orbital remote sensing datasets can identify locations where, at the scale of the instrument resolution, anorthosite is a dominant composition (e.g., Cheek and Pieters, 2014). This includes identifying anorthosite by either the presence of the 1.25 µm absorption band (Adams and McCord, 1971) or a bright featureless spectrum (e.g., Cheek et al., 2013). The results of previous studies can be used to guide the general site selection (Donaldson-Hanna et al., 2014; Ohtake et al., 2009), but investigation of the compositional data and/or derived data products (e.g., Lemelin et al., 2022) is required to find more specific site locations. However, this scale is generally too coarse for detailed traverse planning. Following general identification with compositional datasets, it is recommended that additional higher-spatial resolution datasets, such as image datasets, are employed to identify local geologic regions where exposure of anorthosite rocks is favorable, e. g. high-albedo boulder fields surrounding small fresh craters and contact margins with uplifted peak rings.

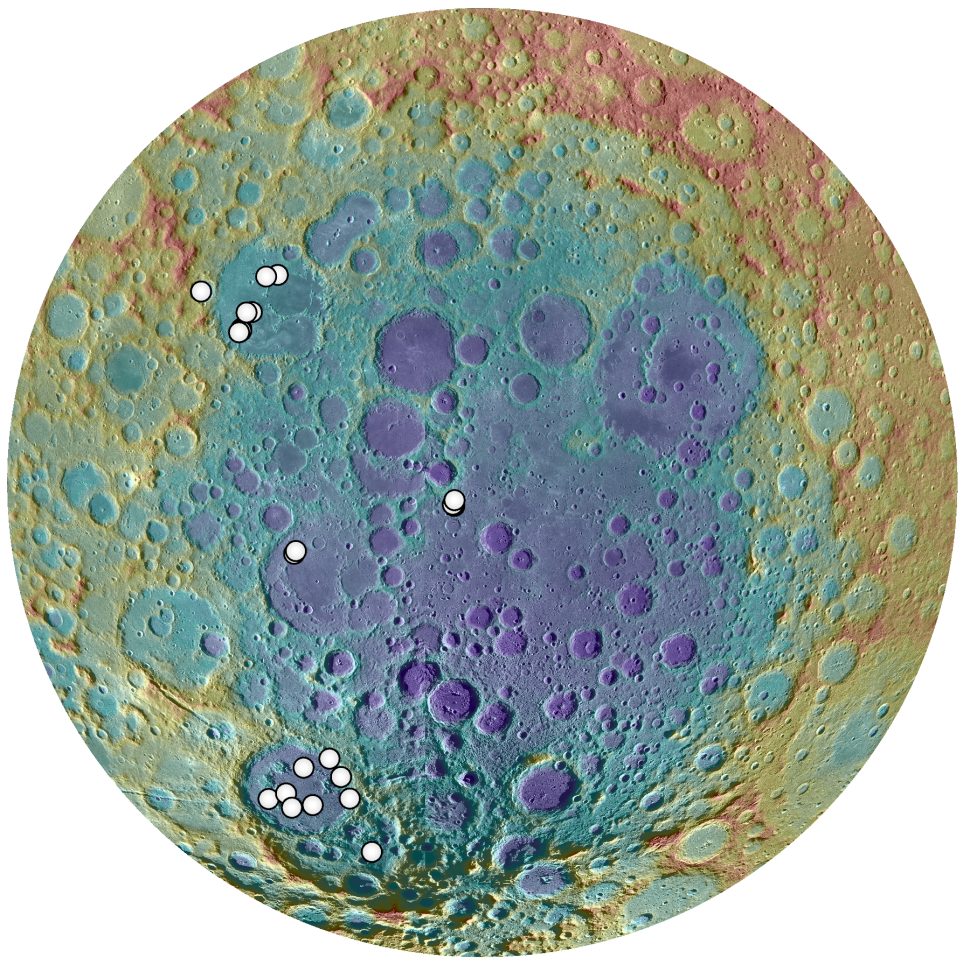

**Figure 4.2.** *Distribution of potential anorthosite sampling locations indicated by >80 wt% plagioclase derived from Kaguya Mineralogy Mapper data (Lemelin et al., 2019).*

#### 4.1.1.5 E1: *In situ* Measurement Requirements

This experiment has two primary *in situ* measurement requirements which consist of the ability to detect anorthosite in the sampling region, and the ability to detect sampleable material consistent with the required sample rock sizes. The detection of anorthosite can be accomplished through a combination of *in situ* imaging via Measurement 1 (M1: Context Imaging) and Measurement 6 (M6: Fine-scale imaging) and compositional measurements via Measurement 2 (M2: Context Mineralogy), Measurement 3 (M3: Fine-scale Mineralogy), and Measurement 4 (M4: Bulk Elemental Chemistry). Additional detail on the spatial resolution requirements for the reconnaissance and sampling phases are provided in Section 4.2.

### 4.1.2 Experiment 2: Determine the Age of SPA Impact Event

The South Pole–Aitken (SPA) basin represents the largest, deepest, and likely oldest preserved impact structure on the Moon, making its absolute age one of the highest-value measurements in lunar science. As one of the earliest retained geologic features on the Moon, the age of SPA is critical for establishing the time by which the lunar surface could retain features—essentially establishing a time zero for impact-based chronologies. Because SPA predates most other visible farside stratigraphy, determining its age is essential for constraining the timing of the earliest basin-forming epoch. Lacking samples from which to derive radiometric ages, the majority of surfaces in the Solar System are dated using crater size-frequency techniques and impact flux models (Neukum et al., 2001; Robbins, 2014; Fassett et al., 2012). Samples returned by the Apollo missions have allowed for the partial calibration of these models for specific regions on the Moon; however, these models currently rely upon unconstrained ages for SPA. Establishing the timing of SPA formation is therefore fundamental for calibrating crater size–frequency chronologies, reconstructing the earliest basin-forming epoch, and constraining models of inner Solar System impact flux (Bottke and Norman, 2017; Cohen et al., 2023; Norman et al., 2010; Fernandes et al., 2013) (e. g. Figure 4.3). Despite its importance for both lunar history and broader Solar System chronologies, the age of SPA remains unconstrained by direct radiometric dating, as no samples with unambiguous provenance from SPA melt sheets or ejecta have yet been collected. While analysis of samples of Apollo basin returned by Chang'e 6 have suggested an age of 4.25 Ga for SPA (Su et al., 2025a), there remain significant uncertainties about the provenance of the dated samples, and broader statistical robustness of the age.

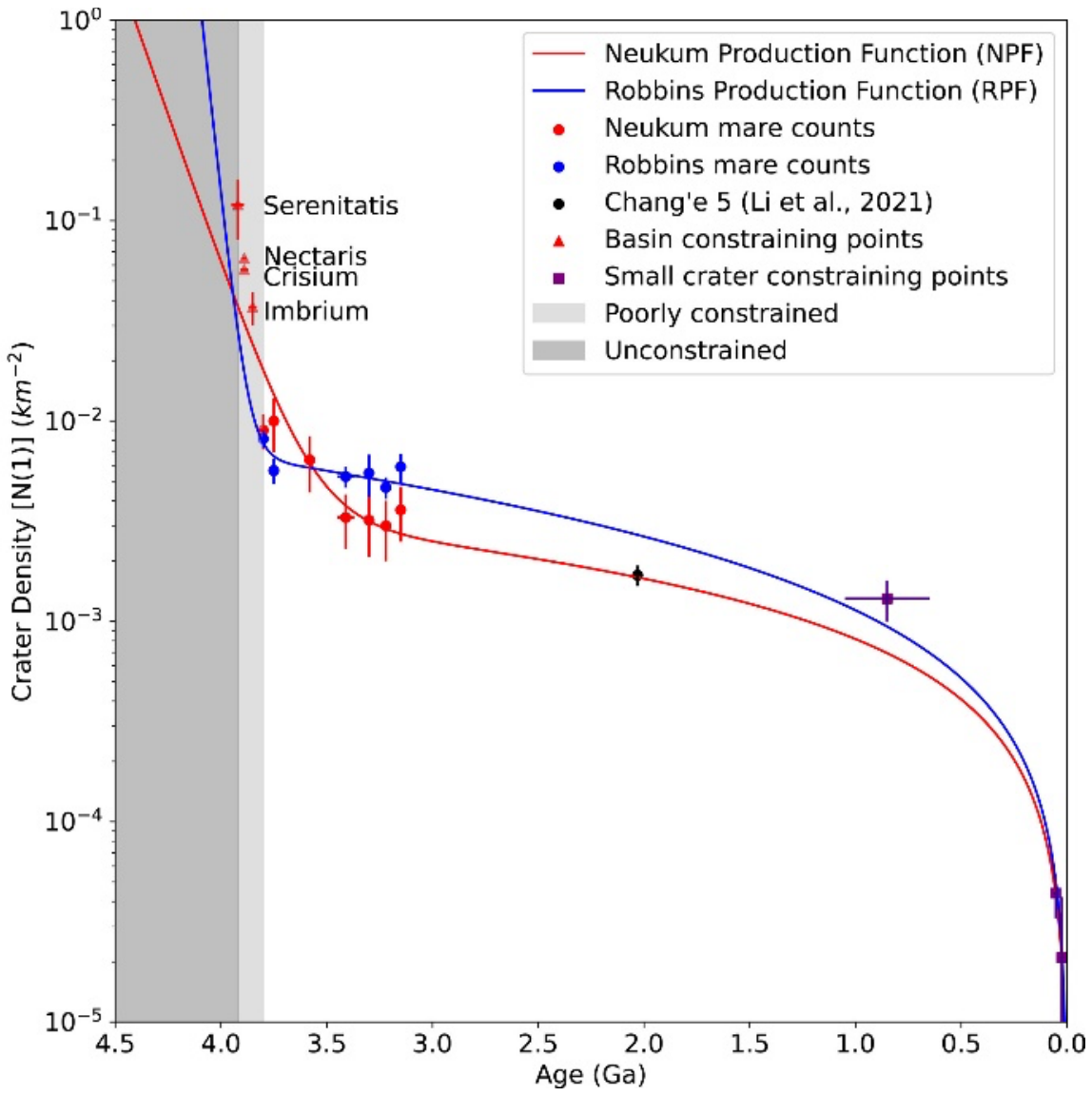

**Figure 4.3.** *Example of two different lunar chronology functions, and the data points that constrain them. Critical differences arise in the functions during the SPA-formation epoch. (From Blevins et al., 2025, Figure 1).*

#### 4.1.2.1 E2: Science Objective Traceability

Determining the age of SPA impact event is directly linked to three mission objectives, including Objectives 1.1, 1.2, and 1.3.

**Objective 1.1** seeks to quantify the age of the Moon to within ±(30) Myr in order to distinguish between competing Early Moon and Late Moon formation models. Following its formation, the Moon may have accreted and differentiated rapidly, consistent with early crystallization models (e.g., Barboni et al., 2017; Nimmo et al., 2024), or may have experienced prolonged or episodic thermal evolution, leading to younger apparent formation ages (e.g., Borg et al., 2011; Gaffney and Borg, 2014; Borg et al., 2023). Resolving these alternatives requires precise and well-contextualized chronologic constraints (e.g., the age of SPA) capable of discriminating among formation scenarios that differ by only tens of millions of years.

**Objective 1.2** seeks to determine whether the Moon cooled rapidly or slowly following its formation. After accretion, the Moon may have undergone a range of cooling and differentiation pathways, including rapid solidification of a global magma ocean, extended crystallization under sustained thermal conditions, or later thermal perturbations such as tidal heating, elevated impact flux, or localized large impacts (e.g., Hartmann et al., 2000; Borg et al., 2015; Maurice et al., 2020; Nimmo et al., 2024; Joy et al., 2025). These scenarios predict distinct distributions of crustal ages, mineral compositions, and isotopic systematics, which can be tested through coordinated geochronologic and geochemical measurements. The age of SPA conclusively establishes a point by which crustal cooling and solidification must have ended.

**Objective 1.3** seeks to determine whether, and when, a giant planet instability produced a sharp change in the inner Solar System impactor flux. The early lunar impact record may reflect either a smoothly declining bombardment from residual planetesimals or a more abrupt episode of enhanced impacts triggered by dynamical rearrangement of the giant planets. These competing scenarios imply fundamentally different histories for Solar System evolution and predict different temporal patterns in basin formation and impact melt ages. Establishing the timing and character of any sharp transition in the lunar impact flux is therefore essential for interpreting the Moon's cratering record and for constraining the dynamical evolution of the early Solar System.

#### 4.1.2.2 E2: Sample Requirements

Impact melt rocks from the South Pole–Aitken (SPA) basin represent one of the most promising archives of early lunar bombardment, but they also present significant challenges for geochronologic interpretation (e.g., Norman et al., 2010; Grange et al., 2013; Mercer et al., 2015). As products of large, high-energy impacts, impact melts may record crystallization ages associated with basin formation; however, they are also susceptible to subsequent thermal overprinting by later impacts, which can partially or fully reset isotopic systems (e.g., Dalrymple & Ryder, 1996; Fernandes et al., 2009; Cohen et al., 2014). This is particularly relevant for SPA, where later basin-forming and regional impacts may have affected melt lithologies over extended timescales. Distinguishing primary crystallization ages from later disturbance therefore requires careful sample selection and the use of multiple complementary chronometers.

In addition, impact melt rocks are commonly compositionally heterogeneous, reflecting mixing of target lithologies and projectile material during impact melting (e.g., Korotev, 2000; Vaughan & Head, 2014; Moriarty et al., 2021a). This heterogeneity can complicate age interpretation but also provides an opportunity to assess source region diversity and melt formation processes. Many impact melts are fine-grained or glassy, limiting mineral separation options and reducing the spread of parent/daughter element ratios needed for high-precision isochron dating (e.g., Borg et al., 2011; Norman & Nemchin, 2014). As a result of these factors, sampling requirements for SPA impact melts must be carefully designed to maximize chronological fidelity and scientific return and fall into three complementary categories.

The sieving and sampling of regolith allows for recovery **of ≥8 mm melt fragments which will be prioritized for collection**. These samples provide the mass, textural context, and mineralogical variability needed to obtain high-precision crystallization ages using Ar-Ar, Pb/Pb and/or Rb/Sr, as well as other isochron systems such as Sm–Nd and Lu–Hf (e.g., Fernandes et al., 2013; Cohen & Coker, 2010; Mercer et al., 2015). These fragments may include crystalline melt clasts as well as glassy materials, offering complementary age constraints on both melt crystallization and later thermal events. Coordinated analyses of multiple fragments per site will help identify age populations and assess the degree of isotopic resetting. Based on Apollo experience, it is anticipated that a substantial fraction of the sieved material will consist of impact melt products (e.g., McKay et al., 1991; Stöffler et al., 2006), making this an efficient strategy for building statistically meaningful age datasets. Collecting multiple large impact melt clasts from distinct SPA localities will permit evaluation of spatial variability in melt ages and compositions, providing critical constraints on the timing and extent of basin-forming and post-basin impact events and allowing the development of a coherent impact chronology for the SPA basin that anchors the early lunar impact record.

Targeting the collection of ≥20 rocklets with a diameter of ≥5 mm diameter is expected to result in **the collection of >100 rocklets of ≥3 mm diameter** based on typical lunar regolith particle size distributions (e.g., Carrier, 2003). These smaller size fragments are sufficient in mass to provide material for thin sections, bulk rock determinations of major, minor, and trace elements, and additional petrologic analyses. These petrologic analyses include determination of crystallization ages of samples via the Pb-Pb system using SIMS, petrological characterization and mineral chemical analyses via electron microprobe/SEM WDS and SDD-EDS, and mineral trace elemental abundances via LA-ICP-MS. These fragments also provide sufficient mass for Ar-Ar dating, which will provide additional information regarding the crystallization age and/or shock history of these samples.

**Bulk regolith samples will be collected to target the fine-grained fraction**, this is expected to contain impact melt glass shards, mineral fragments, and accessory phases such as zircon and baddeleyite (e.g., Nemchin et al., 2009; Grange et al., 2011; Crow et al., 2017; Zhang et al., 2021a). These phases can record both impact-related and inherited ages, enabling investigation of SPA melt formation ages as well as evidence for older crustal components entrained during basin formation.

The SPA basin records a long and complex impact history, such that robust determination of its formation age cannot rely on a single sample or isolated age measurement. Impact melts within SPA are expected to preserve a range of isotopic signatures reflecting primary basin-forming crystallization, partial resetting by subsequent large impacts, and localized reheating events. As a result**, statistically meaningful characterization of the SPA impact event requires analysis of multiple impact melt samples across a range of lithologies, textures, and spatial contexts**. Convergent age populations observed across independent samples and chronometers will provide the strongest evidence for identifying the true SPA basin-forming age, while dispersion in ages will enable quantification of post-SPA thermal modification. This population-based approach is essential for distinguishing the primary SPA signal from later overprinting and for constructing a reliable chronological framework for the basin and the early Moon.

#### 4.1.2.3 E2: Total Baseline Sample Requirements per Sampling Location

- ≥32 rocks ≥8 mm diameter
- ≥400 rocks ≥3 mm diameter
- ≥400 g bulk regolith

The sampling requirements for this experiment also include sample acquisition sufficient to satisfy NASA sample curation requirements. Total sample numbers and overall sample mass reflect the sum of these two sets of requirements.

#### 4.1.2.4 E2: Site and Traverse Requirements

Due to its immense size, the SPA-forming impact would have produced enormous volumes of impact melt. While melt is expected to be present in ejecta, large basins such as SPA retain the majority of their impact melt in a vertically thick central melt pool (e.g., Cintala and Grieve, 1998; Citron et al., 2024).

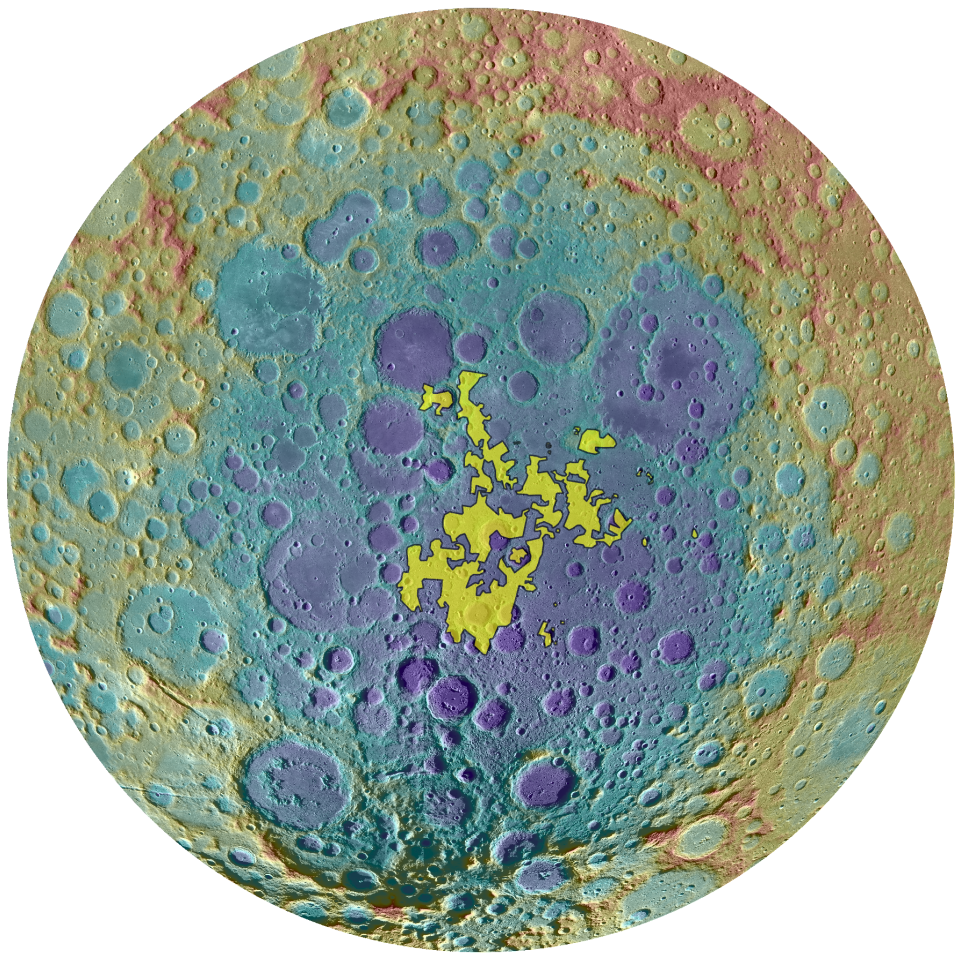

**Figure 4.4.** *Proposed distribution of SPA impact melt sheet materials, adapted from Uemoto et al., 2017, Figure 4.*

Confidently targeting SPA impact melt is a non-trivial challenge due to the basin's ancient age, complex geologic history, and the inability of orbital remote sensing datasets to uniquely identify impact melt. The SPA ejecta deposit is subject to > 4 billion years of reworking through subsequent impact events, deposition of later basin ejecta, and regolith development. As such, it is unclear what fraction of material has retained the age of SPA formation in a form measurable in samples. On the other hand, central SPA, which hosts the thick SPA melt pool, exhibits evidence of significant resurfacing (Moriarty et al., 2021), meaning that SPA impact melt may not be broadly accessible at the contemporary basin surface. As a result, it is recommended that traverse designs focus on targeting the ejecta deposits of strategic large craters that are hypothesized to have excavated into the SPA impact melt sheet, and therefore could contain larger fractions of this impact melt in their own ejecta deposits.

For example, several complex impact craters within central SPA (i.e., Bhabha, Bose, Stoney) reveal a coherent compositional stratigraphy within the uppermost several km of the basin floor, with high-Ca pyroxene-dominated materials in the near surface giving way to low-Ca pyroxenes below ~2-5 km depths. This observed stratigraphy may be the result of differentiation of the impact melt pool (e.g., Hurwitz and Kring, 2014; Uemoto et al., 2017) (e. g. Figure 4.4). **Collecting samples from both low-Ca and high-Ca pyroxene-bearing materials excavated from depth in central SPA would maximize the likelihood of sampling the SPA impact melt pool**. Confidence in the derived ages can be further supported by identifying and **dating SPA impact melt in other locations**, such as within the SPA ejecta re-exposed through subsequent crater- and/or basin-scale impacts, such as the **rims and peak rings of Poincaré basin and Schrödinger basin**. The significant spatial distances between these locations requires long-distance mobility capability.

#### 4.1.2.5 E2: *In situ* Measurement Requirements

No current remote analysis techniques exist for definitively identifying impact melt fragments. The *in situ* measurement requirements for this experiment are Measurement 1 (M1: Context Imaging), Measurement 2 (M2: Context Mineralogy), and Measurement 4 (M4: Bulk Elemental Chemistry) to understand the geology of a region of suitable sampling material within the local region at a rover-scale, placing the sampled material in sufficient context. Although not necessarily needed, Measurement 3 (M3: Fine-scale Mineralogy) and Measurement 6 (M6: Fine-scale Imaging) would be valuable contributors to executing this experiment to increase confidence of melt rock identification while on the Moon prior to sample return.

### 4.1.3 Experiment 3: Determine the Model Age of the Mantle Underlying SPA

Isotopic model ages are generally used to date the formation of isotopically distinct geochemical reservoirs within a planetary body (Section 4.3.3.5). Within the lunar context, model ages have been used previously to constrain the timing of lunar formation as recorded by core-mantle separation ($^{182}$Hf-$^{182}$W system; e.g., Kleine et al., 2008, 2005; Kruijer et al., 2015; Kruijer and Kleine, 2017; Touboul et al., 2007) and to date the timing of mare basalt source region and KREEP formation on the nearside ($^{147}$Sm-$^{143}$Nd, combined $^{146,147}$Sm-$^{142,143}$Nd, $^{176}$Lu-$^{176}$Hf, and $^{207}$Pb-$^{204}$Pb systems; e.g., Borg et al., 2019; Brandon et al., 2009; Carlson and Boyet, 2009; Gaffney and Borg, 2014; McLeod et al., 2014; Nyquist et al., 1995; Snape et al., 2016; Sprung et al., 2013). In the most straightforward sense, mantle model ages derived from the Sm-Nd and Lu-Hf systems date the timing of separation of these mantle sources from a chondritic reservoir that is interpreted as the lunar magma ocean (LMO), as the LMO is thought to have chondritic ratios (e.g., Sprung et al., 2013). Therefore, mantle model ages for mare basalt sources are frequently interpreted as the dating the crystallization of the LMO, which has direct relevance to the timing of lunar formation and the establishment of the LMO.

The goal of Experiment 3 (E3) is to determine the model age of the mantle underlying SPA. While E1 and E2 focus on constraining lunar formation through the age of the crust, E3 seeks to constrain the timing of mantle differentiation, another key process in the formation of the Moon. Specifically, E3 would determine the model age of the mantle, where the model age is derived using multiple isotopic systems to determine the age at which they diverged from an initial parent mantle. Radioactive isotope systems are immensely useful for both the dating of individual samples and as tracers of the evolution of planetary systems, and E3 would utilize ≥3 radiogenic isotope systems (e.g., $^{146,147}$Sm-$^{142,143}$Nd, Lu-Hf, and Pb-Pb) to constrain the timing of lunar formation and evolution.

#### 4.1.3.1 E3: Science Objective Traceability

Determine the model age of the mantle underlying SPA is directly linked to two mission objectives, including Objectives 1.1 and 1.2.

**Objective 1.1** is directly addressed by E3 regarding the age of the Moon by providing model an age(s) for the formation of the mantle sources of ≥3 farside mare basalts.

**Objective 1.2** is directly addressed by E3 regarding models for the duration of magma ocean crystallization. In association with data produced by Experiments 1 and 2 (E1 and E2), and with existing data from the lunar nearside, mantle model ages for mare basalt sources will be used to assess and interpret the duration of magma ocean crystallization. Data from E1 will provide an assessment of the timing of farside crust formation that can be used in tandem with E3 estimates for establishment of mantle sources.

#### 4.1.3.2 E3: Sample Requirements

Mare basalts are among the easier samples to work with from a geochemical and chronological standpoint. Unbrecciated and low-shock samples are typically abundant at mare basalt sampling locations, basalts naturally contain multiple primary minerals, and their formation as mantle partial melts concentrates incompatible elements from the mantle source into the magma. These factors all make chronologic and geochemical measurements easier to perform and interpret as compared to the majority of ancient lunar crustal samples (i.e., anorthosite), which are typically much more strongly affected by brecciation, shock, and thermal annealing, as well as the fact that they contain lower abundances of the elements of interest. There are, however, some constraints that place requirements on the returned samples of the ≥3 spatially and/or temporally distinct mare basalt units.

This experiment requires **≥20 rocklets with a diameter of ≥5 mm diameter to be collected at each site**. Sample analyses and impact mixing models show that between 3–40% of material from mare basalt sampling sites may be exogenous, although this is very dependent on location (e.g., Cao et al. 2025, Su et al. 2025a, Xu et al. 2024, Xie et al. 2020, Huang et al. 2017), so mass requirements have been designed here assuming 20% exogenous materials (Section 5.8.3, Foldout 5.1). Additional factors that will affect the mass requirement for each type of analysis include texture of the basalt samples, abundances of the elements of interest (e.g., REEs, Rb, Sr, Hf, U, Pb) in the samples, compositional homogeneity vs. heterogeneity between individual fragments, and abundance of phases amendable to SIMS Pb-Pb analyses (e.g., zircon, baddeleyite, phosphate, mesostasis). To ensure that large basalt fragments (≥5 mm diameter) with textures and elemental concentrations that enable isochron determinations are collected, ≥80 rocklets (i.e., ≥20 rocklets available for initial analyses by the Science Team) are required.

Rock fragments ≥5 mm diameter will be targeted for Sm-Nd, Rb-Sr, and Lu-Hf isochron dating methods, as these fragments will be of sufficient mass (100s of mg) to enable these methodologies. Isochron age determinations for the basalt sample(s) present at each site are required in order to provide the data necessary for calculation of model ages of the mantle source(s) and for fulfillment of Experiment 8 (E8), which utilizes the same sample set as E3. When rocklet mass, elemental composition, and textures allow, it may be possibly and advantageous to use the same rocklet for isochron ages for multiple isotopic systems, bulk composition determinations, and to make thin sections for subsequent analyses.

Targeting the collection of ≥20 rocklets with a diameter of ≥5 mm diameter is expected to result in **the collection of >100 rocklets of ≥3 mm diameter** based on typical lunar regolith particle size distributions (e.g., Carrier, 2003). These smaller size fragments are sufficient in mass to provide material for thin sections, bulk rock determinations of major, minor, and trace elements, and additional petrologic analyses. These petrologic analyses include determination of crystallization ages of samples via the Pb-Pb system using SIMS, petrological characterization and mineral chemical analyses via electron microprobe/SEM WDS and SDD-EDS, and mineral trace elemental abundances via LA-ICP-MS. These fragments also provide sufficient mass for Ar-Ar dating, which will provide additional information regarding the crystallization age and/or shock history of these samples.

E3 also requires **≥100 g of bulk regolith** to be scooped at these sites, which is expected to contain a fine-grained fraction that includes zircons and other mineral fragments (Barboni et al., 2024, 2017; Crow et al., 2017; Dauphas et al., 2025; Greer et al., 2021; Meyer et al., 1996; Nemchin et al., 2009; Zhang et al., 2021b, 2021a). These zircons will be identified and concentrated to allow chronologic studies. Furthermore, lunar regolith, particularly at mare sites, is expected to constrain volcanic pyroclastic glass beads that will be useful for studies of lunar volcanic history, mantle heterogeneity and volatile contents, and the thermal history of the farside mantle.

#### 4.1.3.3 E3: Total Baseline Sample Requirements per Sampling Location

- ≥80 rocklets ≥5 mm diameter
- ≥400 rocklets ≥3 mm diameter
- ≥400 g bulk regolith

The sampling requirements for this experiment also include sample acquisition sufficient to satisfy NASA sample curation requirements. Total sample numbers and overall sample mass reflect the sum of these two sets of requirements.

#### 4.1.3.4 E3: Site and Traverse Requirements

Experiment 3 requires baseline **sampling of ≥3 spatially and temporally distinct mare basalt units** (Figure 4.5) (≥2 samples for threshold). Sampling locations should use existing studies, such as Pasckert et al., (2018), to identify 3 spatially distinct mare basalt units with the widest spread in temporal ages possible. The requirement of ≥3 spatially and/or temporally distinct mare basalts is necessary so that data for ≥3 distinct mantle source regions can be obtained and compared to assess both the model ages of farside mantle source regions and assess any temporal heterogeneity in those source regions. The threshold requirement for this experiment requires ≥2 samples from spatially and temporally distinct mare basalt units, with the rationale being that at least 2 sample are required to begin to identify any potential heterogeneities in mantle source regions. Multiple samples from a geographically restricted area would not provide sufficient confidence that basalts derived from distinct mantle sources regions are being sampled.

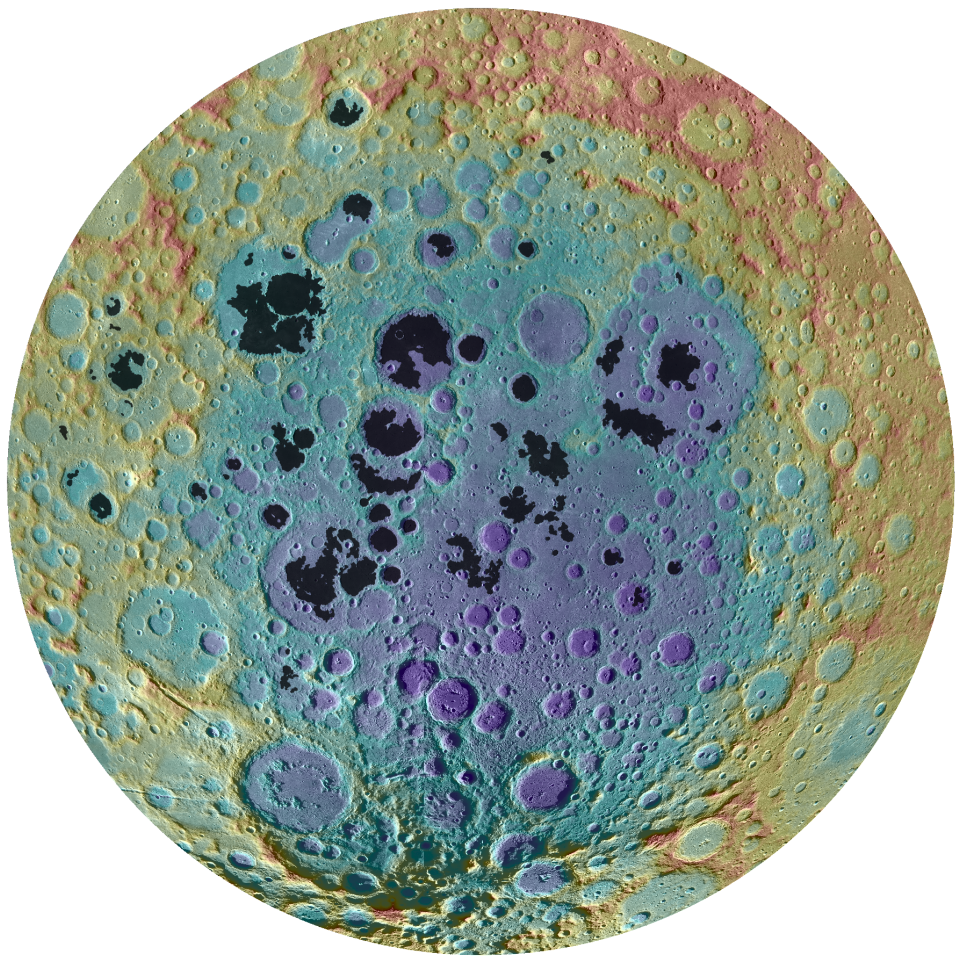

**Figure 4.5.** *Locations of mapped mare units within the SPA region. Derived from LROC global mare product (Nelson et al., 2014).*

#### 4.1.3.5 E3: *In situ* Measurement Requirements

Experiment 3 requires the ability to detect sampleable materials. This will be accomplished by Measurement 1 (M1; Context Imaging). Context imaging via M1 will allow for the assessment of whether the rover workspace contains materials that are suitable for sampling. For basaltic units, past mission experience (e.g., Apollo, Chang'e) has shown that at sampling sites that are clearly within the basaltic unit of interest as mapped from orbit, the regolith will contain rock fragments from the underlying unit of suitable size distribution for the analyses in E3.

### 4.1.4 Experiment 4: Determine the Age of ≥ 1 Farside Imbrian-Aged Impact Crater (D ≥ 80 km) or Basin (D ≥ 300 km) with a Surface Datable by Crater Counts

The geologic history of the Moon has been divided into five main epochs, each of which is anchored by the inferred or calculated formation age of specific features, and an associated impact crater flux (e. g. Wilhelms et al., 1987). The Imbrian period (approximately 3.8-3.2 billion years ago) was particularly important for the lunar nearside, as most of the large basins (including Imbrium) are thought to have formed during the earliest part of this period, around 3.8 Ga. The confluence of so many large basin-forming events around a specific period in the Moon's history led to the hypothesis that the basins were formed in response to a Solar System-wide spike in the impactor flux, called the Late Heavy Bombardment (Section 3.1) (Figure 4.7). The large-scale of the Imbrium impact means that the age is well-constrained by the Apollo and Luna sample collection. In contrast, despite the presence of several Imbrian-aged impact craters and basins on the lunar farside, none have absolute radiometric age constraints. As a result, the Imbrian portion of the lunar chronology remains anchored almost entirely by nearside samples (e.g., Stöffler and Ryder, 2001; Bottke and Norman, 2017), limiting our ability to assess farside impact history and introducing potential bias into global lunar timescales (e.g., Fassett et al., 2012; Robbins, 2014). Determining the absolute age of at least one farside Imbrian-aged impact crater (D ≥ 80 km) or basin (D ≥ 300 km) with a well-characterized crater-count surface is therefore essential for extending and calibrating the lunar chronology beyond nearside-dominated constraints.

#### 4.1.4.1 E4: Science Objective Traceability

Determining the age of at least one (1) farside Imbrian-aged impact crater (D≥80 km) or basin (D≥300 km) with surface dateable by crater counts is directly linked with mission Objective 1.3.

**Objective 1.3** seeks to determine if and when a giant planet instability drove a sharp change in the inner Solar System impactor flux. Large Imbrian-aged farside impact structures sample a critical interval in lunar history during which impact flux models predict either a smooth decline or a temporally concentrated increase associated with dynamical instability among the giant planets. Radiometric dating of impact melt samples from these structures, combined with crater-count–based surface ages, will provide a key calibration point for the lunar cratering chronology and enable quantitative tests of competing bombardment models, thereby constraining the timing and nature of any instability-driven enhancement in the inner Solar System impact rate.

#### 4.1.4.2 E4: Sample Requirements

Impact melt materials exposed within the South Pole–Aitken (SPA) basin provide a rare opportunity to directly calibrate crater size–frequency–derived surface ages for large farside Imbrian impact structures. Several craters and basins within SPA preserve extensive, mappable impact melt deposits whose geomorphic relationships allow their surfaces to be independently dated using crater counts. Obtaining radiometric ages from impact melts associated with one or more of these features will enable a direct comparison between absolute ages and model surface ages, anchoring the farside Imbrian chronology for the first time.

Sampling strategies are therefore focused on collecting impact melt fragments that can be confidently linked, through geologic context and spatial association, to a specific crater (D ≥ 80 km) or basin (D ≥ 300 km). Consistent with experiment 2 description, we expect to recover the same suite of impact melt sample types (i.e., coherent melt clasts, crystalline melt fragments, and glassy impact melt materials) anticipated in the primary sampling strategy. Regolith sieving will target **8 impact melt clasts ≥8 mm in diameter derived from nearby melt-rich units**, maximizing the likelihood that recovered samples represent material produced during formation of the mapped impact structure. These samples are well suited for high-precision Ar–Ar, Pb–Pb, and Rb–Sr dating, with additional isochron constraints provided by Sm–Nd and Lu–Hf systems where appropriate. Analyzing multiple fragments from a single geologic unit will allow identification of coherent age populations that can be robustly tied to a single farside Imbrian impact event.

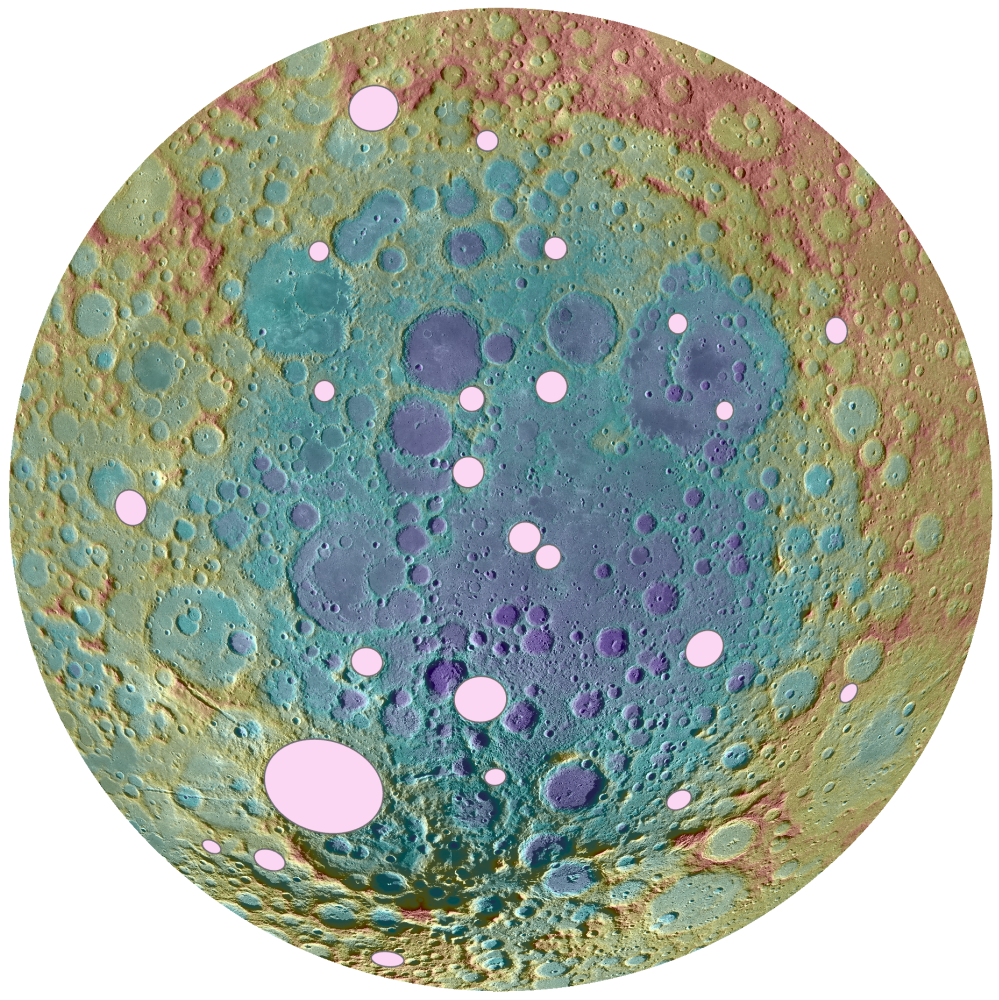

**Figure 4.6.** *Distribution of Imbrian-aged craters and basins with D ≥80 km.*

Targeting the collection of rocklets with a diameter of ≥8 mm diameter is expected to result in **the collection of >100 rocklets of ≥3 mm diameter** based on typical lunar regolith particle size distributions (e.g., Carrier, 2003). These smaller size fragments are sufficient in mass to provide material for thin sections, bulk rock determinations of major, minor, and trace elements, and additional petrologic analyses. These petrologic analyses include determination of crystallization ages of samples via the Pb-Pb system using SIMS, petrological characterization and mineral chemical analyses via electron microprobe/SEM WDS and SDD-EDS, and mineral trace elemental abundances via LA-ICP-MS. These fragments also provide sufficient mass for Ar-Ar dating, which will provide additional information regarding the crystallization age and/or shock history of these samples.

**Bulk regolith samples (≥100 g)** will complement clast-scale analyses by providing access to fine-grained impact melt components and accessory phases that may preserve crystallization ages associated with large impact events. When combined with crater-count–derived surface ages for the same mapped units, these measurements will allow direct calibration of the Imbrian timescale on the lunar farside. Establishing even a single well-constrained absolute age for a farside Imbrian crater or basin will significantly reduce uncertainties in lunar chronology models and improve interpretation of impact flux evolution across the Moon. This calibration is essential for generating the chronological framework of the South Pole–Aitken region and for establishing the broader lunar impact chronology.

#### 4.1.4.3 E2: Total Baseline Sample Requirements per Sampling Location

- ≥32 rocks ≥8 mm diameter
- ≥400 rocks ≥3 mm diameter
- ≥400 g bulk regolith

The sampling requirements for this experiment also include sample acquisition sufficient to satisfy NASA sample curation requirements. Total sample numbers and overall sample mass reflect the sum of these two sets of requirements

#### 4.1.4.4 E4: Site and Traverse Requirements

Within the vicinity of SPA, there is a single Imbrian-aged basin (Schrödinger, D=320 km; Kring et al., 2016) and ~10 Imbrian-aged or younger craters with D=80-300 km (Öhman et al., 2015; Ivanov et al, 2018) (Figure 4.6).  These 11 sites are appropriate for this experiment, though an additional ~13 craters between 50 and 80 km could possibly address this objective with less fidelity. However, larger craters are stratigraphically more significant and, typically, larger landforms have more countable surface area (i.e., broader areas of ejecta with relatively low slope that retain craters), such that crater statistics could be tied to a sample-derived age.

A requirement for this experiment is that **a geologic unit demonstrably associated with the impact-forming event** (e.g., ejecta and/or impact-melt–bearing materials) **is identifiable, laterally mappable, and crater-countable at a scale sufficient to derive a robust CSFD age** (Michael and Liu, 2025). This requirement appears to be met at many of the Imbrian-aged craters in SPA.  Choosing a sample of an impact-derived facies from these Imbrian-aged craters is simpler than some other experiments that require sampling SPA itself (Exp. 2) or a Nectarian/Pre-Nectarian basin (Exp. 5) which have substantial overprinting. Finally, obtaining a datable suite that includes multiple impact-melt lithologies may be enabled by seeking to sample small, fresh craters superposed on the mapped unit. Fresh craters can excavate material from below the upper gardened surface, and enhance the potential for access and return of more coherent melt samples.

#### 4.1.4.5 E4: *In situ* Measurement Requirements

The *in situ* measurement requirements for this experiment are Measurement 1 (M1: Context Imaging), Measurement 2 (M2: Context Mineralogy), and Measurement 4 (M4: Bulk Elemental Chemistry) to understand the geology of the crater/basin at a rover-scale, and place the sampled material in sufficient context.  Although not necessarily needed, Measurement 3 (M3: Fine-scale

Mineralogy) and Measurement 6 (M6: Fine-scale Imaging) would be valuable contributors to executing this experiment to increase confidence of breccia/melt rock identification while on the Moon (prior to sample return).

### 4.1.5 Experiment 5: Determine the Age of ≥ (1) Non-SPA Farside Pre-Imbrian-Aged Impact Basin (D≥300 km)

Pre-Imbrian impact basins on the lunar farside represent some of the oldest preserved large-scale crustal modification events on the Moon and potentially record impact flux during the earliest phases of inner Solar System evolution (Stöffler and Ryder, 2001; Neukum et al., 2001; Bottke and Norman, 2017). As described in Goal 1, and Experiment 4 (Section 3.1, 4.1.4), this period directly precedes the Imbrian period, and allows for a record of the Solar System impact flux at the critical time when proposed spikes in the bombardment history could have taken place. Despite the importance of anchoring this stratigraphic time period, no farside pre-Imbrian basin currently has a direct radiometric age constraint, and the early portion of the lunar chronology remains anchored almost exclusively by nearside Apollo samples and their associated basin interpretations (Fassett et al., 2012; Robbins, 2014). Determining the absolute age of at least one non-SPA farside pre-Imbrian basin will provide a critical calibration point for the earliest preserved impact record and directly test models for early basin formation rates and potential impact flux variations in the pre-Imbrian epoch (Ryder, 1990; Norman et al., 2010).

Establishing a well-constrained pre-Imbrian farside basin age will significantly improve calibration of the early lunar timescale and refine models of early Solar System impact history (Bottke and Norman, 2017; Fassett et al., 2012).

#### 4.1.5.1 E5: Science Objective Traceability

Determining the age of ≥1 non-SPA farside pre-Imbrian-aged impact basin (D≥300 km) is directly linked to mission Objective 1.3.

**Objective 1.3** is to determine if and when a giant planet instability drove a sharp change in the inner Solar System impactor flux. Early lunar history may have been dominated by a relatively steady decline in impacts from leftover planetesimals following accretion, consistent with a smooth decay in bombardment (e.g., Strom et al., 2005; Marchi et al., 2012), or may have included discrete episodes of elevated impact flux triggered by orbital instabilities of the giant planets, resulting in rapid, large-scale basin formation (e.g., Bottke and Norman, 2017; Cohen et al., 2023) (Figure 4.7). Resolving these alternatives requires well-dated impact structures on the lunar farside, including pre-Imbrian and Imbrian basins, that can be linked to crater-count–derived surfaces and impact melt ages. By providing absolute ages for multiple basins using robust chronometers such as Ar-Ar, U–Pb, and Sm–Nd, E5 (and other farside basin experiments, such as E2 and E4) enables the construction of a statistically meaningful lunar impact chronology capable of testing models of early inner Solar System dynamical evolution.

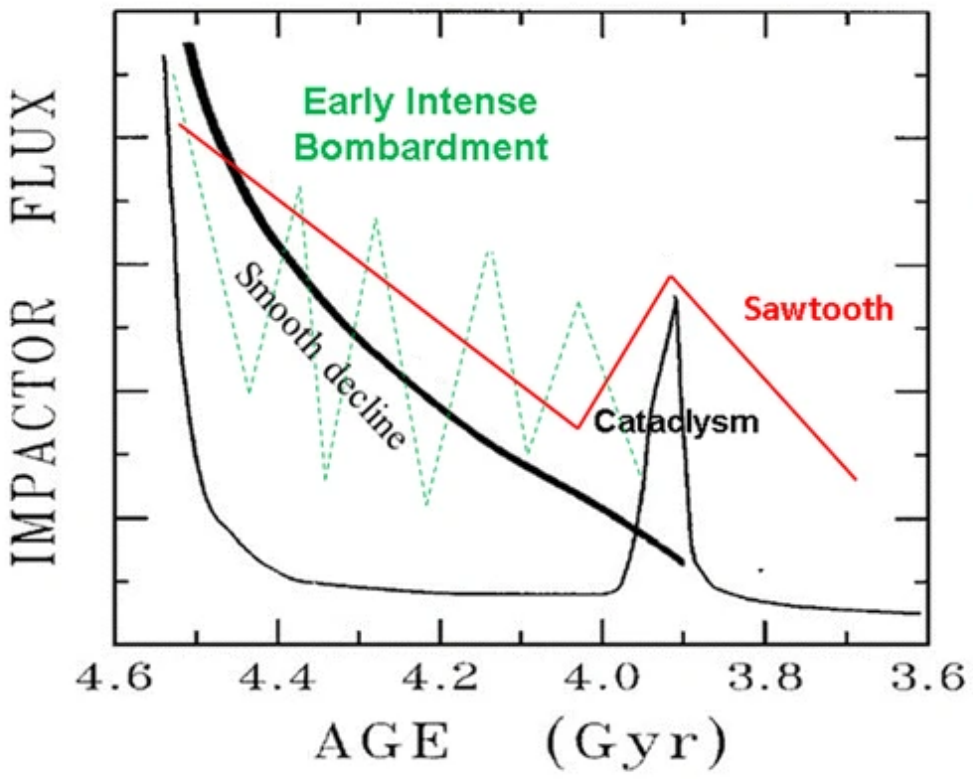

**Figure 4.7.** *Summary of lunar impact scenarios, as presented in Zellner et al., (2017). Graph is by J. W. Delano and modified from Zellner (2001). General flux curves are interpretations of evidence presented in Baldwin (1949, 1964), Hartmann (1965, 1966, 2000), Tera et al. (1974), Neukum et al. (1975 2001), Neukum and Ivanov (1994), and Morbidelli et al. (2012)*

#### 4.1.5.2 E5: Sample Requirements

Impact melt materials associated with large non-SPA farside basins provide the only direct means of establishing absolute ages for pre-Imbrian basin-forming events that are otherwise constrained solely by relative stratigraphy and crater size–frequency distributions. Several farside basins with diameters ≥300 km preserve extensive impact melt exposures whose spatial distribution and stratigraphic relationships indicate formation early in lunar history. Radiometric dating of impact melts linked to one of these basins will provide a critical anchor point for the pre-Imbrian lunar timescale.

Sampling strategies are therefore focused on collecting impact melt materials that can be confidently associated, through geological context and regional mapping, with a specific non-SPA farside basin. Regolith sieving will **target ≥8 mm impact melt clasts derived from basin-related melt deposits and ejecta**, maximizing the likelihood that recovered samples record crystallization ages associated with the basin-forming event. These samples are suitable for high-precision chronologic analyses using Ar–Ar, Pb–Pb, and Rb–Sr systems, with additional constraints from Sm–Nd and Lu–Hf where mineralogy permits. Analyzing multiple melt fragments from a single basin-related unit will enable identification of coherent age populations and improve confidence that measured ages correspond to basin formation rather than later overprinting.

Targeting the collection of rocklets with a diameter of ≥8 mm diameter is expected to result in **the collection of >100 rocklets of ≥3 mm diameter** based on typical lunar regolith particle size distributions (e.g., Carrier, 2003). These smaller size fragments are sufficient in mass to provide material for thin sections, bulk rock determinations of major, minor, and trace elements, and additional petrologic analyses. These petrologic analyses include determination of crystallization ages of samples via the Pb-Pb system using SIMS, petrological characterization and mineral chemical analyses via electron microprobe/SEM WDS and SDD-EDS, and mineral trace elemental abundances via LA-ICP-MS. These fragments also provide sufficient mass for Ar-Ar dating, which will provide additional information regarding the crystallization age and/or shock history of these samples.

**Bulk regolith samples (≥ 100 g)** will complement clast-scale analyses by providing access to fine-grained impact melt components and accessory phases capable of preserving both primary crystallization ages and inherited signatures from target lithologies. These measurements will help assess the degree of crustal excavation during basin formation and evaluate the preservation of early lunar materials within the melt products. When combined with crater-count-derived relative ages and basin stratigraphy, these data will enable calibration of the pre-Imbrian farside basin chronology.

Consistent with Experiment 2, recovery of the same suite of impact melt sample types (coherent melt clasts, crystalline melt fragments, and glassy impact melt materials) is anticipated in the primary sampling strategy. Establishing an absolute age for even a single non-SPA farside pre-Imbrian basin is essential for generating a robust chronological framework for the farside and for anchoring the earliest phases of lunar impact history.

#### 4.1.5.3 E5: Total Baseline Sample Requirements per Sampling Location

- ≥32 rocks ≥8 mm diameter
- ≥400 rocks ≥3 mm diameter
- ≥400 g bulk regolith

The sampling requirements for this experiment also include sample acquisition sufficient to satisfy NASA sample curation requirements. Total sample numbers and overall sample mass reflect the sum of these two sets of requirements.

#### 4.1.5.4 E5: Site and Traverse Requirements

**This experiment requires access to a farside pre-Imbrian impact basin (D ≥ 300 km) (not counting SPA itself).** Specifically, what is most desirable are sites with the best potential for preserving basin-related impact melt materials that can be sampled, thus establishing a confident link between derived sample geochronology and the basin-forming event. Lessons learned from analysis of the Apollo samples has resulted in knowledge of a provenance problem: the sampling site must provide impact melt deposits and/or ejecta whose spatial distribution and stratigraphic relationships support association with a specific basin, because impact melts provide the most direct means to establish absolute ages for pre-Imbrian basin formation where only relative stratigraphy and crater size–frequency constraints currently exist (Wilhelms et al., 1987; Fassett et al., 2012; Orgel et al., 2018). To address these lessons, SPARX requires a basin with diameter ≥300 km, as these basins are sufficiently large to reset a lot of material which can subsequently be sampled, and these basins also have a sufficiently large countable areas to provide a confident assessment of their formation being pre-Imbrium.

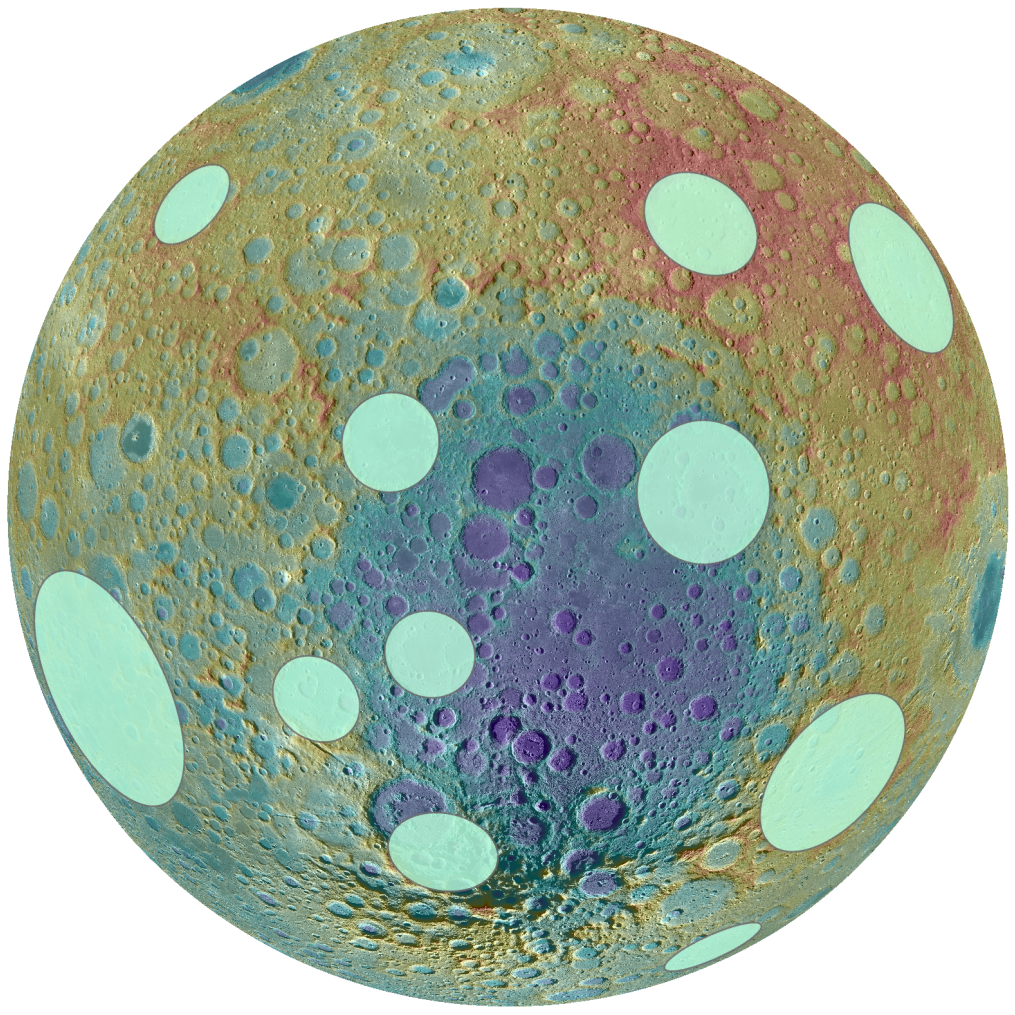

**Figure 4.8.** *Distribution of pre-Imbrian aged basins on the lunar farside. Within SPA, Ingenii, Poincare, and Apollo basins are the most accessible targets for potential sampling.*

Accordingly, the key site requirement is access to a basin-tied unit or source of material. For most basins, the strongest case can be made where access is possible to the **basin's ejecta, peak-ring, or rim.** Parts of a basin's floor not covered by mare also have some potential, though are generally less favorable because these are often covered by younger plains (e.g., Meyer et al., 2020; Giuri et al., 2024). To maximize confidence and reliability of the link to the basin, access to **three geologically distinct settings is required**. These distinct settings maximize the chance of determining a consistent age for the basin and evaluating the extent to which later impacts may have disturbed or mixed the basin melt record.

There are only three candidate pre-Imbrian basins within SPA: Apollo, Ingenii, and Poincare (Figure 4.8). Besides its low latitude, a practical disadvantage of targeting Apollo is that the Chang'e-6 basin has already been sampled and dated as 4.16 Gyr by Chang'e 6 (Chen et al., 2025). However, establishing the chronology of pre-Imbrian basin formation requires measurements from multiple basin settings. A second basin provides the statistical framework and regional geologic context needed to distinguish basin-forming ages from local, or later impact-resetting events, allowing robust interpretation of the observed age populations. Therefore, the most logical targets for SPARX are Ingenii and Poincare. If the traverse extended beyond the rim, the ~355-km Amundsen-Ganswindt (Hostrawser and Osinski, 2025) basin and ~320 km Planck basin are both possible targets, however their considerable distance from other basin target objectives makes these basins less desirable.

#### 4.1.5.5 E5: *In situ* Measurement Requirements

Like Experiment 4, the *in situ* measurement requirements for this experiment are Measurement 1 (M1: Context Imaging), Measurement 2 (M2: Context Mineralogy), and Measurement 4 (M4: Bulk Elemental Chemistry) to understand the geology of the targeted basin at a rover-scale, and place the sampled material in sufficient context. For Experiment 5, particular care should be taken to **use the Context Imaging and Mineralogy to link the sampled material with distinct geologic settings associated with the basin structure**. For example, identifying boulders or regions adjacent to peak ring structures which demonstrate a morphologic or compositional link, or shared provenance. Although not necessarily needed, Measurement 3 (M3: Fine-scale Mineralogy) and Measurement 6 (M6: Fine-scale Imaging) would be valuable contributors to executing this experiment to increase confidence of breccia/melt rock identification while on the Moon (prior to sample return).

### 4.1.6 Experiment 6: Determine the Age and Remnant Magnetization Signatures from ≥3 Returned Samples with Inferred Ages ≥4.0 Ga

Determining whether, when, and how the Moon generated a global magnetic field is central to reconstructing its early thermal and interior evolution, and to establishing the chronology of key processes during the first half-billion years of Solar System history. Because the Moon's deep interior is inaccessible to direct observation, remanent magnetization preserved in lunar rocks provides a critical and unique record of ancient magnetic fields. These magnetic records offer the most direct constraints on the existence, intensity, and longevity of an early lunar dynamo. Accordingly, understanding the origin and evolution of lunar magnetism remains a longstanding scientific priority, as paleomagnetic investigations provide rare insight into dynamo processes operating deep within the Moon that cannot be accessed by other means.

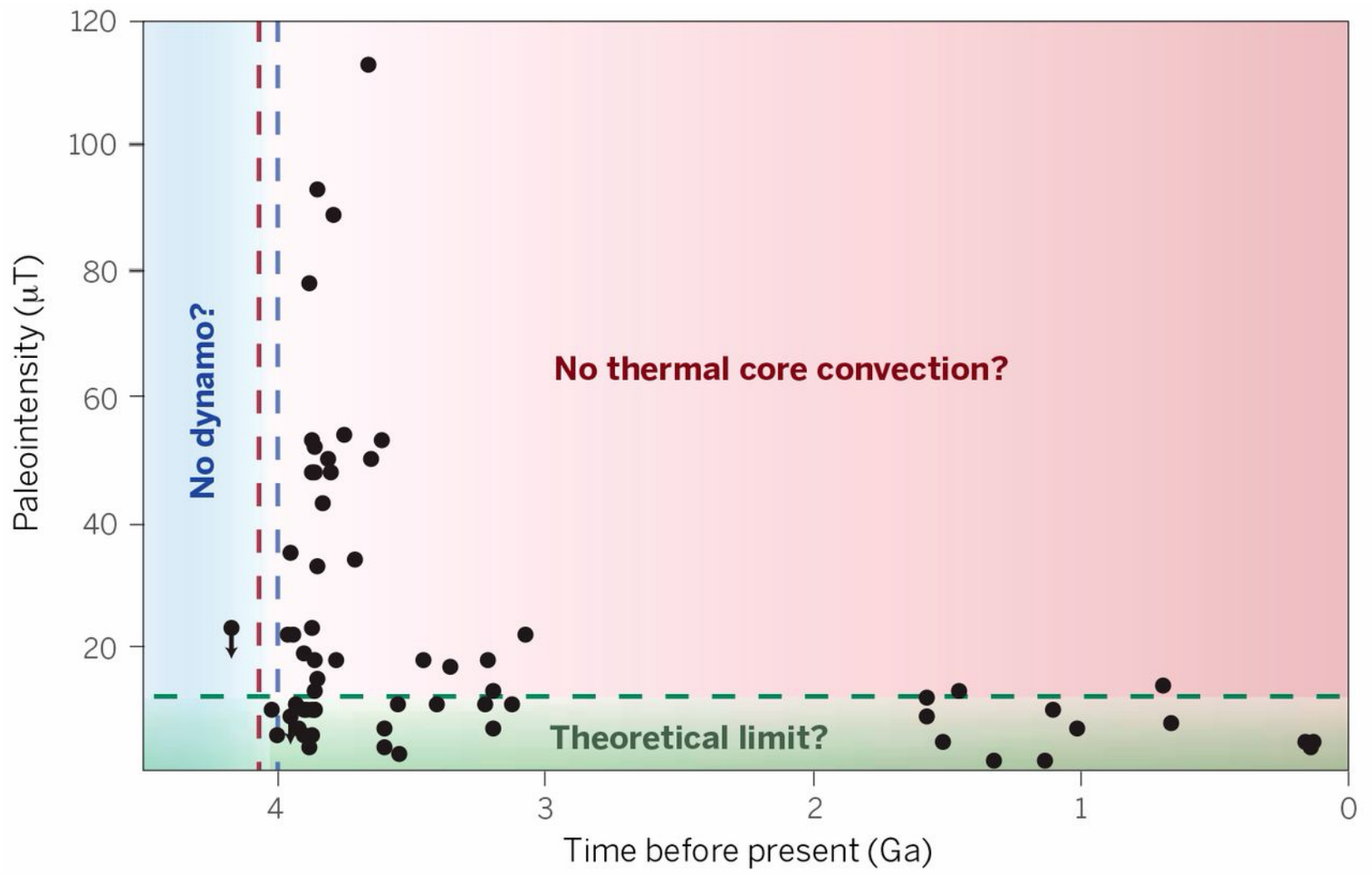

**Figure 4.9.** *Paleointensity measurements of the lunar magnetic field from Apollo samples. The lack of data from samples ≥ 4.0 Ga presents challenges for understanding the presence and strength of a potential lunar dynamo. (adapted from Weiss and Tikoo, 2014, Figure 1, and references therein).*

Progress in constraining the evolution of the lunar dynamo during the Moon's early history is limited by gaps in the existing paleomagnetic record. Nearly all robust lunar paleointensity constraints are derived from Apollo samples, which preferentially sample the lunar nearside and provide limited age coverage prior to ~3.9 billion years ago (Ga) (Figure 4.9). Among these, troctolite 76535 may represent the only Apollo sample yielding a well-constrained paleointensity estimate predating 3.9 Ga, with an age of ~4.25 Ga (Garrick-Bethell et al., 2020).

The next phase of lunar magnetic investigation must therefore prioritize acquisition of high-accuracy paleointensity measurements tied to well-constrained geological context, particularly from the Moon's oldest terrains within the South Pole–Aitken (SPA) basin. Given the basin's estimated age (>4 Ga), samples returned from SPA are uniquely positioned to record magnetic fields present during the Moon's first billion years. Such measurements are essential for resolving the early history of the lunar dynamo and, by extension, constraining the Moon's early interior and thermal evolution. The primary goal of this experiment is to identify and characterize materials older than 4.0 Ga for paleomagnetic analyses; however, confirmation of sample age requires geochronologic analyses that can only be performed on returned materials.

#### 4.1.6.1 E6: Science Objective Traceability

Determining the age and remnant magnetization signatures from returned samples is directly linked to Objective 1.4.

**Objective 1.4** seeks to constrain the evolution of the lunar dynamo in the early epoch of lunar history. Dynamo-driven magnetic field generation represents a fundamental process in early lunar thermal and interior evolution, and returned samples provide direct constraints on the existence, intensity, and temporal evolution of the lunar dynamo during this poorly sampled period. Current paleointensity estimates derived from Apollo samples are overwhelmingly biased toward the post-4.0 Ga epoch, underscoring the need for older samples from pre-4.0 terrain to characterize earlier phases of lunar evolution.

#### 4.1.6.2 E6: Sample Requirements

**Sampling for this objective will be accomplished through coordinated analysis of materials collected to address other mission objectives and returned samples will be made available for paleomagnetic analysis as appropriate.** In addition to constraining the pre–4.0 Ga paleomagnetic record, characterization of the post–4.0 Ga magnetic field history is scientifically valuable and provides essential context for understanding the broader evolution of the lunar magnetic field. Given these complementary objectives, the committee determined that all returned samples should be available for paleomagnetic investigation. Accordingly, this objective does not independently drive sampling strategy but is instead fulfilled through integrated analysis of samples collected to address other scientific goals.

Although this experiment does not impose primary sampling requirements, paleomagnetic investigations do require specific material characteristics for successful implementation. For high-quality paleomagnetic investigation, a total of 6 g per sampling site is recommended, consisting of two discrete analyses of 3 g each (Weiss & Tikoo, 2014; Tikoo et al., 2017; Keane et al., 2021). Of this total mass, approximately 5.4 g is allocated for nondestructive analyses and 0.6 g for destructive analyses. Rock fragments larger than 4 mm in diameter are required to ensure adequate magnetic grain populations and reliable measurements. Individual measurement specimens larger than ~125 mg are required to increase the reliability of paleointensity determinations (Tikoo et al., 2012).

These limitations arise because reliable paleomagnetic measurements require averaging over enough magnetic grains to obtain statistically robust magnetization intensities and directions. Small specimen size may result in magnetic anisotropy effects that bias magnetization directions or introduce demagnetization noise since individual grains are not uniformly magnetized relative to one another (Tikoo et al., 2012).

No maximum mass constraint is imposed from a paleomagnetic perspective. Larger samples are advantageous, as they can be subdivided into mutually oriented specimens to evaluate reproducibility of magnetization direction, intensity, and paleointensity. Additional material may be allocated for complementary analyses—including mineralogy, petrology, shock barometry, geochronology, and thermochronology—to determine the timing of magnetization acquisition and to assess whether the magnetization reflects primary cooling in a dynamo-generated field or subsequent impact-related resetting, thereby strengthening the robustness of paleomagnetic interpretations.

Regolith fines are unsuitable for robust paleomagnetic analysis. In addition to potentially insufficient ferromagnetic grain populations, fines are more likely to have experienced impact processing or shock, increasing the probability that primary magnetization has been modified or reset. Space-weathered exteriors commonly contain nanophase Fe metal, which may complicate interpretation even if such material is predominantly superparamagnetic. Furthermore, fines are too small to permit thin-section preparation and shock barometry, limiting the ability to directly verify that a paleomagnetic specimen is unshocked. In such cases, interpretations would rely on petrographic and age constraints derived from associated regolith materials, introducing uncertainty regarding the equivalence of the analyzed specimen and the characterized lithology.

#### 4.1.6.3 E6: Total Baseline Sample Requirements per Sampling Location

This experiment does not independently drive sampling priorities; however, samples selected for paleomagnetic analysis must meet the following minimum criteria:

- Rocklets > 4 mm in diameter
- ≥ 6 g rocklets

The sampling requirements for this experiment also include sample acquisition sufficient to satisfy NASA sample curation requirements. Total sample numbers and overall sample mass reflect the sum of these two sets of requirements.

#### 4.1.6.4 E6: Site and Traverse Requirements

**This experiment does not independently drive site selection or traverse requirements, as samples will be collected from across the full range of visited locations.** However, sampling from geologically distinct regions and exposures remains important to ensure representative coverage. In particular, targeting terrains aged ≥4.0 Ga is necessary to accomplishing Objective 1.4. These requirements are addressed through traverse planning and sampling strategies established under Goal 1 and associated objectives. Accordingly, this experiment leverages the broader mission traverse without imposing additional site-specific constraints.

#### 4.1.6.5 E6: *In situ* Measurement Requirements

Although this experiment does not drive specific sampling site selection, it does impose *in situ* measurement requirements: the ability to detect sampleable material consistent with the required rock size criteria and the identification of regions within the local area that contain suitable sampling materials. Magnetometer data support identification of sampling regions and interpretation of paleomagnetic data by characterizing the magnetic environment from which viable paleomagnetic samples are collected. In addition, magnetic field measurements enable opportunistic sampling of areas associated with elevated magnetic anomalies, where scientifically informative materials may be concentrated.

Detection of sampleable material can be achieved through a combination of Measurement 1 (M1: Context Imaging), Measurement 2 (M2: Context Mineralogy), Measurement 3 (M3: Fine-Scale Mineralogy), Measurement 4 (M4: Bulk Elemental Chemistry), Measurement 6 (M6: Fine-Scale Imaging), and Measurement 7 (M7: Magnetic Field Measurements). Imaging and mineralogical measurements enable assessment of whether the rover workspace contains materials appropriate for sampling. Spatial resolution requirements for reconnaissance and sampling phases are provided in Section 4.2.

### 4.1.7 Experiment 7: Determine the Composition and Age of Farside Th-Rich Materials

One of the most distinct geochemical signatures of the Moon, KREEP (potassium, rare Earth elements, and phosphorus), is interpreted to form from the last, incompatible-rich dregs of the magma ocean. Although the signature is only observed in a few Apollo basalt samples, orbital remote sensing observations of Th (Gillis et al., 2002) suggest that this material is concentrated in large quantities in the nearside Procellarum KREEP Terrain (Jolliff et al., 2000). As discussed in Goal 2, in particular Objective 2.2 (Section 3.2.2), these incompatible elements play a key role in the initial differentiation of the Moon through crystallization of a lunar magma ocean, and provide insight into (and perhaps even drive) subsequent gravitational restructuring and thermochemical evolution of the lunar interior, as well as hemispherical differences between the lunar nearside and farside. While the presence and nature of these late-stage cumulates on the farside is unknown, Th-bearing materials are an important target. Th is selected as the tracer for these potential materials because it is strongly linked to KREEP signatures on the nearside, and because orbital datasets returned by instruments like Lunar Prospector allow for targeting of regions with potential concentrations.

SPA exhibits a crescent-shaped Th distribution around its rim, with the highest Th abundances observed towards the NW (Figure 4.10) (e.g., Lawrence et al., 1999). Generally, local maxima are associated with ~30-100 km impact structures, while local minima are associated with mare basalts, large basins (including Apollo, Schrodinger, and Ingenii), and the central SPA compositional anomaly (SPACA, see E13 for more detail) (Moriarty et al., 2021). The Th-bearing materials appear to be correlated with enhanced Fe, Ti, K, and Ca, Fe-rich pyroxenes (Moriarty et al., 2021). Although alternative hypotheses (such as antipodal Imbrium ejecta) have been proposed, recent analyses suggest that the Th-bearing materials could be late-stage lunar magma ocean cumulates excavated and ejected during SPA formation. If so, then Th serves as a direct observation of the current distribution of ancient, deep materials revealed by the SPA-forming impact. Collecting such materials would provide invaluable insight into the early Moon.

#### 4.1.7.1 E7: Science Objective Traceability

Collection of a Th sample and *in situ* measurement of Th-rich material is directly relevant to Objective 2.2.

**Objective 2.2** seeks to determine if there was a farside KREEP reservoir or other late-stage magma ocean products prior to the SPA impact. Collection of a Th-rich sample will contribute to knowledge of KREEP materials on the farside and help to constrain hypotheses around lunar evolution.

#### 4.1.7.2 E7: Sample Requirements

A sufficient amount of sample must be collected to measure the age of the sample as well as to measure its geochemical and petrologic properties. **Because the nature of the Th carrier is unknown, the SDT recommends collection of rocklets from all size categories (≥ 8mm, ≥ 3mm, and bulk regolith), to ensure the widest availability of laboratory analyses can be completed.** Rocklets with ≥8 mm diameter provide the mass, textural context, and mineralogical variability needed to obtain high-precision crystallization ages using Ar-Ar, Pb/Pb and/or Rb/Sr, as well as other isochron systems such as Sm–Nd and Lu–Hf (e.g., Fernandes et al., 2013; Cohen & Coker, 2010; Mercer et al., 2015). The smaller size fragments (≥ 3mm diameter) are sufficient in mass to provide material for thin sections, bulk rock determinations of major, minor, and trace elements, and additional petrologic analyses. These petrologic analyses include determination of crystallization ages of samples via the Pb-Pb system using SIMS, petrological characterization and mineral chemical analyses via electron microprobe/SEM WDS and SDD-EDS, and mineral trace elemental abundances via LA-ICP-MS. These fragments also provide sufficient mass for Ar-Ar dating, which will provide additional information regarding the crystallization age and/or shock history of these samples. Finally, bulk regolith samples (≥ 100 g) target the fine-grained fraction, this is expected to contain impact melt glass shards, mineral fragments, and accessory phases such as zircon and baddeleyite (e.g., Nemchin et al., 2009; Grange et al., 2011; Crow et al., 2017; Zhang et al., 2021a). These phases can record both impact-related and inherited ages, enabling investigation of SPA melt formation ages as well as evidence for older crustal components entrained during basin formation.

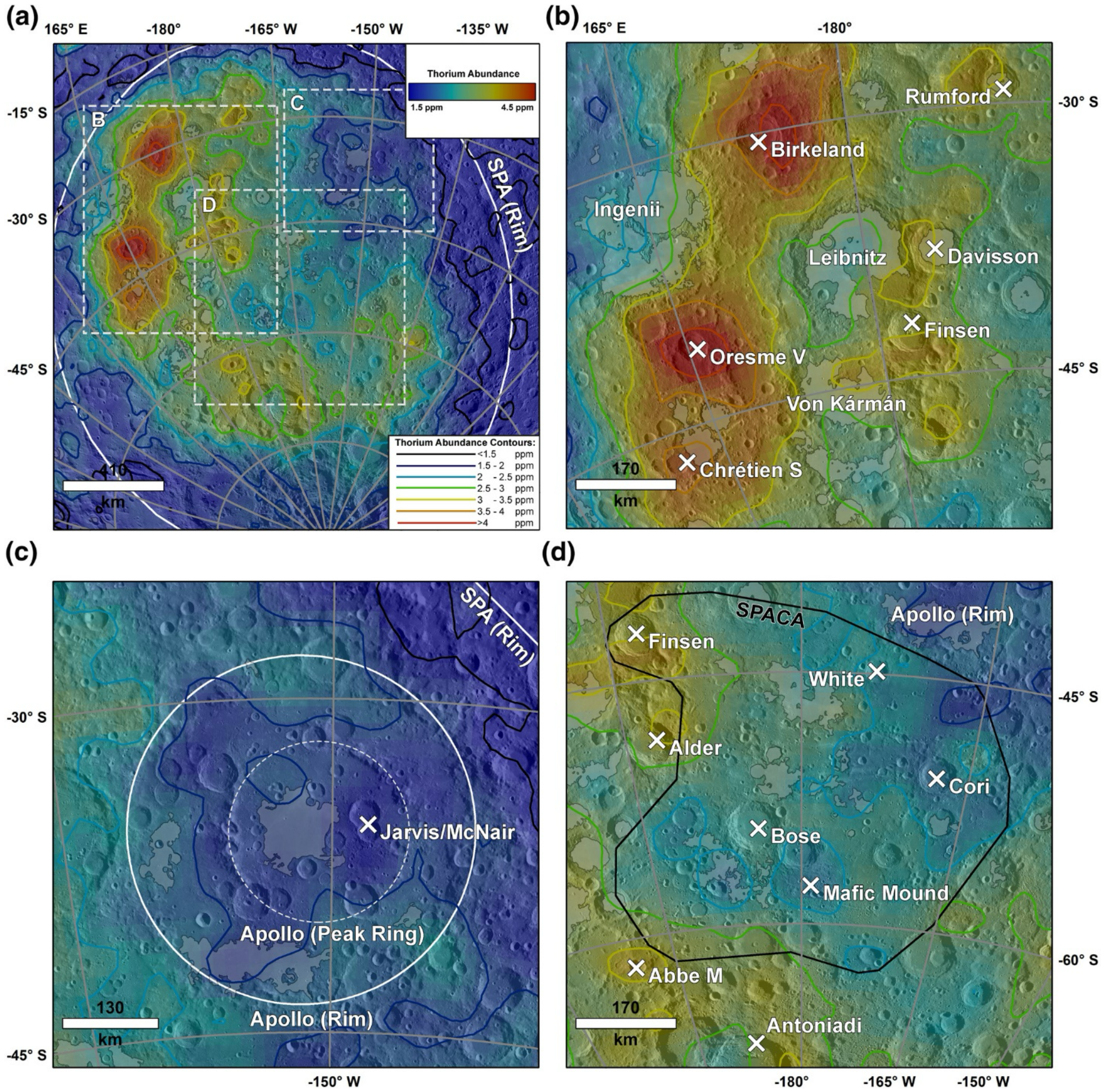

**Figure 4.10.** *Thorium abundance around SPA as revealed by Lunar Prospector (Moriarty et al., 2021).*

#### 4.1.7.3 E7: Total Baseline Sample Requirements per Sampling Location

- ≥32 rocks ≥ 8 mm diameter
- ≥400 rocks ≥ 3 mm diameter
- ≥400.0 g bulk regolith

The sampling requirements for this experiment also include sample acquisition sufficient to satisfy NASA sample curation requirements. Total sample numbers and overall sample mass reflect the sum of these two sets of requirements.

### 4.1.7.4 E7: Site and Traverse Requirements

**This experiment requires access to Th-bearing material**; the baseline available Th content of the desired material (Th > 3 ppm) is based on preexisting orbital datasets (Naito et al 2019) (Figure 4.11). Successful detection methods for Th-bearing material do not use visible imaging but instead rely on high-energy radiation detection (see 4.2.6) and increased proximity to the Th-bearing material. Sample selection of Th-bearing material will likely rely on a meticulous, anomaly-detection driven traverse; identifying Th-bearing material is driven by the radiation signature (i.e., Th gamma-ray signature) count rate corresponding to that of > 3 ppm within the mobility platform workspace (< 1 $m^3$). During E7, the velocity limit of the mobility platform is therefore partially defined by the Th-bearing material and its abundance (i.e., rate of decay). The SDT recommends an initial traverse target selection utilizing preexisting orbital datasets, followed by a detailed "anomaly-detection traverse" using heritage instrumentation is to select a 100 m x 100 m region that has registered a Th signature (Section 4.2.6), then to perform an area sweep in 1 m x 1 m blocks until a "hot spot" is detected within that region (e.g., 3 ppm+ Th signature). Further discussion of this notional ConOps is discussed in Section 6.6.2.

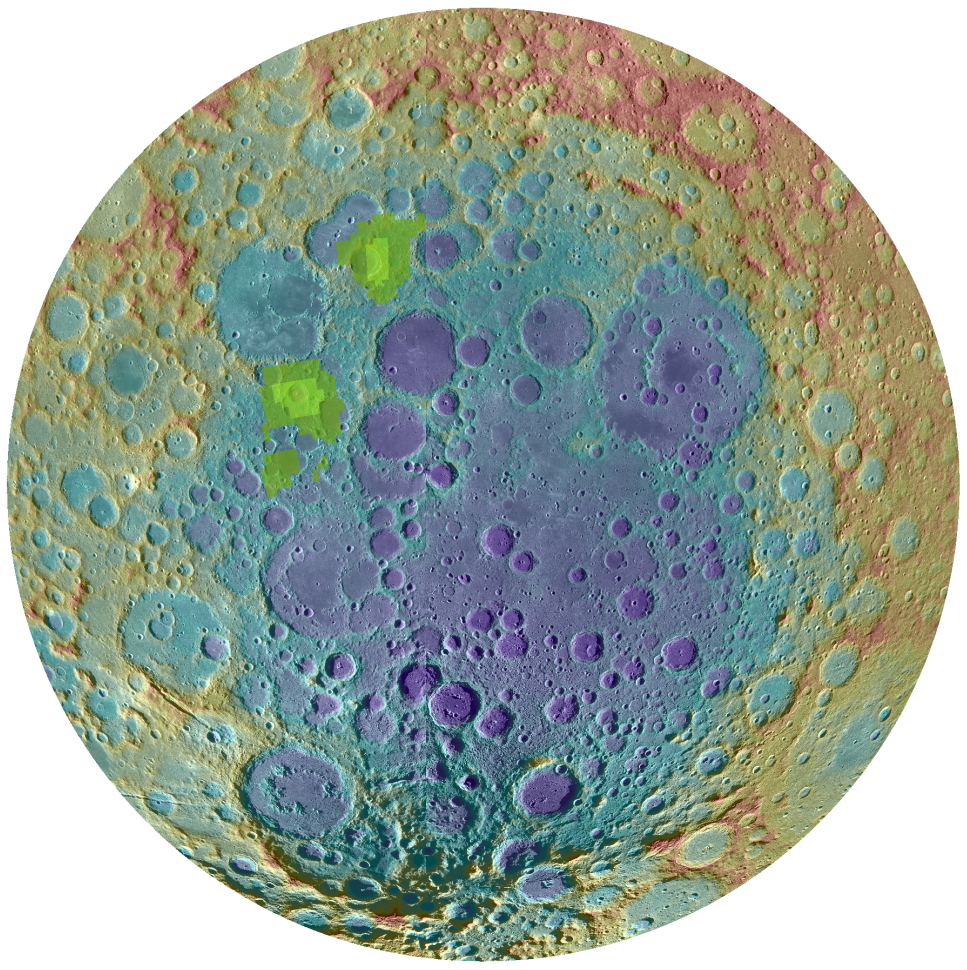

**Figure 4.11.** *Regions with concentrations of Th ≥3 ppm as derived from Lunar Prospector.*

### 4.1.7.5 E7: *In situ* Measurement Requirements

The main *in situ* measurement requirement for this experiment is Measurement 4: Bulk Elemental Chemistry (4.2.6) specifically for measuring the Th content of sample material, context measurements of the surrounding material are also required using Measurement 1 (M1: Context Imaging) and Measurement 2 (M2: Context Mineralogy). The *in situ* measurement will be required to register a Th anomaly (Th > 3 ppm) within a 1 $m^3$ workspace in accordance with the velocity of the mobility platform (approx. 1 kmph). This requirement translates to the instrument detection efficiency and is described in detail in Section 4.2.6: Bulk Elemental Chemistry.

## 4.1.8 Experiment 8: Determine the Composition and Age of Farside Basaltic Units

The Moon presents many asymmetries that can be readily observed from the surface, and perhaps the most visible of these is the distribution of mare basalts. While the familiar nearside of the Moon is marked by a plethora of large mare deposits, the lunar farside hosts almost none. The few mare deposits that are located on the farside are significantly smaller in spatial extant than the nearside, and many of these small deposits are located in craters within South Pole-Aitken basin (Figure 4.12). The volcanic origin of these mare deposits hints that the asymmetric processes which resulted in their formation extend into the lunar interior. The standard lunar magma ocean (LMO) model presents the entire Moon as being molten and solidifying (Wood et al., 1970), and before complete solidification, there is an overturn, or inversion, of the mantle layer. However, this initial LMO model does not necessarily explain the observed asymmetries in surface deposits. Several processes have therefore been proposed to explain why the asymmetry in mantle-derived products exists, including gravitational instabilities associated with mantle overturn (Parmentier et al., 2002), variations in lithospheric thickness (Head et al., 2024), and the formation of SPA (Zhang et al., 2022). By using mare basalt deposits as a tracer of mantle activity, we can investigate the compositional evolution through time.

#### 4.1.8.1 E8: Science Objective Traceability

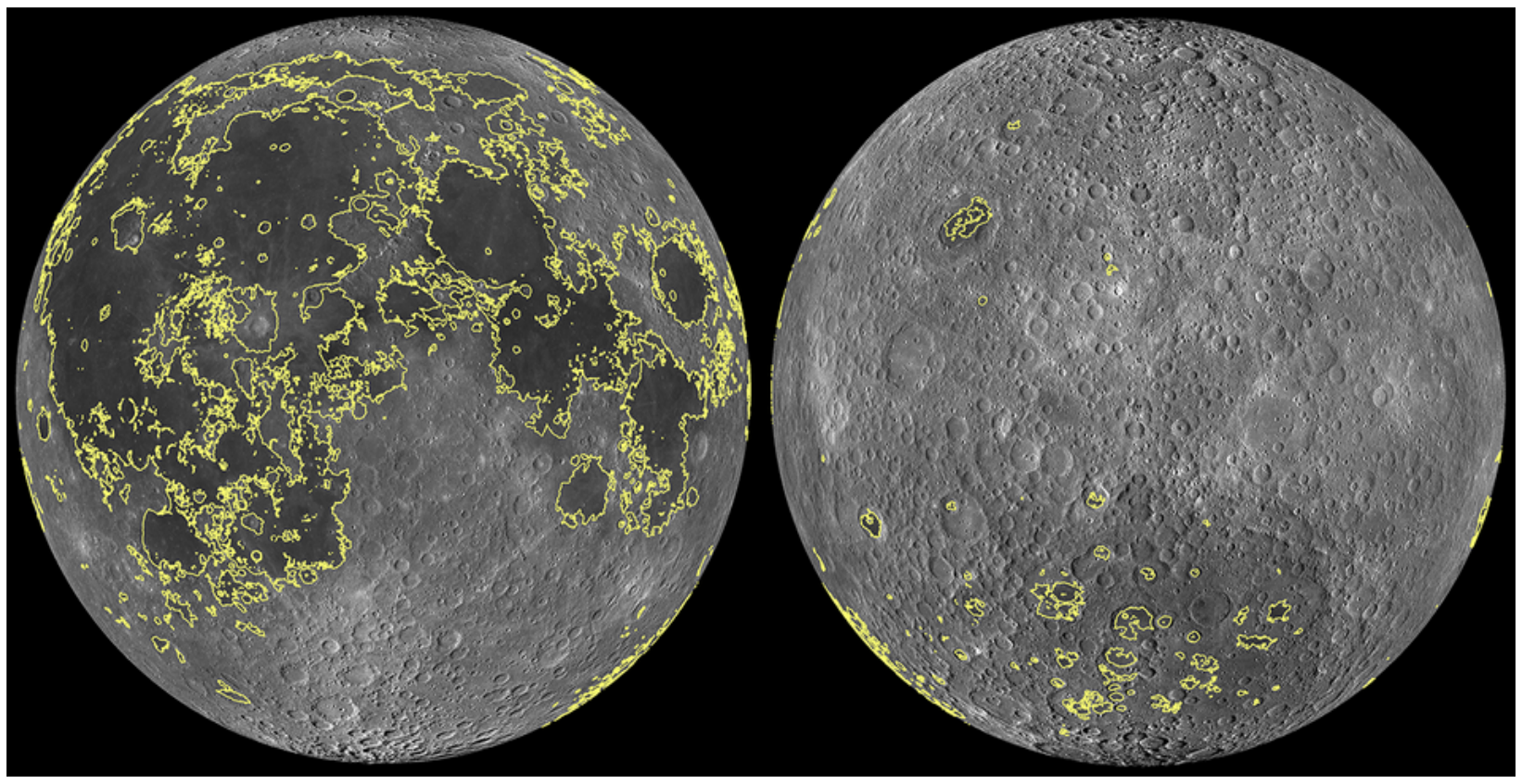

**Figure 4.12.** *Distribution of mapped mare deposits (yellow outline) on the lunar nearside and farside. (Figure credit: LROC/NASA/GSFC/Arizona State University).*

Determining the composition and age of farside basaltic units is directly relevant to Objective 2.3.

**Objective 2.3** requires determining if and how the temporal and compositional evolution of the farside mantle differs from the nearside. Sampling the age and composition of several mare basalts in the South Pole-Aitken Basin will provide critical information for determining how the farside mantle evolved over time, which, in turn, will help to identify the process responsible for the observed asymmetries. These results can then be compared with Apollo and Luna samples to evaluate whether there are substantial differences between nearside and farside basalts – in terms of their compositions and/or the timing of any compositional changes. Similarly, determining the ages of returned samples will provide a temporal view of these compositional changes. These comparisons will begin to address the larger mission goal of whether an asymmetric magma ocean and crustal formation could have caused the observed differences between the lunar nearside and farside.

#### 4.1.8.2 E8: Sample Requirements

Mare basalts are typically among the easier samples to work with from a geochemical and chronological standpoint. Unbrecciated and low-shock samples are typically abundant at mare basalt sampling locations, basalts naturally contain multiple primary minerals, and their formation as mantle partial melts concentrates incompatible elements from the mantle source into the magma. These factors all make chronologic and geochemical measurements easier to perform and interpret as compared to the majority of ancient lunar crustal samples, which are typically much more strongly affected by brecciation, shock, and thermal annealing; as well as, the fact that they contain lower abundances of the elements of interest.

There are, however, some constraints that place requirements on the returned samples of the ≥3 spatially and temporally distinct mare basalt units. This experiment **requires ≥20 rocks with a diameter of ≥5 mm diameter to be collected at each site** (sampling rationale provided in Section 4.1.3.2). As with all sample requirements described in this report, these size requirements are designed to reflect the sample mass required for each recommended type of laboratory analysis (Section 4.3). Additionally, for this experiment, the number of required rocks incorporates impact mixing models which show that between 3–40% of material from mare basalt sampling sites may be exogenous, although this is very dependent on location (e.g., Cao et al. 2025, Su et al. 2025, Xu et al. 2024, Xie et al. 2020, Huang et al. 2017), so final sample requirements have been designed here assuming 20% exogenous materials (Section 5.8.3, Foldout 5.1).

Rock fragments ≥5 mm diameter will be targeted for Sm-Nd, Rb-Sr, and Lu-Hf isochron dating methods, as these fragments will be of sufficient mass (100s of mg) to enable these methodologies. Isochron age determinations for the basalt sample(s) present at each site are required in order to provide the data necessary for calculation of model ages of the mantle source(s) and for fulfillment of Experiment 8 (E8), which utilizes the same sample set as E3. When rock mass, elemental composition, and textures allow, it may be possible and advantageous to use the same rock for isochron ages for multiple isotopic systems, bulk composition determinations, and to make thin sections for subsequent analyses.

Targeting the collection of ≥20 rocks with a diameter of ≥5 mm diameter is expected to result in the **collection of ≥100 rocks of ≥3 mm diameter** based on typical lunar regolith particle size distributions (e.g., Carrier, 2003). These smaller size fragments are sufficient in mass to provide material for thin sections, bulk rock determinations of major, minor, and trace elements, and *in situ* analyses. Laboratory analyses include determination of crystallization ages of samples, petrological characterization, mineral chemical analyses, and mineral trace elemental abundances. These fragments also provide sufficient mass for Ar-Ar dating, which will provide additional information regarding the crystallization age and/or shock history of these samples.

**Bulk regolith (≥ 100 g)** that will be scooped at these sites is expected to contain a fine-grained fraction that will include zircons and other mineral fragments (Barboni et al., 2024, 2017; Crow et al., 2017; Dauphas et al., 2025; Greer et al., 2021; Meyer et al., 1996; Nemchin et al., 2009; Zhang et al., 2021b, 2021a). These zircons will be identified and concentrated to allow chronologic studies. Furthermore, lunar regolith, particularly at mare sites, is expected to contain volcanic pyroclastic glass beads that will be used for Experiment 10 and the study of lunar volcanic history, mantle heterogeneity, volatile contents, and the thermal history of the farside mantle.

#### 4.1.8.3 E8: Total Baseline Sample Requirements per Sampling Location

- 80 rocks ≥5 mm diameter
- 400 rocks ≥3 mm diameter
- 400 g bulk regolith

The sampling requirements for this experiment also include sample acquisition sufficient to satisfy NASA sample curation requirements. Total sample numbers and overall sample mass reflect the sum of these two sets of requirements.

#### 4.1.8.4 E8: Site and Traverse Requirements

This experiment requires samples from **three separate basaltic units**, defined as having **orbitally distinct ages and compositions**. In certain portions of SPA, such as Ingenii, there are multiple adjacent mare deposits with distinct age and compositional characteristics. For most of the basin, however, distinct mare basalts are separated by 100s of kms. Identification of locations for sampling mare basalts should utilize a combination of compositional orbital remote sensing data (e.g., $M^3$, Kaguya), higher-resolution orbital image data (e.g., LROC-NAC), and refer to previous publications that present crater model ages for mare basalts (e.g., Pasckert et al., 2018) (Figure 4.13). Compositional orbital remote sensing datasets can identify locations where, at the scale of the instrument resolution, there are compositionally distinct mare basalts. Crater statistics can be used to identify mare basalts with potentially distinct model ages. These ages and compositional boundaries can either be sourced from the literature (Ivanov et al., 2018) or determined by a future mission team. Within each mare basalt, it is important to identify regions where rocks are likely, such as the ejecta deposits of small fresh craters. This can be achieved using high resolution imagery. Finally, within the scale of basaltic deposits, there is little evidence for compositional variation, therefore, once a unit is selected for sampling, the exact location within the unit can be flexible to broader traverse requirements and path optimization.

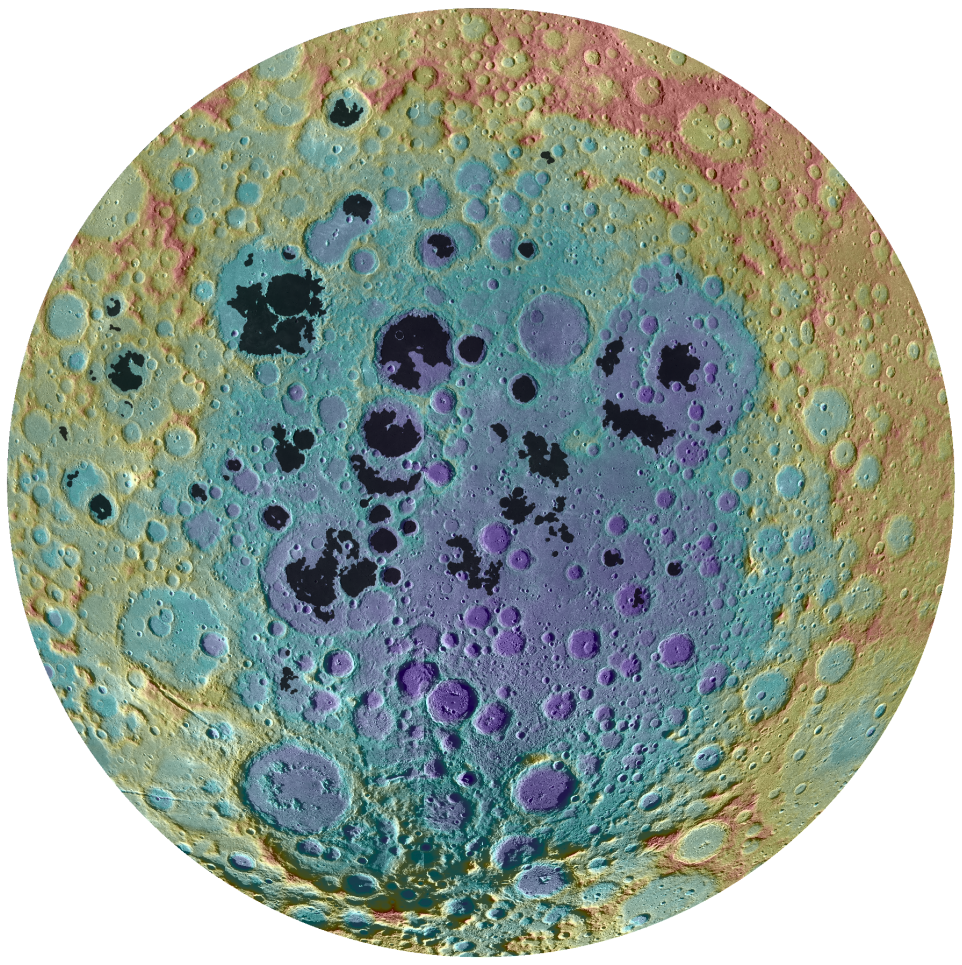

**Figure 4.13.** *Distribution of mare units in SPA from LROC-WAC. (Figure credit NASA/GSFC/ASU).*

#### 4.1.8.5 E8: *In situ* Measurement Requirements

Contextual images of the regions of interest are required to identify a workspace where samples can be collected, requiring Measurement 1 (M1: Context imaging). The team needs to be assured that approximately 20 rocks >5 mm can be collected at each region of interest. These are the same requirements as described in E3 (Section 4.1.3.5).

### 4.1.9 Experiment 9: Sample Hypothesized Excavated Lunar Mantle Material

Despite the wealth of lunar samples returned by the Apollo missions, and samples of lunar meteorites, there are no known samples of the lunar mantle. Our current understanding of the lunar mantle instead derives from geophysical measurements, petrologic investigations of samples derived from mantle melts (e. g. mare basalts), and laboratory experiments designed to synthetically simulate mantle compositions. While all of these processes can provide suggestions about the composition and evolution of the mantle, all are subject to a variety of model-dependent assumptions. The result is that there is no consensus about what the mantle looks like at any specific point in lunar history, including whether the nearside and farside mantle are, or were ever, uniform in composition. Collecting actual mantle material will allow a greater understanding of how the lunar interior evolved and what processes may have contributed to the lunar asymmetry. As the largest, and most ancient impact basin, the South Pole-Aitken basin has the strongest chance to have excavated into the lunar mantle, with the subsequent impact history providing further opportunities for deep mantle material to be excavated in basin ring structures and central peaks (Isaacson et al., 2011) (Figure 4.14).

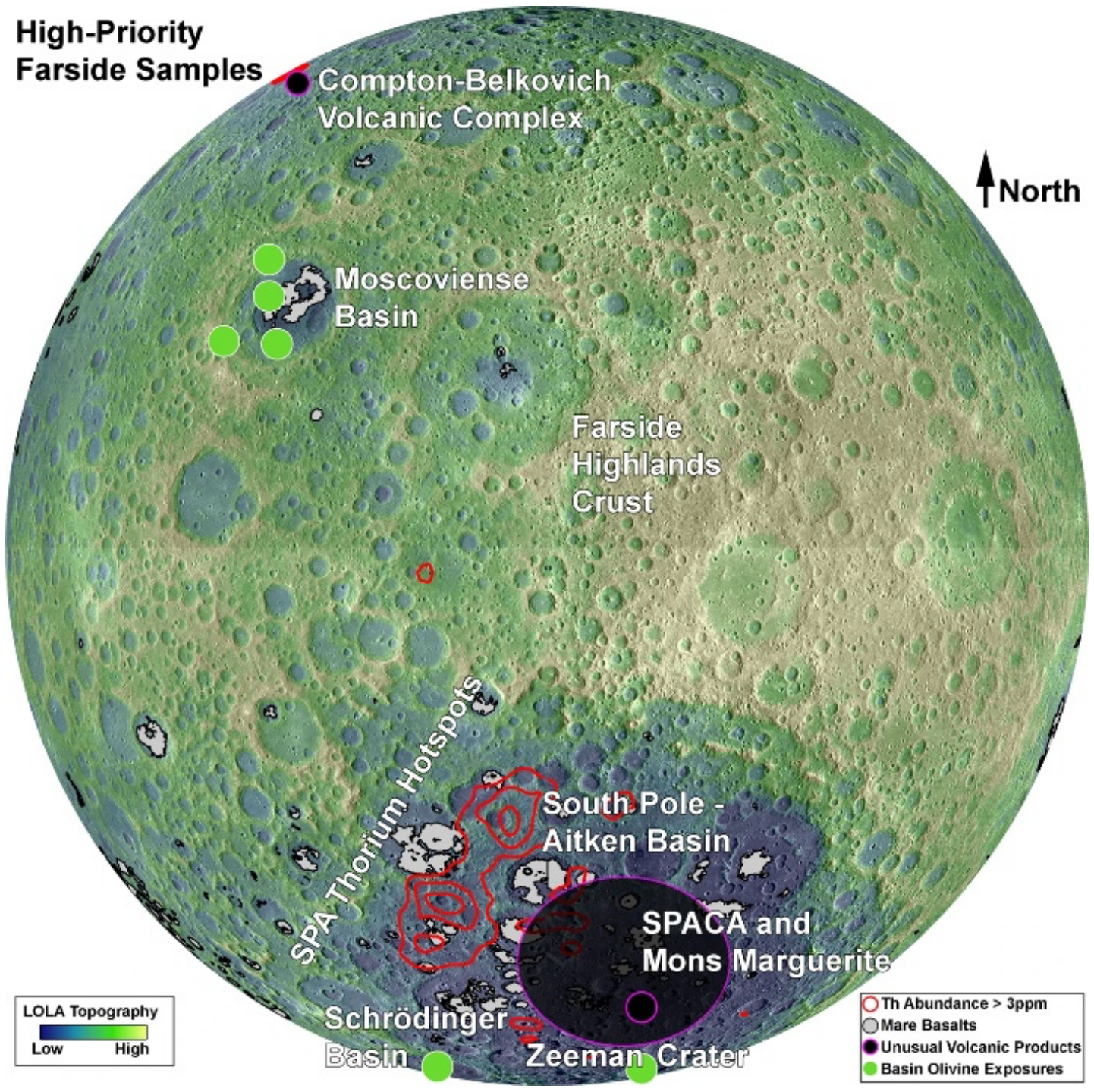

**Figure 4.14.** *SPA hosts a diversity of high-priority sample targets, such as candidate mantle materials including high-Th materials and olivine excavated by large impacts (Moriarty et al., 2021a).*

#### 4.1.9.1 E9: Science Objective Traceability

By seeking to sample hypothesized excavated mantle material, Experiment 9 (E9) is directly relevant to Objective 2.3.

**Objective 2.3** will determine if and how the temporal and compositional evolution of the farside mantle differs from the nearside. The goal of this experiment is to sample locations where the planetary community has identified possible excavated mantle material (e.g., Melosh et al., 2017; Yamamoto et al., 2010; 2023; Andrews-Hanna et al., 2025; Gowman et al., 2026; Moriarty et al., 2021a). The petrologic and chemical measurements made in the lab on SPARX samples will be able to provide insight into whether the returned samples represent mantle material, differences exist between the nearside mantle and the farside mantle, as well as, the implications of those measurements for an asymmetric magma ocean.

#### 4.1.9.2 E9: Sample Requirements

A sufficient amount of sample must be collected to measure the age of the sample as well as to measure its geochemical and petrologic properties, this includes rocks with a range of diameters and also bulk regolith.

**Larger rocks, ≥20 mm in diameter, are required for E9**—as these rocks provide the minimum sufficient mass, mineralogic variation, and sampling options necessary to enable determination of isochrons of highest precision using U-Pb, Rb-Sr, Sm-Nd, and Lu-Hf (e.g., Papanastassiou and Wasserburg, 1975; Sharma et al., 1996; Premo and Tatsumoto, 1993), as well as the geochronometers Pb-Pb and Ar-Ar. From any given location, three such large samples are required to provide a minimum assessment of compositional and chronologic heterogeneity. The requirement for three samples reflects the difficulty in collecting samples that are olivine- and/or low Ca pyroxene-rich compatible with the isotopic age dating process. The large samples will be used for identifying and isolating mineral separates that may provide a spread of parent/daughter element ratios, thus further reducing the age uncertainty.

E9 also requires a large number of smaller rocks sieved from the lunar regolith. These smaller rocks **(≥3 mm in diameter**) would be utilized to determine crystallization and model ages using chronometers (Rb-Sr, Ar-Ar, and U-Pb) (e.g., Papanastassiou and Wasserburg, 1975; Sharma et al., 1996; Cohen and Coker, 2010). Coordinated measurements of **at least 100 particles per site** will yield ages using these complementary chronometers to those used on the larger rocks. Because these samples will potentially contain a mixture of lithologies (and some non-ultramafic) mass estimates are based on our expectations and understanding of Apollo regolith that ~50% of the material sieved will be olivine- or low Ca pyroxene-rich (e.g., McKay et al., 1991; Carrier, 1973).

**At least 100 g of bulk regolith** will be scooped at these sites, and bulk regolith is expected to contain a fine-grained fraction that will include zircons and other mineral fragments (e.g., Nemchin et al., 2009; Meyer et al., 1996; Crow et al., 2017; Barboni et al., 2017; 2024; Zhang et al., 2021, Greer et al., 2023; Dauphas et al., 2025). These zircons will be identified and concentrated to allow chronologic studies, and in particular, to look for ages and compositions that are consistent with origin from an ultramafic rock rather than crustal or basaltic lithology. The number of zircons recovered in Apollo regolith samples has a very small mass compared to the mass of the volume of material searched and thus most of the material remains unused (e.g., Taylor et al., 2009; Crow et al., 2017) and available for future scientific investigations.

#### 4.1.9.3 E9: Total Baseline Sample Requirements per Sampling Location

- ≥12 rocks ≥20 mm diameter
- ≥400 rocks ≥3 mm diameter
- ≥400 g bulk regolith

The sampling requirements for this experiment also include sample acquisition sufficient to satisfy NASA sample curation requirements. Total sample numbers and overall sample mass reflect the sum of these two sets of requirements.

#### 4.1.9.4 E9: Site and Traverse Requirements

The baseline mission requirement of **targeting at least two locations for collecting hypothesized mantle samples** provides an opportunity to interrogate potential mantle diversity, especially if the samples are collected 100's of kilometers apart. Importantly, by targeting two locations for samples, the mission increases the probability of collecting mantle material. While significant efforts have been undertaken to identify candidate mantle materials exposed on the lunar surface using orbital datasets, there is currently no broad consensus in the lunar science community regarding what this mantle material should be (discussion summarized in Moriarty et al., 2021a). For this reason, the hypothesized mantle exposure sites proposed in the literature cannot be definitively identified as mantle materials until they are sampled. Therefore, it is prudent to collect more than one type of candidate mantle material to increase the likelihood of actually sampling the lunar mantle.

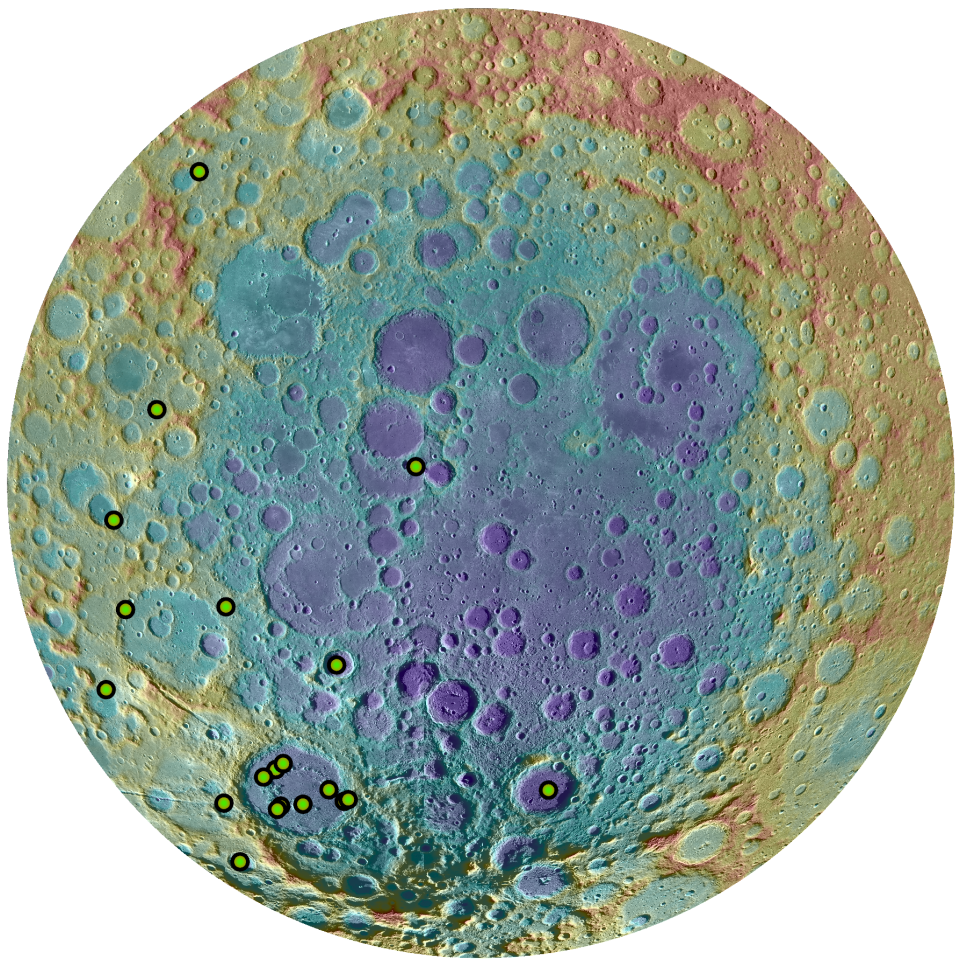

**Figure 4.15.** *Locations of hypothesized exposed mantle materials in the form of Olivine and low-Ca pyroxene from Yamamoto et al. 2023.*

Sample sites must be selected from proposed sites of mantle materials, whether they are specific points on the surface (e.g., Yamamoto et al., 2023) (Figure 4.15) or larger regions of expected mantle material (e.g., Melosh et al., 2017; Moriarty et al., 2021b, Andrews-Hanna et al., 2025; Gowman et al, 2025). Rather than identifying specific lithologies, many of these studies identify locations of specific mineralogies, in particular olivine and low-Ca pyroxene. Thus, identification of potential mantle material sampling locations should utilize compositional remote sensing data (e.g., $M^3$, Kaguya, Lunar Prospector). These datasets can identify regions where, at the scale of the instrument resolution, one of the proposed mantle lithologies may be present. However, this scale is generally too coarse for detailed traverse planning. Following general identification with compositional datasets, additional higher spatial resolution data should be used to refine potential sampling workspaces, such as sites of rock exposures. For proposed mantle materials, these could include boulder fields at the base of central peaks, crater rims, or peaks rings.

#### 4.1.9.5 E9: *In situ* Measurement Requirements

*In situ* measurements for this experiment rely heavily upon identification of hypothesized mantle compositions. Within the broad region, Measurement 2 (M2: Context Mineralogy) should identify the target composition for the traverse site (e. g. olivine or low-Ca pyroxene). Although not necessarily needed, geologic and bulk compositional context measurements from Measurement 1 (M1: Context Imaging) and Measurement 4 (M4: Bulk Elemental Chemistry) are valuable contributors to help understand the geology of the area at a landed vehicle scale and place the sampled material in sufficient context. Once confirmation of one or both materials is established through Context Mineralogy, then imagery needs to be used to select a workspace where samples can be collected. The sampling requirement for rocks ≥ 20mm in diameter, should drive selection of a viable workspace.

### 4.1.10 Experiment 10: Determine the Volatile Content of the Farside Mantle as Recorded by Pyroclastic and Basaltic Materials

Volatile elements and compounds (e.g., H, $H_2O$, $CO_2$, S, Cl, F) serve as critical tracers for the formation, thermal evolution, and volcanic history of the Moon. These elements and compounds, which preferentially exist in the gaseous form at the lunar surface, are sequestered in the lunar mantle during the Moon's formation, and provide critical information about how the Moon formed, and from what material. While the discovery of indigenous water in Apollo pyroclastic glass beads overturned the decades-long "bone-dry" lunar paradigm (e.g., Taylor et al., 2006; Saal et al., 2008), our current understanding of these elements is primarily limited to the Moon's nearside. Experiment 10 investigates the hemispheric dichotomy of the Moon by determining if the farside mantle is geochemically distinct from the nearside. Recent analyses of the water abundances and H isotopic compositions of apatite and melt inclusions in a basalt returned from the Chang'e-6 mission suggest the farside mantle is drier than the lunar nearside, potentially supporting the hemispheric dichotomy of the Moon (He et al., 2025) (Figure 4.16). However, additional data are required to understand if this measurement is indeed representative of the farside mantle, or if it is anomalous. Experiment 10 is essential for understanding the Moon's abundance, distribution, and origins of lunar magmatic volatiles (e.g., McCubbin et al., 2023).

#### 4.1.10.1 E10: Science Objective Traceability

Determining the volatile content of the farside mantle as recorded by pyroclastic and basaltic materials is directly linked to Mission Objective 2.3.

**Objective 2.3** seeks to determine if and how the temporal and compositional evolution of the farside mantle differs from the nearside. Measuring the concentration of volatile species ($H_2O$, S, Cl, F) in farside samples establishes the chemical baseline of the mantle source, allowing for a direct comparison with the reservoirs observed on the nearside. By analyzing both basaltic flows and pyroclastic deposits, the experiment tracks how the mantle's volatile inventory changed over geologic time, revealing the thermal cooling history of the farside. This data set acts as the primary counterpoint to Apollo-era nearside data, enabling the first comprehensive, global model of lunar mantle heterogeneity.

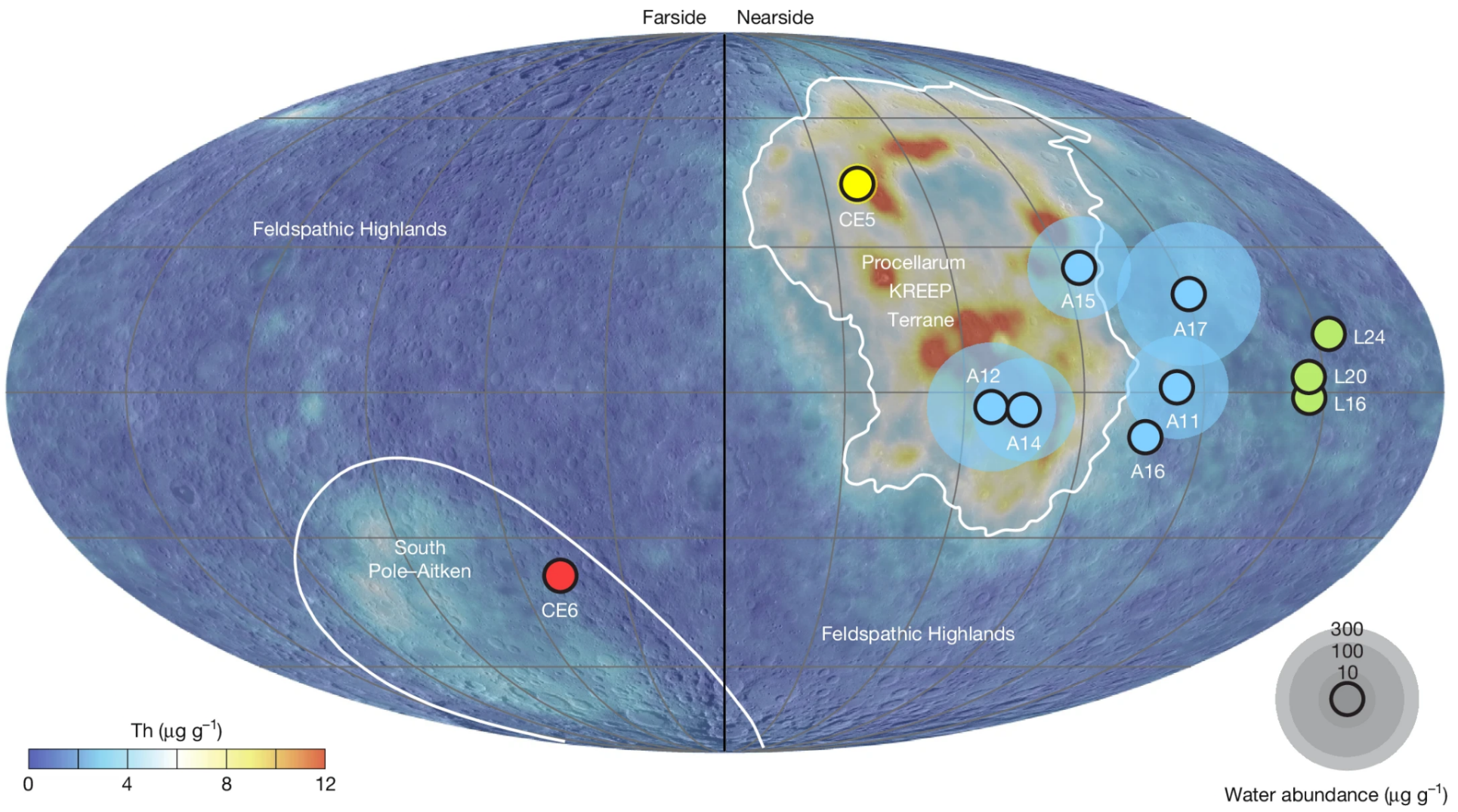

**Figure 4.16.** *The Apollo (prefix A), Luna (prefix L), and Chang'e-5 (CE5) missions collected samples in and around the Procellarum KREEP Terrane which exhibits notably higher thorium (Th) than the Feldspathic Highlands and the SPA Basin. The Chang'e-6 (CE6) mission returned the first lunar samples from the lunar farside from the SPA Basin. The terrane boundaries are from ref. 5 and the Th base map is adapted from ref. 45. The sizes of the circles are scaled to the estimated maximum water abundance1,9,33 in the mantle source beneath the sampling sites. (Figure credit He et al. (2025).*

#### 4.1.10.2 E10: Sample Requirements

To constrain the volatile content of the farside mantle, **two sets of samples are required**: (1) **sieved rock samples from farside mare basalt deposits** (E3, E8), and (2) **bulk regolith collected from a farside pyroclastic deposit**. The sieved rock sample requirements are 20 rocks ≥ 5 mm in diameter, and 100 rocks ≥ 3 mm in diameter. The slight increase in sample number requirement (compared with E2, for example) reflects that these materials contribute to both Experiment 3 and Experiment 10. To address Experiment 10, pyroclasts will be separated from ≥ 100 g of bulk regolith. Pyroclasts represent the deepest-sourced and most primitive magmas on the Moon (Green et al., 1975). Therefore, targeting and collecting 100 g of bulk regolith from a farside pyroclastic deposit, sieving said regolith material for pyroclasts (which typically range 50 to 100 µm in maximum axis), and analyzing the pyroclasts in terrestrial laboratories will be critical for constraining the farside mantle volatile budget.

All returned samples will be characterized petrologically, and analyzed for their elemental chemistry, geochronology, and volatile abundances. The returned samples will undergo petrologic characterization to identify mineral phases (i.e., apatite, a mineral that hosts OH, Cl, and F), mineral textures, and the presence of melt inclusions in major minerals (e.g., olivine). It is necessary to identify melt inclusions, which are trapped magma inside crystals, to understand the initial (undegassed) mantle volatile signatures. These analyses can be conducted on rocklets, polished sections, and suitable glassy pyroclasts separated from the bulk regolith. The samples will also be analyzed for their major element (Fe, Mg, Ca, K, Si, Ti, Al, Cr, Mn) abundances to understand the origin and evolution of the melt and to classify the basaltic suites; and trace element (Ni, Co, highly-siderophile element) abundances to identify depth of melting and mantle differentiation processes. These analyses can be conducted on rocklets, polished sections, and sieved pyroclasts. Geochronologic measurements are needed to determine the cosmic ray exposure (CRE) age of the samples to ±10 Myr (Section 4.3.3.7). This CRE age will be used to constrain the isotopic composition of volatiles in the samples. CRE ages are necessary to correct for potential solar-wind implanted H in samples, applicable to measurements of water in apatite. These analyses can be conducted on rocklets. Finally, samples will be analyzed for the abundances of volatiles (e.g., $H_2O$, Cl, F, S) and the isotopic composition of volatiles (e.g., D/H) to understand the volatile inventory in the farside interior and the history of degassing in the samples (e.g., Tartese et al., 2013)(Section 4.3.4). These analyses can be conducted on polished samples and suitable glassy pyroclasts.

#### 4.1.10.3 E10: Total Baseline Sample Requirements per Sampling Location

- ≥80 rocks ≥ 5 mm diameter (required for Experiment 3)
- ≥400 rocks ≥ 3 mm diameter (required for Experiment 3)
- ≥400 g bulk regolith from a pyroclastic deposit

Note that mare basalts are sampled as part of Experiment 3, so no additional mare basalt samples are required. The sampling requirements for this experiment also include sample acquisition sufficient to satisfy NASA sample curation requirements. Total sample numbers and overall sample mass reflect the sum of these two sets of requirements.

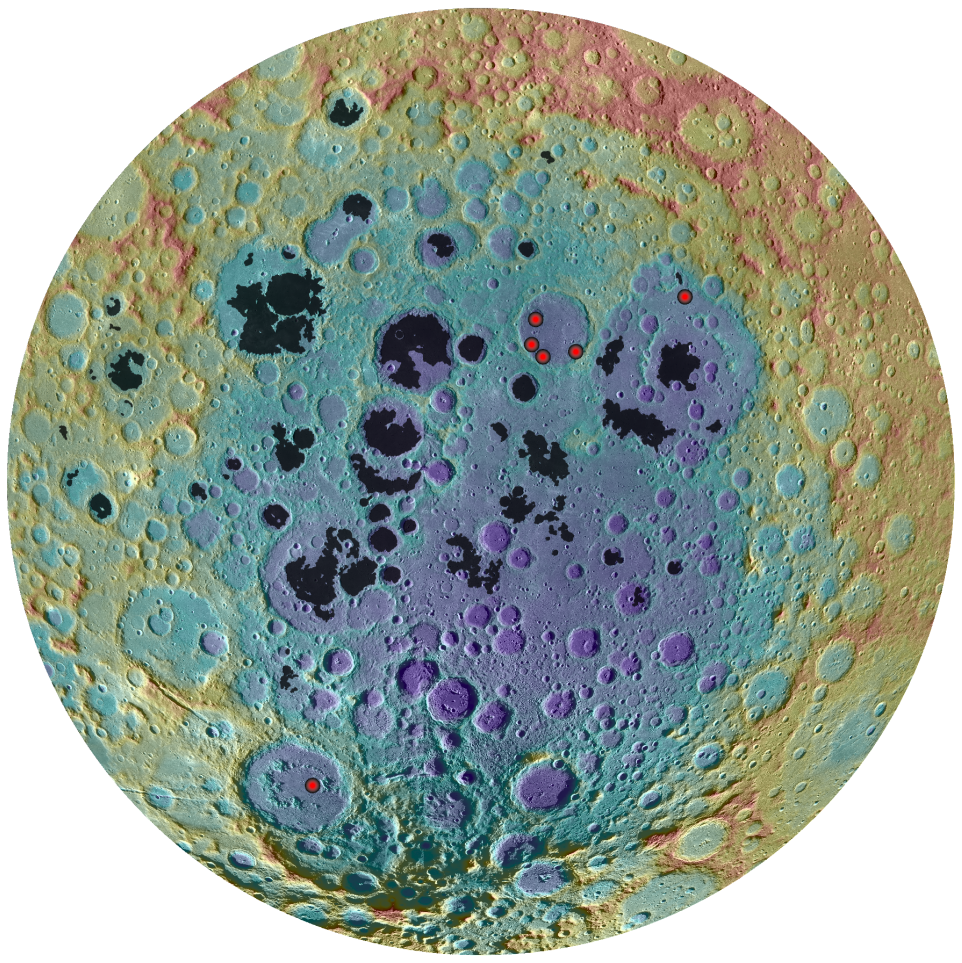

**Figure 4.17.** *Distribution of mare deposits (black) and pyroclastic deposits (red) within SPA. Pyroclastic deposit locations from Gaddis et al. (2003).*

#### 4.1.10.4 E10: Site and Traverse Requirements

As basaltic materials will be collected as part of Experiment 3, **no additional mass or sites to obtain sieved basaltic materials** are required for Experiment 10. In addition to the basaltic rocks, Experiment 10 requires **bulk regolith samples from ≥1 pyroclastic deposit(s)** to address baseline mission requirements. Lunar pyroclastic deposits have been well-characterized using orbital remote sensing datasets (e.g., Gaddis et al., 2000; Gaddis et al., 2003; Besse et al., 2014), and the selection of a potential traverse destination should come from a recognized database (e.g., Gaddis et al., 2003). Following the identification of a specific deposit, potential traverse and sample locations should be refined using a combination of compositional orbital remote sensing data (e.g., $M^3$, Kaguya), and higher- resolution orbital image data (e.g., LROC-NAC).

#### 4.1.10.5 E10: *In situ* Measurement Requirements

*In situ* measurement requirements for Experiment 10 require only measurements to support sample selection and characterization. Within a well-mapped pyroclastic deposit (e.g., Schrödinger), pyroclastic material is expected to dominate the regolith. Context imaging via Measurement 1 (M1: Context Imaging) will allow for the assessment of whether the rover workspace contains materials that are suitable for sampling. Although not necessarily needed, compositional context measurements from Measurement 2 (M2: Context Mineralogy) and Measurement 4 (M4: Bulk Elemental Chemistry) are valuable contributors to help understand the geology of the area at a landed vehicle scale and place the sampled material in sufficient context.

### 4.1.11 Experiment 11: Determine the Crustal Structure (e.g., Thickness, Density, Remnant Magnetic Field Strength) Across the SPA Basin (from the Center to the Putative Rim) to Constrain Mega-Basin Formation and Evolution Models

As the Moon's largest and oldest impact structure, the South Pole-Aitken (SPA) basin preserves a record of deep crustal excavation and large-impact processes that cannot be accessed elsewhere on the Moon. With a diameter in excess of 2500 km and a depth of up to 8 km, the SPA impact event forever altered the Moon's structure. Various compositional (e. g. Pieters et al., 2001) and geophysical studies (e. g. Wieczorek et al., 2012) suggest that the impact event excavated well into the lower crust, and possibly even the upper mantle of the Moon, resulting in a basin today with some of the thinnest remaining crust. While scars of the impact are visible on the surface, the larger structural effects, including how the crust responded to, and evolved from, such a large impact event are hidden beneath the surface (Figure 4.18). Geophysical investigation techniques are therefore required to study the subsurface structure of the basin, and provide valuable data that can be used to understand a whole host of basin properties resulting from, or responding to, the formation of the SPA mega-basin.

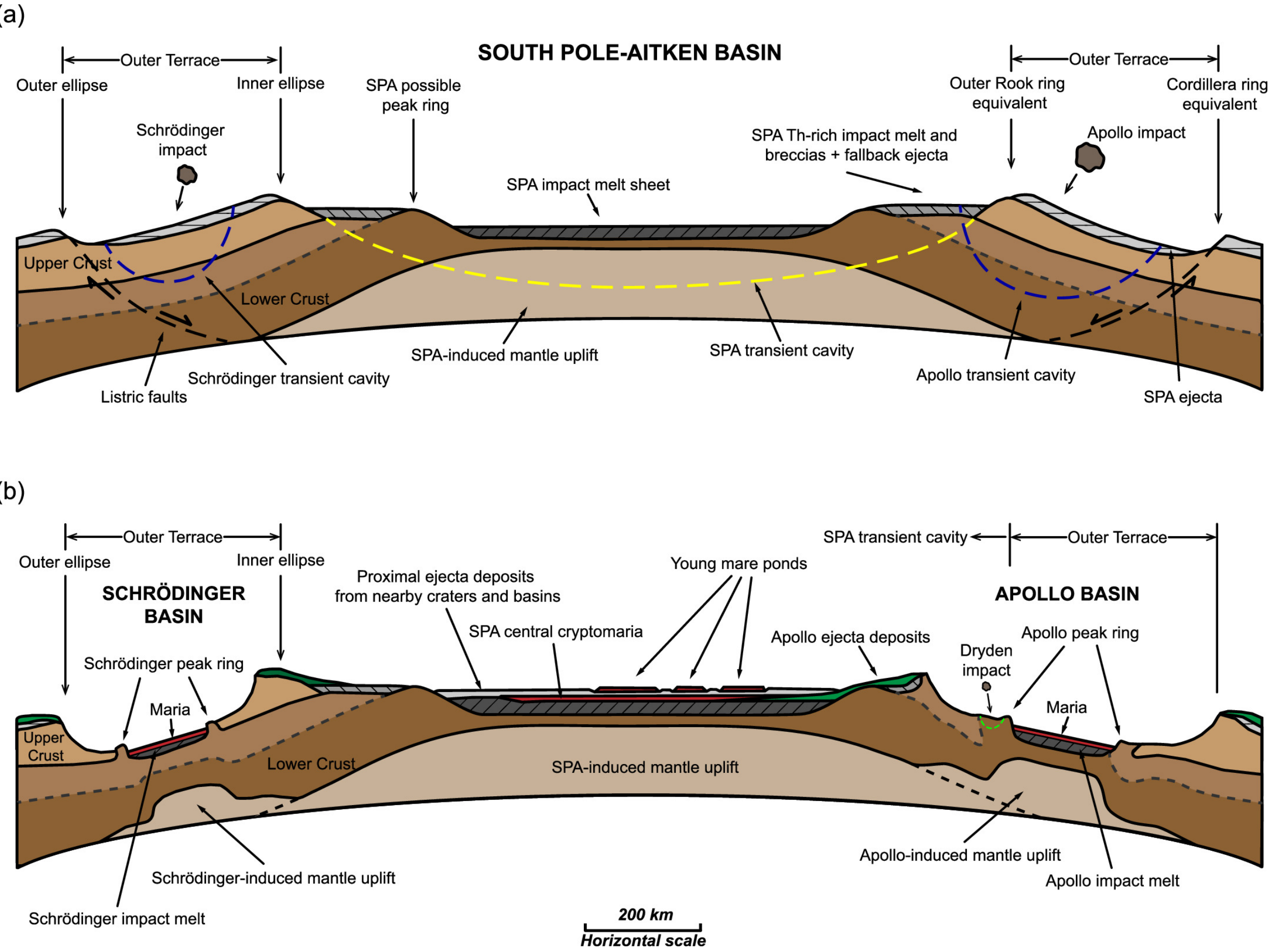

**Figure 4.18.** *Cross-sectional diagrams of SPA basin structure (a) before and (b) after Apollo and Schrodinger impact basin formation. A 650 km traverse from central SPA would sample multiple transitions in crustal thickness/composition. (Figure from Wang et al. 2024).*

A basin-scale (≥ 650 km) geophysical traverse across SPA basin will directly characterize variations in crustal properties, such as thickness, density, and remnant magnetization, from the basin center toward the putative rim, providing critical constraints on mega-basin formation and evolution models. Variations in crustal density constrain models of mega-basin formation and collapse by quantifying the relative contributions of impact melt, crustal backfill, and impact ejecta to the SPA interior. Measurements of crustal thickness gradients within the basin will further constrain the orientation and geometry of the SPA impact (e.g., Andrews-Hanna et al., 2025). Gravity measurements collected from the basin interior toward the rim will distinguish basin rim structure and constrain the overall shape of SPA, a critical parameter for narrowing the range of viable impact origin and post-impact evolution scenarios. Magnetic measurements acquired along the traverse will constrain the depth, strength, and origin of magnetized crustal materials, and will help determine whether observed magnetic anomalies are directly associated with basin formation or were emplaced by subsequent processes.

Geophysical measurements acquired along a traverse provide essential ground truth for models of basin formation, and provide critical inputs for refining the next generation of models at a scale that is inaccessible from orbital assets.

#### 4.1.11.1 E11: Science Objective Traceability

Determining the crustal structure across the SPA basin directly addresses Objective 3.1.

**Objective 3.1** seeks to identify and characterize discontinuities in crustal thickness, density, and magnetic properties from the basin interior to the rim, which are diagnostic of the processes that governed mega-basin formation and modification.

#### 4.1.11.2 E11: Sample Requirements

This experiment consists entirely of *in situ* measurements and **does not require returned samples**.

#### 4.1.11.3 E11: Total Baseline Sample Requirements per Sampling Location

None; this experiment does not require returned samples.

#### 4.1.11.4 E11: Site and Traverse Requirements

This experiment **requires a traverse of over 650 km distance from the central SPA region radially outward towards the basin rim**. Although SPA has a diameter of approximately 2,400—2,600 km, a 650 km radial traverse distance (Figure 4.19) encompasses the critical transition from the center of the excavated melt cavity across the proposed peak-ring transition (Gowman et al., 2026). Making geophysical measurements across this transition enables direct examination of variations in crustal density and thickness across the basin interior. If combined with Artemis measurements obtained near the basin rim at the South Pole, these data will provide a more comprehensive picture of spatial variations in SPA crustal structure. A traverse from the central SPA region across the crustal thickness transition zone will also provide access to regions where crustal magnetic anomalies are poorly constrained, and where the presence and distribution of magnetic carriers could provide critical information about SPA formation or subsequent basin modification. The successful execution of this experiment requires acquiring the transition from central SPA to the outer basin (650 km); however, **the traverse path that is taken to meet this total radial distance is flexible**; indeed, longer path lengths that include variable topographic units or extend past the initial basin transition zone can provide even more compelling data.

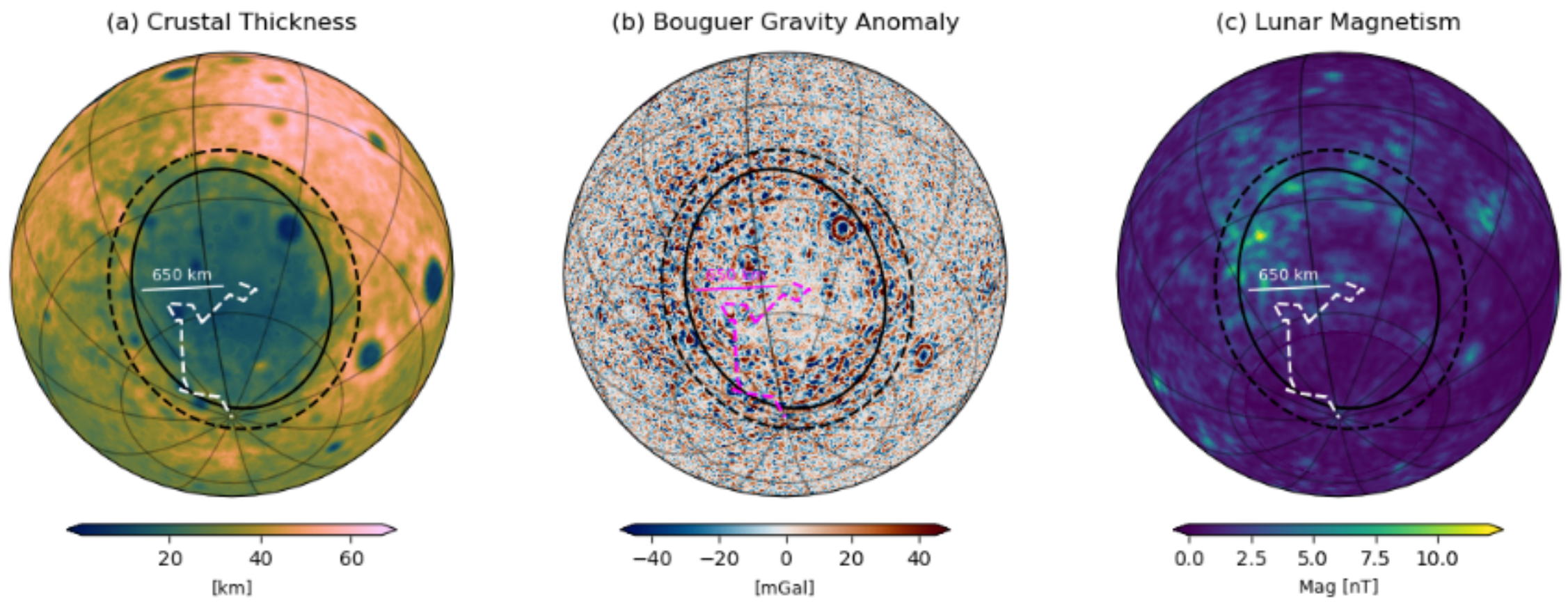

**Figure 4.19.** *A basin-scale (≥ 650 km) geophysical traverse across SPA basin will directly characterize variations in crustal properties, such as thickness, density, and remnant magnetization. Lunar maps of SPA showing (a) crustal thickness, (b) high-passed Bouguer gravity anomaly, and (c) magnetism, in an orthographic projection centered on 190°E, -50°S. A 650 km radial track is shown in white (magenta), extending from central SPA towards the rim, and the example baseline mission traverse is shown with a dashed white (magenta) line. Solid and dashed black ellipses denote the inner and outer basin rims, respectively, after James et al. (2019). Crustal thickness is from Model 1 of Wieczorek et al. (2013), high-passed Bouguer gravity is from Gowman et al. (2026), and lunar magnetism is from the 30-km map of Hood et al. (2021).*

#### 4.1.11.5 E11: *In situ* Measurement Requirements

Measurements for this experiment include Measurement 7 (M7: Magnetic Field) and Measurement 5 (M5: Gravity Field). Gravity field measurements enable an estimate of changes in crustal density and thickness, and magnetic field measurements enable an estimate of the distribution, strength, and depth of magnetic anomalies within the lunar crust.

Characterizing gravity anomalies across SPA basin is essential for constraining crustal structure, density heterogeneity, and the impact processes that shaped the basin; *in situ* gravity measurements therefore require sufficient sensitivity and spatial resolution to resolve both regional- and basin-scale density variations. Background gravity anomalies in the farside highlands related to heterogeneities in the density/porosity of the upper crust are typically about 10 mGal in amplitude, while anomalies associated with the SPA rim and inner ejecta annulus are typically ~20-50 mGal (Gowman et al., 2026). A **relative measurement accuracy of ≤1 mGal** is sufficient to resolve these anomalies and could distinguish density contrasts of ≤200 kg $m^{-3}$—comparable to the density contrast between different major lunar lithologies—over an elevation change of ≥100 meters. The width of gravity anomalies within the SPA rim annulus/ejecta zone is between ~ 17 and 60 km (Gowman et al., 2026), although a surface campaign could detect smaller-scale structures than those observable from orbit. A **distance between measurements of ≤ 2 km** would allow for characterization of anomalies of ~4 km in wavelength, comparable to the scale of local crustal blocks, dikes, faults, and intrusions and more than sufficient to resolve 10s to 100s of km scale features associated with complex craters and larger impact basins. Such a cadence would allow for a characterization of regional anomaly variations ensuring that the traverse captures large-scale structural changes within the SPA basin.

Magnetic field measurements acquired along such a traverse will enable characterization of surface magnetic features that are unresolvable from orbital observations, including their depth, spatial distribution, and field strength. Examining magnetic anomalies from central SPA outward enables assessment of whether the distribution of magnetic carriers is linked to SPA formation (e.g., remnants of the SPA impactor) or subsequent geological processes (e.g., basaltic dike intrusion; Purucker et al., 2012), thereby constraining the sequence of events that shaped the evolution of SPA. Correlations between magnetic field strength, source depth, and surface geology further permit discrimination between magnetization mechanisms, which allows us to distinguish between regional-scale and localized processes that have operated on the Moon. For example, dynamo-related fields are expected to produce broader, deeper-seated signatures, whereas impact-related magnetization is typically shallow, spatially localized, and correlated with surface topography, crater interiors, melt deposits, or fill material (e.g., Yang and Wieczorek, 2024).

Apollo surface magnetic field measurements provide an *in situ* reference for defining magnetometer performance requirements for SPARX. Apollo surface measurements show substantial variability in remanent magnetic field strength, ranging from as low as ~3 nT at the Apollo 15 site up to ~330 nT at Apollo 16 (Dyal et al., 1974). Given that the Apollo 15 landing site exhibits surface field strengths up to an order of magnitude lower than those inferred for portions of SPA basin (e.g., Dyal et al., 1974; Tsunakawa et al., 2015), **a measurement resolution of ≤1 nT is sufficient** to resolve even the weakest lunar surface fields measured to date. Orbital measurements indicate total surface field strengths of ≲1,000 nT within SPA basin (Tsunakawa et al., 2015). However, *in situ* measurements conducted at smaller spatial scales may detect stronger localized anomalies that are not fully resolved from orbit. Accordingly, **a dynamic range of ≥2,000 nT is recommended** to ensure adequate margin for resolving localized, high-amplitude anomalies.

The ability to constrain source depth is governed primarily by the spatial wavelength of magnetic anomalies. Shallower sources produce shorter-wavelength (smaller-scale) anomalies, and deeper sources produce longer-wavelength (larger-scale) anomalies (e.g., Alldredge and Van Voorhis, 1961; Purucker and Clark, 2011). Therefore, to distinguish spatial variations in magnetic field strength and constrain source depth, measurements with **≤1 nT resolution and a sampling rate of ≥1 Hz are recommended**. A ≥1 Hz sampling rate satisfies most geophysical investigations, while preserving flexibility to increase sampling rates for improved resolution of short-wavelength anomalies, and is consistent with sampling rates implemented on other lunar and Martian stationary and mobile platforms (e.g., Dyal et al., 1970; Du et al., 2020). Because the depth of SPA magnetic anomalies is not yet known, resolving deeper crustal sources would require observations spanning spatial scales of tens of kilometers or more. The planned basin-scale geophysical traverse (>650 km) is therefore expected to provide sufficient spatial coverage to constrain the depth and distribution of magnetic sources within SPA.

#### 4.1.11.6 E11: Additional Measurements

Pairing geophysical magnetometer measurements with paleomagnetic analyses enables collected samples to be placed in a regional crustal context. Although magnetometer measurements are not required for initial sample acquisition—since all collected samples will be available for paleomagnetic studies (E6; Section 4.1.6) and such analyses would be performed even in the absence of a magnetometer—these contextual magnetometer measurements are highly complementary, as they improve interpretation of sample-scale magnetization in terms of source depth and emplacement mechanism and can guide opportunistic sampling by identifying regions of elevated magnetic anomaly that might otherwise be unsampled.

### 4.1.12 Experiment 12: Determine if the Massifs in the Southwestern Portion of the Basin Constitute the Diffuse Rim of SPA

Despite, or perhaps because of, its immense size, the actual physical extent of SPA remains poorly defined. Various studies have defined the extent of SPA based alternately on variations in topography, crustal thickness, composition, or the distribution of massifs interpreted as remnants of the basin rim (James et al. 2019) (Figure 4.20). Determining the actual extent of the basin is further complicated by a complex, diffuse rim structure consisting of an inner rim, outer rim, and incomplete exterior scarp (Garrick-Bethell and Zuber 2009; James et al. 2019; Bernhardt et al. 2025); in addition to an inner annulus of material potentially representative of rim collapse and/or ejecta deposits (Gowman et al., 2026). Whereas some portions of the north and east rims of the SPA basin are well-defined topographically, diffuse archipelagos of massifs sit where we expect to find the south and west rims of the basin. Whether or not these massifs represent a portion of SPA's rim is critical to linking massif distributions to other rim markers such as transitions in topography and crustal thickness.

This experiment uses bulk density and compensation state of the massifs to determine their origin. While a massif representing a portion of the basin rim might have a bulk density consistent with that of the feldspathic highlands, a massif representing dense deposits in the lower crust / upper mantle material excavated by the SPA impact may instead have a relatively higher mafic or ultramafic bulk density, and a partially compensated state. Determining the true extent and location of the SPA rim in the western and southern portions of the basin is critical for constraining models of basin formation, including resolving on-going questions about the direction of the SPA impact itself (Andrews-Hanna et al, 2026; Bernhardt et al., 2025; Wieczorek et al. 2012).

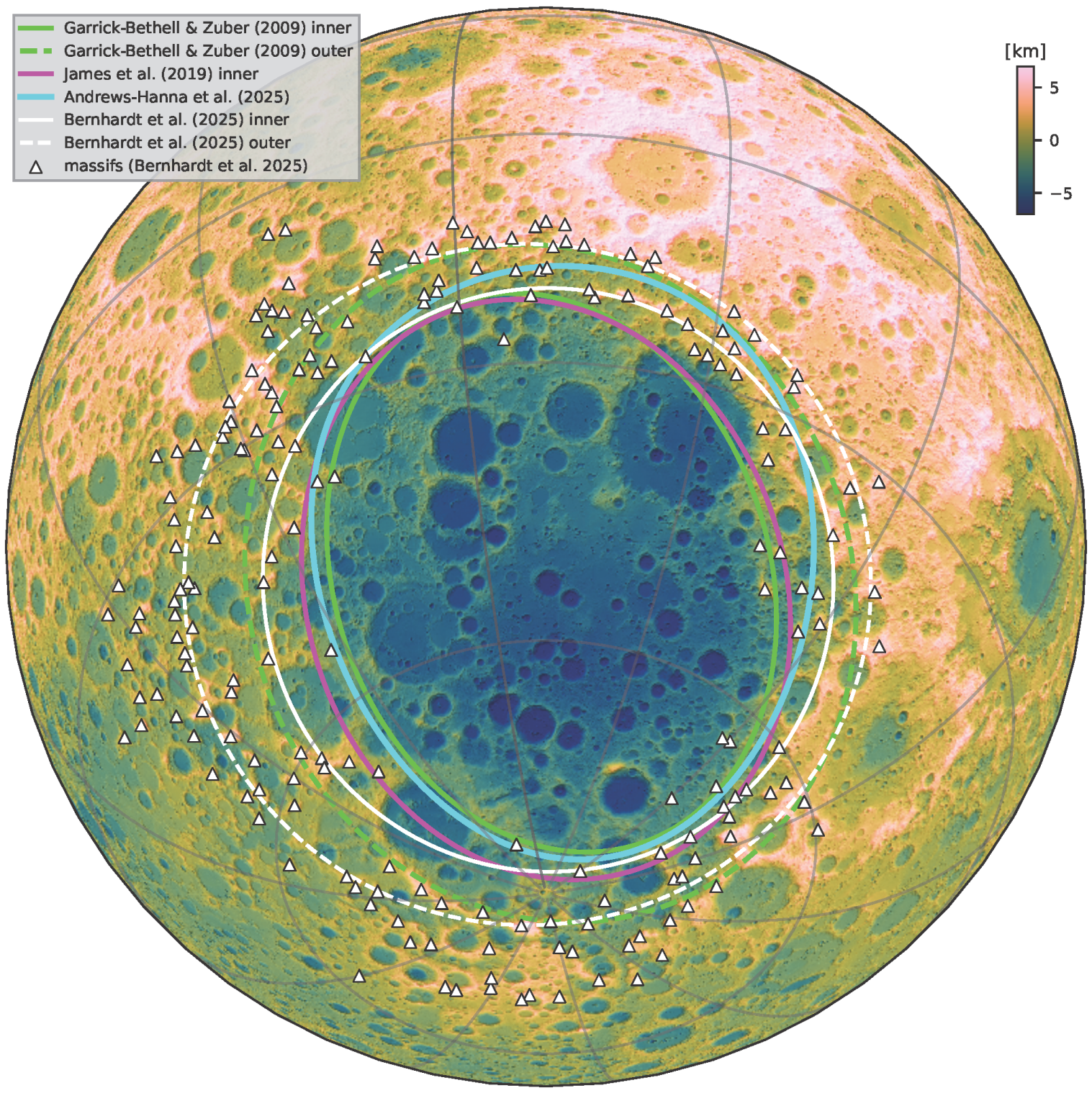

**Figure 4.20.** *Lunar topography (Smith et al. 2010) centered on SPA with various proposed SPA inner and outer basin rims. Proposed rim massifs are from Bernhardt et al. (2025). Proposed rim locations vary greatly depending on if they are based on topography (Garrick-Bethell & Zuber 2009), crustal thickness (James et al. 2019), gravity (Andrews-Hanna et al. 2025), or massifs (Bernhardt et al. 2025).*

#### 4.1.12.1 E12: Science Objective Traceability

Constraining the nature of SPA's diffuse rim directly addresses **Objective 3.1** "Determine whether there are discontinuities in the basin structure (crustal thickness, density, magnetic field) across SPA to constrain models for the formation of a mega-basin".

By determining whether massifs in the southwestern portion of the basin represent crustal rim material, we will be able to better fit the outline of the basin structure, and provide critical constraints for models of SPA formation.

#### 4.1.12.2 E12: Sample Requirements

None; **this experiment does not require returned samples**.

#### 4.1.12.3 E12: Total Baseline Sample Requirements per Sampling Location

None; this experiment does not require returned samples.

#### 4.1.12.4 E12: Site and Traverse Requirements

This experiment requires traversing ≥1 km in elevation up a hypothesized SPA rim massif. Several massifs have been proposed to be part of the SPA rim structures (Bernhardt et al. 2025) (Figure 4.21). Traverse designs should target a **massif with rover-accessible slopes, and a traversable height of at least 1 km**. A traverse height in excess of 1 km is required to achieve bulk density estimates with a precision of <200 kg/m$^3$, which is required to distinguish between a feldspathic or mafic origin for the massif material. This assumes a relative gravity accuracy of 10 mGal; better precision would proportionally reduce the required traverse height (as shown in Figure 4.22) and/or facilitate the inference of depth-dependent density within the massif.

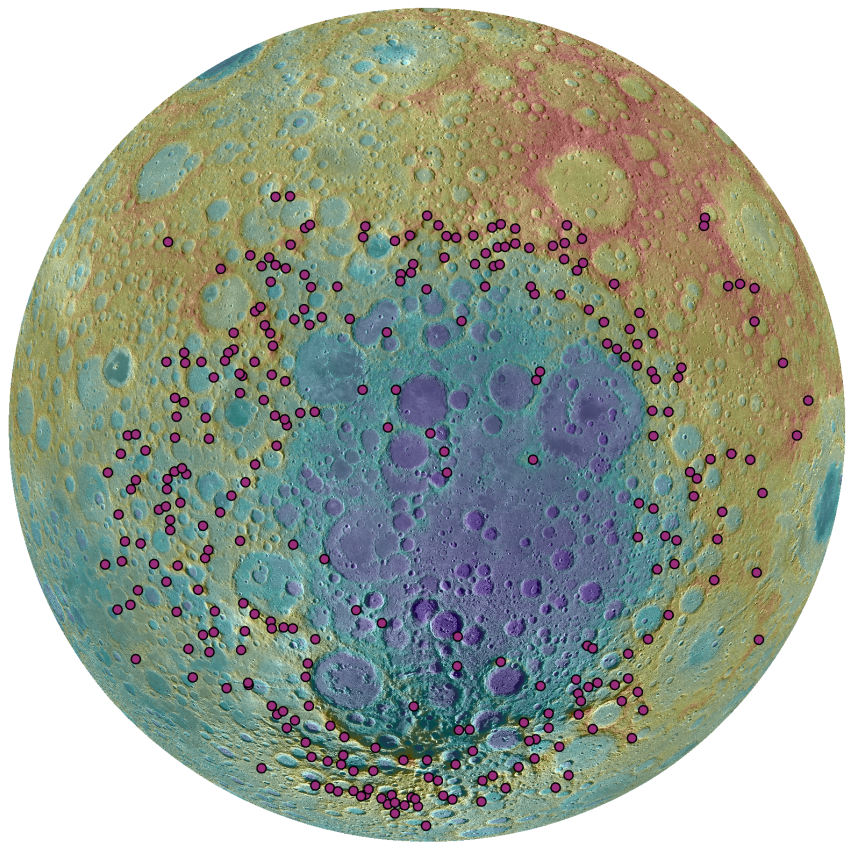

**Figure 4.21.** *Locations of proposed SPA rim massifs from Bernhardt et al., 2025.*

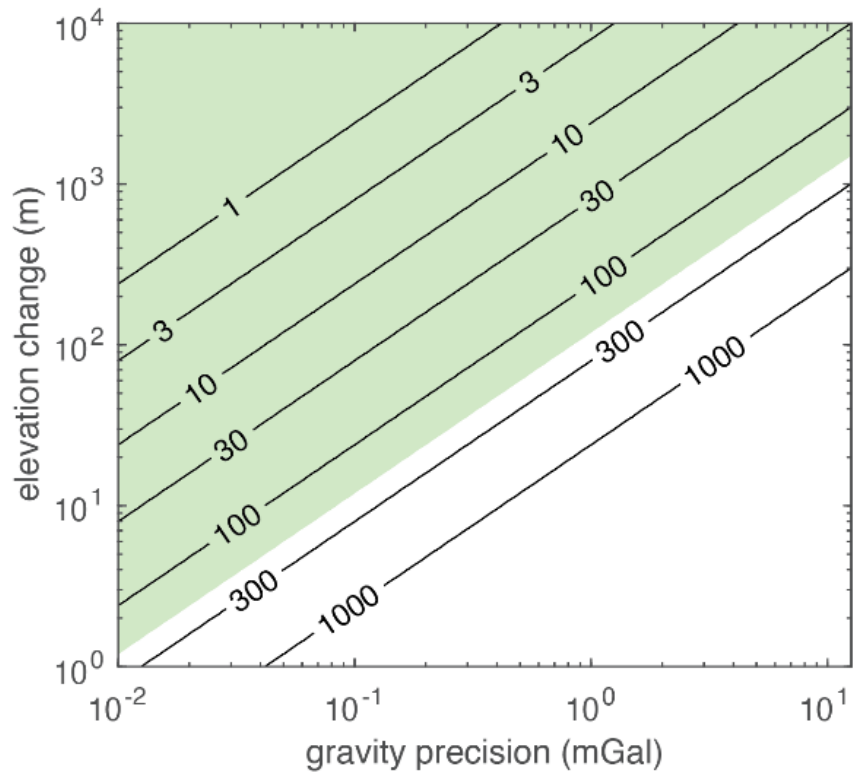

**Figure 4.22.** *Density uncertainty associated with gravity measurements of various precision and elevation change. The green zone indicates a density uncertainty of <200 kg/m$^3$, which should be sufficient to rule out some lithologies.*

#### 4.1.12.5 E12: *In situ* Measurement Requirements

Measurement 5 (M5: Gravity Field) is the required *in situ* measurement to determine the density of the massif structure and regional crustal thickness. Measurements at the peak and base of a topographic feature are most crucial for bulk density determination, but measurements along the flanks may additionally facilitate the inference of depth-dependent density. Based on the 1km elevation gain, and assuming a safe traverse slope requirement of 6°, the baseline distance for this requirement is 10 km. The standard SPARX baseline ConOps defines a 2 km minimum measurement spacing for Science Interval Stops, and the rover should be mechanically quiet during these gravity measurements. Combining these two measurement requirements results in a derived requirement for **5 gravity measurements across the approach, ascent, peak, and descent from the massif**. In addition, these requirements implicitly assume a knowledge of the elevation at which the measurement is taken; an error of 0.1 mGal will be introduced to the inferred free air anomaly with every meter of elevation error. The gravity uncertainty associated with elevation should be less than that of the reading, as a recommendation. **For a gravimeter with 10 mGal accuracy, this yields a requirement of ≤100-meter elevation accuracy. More precise gravimeters will require proportionally more precise elevation**.

### 4.1.13 Experiment 13: Determine the Stratigraphy and Composition of the Unique Deposits in Central SPA

The SPA forming impact dramatically altered the lunar surface, producing not only structural changes in the crust (Sections 3.3, 4.1.11, 4.1.12), but also widespread regional geochemical, geothermal, and geophysical changes that influenced subsequent crustal evolution. In addition to excavating several kilometers of crustal material (e.g., Wieczorek et al., 2013), the SPA impact would have also imparted a phenomenal amount of energy to the surface, melting several kilometers of the remaining central basin material (e.g., Citron et al., 2024). The material in the central portion of the basin must then evolve in a new way, subject to a range of different processes, all of which could alter the composition and distribution of materials observed in the basin today (Figure 4.23). Some of these processes include basin modification (cavity collapse) (e.g., Trowbridge et al., 2020; Citron et al., 2024), volcanic resurfacing (e.g., Whitten and Head, 2013; 2015; Moriarty and Pieters, 2018; Izquierdo et al., 2024; Chuang et al., 2025), deposition of ejecta from subsequent impacts (Petro and Pieters, 2008), or differentiation of a massive impact melt pool (e.g., Hurwitz and Kring, 2014; Vaughan and Head, 2014; Uemoto et al., 2017). To unravel the history and contributions of these different processes to SPA's interior, targeted sampling across diverse geologic units is required.

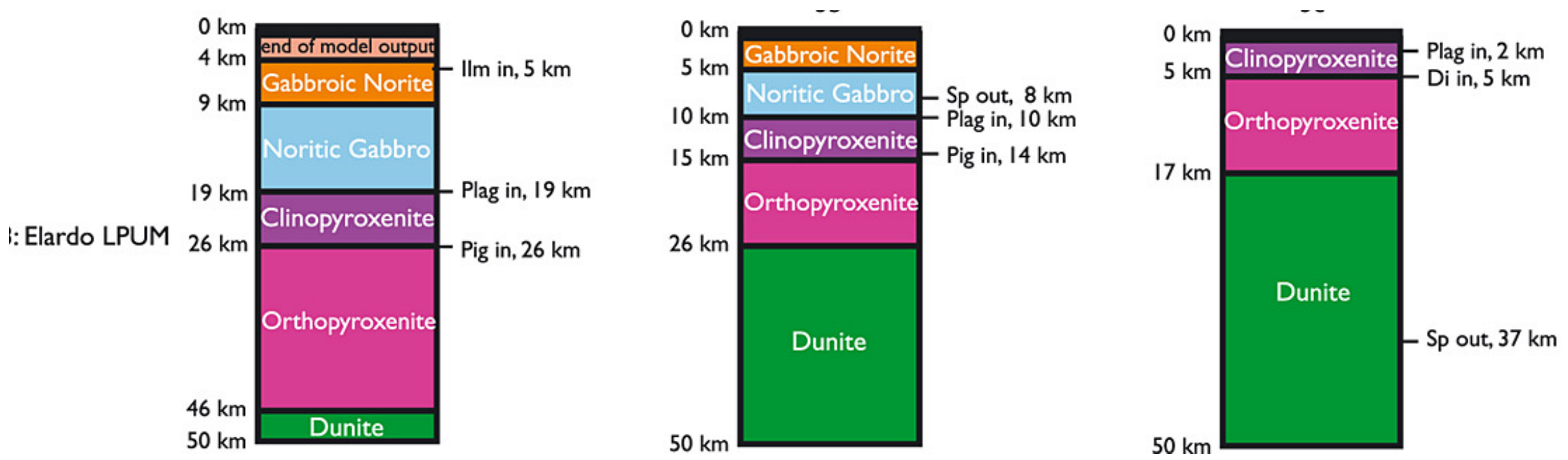

**Figure 4.23.** *An example of several possible subsurface stratigraphies produced by differentiation of the SPA impact melt pool. (Figure from Hurwitz and Kring, 2014).*

Experiment 13 tackles the perennial geologic challenge of using surficial samples and deposits to understand complex stratigraphy at depth. Generally, the surface of central SPA exhibits a lithology spectrally dominated by Ca,Fe-rich pyroxenes (Moriarty and Pieters, 2018, Moriarty et al., 2026). These Ca,Fe-rich pyroxenes are observed in a wide range of plains units (mare, cryptomare, and non-mare), and are also observed in the walls, rims, and ejecta of large (~50-100 km) complex craters, indicating that they persist to depths of several km. However, the deepest material excavated in central SPA (in the central peaks and some walls/rims/ejecta of the largest craters) exhibit a very different composition, one dominated by Mg-rich pyroxenes. The exposure of this distinct lithology in the deepest-excavated materials directly reflects the compositional stratigraphy of the basin as a whole. Therefore, **by targeting the ejecta of craters which have excavated strategically relevant depths, this experiment seeks to probe different parts of the subcrustal stratigraphy**, and compare the results with a range of proposed compositional models (e.g., Hurwitz and Kring, 2014, Vaughan and Head, 2014). These strategic spot measurements can then be paired with surface observations made at regular intervals across the basin, which will allow for the tracking of important compositional contributors like volcanic resurfacing and deposition of ejecta from other non-SPA impact events. The combination of these approaches will provide critical insight into the current structure and composition of the basin interior, and providing essential context for understanding the evolution of these large-scale impact processes across the Solar System.

#### 4.1.13.1 E13: Science Objective Traceability

Experiment 13 is directly relevant to Science Objective 3.2 for determining the stratigraphy and composition of unique deposits in SPA.

**Objective 3.2** seeks to distinguish the relative contributions of impact melt differentiation, crustal backfilling, and volcanic infilling to the interior composition and stratigraphy of SPA. As discussed in Section 3.3.2, each of these processes produces variations in compositional stratigraphy, and distribution of deposits around the basin. In order to determine their relative contributions, samples must be acquired from critical stratigraphic horizons, and different units.

#### 4.1.13.2 E13: Sample Requirements

Distinguishing between primordial crust, SPA impact melt, foreign basin ejecta, and volcanic materials through analyses of sample age, geochemistry, and other properties is required for this experiment. These materials are mafic-bearing and should include multiple mineral systems and relatively higher abundances of elements relevant to radiometric dating than highlands crustal rocks. However, some of these materials may be significantly shocked and brecciated, which complicates these measurements.

This experiment requires **≥8 rocks with a diameter of ≥8 mm diameter** to be collected at each site. As with all sample requirements described in this report, these size requirements are designed to reflect the sample mass required for each recommended type of laboratory analysis (Section 4.3). Rock fragments ≥8 mm diameter will be targeted for Sm-Nd, Rb-Sr, and Lu-Hf isochron dating methods, as these fragments will be of sufficient mass (100s of mg) to enable these methodologies. When rock mass, elemental composition, and textures allow, it may be possible and advantageous to use the same rock for isochron ages for multiple isotopic systems, bulk composition determinations, and to make thin sections for subsequent analyses.

Targeting the collection of ≥8 rock with a diameter of ≥8 mm diameter is expected to result in the collection of **≥100 rock of ≥3 mm diameter** based on typical lunar regolith particle size distributions (e.g., Carrier, 2003). These smaller size fragments are sufficient in mass to provide material for thin sections, bulk rock determinations of major, minor, and trace elements, and *in situ* analyses. *In situ* analyses include determination of crystallization ages of samples, petrological characterization, mineral chemical analyses, and mineral trace elemental abundances. These fragments also provide sufficient mass for Ar-Ar dating, which will provide additional information regarding the crystallization age and/or shock history of these samples.

**Bulk regolith (≥100 g)** that will be scooped at these sites is expected to contain a fine-grained fraction that will include zircons and other mineral fragments (Barboni et al., 2024, 2017; Crow et al., 2017; Dauphas et al., 2025; Greer et al., 2021; Meyer et al., 1996; Nemchin et al., 2009; Zhang et al., 2021b, 2021a). These zircons will be identified and concentrated to allow chronologic studies.

#### 4.1.13.3 E13: Total Baseline Sample Requirements per Sampling Location

- ≥ 32 rocks ≥8 mm diameter
- ≥ 400 rocks ≥3 mm diameter
- ≥400 g bulk regolith

The sampling requirements for this experiment also include sample acquisition sufficient to satisfy NASA sample curation requirements. Total sample numbers and overall sample mass reflect the sum of these two sets of requirements.

#### 4.1.13.4 E13: Site and Traverse Requirements

To address this experiment, it is necessary to **collect samples from a representative sampling of surface units and origin depths.** Because of the requirement to test material excavated from specific depths, the traverse and sampling locations for E13 are more constrained than other experiments. At minimum, it is required to collect samples from four distinct sites: **≥ (1) sample of low-calcium pyroxene (LCP) material** excavated by Bose crater; **≥ (1) sample of high-calcium pyroxene (HCP) material** excavated by Bhabha crater; **≥ 1 sample from the ejecta of a morphologically fresh ~10-km diameter crater** within a plains unit; **≥ 1 sample of light plains material** in central SPA are suggested for E13. The craters of Bose and Bhabha are specifically targeted because remote sensing dataset identifications of the critical lithologies in the central peaks and associated ejecta deposits (Figure 4.23), and the close proximity of the two craters optimizes sample acquisition. Samples acquired from both a LCP crater and a HCP crater should target parts of the ejecta deposit with material representative of the ejecta deposit as a whole. Samples from the fresh 10-km diameter crater and light plains material are more flexible, and can be selected from regions that optimize the traverse path. Candidate fresh 10-km craters with well-defined ejecta should be identified using LROC data, and candidate light plains deposits should be identified from existing geologic databases (e. g. Meyer et al., 2020) and confirmed with LROC-NAC data.

#### 4.1.13.5 E13: *In situ* Measurement Requirements

Experiment 13 requires **the ability to detect sampleable materials and to make continuous measurements of geologic context throughout the central SPA region**. This will be accomplished by Measurement 1 (M1; Context Imaging), Measurement 2 (M2: Context Mineralogy), and Measurement 4 (M4: Bulk Elemental Chemistry) to identify anomalous materials and to understand the geology of the area at a rover-scale, placing the sampled material in sufficient context. Additionally, regular interval stops for M1, M2, and M4 every ~2 km will further complete a continuous image and compositional data set that can help to identify potentially anomalous features in stratigraphy throughout the traverse, and at a spatial scale that significantly exceeds current orbital remote sensing datasets.

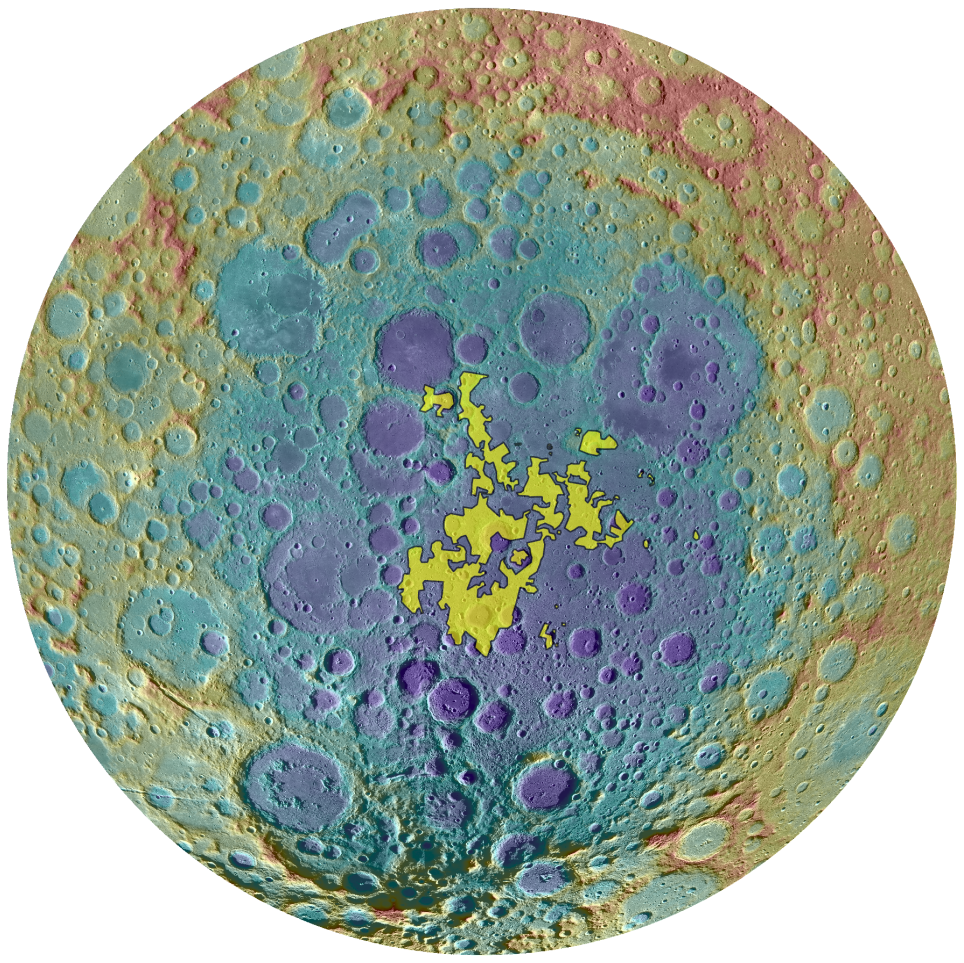

**Figure 4.24.** *Distribution of predicted exposure of SPA impact melt sheet materials including LCP and HCP. Adapted from Uemoto et al., 2017, Figure 4.*

### 4.1.14 Experiment 14: Determine the Origin and Extent of the Low-Calcium Pyroxene Annulus Enclosing Central SPA

Outside of Central SPA, which shows a high Ca,Fe pyroxene-bearing composition (likely resulting from impact melt differentiation and/or volcanic resurfacing, see E13), the SPA interior is dominated by a different lithology-- Mg-rich low-Ca pyroxenes (LCP). These units are observed in a fairly uniform annulus exterior to the central HCP region (Figure 4.25) (Moriarty and Pieters, 2018; Moriarty et al., 2026). These LCP signatures are observed in the ejecta and central peaks of large craters in the HCP region, suggesting excavation from a deep LCP layer (Moriarty and Pieters, 2018; Moriarty et al., 2021; 2026). The distribution of these LCP units suggests that they represent an important part of SPAs evolutionary history. Therefore, in order to fully understand the compositional contributions and evolution of central SPA, it is necessary to determine the origin and extent of these LCPs.

#### 4.1.14.1 E14: Science Objective Traceability

This experiment directly addresses Objective 3.2 by constraining LCP-bearing materials.

**Objective 3.2 aims to distinguish the relative contributions** of impact melt differentiation, crustal backfilling, and volcanic infilling to the interior of SPA. By determining the origin of LCP-bearing materials within SPA, this experiment will constrain the source of a major component of SPA constituting an annulus interior of the SPA rim and surrounding central SPA (SPACA).

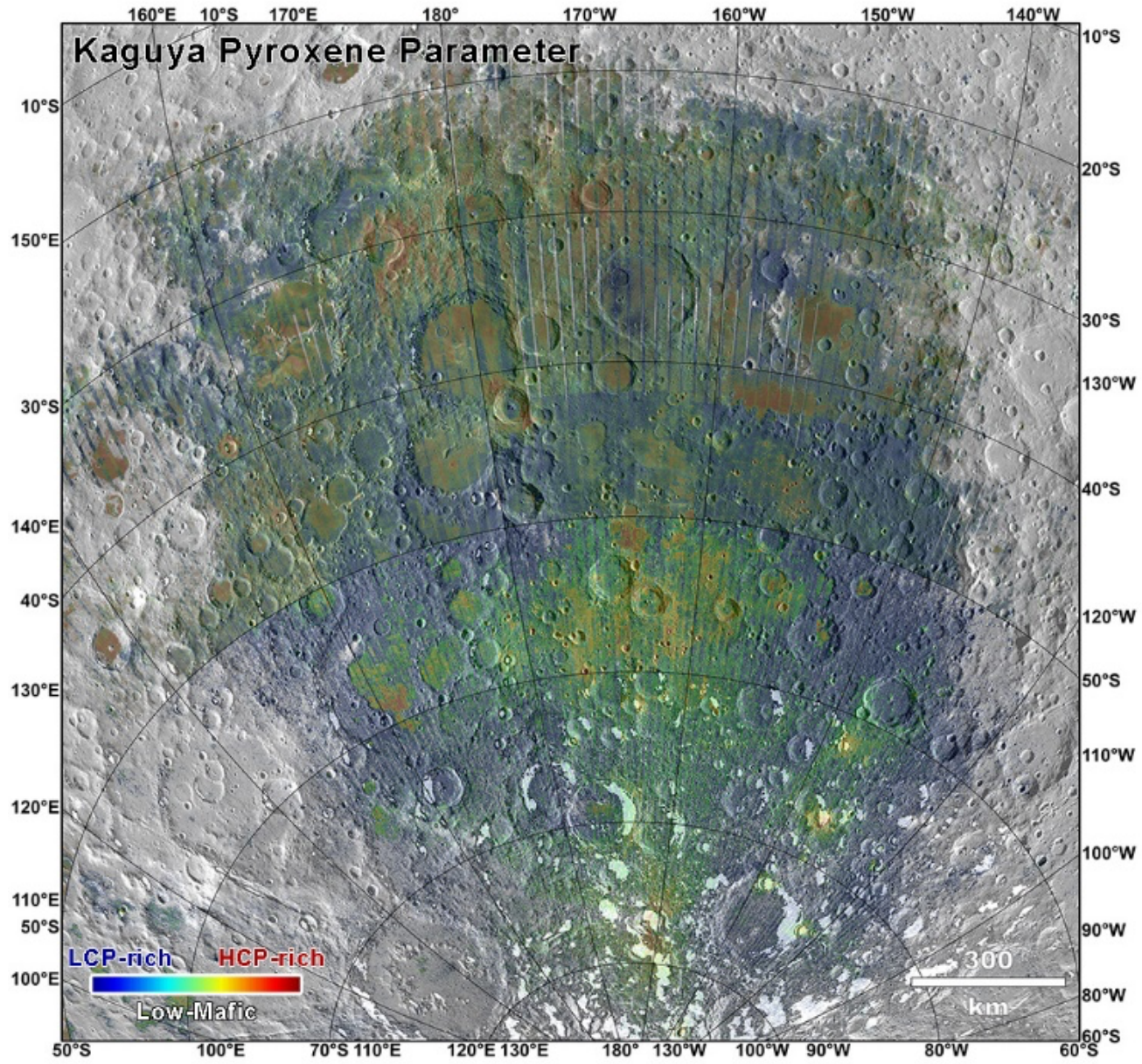

**Figure 4.25.** *A pyroxene parameter map derived from Kaguya mineral abundance maps highlighting the distribution of high-Ca,Fe pyroxene (HCP), Mg-rich low-Ca,Fe pyroxene (LCP), and material with < 30 wt% total pyroxene abundance (low mafic) (Moriarty et al., 2026).*

#### 4.1.14.2 E14: Sample requirements

Comparison of Mg-pyroxene samples collected from throughout SPA, including at least one from the Mg-pyroxene annulus, by measuring sample age, geochemistry, and other properties is required for this experiment. This experiment requires **≥8 rocks with a diameter of ≥8 mm diameter** to be collected at each site. As with all sample requirements described in this report, these size requirements are designed to reflect the sample mass required for each recommended type of laboratory analysis (Section 4.3). Rock fragments ≥8 mm diameter will be targeted for Sm-Nd, Rb-Sr, and Lu-Hf isochron dating methods, as these fragments will be of sufficient mass (100s of mg) to enable these methodologies. When rock mass, elemental composition, and textures allow, it may be possible and advantageous to use the same rock for isochron ages for multiple isotopic systems, bulk composition determinations, and to make thin sections for subsequent analyses.

Targeting the collection of ≥8 rock with a diameter of ≥8 mm diameter is expected to result in the collection of **≥100 rock of ≥3 mm diameter** based on typical lunar regolith particle size distributions (e.g., Carrier, 2003). These smaller size fragments are sufficient in mass to provide material for thin sections, bulk rock determinations of major, minor, and trace elements, and *in situ* analyses. *In situ* analyses include determination of crystallization ages of samples, petrological characterization, mineral chemical analyses, and mineral trace elemental abundances. These fragments also provide sufficient mass for Ar-Ar dating, which will provide additional information regarding the crystallization age and/or shock history of these samples.

**Bulk regolith (≥100 g)** that will be scooped at these sites is expected to contain a fine-grained fraction that will include zircons and other mineral fragments (Barboni et al., 2024, 2017; Crow et al., 2017; Dauphas et al., 2025; Greer et al., 2021; Meyer et al., 1996; Nemchin et al., 2009; Zhang et al., 2021b, 2021a). These zircons will be identified and concentrated to allow chronologic studies.

#### 4.1.14.3 E14: Total Baseline Sample Requirements per Sampling Location

- ≥32 rocks ≥8 mm diameter
- ≥100 rocks ≥3 mm diameter
- ≥400 g bulk regolith

The sampling requirements for this experiment also include sample acquisition sufficient to satisfy NASA sample curation requirements. Total sample numbers and overall sample mass reflect the sum of these two sets of requirements.

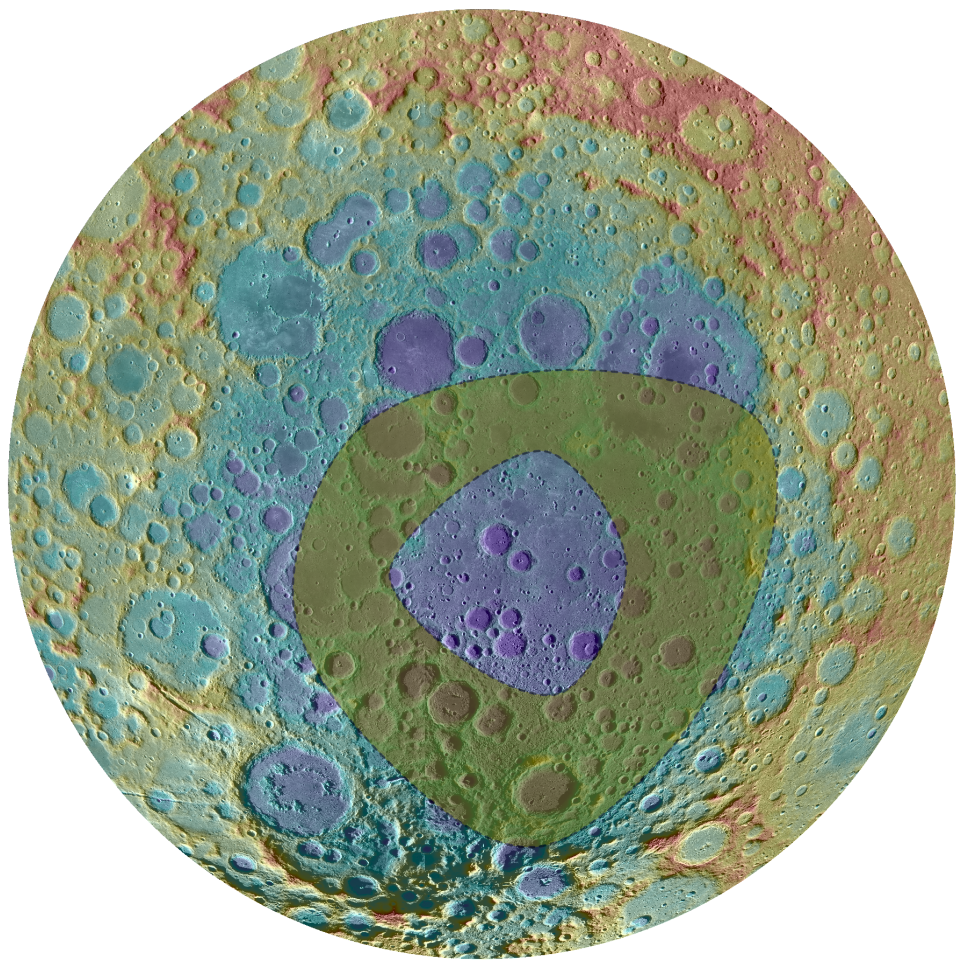

**Figure 4.26.** *Approximate regions of low-Ca pyroxene annulus observed outside the central SPA region.*

#### 4.1.14.4 E14: Site and Traverse Requirements

This experiment requires acquiring samples from **at least 1 location in the Mg-pyroxene annulus enclosing central SPA.** Sites that have low-CA pyroxene detections from orbit have been identified (e.g., Moriarty and Pieters, 2018; Yamamoto et al. 2023), and some of these sites further correspond with an inter-basin annulus of high-density material that may represent dense materials (lower crust and upper mantle) excavated during the SPA impact (Gowman et al. 2026) (Figure 4.26). When selecting a sample site location, **prioritization should go to these identified collocated LCP and high-density material sites**. If traverse design does not allow for accessing any of these sites, priority should be given to LCP detections (e.g., Moriarty and Pieters, 2018; Yamamoto et al. 2023) that are compliant with traverse constraints.

#### 4.1.14.5 E14: *In situ* Measurement Requirements

This experiment has two *in situ* measurement requirements: the ability to detect material of key lithologies, including high- and low-Ca pyroxene, olivine, and feldspar, and the ability to detect sampleable material. The detection of key lithologies can be accomplished through a combination of Measurement 1 (M1: Context Imaging) and Measurement 2 (M2: Context Mineralogy), with bulk context from Measurement 4 (M4: Bulk Elemental Chemistry). Additional details on the spatial resolution requirements for the reconnaissance and sampling phase are provided in Section 4.2.

## 4.2 *IN SITU* MEASUREMENT REQUIREMENTS AND STRAWMAN INSTRUMENT SUITE

### 4.2.1 Introduction

The SPARX SDT evaluated and defined the *in situ* measurement requirements necessary to accomplish the mission's science objectives. *In situ* measurements will broadly serve three purposes: (1) to ensure that collected samples intended for Earth return meet science requirements, (2) to place returned samples within their geologic context, and (3) to directly address mission science objectives. These measurements fall into three broad categories. (1) Landscape-scale contextual data (Sections 4.4.2, 4.4.3, and 4.4.6) support identification of candidate sampling targets at a distance and guide traverse planning. (2) Fine-scale measurements (Sections 4.4.4 and 4.4.5) characterize the size, texture, and composition of materials within a specific sampling workspace. (3) Geophysical measurements (Sections 4.4.7 and 4.4.8) characterize the basin's subsurface structure and magnetic field environment. A summary traceability between all *in situ* measurements and the SPARX Objectives and Experiments is provided in Table 4.1, and traceability for payload measurement requirements and strawman payload instruments can be found in Section 4.2.9, Foldout 4.1.

**Table 4.1.** *In situ Measurement Traceability to SPARX Science.*

| IN SITU MEASUREMENTS | | |
|---|---|---|
| MEASUREMENT | APPLICABLE EXPERIMENTS | APPLICABLE OBJECTIVES |
| M1 Context Imaging | E1 - E3, E4¢ - E7¢, E8 - E10, E13, E14 | 1.1, 1.2, 1.3, 1.4¢, 2.1, 2.2¢, 2.3, 3.2 |
| M2 Context Mineralogy | E1, E2¢, E4¢ - E7¢, E9¢, E10¢, E13, E14 | 1.1, 1.2, 1.3¢, 1.4¢, 2.1, 2.2¢, 2.3¢, 3.2 |
| M3 Fine-scale Mineralogy | E1, [E4], [E5], E6¢ | 1.1, 1.2, [1.3], 1.4¢, 2.1 |
| M4 Bulk Elemental Chemistry | E1¢, E2¢, E4¢ - E6¢, E7, E9¢, E10¢, E13, E14¢ | 1.1¢, 1.2¢, 1.3¢, 1.4¢, 2.1¢, 2.2, 2.3¢, 3.2 |
| M5 Gravity Field | E11, E12 | 3.1 |
| M6 Fine-scale Imaging | E1, [E4], [E5], E6¢ | 1.1, 1.2, [1.3], 1.4¢, 2.1 |
| M7 Magnetic Field | E6¢, E11 | 1.4¢, 3.1 |
| ¢ denotes applicable items associated with context measurements; bracketed items are associated with baseline measurements | | |

The first purpose for *in situ* data is to ensure that the samples collected for Earth return are, to the best ability of the mission, appropriate and sufficient for addressing the mission's science objectives. As a sample return mission, the majority of science objectives defined by the SPARX SDT require the acquisition of specifically designated samples (both in size and lithology) for later analysis in terrestrial laboratories. These *in situ* measurements provide the information necessary for the science team to guide sampling decisions at the rover scale within the broader orbitally defined traverse regions. They also enable verification that materials within a given work volume meet established science criteria, including minimum size thresholds and required mineralogical or elemental compositions (Section 4.2.9).

Secondly, the scientific value of returned samples depends on a robust understanding of their geologic context, including their provenance, formation processes, and relative age. Establishing this context on the Moon is complicated by the character of the lunar surface — bedrock exposures are rare, and regolith has been extensively reworked by impact-induced mixing. Nevertheless, orbital remote sensing data demonstrate that distinct lithologies dominate different localities on the Moon, indicating that coherent geologic units remain identifiable at regional scales. *In situ* measurements are therefore required to characterize lithologies encountered both along the traverse, and within sampling regions. These data provide insight into whether collected samples are representative of broader units or reflect localized, potentially unique materials. Where possible, *in situ* observations will also document stratigraphic relationships of units and link sampled materials to identifiable geologic features, such as crater rims and peak rings, boulders, fractures, lava flows, and pyroclastic vents. Collectively, this contextual information anchors laboratory analyses within a coherent framework of the Moon's geologic evolution. **The SPARX SDT emphasizes that the science objectives cannot be fully achieved without this contextual sample information**. The SDT finds that geologic context measurements, consisting of partial panoramic context imaging, partial panoramic context mineralogy, bulk elemental chemistry, gravity field, and magnetic field **measurements be made from a stationary position approximately every 2 km** while traversing, or between sampling locations. This distance represents the approximate distance to the horizon, and therefore is required to provide a near-continuous record of surface properties across the duration of the mission. Further description of these "Science Interval Stops" can be found in Section 6.6.1.

Finally, we note that, SPARX objective 3.1 is achieved solely through *in situ* measurements. Determining the broad structure of the SPA basin, and how it reflects the response of the Moon to the impact event requires geophysical and magnetic field measurements to be carried out across a wide traverse of the basin. These measurements are described more fully in Experiment 11 and 12 (Section 4.1.11–4.1.12).

**The instruments identified below are recommended because they provide the minimum measurement capabilities required to achieve the mission's science objectives**. While additional *in situ* capabilities increase the mission's discovery potential—particularly in the event that samples are not returned to Earth due to technical or programmatic issues— such additions would also increase mission complexity, timeline, and cost. The SPARX SDT therefore recommends only measurements necessary to collect and contextualize scientifically optimal samples while minimizing impacts on vehicle complexity, time, mass, and overall mission resources. Further discussion of the balance between *in situ* measurement capabilities and return sample science can be found in Section 7.4.

### 4.2.2 Context Imagers

#### How measurement contributes to experiments and goals:

To first order, context imaging will guide SPARX to workspaces that contain materials that fulfill the physical property requirements for sampling. The requirements for this are set by the ability to find rocks for Experiment 1 (≥40 mm in diameter): "Determine the composition and age of the anorthositic farside highlands crust" and Experiment 9 (≥20 mm in diameter): "Sample hypothesized excavated lunar mantle material." In addition to the far-field resolution requirements that derive from these experiments, these experiments also set the requirement for color context imagery, as color can be a critical discriminator in identifying these two lithologies (i.e., anorthosite and hypothesized olivine).

Even on the predominantly grey Moon, color can be an important tool to distinguish materials (Figure 4.27). The discovery of orange glass by the Apollo 17 astronauts is a dramatic example where visible color revealed new geologic materials that would not have otherwise been visible in gray scale images. In addition to composition, color can also be used to help differentiate maturity of lunar materials (Blewett et al., 2023a). The specific wavelengths of the RGB color bands should be chosen such that they are sensitive to lunar parameters like optical maturity, titanium, and iron content. When combined with additional context imaging information, like rock and regolith texture, context imaging can provide a first-order assessment of lithology. This type of integrated color context imaging data should also be used to generate geologic maps of the traverse to place the samples collected by the mission within their geologic context.

#### Measurement requirements and justification:

The spatial resolutions required for color context imaging are set by the requirement to identify rocks ≥20 mm in size that are just beyond the rover's workspace, which equates to a spatial resolution requirement of **≤5 mm/pixel (0.5 cm /pixel) at ≥5m distance**. Images for regional context imagine (≥100 m) should be able to resolve larger boulders (≥0.5 m diameter) that could be sources of sampleable material. These images will also be used for geologic mapping and interpretation, so they should also show geomorphic features ≥1 m in size (e.g., stratigraphic

### Balancing the scope of *in situ* compositional characterization

The SPARX SDT discussed at length the appropriate balance of *in situ* compositional characterization of prospective sample material. The Mars rover missions have shown that "characterization" can range from simple monochrome imaging to advanced geochemical and elemental characterization with techniques such as LIBS, XRF, and Raman spectroscopy. The SDT ultimately decided that the for the science objectives defined in the SPARX STM, compositional characterization that emphasized the identification of specific bulk mineralogic characteristics was sufficient. The rationale for this includes the observation that none of the other listed techniques (LIBS, XRF, Raman) could provide data that would either allow for better triage or selection of samples for return, or completion of certain science objectives completely *in situ*. For example, several SPARX objectives require the acquisition of impact melt material, for which there is no unique geochemical signature; instead, measurements should focus on identifying material that is broadly representative of the targeted host deposit. In the case of a discrete required lithology, such as anorthosite, after bulk lithology, the critical sample characteristic which required for geochronology measurements is the shock state of the minerals, a parameter that is not possible to characterize *in situ* with a useful degree of precision. Therefore, while more robust *in situ* geochemical characterization techniques should not be excluded from future consideration, the SPARX SDT is not requiring them or recommending them within the strawman payload.

layers, ejecta rays, deposits with distinctly different albedos). This sets a spatial resolution requirement in the far field to be **≤10 cm/pixel at ≥100 m distance**. For both workspace and regional context imagining, "resolve" is defined as a minimum of 3 pixels covering the feature.

Beyond the broad **requirement for color images**, there are no requirements for the number and wavelength ranges used to generate color; either approximate true color or enhance true color, with wavelengths chosen to also be sensitive to properties like optical maturity (e. g. Blewett et al., 2023a) and iron content would fulfill this need. Preference should be given to color ranges that approximate or overlap with existing remote sensing datasets (e. g. LROC WAC, Clementine, or Kaguya TC), as this would allow for cohesive mapping from the rover-scale to orbital scale, and enhance return of the geologic context maps. Context images must include much of the area around the vehicle, and imagers should be able to achieve this field of view through either their own internal mechanisms, via an external scan platform, or through utilizing vehicle movement. The field of view requirements are an **azimuthal field of view ≥180˚**, and images from the **sampling workspace to at least +20˚ above the horizon** in elevation. Context images must also be able to be used to generate 3D reconstructions of the surrounding terrain to aid in interpretation of geologic features and stratigraphic relationships, and are directly needed to address E11 and E12. The SDT notes that **stereo image acquisition is not required to** achieve SPARX requirements, although such a capability might be desirable science or engineering operations.

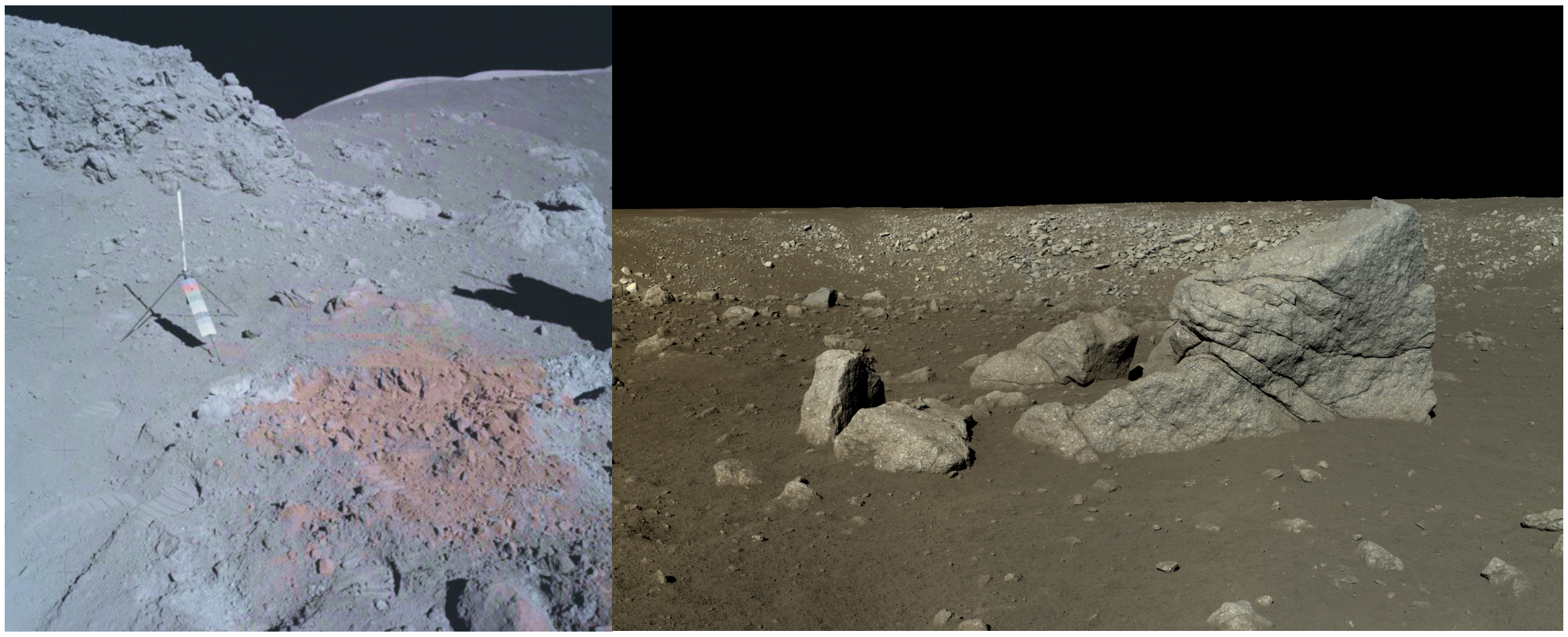

**Figure 4.27.** *Two color images from the surface of the Moon. Left: NASA photo AS17-137-20990 showing "Orange glass" spotted by the Apollo 17 astronauts. Contrast- and color-balancing the images significantly enhances out the orange color. Right: A color image from the Yutu rover shows color and textural variations similar to what may be present along the SPARX traverse. Credit: National Astronomical Observatories of Chinese Academy of Sciences.*

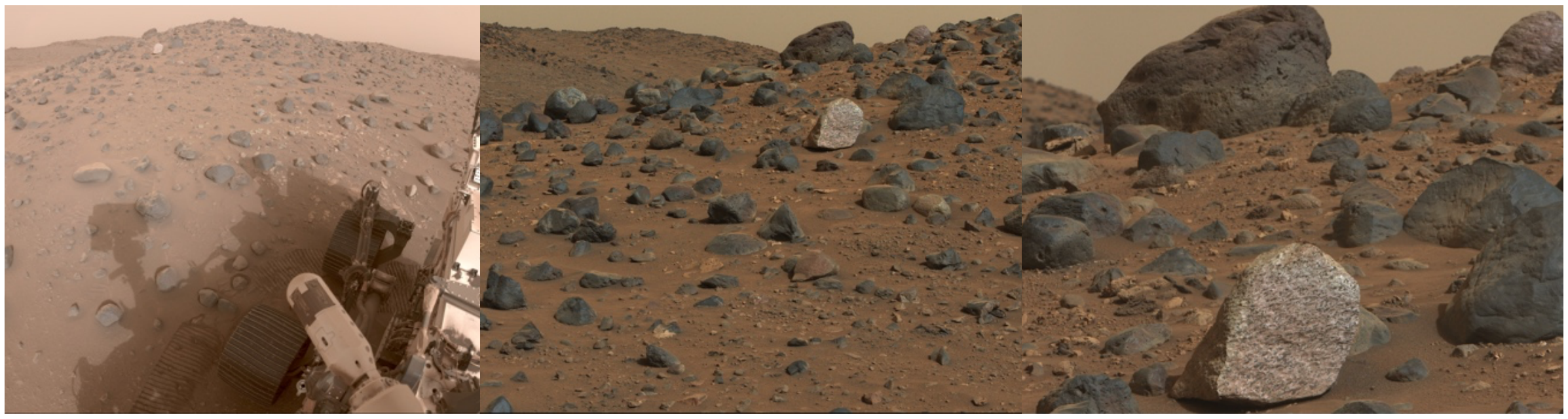

**Figure 4.28.** *Example radiometrically corrected images from context cameras on the Perseverance Mars rover. The light toned rock, nicknamed "Atoko Point," is ~45 cm wide and 35 cm tall and was found to be pyroxene by the rover's compositional instruments. Left: Left Navcam (EECAM) image from sol 1162, ID: nlf_1162_0770105134_737rad_n0524312ncam03162_0a0095j. Middle: Left Mastcam-Z image at 34mm focal length acquired on sol 1164, ID: zrf_1164_0770275563_910rad_n0524312zcam03911_0340lmj. Right: Right Mastcam-Z image at 110mm focal length acquired on sol 1164, ID: zrf_1164_0770274620_848rad_n0524312zcam03910_1100lmj. All images downloaded from Perseverance Analyst's Notebook, https://an.rsl.wustl.edu/m20/AN/an3.aspx.*

### Example instruments capable of context imaging (non-exhaustive list):

There are a number of high-heritage, space qualified color imagers capable of supporting this measurement requirement. This includes the color Enhanced Engineering Camera (EECAM) on the Mars 2020 rover (Figure 4.28, Maki et al., 2020) and MahinaCam, an instrument being developed specifically for a future lunar surface mission (Blewett et al., 2023b), both provide color information without the need for moving parts. The Mastcam and Mastcam-Z color cameras on the Mars Science Laboratory and Mars 2020 rover respectively (Malin et al., 2017; Bell et al., 2021) also provide RGB color with the option to apply filters that enable even more compositional information across additional wavelengths. The Integrated Vision System (IVS) concept is being developed for a lunar polar mission and combines a multispectral imager with a LIDAR system for active spectral sensing and ranging that would provide compositional information beyond three color RGB alongside stereo information (Osinski et al., 2023). Several commercial lunar lander payloads (CLPS) also carry color cameras that may meet the SPARX requirements.

### 4.2.3 Compositional Context Imagers

#### How measurement contributes to experiments and goals:

Similar to context imaging, the ability to map minerals and identify rock types at the landscape scale is necessary both to find where scientifically valuable samples are located, and also to provide additional information about their geologic context. SPARX must specifically be able to find anorthosites that will be used to address Experiment 1: "Determine the composition and age of the anorthositic farside highlands crust", and olivine and/or low-calcium pyroxene to address Experiment 9: "Sample hypothesized excavated lunar mantle material". SPARX must also be able to map the distribution of low-calcium pyroxene along its traverse near central SPA to fulfill Experiment 14: "Determine the origin and extent of the low-calcium pyroxene annulus enclosing central SPA," and to more broadly map common lunar minerals along the entire traverse to address Experiment 13: "Determine the stratigraphy and composition of the unique deposits in central SPA". Mineral maps across the SPARX traverse, and especially within the areas where samples are collected, will also contribute to the ability to define geologic units and provide additional geologic context for collected samples.

#### Justification for spatial resolution, mineral detection, and field of view requirements:

The spatial resolution requirements for context mineralogy imagers are coarser than context imagers because compositional information may be interpretable even if rocks and rocklets fill a single pixel or less. For midfield images (≥5 m distance), SPARX must be able to determine the mineralogy of large rocks (≥20 mm) just beyond the workspace, leading to a requirement of **≤10 mm / pixel at ≥5 m distance**. In the farfield (≥100 m), resolution should be sufficient to determine the mineralogy of large boulders (≥0.5 m scale) that might shed sampleable material, and to determine mineralogy of a ≥1 m scale geomorphic features, leading to a requirement of **≤1 m / pixel resolution at ≥100 m distance**. Compositional imagers must be able to identify **high- and low-calcium pyroxene, olivine, and feldspar** when those minerals are present in abundances that fill a pixel. The field of view requirements for a compositional imager match those of context imaging, which is the ability to capture the broad context around the vehicle. These requirements are **≥180˚ azimuthal imaging**, and image at least from the **sampling workspace to +20° above the horizon in elevation**. Compositional imagers should be able to demonstrate that the stated field of view requirements can be meet using, for example, internal design mechanisms, an external scan platform, or mobility offered by the SPARX platform (e. g. rover turning).

The SPARX SDT is additionally strongly recommending context mineralogy instruments that acquire data in an **imaging framework**, as opposed to point acquisition. The key requirement driving this recommendation is the scientific requirement to understand the mineralogy across the entire scene (≥180° azimuthal, and +20° above the horizon in elevation) in a reasonable amount of time for both the Science Interval Stops (~every 2 km) (Section 6.6.1), and also within the sampling region. The acquired compositional information is used to both find key targets for sampling, and for characterizing the local stratigraphy. Given the projected mission duration of SPARX, the SDT is recommending a large number of autonomous rover operations, which additionally favor an imaging framework for the acquisition of mineralogic information. While the SPARX SDT does not exclude point spectrometers, future investigators should demonstrate that these systems can acquire the required contextual information within a duration that is commensurate with the currently baselined ConOps operational envelop, and that operations can be carried out in an autonomous manner.

### Example instruments capable of context mineralogy (non-exhaustive list):

A number of compositional imagers in development for lunar surface mission could partially or fully fulfill the requirements above (Figure 4.29). For example, the Ultra-Compact Imaging Spectrometer for the Moon (UCIS-Moon) is a short wavelength infrared (SWIR) imaging spectrometer that measures radiance from 0.6 – 3.6 µm and could map olivine, low- and high-calcium pyroxene, and Fe-bearing feldspars (Fraeman et al., 2023). The Lunar Intermediate Infrared Imaging Spectrometer (LIIRIS) observes in the 4-8 µm intermediate infrared (IMIR) wavelength range, which allows it to map olivine, pyroxenes, and determine their Mg# (Kremer et al., 2023). There are a wide number of thermal imagers that are sensitive to bulk silicate mineralogy and are particularly favorable for identifying Si-rich, low-Fe material, alongside thermophysical properties and thermal environments (Greenhagen et al., 2023). These include the Lunar Advanced Filter Observing Radiometer for Geologic Exploration (LAFORGE), a multispectral thermal infrared imaging radiometer slated to fly on the Canadian Space Agency’s (CSA) Lunar Rover Mission (LRM) that is designed to maximize temperature accuracy at temperatures below 100K and also has the ability characterize the abundance of silicates and oxides using multiband filter capabilities (Greenhagen et al., 2025).

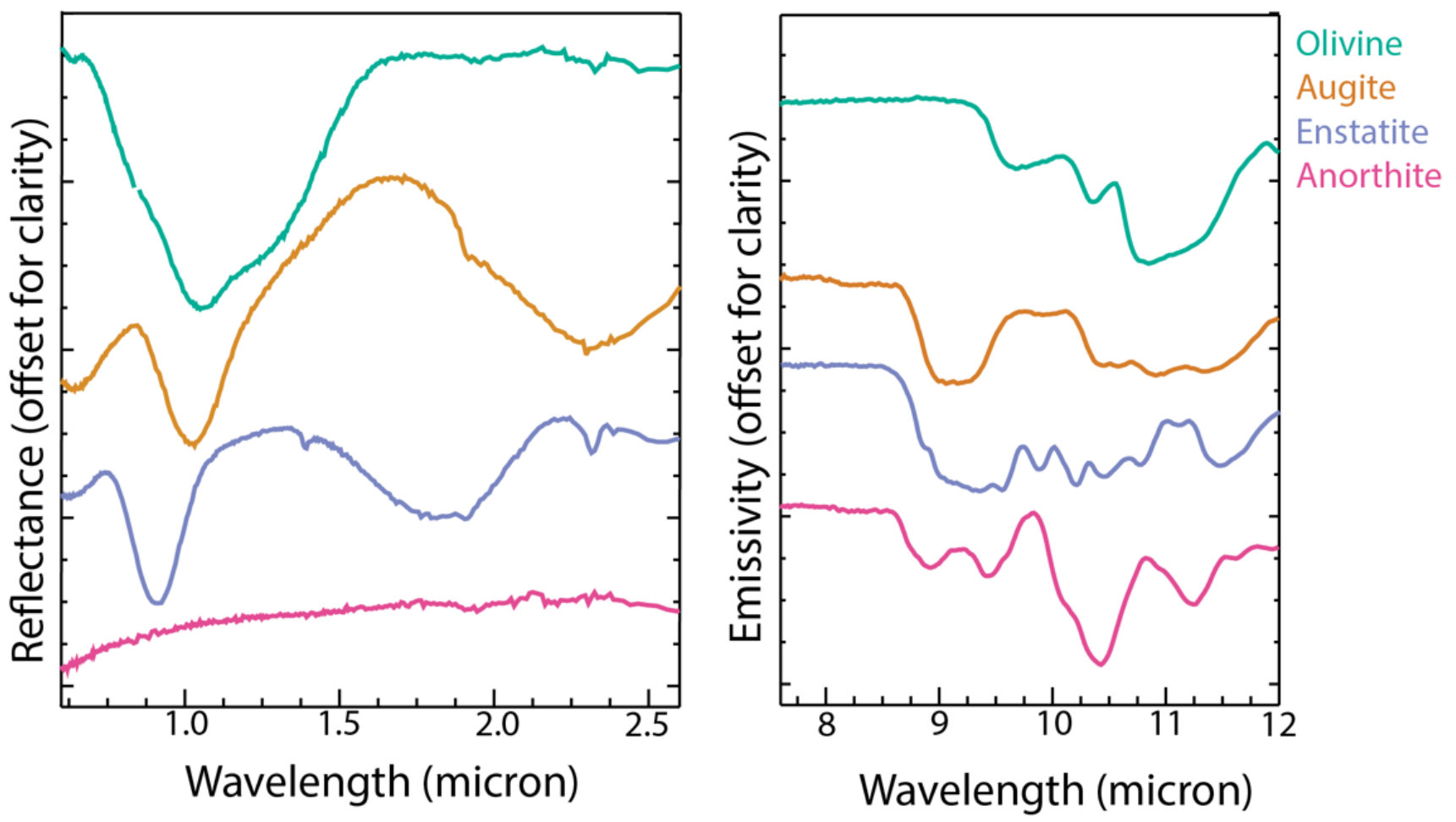

**Figure 4.29.** *Laboratory spectra showing visible shortwave infrared (VSWIR, left) and thermal infrared (TIR, right) wavelengths can be used to identify common lunar rock forming minerals.*

## 4.2.4 Color Fine-Scale Imaging

### How measurement contributes to experiments and goals:

Fine-scale imaging is defined as imaging data that are used to identify and contextualize the material within a rover work volume. This is distinguished from context imaging, which has the primary purpose of finding rocks at the landscape scale and providing geologic context. Fine scale imaging will reveal small scale textural features within a work volume, such as the presence and size of vesicles, large crystals, or glass beads. It will also show the size distribution of material in a work volume, which is critical for the projected sampling efficiency of the SPARX operations. The majority of science objectives and experiments where samples are acquired require specific numbers of grains that are 3 – 8 mm in diameter, and fine-scale imaging assists in verifying these requirements have been met.

#### Justification for spatial resolution, mineral detection, and field of view requirements:

This imager should be able to identify when grain size distributions in a workspace are coarser than typical, in order to maximize sampling efficiency. This translates to **≤0.3 mm / pixel in the sampleable workspace. Three band color** is also desired for similar reasons as described in context imaging.

#### Example instruments capable of fine-scale imaging (non-exhaustive list):

Ideally this requirement can be met by the same instrument responsible for context imaging. Color arm mounted cameras with high spatial resolution, similar to the Mars Hand Lens Imager (MAHLI) on Curiosity or the Wide-Angle Topographic Sensor for Operations and eNgineering (WATSON) on Perseverance, could also meet this requirement, although the addition of an instrument on an arm would add design and operational complexity to the mission that may not be necessary to meet requirements.

#### Baseline vs. threshold determination:

The fine-scale imaging measurement is considered baseline only. In the event that a dedicated fine-scale imaging instrument is not selected, the SDT notes that many currently available imagers are capable of meeting both the Context Imaging (4.2.1) and Fine-Scale Imaging (4.2.4) requirements, and NASA should consider prioritizing a payload selection that can meet both.

### 4.2.5 Fine-Scale Mineralogy

#### How measurement contributes to experiments and goals:

In contrast to context mineralogy, which will be used to find locations to sample, fine-scale mineralogy will reveal the mineralogy of individual sampleable rocks (≥2 cm in diameter) in the rover workspace, to ensure the material that will be / has been sampled has the correct composition.

#### Justification for spatial resolution, mineral detection, and field of view requirements:

Similar to context mineralogy, a fine-scale mineralogy instrument should be able to **identify olivine, low calcium pyroxene, and anorthosite** but at a spatial resolution sufficient to determine the mineralogy of rocks ≥2 cm in the workspace, e. g. **≤2 cm / pixel**.

#### Example instruments capable of context mineralogy (non-exhaustive list):

Similar to context vs. fine-scale imagery, this requirement is ideally met by a single instrument. In addition to instruments described above, there are several high heritage instruments that can assess mineralogy at the centimeter scale only. These include the Advanced Multispectral Infrared Microimager (AMIM), a compact microscopic imager that provides spatially-correlated mineralogical and microtextural information at up to 30 micrometers / pixel with spectral coverage from 0.45 - 3.6 µm (Núñez et al., 2023). Elemental chemistry instruments may also provide a proxy for mineralogy at the centimeter scale and below, if areas that are analyzed are monomineralic. High heritage instruments that acquire chemistry *in situ* include the ChemCam (Maurice et al., 2012) and SuperCam (Maurice et al., 2021) instruments on Curiosity and Perseverance respectively, or an Alpha-Particle X-Ray Spectrometer (APXS), versions of which have been carried by multiple Mars and lunar missions (e. g. Squyers et al., 2003; Gellert, et al., 2013, Shanmugam, et al., 2014).

### 4.2.6 Bulk Elemental chemistry

#### How measurement contributes to experiments and goals:

One of the highest priority science objectives for the SPARX mission is collecting sample(s) of Th-rich material in order to determine whether these materials originated from a farside KREEP reservoir (or other late-stage mantle cumulate) (Objective 2.2, Section 3.2.2). This is captured in E7 "Determine the composition and age of farside Th-rich materials". As described in section 4.1.7 and shown in Figure 4.10, Th-rich material exists in several places within SPA, including a prominent regional distribution in the northwestern portion of the basin, with additional discrete exposures around the crater Abbe M, located to the southeast of Poincare basin. Although colloquially described as exposures here, the actual surficial nature, and host material for this Th-rich material is unknown. As described in Section 3.2.2, on the lunar nearside, Th-rich materials are associated with the signature of KREEP, which is most commonly associated with basaltic lithologies, both in extensive surficial deposits, and also in grains that are well-mixed within the lunar regolith (e.g. Shearer et al., 2006). As a result of the uncertainties surrounding the lithological location of Th-rich material, the SPARX SDT is recommending measurement strategies that emphasize broad spatial surveys in the regions where Th-rich samples are to be acquired.

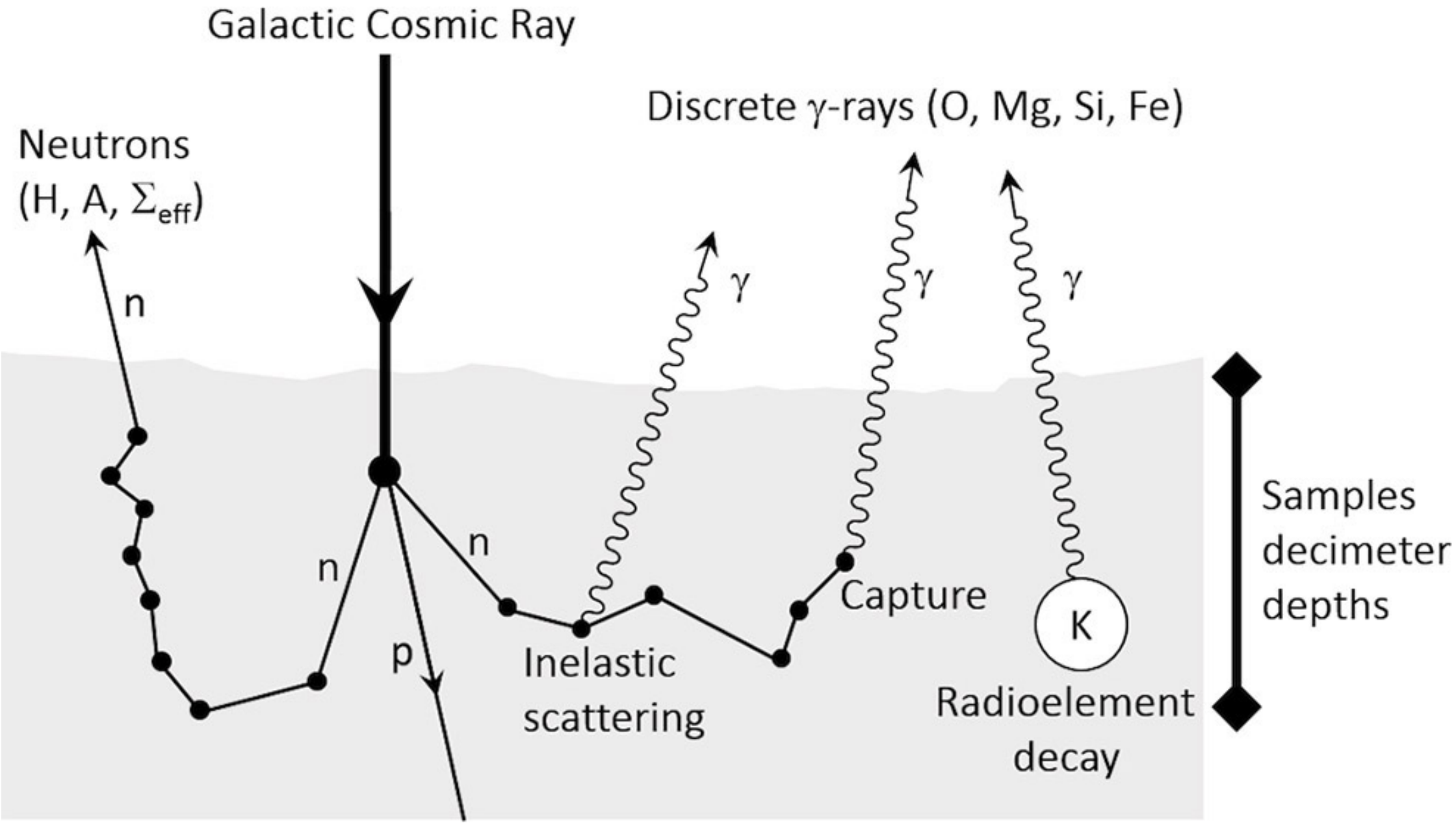

**Figure 4.30.** *From Prettyman et al (2015). Cartoon of gamma-ray production processes in the surface of planetary bodies. Galactic cosmic rays—mainly high-energy protons (p)—smash apart atoms in the top surface of airless bodies. This produces a shower of secondary particles, which include neutrons (n). These neutrons then undergo successive collisions with atoms in the surface and produce gamma rays (γ) via inelastic scattering and radiative capture. Gamma rays can also be produced by the decay of natural radioelements, such as potassium (K), thorium (Th), and uranium (U). The spectrum of the gamma rays can be analyzed to determine the concentration of the detected elements. Important rock-forming elements include oxygen (O), magnesium (Mg), silicon (Si) and iron (Fe). In addition, the escaping neutrons convey information about the hydrogen content (H), average atomic mass (A), and the neutron macroscopic capture cross section ($\Sigma_{e\#}$) of the surface materials.*

**Table 4.2.** *Proposed GRNS instrument sensitivity requirements.*

| Proposed GRNS Instrument Requirement 1 | Proposed GRNS Instrument Requirement 2 |
|---|---|
| The GRNS instrument shall detect Th with a detectability critical level (Lc) corresponding to 3ppm (*see Knoll 2010 equation 3.61*) in under 30 minutes either at a single stationary location or integrated across a ~500 m rover "noodle" track. | The GRNS instrument shall detect Fe, Mg, Si, Al at a minimum threshold of 5 wt% +/- 1 wt% 1$\sigma$ , with a FWHM of ≤ 10% at 662 keV. |
| **Rationale:** For sample selection purposes in finding *a sample that contains Th in the SPA region* in accordance with Objective 2.2 and constrained by ConOps sampling duration time. This requirement is intended to initially "dowse" for Th, resulting in a longer duration measurement after a peak abundance of Th in an area is located. | **Rationale:** For distinguishing between major elemental lines during bulk composition context measurements and tracking of Fe/Si and Mg/Si ratios along the rover traverse. These context measurements are meant to be done over significantly longer integration times (1 – 4 hours). |

In addition to the specific requirements to measure Th for E7, the SPARX SDT also requires measurements of bulk elemental chemistry as a part of the standard suite of geological context measurements (e.g., Section 4.1). These measurements should focus on identifying major rock forming elements including, at minimum, Fe, Mg, Si, and Al. These measurements are required to track changes in the geologic evolution of the basin (Objective 3.2, Section 3.3.2), and provide further context for sampling regions.

### Justification for measurement requirements:

Detection of Th-rich material can be accomplished through the use of gamma-ray spectrometers (GRS) (Reedy 1978) or could be inferred from microwave wavelength measurements (Siegler et al 2023). Publications by (Prettyman et al 2006, Lawrence et al 2007, Yamashita et al 2010, and Kobayashi et al 2012) using orbital gamma-ray and neutron spectrometer (GRNS) data sets, such as Lunar Prospector, suggest that the abundance of Th on the Moon ranges from 0 – 20 ppm, with ~3 ppm being the average in SPA (30 km footprint in polar orbit). However, whole rock abundance data from Apollo samples show Th abundances ranging from 0 ~ 50 ppm (Heiken et al 2005), suggesting that the laboratory derived Th abundance for any particular sample could exceed the orbital detection level. This is largely reflective of the large spatial footprint of orbital GRNS data (~1 km/px), compared with the often heterogenous distribution of Th-rich materials on the surface. **The 3-ppm Th used as the minimum threshold measurement requirement for Experiment 7 reflects the current orbital datasets and was chosen specifically to emphasize sample return from locations within SPA with Th enhancement above the background.** The SPARX SDT also emphasizes that, based on evidence from the comparison of nearside orbital vs. laboratory measurements of Th, it is highly

#### Lunar volatiles investigations enabled by SPARX

Although many of the sites selected by the SPARX SDT do not indicate significantly increased hydrogen abundances (≲0.01 wt%, Wilson et al., 2018); orbital measurements of the lunar South Pole are based on hydrogen map footprint measurement sizes of 30-45 km for bulk composition (Lawrence et al., 2006) often re-analyzed using forward modeling assumptions (e.g., Wilson et al., 2018) such that surface *in situ* measurements may reveal smaller more H-rich deposits. Instruments selected for the SPARX rover (e.g., mineralogy and bulk geochemistry instruments) may enable studies of polar volatiles at higher fidelity field of view (< 1 $m^3$).

likely that these localized Th "hot spot" regions should have materials with Th abundances higher than 3 ppm.

The 3-ppm Th abundance sensitivity requirement set by the STM (Section 3.6) is challenging, though attainable by most GRNS systems, depending highly on integration time and crystal characteristics. Overall, the ConOps time-limit (i.e., integration time) is what will ultimately drive the GRNS instrument design for detection of Th. To better define the need for Th detection, draft requirements are suggested in Table 4.2. For the **required major rock-forming elements** that are not naturally decaying (e.g. **Fe, Si, Mg, Al**), the galactic cosmic ray (GCR) flux is required. Knowledge of the surrounding nuclear environment is typically inferred through neutron spectroscopy measurements (Prettyman 2006). Therefore, **the SDT recommends the completion of a full GRNS suite (i.e., GRS and NS) to provide the required composition measurements.**

### Heritage Instruments Capable of Measuring Th and Major Rock-Forming Elements (non-exhaustive list):

There are a number of heritage gamma-ray and neutron spectrometer instruments capable of measuring Th and major rock-forming elements. Figure 4.32 depicts an ideal example of lunar count rates for a variety of heritage detector efficiencies. Figure 4.31 depicts an example comparison of energy resolution for several GRS instruments. The GRS on board NASA Lunar Prospector consisted of a cylindrical bismuth germinated oxide (BGO) crystal surrounded by an anticoincidence shield (ACS) of borated plastic, while the NS consisted of a pair of He-3 tubes (Feldman et al., 2004). BGO is highly efficient at interacting with and measuring gamma-rays, making it a potential candidate for "Th-dowsing", although it typically has an energy resolution (~10% FWHM) which makes it less suitable for measurement of major rock-forming elements. The GRS onboard the Near-Earth Asteroid Rendezvous (NEAR) spacecraft (targeting asteroid 433 Eros) was a NaI(Tl) scintillator, combined with a BGO ACS (Goldsten, 1998). NaI(Tl) typically has a decent energy resolution (~7% FWHM) for detection of gamma-rays, making it and other similar-resolution GRS scintillators suitable for measurement of rock-forming elements. However, NaI(Tl) is less efficient than BGO, potentially requiring a larger crystal size to meet the same efficiency as BGO for a limited mission time ConOps. High-purity germanium (HPGe) spectrometers have flown on the NASA Mars Odyssey spacecraft (Boynton et al., 2004), the NASA MErcury Surface, Space ENvironment, GEochemistry, and Ranging (MESSENGER) spacecraft (Goldsten et al., 2007), the NASA Psyche spacecraft (Lawrence et al., 2025), and on astrophysics missions. HPGe GRS have a superior energy resolution (< 1%) but comparable efficiency when compared to scintillators. These instruments may allow for measurements of additional rock-forming elements, such as Ti, Ca, Na, and others, due to their precise energy resolution. However, due to the nature of the germanium crystal sensor head they must be operated at cryogenic temperatures (~80 K) which can cause complexity in thermal design and interface requirements (e.g., increased power requirements due to cryocooling or stringent thermal interface requirements if passive techniques are used).

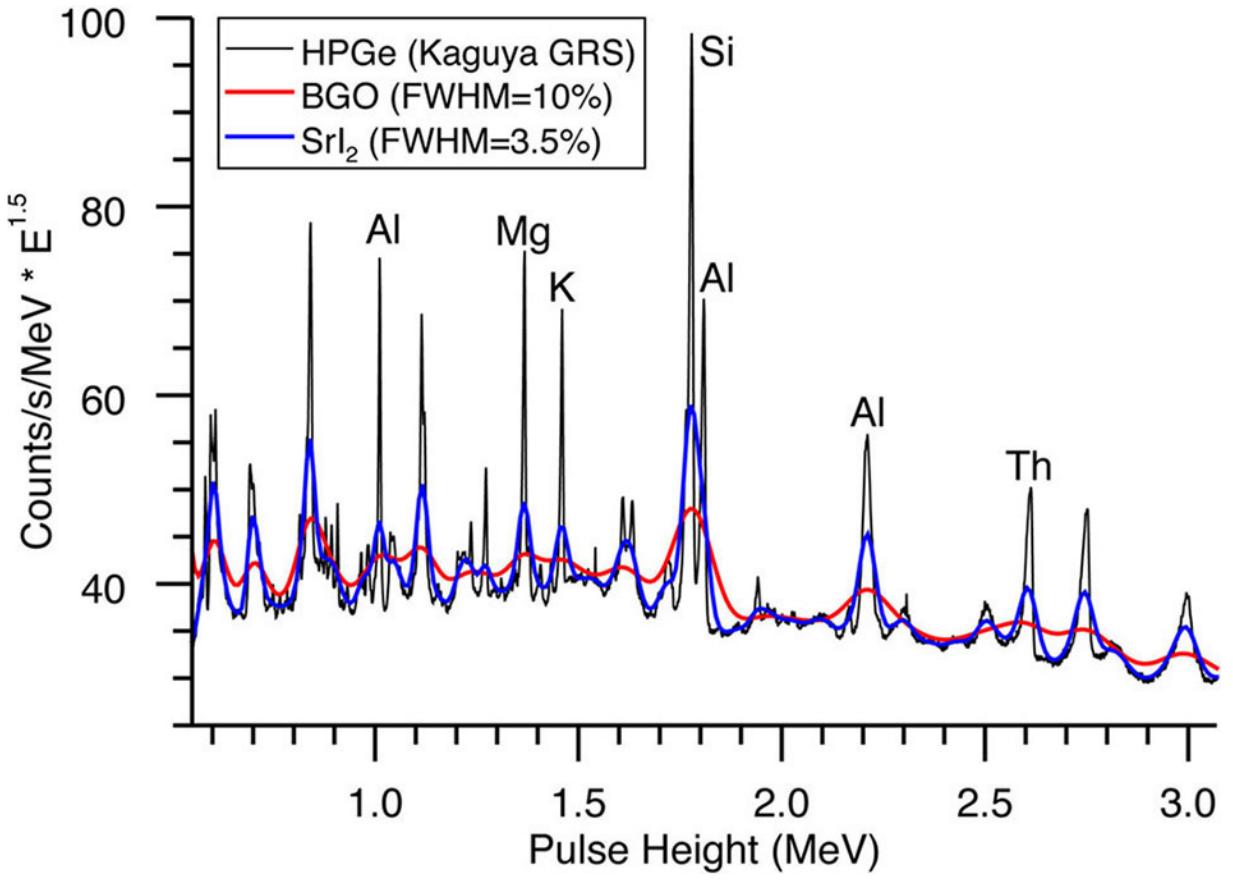

**Figure 4.31.** *Illustration showing how energy resolution (i.e., peak sharpness) affects the ability to extract chemical information from a lunar gamma-ray spectrum. The whole-Moon spectrum acquired by the Kaguya gamma-ray spectrometer (GRS), which was equipped with a HPGe detector, is shown here. For comparison, the measurement has been degraded using a model to match the performances of BGO and strontium iodide ($SrI_2$) scintillators. Figure from Prettyman et al (2015) SPIE.*

Other GRS options include CdZnTe (CZT) with a resolution of ~2%, part of the GRaND instrument which flew on NASA Dawn (Prettyman et al., 2003); CeBr3(Ce) with a resolution of ~4%, which is currently onboard ESA/JAXA BepiColombo (Mitrofanov et al., 2010; Benkhoff et al., 2021); Cs2YLiCl6:Ce (CLYC) a dual detection scintillator capable of detecting neutrons with a resolution of ~5%, which has flown on NASA LunaH-Map and is manifested on NASA Lunar-VISE (Hardgrove et al., 2020; Hardgrove et al 2025).

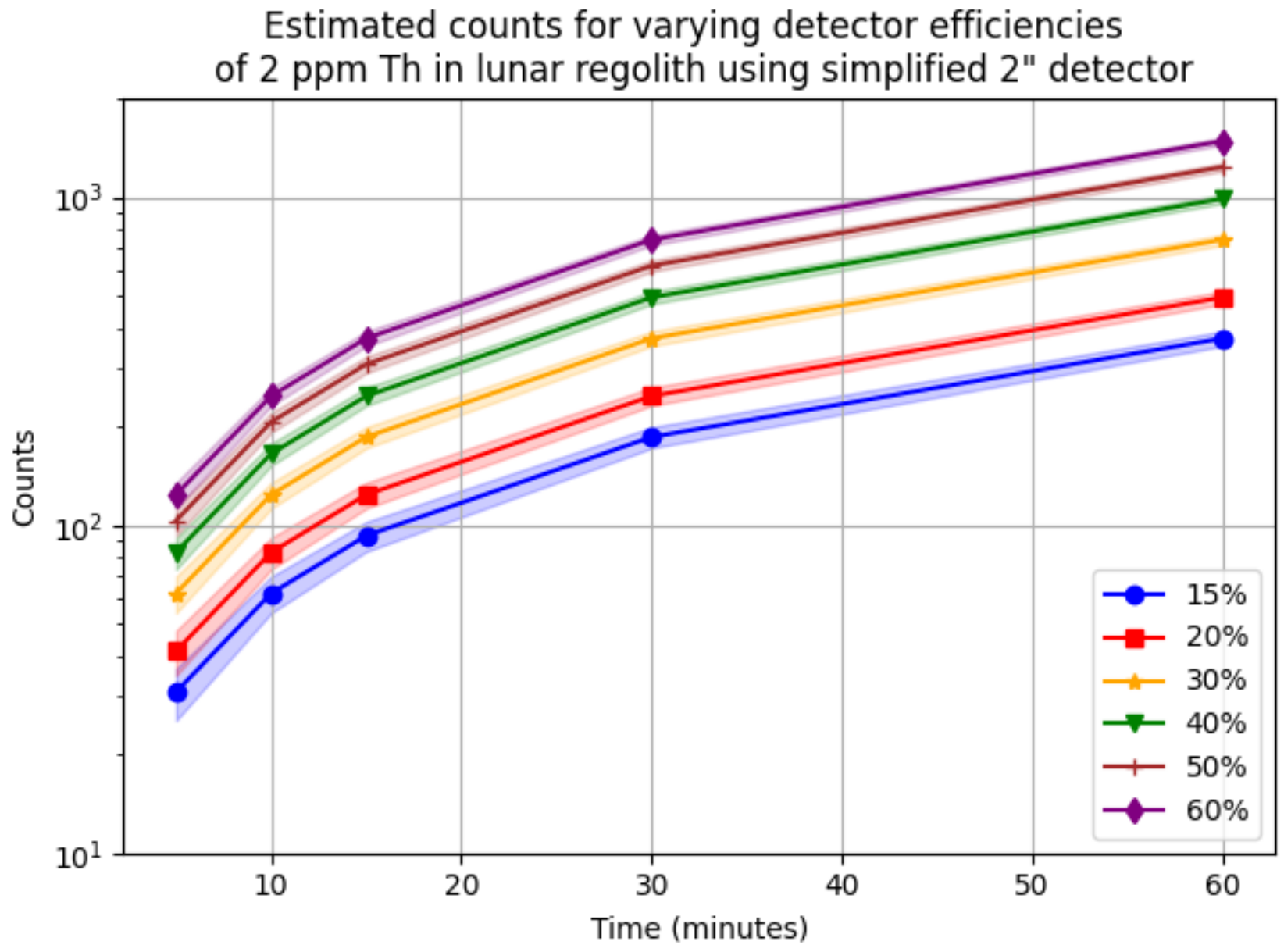

**Figure 4.32.** *Results of count (N) calculations done for a 2-inch diameter cylindrical detector with flat-face perpendicular to a 1-meter diameter emission surface at a distance of 50 cm using the approximated solid angle estimation from Knoll (2010). Intrinsic detector efficiencies ranging from 15 – 60% are depicted, with Poisson statistical uncertainty ($\sqrt{N}$) applied as the lighter contours around each efficiency trace. The 60% efficiency line corresponds well to a BGO scintillator, similar to that on Lunar Prospector; the 15%-20% lines correspond well to a HPGe spectrometer, similar to that on Psyche; and the 20%-30% lines correspond well to that of a Lunar-VISE-like instrument. Note that these results are ideal, and do not account for the energy resolution which can be advantageous for truly identifying radionuclides within a gamma-ray spectrum, nor the contribution to uncertainty from background sources or detector degradation/noise.*

### Nuclear imaging technologies from national defense initiatives

Nuclear imaging technologies have been around for decades, used in nuclear nonproliferation monitoring and nuclear plant materials transport monitoring (Kroeger et al, 2002). In more recent years, miniaturized, hand-held gamma-ray and neutron imaging systems have been developed for use with the International Atomic Energy Agency for detection of nuclear threats and radiation leaks (Dreyer et al, 2014). Compton imagers in particular could be utilized for finding Th-rich individual rock materials in limited integration time scenarios. Alpha back-scatter imaging techniques could also potentially be used in rapidly locating buried H-rich deposits (Unzueta et al., 2025). These types of nuclear imaging techniques could help astronauts rapidly assess and locate critical geologic materials under highly constrained mission time-limits (i.e., a lunar day) for mission architectures that combine SPARX rover activities with Artemis astronaut rendezvous (Section 7.6.2).

### 4.2.7 Magnetic Field

#### How measurement contributes to experiments and goals:

As described in Objective 1.4 (Section 3.1.4), constraining the evolution of the early lunar dynamo is the highest priority study for lunar paleomagnetism in SPA. This objective requires high-accuracy paleointensity measurements from returned samples to be tied to well-constrained geological context measurements of surface geology, composition, and present-day magnetic environment (E6, Section 4.1.6). Characterization of the present-day magnetic environment is required to both understand spatial variations in the strength, and also source depth of remanent crustal magnetism across the SPA basin. These measurements are themselves required to constrain competing hypotheses for the origin of magnetic crustal material, and assess whether magnetic anomalies are related to the formation of the SPA basin, or an evolving lunar dynamo. In addition to the focused paleomagnetism studies, a magnetometer survey conducted within SPA also contributes to Objective 3.1 to “Determine the crustal structure (thickness, density, remnant magnetic field strength) across the SPA basin (from the center to the putative rim) to constrain mega-basin formation and evolution models” (Section 3.3.1). Magnetic field data will constrain the spatial structure and source depths of crustal remanent magnetization, providing the context needed to link surface magnetization to paleofield intensity, dynamo evolution, and mega-basin formation processes (E11, Section 4.1.11).

#### Justification for Measurement Requirements:

Apollo surface measurements demonstrate the significant variability in the surface strength of the lunar magnetic field; ranging from 3 nT at the Apollo 15 exploration region to 330 nT at Apollo 16 (Dyal 1974). Accordingly, the SPARX SDT **requires a magnetometer that can accommodate a large dynamic range (≥2,000 nT)** to capture the full range of potential surface field intensities. To resolve the lowest measured surface magnetic field value (based on the observed 3 nT fields at Apollo 15), a **resolution of ≤1 nT is sufficient**. While not required for SPARX objectives, the SDT notes that equipping a magnetometer with a higher field sensitivity resolution of 0.25 nT would enable additional space weather applications of the measurements, and is comparable to the Mars Atmosphere and Volatile EvolutioN (MAVEN) magnetometer’s requirements and capabilities (Connerney et al., 2015). A **sampling rate of ~10 Hz is adequate** **for most geophysical applications** (Banfield et al, 2019) including those outlined by SPARX. However, for investigations that seek to include additional space weather investigations will re quire a magnetometer with a sampling rate ≥20 Hz (Loto'aniu et al, 2019). **Both the static (DC) and time varying (AC) magnetic fields must be measured**. Magnetic field measurements can be conducted passively, provided that the rover's electromagnetic (EM) state is well-characterized. Close coordination between the platform and magnetometer teams will be required to assess rover-generated EM noise and, where necessary, modify circuitry or operational models to minimize interference.

### Magnetometers enable better science for other instruments

Solar energetic particle events (SEPs) can significantly increase particle flux in the vicinity of airless bodies when those bodies lie within the SEP propagation trajectory. Elevated particle flux can affect sensitive science instruments (e.g., semi-conductors, scintillators) by inducing baseline noise shifts or transient artifacts in electronic systems and instruments. The effects of a SEP event can persist for ~ a day up to a month, during which affected instrument datasets are often excluded from analysis (Peplowski 2016). Magnetometers can detect and characterize SEP-related disturbances, providing independent means of identifying periods of elevated solar activity and determining when the local radiation environment has returned to baseline conditions. Improved characterization of the magnetic field in the vicinity of the rover may enable correction of instrument datasets and recovery of data that would otherwise be discarded.

### Example instruments capable of magnetic field measurements (non-exhaustive list):

Fluxgate magnetometers, such as those used on the NASA MAVEN, InSight, Europa Clipper

missions, measure the full vector magnetic field (Bx, By, Bz). Fluxgate magnetometers can be designed with mission-specific dynamic ranges, allowing optimization for anticipated field strengths. Search-coil (induction) magnetometers are better suited for time-varying (AC) magnetic fields and are commonly used for space weather investigations. Overhauser and scalar magnetometers are useful for high-precision measurements of total magnetic field magnitude and are often used to calibrate fluxgate magnetometers (Kellock et al, 1996). For determining the magnetic context of geologic samples, measurement of the full magnetic field vector is preferred. Accordingly, a fluxgate magnetometer, or comparable instrument capable of vector field measurements, is well suited to meet mission requirements.

**Baseline vs. Threshold Determination**: The magnetic field measurement is considered baseline only. The reflects the baseline only nature of Obj. 1.4. The baseline only distinction also reflects the increased design and cost burdens of incorporating a magnetometer and magnetic cleanliness requirements into a rover design. While the SDT feels that these design requirements could be met by sufficient early characterization of these requirements, the overall final architecture may ultimately drive the practicality of inclusion. Despite this, the SDT emphasizes the significant scientific value in including magnetic field measurements.

### Heliophysics and space weather investigations enabled by SPARX

The SPARX mission is intended to survive multiple lunar day-night cycles, with a total duration expected to enable observations across several lunar orbits and repeated passages of the Moon through Earth's magnetotail (~20% of the Moon's orbit about Earth resides in the magnetotail). The inclusion of instruments on SPARX, such as a Gamma-Ray and Neutron Spectrometer (GRNS; sensitive to particles in the ~20keV–10MeV range) and a magnetometer enable monitoring of the solar wind, galactic cosmic ray (GCR) flux, solar energic particle (SEP) events, and the local electromagnetic environment as a function of time on the lunar surface (Shang, et al 2026). When combined with complementary observation heliophysics assets (e.g., multi-spacecraft measurements of SEPs within the Heliophysics Solar Observatory), these data will contribute to improved understanding particle transport, solar-planetary interactions, and the plasma radiation environment throughout the Solar System. Characterizing the lunar radiation and electromagnetic environment is also essential for enabling sustained lunar surface operations and long-term human presence.

### 4.2.8 Gravity Field

#### How measurement contributes to experiments and goals:

Gravity anomalies are a function of the distribution of bulk density in the lunar interior, including shallow crustal heterogeneities, relief on the crust–mantle interface, and mantle density anomalies. Gravimeters measure the strength and variation of gravity fields at specific locations on a planetary surface, and these changes can then be inverted to model variations in subsurface density representative of changes in crustal structure. Therefore, measurement of gravity anomalies within a broad portion of SPA is required to address Obj. 3.1 "Determine the crustal structure (thickness, density, remnant magnetic field strength) across the SPA basin (from the center to the putative rim) to constrain mega-basin formation and evolution models" (Section 3.3.1)". The investigation of this objective has two constituent experiments, both of which require measuring gravity anomalies within SPA. First, measurements of gravity fields taken across a traverse from the center of SPA to at least the edge of the model hypothesized inner rim (650 km) are required to constrain the subsurface structure of the mega-basin impact event (E11, Section 4.1.11). Second, gravimeter measurements taken while traversing over a hypothesized SPA rim massif to determine if the bulk density of the massif is consistent with uplifted crustal rim material, or with a different provenance (E12, Section 4.1.12).

#### Justification for Measurement Requirements:

The gravity instrument must be able to determine the amplitude of the gravitational acceleration vector at the rover with a relative accuracy **Baseline measurement of ≤1 mGal (Threshold of ≤10 mGal)**. This measurement requirement is sufficient to resolve density contrasts of ≤200 kg/m$^3$ when the rover experiences an elevation change of ≥100 meters; this is comparable to the density contrast between different major lunar lithologies (e. g., Kiefer et al., 2012). While an elevation change of ≥100 meters is a measurement minimum requirement, the SPARX SDT recommends a baseline requirement of 1000 meters of elevation change to ensure that the targeted location is consistent with hypothesized SPA rim massif material (e.g. Figure 4.33). **The instrument's relative accuracy must be stable between measurements made at** **sequential locations over a ≥10 km range (stops every 2 km), with the mission vehicle remaining mechanically quiet during gravity measurements**. The ≥10 km range is derived from the Experiment 12 requirement of a 1km elevation gain with an average slope of 6°, which is below the traverse slope safety requirement of ≤10° (further detail 4.1.12).

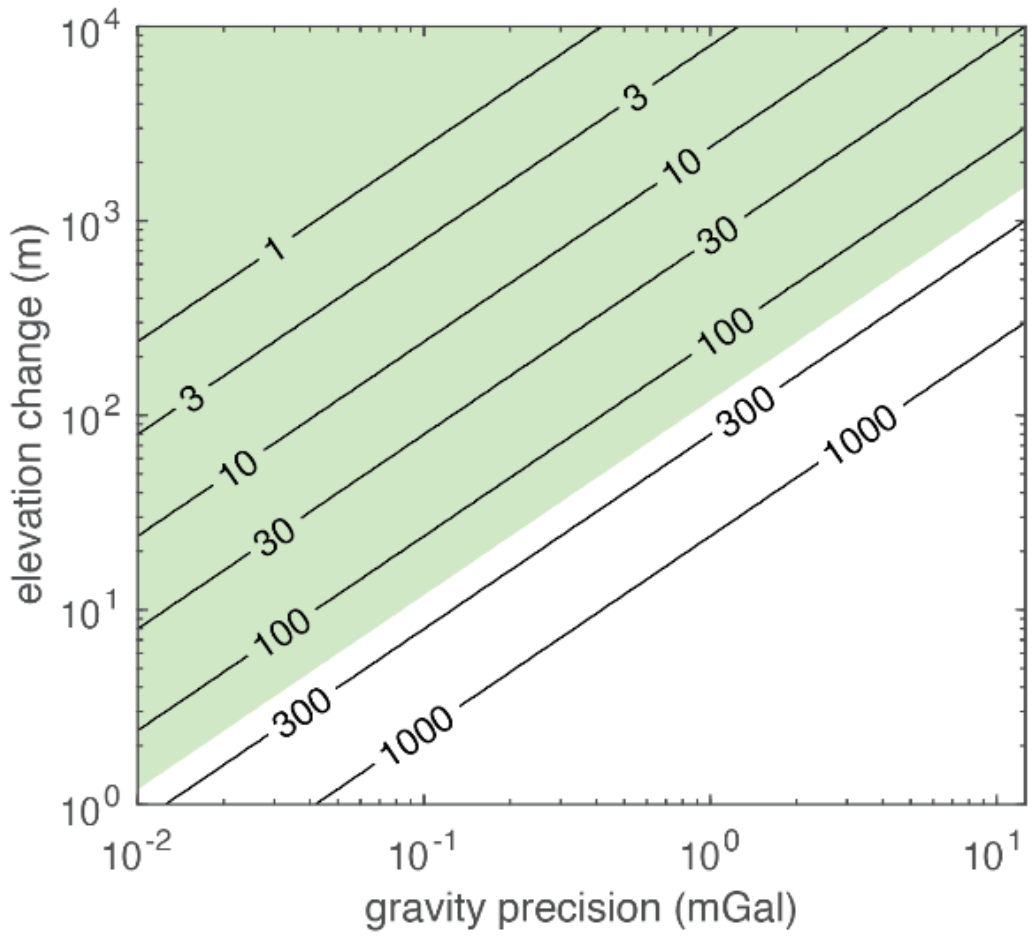

**Figure 4.33.** *density uncertainty associated with gravity measurements of various precision and elevation change. The green zone indicates a density uncertainty of <200 kg/m³, which should be sufficient to rule out some lithologies.*

## Example instruments capable of gravity measurements (non-exhaustive list):

Radio-based instruments, such as the Ka-band systems used for GRAIL, require the use of multiple spacecraft to measure tiny distance changes as a result of gravitational changes (Zuber et al 2013). Superconducting gravimeters, such as Superconducting QUantum Interference Devices (SQUID), offer extremely high precision measurements but require the use of extreme cryogenic systems (4 K) (Huang et al, 2022). Spring or vibrating-string gravity instruments are ideal for lunar surface measurements. The lunar Traverse Gravimeter Experiment (TGE), carried by the Apollo 17 astronauts, is currently the only *surface-based traverse* gravity measurement to have been collected on the Moon (Talwani et al, 1976). The Mars Science Laboratory (MSL) Curiosity rover is equipped with dual redundant Rover Inertial Measurement Units (RIMUs) which are used for obtaining rover attitude and position information, particularly during drives, and are located in the rover wheels. Each RIMU includes a set of three-axis micro-electromechanical (MEMS) accelerometers and fiber-optic gyros. When using these MEMS, final precision in the relative determination of Martian surface gravity is approximately 10 mGal. However, these measurements are based on model density, relying on assumptions of the Martian near-subsurface properties (Lewis et al 2019). Either a technologically updated Apollo-era-based gravimeter or MEMs-based accelerometer instrument is sufficient for SPARX gravity measurements.

### 4.2.9 Table of *In Situ* Measurement Requirements

Foldout 4.1 displays the *In Situ* Measurement requirements.

**Finding: The SDT finds that a baseline set of 7 in situ measurement classes (context imaging, context mineralogy, fine-scale imagine, fine-scale mineralogy, bulk elemental chemistry, gravity field, and magnetic field) are required to achieve the SPARX objectives, and these measurements can be accomplished by a minimum payload of 5 instruments.**

**Finding: The SDT finds that in situ measurements taken along the prescribed intervals of the SPARX mission, and in particular, in the sample acquisition regions are required to meet the science objectives. The SDT further notes that in situ measurements taken in a manner that are not coincident with paired sample activities (e.g., measurements taken at different locations on the Moon or as part of different missions) do not successfully meet the required criteria, and risk not achieving SPARX science objectives.**

| | MEASUREMENT TYPE | DRIVING FUNCTIONAL NEEDS | REQUIREMENTS PARAMETER | VALUE | JUSTIFICATION AND NOTES | STRAWMAN PAYLOAD | NOTES | SPARX INSTRUMENT GUIDING PRINCIPLES |
|---|---|---|---|---|---|---|---|---|
| THRESHOLD | (M1) Context imaging | Enable geologic mapping and topographic reconstruction of sample sites (≥100 m range)<br>———<br>Show if rover workspace contains sampleable material (rocks ≥5 mm in diameter and regolith fines)<br>———<br>Reveal waypoints of interest for future drives | Spatial scale (midfield): | ≤0.5 cm/pixel at ≥5 m distance | Sufficient to identify larger sampleable rocks (≥2 cm in size) just beyond the rover workspace (≥5 m), and map sites. | MastCam-Z | | Below is a series of guiding principles for SPARX science instruments. These are not immutable, but are suggested constraints to ensure that this mission, with its unique requirements and concept of operations, is able to succeed.<br><br><u>Accomodation:</u> Instruments are permitted to be body or mast-mounted. No instruments shall be on the arm, or require close contact to the lunar surface.<br><br><u>Time To Acquire Observations:</u> Instruments shall not require prolonged operations that inhibit mission progress. It is preferable to have instruments that can operate while roving. If instruments require stops, then shorter stops are greatly preferred. The current baseline assumes interval stops are currently allocated ≤1 hour every 2 km (and ideally much shorter times is preferable).<br><br><u>Time of Day Sensitivities:</u> Instruments shall be operable during any time of the lunar day. If instruments require additional illumination, they must provide it.<br><br><u>Data Volume:</u> The total data volume collected across the payload must meet SPARX ConOps scenarios for onboard data storage and downlink. |
| | | | Spatial scale (farfield): | ≤10 cm/pixel at ≥100 m distance | Sufficient to detect larger boulders (≥0.5 m scale) that might shed sampleable material, or detect a ≥1 m scale geomorphic features (e.g., stratigraphic layers, ejecta rays, melt deposits) at a range of ≥100 m (comparable to the scale of many sample sites, and the distance between "snapshots") | | | |
| | | | Color: | ≥3 color bands | Look for compositional and maturity variation (e.g., Blewett et al. 2022) with less resource requirements than mineralogy instrument. | | | |
| | | | Topographic reconstruction | Nominally stereo, proposer to demonstrate sufficiency | 3D reconstructions of terrain aids in interpretation of geologic features and stratigraphic relationships, directly addresses E11 and E12 | | | |
| | | | Field of view: | Azimuthal: ≥180° panoramic capability<br>Elevation: from the workspace to +20° above the horizon | Panoramic capability may be accomplished either with the instrument and/or rover mobility | | | |
| | (M2) Context mineralogy | Finding key mineralogies (including anorthosites, olivines, and high/low-Ca pyroxene) for sampling<br>———<br>Enable geologic mapping and determination of mineralogy at the scale of the rover within a sample site (≥100 m range) | Spatial scale (midfield): | ≤1 cm/pixel at ≥5 m distance | Sufficient to determine the mineralogy of sampleable rocks (≥2 cm—relevant for the "rock hunting" sampling modes) just beyond the rover workspace (≥5 m). | UCIS | | |
| | | | Spatial scale (farfield): | ≤1 m/pixel at ≥100 m distance | Sufficient to determine the mineralogy of larger boulders (≥0.5 m scale) that might shed sampleable material, or determine mineralogy of a ≥1 m scale geomorphic features (e.g., stratigraphic layers, ejecta rays, melt deposits) at a range of ≥100 m (comparable to the scale of many sample sites, and the distance between "snapshots") | | | |
| | | | Mineralogy: | Ability to detect material of key lithologies, such as high- and low-Calcium pyroxene, olivine, and feldspar when they fill a pixel | Proposer must justify why spectral range, sampling, and SNR of their proposed instrument is sufficient to meet these requirements and the ability to identify major rock-forming minerals on the Moon. | | | |
| | | | Field of view: | Azimuthal: ≥180° panoramic capability<br>Elevation: from the workspace to +20° above the horizon | Panoramic capability may be accomplished either with the instrument and/or rover mobility. Understanding the mineralogy across the <u>entire scene</u> in a reasonable amount of time will be critical for finding key targets to sample and characterizing local stratigraphy. | | | |
| | (M3) Fine-scale mineralogy | Determine mineralogy of <u>individual</u> sampleable rocks (≥2 cm in diameter) in the rover workspace, including the scoop | Spatial scale (nearfield): | ≤2 cm/pixel in the rover workspace | Sufficient to determine the mineralogy of sampleable rocks (≥2 cm—relevant for the "rock hunting" sampling modes) within the rover workspace. | UCIS | | |
| | | | Mineralogy: | Ability to detect material of key lithologies, such as high- and low-Calcium pyroxene, olivine, and feldspar when they fill a pixel | Proposer must justify why spectral range, sampling, and SNR of their proposed instrument is sufficient to meet these requirements and the ability to identify major rock-forming minerals on the Moon. | | | |
| | | | Field of view: | Capability to characterize the mineralogy of the sampleable rocks in the rover workspace in the scoop | - | | | |
| | (M4) Bulk elemental chemistry | Determine abundance of major rock forming elements in lunar regolith in the vicinity of the rover<br>---------<br>Identify Th-rich regions ≥3ppm | Elemental identification: | Threshold: Th<br>Baseline: Fe, Si, Al, Mg, Ti, Na, K, Ca | Identification of high / low Th regions are required to meet science objectives. | GRNS (Lunar-VISE) | | |
| | | | Sensitivity: | Th (2.6MeV) with a detectability critical level (Lc) corresponding to 3ppm (see Knoll 2012 equation 3.61) in under 30 minutes<br><br>Fe (846keV), Si (1.78MeV), Al (2.2MeV), Mg (1.4MeV), K (1.46MeV threshold), Ca (3.7MeV threshold), Ti (6.67MeV threshold) at a minimum Lc corresponding to 5 wt%, with a FWHM of ≤10% at 662 keV | | | | |
| | | | Field of View: | Ability to resolve a local Th hotspot (counts at a minimum of Lc + 10% or higher) within a 10 m by 10 m grid area | | | | |
| | | | Energy resolution: | ≤10% FWHM | | | | |
| | (M5) Gravity field | Determine the amplitude of the gravitational acceleration vector at the rover | Relative accuracy: | Baseline: ≤1 mGal<br>Threshold: ≤10 mGal | ≤1 (≤10) mGal relative accurary is sufficient to resolve density contrasts of ≤200 kg/m^3—comparable to the density contrast between different major lunar lithologies—over an elevation change of ≥100 (≥1,000) meters. | IMU | Vehicle must be mechanically quiet during gravity measurements. | |
| | | | Stability: | Proposer to demonstrate that the instrument's relative accuracy is stable between between measurements made at sequential locations over a ≥10 km range. | The ≥10 km range was estimated based on assuming the rover must gain 1 km elevation, which for plausible slopes (≤10°) would be, conservatively, 10 km in distance. | | | |
| | | | Distance between measurement locations: | Baseline: ≤2 km<br>Threshold: ≤10 km | The mission should exceed the resolution limits afforded by GRAIL (approximately 10 km). | | | |
| BASELINE | (M6) Fine-scale imaging Baseline, not threshold | Identify and contextualize sampleable material in the rover workspace | Spatial scale (nearfield): | ≤[0.3] mm/pixel in the rover workspace | Sufficient to identify coarser than typical grain size distributions to maximize sampling efficiency (for ≥3 mm diameter rocks in the rover workspace). | MastCam-Z | *This requirement may be met by engineering cameras. | |
| | | | Color: | ≥[3] color bands | Determining whether the sampleable material are representative of the surrounding terrains. | | | |
| | | | Field of view: | Capability to image the entire rover workspace (including the scoop) | - | | | |
| | (M7) Magnetic field Baseline, not threshold? | Determine the magnetic field strength and direction in the vicinity of the rover | Dynamic range: | ≥2,000 nT | There have been limited magnetic field measurements from the surface of the Moon, but they are expected to greatly exceed what is measured from orbit. From Apollo, there is substantial variability in the measured surface magnetic field strength, from 3 nT at Apollo 15 to 330 nT at Apollo 16 (Dyal 1974), and thus it is appropriate to have large dynamic range. | IFG (InSight) | Magnetic field measurement requirements are very sensitive to the choice of traverse, and in particular, whether the rover will visit mapped magnetic anomalies (e.g., near Ingenii). | |
| | | | Resolution: | ≤1 nT | ≤1 nT would easily resolve the surface magnetic field of even the weakest Apollo landing site. | | | |
| | | | Noise level and stability: | Adequacy to be justified by proposer | - | | | |
| | | | Sampling rate: | ≥1 Hz | ≥1 Hz is appropriate for most geophysical applications (although ≥10 Hz is preferable for other applications, including space weather). | | | |

**Foldout 4.1.** SPARX in situ measurement requirements. The full SPARX science traceability matrix can be found in Foldout 3.2. Rows in gray are baseline (not threshold) measurement types.

## 4.3 LABORATORY MEASUREMENT REQUIREMENTS

The SPARX Experiments (Section 4.1) define a range of required rock sizes and lithologies to be collected (e.g., Figure 4.34) and examined using terrestrial laboratory facilities. The measurements required to address Goals and experiments defined in the SPARX STM (Section 3.6), and Table: Sample Requirements (Section 4.3.6) fall into five broad categories: (1) petrologic sample characterization, (2) bulk elemental chemistry, (3) geochronologic measurements, (4) volatile measurements, and (5) remanent magnetization measurements. The following sections include discussion of scientific rationale for different investigations, as well as, noting places of suggested analysis synergy (e.g., places where a logical progression of investigations can be performed on a single sample). The recommendations in this section reflect best practices developed by the lunar petrology community and are designed to maximize scientific return from the collected samples.

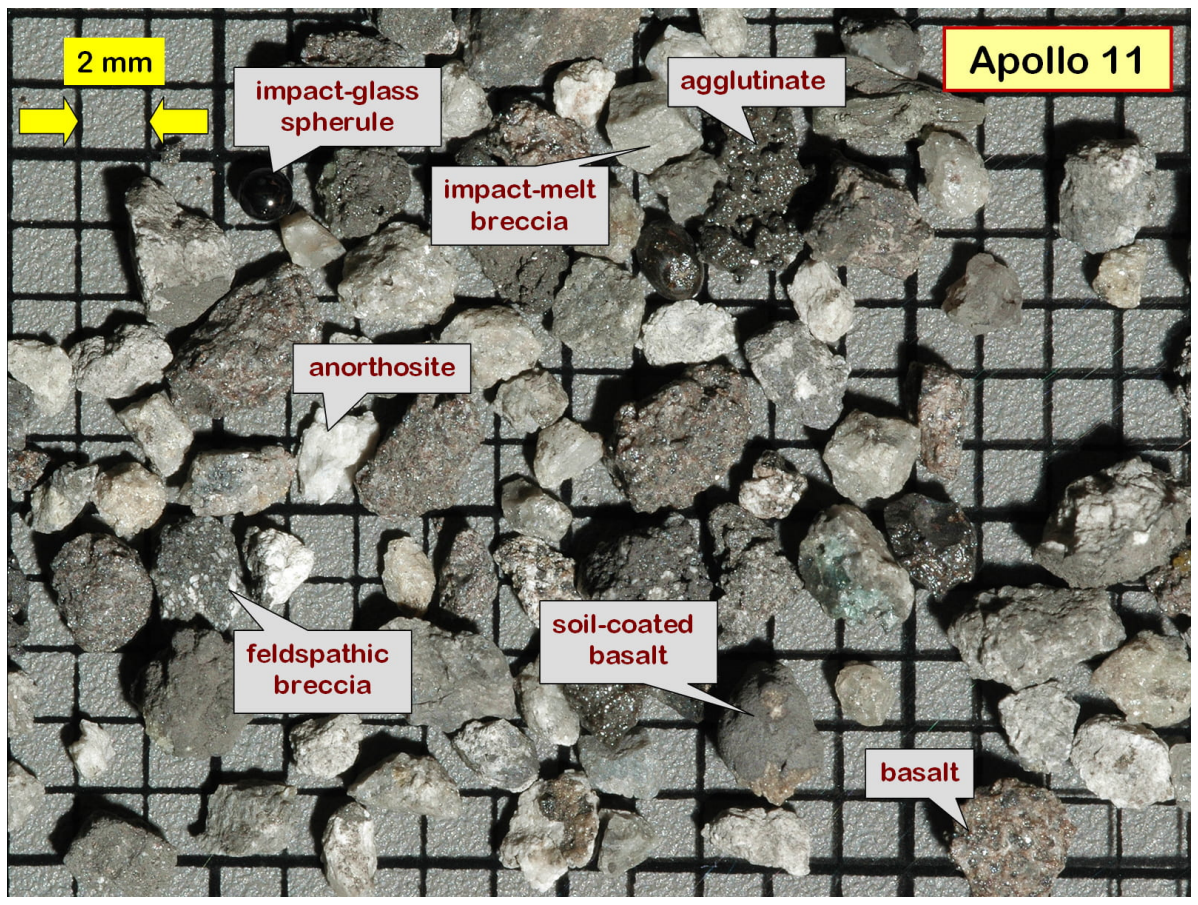

**Figure 4.34.** *Lithic fragments from the 2-4 mm grain-size fraction of lunar regolith collected during Apollo 11. The sample contains basalts (and basalts coated in lunar dust), impact glass spherules, anorthosites, regolith breccias, and agglutinates. Photo credits: Randy Korotev.*

The SDT notes that the measurements described here will require the work of many dozens of researchers and laboratories. While the full discussion of these analyses is beyond the scope of the TOR, the SDT recommends NASA pursue a strategy similar to the Apollo sample analysis teams and the planned Artemis sample analysis teams with early community solicitations, and integration of key team personnel with the sample preparation and characterization processes. The SDT further notes that the findings of the LEAG/ExMag Artemis Sample Return Specific Action Team (ASR-SAT) report are highly applicable to future SPARX returned samples, and the SDT encourages NASA to implement key findings of this report in advance of SPARX sample return. In particular, the SDT highlights the ASR-SAT finding of an impending critical lack of both laboratory facilities and trained personnel in the United States to carry out many of the highest-priority SPARX investigations. The SDT strongly recommends NASA address the findings of this report, particularly for the critical laboratory facilities required for SPARX-related investigations.

### 4.3.1 Petrologic Sample Characterization

#### 4.3.1.1 Recommendations for Initial Characterization

Once samples are returned from the Moon, they should undergo initial triage in coordination with the NASA Johnson Space Center (JSC) curation team, which involves mostly non-destructive characterization to preserve sample integrity. The following recommendations should be carried out, in addition to, the standard curation characterization requirements.

A standard portion (~1-3 g) of larger samples will be prepared for thin sectioning. Although a small amount of material is lost making any thin section, numerous sections can be made from one piece or billet (~0.3 g of material for each thin section (Lunning et al., 2021)), and the sections are a long-term resource that can be utilized by numerous researchers over many decades. **For initial characterization, the SPARX SDT recommends the use of several techniques** **including: visible light microscopy, structured light scanning, micro-X-ray computed tomography (Micro-XCT)** (Figure 4.35), **and micro-X-ray fluorescence (Micro-XRF).** These techniques provide a first order understanding of the type of sample present (e. g. basalt, impact melt, breccia), supporting verification of successful sample acquisition, and facilitating categorization of each sample for future laboratory analyses. The listed techniques represent the current standard for non-destructive sample characterization, and were utilized by the recent Apollo Next Generation Sample Analysis (ANGSA) team (Shearer et al., 2024) as part of an initiative to prepare the lunar sample community for the next generation of returned lunar samples. The initial petrologic characterization techniques have been selected to provide information on sample type, structure, and initial compositional classification. Additionally, two of the recommended techniques (structured light scanning and micro-XCT) provide digital 3-dimensional records of the sample surface and interior structure, which will serve as a vital resource for understanding initial sample context once destructive sample analysis techniques begin. A brief description of each of the recommended petrologic characterization techniques is provided below, with additional information provided in Table 4.3 and Appendix 10.3.

1. Visible Light Microscopy: rough samples studied with binocular light microscopes, and polished thin sections studied with petrographic microscopes. Provides information on overall sample classification, also provides information on sample cooling history.
2. Structured Light Scanning: Produces high-fidelity 3-d digital photograph of samples through the projection of patterned light.
3. Micro-XCT: produces 3-d density map of sample interior allowing for non-destructive identification of interior sample heterogeneities and sample type. (Figure 4.35)
4. Micro XRF: provides elemental maps of sample surface, allowing for triage of major sample mineralogic phases and bulk compositional trends.

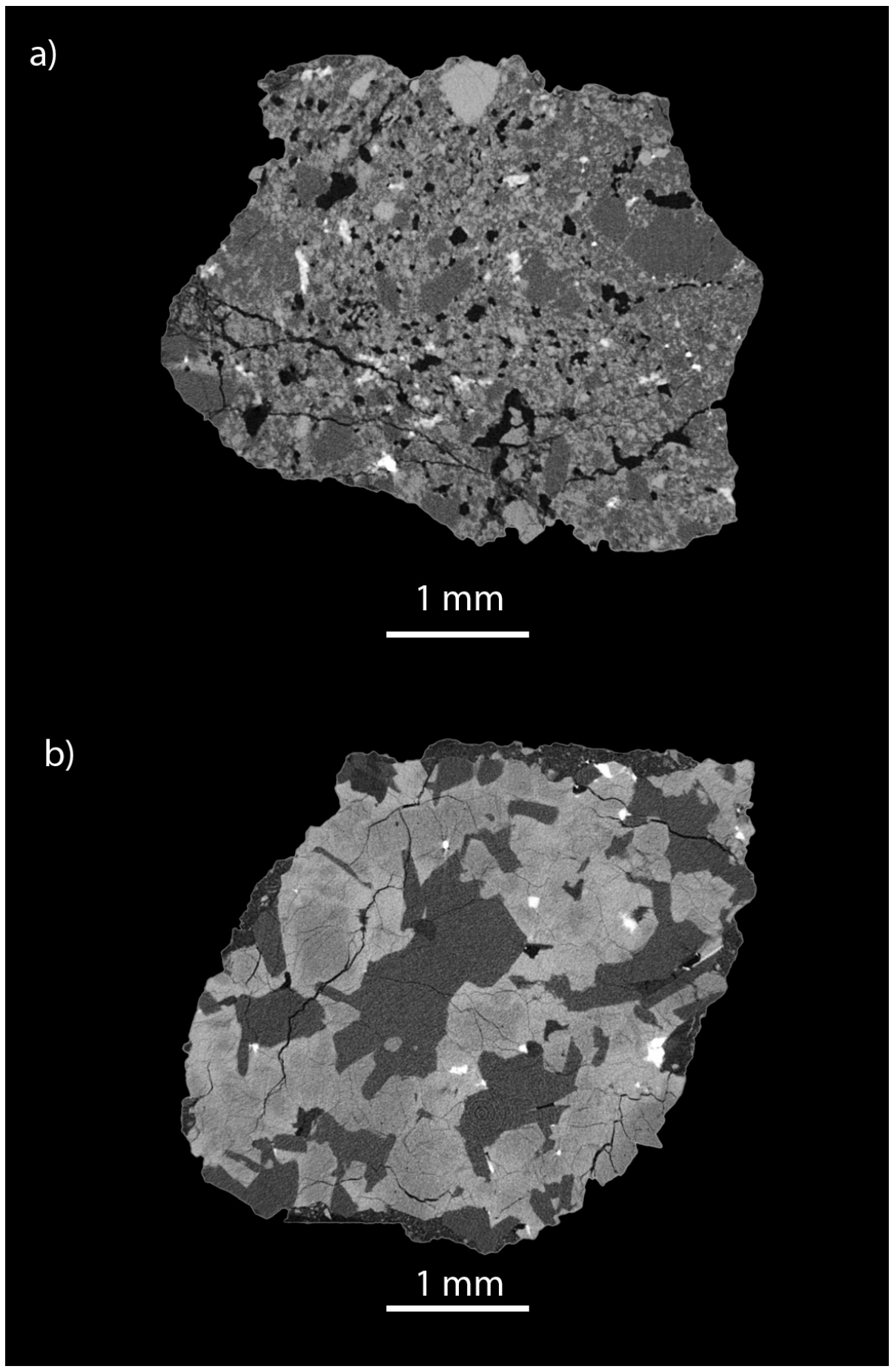

**Figure 4.35.** *Grayscale XCT slices of Apollo samples scanned during the ANGSA program. (a) 73002,1068A; impact melt breccia. (b) 73001,1234B; very-low Ti basalt. Sources: https://ada-prod.prod.astromat.org/doi/10.60707/ah0m-p611 and https://ada-prod.prod.astromat.org/doi/10.60707/4443-4366*

### 4.3.1.2 Additional Characterization Requirements

Following the initial triage and characterization of the samples, and curation allocation, samples shall be prepared for analysis by the lunar community. Prior to the STM defined laboratory analysis (Sections 3.6, 4.3.6), the SPARX SDT recommends that researchers perform an additional intermediary characterization of the sample mineralogy and petrology. One or more of the following techniques should be used to provide detailed textural and chemical information. The results of these investigations are required for both contextual understanding of the sample, and for successfully addressing the proposed investigations.

**Table 4.3.** *Petrologic sample characterization techniques. Sample Form abbreviations: CP (Chip); TNS (Thin Section); TKS (Thick Section); TF (Thin Foil); PWD (Powder); FIB (Focused Ion Beam).*

| Method | Measurement | Measurement capability | Application | Sample Form | Reference |
|---|---|---|---|---|---|
| Optical Light Microscopy | mineralogy, mineral textures, lithologic descriptions | 2% precision on modal mineralogy for major mineral phases | basic characterization of all samples | CP, TNS, TKS | Shearer and Borg (2006) |
| Structured Light Scanning | 3D geometry of rocks and clasts | -sub micron resolution<br>-3D geometry and shape information | basic characterization of all samples | CP | Ryan et al. (2026) |
| Micro-XCT (Micro-Computed Tomography) | interior imaging of rocks for textural and compositional information | micron scale 3D compositional imaging | basic characterization of all samples | CP | O'Neal et al. (2025) |
| Micro-XRF (X-ray Fluorescence) | major and minor elements | non-destructive, qualitative and rapid compositional information at micron scale | characterization of sample in support of E1-E10, and E13-E14 objectives | CP, TNS, TKS | Cohen et al. (2025); Crow et al. (2024) |
| Scanning Electron Microscopy (SEM) | major and minor elements, BSE elemental mapping | quantitative mineral chemistry and micrometer-scale textural characterization | characterization of sample in support of E1-E10, and E13-E14 objectives | CP, TNS, TKS | Seddio et al. (2014) |
| Electron Microprobe Analysis (EMPA) | quantitative analysis | Quantitative major and minor element mineral chemistry with 0.05% analytical preceision and ~100 ppm detection limits | characterization of sample in support of E1-E10, and E13-E14 objectives | TNS, TKS | Cohen et al. (2025) |
| Raman Spectroscopy | mineral identification | non-destructive and rapid mineralogical information at micron scale | characterization of sample in support of E1-E10, and E13-E14 objectives | CP, TNS, TKS, PWD | Ling et al. (2011); Cao et al. (2022) |
| Electron Backscatter Diffraction (EBSD) | deformation textures | non-destructive and rapid mineralogical information at micron scale | characterization of sample in support of E1-E10, and E13-E14 objectives | TNS, TKS | Cernok et al. (2019); Timms et al. (2012) |
| Transmission Electron Microscopy (TEM) | mineral identification | mineralgoic and compsoitional information at nm scale | characterization of sample in support of E1-E10, and E13-E14 objectives | TF, FIB | McFadden et al. (2024); Keller et al. (2021) |

Recommended characterization techniques:

Scanning Electron Microscopy (SEM), Electron Microprobe Analysis (EMPA), Raman spectroscopy, Electron Backscatter Diffraction (EBSD), Transmission Electron Microscopy (TEM).

Additional details about each of these techniques can be found in Table 4.3 and Appendix 10.3. The SPARX SDT recommends that whenever possible, the thin sections prepared during the initial triage stages be used during these additional characterization investigations as a means of efficiently utilizing the returned sample mass.

### 4.3.2 Elemental Chemistry

#### 4.3.2.1 Whole Rock Composition Analysis (Major and Trace Elements)

The origin and evolution of the Moon is understood through some of its unique geochemical features, and some groups of elements are particularly useful. For example, K, the rare earth elements (REE), P, U, and Th are all concentrated in the nearside Procellarum KREEP terrane. Whereas, basalt compositional groups are defined by their Ti (e.g., high, low, and very low) and K contents, and/or the extent of their negative Eu anomalies. Meanwhile, anorthosites can be ferroan or magnesian, and impactor geochemical signatures in impact melt rocks are sleuthed using siderophile (Fe, Ni, Co, Mo, W, Ge) and highly siderophile (Au, platinum group elements, and Re) elements. Application of geochronometers such as Rb-Sr, Sm-Nd, U-Pb, and Lu-Hf systems require understanding of these trace element concentrations and locations in the rocks. **For a majority of the experiments the major and trace element data will be used in combination with petrologic and isotopic data to provide bulk compositional context, ensure the proper materials are identified to address the objectives, and to help interpret results.**

A wide range of analytical approaches can be used to determine bulk chemical composition of lunar samples. The choice of technique depends somewhat on the lithology and the elements of interest, and available laboratory instrumentation. The types of bulk chemical analyses and recommended precision are related to the specific experiment goals (Section 4.1) and overall science objective (Section 3.1–3.3). Here we define four classes of investigation, and the concerns which drive their bulk chemical composition analysis:

1. Samples for compositional comparison: Many SPARX experiments seek to understand the evolving composition of the Moon through comparisons with the existing sample collection. For these analysis, major and trace element analysis should be of comparable precision and speciation to contemporary Apollo sample analysis (i.e., ANGSA). Relevant experiments: E1 - E10, E13, E14.
2. Sample for geochronology: Determining the age of specific samples is central to multiple SPARX experiments and objectives. Three kinds of analysis will be needed to meet the geochronologic requirements: bulk Ar-Ar dating of fine-grained rocks (e.g., impact melts), mineral isochrons from larger coarser rocks (e.g., anorthosites), and high-resolution laser or ion microprobe analysis of targeted mineral grains (e.g., zircons). For samples dated by mineral isochrons, the SDT recommends laser ablation analyses to help identify the widest range of parent-daughter ratios in minerals within each rock, followed by TIMS analyses of mineral separates. This combined approach will lead to the highest precision isochrons. Relevant experiments: E1–9, E13, E14.
3. Samples for Th analysis: These samples require the highest precision; therefore, the SDT specifically recommends bulk chemistry characterization through OES- or ID-ICP-MS, both of which allow Th determinations to better than 5% (e.g., Li et al., 2024; Ji et al., 2025; Wang et al., 2026). Relevant experiments: E7

4. Micro sample analysis: Some samples will have limited material available for sample preparation and analysis. In addition, some samples will be comprised of brecciated material, including small clasts and mineral fragments that will need to be analyzed for major and trace element content. In these cases, the bulk analysis techniques described above cannot be utilized and some more focused and higher spatial resolution techniques must be utilized such as micro-XRF, SIMS, or LA-ICP-MS. Typical uncertainties for SIMS or LA-ICP-MS measurements are slightly higher than those for bulk analyses, but still ~10% or better depending on concentration levels, number and type of standards used, and counting times. Micro-XRF is more qualitative, providing elemental mapping over broader areas than SIMS or LA-ICP-MS. Relevant experiments: E1–10, E13, E14.

There are multiple techniques available which could provide the required bulk chemistry information. Investigating teams should demonstrate that their proposed techniques are commensurate with the class of information required. Listed below, and provided in Table 4.4 are techniques that are suitable for these analyses. OES and ICP-MS require 50-100 mg of material for analysis, while XRF typically requires more material: 0.5 to 1 g (fused beads) or 5 to 8 g (pressed powder pellet). Coarser grained rocks require more mass to attain a representative sample for bulk chemical analysis (several g of material), and typically more than the minimum required for analysis (~100 mg), which is also an important driver of collected mass. Full descriptions of the techniques, any specific advantage for particular elements, historical context, and examples of analyses from some recent publications are provided in Appendix 10.3.

#### 4.3.2.2 Recommended Bulk Chemistry Analysis Techniques

Optical Emission Spectroscopy (OES), X-Ray Fluorescence Spectroscopy (XRF), Inductively Coupled Plasma Mass Spectroscopy (ICP-MS) (includes ID-ICP-MS, MC-ICP-MP, and Quadrupole ICP-MS), Instrumental neutron activation analysis (INAA), Micro-X-ray Fluorescence Spectroscopy (µXRF), Secondary Ion Mass Spectrometry (SIMS) (includes SHRIMP and NanoSIMS), Laser-ablation Inductively Coupled Plasma Mass Spectroscopy (LA-ICP-MS).

**Table 4.4.** *Elemental chemistry techniques. Sample Form abbreviations: CP (Chip); TNS (Thin Section); TKS (Thick Section); TF (Thin Foil); PWD (Powder); FIB (Focused Ion Beam).*

| Method | Measurement | Measurement capability | Application | Sample Form | Reference |
|---|---|---|---|---|---|
| Optical Emission Spectroscopy (OES) | major and trace elements | 2-5% precision; detection limits vary with element; optimal for Th, REE, metals | characterization of sample in support of E1-E10, and E13-E14 objectives | PWD | Wang et al. (2026); Ji et al. (2025) |
| X-ray Fluorescence Spectroscopy (XRF) | major and trace elements | 2-5% precision; detection limits vary with element | characterization of sample in support of E1-E10, and E13-E14 objectives | PWD | Ryder and Schuraytz (2001) |
| Inductively Coupled Plasma Mass Spectroscopy (ICP-MS) | major and trace elements | 2-5% precision; detection limits vary with element; optimal for Th, REE, metals | characterization of sample in support of E1-E10, and E13-E14 objectives | PWD | Neuman et al. (2025) |
| Instrumental neutron activation analysis (INAA) | major and trace elements | historical choice for Th, many REE, some metals | characterization of sample in support of E1-E10, and E13-E14 objectives | CP, PWD | Korotev (2022) |

## 4.3.3 Required Geochronologic Measurements

### 4.3.3.1 Age Determinations

Determination of crystallization ages for various lithologies is required in multiple experiments, including the highest priority SPARX objectives. Age determinations for most sample types (anorthosites, impact melts, basalts, and other crustal and mantle rocks) require broadly similar laboratory techniques, and can utilize multiple geochronometers when samples are both sufficiently crystalline, and also contain adequate parent and daughter element abundances. The isotopic systems most commonly used and most amenable to dating crystalline igneous rocks and impact melts are $^{147}$Sm-$^{143}$Nd, Rb-Sr, U-Pb, Pb-Pb, Lu-Hf, $^{40}$Ar–$^{39}$Ar, and in the case of some very ancient samples, $^{146}$Sm-$^{142}$Nd. Owing to its broad applicability, robustness, and resistance to thermal-shock resetting (Borg et al., 2015), the $^{147}$Sm-$^{143}$Nd system is required for experiments involving the dating of anorthositic samples (E1; discussed separately below), SPA impact melt (E2), and mare basalts (E3, 8, 9). Additional systems, including Rb-Sr, U-Pb, Pb-Pb, Lu-Hf, $^{40}$Ar–$^{39}$Ar, and some *in situ* techniques for specific phases such as zircon, baddeleyite, and phosphates, can be used as well, and are described in detail for different rocks types below. Mass and sample form requirements are described in the Appendix: Samples, and images of the analysis process can be seen in Figure 4.36

### 4.3.3.2 Age Determinations for Anorthosites

Experiment 1 requires the geochronologic dating of anorthosite samples collected from multiple locations (4.1.1). Lunar anorthosites have proven to be among the most challenging materials to robustly date for multiple reasons (see discussion in Borg et al., 2015). First, by definition, anorthosites have limited mineralogical diversity, which hampers isochron dating methods. They generally lack minerals such as zircon and baddeleyite that are amenable to individual mineral-scale dating. Second, anorthosites have low abundances of elements that are used in chronology (e.g., Sm, Nd, Lu, Hf, Rb, K). Both of these factors significantly increase the amount of sample mass required to obtain a precise and interpretable age. Third, lunar anorthosites are some of the oldest igneous rocks on the Moon, and have almost ubiquitously been subjected to impact heating and metamorphism, which frequently causes disturbance of chronometers involving particularly mobile elements (e.g., Rb, Pb, Ar). Finally, the effects of thermal and epithermal neutron capture on isotopes of Sm, Nd, and Hf require complex, composition-dependent corrections to remove the effects of irradiation on isotopic compositions (e.g., Borg et al., 2019, 2015; Brandon et al., 2009; Gaffney and Borg, 2014; Nyquist et al., 1995; Rankenburg et al., 2006; Sprung et al., 2013). Importantly, none of these four factors can be robustly evaluated on the lunar surface, limiting the ability to triage samples based on their likelihood of yielding reliable crystallization ages. Despite these challenges, dating of lunar anorthosites is possible, and the results from such analyses are required to address some of the highest priority SPARX objectives, including Obj. 1.1 and 1.3.

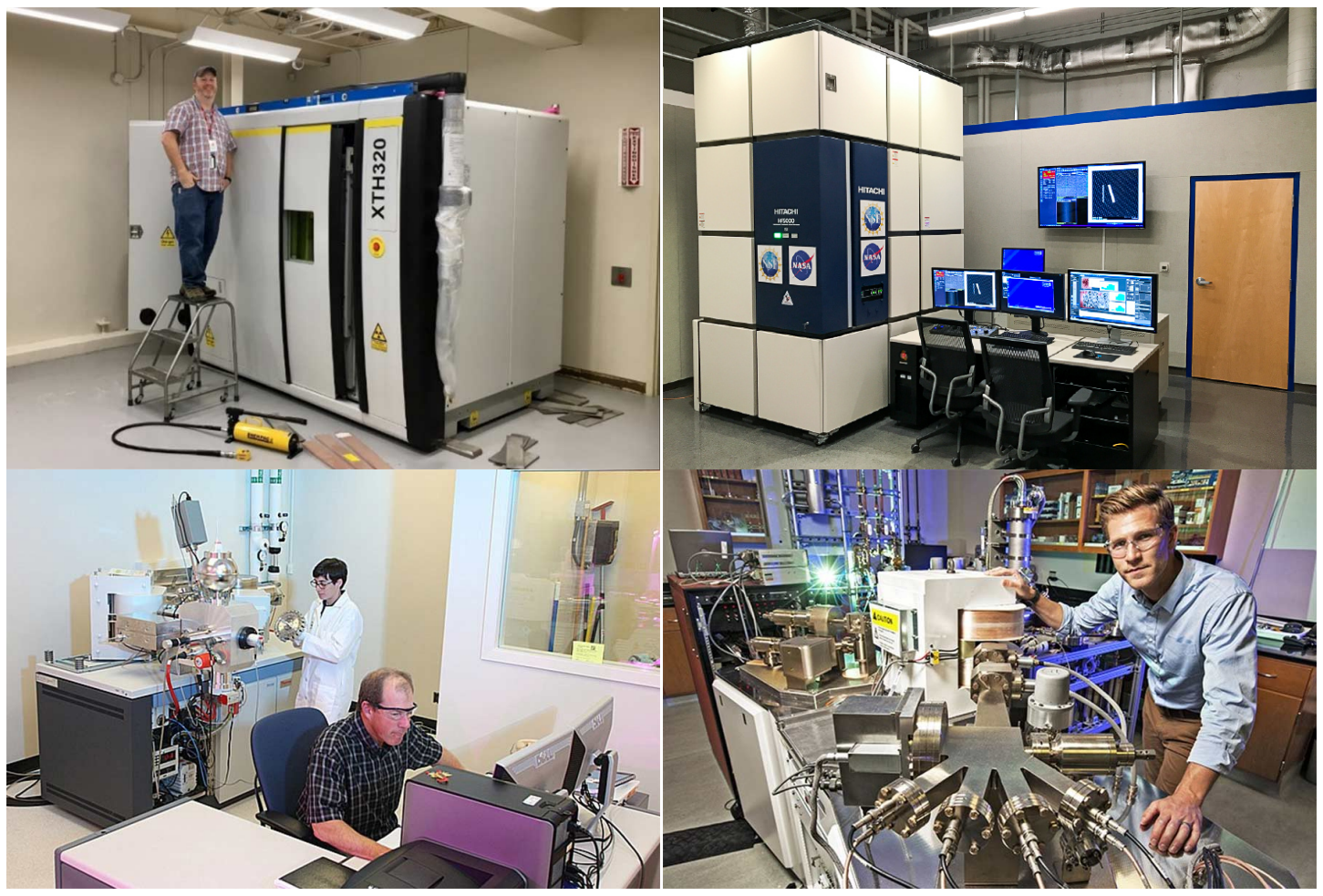

**Figure 4.36.** *Laboratory instrumentation recommended for the study of returned samples. (a) Researchers preparing to use geochronologic techniques to date Apollo 17 highlands samples. (b) The Noble Gas lab at Lawrence Livermore National Laboratory used for Ar-Ar dating of Apollo samples. (c) The X-ray computed tomography (XCT) scanner at NASA Johnson Space Center used to characterize lunar samples. (d) The 12-foot-tall transmission electron microscope (TEM) at the University of Arizona used for atomic-resolution imaging of lunar materials. Sources: https://www.lpi.usra.edu/ANGSA/news/angsa-in-action/, https://kalfaa.lpl.arizona.edu/facility/tem.*

Following the methodologies of successful anorthosite dating efforts (Borg et al., 1999, 2011; Marks et al., 2019; Sio et al., 2020), for E1, the SDT **requires the dating of anorthosite samples using the 147Sm-143Nd isochron system**, as this system is the most resistant to thermal metamorphism. Additionally, **≥2 geochronometers other than 147Sm-143Nd (e.g., Rb-Sr, 146Sm-142Nd, U-Pb, Pb-Pb, Lu-Hf, and Ar-Ar) are required and will be dictated by the composition and mineralogy of the returned samples**. The combined application of multiple chronometers is essential for identifying isotopic disturbance and establishing a robust chronology for lunar crust formation.

#### 4.3.3.3 Age Determinations for Impact Melts

In the case of **E2, the geochronology measurement requirement is stipulated as "40Ar–39Ar ± the 147Sm-143Nd isochron system** (4.1.2), reflecting the expectation that large-basin SPA impact melt samples will be predominantly crystalline due to prolonged cooling of thick melt sheets, thereby enabling application of both Ar–Ar and additional isochron systems; however, the degree of crystallinity cannot be guaranteed prior to sample return. Establishing a confident age for the SPA impact event in E2 will require convergence of results from multiple chronometers applied to multiple samples, which ultimately drives the total mass requirements for this experiment (see Appendix). **The application of Ar–Ar to multiple clasts will also enable statistical evaluation of age populations and discrimination between primary SPA melt and later thermally reset materials (e. g. Figure 4.37).**

Experiments 4 and 5 require the dating of two additional non-SPA craters or basins, where the target lithology is impact melt deposits from targeted craters or basins. Age determinations via the Ar-Ar system for samples in E4 and 5 will provide direct information regarding the ages of the craters or basins sampled by those experiments which are expected to consist of crystalline impact melt sheets, melt breccias, or shock-melted lithologies formed during basin-scale or large-crater-forming events. Even in cases where returned materials are brecciated or partially recrystallized, they are still likely to contain melt-derived components that record the timing of impact-related melting and cooling. Ar-Ar is a commonly and extensively used system for addressing basin chronology on the Moon due to the ease with which the Ar system is reset by impact processes, effectively establishing a new radiometric "clock" at the time of melt crystallization (e.g., Cohen et al., 2023, 2005; Dalrymple and Ryder, 1996; Norman et al., 2006). This resetting behavior makes impact melts among the most direct recorders of basin- and crater-forming events.

Thus, for samples in E4, 5, dating via the Ar-Ar system is required, as this system is widely applicable, and depending on the compositional and mineralogical characteristics of the samples, ≥2 independent geochronometer systems are required where permitted by the nature of the returned materials.

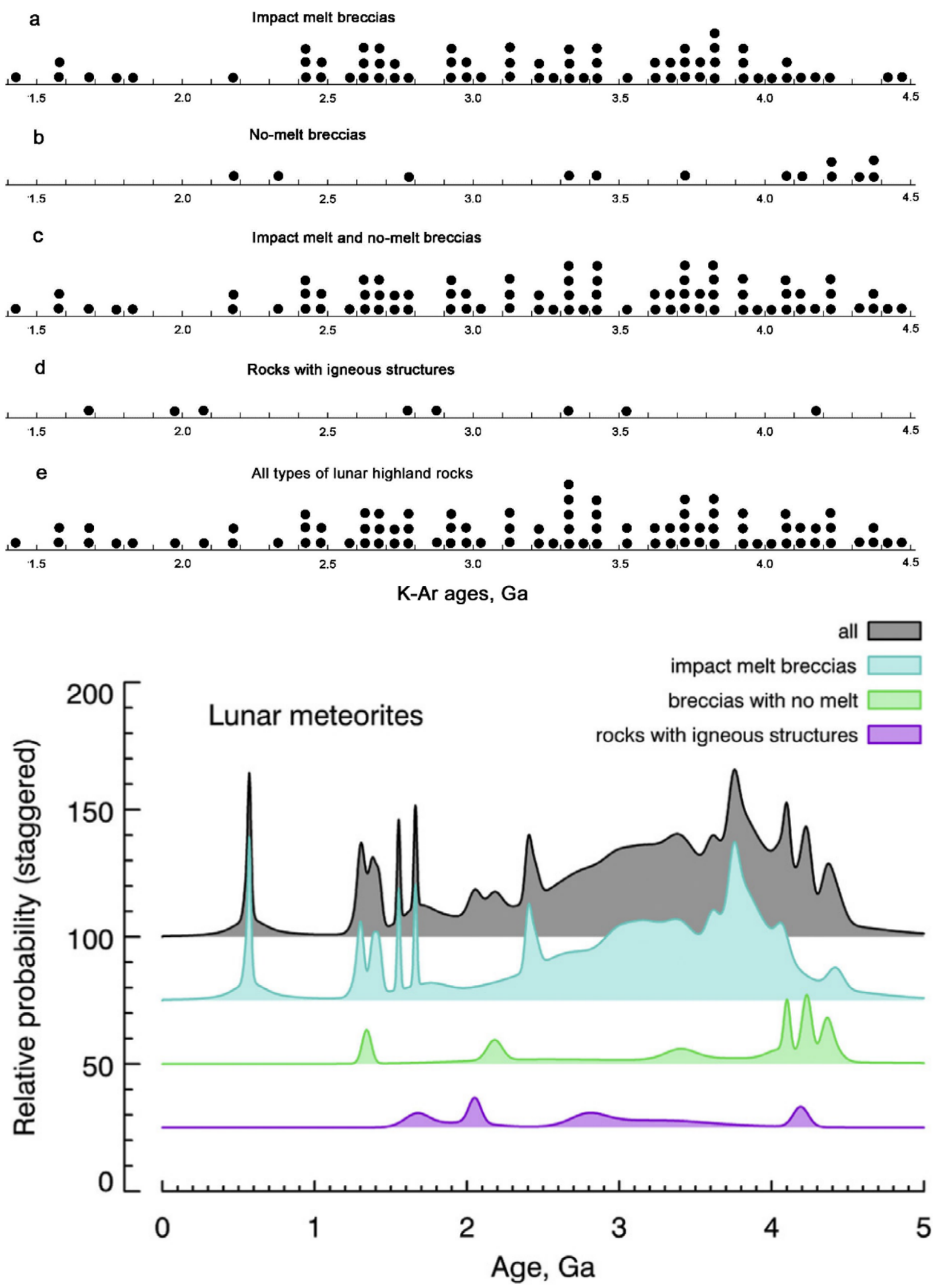

**Figure 4.37.** *(a) Histogram of K-Ar ages for lunar meteorites, divided into impact melt breccia, breccia with no melt, rocks with igneous structures, and all lunar highland rocks. (b) Relative probability plots of K-Ar age distribution for lunar meteorites, divided into the same lithological categories as (a). These figures demonstrate how suites of K-Ar ages are used to understand the statistical probability of age distributions for multiple key resetting events experienced by each rock. (Figure 9 and 10, Michael et al. 2018).*

#### 4.3.3.4 Age Determinations for Mare Basalts

**For E3, 8, and 9, ≥2 additional geochronometers other than 147Sm-143Nd are required for the dating of mare basalt samples and will be dictated by the composition and mineralogy of the returned samples**; examples of adequate chronometers include: Rb-Sr, U-Pb, Pb-Pb, Lu-Hf, and Ar-Ar. Although mare basalts are typically some of the most straightforward lunar samples to date due to their relatively high abundances of parent and daughter elements, diverse mineralogy, and typically low shock states, the application of multiple chronometers is essential for evaluating potential disturbance, and confirming crystallization ages. **Different**

**isotopic systems possess distinct closure temperatures and sensitivities to impact-related thermal resetting, and concordant ages among systems provide robust validation that the measured age reflects primary magmatic crystallization rather than partial resetting or later thermal overprinting**. Furthermore, the Lu-Hf, Pb-Pb, and $^{146}$Sm-$^{142}$Nd isotopic compositions of mare basalts can be used to calculate reservoir model ages (see below), directly addressing E3 objectives related to the evolution of the mantle underlying SPA. Such integrated chronologic and isotopic approaches have been successfully applied to Apollo, Chang'e, and meteoritic mare basalts (e.g., Borg et al., 2019; Brandon et al., 2009; Gaffney and Borg, 2014; McLeod et al., 2014; Nyquist et al., 1995; Rankenburg et al., 2006; Snape et al., 2016; Sprung et al., 2013).

### 4.3.3.5 Isotopic Model Ages

Isotopic model ages are calculated to determine the formation of geochemical reservoirs within the Moon based on the nature of the parent-daughter fractionation mechanism (e.g., Hf-W via core formation, Sm-Nd and Lu-Hf via silicate differentiation). For the Moon, previous workers have primarily applied the Hf-W, $^{147}$Sm-$^{143}$Nd, $^{146}$Sm-$^{142}$Nd, Lu-Hf, and U-Pb systems to lunar samples in order to calculate model ages for lunar formation, silicate differentiation, and the formation of the KREEP reservoir (see review in Gaffney et al., 2023).

Experiment 3 specifically requires determination of the model age of the mantle underlying SPA, although we note the measurements required in order to calculate such an age factor into other Experiments as well. Specifically, E3 **requires the determination of the crystallization ages of the mare basalts collected using the 147Sm-143Nd isochron system and then ≥2 additional systems, including Rb-Sr, U-Pb, Pb-Pb, Lu-Hf, and Ar-Ar**. The $^{147}$Sm-$^{143}$Nd isochron system is required because it is commonly used in the dating of lunar basalts, is resistant to resetting and disturbance from impacts, and is often used in the calculation of reservoir model ages. In addition to age determinations with $^{147}$Sm-$^{143}$Nd and ≥2 additional systems, E3 requires the use of ≥2 systems including $^{147}$Sm-$^{143}$Nd, $^{146}$Sm-$^{142}$Nd, Lu-Hf, and Pb-Pb to calculate model ages of the underlying mantle. The laboratory measurements required for $^{147}$Sm-$^{143}$Nd isochron age determination provide the necessary information for calculating a $^{143}$Nd model age for the SPA mantle for each basalt, thereby linking crystallization ages directly to mantle source evolution.

Depending on which dating systems are selected, the requirement for additional model age information may require additional laboratory measurements. Although the Lu-Hf system can be used for isochron age determinations for lunar samples (e.g., Carlson et al., 2014), more commonly this system has been used for model age determinations based on the Hf isotopic compositions of samples and ages determined via other radiogenic systems (e.g., Barboni et al., 2017; Gaffney and Borg, 2014; Sprung et al., 2013). Measurements of the $^{146}$Sm-$^{142}$Nd system can be used in conjunction with the $^{147}$Sm-$^{143}$Nd system to calculate a $^{142}$Nd-$^{143}$Nd source model age (e.g., Borg et al., 2019; Brandon et al., 2009; McLeod et al., 2014; Nyquist et al., 1995). Although this calculation falls under the purview of E3 where mantle model ages are the focus, it is possible to include $^{142}$Nd-$^{143}$Nd data from other lithologies such as anorthosites in such a model age calculation (Borg et al., 2019), which would be useful in answering questions related to O1.1 and O1.2.

### 4.3.3.6 Ages for Other Sample Types

For E6, 7, 9, 13, and 14, the expected lithologies are more uncertain. For example, E7 is designed to determine the composition and age of farside Th-rich materials. From orbital observations, it is difficult to determine the lithologic nature of these materials (e.g., crystalline impact melt, poorly crystalline impact melt, an igneous lithology, reworked brecciated materials, etc.) and therefore is difficult to prescribe specific chronometers. In almost any case, however, the Ar-Ar system will be applicable to any lithology returned as part of E6, 7, 9, 13, and 14. Depending on the nature of the samples, the $^{147}$Sm-$^{143}$Nd, Rb-Sr, U-Pb, Pb-Pb, and Lu-Hf systems may also be applicable and may serve as drivers for returned sample mass requirements and allocation strategies.

#### 4.3.3.7 Cosmic Ray Exposure Age Dating

Experiment 10 is to determine the volatile content of the farside mantle as recorded by pyroclastic and basaltic materials. This requires quantification of the surface exposure, and regolith residence histories of returned samples to ensure that measured volatile abundances and isotopic compositions reflect primary mantle signatures rather than secondary cosmogenic or solar wind contributions. Cosmic ray exposure (CRE) age dating using noble gases will therefore be applied, where appropriate, to constrain the near-surface residence history of returned materials and to distinguish primary magmatic volatile inventories from exposure-related modifications. CRE ages are determined from concentrations of cosmogenic noble gas isotopes (e.g., $^{3}$He, $^{21}$Ne, and $^{38}$Ar) produced by spallation reactions induced by galactic cosmic rays in target elements such as O, Mg, Al, Si, Ca, and Fe (e.g., Eugster, 2003; Wieler, 2002). Because cosmogenic production rates depend on chemical composition, shielding depth, and sample geometry, CRE age calculations require careful characterization of bulk composition and, where possible, multiple cosmogenic isotope systematics (e.g., helium, neon, argon, krypton and xenon) to evaluate shielding conditions and identify complex or multi-stage exposure histories (Leya and Masarik, 2009; Hohenberg et al., 1978).

Application of noble gas CRE dating to pyroclastic glasses, mare basalts, impact melt clasts, and regolith-derived lithologies will allow determination of excavation depths, regolith residence times, and potential multi-stage exposure histories, as demonstrated for Apollo samples and lunar meteorites (e.g., Eugster et al., 2001; Fogg et al., 2012; Joy et al., 2011; Meier et al., 2014, Curran et al., 2019; Cohen et al., 2025). For pyroclastic and basaltic materials in particular, determining CRE ages enables correction for cosmogenic noble gas production, evaluation of solar wind implantation effects, and assessment of diffusive loss processes that may modify volatile inventories during prolonged surface exposure. In this way, **CRE dating provides essential context for interpreting mantle volatile signatures and ensuring that volatile concentrations and isotopic ratios are accurately attributed to farside mantle source regions rather than secondary surface processes**.

#### 4.3.3.8 Required Techniques for Geochronologic Measurements

The SPARX SDT recommends two primary analytical techniques for performing the required geochronologic measurements.

1. Thermal Ionization Mass Spectrometry (TIMS): a technique for high-precision measurement of isotope ratios. TIMS is the optimal method for measuring specific elements, including some chronometers (e.g., Sr, Nd, Pb).
2. Noble Gas Mass Spectrometry: technique used for the elemental and isotopic measurements of chemically inert noble gases. This technique is used for measurements involving the $^{40}$Ar/$^{39}$Ar chronometer, and certain cosmic-ray exposure age measurements.

Additional details on both of these techniques can be found in Appendix 10.3 and Table 4.5.

**Table 4.5.** *Techniques for geochronologic measurements. Sample Form abbreviations: CP (Chip); TNS (Thin Section); TKS (Thick Section); TF (Thin Foil); PWD (Powder); FIB (Focused Ion Beam).*

| Method | Measurement | Measurement capability | Application | Sample Form | Reference |
|---|---|---|---|---|---|
| Thermal Ionization Mass Spectrometry (TIMS) | Rb-Sr, Sm-Nd, U-Th-Pb, Lu-Hf, W-Hf isotopes and ages | 5-10 ppm, giving age precision of +/- 10-20 Ma or 0.25-0.5% for 4 Ga old samples | chronologic information for E1-E10, and E13-E14 objectives | CP, PWD | Carlson et al. (2014); Borg et al. (2015) |
| Noble gas mass spectrometry | 40Ar-39Ar ages; 37Ar/38Ar exposure ages | 0.5-1% precision, but dependent upon thermal and shock disturbance history | chronologic information for E1-E10, and E13-E14 objectives | CP, PWD | Norman et al. (2006); Cohen et al. (2023) |
| Secondary Ion Mass Spectrometry (SIMS) (also SHRIMP and NanoSIMS) | U-Th-Pb isotopes and ages; trace elements and H,S,P, F, Cl, O, Li elemental and isotopic | +/- 1% for isotopic analyses; 10% precision for ppm to sub-ppm detection | chronologic information for E1-E10, and E13-E14 objectives | CP, TNS, TKS | Arai et al. (2025); Zhang et al. (2025); Dottin et al. (2025) |
| Laser-ablation Inductively Coupled Plasma Mass Spectroscopy (LA-ICP-MS) | major and trace elements; some isotopic systems | +/- 1% for isotopic analyses; 2-5 % precision for ppm to sub-ppm detection | chronologic information for E1-E10, and E13-E14 objectives | TNS, TKS | McIntosh et al. (2024); Nottingham et al. (2024) |

### 4.3.4 Volatile Abundance

Constraining the distribution and behavior of volatiles in the lunar interior is fundamental to understanding the mechanisms, timing, and extent of the Moon's thermochemical evolution. To measure the volatile content of the farside mantle for Experiment 10, returned samples will be characterized petrologically, and analyzed for the abundances, isotopic signatures, and speciation of volatile elements. To constrain lunar volatiles, previous studies focused on magmatic volatiles, such as F, Cl, H, C, and S (e.g., McCubbin et al., 2023 and references therein). Investigations frequently focused on the study of apatite (e.g., Boyce et al., 2014; McCubbin et al., 2015; Tartese et al., 2014; Treiman et al., 2016) and melt inclusions (e.g., Hauri et al., 2011; Saal et al., 2013; Chen et al., 2015; Wetzel et al., 2015; Ni et al., 2017; Singer et al., 2017; McCubbin and Barnes, 2020). Lunar apatite contains F, Cl, OH, and vacancies in its crystal structure and is a late-crystallizing phase in lunar rocks. In contrast, melt inclusions are trapped melt in growing crystals and could potentially record the earliest conditions of the magma from which the minerals crystallized. However, lunar melt inclusions often become fully crystallized, showing post-entrapment loss or addition of volatiles and Fe and Mg (e.g., Roedder and Weiblen, 1971). After correcting for this effect, their measurements offer insight into the earliest volatile signatures of the pre-eruptive magma. Volatile elements have also been analyzed in pyroclastic glass beads (e.g., Saal et al., 2008, 2013; Hauri et al., 2011, 2015; Chen et al., 2015; Wetzel et al., 2015; Su et al. 2023; 2025a; 2025b), mesostasis glass (Greenwood et al., 2017), and nominally anhydrous minerals (NAMs) (e.g., Hui et al., 2013, 2017; Mills et al., 2017; Simon et al., 2020). Further studies of bulk samples have focused on moderately volatile elements (MVEs) and chalcophiles (e.g., Zn, K, S, Se, Te) to understand depletion during the Moon-forming impact and potential impactor additions.

Volatile elements are analyzed using a variety of techniques, namely EPMA, SIMS, NanoSIMS, TOF SIMS, FTIR, micro-FTIR, XANES, LA-ICP-MS, MC-ICP-MS, and stepped combustion (these are described in Table 4.6 and in APPENDIX: SAMPLES). Some studies involve the use of experimental petrology prior to *in situ* volatile element analyses; whereby crystalline melt inclusions are homogenized by heating them and their host olivine crystals (e.g., Ni et al., 2019). These geochemical analyses will be coupled with geochronologic measurements (e.g., Ar-Ar, Sm-Nd) and cosmic ray exposure (CRE) age of the samples to ±10 Myr. These CRE ages are necessary to correct for potential solar-wind implantation and spallation effects, ensuring that the signatures measured represent the pristine lunar interior rather than secondary surface processes.

**Table 4.6.** *Volatile techniques. Sample Form abbreviations: CP (Chip); TNS (Thin Section); TKS (Thick Section); TF (Thin Foil); PWD (Powder); FIB (Focused Ion Beam).*

| Method | Measurement | Measurement capability | Application | Sample Form | Reference |
|---|---|---|---|---|---|
| Nanoscale Secondary Ion Mass Spectrometry (NanoSIMS) | abundances and isotopic composition of volatile systems (H, Cl, C, O, F, S) | ±1–10 ppm for abundances of volatiles and ±20–50‰ for isotopic systems | volatile information for E10 | TNS, TKS, FIB | McCubbin et al., 2015; Barnes et al., 2013; Stephant et al., 2019 |
| Secondary Ion Mass Spectrometry (SIMS) | abundances and isotopic composition of volatile systems (H, Cl, S, N, C) and moderately volatile systems | ±2–20 ppm for abundances of volatiles and ±1–50‰ for isotopic systems | volatile information for E10 | CP, TNS, TKS | Saal et al., 2008; Hauri et al., 2011; Greenwood et al., 2011; Boyce et al., 2015; Wetzel et al., 2015 |
| Electron Microprobe Analysis (EMPA) | abundances of volatiles | ±50–100 ppm | volatile information for E10 | TNS, TKS | McCubbin et al., 2023 |
| MC-ICP-MS | High-precision bulk analysis of moderately volatile elements (K, Zn, Fe, Cu, Cl) | < 1 ppb in solution; ± 0.02‰ | volatile information for E10 | PWD | Wang and Jacobsen, 2016; Marty, 2003; Furi et al., 2012 |
| LA-ICP-MS | S, Cl, Zn | 10 ppm; ± 0.3‰ | | TKS | Jackson et al., 2021; Lin et al., 2017 |
| Stepped combustion | H, C, noble gases | 0.1 ng; ± 0.5‰ | volatile information for E10 | CP, PWD | Mortimer et al., 2016; Marty, 2003; Furi et al., 2012 |
| TIMS | Cl, Pb, Cr | ± 0.1‰ | volatile information for E10 | CP, PWD | Sharp et al., 2010; Gargano et al., 2020; Borg et al., 2011 |

**Do We Really Need Sample Return? Why Can't These Measurements Be Made *In situ*?**

Many of the Science Goals of the SPARX mission rely on compositional and age data for the samples in question in order to achieve their objectives. Since there are existing *in situ* instruments that provide rock chemical and chronologic data, a natural question would be, Why is it necessary to return samples to Earth instead of collecting these data on the Moon? In short, existing *in situ* instrumentation is simply not capable of collecting the necessary geochemical and chronologic data, either at all and/or to the precision necessary,

The geochemical composition of bulk rock samples, their constituent minerals, and textural relationships are required for many of the SPARX Science Objectives. Some exiting flight instrumentation, such as the Planetary Instrument for X-Ray Lithochemistry (PIXL) instrument, for example, are capable of collecting geochemical data, including at the micro-scale. Although PIXL and other *in situ* geochemical instruments are able to collect major elements abundances (e.g., >1 wt. %) and some minor and trace (e.g., ≤ 1000 ppm) elements, they are not capable of quantifying the abundances of all necessary elements to the required precision and at all required spatial scales (bulk rock and individual minerals), and in most cases, cannot detect the necessary elements at all. Earth-based geochemical laboratory techniques such as ICP-MS rely on the dissolution of rock samples, dilution of the sample, and calibration against standards during the analysis. Laboratory imaging techniques rely on analyzing highly polished surfaces to provide textural and chemical information at scales ~100x smaller than PIXL, for example.

Similarly, determining the ages of samples is necessary for many of the SPARX Science Objectives, but *in situ* radiometric dating techniques are incapable of determining rock and mineral ages with the necessary isotopic systems and to the precision required. The most relevant *in situ* instrument is the Chemistry, Organics, and Dating EXperiment (CODEX), which was selected as a payload for the Dating an Irregular Mare Patch with a Lunar Explorer (DIMPLE) CLPS investigation (Anderson et al., 2025). CODEX can achieve a precision of about 300 Myr for the Rb-Sr isotopic system, which is significantly worse than what is needed for SPARX. Laboratory dating techniques can routinely provide precision 10x – 100x better, and no current *in situ* instruments can determine ages for the Lu-Hf, U-Pb, Pb-Pb, and both Sm-Nd systems. Such ages determinations to the precision needed require months of detailed laboratory work. Therefore, it is impossible for SPARX to achieve its Science Objectives with *in situ* instrumentation. Sample return is required.

### 4.3.5 Remanent Magnetization Measurements

Constraining the timing and intensity of magnetization recorded in lunar samples is required to evaluate the mechanisms, duration, and variability of the lunar dynamo (Objective 1.4; Experiment 6). Paleomagnetic investigations determine the presence, origin, age, and paleointensity of stable natural remanent magnetization (NRM). A key investigation is to assess whether the observed remanence represents thermoremanent magnetization (TRM) acquired during cooling in an ambient magnetic field. TRM is acquired during cooling through magnetic blocking temperatures in the presence of an ambient field and would therefore directly record the intensity of the global magnetic field generated by a lunar core dynamo at the time of crystallization (Weiss and Tikoo, 2014). However, lunar samples, particularly breccias, may record complex magnetization histories. Other remanence mechanisms (e.g., shock remanent magnetization (SRM), viscous remanent magnetization (VRM), or crystallization remanent magnetization (CRM)) may form during transient impact processes, a spacecraft's artificial magnetic field, crystallization, or later alteration and do not uniquely require a sustained global dynamo (e.g., Pearce et al., 1973; Sugiura and Strangway, 1980; Gattacceca et al., 2010). Establishing the mechanism of remanence acquisition is therefore required prior to interpreting any paleointensity estimate as evidence for a dynamo field. Determination of paleofield vector direction is not required. Recovery of directional information would necessitate oriented sampling and preservation of sample orientation throughout collection, curation, and return. Given limited access to coherent bedrock within the SPA environment and the substantial engineering and curatorial requirements associated with oriented coring and storage, such measurements were determined to be not commensurate with the broader programmatic scope of SPARX. Accordingly, Experiment 6 focuses on magnetization timing and paleointensity.

Returned samples must undergo coordinated petrologic, geochronologic, and rock magnetic characterization. Petrographic analyses are required to identify magnetic carriers and assess their stability. Geochronologic and/or thermochronologic measurements are required to establish crystallization ages and evaluate potential impact-related resetting. Both of these analyses will be carried out on the broader SPARX returned sample collection first. Following these investigations, samples compatible with the E6 paleomagnetic investigation (ie. samples hosting magnetic carriers and with inferred age ≥ 4.0 Ga) will be identified for further analysis. Paleomagnetic analysis begins with measurement of the NRM, followed by progressive removal of magnetic overprints using alternating field (AF) demagnetization (nondestructive) or thermal demagnetization (destructive). After isolation of the primary remanent component, laboratory-induced magnetizations in known fields are used to estimate paleointensity.

Thermal methods (e.g., Thellier and Thellier, 1959; Yu et al., 2004) provides the most robust paleointensity determinations when TRM acquisition is demonstrated, as it most closely replicates the mechanism in which rocks initially acquired their magnetism (e.g., Apollo regolith breccia 15498; Tikoo et al., 2017). However, thermal treatment may induce chemical alteration in samples containing phases such as troilite or, at elevated temperatures, ilmenite. In such cases, nonheating approaches are preferred, such as acquisition of saturation isothermal remanent magnetization (sIRM; e.g., Stephenson et al., 1977; Gattacceca and Rochette, 2004) or anhysteretic remanent magnetization (ARM; e.g., Stephenson and Collinson, 1974) as proxies for TRM. IRM- and ARM-based methods typically yield paleointensity estimates with uncertainties on the order of a factor of ~5 (Weiss and Tikoo, 2014). Selection of destructive versus nondestructive techniques will depend on mineralogy and curatorial constraints.

Sample mass directly affects the robustness of paleomagnetic measurements. Specimens smaller than ~125 mg may be unreliable due to magnetic anisotropy effects and insufficient averaging over magnetic grain populations (e.g., Tikoo et al., 2012; Bellon et al., 2025). Larger samples are preferred and may be subdivided into mutually oriented specimens to evaluate reproducibility of remanent intensity and paleointensity estimates, while also allocating material for mineralogic characterization, shock assessment, and age determination. Newer techniques, such as quantum diamond microscopy (QDM; e.g., Glenn et al., 2017; Yang et al., 2025) and superconducting quantum interference device (SQUID) microscopy (e.g., Weiss et al., 2007), provide high-spatial-resolution magnetic field imaging and are particularly valuable for identifying ferromagnetic carriers and assessing magnetization heterogeneity at submillimeter to submicrometer scales. These approaches are especially useful for weakly magnetized lunar samples (e.g., Weiss et al., 2007) and can be applied even to individual grains (e.g., Yang et al., 2025). However, because thermoremanent magnetization represents a statistical bias in the alignment of many magnetic grains, robust paleointensity determinations still require sufficient averaging over an adequate grain population (e.g., Tikoo et al., 2012; Bellon et al., 2025). In lunar materials with low magnetic mineral abundances, magnetization may be dominated by a limited number of grains, increasing uncertainty in both bulk and spatially resolved measurements.

**A range of analytical approaches may be employed to characterize remanent magnetization and evaluate paleointensity, depending on sample lithology, magnetic mineralogy, available instrumentation, and curatorial constraints**. The techniques listed in Table 4.7 are representative of established paleomagnetic methods but are not exhaustive. Investigating teams should demonstrate that proposed analytical strategies are appropriate for the returned materials and commensurate with the objectives of Experiment 6.

**Table 4.7.** *Techniques to estimate paleointensity.*

| Method | Measurement | Measurement capability | Application | Sample Form | Reference |
|---|---|---|---|---|---|
| Thellier–Thellier | Thermoremanent magnetization (TRM) acquisition and paleointensity determination | Destructive; highest-fidelity paleointensity method when TRM acquisition is demonstrated | Determine remanence mechanism and estimate paleointensity for E6 | rocklets | Thellier and Thellier, 1959 |
| In-field Zero-field Zero-field In-field (IZZI) | Thermoremanent magnetization (TRM) acquisition and paleointensity determination | Destructive; highest-fidelity paleointensity method when TRM acquisition is demonstrated | Determine remanence mechanism and estimate paleointensity for E6 | rocklets | Yu et al., 2004 |
| Anhysteretic Remanent Magnetization (ARM) | Nondestructive; paleointensity proxy with typical uncertainties of up to a factor of ~5 | nondestructive, yield paleointensity estimates within a 2 sigma uncertainty | Determine remanence mechanism and estimate paleointensity for E6 | rocklets | Stephenson and Collinson, 1974 (ARM method); Weiss & Tikoo, 2014 (Uncertainty analysis) |
| Saturation Isothermal Remanent Magnetization (sIRM) | Nondestructive; paleointensity proxy with typical uncertainties of up to a factor of ~5 | nondestructive, yield paleointensity estimates within a 2 sigma uncertainty | Determine remanence mechanism and estimate paleointensity for E7 | rocklets | Gattacceca and Rochette, 2004; Stephenson et al., 1977 (IRM method); Weiss & Tikoo, 2014 (Uncertainty analysis) |
| Superconducting Quantum Interference Device (SQUID) microscopy | magnetic field imaging | moment resolution of 1e-15 Am2; spatial resolution of ~100 µm | Determine remanence mechanism and estimate paleointensity for E8 | rocklets; clasts | Weiss et al., 2007 |
| Quantum Diamond Microscopy (QDM) | magnetic field imaging | spatial resolution of ~2–4 µm; magnetic field sensitivity (~2.09–3.39 to 20 µT·Hz−1/2) | Determine remanence mechanism and estimate paleointensity for E9 | rocklets, clasts, individual grains | Glenn et al., 2017; Yang et al., 2025 |

### 4.3.6 Table of *Sample Requirements*

Foldout 4.2 displays the SPARX sample requirements.

Figure 4.38 summarizes how these requirements would translate to the analytical flow for an example sample type: anorthosites.

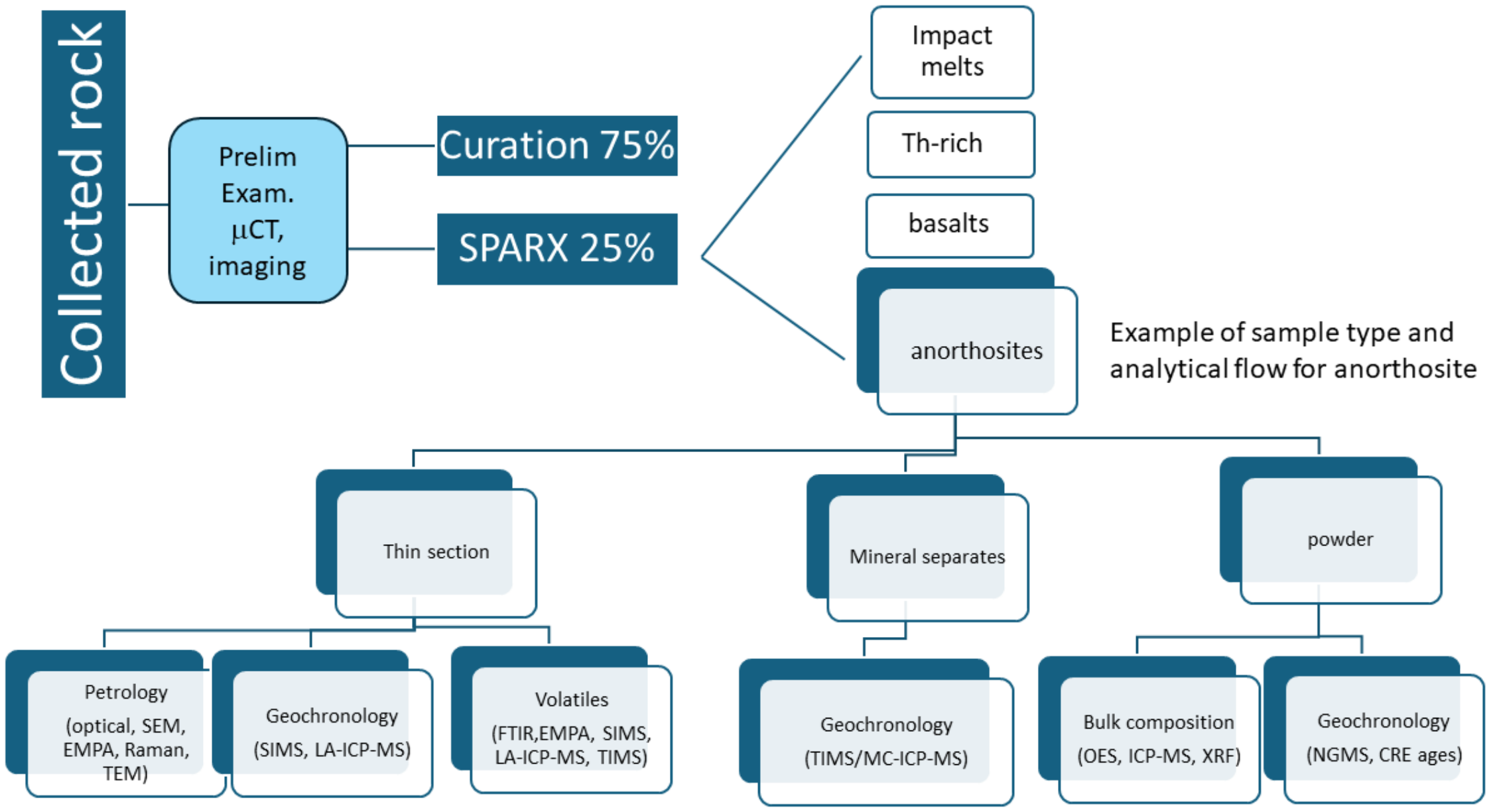

**Figure 4.38.** *This flow demonstrates the recommended preliminary examination and petrologic characterization analyses to be conducted on all received samples. The flow then demonstrates the different preparations for samples, and the classes of experimental investigations for each preparation type.*

**Finding: The SDT finds that the measurements required to address the defined SPARX objectives cannot be achieved using available in situ analysis methods, and therefore the SPARX mission requires sample return to Earth to successfully address mission science objectives.**

**Finding: The SDT finds that SPARX science objectives require the return of specifically targeted lithologies in a range of prescribed rock sizes (diameters ≥ 3mm, 5mm, 8mm, 20, and 40mm) and unsieved bulk regolith. The SDT notes that returning fewer, larger mass samples risks not meeting the defined measurement requirements and significantly increases scientific risk.**

**Finding: The SDT finds that the results of the LEAG/ExMAG Artemis Sample Return Specific Action Team (ASR-SAT) report are highly applicable to future SPARX returned samples, and the SDT encourages NASA to implement key findings of this report in advance of SPARX sample return. In particular, the SDT highlights the ASR-SAT finding of an impending critical lack of both laboratory facilities and trained personnel in the United States to carry out many of the highest-priority SPARX investigations. The SDT strongly recommends NASA address the findings of this report, particularly for the critical laboratory facilities required for SPARX-related investigations.**

| Experiments | | In situ measurement requirements for selecting samples | Laboratory analyses of returned samples | | Required sample components | Desired sample lithology (Note: no requirement to verify that sample this lithology) | Sample requirements – Required sample characteristics: Assumed particle density (Kiefer et al. 2015) | Particle size | Particle mass (or bulk mass) (assuming cubical particles) | Number of particles | Sample curation requirements: Mass | Number of particles | Required sample mass – Components: Baseline | Components: Threshold | Total: Baseline | Total: Threshold |
|---|---|---|---|---|---|---|---|---|---|---|---|---|---|---|---|---|
| E1 | Determine the composition and age of the anorthositic farside highlands crust | **Ability to detect anorthosite**<br>**Ability to detect sampleable material** | **Geochronology:** Determine the age and isotopic composition of anorthositic highlands crust utilizing the 147Sm-143Nd isochron system and ≥[2] additional independent geochronometer systems, including Rb-Sr, 146Sm-142Nd, U-Pb, Pb-Pb, Lu-Hf, and Ar-Ar<br>**Petrologic characterization** of returned samples | ROCK HUNTING | Rocks (≥40 mm) | Anorthosite | 2.2 g/cc | ≥[40] mm | ≥[140.8] g | ≥[3] | – | ≥[12] | ≥1,689.6 g | ≥[1,689.6] g | ≥2,107.6 g | ≥[1,707.6] g |
| | | | | | Rocklets (≥8 mm) | | | ≥[8] mm | ≥[1.1] g | ≥[4] | – | ≥[16] | ≥18.0 g | ≥[18.0] g | | |
| | | | | | Bulk regolith | No requirement | – | – | ≥100.0 g | – | ≥400.0 g | – | ≥400.0 g | – | | |
| E2 | Determine the age of the SPA impact event | **Ability to detect sampleable material** | **Geochronology:** Determine the age of SPA impact melt utilizing the Ar-Ar ± the 147Sm-143Nd isochron system, and ≥[1] additional independent geochronometer system, including Rb-Sr, U-Pb, Pb-Pb, 146Sm-142Nd, Lu-Hf<br>**Petrologic characterization** of returned samples | STANDARD | Rocklets (≥8 mm) | Impact melt | 2.7 g/cc | ≥[8] mm | ≥[1.4] g | ≥[8] | – | ≥[32] | ≥44.2 g | ≥[44.2] g | ≥473.4 g | ≥[473.4] g |
| | | | | | Rocklets (≥3 mm) | | | ≥[3] mm | ≥[0.1] g | ≥[100] | – | ≥[400] | ≥29.2 g | ≥[29.2] g | | |
| | | | | | Bulk regolith | No requirement | – | – | ≥[100.0] g | – | ≥[400.0] g | – | ≥400.0 g | ≥[400.0] g | | |
| E3 | Determine the model age of the mantle underlying SPA | **Ability to detect sampleable material** | **Geochronology:** Determine the ages and isotopic compositions of mare basalts utilizing the 147Sm-143Nd isochron system and ≥[2] additional independent geochronometer systems, including Rb-Sr, U-Pb, Pb-Pb, Lu-Hf, and Ar-Ar, and calculate the model age of the underlying mantle source utilizing ≥[2] systems including 147Sm-143Nd, 146Sm-142Nd, Lu-Hf, and Pb-Pb<br>**Petrologic characterization** of returned samples | STANDARD-ish | Rocklets (≥5 mm) | Basalt | 3.2 g/cc | ≥[5] mm | ≥[0.4] g | ≥[20] | – | ≥[80] | ≥32.0 g | ≥[32.0] g | ≥466.6 g | ≥[466.6] g |
| | | | | | Rocklets (≥3 mm) | | | ≥[3] mm | ≥[0.1] g | ≥[100] | – | ≥[400] | ≥34.6 g | ≥[34.6] g | | |
| | | | | | Bulk regolith | No requirement | – | – | ≥[100.0] g | – | ≥[400.0] g | – | ≥400.0 g | ≥[400.0] g | | |
| E4 | Determine the age of ≥1 farside Imbrian-aged impact crater (D≥80 km) or basin (D≥300 km) with surface dateable by crater counts | Ability to detect sampleable material | **Geochronology:** Determine the age of impact melt utilizing Ar-Ar, and when sample characteristics allow, ≥[2] independent geochronometer systems, including, U-Pb, Pb-Pb, 147Sm-143Nd, Rb-Sr, and Lu-Hf<br>**Petrologic characterization** of returned samples | STANDARD | Rocklets (≥8 mm) | Impact melt | 2.7 g/cc | ≥8 mm | ≥1.4 g | ≥8 | – | ≥32 | ≥44.2 g | – | ≥473.4 g | – |
| | | | | | Rocklets (≥3 mm) | | | ≥3 mm | ≥0.1 g | ≥100 | – | ≥400 | ≥29.2 g | – | | |
| | | | | | Bulk regolith | No requirement | – | – | ≥100.0 g | – | ≥400.0 g | – | ≥400.0 g | – | | |
| E5 | Determine the age of ≥[1] non-SPA farside pre-Imbrian-aged impact basin (D≥300 km) | Ability to detect sampleable material | **Geochronology:** Determine the age of impact melt utilizing Ar-Ar, and when sample characteristics allow, ≥[2] independent geochronometer systems, including, U-Pb, Pb-Pb, 147Sm-143Nd, Rb-Sr, and Lu-Hf<br>**Petrologic characterization** of returned samples | STANDARD | Rocklets (≥8 mm) | Impact melt | 2.7 g/cc | ≥[8] mm | ≥[1.4] g | ≥[8] | – | ≥[32] | ≥44.2 g | ≥[44.2] g | ≥473.4 g | ≥[473.4] g |
| | | | | | Rocklets (≥3 mm) | | | ≥[3] mm | ≥[0.1] g | ≥[100] | – | ≥[400] | ≥29.2 g | ≥[29.2] g | | |
| | | | | | Bulk regolith | No requirement | – | – | ≥[100.0] g | – | ≥[400.0] g | – | ≥400.0 g | ≥[400.0] g | | |
| E6 | Determine the age and remnant magnetization signatures from ≥3 returned samples with inferred ages ≥4.0 Ga | Ability to detect sampleable material | **Geochronology:** Determine the age of returned samples utilizing Ar-Ar, and, including, U-Pb, Pb-Pb, 147Sm-143Nd, Rb-Sr, and Lu-Hf. when sample characteristics allow, ≥2 independent geochronometer systems<br>**Petrologic characterization** of returned samples<br>**Determine the remnant magnetic field strength of returned samples** | | **Does not drive sampling requirements**; this objective is completed as part of the coordinated analysis of samples collected for other objectives | | **Does not drive sampling requirements;** this objective is completed as part of the coordinated analysis of samples collected for other objectives | | | | | | | | | |
| E7 | Determine the composition and age of farside Th-rich materials | **Ability to detect Th abundance**<br>Ability to detect sampleable material | **Geochronology:** Determine the age of returned samples utilizing Ar-Ar, and when sample characteristics allow, ≥2 independent geochronometer systems, including, U-Pb, Pb-Pb, 147Sm-143Nd, 146Sm-142Nd, Rb-Sr, and Lu-Hf<br>**Petrologic characterization** of returned samples<br>**Elemental chemistry** including major elements (Fe, Mg, Ca, K, Si, Ti, Al, Cr, Mn) and trace elements (Th, U, Hf, Zr, Sc, Rb, Sr, Pb, K, P, W, and rare-Earth elements) | STANDARD | Rocklets (≥8 mm) | **No requirement** (the Th-carrier is unknown) | 2.7 g/cc | ≥[8] mm | ≥[1.4] g | ≥[8] | – | ≥[32] | ≥44.2 g | ≥[44.2] g | ≥473.4 g | ≥[473.4] g |
| | | | | | Rocklets (≥3 mm) | | | ≥[3] mm | ≥[0.1] g | ≥[100] | – | ≥[400] | ≥29.2 g | ≥[29.2] g | | |
| | | | | | Bulk regolith | | – | – | ≥[100.0] g | – | ≥[400.0] g | – | ≥400.0 g | ≥[400.0] g | | |
| E8 | Determine the composition and age of farside basaltic units | Ability to detect sampleable material | **Geochronology:** Determine the ages and isotopic compositions of mare basalts utilizing the 147Sm-143Nd isochron system and ≥[2] additional independent geochronometer systems, including Rb-Sr, U-Pb, Pb-Pb, Lu-Hf, and Ar-Ar, and calculate the model age of the underlying mantle source utilizing ≥[2] systems including 147Sm-143Nd, 146Sm-142Nd, Lu-Hf, and U-Pb<br>**Petrologic characterization** of returned samples<br>**Elemental chemistry** including major elements (Fe, Mg, Ca, K, Si, Ti, Al, Cr, Mn) and trace elements (Th, U, Hf, Zr, Sc, Rb, Sr, Pb, K, P, and rare-Earth elements) | | **Returned sample requirements are the same as (E3)**; no additional samples are required | | **Returned sample requirements are the same as (E3)**; no additional samples are required | | | | | | | | | |
| E9 | Sample hypothesized excavated lunar mantle material | **Ability to detect hypothesized mantle material, such as olvine and low-calcium pyroxene**<br>Ability to detect sampleable material | **Geochronology:** Determine the ages and isotopic compositions of returned samples utilizing Ar-Ar, and when sample characteristics allow, ≥2 independent geochronometer systems, including, U-Pb, Pb-Pb, 147Sm-143Nd, 146Sm-142Nd, Rb-Sr, and Lu-Hf<br>**Petrologic characterization** of returned samples<br>**Elemental chemistry** including major elements (Fe, Mg, Ca, K, Si, Ti, Al, Cr, Mn) and trace elements (Th, U, Hf, Zr, Sc, Rb, Sr, Pb, K, P, and rare-Earth elements) | ROCK HUNTING | Rocks (≥20 mm) | Course grained, low-Ca pyroxene and/or olivine | 3.4 g/cc | ≥20 mm | ≥27.2 g | ≥3 | – | ≥12 | ≥326.4 g | – | ≥763.1 g | – |
| | | | | | Rocklets (≥3 mm) | | | ≥3 mm | ≥0.1 g | ≥100 | – | ≥400 | ≥36.7 g | – | | |
| | | | | | Bulk regolith | No requirement | – | – | ≥100.0 g | – | ≥400.0 g | – | ≥400.0 g | – | | |
| E10 | Determine the volatile content of the farside mantle as recorded by pyroclastic and basaltic materials | Ability to detect sampleable material | **Geochronology:** Determine the cosmic ray exposure age to ±10 Myr<br>**Petrologic characterization** of returned samples<br>**Elemental chemistry** including major elements (Fe, Mg, Ca, K, Si, Ti, Al, Cr, Mn) and trace elements (Ni, Co, highly-siderophile elements)<br>**Volatile abundance** (H2O, CO, Cl, F, S) | | **Mare basalts are sampled as part of (E3)**; no additional mare basalt samples are required | | **Mare basalts are sampled as part of (E3);** no additional mare basalt samples are required | | | | | | | | | |
| | | | | BULK | Bulk regolith | Pyroclastic material | – | – | ≥100.0 g | – | ≥400.0 g | – | ≥400.0 g | ≥400.0 g | ≥400.0 g | – |
| E11 | Determine the crustal structure (thickness, density, remnant magnetic field strength) across the SPA basin (from the center to the putative rim) to constrain mega-basin formation and evolution models | **None**; this experiment does not require returned samples | | | **None**; this experiment does not require returned samples | | **None**; this experiment does not require returned samples | | | | | | | | | |
| E12 | Determine if the massifs in the southwestern portion of the basin constitute the diffuse rim of SPA | **None**; this experiment does not require returned samples | | | **None**; this experiment does not require returned samples | | **None**; this experiment does not require returned samples | | | | | | | | | |
| E13 | Determine the stratigraphy and composition of the unique deposits in central SPA | **Ability to detect material of key lithologies, including high- and low-Calcium pyroxene, olivine, and feldspar**<br>Ability to detect sampleable material | **Geochronology:** Determine the ages and isotopic compositions of returned samples utilizing Ar-Ar, and when sample characteristics allow, ≥[2] independent geochronometer systems, including, U-Pb, Pb-Pb, 147Sm-143Nd, 146Sm-142Nd, Rb-Sr, and Lu-Hf<br>**Petrologic characterization** of returned samples<br>**Elemental chemistry** including major elements (Fe, Mg, Ca, K, Si, Ti, Al, Cr, Mn) and trace elements (Th, U, Hf, Zr, Sc, Rb, Sr, Pb, K, P, W, and rare-Earth elements) | STANDARD | Rocklets (≥8 mm) | **Site dependent** (including impact melt and high- and low-Ca pyroxene) | 2.7 g/cc | ≥[8] mm | ≥[1.4] g | ≥[8] | – | ≥[32] | ≥44.2 g | ≥[44.2] g | ≥473.4 g | ≥[473.4] g |
| | | | | | Rocklets (≥3 mm) | | | ≥[3] mm | ≥[0.1] g | ≥[100] | – | ≥[400] | ≥29.2 g | ≥[29.2] g | | |
| | | | | | Bulk regolith | No requirement | – | – | ≥[100.0] g | – | ≥[400.0] g | – | ≥400.0 g | ≥[400.0] g | | |
| E14 | Determine the origin and extent of the low-calcium pyroxene annulus enclosing central SPA | **Ability to detect material of key lithologies, including high- and low-Calcium pyroxene, olivine, and feldspar**<br>Ability to detect sampleable material | **Geochronology:** Determine the ages and isotopic compositions of returned samples utilizing Ar-Ar, and when sample characteristics allow, ≥[2] independent geochronometer systems, including, U-Pb, Pb-Pb, 147Sm-143Nd, 146Sm-142Nd, Rb-Sr, and Lu-Hf<br>**Petrologic characterization** of returned samples<br>**Elemental chemistry** including major elements (Fe, Mg, Ca, K, Si, Ti, Al, Cr, Mn) and trace elements (Th, U, Hf, Zr, Sc, Rb, Sr, Pb, K, P, W, and rare-Earth elements) | STANDARD | Rocklets (≥8 mm) | Low-Ca pyroxene | 2.7 g/cc | ≥8 mm | ≥1.4 g | ≥8 | – | ≥32 | ≥44.2 g | – | ≥473.4 g | – |
| | | | | | Rocklets (≥3 mm) | | | ≥3 mm | ≥0.1 g | ≥100 | – | ≥400 | ≥29.2 g | – | | |
| | | | | | Bulk regolith | No requirement | – | – | ≥100.0 g | – | ≥400.0 g | – | ≥400.0 g | – | | |

| Experiment | Required sample components | Design reference mission: sample requirements (baseline only) – Sample acquisition mode | Scale factor for sample quality: Scale factor | Explanation | Achieve sample: Number (of required size) | Mass (of required size) | Scale factors for scooping and sieving operations – Incidental samples collected (adapted from Carrier 2003) – Rocklets (≥3 mm): Scale factor | Number (≥3 mm) | Scale factor | Mass (≥3 mm) | Total mass collected per acquisition mode (including untargeted smaller particles) | Total mass collected per experiment per site | Total mass collected per experiment for all sites* |
|---|---|---|---|---|---|---|---|---|---|---|---|---|---|
| E1 | Rocks (≥40 mm) | Targeted rock picking | 1.0× | ≥40-mm rocks are targeted individually using remote sensing data. We assume all targeted rock will meet requirements, and no "sample quality" factor is required. | ≥12 | ≥1,689.6 g | **N/A:** We do not assume incidental rocks and regolith will be collected when performing targeted rock picking. | | | | ≥1,689.6 g | **≥2,244.6 g** | **≥6,7338 g** |
| | Rocklets (≥8 mm) | Targeted sieving | 2.0× | Anorthosite material is expected to be rare, but the exact abundance is uncertain. With this uncertainty, we assume a conservative 50%. | ≥32 | ≥36.0 g | 71.0× | ≥2,272 | 4.3× | ≥155.0 g | ≥155.0 g | | |
| | Bulk regolith | Untargeted scooping | 1.0× | There are no lithology requirements for bulk regolith, so no additional "sample quality" scale factor is required. | – | ≥400.0 g | **N/A:** We do not assume incidental rocklets in bulk regolith samples will count towards rocklet requirements. | | | | ≥400.0 g | | |
| E2 | Rocklets (≥8 mm) | Targeted sieving | 3.0× | Impact mixing models suggest that 14–86% of material in central SPA corresponds to impact melt, with 50–80% being the preferred range (e.g., Petro & Pieters 2004). Also note that not all impact melt breccias will have dateable impact melt material (50% is common, e.g., Cohen et al. 2005). We assume a conservative 3× factor, equivalent to assuming 33.3% of material is impact melt. | ≥96 | ≥132.7 g | 71.0× | ≥6,816 | 4.3× | ≥570.7 g | ≥570.7 g | **≥970.7 g** | **≥2,9121 g** |
| | Rocklets (≥3 mm) | Targeted sieving | 3.0× | | ≥1,200 | ≥87.5 g | **Requirement met by incidental sample collection** | | | | | | |
| | Bulk regolith | Untargeted scooping | 1.0× | There are no lithology requirements for bulk regolith, so no additional "sample quality" scale factor is required. | – | ≥400.0 g | **N/A:** We do not assume incidental rocklets in bulk regolith samples will count towards rocklet requirements. | | | | ≥400.0 g | | |
| E3 | Rocklets (≥5 mm) | Targeted sieving | 1.25× | Sample analyses and impact mixing models show that between 3–40% of material from mare basalts may be exogenous, although this is very dependent on location (e.g., Cao et al. 2025, Su et al. 2025, Xu et al. 2024, Xie et al. 2020, Huang et al. 2017). 5–20% is common. We designed to 20%. | ≥100 | ≥40.0 g | 9.0× | ≥900 | 2.1× | ≥84.0 g | ≥84.0 g | **≥484.0 g** | **≥1,452 g** |
| | Rocklets (≥3 mm) | Targeted sieving | 1.25× | | ≥500 | ≥43.2 g | **Requirement met by incidental sample collection** | | | | | | |
| | Bulk regolith | Untargeted scooping | 1.0× | There are no lithology requirements for bulk regolith, so no additional "sample quality" scale factor is required. | – | ≥400.0 g | **N/A:** We do not assume incidental rocklets in bulk regolith samples will count towards rocklet requirements. | | | | ≥400.0 g | | |
| E4 | Rocklets (≥8 mm) | Targeted sieving | 3.0× | The exact quality factor for this investigation will depend sensitively on the specific crater and its location. We assume 3× to be consistent with SPA impact melt assumptions. | ≥96 | ≥132.7 g | 71.0× | ≥6,816 | 4.3× | ≥570.7 g | ≥570.7 g | **≥970.7 g** | **≥970.7 g** |
| | Rocklets (≥3 mm) | Targeted sieving | 3.0× | | ≥1,200 | ≥87.5 g | **Requirement met by incidental sample collection** | | | | | | |
| | Bulk regolith | Untargeted scooping | 1.0× | There are no lithology requirements for bulk regolith, so no additional "sample quality" scale factor is required. | – | ≥400.0 g | **N/A:** We do not assume incidental rocklets in bulk regolith samples will count towards rocklet requirements. | | | | ≥400.0 g | | |
| E5 | Rocklets (≥8 mm) | Targeted sieving | 3.0× | The exact quality factor for this investigation will depend sensitively on the specific crater and its location. We assume 3× to be consistent with SPA impact melt assumptions. | ≥96 | ≥132.7 g | 71.0× | ≥6,816 | 4.3× | ≥570.7 g | ≥570.7 g | **≥970.7 g** | **≥2,912.1 g** |
| | Rocklets (≥3 mm) | Targeted sieving | 3.0× | | ≥1,200 | ≥87.5 g | **Requirement met by incidental sample collection** | | | | | | |
| | Bulk regolith | Untargeted scooping | 1.0× | There are no lithology requirements for bulk regolith, so no additional "sample quality" scale factor is required. | – | ≥400.0 g | **N/A:** We do not assume incidental rocklets in bulk regolith samples will count towards rocklet requirements. | | | | ≥400.0 g | | |
| E6 | | Does not drive sampling requirements; this objective is completed as part of the coordinated analysis of samples collected for other objectives | | | | | **Does not drive sampling requirements;** this objective is completed as part of the coordinated analysis of samples collected for other objectives | | | | | **None**; accomplished by coordinated analysis of all returned samples | **None**; accomplished by coordinated analysis of all returned samples |
| E7 | Rocklets (≥8 mm) | Targeted sieving | 2.0× | The carrier of the Th-rich material is unknown; thus, it is challenging to estimate the quality factor. We assume 50%. | ≥64 | ≥88.5 g | 71.0× | ≥4,544 | 4.3× | ≥380.4 g | ≥380.4 g | **≥780.4 g** | **≥1,560.8 g** |
| | Rocklets (≥3 mm) | Targeted sieving | 2.0× | | ≥800 | ≥58.3 g | **Requirement met by incidental sample collection** | | | | | | |
| | Bulk regolith | Untargeted scooping | 1.0× | There are no lithology requirements for bulk regolith, so no additional "sample quality" scale factor is required. | – | ≥400.0 g | **N/A:** We do not assume incidental rocklets in bulk regolith samples will count towards rocklet requirements. | | | | ≥400.0 g | | |
| E8 | | Returned sample requirements are the same as (E3); no additional samples are required | | | | | **Returned sample requirements are the same as (E3)**; no additional samples are required | | | | | **Returned sample requirements are the same as (E3)**; No additional samples are required | **Returned sample requirements are the same as (E3)**; No additional samples are required |
| E9 | Rocks (≥20 mm) | Targeted rock picking | 1.0× | ≥20-mm rocks are targeted individually using remote sensing data. We assume all targeted rock will meet requirements, and no "sample quality" factor is required. | ≥12 | ≥326.4 g | **N/A:** We do not assume incidental rocks and regolith collected either during targeted rock picking or bulk regolith sampling will count towards rocklet requirements. | | | | ≥326.4 g | **≥799.8 g** | **≥1,580.2 g** |
| | Rocklets (≥3 mm) | Targeted sieving | 2.0× | Mantle material is expected to be rare, but the exact abundance is uncertain. With this uncertainty, we assume a conservative 50%. | ≥800 | ≥73.4 g | | | | | ≥73.4 g | | |
| | Bulk regolith | Untargeted scooping | 1.0× | There are no lithology requirements for bulk regolith, so no additional "sample quality" scale factor is required. | – | ≥400.0 g | | | | | ≥400.0 g | | |
| E10 | | Mare basalts are sampled as part of (E3); no additional mare basalt samples are required | | | | | **Mare basalts are sampled as part of (E3);** no additional mare basalt samples are required | | | | | **≥400.0 g** | **≥400.0 g** |
| | Bulk regolith | Untargeted scooping | 1.0× | When in a well-mapped pyroclastic deposit (e.g., Schrödinger), pyroclastic material is expected to dominate the regolith, so no additional "sample quality" scale factor is required. | – | ≥400.0 g | **N/A:** We do not assume incidental rocklets in bulk regolith samples will count towards rocklet requirements. | | | | ≥400.0 g | | |
| E11 | | None; this experiment does not require returned samples | | | | | **None**; this experiment does not require returned samples | | | | | **None**; this experiment does not require returned samples | **None**; this experiment does not require returned samples |
| E12 | | None; this experiment does not require returned samples | | | | | **None**; this experiment does not require returned samples | | | | | **None**; this experiment does not require returned samples | **None**; this experiment does not require returned samples |
| E13 | Rocklets (≥8 mm) | Targeted sieving | 3.0× | The exact abundance of these materials in these regions is uncertain. As described in (E3), estimates for the abundance of impact melt range from 14–86%. As many of the samples collected here are also utilized to address (E3), we assume the same quality factor. | ≥96 | ≥132.7 g | 71.0× | ≥6,816 | 4.3× | ≥570.7 g | ≥570.7 g | **≥970.7 g** | **≥1,941.4 g** |
| | Rocklets (≥3 mm) | Targeted sieving | 3.0× | | ≥1,200 | ≥87.5 g | **Requirement met by incidental sample collection** | | | | | | |
| | Bulk regolith | Untargeted scooping | 1.0× | There are no lithology requirements for bulk regolith, so no additional "sample quality" scale factor is required. | – | ≥400.0 g | **N/A:** We do not assume incidental rocklets in bulk regolith samples will count towards rocklet requirements. | | | | ≥400.0 g | | |
| E14 | Rocklets (≥8 mm) | Targeted sieving | 2.0× | The exact abundance of these materials in these regions is uncertain. With this uncertainty, we assume a conservative 50%, consistent with other experiments described here. | ≥64 | ≥88.5 g | 71.0× | ≥4,544 | 4.3× | ≥380.4 g | ≥380.4 g | **≥780.4 g** | **Returned sample requirements acquired as part of E9; no additional samples required** |
| | Rocklets (≥3 mm) | Targeted sieving | 2.0× | | ≥800 | ≥58.3 g | **Requirement met by incidental sample collection** | | | | | | |
| | Bulk regolith | Untargeted scooping | 1.0× | There are no lithology requirements for bulk regolith, so no additional "sample quality" scale factor is required. | – | ≥400.0 g | **N/A:** We do not assume incidental rocklets in bulk regolith samples will count towards rocklet requirements. | | | | ≥400.0 g | | |

**Foldout 4.2.** SPARX sample requirements. The full SPARX science traceability matrix is in Foldout 3.2. Rows in gray are baseline (not threshold) requirements. The number and location of sample sites can are in Foldout 5.1

## 4.4 RETURNED SAMPLE MASS REQUIREMENTS

In Section 4.1and 4.3, we described the numbers and types of samples required for each of the Experiments to achieve the Science Objectives. In this section, we will clarify the curation requirements and assumptions of sample properties used to derive the total sample mass requirements for each of the Experiments. A discussion of how the required sample masses and overall sample properties affect the recommended sampling strategy, and the baseline sampling concept of operations can be found in Section 5.5 and 6.6, respectively.

### 4.4.1 Sample Curation

While the SPARX SDT was not tasked with discussing curation analyses or cost (APPENDIX: SPARX TOR), the effects of curation activities on returned sample mass were important for the SDT to consider. The NASA standard for sample curation requires that no more than 25% of returned materials are released to the mission science team, while the remaining 75% of materials are curated for future investigations (NPD7100.10E). In order to accommodate this requirement, the SDT first defined the number and masses of samples required to achieve the listed Experiments and Objectives. Where applicable, the sample masses were then forward calculated to sample rock sizes (see 4.4.2 below). This total was then multiplied by a factor of 4 to accommodate the sample curation requirements. These totals are then used to calculated the total baseline and threshold sample mass requirements. These sample mass requirements are considered implementation agnostic, and can be viewed in Section 4.3.6 (Table: Sample Requirements).

### 4.4.2 Sample Size vs. Sample Mass Requirements

The SPARX SDT had several discussions about the most useful way to both define and derive mission sample requirements. The Sample Requirements sections for each of the 14 Experiments (Section 4.1.1-4.1.14) describe the scientific rationale for the required samples, including the required lithology, and the required sample sizes and masses for the flow-down terrestrial laboratory analyses, including curation requirements (Sections 4.1.x.3). To provide full traceability between sample size characteristics and sample masses, the SDT used the published densities for each relevant lithology (Kiefer et al., 2012; 2015) multiplied by the volume of a spherical particle with the required diameter. The resulting number represents a likely upper bound on the required sample mass, recognizing that most lunar rocks are not fully spherical in shape, however, they might also host small contributions from denser minerals. The majority of Experiments require ~500 g of material per location to meet the Science Objectives and curation requirements. The exception to this is the sample material required for E1, which targets anorthositic rocks, and requires > 2000 g of sample material per location to meet science and curation requirements. The full traceability of this process is shown Section 4.3.6 (Table: Sample Requirements).

### 4.4.3 Scaling Factors for Sample Requirements

The straightforward “Sample Requirement” numbers, however, are agnostic of the realities of sample acquisition on the Moon and assume that 100% of sampled material acquired at a sampling location is applicable for the described investigations. This is an unrealistic expectation. The SDT therefore included a number of additional sample scaling factors to more realistically represent the total sample mass requirements. The scaling factors, are designed to address two main aspects of sample collection (1) the degree to which a targeted lithology is likely to be represented in a given sample acquisition, and (2) the relative abundance of the relevant rock size fractions in any given sample acquisition. Although this second item is directly reflective of the baseline sample acquisition method described in the SPARX Design Reference

Mission (Section 6.5.4), we will describe the rationale here, as any mission design that prioritizes smaller rock size fractions will need to account for the variability of the lunar regolith.

### Scaling Factor for Sample Quality:

This scaling factor captures the expected degree to which the sampled material is likely to reflect the targeted lithology. This scaling factor ranges from 1x to 3x. For lithologies that require highly targeted acquisition (e.g., anorthosite), the scaling factor is low (1x), reflecting that samples will be individually targeted, and therefore have a high probability of reflecting the required lithology. For other sample types (e.g., impact melt materials), the scaling factor is significantly higher (x3). This higher scaling factor reflects the difficulty in targeting the required lithology using *in situ* instrumentation (Section 4.2), the unknown mixing fraction of the target lithology in the local regolith, and the fraction of the regolith that is comprised of agglutinates.

### Scaling Factors for Scooping and Sieving:

As the title implies, this scaling factor accounts for the specific sample acquisition mode baselined by the SPARX Design Reference Mission. The SPARX sample requirements define several specific size classes for acquired samples, which trace to sample masses and properties required for terrestrial laboratory analyses. These relevant size fractions must be sieved from the lunar regolith, which itself has a particle size frequency distribution (e. g. Carrier et al., 2003). The factors shown in these columns scale for the amount of incidental sample that must be acquired in order to meet the number of required rocks ≥3 mm in diameter. Section 6.5.4 describes the derivation of these scaling factors in more detail.

## 4.4.4 Total Baseline and Threshold Sample Mass Requirements

The total baseline and threshold sample mass requirements reflect the amount and type of samples required to address the SPARX Science Objectives, the number of total locations from which samples must be acquired to address the Science Objectives, the fraction of sample that must be additionally acquired to satisfy future curation requirements while still allowing for the described science investigations, and the factored scaling reflecting the lithological and size fraction variability of the lunar regolith.

- **Total Baseline Sample Mass Requirements: ≥20.5 kg**
- **Total Threshold Sample Mass Requirements: ≥ 9.8 kg**

The baseline and threshold mission designs will be discussed in detail in Section 5.8.

**The sample mass requirements quoted above reflect the minimum mass of specified samples required to achieve the science defined in the SPARX STM, as such, they should be considered "unmargined".** The actual mass returned from the Moon in a sample return mission is anticipated to be higher due to a number of factors including, but not limited to, additional sample material acquired during scooping that is ancillary to the proposed investigations, additional material required or desired to fill the sample container volume to prevent mechanical abrasion of the samples, unplanned science of opportunity samples, and sample container mass.

Although not required by the TOR, the SDT has provided a first order estimate of anticipated total return mass as a function of required sample mass (Figure 4.39). In this plot "tare" refers to the mass contributions of all elements that are not the explicitly required samples, here dominated by a simple sample container system derived from the SPARX design reference mission (Section 6.5.4). **For the SPARX baseline mission, the anticipated total sample return mass is ≥ 38 kg. Both the baseline sample mass and total return mass for SPARX** **are neatly in-line with the sample mass returned by Apollo 11.** When placed in this context, it demonstrates the extremely focused nature of the SPARX required samples.

The SDT also highlights the importance of considering non-sample mass contributions for the threshold mission design. **The SPARX threshold mission science requires ≥ 9.8 kg of sample material, and while this is within the TOR-defined robotic sample return NTE limit of 10 kg, the resulting anticipated total return mass of ≥25 kg is not.** The SDT therefore strongly advises NASA to consider the total return mass when considering the mission scope, and the mission implementation architectures presented in Section 5.

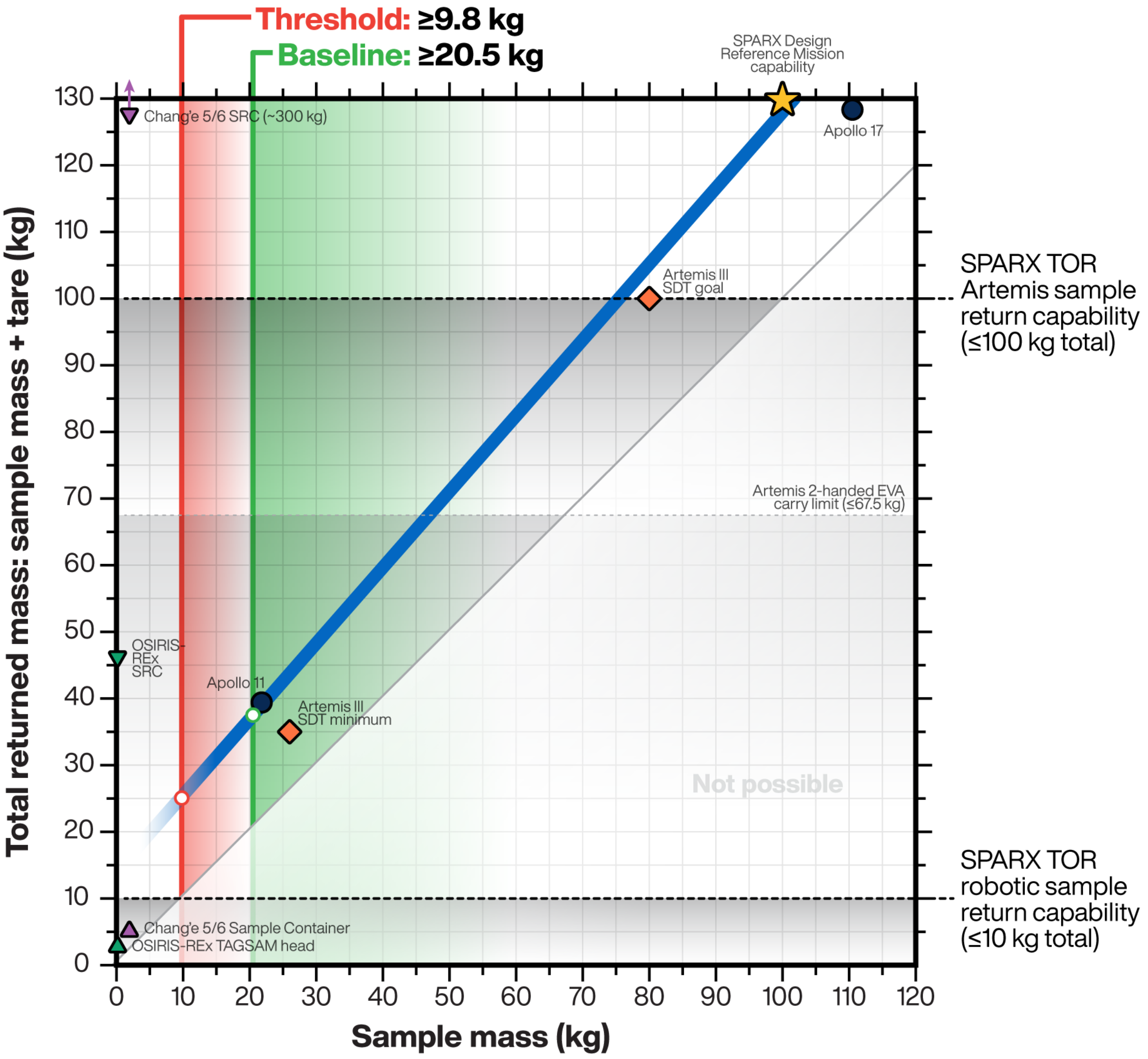

**Figure 4.39.** *First order relationship of scientific sample mass and total returned sample mass. The SPARX baseline and threshold estimates are shown, as are examples of recent sample return missions, including OSIRIS-REx, and Chang'e 5 and 6.*

### Why are the SPARX sample mass requirements so much higher than other recent missions, i.e., OSIRIS-REx?

Sample return aficionados will realize that the mass of material proposed for return by SPARX is much higher than, for example, the recent OSIRIS-REx sample return mission from asteroid Bennu. There are many differences between SPARX and OSIRIS-REx that contribute to this difference, with some being operational, some scientific, and others sample-related. Below some of these factors are compared and assessed for these two cases.

**Knowledge base:** SPARX is building on decades of research and exploration of the Moon, including previously returned lunar samples, and has numerous specific objectives that have been developed based on previous work and outstanding questions in lunar science. OSIRIS-REx in comparison, had the bare minimum data required to plan a mission, but had only limited detailed information about mineralogy, surface lithology and rock sizes. In fact, the sampling head was designed with the assumption that there would be plentiful < 2 cm materials to sample, whereas imaging of Bennu upon initial arrival revealed the very coarse and meter-sized boulders that littered the surface – indeed the possibility of sampling at all became an immediate concern.

**Sampling head and mechanism:** The OSIRIS-REx sampling mechanism was designed for < 2 cm materials and had a capacity near 1 kg. In fact, even though the sampling event was highly successful, the flap did not close tightly, resulting in loss of substantial material before stowage which ultimately led to collection of a smaller mass (even despite this sample loss the 120 g was much higher than the baseline 15 g needed to address the missions science goals). SPARX does not have the sampler collection or storage limitations that OSIRIS-REx did, and furthermore the higher masses are required for the SPARX sample measurements (see below).

**Sample grain size:** All volatile-rich chondritic meteorite material in worldwide collections is comprised of very fine grained highly altered material, and thus the expectation for OSIRIS-REx samples. Smaller masses of material of Bennu should thus be required to attain representative sample of the asteroid. In comparison, SPARX sampling campaign is targeting multiple and specific lithologies, while also testing hypotheses regarding differences in impact structures and events across the lunar farside.

**Chronology approaches:** Many SPARX goals involve chronology and dating of specific lithologies and rock types, and require specific mineral types, masses, and compositional ranges. In comparison, because of the fine-grained nature and the ancient ages of Bennu material, micro-analytical techniques are used to determine ages, and age dating of rocks was not a high priority of the OSIRIS-REx mission; their ancient nature was already anticipated and expected.

**Science questions:** The Moon is a differentiated body with a core, mantle, and crust all of which have evolved over 4.5 billion years. Thermal events have transformed lunar materials due to melting, impact events, and volcanic activity. As a result, the Moon has wide range of rock types that span micron-scale to cm-scale in textures. The science goals of SPARX require multiple lithologies and grain sizes for their resolution. Bennu on the other hand is an ancient undifferentiated body that has not experienced the heating, melting, or mega-impact events of the Moon, and thus the lithologies needed to address questions of its history and geologic evolution are of a much more limited scope.

**Heterogeneity of lithologies:** The lunar samples are known to be polymict breccias that contain a wide variety of rock types in clasts, many due to impact dynamics which in some cases were transported 100s-1000s of kilometers from the original site to be included in regolith with other unrelated materials. Because of this, and our relatively mature understanding of lunar breccias, sampling plans must account for this heterogeneity and typically include collection of more mass than is required so that ample mass of the targeted lithology is collected. For Bennu, the heterogeneity is not as severe, and in fact most of the rock types collected – even the two most common and voluminous lithologies – the angular and hummocky – have nearly identical overall mineralogy and similar textures. This, coupled with the fine grain sizes, means that the 120 g of Bennu material has allowed all of the scientific goals of the OSIRIS-REx mission to be addressed.

**Contamination concerns:** Because of focus on organic geochemistry and desire to understand aspects of Bennu that relate to the origin of life, OSIRIS-REx mission dedicated significant resources to understanding the sample exposure to various potential contaminants. Elimination of as many possible contaminants, coupled with the limitation of materials that would ultimately touch or come into contact with the sample influenced the design and implementation of nearly every aspect of the mission – spacecraft environment, sample collection, sample storage and transport, sample recovery, and sample curation. SPARX builds on decades of understanding effects of sample handling and sample curation of lunar materials, and has more flexibility in how samples are stored, prepared, allocated, analyzed, than the OSIRIS-REx Bennu samples.

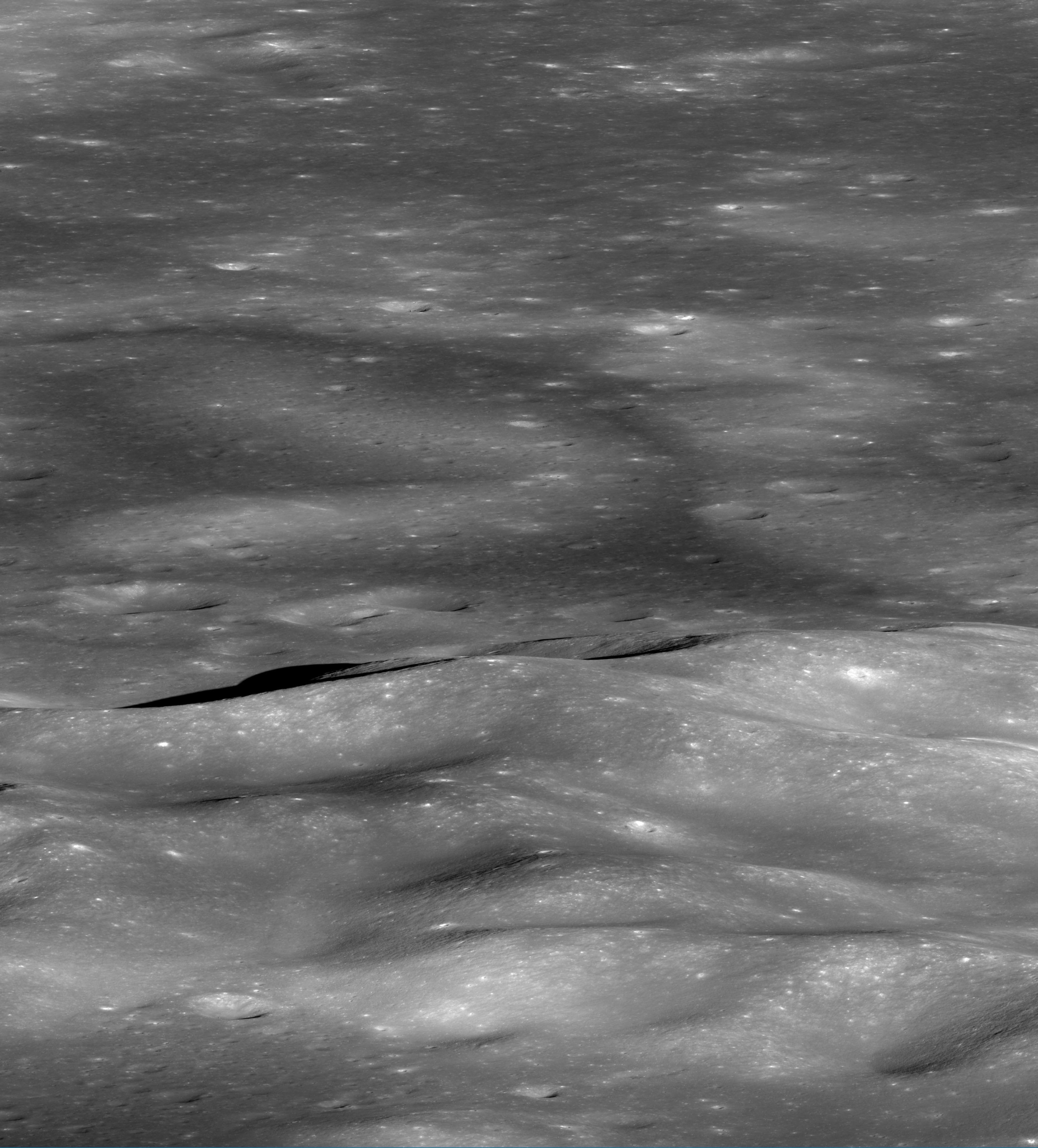

**Magnetic swirls in Mare Ingenii, in the northeastern region of South Pole-Aitken basin.**
Image credit: LROC NAC M159017963R / NASA / GSFC / Arizona State University.

# 5 MISSION ARCHITECTURES

## 5.1 INTRODUCTION

Since the 2003 Decadal Survey ("New Frontiers in the Solar System"), SPA sample return has been recognized as the highest priority lunar science mission. This ultimately resulted in placing SPA sample return on the shortlist for NASA's New Frontiers program of medium sized (~$1B) competed planetary science missions (NASEM 2003, NASEM 2011). Over the years, SPA sample return mission concepts have been proposed to NASA's competed programs multiple times. Examples, such as MoonRise would have placed a static lander in central SPA and returned ~2 kilograms of material robotically (e.g., Jolliff et al. 2012, 2021). Despite multiple proposals from different organizations, NASA did not select an SPA sample return mission for flight during any of those solicitation rounds. Recurring weaknesses for past SPA sample return mission concepts included concerns about the ability to achieve all of the science objectives with scooping/sieving material at a single site, and concerns of the cost and complexity of a sample return architecture that often-required multiple independent spacecraft (lander, communications relay, Earth return vehicle, etc.) within the prescribed cost envelope of a New Frontiers mission.

In the most recent Decadal Survey ("Origins, Worlds, and Life"), the committee revisited the approach to accomplishing SPA sample return utilizing a different mission architecture. Inspired by the Intrepid mission concept for a long-range lunar rover (Robinson et al. 2020), the committee studied two mission concepts for accomplishing SPA sample return: Endurance-A and Endurance-R (Keane et al. 2021). Both concepts would utilize a highly autonomous robotic rover to traverse ≥2,000 kilometers of the diverse terrains within SPA, while collecting *in situ* measurements and caching samples for return to Earth for detailed analysis in ground-based laboratories. The primary architectural difference between the two concepts was how the samples were returned to Earth: Endurance-A ("A" is for "Astronaut") would rendezvous with Artemis astronauts and transfer a large cache (up to 100 kilograms) for return to Earth; Endurance-R ("R" is for "Robotic") would rendezvous with a separate robotic sample return vehicle and transfer a small cache (up to 2 kilograms) for return to Earth. Ultimately, the Decadal Survey prioritized Endurance-A for a variety of reasons, including: (1) a compelling synergy with Artemis, (2) higher returned sample mass, and (3) a lower total cost to NASA's Science Mission Directorate ("producing flagship-level science at a fraction of the cost"). Since only Endurance-A was recommended for implementation, the concept name was eventually simplified to "Endurance" after the Decadal Survey (Keane et al. 2024). Note that the Endurance concept defined by the Decadal Survey does not have any relevance to Blue Origin's Blue Moon Mark 1 lander which was recently named "Endurance."

In response to the Origins, Worlds, and Life Decadal recommendations, NASA convened the SPARX SDT to provide an analysis to the Deputy Associate Administrator for Exploration (DAAX) on prioritized science objectives and requirements and different implementation approaches for a SPA sample return and exploration mission. The SPARX TOR tasked the SDT with assessing an undefined range of implementation approaches for achieving different hierarchies of the proposed SPARX objectives, and from this range of approaches, to designate a singular baseline mission architecture (Section 2, APPENDIX: TOR). In this section, we discuss the locations within SPA where SPARX objectives could be achieved (Section 5.2), the Figure of Merit defined by the SDT for evaluating different implementation architectures (Section 5.3), modes of sample return (Section 5.4), methods of sample acquisition (Section 5.5), modes of surface mobility (Section 5.6), investigated mission architectures (Section 5.7), and the final baseline and threshold mission architecture (Section 5.8).

**Ultimately, the SDT concluded that the most compelling and practical approach for addressing SPA science objectives is a long-range rover that would scoop, sieve, and store lunar samples before rendezvousing with Artemis astronauts for sample return. All other concepts investigated by the SDT, including static landers, hoppers, and short-range missions fell below the science threshold.** This concept is the SPARX baseline that serves as the basis for the design reference mission described in Section 6.

## 5.2 REGIONS OF INTEREST ACROSS SOUTH POLE-AITKEN BASIN

As described in Section 4, the SPARX science objectives require *in situ* measurements and samples acquired from a diverse range of locations and lithology types within SPA. In this section, we describe the main "regions of interest" where science objectives could be met. The minimum number of locations required to achieve the SPARX objectives is defined in Section 5.3.

### 5.2.1 Central Plains of the South Pole-Aitken Basin (CSPA)

#### 5.2.1.1 Overview of CSPA

The interior of SPA is one of the most compositionally and stratigraphically complex, and scientifically compelling regions on the Moon. Situated within an extensive topographic depression (Wieczorek et al., 2013), a relic of the initial excavation cavity, the central region of SPA is characterized by widespread smooth plains, unusual volcanic features, a distinctive high-Ca,Fe pyroxene-bearing surface composition, and the signature of excessive mass in the subsurface (Figure 5.1) (e.g., Moriarty and Pieters, 2016; 2018; James et al., 2019). The complex compositional stratigraphy of this region is observed in the central peaks of craters, which reveal a more Mg-rich mineralogy. The unusual composition of this region has been referred to as the South Pole-Aiken Compositional Anomaly (SPACA), and has been hypothesized to reflect the broadly unconstrained history of impact melt differentiation, crustal backfilling, and volcanic infilling that occurred in the wake of the SPA impact event (Trowbridge et al., 2020; Citron et al., 2024, Whitten et al., 2015; Moriarty and Pieters, 2016; 2018; Izquierdo et al., 2024; Wang et al., 2024; Chuang et al., 2025, Petro and Pieters, 2008, Hurwitz and Kring, 2014; Vaughan and Head, 2014; Uemoto et al., 2017).

#### 5.2.1.2 Key Science Objectives for CSPA

The unique compositions and geologic setting of the central SPA region make studying this region essential for reconstructing SPA's geologic history and understanding the wider processes of impact basin formation and lunar evolution. Several of the SPARX objectives can only be accomplished by leveraging the excavated stratigraphy of craters within CSPA. The craters Bose and Bhabha are located in the center of the SPACA anomaly, and their ejecta deposits reveal distinct compositions that are required to investigate the hypotheses for post-impact SPA modification as described in E13. The deep crustal excavation provided by these craters also provides an important location to potentially access SPA impact melt, which is required for E2. The CSPA region also hosts a number of small mare deposits with a range of proposed crater-retention ages (Pasckert et al., 2018), making it an ideal location to acquire basalt samples for E3 and E8. Finally, the CSPA region covers, by definition, the center of the basin, and extends for several hundred kilometers, making this region the dominant traverse target for the geophysical surveys proposed by E11, and the distal portions of the central SPA region are ideal for addressing the origin of the compositional anomaly at the edge of the central region (E14).

The SPA impact event would have stripped away the original highlight floatation crust, and therefore this region is non-ideal for acquiring anorthosite samples (E1). Similarly, Lunar Prospector data do not show Th enrichment (>3 ppm) within the broad CSPA region (E7). Finally, data returned by remote sensing datasets has not led to hypothesized excavations of mantle lithologies (E9).

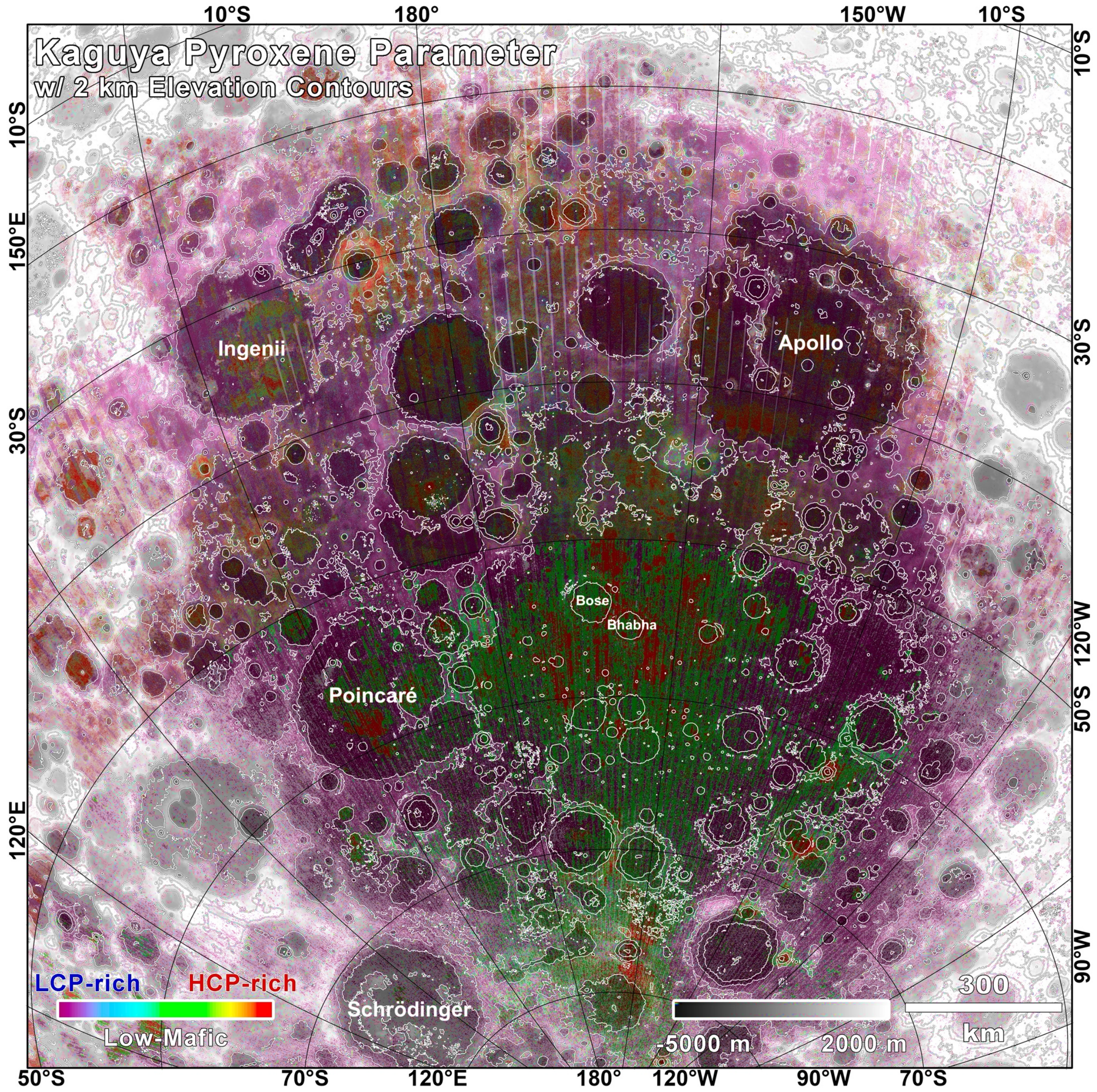

**Figure 5.1.** *LOLA topography integrated with Kaguya mineral abundance data highlights a relatively smooth low-topography zone in central SPA with a surface composition enriched in high-Ca,Fe pyroxenes (SPACA). Figure modified from Moriarty et al., 2026.*

## 5.2.2 Poincaré Basin

### 5.2.2.1 Overview of Poincaré

Poincaré is a D~340 km impact structure in the western part of SPA (56.7°S, 163.6°E) (Figure 5.2). Based on the number of superposed impact craters and its degradation state, it is thought that Poincaré is Pre-Nectarian in age (Wilhelms et al., 1987; Fassett et al., 2012). Indeed, Poincaré is inferred to be the oldest post-SPA basin of its size, as it has more superposed craters and is more degraded than Apollo basin (e.g., Chen et a., 2025). Poincaré's western rim and western inner ring are preserved, while the eastern portion of the basin has been reworked by several 70–100 km craters (Abbe, Hess, Poincaré E and F, and two large unnamed craters). There is widespread mare on Poincaré's floor and on the floor of Poincaré E and F.

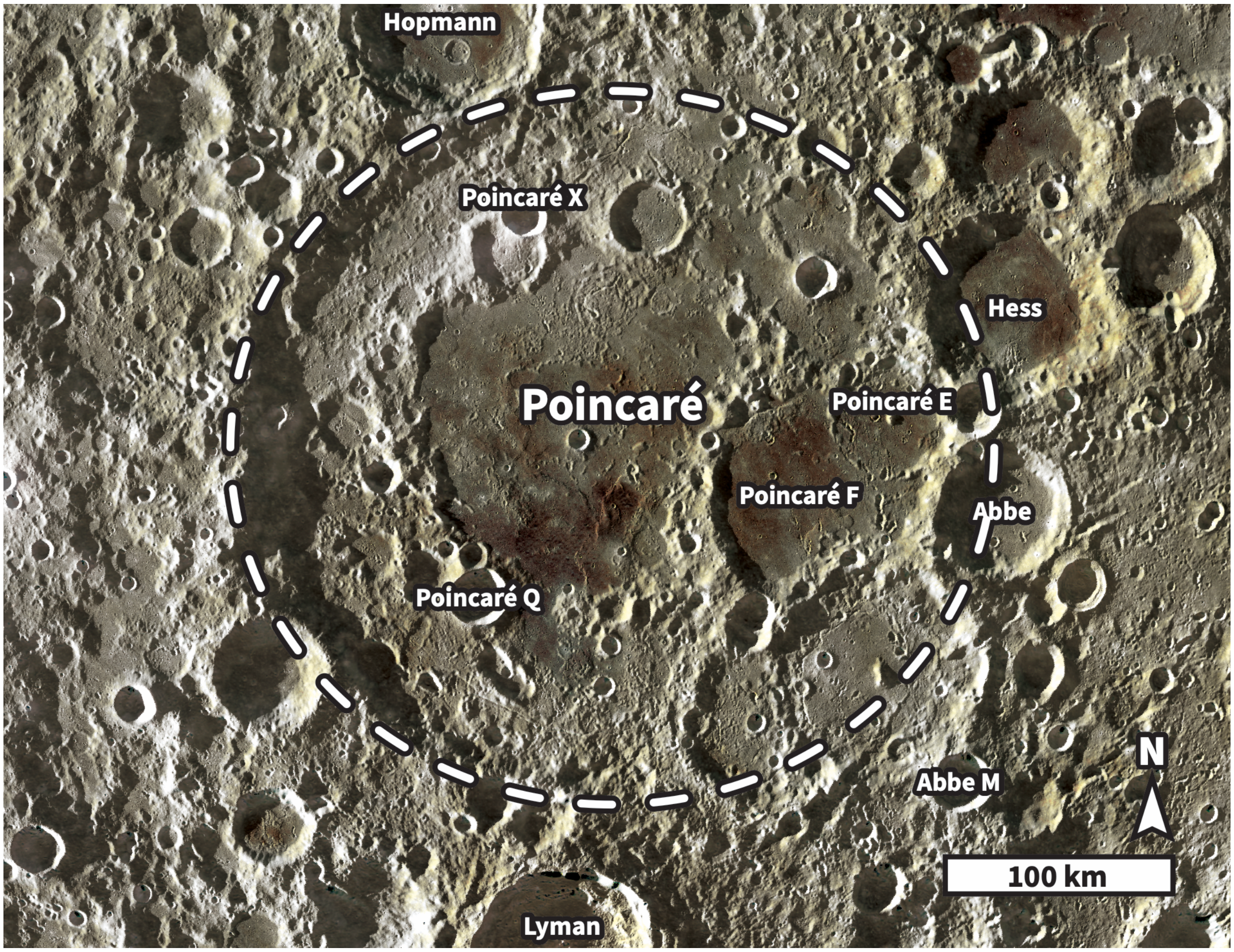

**Figure 5.2.** *Annotated LROC WAC mosaic of the Poincaré impact basin combining the three-color Hapke-normalized mosaic of 689 nm (red channel), 415 nm (green channel) and 321 nm (blue channel), with the low-sun (big shadows) mosaic. Images rendered by ACT/LROC Quickmap and composited by James Tuttle Keane.*

### 5.2.2.2 Key Science Objectives for Poincaré

Poincaré is an optimal location inside SPA for addressing chronology objectives (Goal 1, Section 3.1), because samples have the potential to include material that record the age of the SPA (E2) and Poincaré itself (E5). One of the areas where this could be best pursued is on Poincaré's inner ring. Numerous Imbrian craters are potentially available for direct sampling to pursue experiment E4; there are also large and distinct secondaries from Antoniadi crater that potentially could enable dating that large Imbrian impact. The range of mare units (E3, E8, E10) in Poincaré range in age from at least 3.4–3.6 Gy (Pasckert et al., 2018).

For exploring the Moon's flotation crust/anorthosite, there are small areas along the northern inner ring (near Poincaré X) with high (>80%) inferred abundances of plagioclase based on Kaguya data (Lemelin et al., 2022) that could be prime targets for anorthosite (E1). In the areas where the inferred plagioclase abundance is high, it is not as high as in some of the other portions of SPA.

Poincaré is not a preferred location to target either coherent SPA impact melt-sheet materials, which likely would have been destroyed by the formation of Poincaré basin itself, or several other central-SPA-specific geomorphic targets. Additionally, there are no obvious pyroclastic deposits in the basin, though there are several floor-fractured craters (often associated with pyroclastic eruptions elsewhere) in the Poincaré interior (e.g., Jozwiak et al., 2012, 2015). Based on Lunar Prospector, the thorium abundance is relatively low inside Poincaré. Thorium values are relatively higher on Poincaré's eastern third, and reach a local maximum of 3.1 ppm on the 0.5 ppd data just outside the southeast rim of Poincaré, near Abbe M, making this location a viable regional target for E7.

### 5.2.3 Schrödinger Basin

#### 5.2.3.1 Overview of Schrödinger

Schrödinger Basin is a well-preserved impact structure located along the southwestern margin of SPA (center: 75°S, 132.5°E) measuring ~315 km in diameter. It is one of the youngest basins on the Moon, with formation in the early Imbrian and an estimated of age of ~3.8 billion years (Wilhelms et al., 1987; Fassett et al., 2012). Schrödinger's most striking feature is its nearly pristine central peak ring that stretches ~150 km in diameter and rises 1 to 2.5 km above the surrounding basin floor (Figure 5.3). This ring is proposed to have uplifted deep-seated crystalline rocks (Kring, 2016). Spectral remote sensing data show that the composition of the peak ring is distinct from its surroundings and that it contains pure mineral assemblages of plagioclase (attributed to anorthosite rocks), low calcium pyroxene, and olivine (Kramer et al., 2013; Kring, 2016).

Schrödinger's floor is predominantly covered by impact melt that displays multiple textures and lithologies (Shoemaker et al., 1994). The basin's floor also hosts several distinct volcanic deposits, including multiple small mare patches and a pyroclastic deposit. The pyroclastic deposit is located on east side of the basin and forms a conical apron of material surrounding a large volcanic vent (e.g., Kring et al., 2025). Schrödinger's floor also contains at multiple smooth flat plains that have been mapped as mare basalts (Kramer et al., 2013). These features are located in the north and northeastern section of the basin floor and are significantly younger than the basin itself.

#### 5.2.3.2 Key Science Objectives for Schrödinger

Schrödinger provides the opportunity to collect samples that address five different experiments. It is the only location in the SPA region where SPARX could confidently collect sample from a pyroclastic deposit (E10), while the volcanic features on the crater's floor would give access to a mare basalt unit (E3, E8, E10), and give SPARX an opportunity to investigate a volcanic vent and fracture feature. The basin itself is an Imbrian impact, so samples from the area would also provide definitive age dates from this period (E4). SPA impact melt may be present in the southeast portion of Schrödinger (E2) (Hurwitz & Kring, 2014), although regions closer to the center of SPA provide higher priority areas to sample SPA melt based on assumed geologic context. Similarly, the observed plagioclase material in the peak ring could provide samples of the anorthosite highlands crust (E1); however, accessing these exposures is challenging, and sampling activities would rely on identifying boulders or other material that has been shed at the base of the peaks. Remote sensing observations of the peak rings of Schrödinger have also suggested exposures of olivine, which has been hypothesized to represent mantle material (E9) (Yamamoto et al., 2010), although the same caveats about sample accessibility apply.

Schrödinger is not a preferred location for acquiring samples of thorium rich material, as Lunar Prospector data suggest only trace (~1ppm) Th abundance in the region. Additionally, due to its location at the far southern and western limb of SPA, Schrödinger is a non-ideal location for investigating lithologies and morphologies associated with central SPA basin formation and evolution processes. However, the location of SPA at the edges of SPA does make it a viable regional target for investigating hypothesized SPA rim massifs (E12).

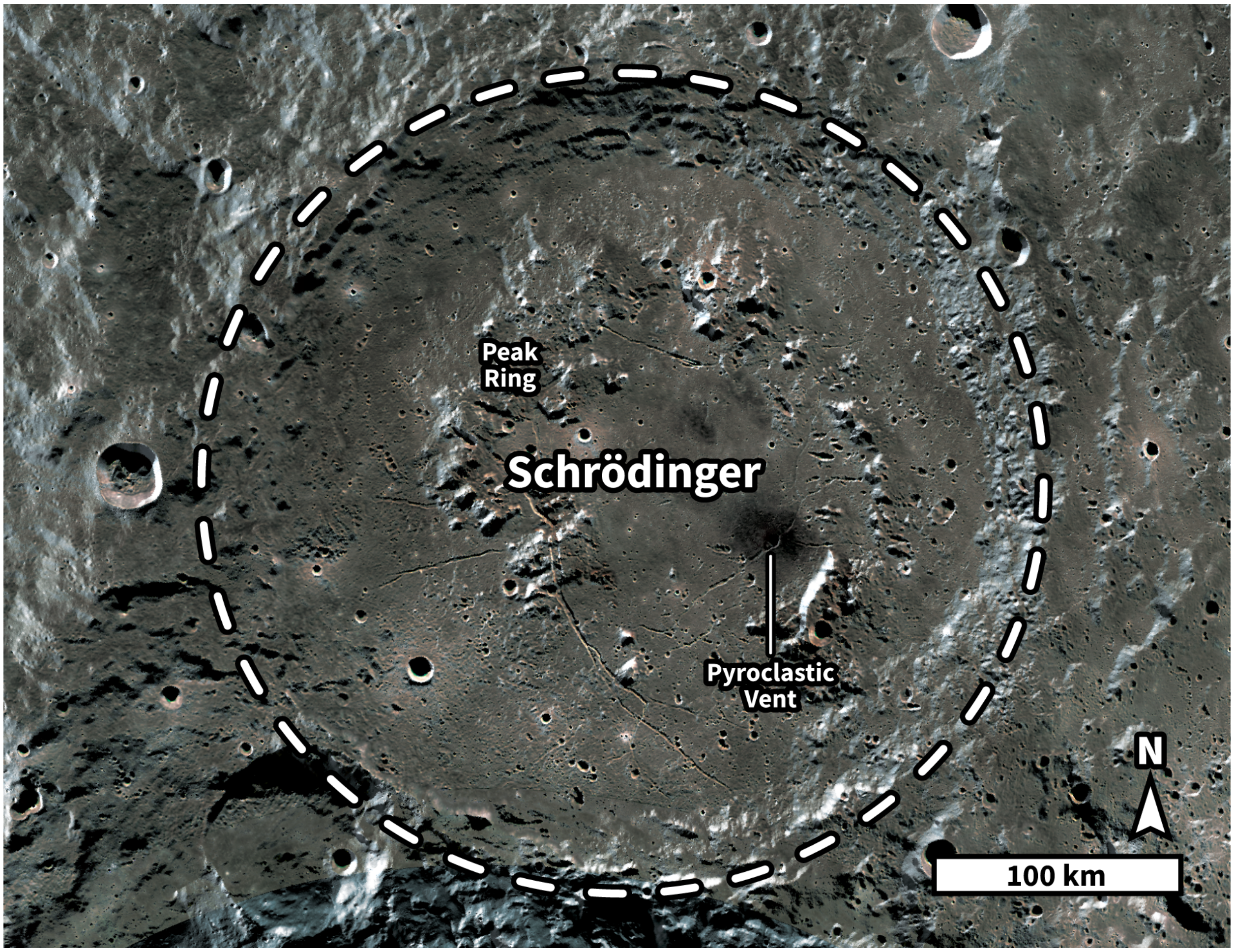

**Figure 5.3.** *Annotated LROC WAC mosaic of the Schrödinger impact basin combining the three-color Hapke-normalized mosaic of 689 nm (red channel), 415 nm (green channel) and 321 nm (blue channel), with the large solar incidence WAC mosaic. Images rendered by ACT/LROC Quickmap and composited by James Tuttle Keane.*

### 5.2.4 Ingenii Basin

Given the balance of science objectives, and the architectural constrains imposed by the latitude of the basin, Ingenii is not considered a destination for either the baseline or threshold mission. However, the SDT recognizes that a mission profile that includes Ingenii basin could accomplish a significant number of SPARX Objectives and thus warrants special consideration.

### 5.2.4.1 Overview of Ingenii

Ingenii Basin (~320 km diameter) is located just interior to the northwestern rim of the South Pole–Aitken (SPA) basin (33.2°S, 164.8°E), and contains a unique combination of geologic, volcanic, geochemical, and magnetic features that make it a compelling science target for an SPA robotic sample return mission (Figure 5.4). It is an ancient, Pre-Nectarian impact basin that has been heavily modified by subsequent geologic processes, including superposed craters (e.g., Thomson and Thomson M) and pronounced furrowing of its basin walls. Ingenii lies near the antipode of the Imbrium basin, and its wall furrows have been hypothesized to result either from focused seismic waves generated during the Imbrium impact (Schultz & Gault, 1975) or from the concentrated emplacement of ejecta at the basin's antipode (Wilhelms et al., 1987). South of O'Day crater, along the western rim of Ingenii, a region of elevated terrain has been proposed to represent an inner ring massif of SPA (Bernhardt et al., 2025).

The floor of Ingenii is largely filled with mare basalts, which also completely flood the floors of Thomson and Thomson M craters. In addition to mare materials, Ingenii also contains several exposures of plagioclase-rich lithologies in basin floor regions that are not mantled by mare basalts, providing potential access to crustal materials. Ingenii is also magnetically distinctive within SPA, hosting multiple magnetic anomaly hotspots along its rim and interior (Figure 5.5). Many of the strongest anomalies spatially coincide with lunar swirls—enigmatic, high-albedo, sinuous surface features (Blewett et al., 2011) that can be prominently seen on the basin floor (Figure 5.4). While not located within Ingenii itself, two thorium (Th) hotspots occur just exterior to the basin: one associated with Birkeland crater to the east and another near Oresme V to the south.

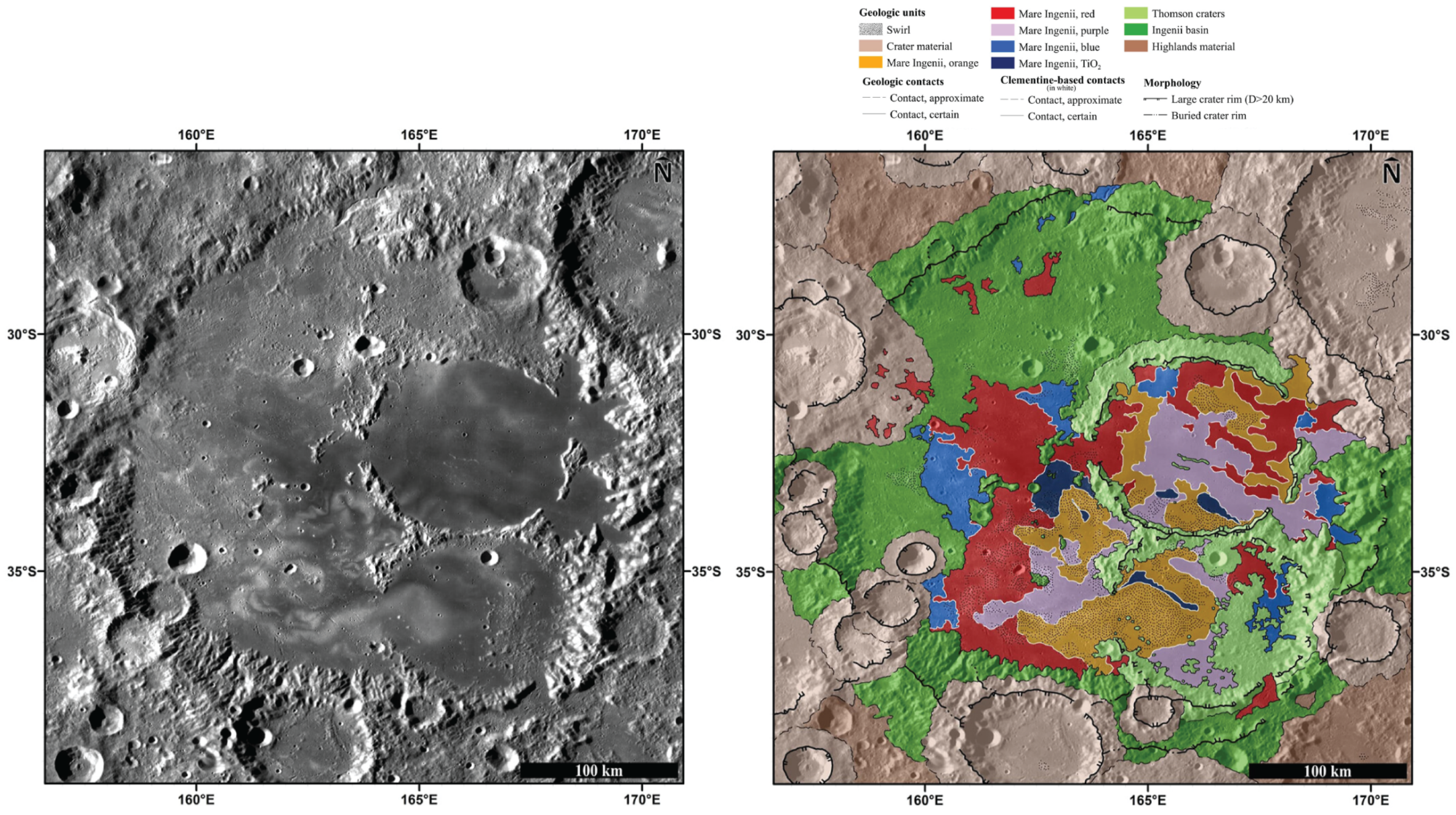

**Figure 5.4.** *Top left: LROC WAC image mosaic of the Ingenii basin (adapted from Salari et al., 2023). Top right: Morpho-stratigraphic map of the Ingenii basin region visualized on top of the Clementine UVVIS Color Ratio mosaic (adapted from Salari et al., 2023). Bottom: Lambert azimuthal equal-area projections of the lunar farside centered on South Pole-Aitken basin showing the magnetic field strength (left) and topography (right), adapted from Wieczorek et al. (2013).*

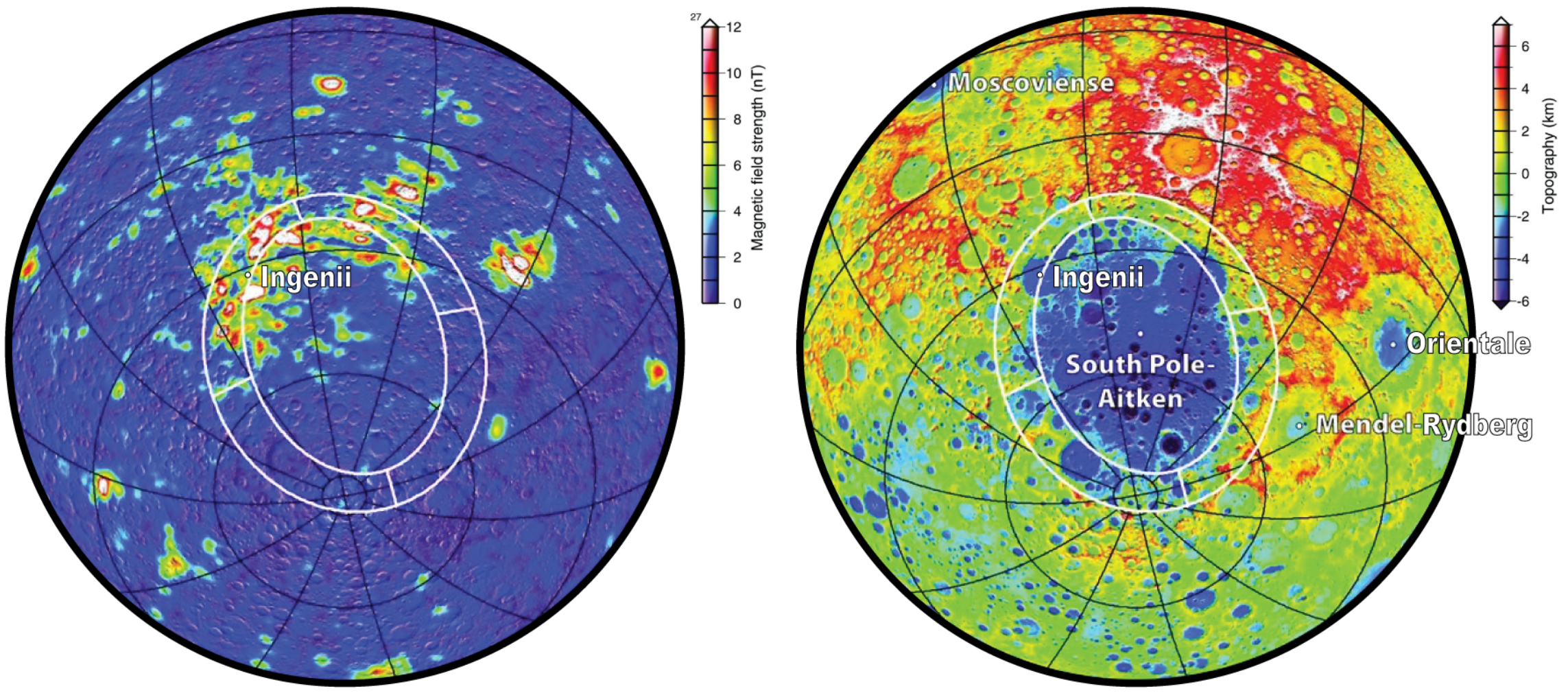

**Figure 5.5.** *Lambert azimuthal equal-area projections of the lunar farside centered on South Pole-Aitken basin showing the magnetic field strength (left) and topography (right), adapted from Wieczorek et al. (2013). Ingenii basin lies within the "crescent" of strong magnetic anomalies NW of South Pole-Aitken that have been interpreted as potential impactor core material.*

### 5.2.4.2 Key Science Objective for Ingenii

Ingenii basin has the ability to address multiple SPARX science objectives and presents the opportunity to address specific science questions that cannot be addressed in other parts of SPA. The diverse lithologies present within Ingenii, including exposed plagioclase rich lithologies contribute to E1. The mare basalts within Ingenii are particularly notable because multiple flow units can be distinguished using orbital compositional datasets, and these units and exhibit a wide range of emplacement ages. Crater model ages for 11 mapped mare units range from 3.22 to 3.67 Ga (Pasckert et al., 2018), representing the largest age spread of mare volcanism documented within a single impact crater in SPA, and highly conducive to addressing E3 and E8.

Experiment E7, which focuses on sampling Th-rich material, could be addressed by sampling the observed Th anomalies near the outside of the basin rim (Figure 5.6) (e.g., Birkeland and Oresme V. craters). Lunar Prospector data suggests that the Th in these regions exceeds 5 ppm, making these anomalies the most suitable within SPA for addressing the potential for KREEP exposures within SPA.

There are several large, Imbrian-aged craters within the Ingenii region, suitable for E4; however, the location of Ingenii at the Imbrian antipode does raise the potential for contamination/overprinting by the Imbrian impact itself. While Ingenii is chronostratigraphically associated with the Pre-Nectarian period; the combination of its location at the Imbrian antipode, combined with the significant modification to the basin rim and peak ring structures, has led to the concern about the ability to successfully acquire samples of the basin for dating, as required by E5. Similarly, Ingenii is not considered an ideal location for dating the SPA impact event (E2), as the formation of Ingenii likely would have destroyed or reset any SPA impact melt in the region.

### 5.2.4.3 Alternative Science Objectives for Ingenii

While Ingenii was not prioritized for the baseline or threshold mission, the SDT recognizes that a mission profile that includes Ingenii basin could accomplish a significant number of SPARX Objectives. Additionally, if future mission profiles are to consider Ingenii, the SDT recommends the inclusion of an additional experiment, in support of SO 3.1 "Determine whether discontinuities in crustal structure (e.g., crustal thickness, density, and magnetic field) across SPA constrain models for mega-basin formation". This proposed experiment could be to "Constrain the source depth of magnetic anomaly materials in the northwestern portion of the basin to determine whether they are derived from impactor materials.". This experiment would leverage the magnetometer that is already part of the recommended SPARX baseline payload and would investigate the nature of the unique magnetic environment around Ingenii basin.

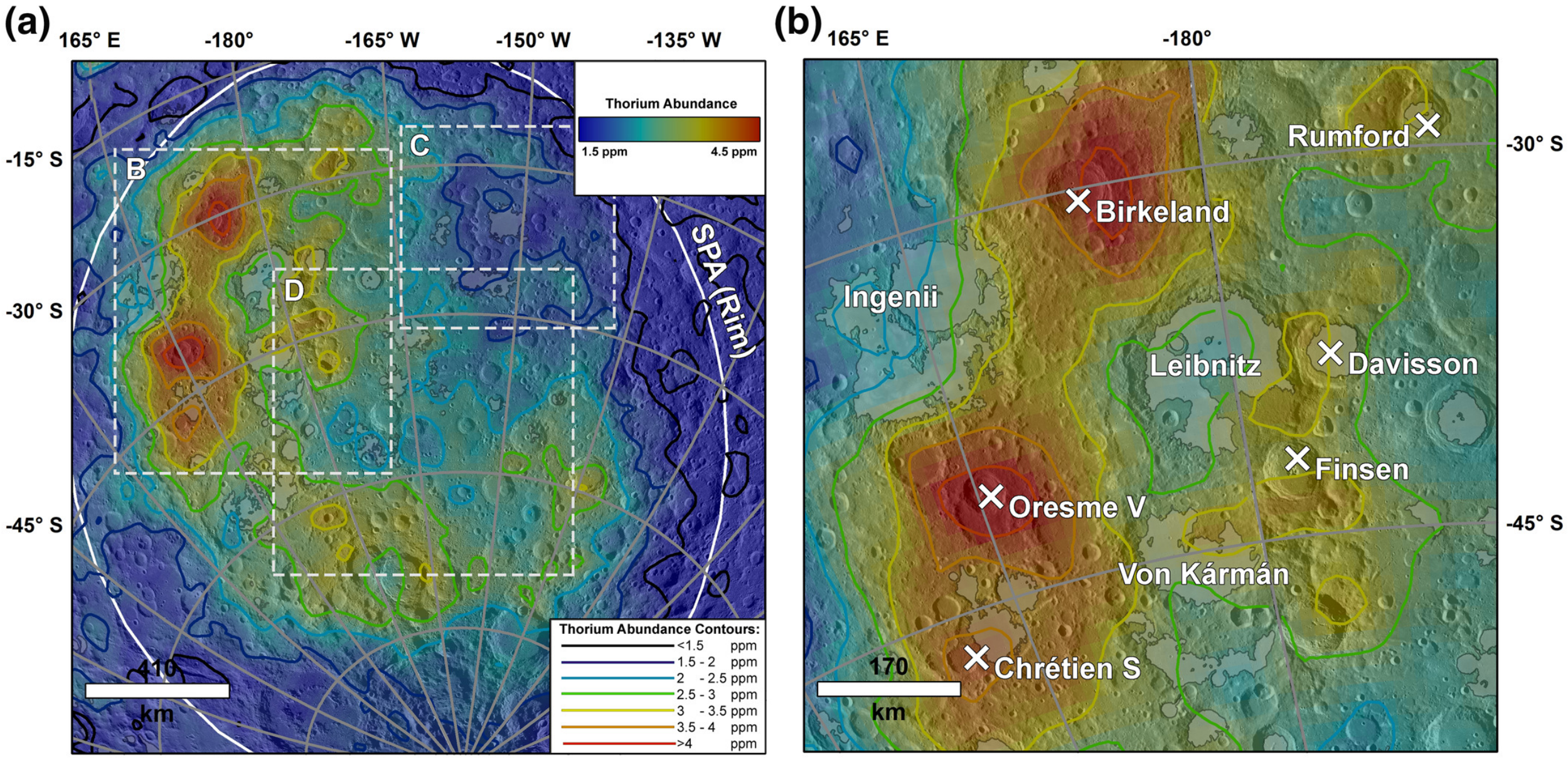

**Figure 5.6.** *Thorium concentrations within SPA (a) and in the NW region (b) show that Ingenii is situated close to a crescent of high Th concentrations. Figure modified from Moriarty et al. (2021).*

### 5.2.5 Other Regions of Interest Not Prioritized

Two other regions of interest within SPA were considered but not prioritized by the SDT:

- Apollo basin: Apollo represents a large pre-Imbrian basin that may meet several of the SPARX SDT Science Objectives. However, this region was not prioritized by the SDT as samples of mare material from this basin have been returned by the Chang'e 6 mission. The SDT therefore prioritized locations from which no recognized materials have been returned.
- The South Pole Artemis Exploration Zone (Poleward of 84°S): While the Artemis exploration zone has value for other science investigations (e.g., polar volatiles), the SDT did not prioritize this region for SPARX, owing to the lack of relevant geologic terrains for the SPARX science objectives.

## 5.3 FIGURES OF MERIT FOR ASSESSING MISSION ARCHITECTURES

The science objectives of SPARX are bold and ambitiously address multiple Decadal-level science questions (Section 3.5). The SDT has defined a series of 14 Experiments (Section 4.1.1 – 4.1.14) that capture the measurements (*in situ* and terrestrial), mission requirements (sample lithologies, sizes, numbers), and locations within SPA where the science objectives can be met. The resulting science requirements for the overarching SPARX mission can be distilled into a series of Figures of Merit that can be used to evaluate (quantitatively or qualitatively) different mission architectures. In this section, we summarize these figures of merit, which we then use to evaluate different mission architectures in the following sections (Sections 5.4-5.7).

### 5.3.1 Total Number of Science Objectives - Traverse

**The most significant figures of merit for SPARX—and largest discriminator between mission architectures—is the number and hierarchy of science objectives that can be feasibly achieved by a given architecture.** This figure of merit has a broad scope and can be influenced by many aspects of a potential mission design including the traverse, the size of the surface exploration asset, and the sample return mechanism. We will explore the influence of the latter two in subsequent sections, and in this section will focus on identifying the number of science objectives that can be feasibly accomplished by various combinations of locations within SPA.

The enormous scale of SPA contributes to its scientific importance, but it also presents a challenge for exploration. To visualize these scales, Figure 5.7 shows the SPA region reprojected onto North America. SPA is >2,500 kilometers (>1,550 miles) in diameter, comparable in scale to the United States west of the Mississippi River. SPARX's Science Objectives center on the analysis of returned samples from diverse sites across SPA. Of the 14 Science Experiments defined by the SDT, 12 require the collection and return of samples, and they each flow to different requirements in terms of terrain type and lithology as summarized in FO: Site and Traverse Requirements (Section 5.8.3) and described in much more detail in Section 4.

**Table 5.1.** *Addressing SPARX's science objectives and experiments requires a diverse set of samples with a range of lithologies. This table provides a greatly simplified summary of the true requirements, which are more completely described in Section 4. Numbers in brackets are threshold requirements, whereas numbers outside of brackets are baseline requirements. The total number of distinct sites is identified at the bottom of the table, although it is important to note that some sample sites address more than one experiment, so these values do not represent the sum of the numbers in the column above. The traceability to number of sample sites is better illustrated in Section 5.8.3, FO: Site and Traverse Requirements.*

| TYPES OF SAMPLES: (SIMPLIFIED FROM THE STM) | NUMBER OF SAMPLE SITES: | EXPERIMENTS: |
|---|---|---|
| Anorthositic highlands material | ≥3 [≥1] | E1 |
| Impact melt from SPA | [≥3] | E2 |
| Impact melt from pre-Imbrian-aged impact basins (e.g., Poincaré, Ingenii) | [≥3] | E5 |
| Impact melt from Imbrian-aged impact craters | ≥1 | E4 |
| Mare basalts | ≥3 [≥2] | E3, E8, E10 |
| Pyroclastic deposits | ≥1 | E10 |
| Material from Th-rich regions | ≥2 [≥1] | E7 |
| Hypothesized mantle material | ≥2 | E9 |
| Material sampling the stratigraphy in/around central SPA | ≥5 [≥2] | E13, E14 |

| | |
|---|---|
| **Baseline number* of distinct sample sites:** | ≥20 |
| **Threshold number* of distinct sample sites:** | [≥10] |

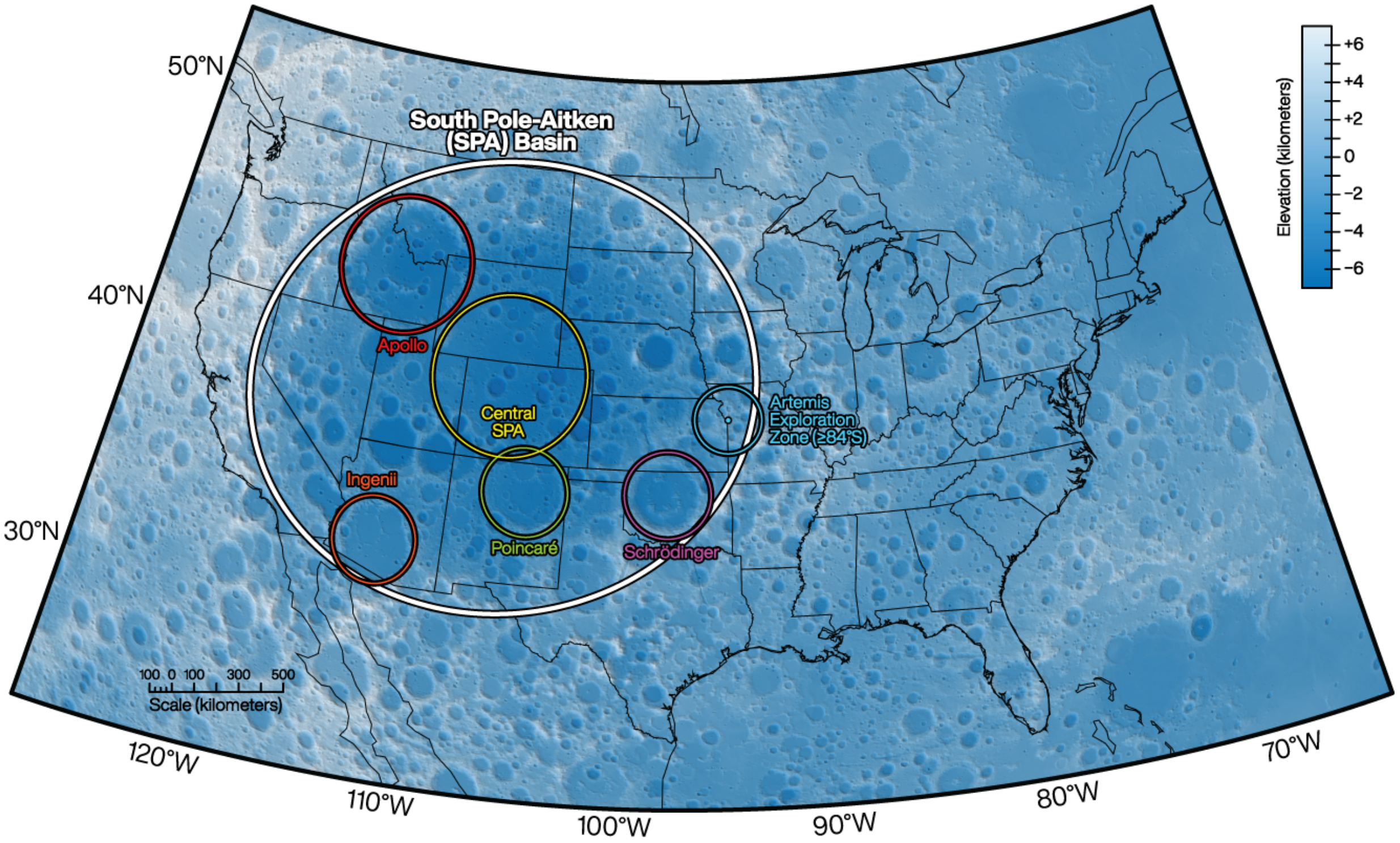

**Figure 5.7.** *Exploration of South Pole-Aitken is akin to exploring a continent. This map shows the continental United States, including the outlines of country and state borders in black, overlain by lunar topography which has been reprojected to the same scale. In this map, the lunar South Pole is centered on Kansas City, and oriented with the lunar farside towards the left (western United States) and lunar nearside towards the right (eastern United States). Distances are accurately depicted in the radial direction from Kansas City, although the lunar topography is stretched in the azimuthal direction towards the distal edges of the map due to the reprojection. The map is projected in a conic equal area projection.*

Orbital remote sensing data from a variety of lunar orbiters (e.g., NASA's Lunar Reconnaissance Orbiter and Lunar Prospector, ISRO's Chandrayaan-1, and JAXA's SELENE), have provided sufficient data to determine where on the lunar surface these materials may be accessible. The spatial distributions for each of these materials is shown in snapshot for each Experiment (Section 4.1.1-4.1.14). Unfortunately, these locations are widely distributed across SPA. FO: Site and Traverse Requirements (Section 5.8.3) shows the SPARX Science Experiments and their individual mission requirements, and how they trace to specific SPA Regions of Interest (Section 5.2). **There is no single Region of Interest within SPA where all of highest priority science objectives can be confidently met.** Confidence here is defined as a reasonable degree of scientific certainty stemming from existing orbital remote sensing data and geologic setting that the desired lithology will be present, and sampleable, in abundances sufficient to meet the sample measurement requirements. While all of the regions of interest can confidently address some of the SPARX objectives, none can address a preponderance of the objectives. Of the regions of interest, the most "science rich" is Poincaré—although even this region of interest only confidently and completely addresses 7 of the SPARX Science Experiments, partially addresses 5 Experiments, but does not address E2, which is a key component of the highest priority SPARX objective O1.3 "Determine if and when a giant planet instability drove a sharp change in the inner Solar System

| EXPERIMENT: | | MISSION REQUIREMENT: | SPA REGIONS OF INTEREST: CENTRAL SOUTH POLE-AITKEN | POINCARÉ | SCHRÖDINGER | INGENII | ARTEMIS EXPLORATION ZONE |
|---|---|---|---|---|---|---|---|
| E1 | Determine the composition and age of the anorthositic farside highlands crust | Sample ≥3 [≥1] geologically distinct exposures of farside anorthositic highlands material | X | ✓✓ | ✓✓ | ✓✓ | ✓ |
| E2 | Determine the age of SPA impact event | Sample [≥3] geologically distinct units hypothesized to be excavated or deposited SPA impact melt | ✓ | X | ✓ | X | X |
| E3 | Determine the model age of the mantle underlying SPA | Sample ≥3 [≥2] spatially and/or temporally distinct mare units as mapped from orbital datasets | ✓✓ | ✓✓ | ✓ | ✓✓ | X |
| E4 | Determine the age of ≥1 farside Imbrian-aged impact crater (D≥80 km) or basin (D≥300 km) | Sample ≥1 farside Imbrian-aged impact crater or basin with a dateable surface (e.g., peak ring or ejecta blanket) | ✓✓ | ✓✓ | ✓✓ | ✓✓ | ✓ |
| E5 | Determine the age of ≥1 non-SPA farside pre-Imbrian-aged impact basin (D≥300 km) | For [≥1] non-SPA farside pre-Imbrian impact basins (either Poincaré or Ingenii), sample [≥3] geologically distinct units associated with the crater likely to retain a basin age (e.g., peak ring, rim) | X | ✓✓ | X | ✓✓ | X |
| E6 | Determine the age and remnant magnetization signatures from ≥3 returned samples with inferred ages ≥4.0 Ga | (E6) does not drive sample specific sample site selections; this experiment is completed as part of the coordinated analysis of samples collected for other experiments | ✓ | ✓ | ✓ | ✓ | ✓ |
| E7 | Determine the composition and age of farside Th-rich materials | Sample ≥2 [≥1] geographically distinct Th-rich regions identified by preexisting orbital datasets to have Th-abundances ≥3 ppm | X | ✓✓ | X | ✓✓ | X |
| E8 | Determine the composition and age of farside basaltic units | Same as (E3); no additional sites are required | ✓✓ | ✓✓ | ✓ | ✓✓ | X |
| E9 | Sample hypothesized excavated lunar mantle material | Sample ≥2 geologic unit that is hypothesized to be excavated mantle, including hypothesized low-calcium pyroxene and olivine exposures proposed to trace mantle ejecta | X | ✓ | ✓ | X | X |
| E10 | Determine the volatile content of the farside mantle as recorded by pyroclastic and basaltic materials | Basaltic materials are collected as part of (E3); no additional sites are required | ✓✓ | ✓✓ | ✓ | ✓✓ | X |
| E11 | Determine the crustal structure (thickness, density, remnant magnetic field strength) across the SPA basin (from the center to the putative rim) to constrain mega-basin formation and evolution models | Perform a traverse of SPA that spans from central SPA out radially [≥650] km towards the hypothesized rim region | ✓ | ✓ | ✓ | ✓ | ✓ |
| E12 | Determine if the massifs in the southwestern portion of the basin constitute the diffuse rim of SPA | Traverse [≥1] km in elevation up a hypothesized SPA rim massif | X | ✓ | ✓ | ✓ | ✓✓ |
| E13 | Determine the stratigraphy and composition of the unique deposits in central SPA | Sample the geologic diversity of central SPA including: [≥1] sample of low-calcium pyroxene material excavated by Bose crate; [≥1] sample of high-calcium pyroxene material excavated by Bhaba crater; ≥1 sample from the ejecta of a morphologically fresh ≥10-km diameter crater with a well-defined ejecta blanket (plausibly excavating through mare basalts); ≥1 sample of light plains material in central SPA | ✓✓ | X | X | X | X |
| E14 | Determine the origin and extent of the low-calcium pyroxene annulus enclosing central SPA | Acquire ≥1 sample from the low-calcium pyroxene annulus enclosing central SPA, but excluding mare basalts and non-SPA impact basin material | ✓✓ | ✓✓ | X | X | X |

**Table 5.2.** *Traceability between SPARX's Science Experiments (Section 4.1) and SPA Regions of Interest (Section 5.2). Two check marks (✓✓) and a green color indicate that samples and measurements from a campaign within that region completely address the experiment; One check mark (✓) and yellow that samples and measurements from a campaign within that region contribute to addressing the experiment, but that cannot alone completely address it; a "X" and red*

*indicates that samples and measurements from a campaign within that region cannot address the experiment. Requirements in brackets are threshold requirements, whereas requirements are outside of brackets are baseline requirements. Note that a "campaign" within a single region of interest may still require long-range mobility (100s of kilometers).*

| | Science Value (normalized to 100): | |
|---|---|---|
| **Central SPA – Poincaré – Schrödinger** | 100 (Highest) | ↑ SPARX Baseline |
| **Central SPA – Poincaré – Ingenii** | 89 | |
| **Central SPA – Ingenii** | 89 | |
| **Central SPA – Poincaré** | 78 | |
| **Central SPA – Schrödinger** | 70 | ↑ SPARX Threshold |
| **Central SPA only** | 52 | |
| **Poincaré only** | 41 | |
| **Ingenii only** | 41 | |
| **Schrödinger only** | 37 (Lowest) | |
| **South Pole Artemis Exploration Zone** | Not prioritized<br>(lower SPARX science value than Schrödinger) | |

**Table 5.3.** *A summary of science value for different regions of interest and traverse between them. Science value was scored by the SDT based on the extent to which each region (or sites therein) could address the motivating science objectives and goals. These science values were normalized to 100 for the highest-value option (a traverse from Central SPA to Poincaré to Schrödinger).*

impactor flux" (Section 3.4). Therefore, mission architectures that explored only Poincare would be a significant risk of not meeting the SPARX highest priority science objectives.

The Experiments that flow from the highest priority SPARX science objectives (O1.3, O1.1, and O2.2 (Section 3.4), are E2, E4, E5, and E7. It is clear from Table 5.2 that, **in order to address the highest priority SPARX science objectives, prospective mission architectures must combine exploration of at least 2 regions of interest**.

**To address the SPARX baseline science Experiments, the SDT recommends an exploration architecture that includes 3 regions of interest: the unique central SPA region, and two of the three large basins.** The rationale for including two large basins in the baseline investigation is motivated by the scientific desire to sample critical lithologies (including anorthosite and SPA impact melt) from as diverse a range of geologic settings as possible, thereby increasing scientific confidence is the viability of the returned sample suite.

Given the above parameters, the SDT evaluated the relative science value of traverses between two or more Regions of Interest. Traversing between Regions of Interest combines the science targets of individual Regions of Interest and also enables science targets between the two Regions of Interest that otherwise may not be accessible. For example, some SPARX Science Experiments, particularly E11, drive a requirement for acquiring measurements along transects of SPA, necessitating investigations outside single Regions of Interest. Table 5.3 shows a comparison of the relative science value between different Region of Interest combinations. Of the traverses considered, a Central SPA to Poincaré to Schrödinger traverse (similar to the original Endurance-A traverse in the Decadal Survey concept study report) was deemed the highest scientific value, fully addressing all of the Science Experiments. The next three highest priority traverses included Central SPA to either Poincaré or Ingenii basin. Poincaré and Ingenii prioritized higher than Schrödinger for a variety of factors, most notably that Poincaré and Ingenii are older (pre-Imbrian), and therefore able to address the lithology target requirements of E5 (Section 4.1.5), which are required to address O1.3, the highest priority SPARX objective. Traverses that focused on individual SPA Regions of Interest were determined not to meet the SPARX threshold science requirements, with the Artemis Exploration Zone being the lowest science value overall.

Another way to visualize the prospective mission implementation and establish whether there is a preferential direction of investigation, is to consider science return as a function of distance. Figure 5.8 shows an illustrative example of the science "value" as a function of distance for both the baseline and threshold traverses (described further in Section 5.8 and Section 6.7). In this analysis, cumulative science value is determined by how each site contributes to each objective (acknowledging that some sites address more than one objective). While this analysis is incredibly simplified, it nevertheless highlights two important details: (1) Most of the science value for these long-range rover architectures is in the first half of the traverse in central SPA and Poincaré, and the entire threshold mission could be accomplished just in these two regions. (2) Traverses starting at the South Pole Moon base and traveling north (i.e., doing the reverse of these plots) must travel ≥1,000 kilometers before they would reach the highest priority science destinations; therefore, suggesting traverses should start in central SPA.

**Finding: The SDT recommends baseline mission implementation architectures that can support exploration of the Central SPA region, Poincaré basin and surrounding region, and the Schrödinger basin and surrounding region. The SDT recommends a threshold mission implementation architecture that supports exploration of the Central SPA region through the Poincaré basin and surrounding region.**

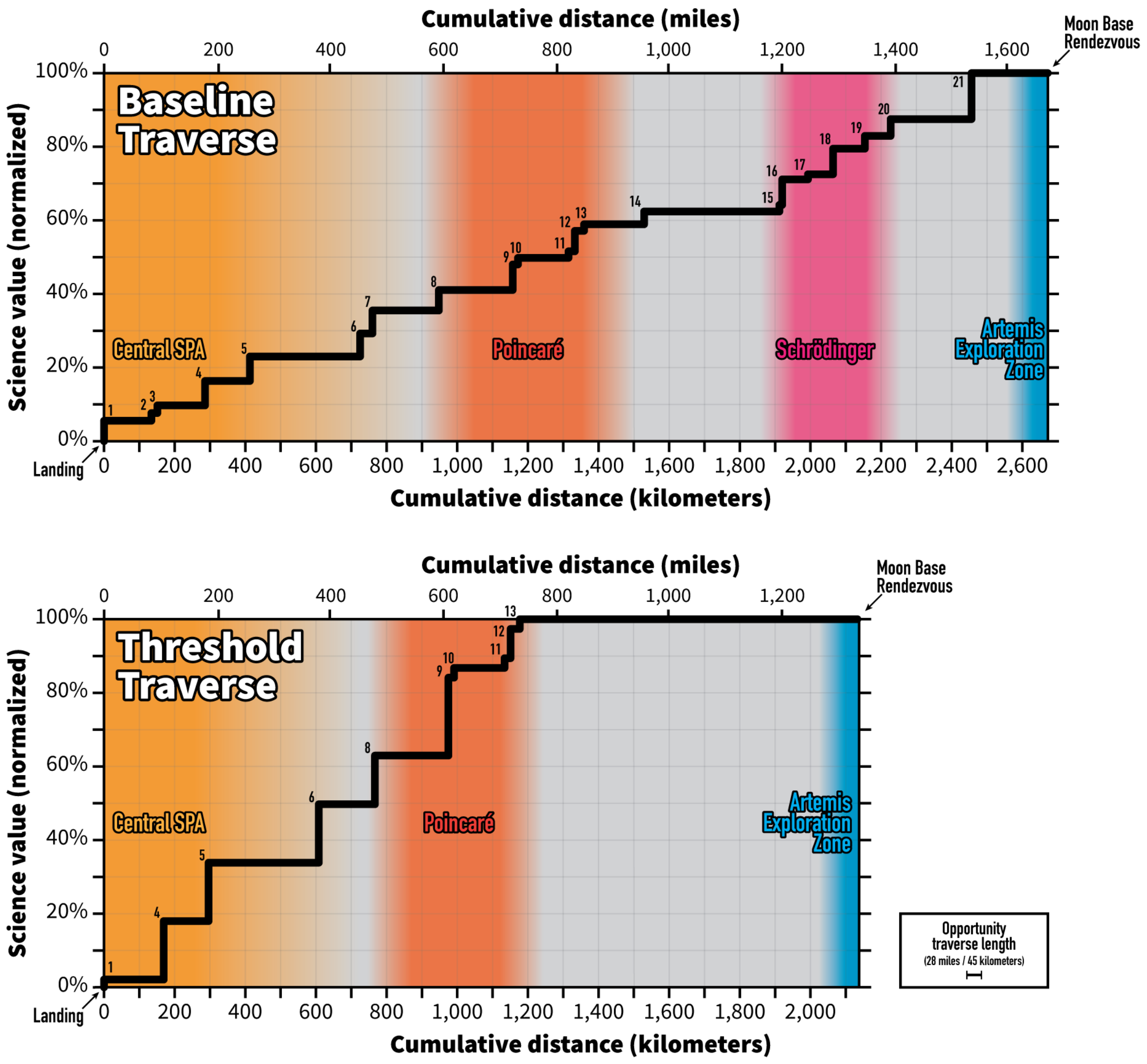

**Figure 5.8.** *Illustration of notional science value as a function of distance for the baseline (top) and threshold (bottom) traverses. Cumulative science value was determined by assessing how each site contributed to each objective. Some sites have more science value in this context, as analysis of samples from those sites may address more than one objective. This is chart is intended only for illustrative purposes and should not be used to make mission design decisions—which instead require evaluation of individual sites and traverses compared to the SPARX requirements.*

### 5.3.2 Sample Mass and Characteristics

At its core, SPARX is a sample return mission, and mission success is only achieved if SPARX delivers its samples to Earth for analysis in terrestrial laboratories. As shown in Table 5.4, all three SPARX Science Goals flow to required analyses of returned samples. Of the 14 SPARX Science Experiments defined by the SDT, 12 require the collection and return of samples. The two Science Experiments that do not require sample return are largely geophysical in nature, and even in those cases, sample return is still enhancing for the interpretation of the geophysical measurements. Consequently, the SDT considered several Figures of Merit reflecting the return mass and characteristics (size, lithology) of the potential samples.

| Goals | Objectives | Experiments | Are Samples Required? |
|---|---|---|---|
| (G1) What was the chronology of key processes and events during the first half billion years of solar system history? | (O1.1) Quantify the age of the Moon to distinguish between early- and late-Moon formation models | (E1) Determine the composition and age of the anorthositic farside highlands crust | Yes |
| | | (E2) Determine the age of SPA impact event | Yes |
| | | (E3) Determine the model age of the mantle underlying SPA | Yes |
| | (O1.2) Distinguish between fast- and slow-cooling models for lunar magma ocean crystallization | (E1) Determine the composition and age of the anorthositic farside highlands crust | Yes |
| | | (E2) Determine the age of SPA impact event | Yes |
| | | (E3) Determine the model age of the mantle underlying SPA | Yes |
| | (O1.3) Determine if and when a giant planet instability drove a sharp change in the inner Solar System impactor flux | (E2) Determine the age of SPA impact event | Yes |
| | | (E4) Determine the age of ≥1 farside Imbrian-aged impact crater (D≥80 km) or basin (D≥300 km) with surface dateable by crater counts | Yes |
| | | (E5) Determine the age of ≥1 non-SPA farside pre-Imbrian-aged impact basin (D≥300 km) | Yes |
| | (O1.4) Constrain the evolution of the lunar dynamo in the pre-4.0 Ga epoch of lunar history | (E6) Determine the age and remnant magnetization signatures from ≥3 returned samples with inferred ages ≥4.0 Ga | Yes |
| (G2) Does the lunar farside preserve evidence of asymmetric magma ocean evolution and planetary crustal formation processes? | (O2.1) Determine if the farside highlands crust is similar to nearside crust in age and/or composition | (E1) Determine the composition and age of the anorthositic farside highlands crust | Yes |
| | (O2.2) Determine if there was a farside KREEP reservoir, or other late-stage magma ocean products, prior to the SPA impact | (E7) Determine the composition and age of farside Th-rich materials | Yes |
| | (O2.3) Determine if and how the temporal and compositional evolution of the farside mantle differs from the nearside | (E8) Determine the composition and age of farside basaltic units | Yes |
| | | (E9) Sample hypothesized excavated lunar mantle material | Yes |
| | | (E10) Determine the volatile content of the farside mantle as recorded by pyroclastic and basaltic materials | Yes |
| (G1) What are the effects of mega-basin impacts on terrestrial bodies? | (O3.1) Determine whether there are discontinuities in the structure (crustal thickness, density, magnetic field) across SPA to constrain models for the formation of a mega-basin | (E11) Determine the crustal structure (e.g., thickness, density, remnant magnetic field strength) across the SPA basin (from the center to the putative rim) to constrain mega-basin formation and evolution models | No (enhancing) |
| | | (E12) Determine if the massifs in the southwestern portion of the basin constitute the diffuse rim of SPA | No (enhancing) |
| | (O3.2) Distinguish the relative contributions of impact melt differentiation, crustal backfilling, and volcanic infilling to the interior composition and stratigraphy of SPA | (E13) Determine the stratigraphy and composition of the unique deposits in central SPA | Yes |
| | | (E14) Determine the origin and extent of the low-calcium pyroxene annulus enclosing central SPA | Yes |

**Table 5.4.** *Simplified version of the SPARX Science Traceability Matrix (STM) demonstrating that sample return is required to address SPARX's science. More detailed STMs are in Foldouts 3.2, 4.1, 4.2, and 5.1.*

### 5.3.2.1 Total Sample Mass

For a sample return mission, the total mass of required samples is singularly driving. Table 5.5, and FO: Sample Requirements (Section 4.3.6) provides a summary of the total sample mass requirements for SPARX, which are justified further in Section 4. **For SPARX to achieve mission success, a baseline of ≥20.5 kilograms (threshold ≥9.8 kilograms) of lunar samples must be returned to Earth. Mission architectures that cannot meet these requirements were deemed not viable.** These sample masses are minimum values, and considered unmargined, as discussed in Section 4.4. The SDT recognizes and emphasizes that real world sample containment and return systems should margin for a greater return mass, approximate scaling for which is shown in Section 4.4, Figure 4.38. The SPARX design reference mission (Section 6) accounts for this by designing for a return mass >100 kg. However, the SDT notes that sample masses consistent with this return mass would be a "stretch" goal, and this sample mass is not required to achieve the SPARX baseline science, nor does this report discuss the strategies for acquiring this additional sample material either through additional traverse stops or non-targeted "science of opportunity" stops.

| | | |
|---|---|---|
| **SPARX Baseline:** | **≥20.46 kg** | Total from across 20 sample campaigns |
| **SPARX Threshold:** | **≥9.82 kg** | Total from across 10 sample campaigns |

**Table 5.5.** *Summary of SPARX sample mass requirements, which are described further in Section 4. These sample mass requirements are exclusive of sample containers (i.e., tare).*

**Finding: the SPARX baseline mission requires an implementation that can return ≥20.5 kg of lunar samples, and the threshold implementation must be able to return ≥9.8 kg of sample material.**

### 5.3.2.2 Sample Mass, Sizes, and Lithologies, and Per Site

As described in the previous figure of merit (Section 5.3.1), SPARX's samples must be returned from distinct locations across SPA. The baseline number of sample sites is 20 (threshold 10), and the sample requirements for each site are variable, driven by the Experiment requirements. All 20 required sample sites can be distilled into 6 distinct "classes" (A–F) summarized in Table 5.6, with classes reflecting a distinct combination of sample collection mode, and the numbers and sizes of rocks required to meet the Experiment requirements. The total sample mass ranges from between ≥0.48 to ≥2.24 kilograms for each sample site.

**The SDT emphasizes that simply collecting a large mass of lunar samples is not sufficient to meet SPARX's Science Objectives; the samples must be of the correct size, lithology, and from the appropriate geologic setting.** These requirements vary between each SPARX Science Experiment as described in Section 4.1, FO: Sample Requirements (Section 4.3.6), and are captured in a simplified form in Table 5.6. The information defined by each of the sample classes in Table 6 can be used to outline the requirements of a successful sample collection system and ConOps. Two examples of this are provided below:

- Science Experiment E1 (Sample Class A): This experiment is focused on understanding the early anorthositic crust of the Moon; therefore, unsurprisingly, this experiment requires the collection of anorthositic rocks. Past experience with laboratory analyses of lunar anorthosites has proven to be extremely challenging (Section 4.3.3.2), and the best results have been achieved with a small number of relatively large samples (≥40 millimeters in diameter). In this case, the size and lithology of the rocks are driving ("quality over quantity"). This combination of factors also drives requirements for *in situ* measurements to identify the correct rocks (Section 4.2) and some mobility within a given sample site (described further in Section 6.6) in order to pick individual rocks (i.e., "targeted" sampling).
- Science Experiment E2 (Sample Class C): This experiment is focused on determining the age of the SPA impact, requiring the collection of SPA impact melt. Unfortunately, it is difficult to confidently identify impact melt on the surface of the Moon—even with *in situ* remote sensing

instruments—as impact melt can appear similar to other lithologies (e.g., other igneous rocks) (Section 4.2–4.3). Past experience has demonstrated that in this case, it is less important to target certain lithologies, and more important to collect large numbers of smaller rocks from the regolith in strategically chosen geologic settings (i.e., "quantity over quality"). By collecting large numbers of smaller rocks (hundreds to thousands, as noted in Table 5.6) terrestrial laboratories can leverage statistics to address the motivating science experiment (Section 4.3.3.3). Furthermore, the types of laboratory methods to be implemented here are more amenable to analyses of small samples than other experiments (such as E1).

**Table 5.6.** *Simplified table of SPARX sample requirements. For more detail, refer to the Science Traceability Matrices in Foldouts 3.1 and 4.2, and Science Experiment descriptions sections (4.1.1–4.1.14). These values already account for factors associated with curation and sample quality.*

| | | Class A | Class B | Class C | Class D | Class E | Class F |
|---|---|---|---|---|---|---|---|
| | **Number of samples sites of this class (baseline)** | 3 | 2 | 9 | 2 | 3 | 1 |
| | **Science experiments** | E1 | E9 | E2, E4, E5, E13 | E7, E14 | E3 | E10 |
| | **Driving lithology** | Anorthosite | Low-Ca Pyroxene, Olivine | Impact Melt | (varies) | Basalt | Pyroclastics |
| | **Sample collection modality** | Targeted scooping and sieving | | Untargeted scooping and sieving | | | |
| **Requirements per site** | **Number of ≥40 mm rocks** | ≥12 | — | — | — | — | — |
| | **Number of ≥20 mm rocks** | — | ≥12 | — | — | — | — |
| | **Number of ≥8 mm rocks** | ≥32 | — | ≥96 | ≥64 | — | — |
| | **Number of ≥5 mm rocks** | — | — | — | — | ≥100 | — |
| | **Number of ≥3 mm rocks** | — | ≥800 | ≥1,200 | ≥800 | ≥500 | — |
| | **Bulk regolith (including fines)** | ≥400 g | ≥400 g | ≥400 g | ≥400 g | ≥400 g | ≥400 g |
| | **Total mass** | ≥2,245 g | ≥800 g | ≥971 g | ≥780 g | ≥484 g | ≥400 g |

**Finding: The SPARX sample acquisition and caching system must be able to: (1) Acquire samples specifically targeted from *in situ* measurements, (2) acquire specified numbers of samples within certain size ranges, (3) acquire un-sieved regolith samples.**

#### 5.3.2.3 Sample Integrity, Contamination, and Cross-Contamination

Preserving the scientific value of SPARX's returned samples requires adherence to guidelines for maintaining the sample's integrity, minimizing contamination from the spacecraft, and minimizing cross-contamination between samples.

In terms of contamination, a driving architectural concern is avoiding chemical and physical contamination from any spacecraft thrusters. During landing (or launching), rocket exhausts can scour the lunar surface and deposit chemical byproducts which may impact the scientific analysis or even be hazards for humans handling the samples. In general, it is desirable to collect samples at some distance from landers. The SDT did not define a specific distance threshold between a lander and sample sites, as the details depend strongly on the details of the lander and the landing scenario. The size of lander blast area may be approximate by comparing lander size and blast zone relationships from past studies (e.g., Clegg-Watkins et al. 2016). Additional sources of contamination could include lubricants in gears and motors.

In terms of cross-contamination between sample sites (i.e., the amount of sample material from one site that can be mixed into another site's sample material), the SPARX requirements are generally minimal. The lunar regolith is, itself, already partially cross-contaminated by impact gardening, a process that can mix materially laterally and vertically over great distances. Therefore, the SDT did not define specific inter-container contamination thresholds, although ≤1% by volume was assumed to be an appropriate criterion that would not substantially drive the mission architecture, while still preserving the unique character of each sample. Some cross-contamination can be tolerated especially if the cross-contamination is documented during the mission (e.g., the vehicle sees a rock bouncing into the wrong sample container). Finally, although not a contamination concern, the SDT notes that depending on container design, there is a potential for sample degradation and "rock tumbling" to occur as a result of mission activities on the surface; therefore, the identified requirement for the collection of unsieved regolith at each sample location is strongly recommended to preserve the sieved rock components. This is in addition to the scientific requirements for unsieved regolith.

While SPARX may not have the most stringent sample integrity and contamination requirements compared to other missions, it is important to note that the samples would ultimately be curated on Earth alongside other lunar and planetary samples. For SPARX's samples to be curated, they should meet the same sample integrity and contamination requirements as other lunar samples.

**Finding: SPARX requires that samples be acquired outside the contamination zone of landing rocket exhaust. Cross-contamination between acquired samples should be kept to ≤ 1% by volume, and all mission-level sample integrity and contamination requirements should be commensurate with NASA lunar sample curation requirements.**

### 5.3.3 *In situ* Measurements and Geologic Context

While sample return is an essential component of SPARX, acquiring *in situ* measurements is essential for providing geologic context, sample region identification, and for addressing Science Experiments that do not strictly require samples (such as E11 and E12). As outlined by the SDT in Section 4.2, *in situ* measurements fall into two broad categories: sample site characterization and traverse science.

#### 5.3.3.1 Sample Site Characterization

Anytime a geologist is handed a rock and asked to identify it, one of the first questions the geologist might ask is "where did it come from?" Geologic context—the observations and understanding of the geology and environment from which samples are acquired—is crucial for properly identifying, characterizing, and studying returned samples. Without geologic context, the ability to understand the scientific provenance of a sample is severely hindered.

The SPARX mission includes requirements to characterize sample sites with a range of *in situ* measurements, including context imaging (at both regional (M1) and fine-scale (M6)), mineralogy (at both regional (M2) and fine-scale (M3)), and determination of the bulk elemental chemistry of the (M4), the gravity field at the site (M5) and the magnetic field (M7).

> **Finding: SPARX implementations must acquire sufficient *in situ* measurements at each site to adequately characterize the geologic environment.**

#### 5.3.3.2 Traverse Science

Several SPARX Experiments require non-sampling *in situ* measurements to be taken at specific spatial intervals within the region. Notionally described as "along a traverse", these measurements are required to characterize the changes in mineralogy, elemental chemistry, gravity field, and magnetic field that are occurring at scales finer than have been resolved by orbital remote sensing instruments. Additionally, these regional scale measurements provide further geologic context for individual sample sites, and significant benefits to opportunistic and future science investigations that cannot be conceived of prior to the mission.

> **Finding: SPARX science objectives require the periodic acquisition of *in situ* geologic context measurements in-between major regions of interest and designated sampling sites.**

### 5.3.4 Cost, Complexity, Risk, and other Programmatic Factors

Finally, any viable mission implementation must fit within programmatic constraints, including cost, schedule, and a NASA-defined risk posture. The SPARX TOR (Section 2, APPENDIX: TOR) describes many of these constraints. In general, architectures that minimize the number of distinct spacecraft and spacecraft interfaces are preferred, as are architectures that leverage existing or planned infrastructure. In some cases, combinations of architectures can provide uniquely high science value per dollar, such as pairing robotic mission with planned Artemis crewed missions. In some cases, there are trades between technical risk and programmatic risk that must be evaluated. For example, utilizing astronaut-facilitated sample return trades the cost/technical risk associated with robotic sample return for programmatic risk associated with linking with Artemis's schedule, landing site selection, and payload capacities of future Artemis architecture. The SDT evaluated these figures of merit using both past experience and quantitative analysis where possible.

**Finding: The SPARX implementation architecture should optimize science return within a framework of minimized cost, complexity, and overall mission risk.**

## 5.4 MODES OF SAMPLE RETURN

**A core aspect of SPARX is sample return, and mission success is only achieved if SPARX delivers its samples to Earth for analysis in terrestrial laboratories.** Therefore, the details of how the samples are returned to Earth is a critical component of the overall mission architecture. The SDT was tasked with investigating two potential avenues of sample return using either Artemis astronauts at the South Pole, or commercial-derived robotic sample return capabilities. During the course of the study, the scope was expanded to include Artemis astronaut return from a non-polar destination. Per the SPARX TOR (APPENDIX: TOR), astronaut-assisted sample return implementations shall not exceed 100 kg of returned samples (inclusive of sample containers), while commercially derived (robotic) sample return capabilities shall not exceed 10 kg of returned samples.

Over the past half-century, many missions have performed extraterrestrial sample return (as summarized in Figure 5.9), all utilizing vastly different mission architectures. In the following sections, we examine the modes of sample return in more detail, including examining past missions and concepts, and examining the viability of the assumptions defined in the SPARX TOR.

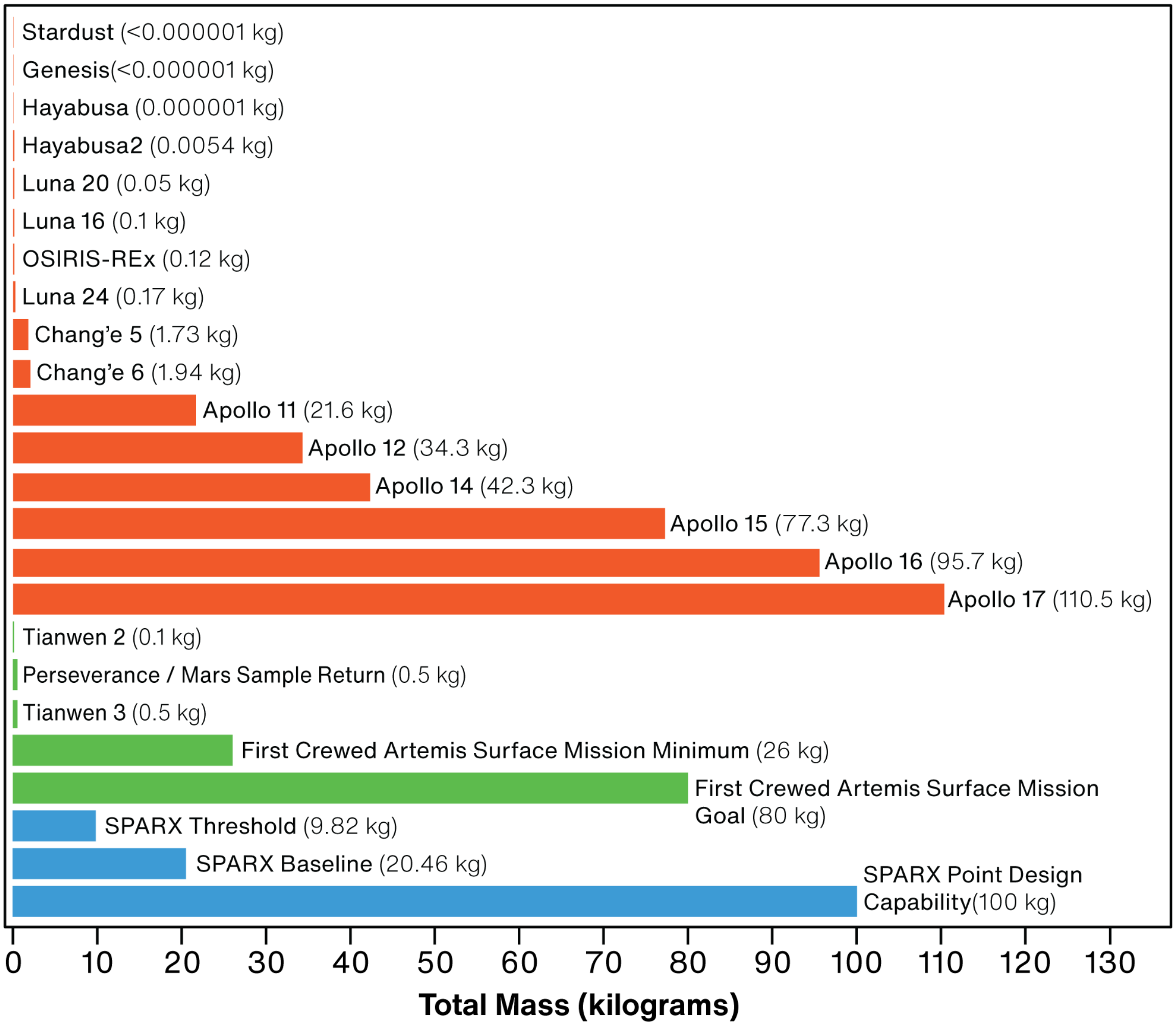

**Figure 5.9.** *Summary of the total returned sample mass from past, current, and future missions. Completed missions are shown in orange; active missions or missions in development are shown in green; SPARX's requirements are shown in blue.*

### 5.4.1 Robotic Sample Return

One approach to sample return is via uncrewed, robotic missions. This has been accomplished a small number of times on the Moon by the USSR in the 1970s (Luna 16, 20, and 24) and China in the 2020s (Chang'e 5 and 6, Figure 5.10).

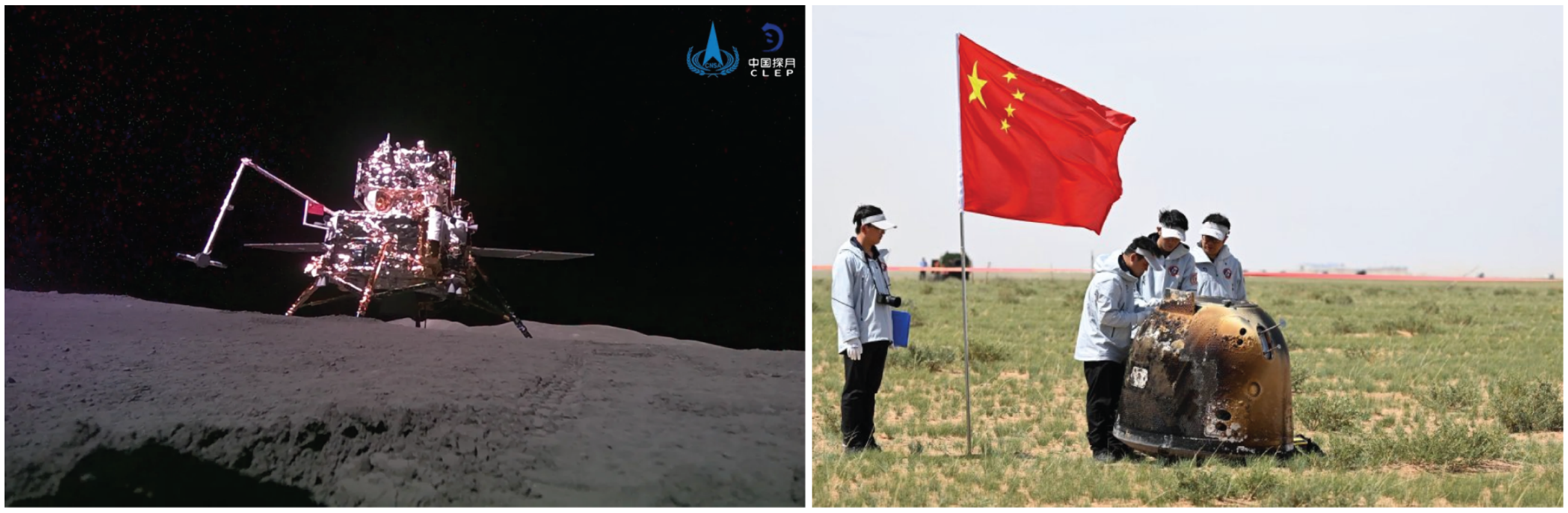

**Figure 5.10.** *Chang'e 6 was the most recent, successful robotic lunar sample return mission. The image on the left shows the Chang'e 6 lander on the surface of the Moon, while the image on the right shows the Chang'e 6 sample return capsule landed on Earth. Images by CNSA.*

The United States has never attempted robotic sample return from the surface of the Moon; nevertheless, robotic sample return concepts have been studied by NASA for decades. Two notable examples include the MoonRise concept for SPA sample return which has been proposed multiple times for NASA's competitive New Frontiers program (Figure 5.11; Jolliff et al. 2012, 2021), and the Endurance-R concept developed for the most recent Planetary Science and Astrobiology Decadal Survey (Figure 5.11; Keane et al. 2022). Despite variable science goals, most architectures for lunar sample return have common elements, including: (1) a lunar lander; (2) *in situ* instruments to select and contextualize samples; (3) a sample collection system (e.g., arm with a scoop or sieve); (4) a sample stowage system; (5) a lunar ascent vehicle to launch the samples either into lunar orbit (where it may rendezvous with another spacecraft) or directly to Earth; and (6) a sample return capsule (SRC) designed for Earth entry and landing. The large number of elements, distinct spacecraft, and interfaces usually results in a complex and costly system.

The SPARX SDT was not tasked with explicitly investigating or developing the elements of a robotic return architecture beyond assessing the sample return viability given the SPARX requirements. Therefore, the SDT directs those interested in the nuances of how such a robotic return system might interface with a SPARX-like mission to the Endurance-R concept study. The Endurance-R concept study presents a point design that accomplishes aspects of a SPARX-like mission, but highlights some of the challenges with robotic sample return (Keane et al. 2021) observed in previous generation concepts, such as MoonRise. The concept study report included a detailed reference design for a CLPS-facilitated, robotic sample return architecture. In this concept, a CLPS lander delivered elements to the lunar surface for collecting the sample cache from the rover (the "sample transfer link") and transferring them to an Earth Return Vehicle (ERV) and Sample Return Capsule (SRC). While this concept focused on collecting a cache of samples from a rover, the sample transfer link could be replaced with a slightly more sophisticated arm and sample collection system, similar to the system initially proposed for Mars Sample Return. The Endurance-R concept study highlights several elements that are relevant for investigating the viability of commercially derived robotic sample return implementations for SPARX.

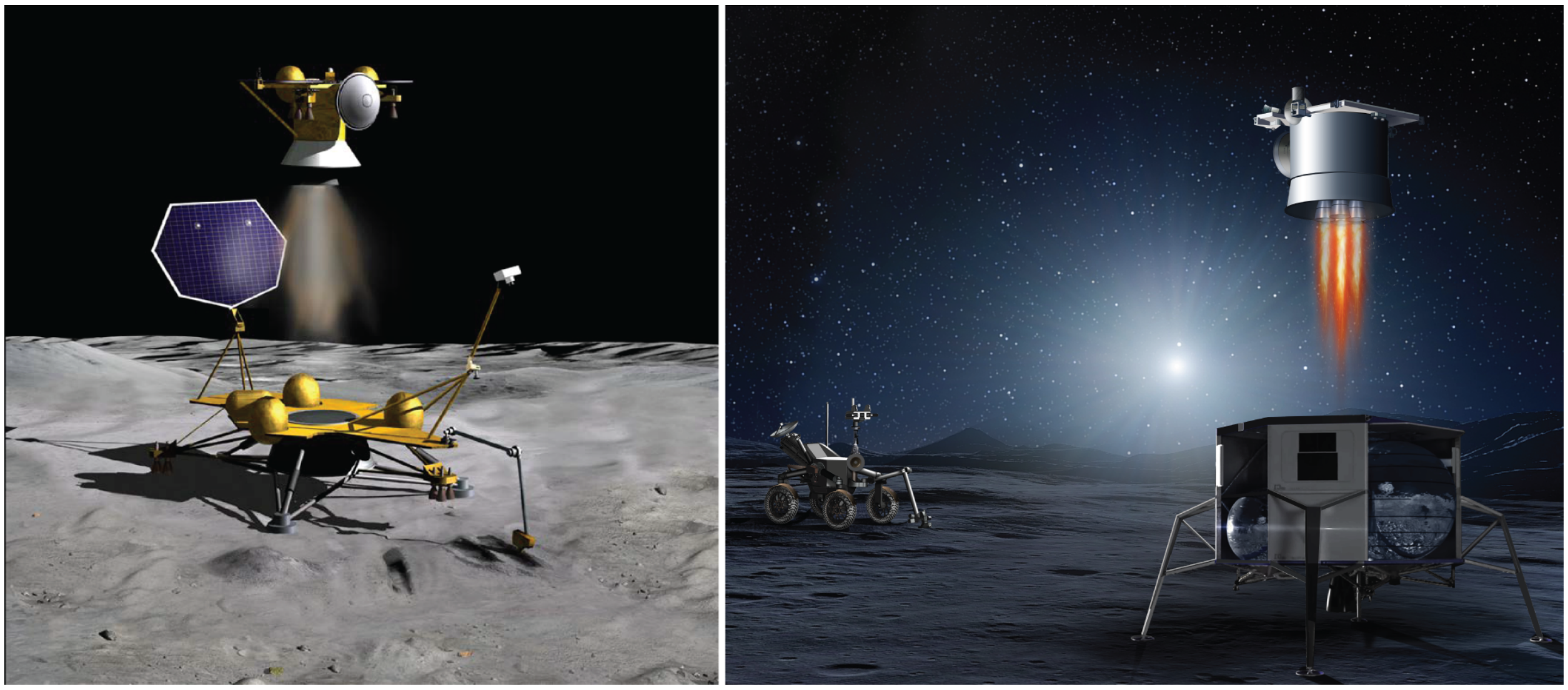

**Figure 5.11.** *Two past concepts for robotic sample return from SPA: MoonRise and Endurance-R. Left: Artist concept of the New Frontiers MoonRise concept for robotically collecting approximately 2 kilograms of samples from SPA and returning them to Earth (Jolliff et al. 2012). The MoonRise concept evolved over the course of many proposals, and this artist rendering shows an incarnation that included a robotic lander with sample collection arm, cameras, and Earth Return Vehicle (shown launching off the surface of the Moon), plus a Sample Return Capsule that would ultimately enter Earth's atmosphere and land the samples. Right: Artist concept of the Endurance-R mission concept developed by the Planetary Science and Astrobiology Decadal Survey (Keane et al. 2021).*

It is not within the purview of the SDT to prognosticate on future commercial capabilities; however, we will briefly highlight three areas of SDT concern surrounding robotic sample return implementations for SPARX.

1. **Sample Return Capsule (SRC) Capabilities**: Samples being returned to Earth robotically must reenter the Earth's atmosphere in some sort of SRC. Current generation SRC, like those flown on OSIRIS-REx (Figure 5.12), have limited size, and a payload capacity for approximately 2 kg of samples. At present, no produced SRC meets either the SPARX TOR assumptions for robotic sample return (10 kg), nor the unmargined SPARX science requirements for sample return (threshold ≥9.82 kg, baseline: ≥20.46 kg; Section 4.4). While it is possible to develop larger SRCs, this would require new technology development, and the increased mass would ripple into the rest of the sample return system design (e.g., exponentially increasing the size of the lunar ascent vehicle). Although such considerations are not within the costing purview of the SDT, it is nevertheless strongly noted.
2. **Demonstration of commercial lunar launch capabilities:** While significant strides have been made in commercial lunar exploration, and many commercial companies are anticipating and developing systems to return materials from the surface of the Moon, none have currently done so. The advent of NASA's Ignition and Moon Base programs may have the ability to rapidly grow and support this capability. The SDT strongly urges NASA to consider viable demonstrations of commercial lunar launch and return for any proposed system prior to the SPARX sample return owing to the critical nature of sample return to SPARX mission success.

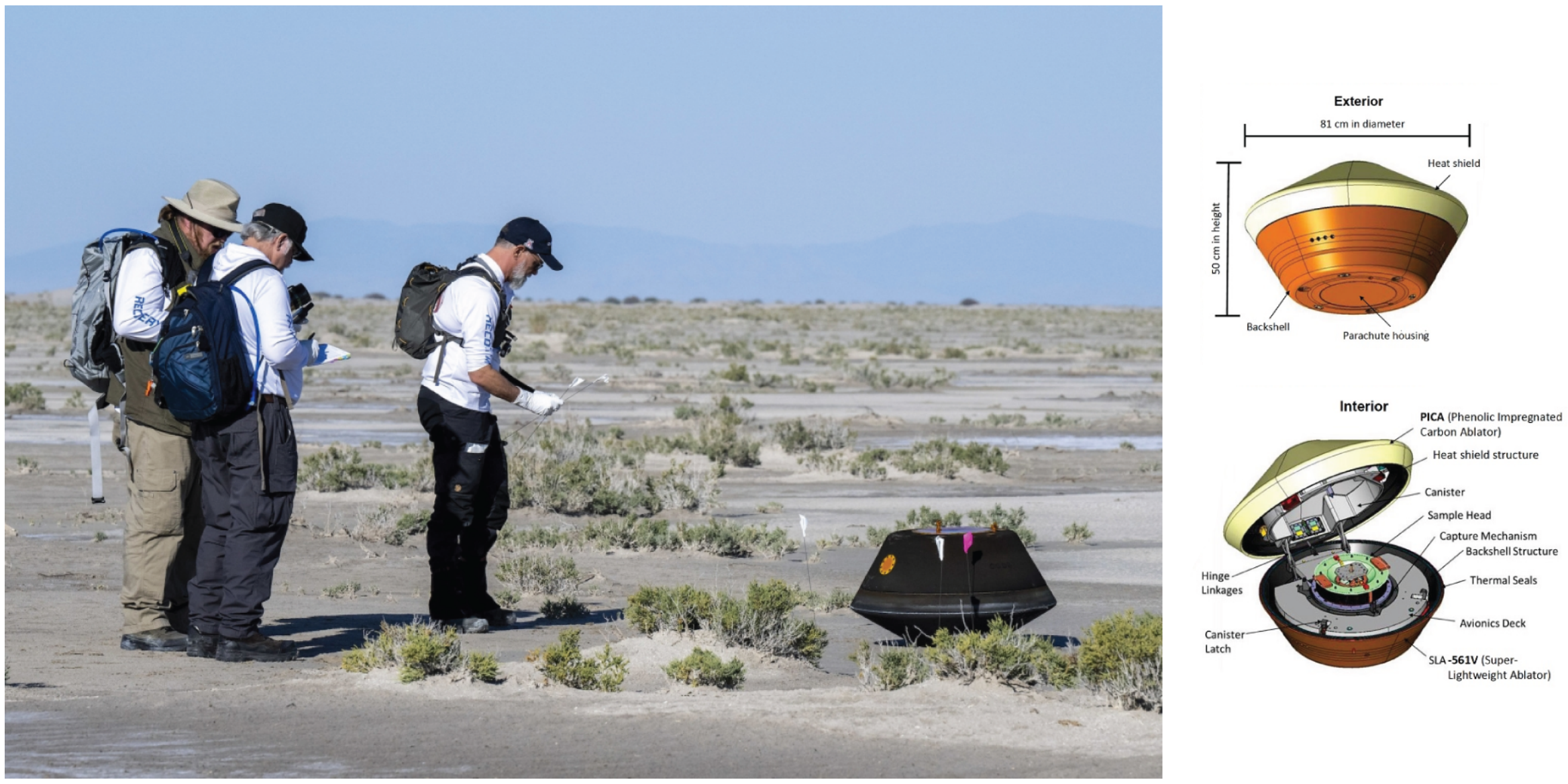

**Figure 5.12.** *The OSIRIS-REx Sample Return Capsule (SRC) represents the current paradigm for sample return. Left: image of the OSIRIS-REx team retrieving the SRC in the Utah Test and Training Range. Image credit: Lonnie Shekhtman (NASA) and Daniel Stolte (University of Arizona). Right: schematic diagram of the OSIRIS-REx SRC. Image credit: Lockheed Martin.*

3. **Historic cost and cost growth of robotic sample return missions:** Historically, sample return from SPA has been considered as a New Frontiers (~$1B cost cap) level mission. Endurance-R provides the most current cost estimate for a robotic sample return that partially leverages CLPS (Keane et al. 2021). At the time of the study, the CLPS program office reported an estimated cost for such a CLPS lander at $200M (FY25). The JPL Team-X study for the other sample return elements estimates a total cost for the sample transfer link, ERV, SRC, and operations at $716M (FY25). There are several important caveats with both of these costs: this costing was performed early in the CLPS program history, the sample return elements were designed and costed as Class-B components (appropriate for a New Frontiers class mission), sample transfer was accomplished via a 1-degree-of-freedom transfer link, the returned sample mass was ~2 kg. Consequently, it is anticipated that costs for sample return of the magnitude required by SPARX would significantly exceed the Endurance-R values. The SDT notes that if NASA, industry, or international partners developed high-mass (≥10 kg, exclusive of sample containers) robotic sample return capabilities at low cost, this would change the paradigm and open up more viable, robotic sample return options for accomplishing SPARX's science objectives.

**Finding: The SDT finds that while the unmargined SPARX threshold sample return mass (9.8 kg) is technically below the TOR recommended commercially derived robotic NTE value of 10 kg, many practical challenges remain before recommending this implementation system. The SDT chiefly suggests that NASA consider the current and near-future capabilities, and capability demonstration of commercially derived sample return systems against the SPARX requirements.**

### 5.4.2 Astronaut-Facilitated Sample Return

NASA's Artemis program will return astronaut explorers to the surface of the Moon for the first time in over half a century. The return of crew to the lunar surface affords the opportunity for crew to collect large sample masses and return them to Earth. At the time of the writing of this report, the details of Artemis infrastructure are not fixed, but there are two key elements (Figure 5.13):

- Human Landing System (HLS): The HLS is the vehicle that will land astronauts on the lunar surface and safely return them to lunar orbit. At present, two companies are developing HLS spacecraft: SpaceX (Starship) and Blue Origins (Blue Moon). The final capabilities of these vehicles for Artemis missions are uncertain; however, per the SPARX TOR, a NTE sample mass (inclusive of sample containers) of 100 kg is proposed.
- Orion: Orion is the spacecraft that takes astronauts from Earth to lunar orbit, where they can transfer back and forth between HLS, before ultimately returning to Earth in Orion. At present, Orion has the capability of returning payloads (e.g., samples) ≤100 kilograms (e.g., NASA 2023). In the current architecture, Orion is the only mechanism by which samples are returned to Earth and thus represents the near-term "bottleneck" in total sample mass. This may evolve as Orion and the overall Artemis infrastructure evolves.

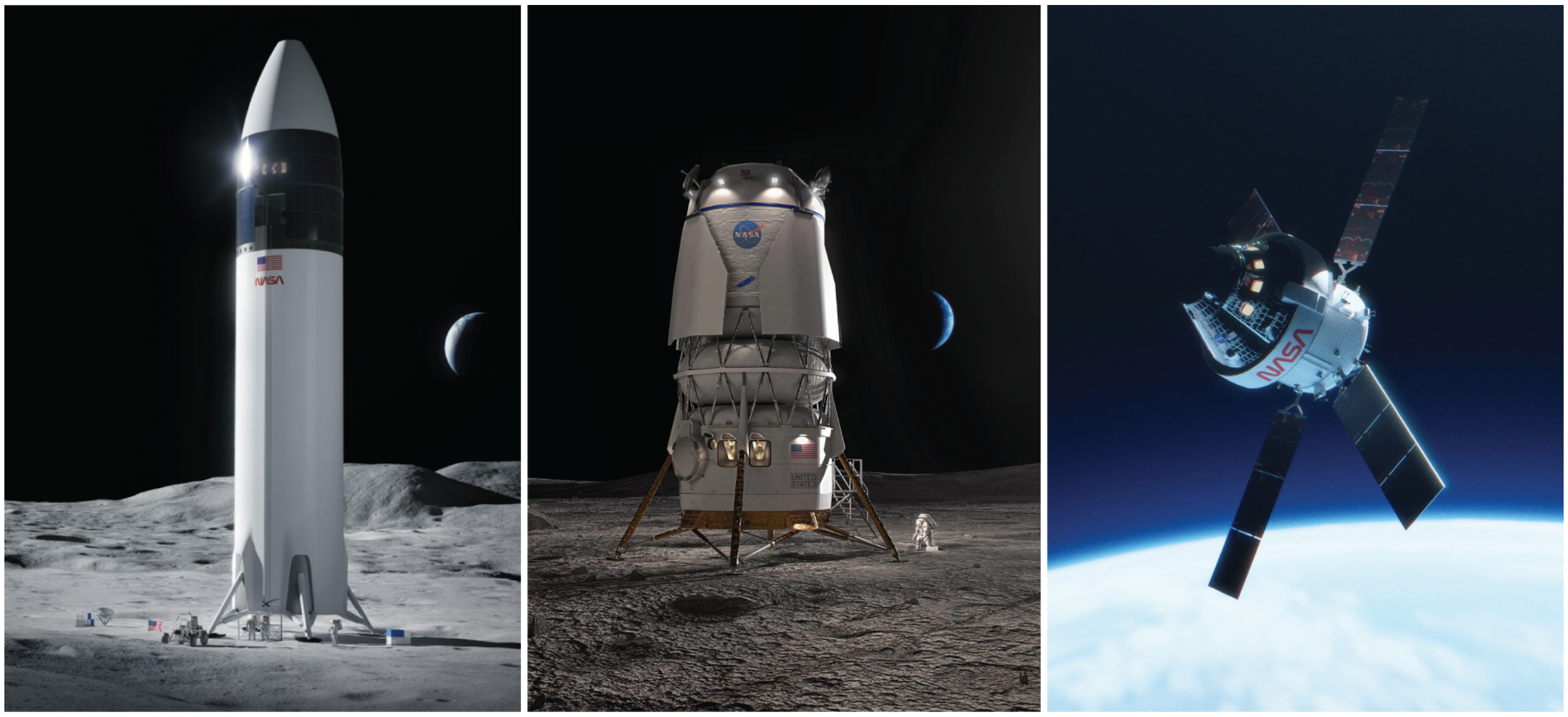

**Figure 5.13.** *The current Artemis infrastructure under development by NASA. Left: Concept art of the SpaceX Starship Human Landing System (NASA / SpaceX). Middle: Blue Origins Blue Moon Human Landing System (NASA / Blue Origins). Right: Orion spacecraft (NASA). None of the images are to scale.*

**While it is important to emphasize the uncertainty with the Artemis architecture, what is clear is that astronaut-facilitated sample return affords an order-of-magnitude increase in the total returned sample mass compared to robotic sample return approaches. The SPARX baseline sample mass requirement is within the perceived and listed capabilities of a single near-term Artemis mission.** SPARX's baseline (20.5 kg) and threshold (9.8 kg) sample mass requirements both fit within the HLS payload requirements and are consistent with reasonable first-order assumptions for possible up-mass when accounting for sample container and other associated masses (i.e., tare, described further in Section 4.4). The SDT notes that larger sample return masses comparable to, if not exceeding, what was achieved by

later Apollo missions like Apollo 17 which returned ≥110 kg of samples is strongly desired if the availability of return architecture assets supports it. This capability would allow greater flexibility in returning Artemis-specific samples, in addition to, the SPARX samples in a single mission architecture. Decisions surrounding the acquisition of additional samples, or sample mass beyond the SPARX baseline requirements should be explored by the SPARX Science Team at the time when NASA makes conclusive decisions about the mission architecture.

Astronaut-facilitated sample return also provides the opportunity for astronauts to enhance the science return of the samples, whether through synergistic exploration or sample triage (discussed additionally in Section 7). Astronauts are uniquely poised to perform complex scientific activities while on the surface of the Moon that may be synergistic with the SPARX surface platform. For example, astronauts on extravehicular activities (EVAs) could collect samples or explore regions that are inaccessible by robots. Astronauts could also utilize human perception and *in situ* instruments to triage or high-grade samples (i.e., selecting the high-priority or high-value samples from larger volume of samples), which may be beneficial if Artemis infrastructure cannot return the full volume of samples.

While astronaut-facilitated sample return has many advantages, it is not without its drawbacks. Astronaut-facilitated sample return means that the samples and sample container must be rated for handling by astronauts. Relying on astronaut-facilitated sample return also introduces programmatic risks, as mission success is dependent on future Artemis missions and their capabilities (described further in Section 7), and syncing the sample transfer may introduce schedule pressure. Finally, at present, all Artemis astronaut missions are targeting the lunar South Pole (the Artemis exploration zone, poleward of 84°S), which is far from the SPARX Regions of Interest. Utilizing astronaut-facilitated sample return either requires samples to be delivered to the South Pole, or a non-polar Artemis mission.

**Finding: Sample return by Artemis astronauts is commensurate with the baseline and threshold sample mass required by SPARX. Astronaut architectures have the opportunity to provide additional benefits, however, are dependent on concordant schedules between SPARX and Artemis launches, and permissible astronaut exploration zones.**

### 5.4.3 Approaches Without Sample Return (i.e., *In situ* Methods Only)

As a defined sample return mission (see APPENDIX: TOR), the SDT did not consider mission implementations that did not require at least some level of sample return. Despite the emphasis on sample return, the SDT did rigorously consider non-sample science objectives in developing SPARX. In accordance with these discussions, the SDT defined one science objective and two science experiments that could be accomplished solely through *in situ* measurements. The SDT considered *in situ* measurement techniques that would allow a greater degree of sample analysis on the surface of the Moon; however, these methods were determined to be not sufficient for fully achieving SPARX objectives. Additional discussion of this topic can be found in Sections 4.2, 4.3 and 7.4.

Although not within the purview of the SPARX SDT TOR, it is worth briefly discussing the possibility of expanded sample analysis techniques accomplished via infrastructure as part of a future Moon

Base. While, in principle, the Moon Base may afford more mass and power for the large laboratory instruments, plus crew availability for more sophisticated sample handling, preparation and analysis, there is insufficient information to properly evaluate what fraction (if any) of SPARX's terrestrial laboratory experiments could be achieved at a hypothetical Moon Base. Indeed, the large, complex operations and maintenance of terrestrial laboratory facilities suggests that these types of facilities would not be a near-term priority for the developing Moon Base.

**Finding: The preponderance of science objectives and laboratory measurements required by SPARX necessitate implementation architectures with terrestrial sample return.**

### 5.4.4 Summary

The SDT investigated two broad categories of sample return for the SPARX samples, with TOR guidance focusing heavily on constraints provided by return sample mass. Table 5.7 compares the SPARX total sample mass requirements against the total sample mass capabilities of the different sample return architectures described in this section. This table highlights the main conclusions of the SDT regarding sample return architecture:

(1) Astronaut-facilitated sample return is the only investigated architecture option capable of returning the SPARX baseline sample mass, with additional mass margin available.

(2) Commercially derived robotic sample return, per the TOR guidelines, is potentially viable for the SPARX threshold mission. However, additional work would need to be completed to determine the total return mass (including sample containers, sample return capsule, etc.), and further demonstration of commercial lunar surface launch capabilities for this option to prove viable.

| | | | **Robotic Sample Return** (past precedent ≤2 kg) | **Robotic Sample Return** (SPARX TOR guidelines: ≤10 kg) | **Astronaut-Facilitated Sample Return** (SPARX TOR guideline: ≤100 kg) |
|---|---|---|---|---|---|
| **SPARX Baseline:** | **≥20.46 kg** | Total from across 20 sample campaigns | **X** (≥11 missions required) | **X** (≥3 missions required) | ✓✓ (≥1 mission required) |
| **SPARX Threshold:** | **≥9.82 kg** | Total from across 10 sample campaigns | **X** (≥10 missions required) | ✓ (≥1 mission required) | ✓✓ (≥1 mission required) |

**Table 5.7.** *Summary of SPARX's* <u>*total*</u> *sample mass requirements (described further in Section 4.4), compared to the capabilities for different sample return architectures described here. Two check marks (✓✓) and green indicates that the mode of sample return can return the total SPARX sample mass requirements; one check mark (✓) and yellow that the mode of sample return might be able to return the total SPARX sample mass requirements, depending on the mass of the sample container (i.e., tare); an "X" and red indicates that the mode of sample return cannot meet the total SPARX sample mass requirements.*

## 5.5 Modes of Sample Acquisition

Samples of planetary surface can be acquired in a number of different ways, including drilling, scooping, targeted grabbing, and lofting. Here, we discuss the acquisition system requirements imparted by the SPARX sample requirements and identify Figures of Merit for a SPARX sampling system. As discussed in Section 5.3, the SPARX sample acquisition system must be able to: (1) Acquire samples specifically targeted from *In situ* Measurements; (2) Acquire specified numbers of samples with a range of minimum diameter requirements; (3) Acquire un-sieved regolith samples. With these requirements in mind, we briefly discuss the available modes of sample acquisition.

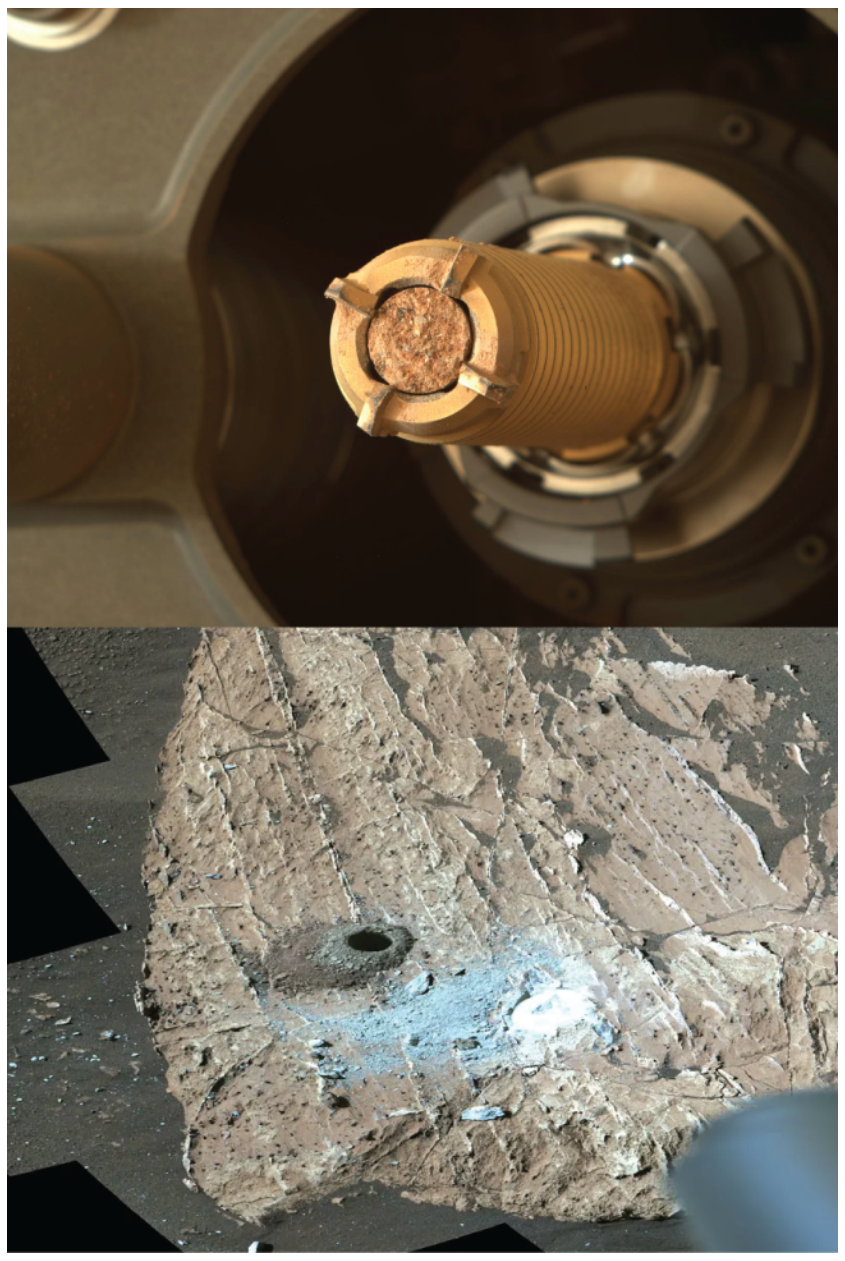

**Figure 5.14.** *Top: Mastcam-Z image of the drill after collecting the "Swift Run" core on the surface of Mars (NASA / JPL-Caltech / ASU). Bottom: Mastcam-Z image of the "Cheyava Falls" workspace, showing a dark hole where the rover acquired a drill sample (PIA26401: NASA / JPL-Caltech / ASU / MSSS).*

Drilling (or coring) has been used as a primary mode of sample acquisition on multiple planetary missions, including Luna 16, 20, 24, Chang'e 5 and 6, as well as NASA's Mars 2020 Perseverance rover (Figure 5.14). The benefits of drilling are that it returns a continuous, or near continuous, sample of material while largely preserving stratigraphy and allowing access to material's not directly exposed to the space environment. However, the exact nature of these materials, including the sizes and compositions of rocks and clasts embedded in the sample core are often unknown until the sample is examined on Earth. The majority of these systems are also only able to acquire a single core. The Perseverance rover uses a specially designed coring and storing system to allow for the acquisition of multiple sample tubes and also uses a complex interior suite of instruments to thoroughly characterize the geologic nature of each sample. Neither of these drilling styles are commensurate with the SPARX sample requirements. Most specifically, the SPARX requirements for specific sizes and numbers of rocks and rocklets to be acquired at each sample site would prove extremely challenging to either meet or verify using a nominal drilling system.

### 5.5.1 Targeted Grabbing

Targeted grabbing, or, the collection of hand-sized rocks, was used effectively by the Apollo astronauts. The large rocks collected often contained a multitude of clasts, allowing for a large number of analyses in terrestrial laboratories. However, much like with drilling, it is almost impossible to verify the presence and nature of these internal clast components prior to examination on Earth. The size and generally irregular conformation of hand-size samples also pose challenges for sample storage systems, and the anticipated volume constraints force a tradeoff between fewer larger samples or multitudes of smaller samples. The nature of the proposed SPARX measurements (Sections 4.1, 4.3) favor multiple investigations of smaller samples, which allow for greater statistics and more robust analyses required for meeting SPARX objectives.

### 5.5.2 Pneumatic Sampling or Lofting

Another method of sample acquisition is pneumatic sampling—the use of pressurized gas to disrupt a regolith surface, lofting grains which can then be captured by a containment system. This method has been used primarily for sample collection on asteroids, for example, by OSIRIS-REx on the asteroid Bennu. The method has also recently been successfully demonstrated on the Moon with the Lunar PlanetVac (LPV) system on Blue Ghost Mission 1 (Figure 5.15) (Zacny et al., 2026). The SDT received a briefing on the LPV system in April 2025, including significant discussion about the system capabilities for post-acquisition sample verification and sample sorting technologies. Ultimately, the SDT determined that the large number of individual sample clasts that must be collected, combined with the multiple different diameter requirements were not well-matched to the LPV or other more traditional lofting methods.

### 5.5.3 Scooping and Sieving

Scooping and sieving offers a low-risk, mechanically simple and robust approach to collecting samples from regolith-rich environments. Scooping and sieving has been implemented on many planetary science missions, including Surveyor, the Mars Science Laboratory rover, Curiosity, the Chang’e 5 and 6 missions, and the Apollo missions (Figure 5.16). The use of sieve gratings can allow for the selection of specific rock sizes if required.

Scooping and sieving is particularly effective for sampling on the Moon. The vast majority of the lunar surface is dominated by regolith—a combination of rocks and fine particles that are thoroughly broken and well-mixed by impact processes. The Apollo surface operations demonstrated that simple scoops, rakes, and sieves allowed the astronauts to rapidly harvest diverse materials that captured the varied lithologies at each site required for many scientific investigations. **The SPARX sample requirements, particularly the requirements concerning the sizes and numbers of rocks (Section 4.3.6, FO Sample Requirements), are readily met by scooping and sieving techniques.**

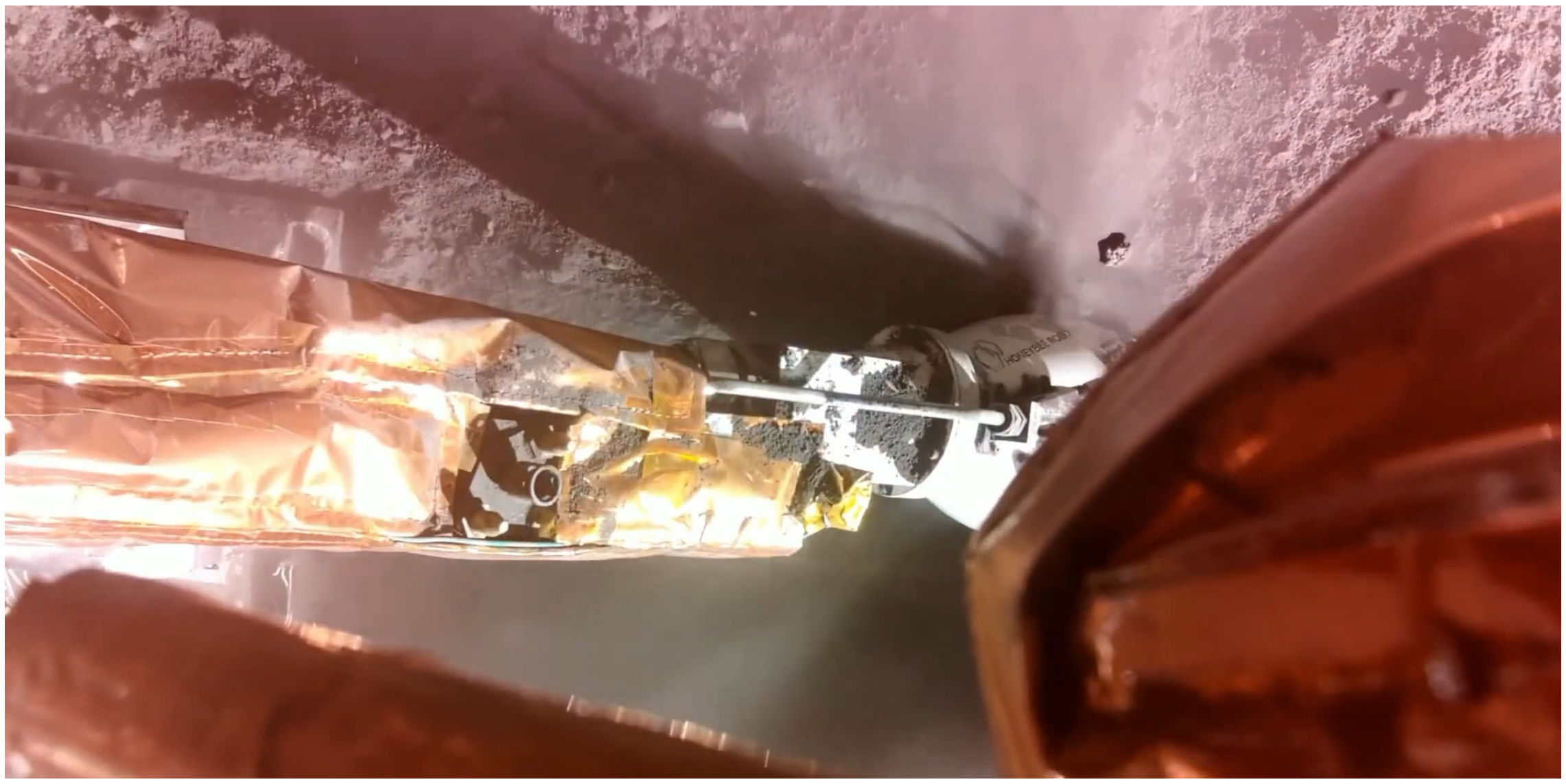

**Figure 5.15.** *Lunar PlanetVac (Zacny et al., 2026) mid sample acquisition during Firefly Aerospace Blue Ghost Mission 1. Dust can be seen entrained in gas emerging from the pneumatic sampler, along with lofted rocks. Image credit: NASA / Firefly Aerospace / Honeybee Robotics, a Blue Origin Company.*

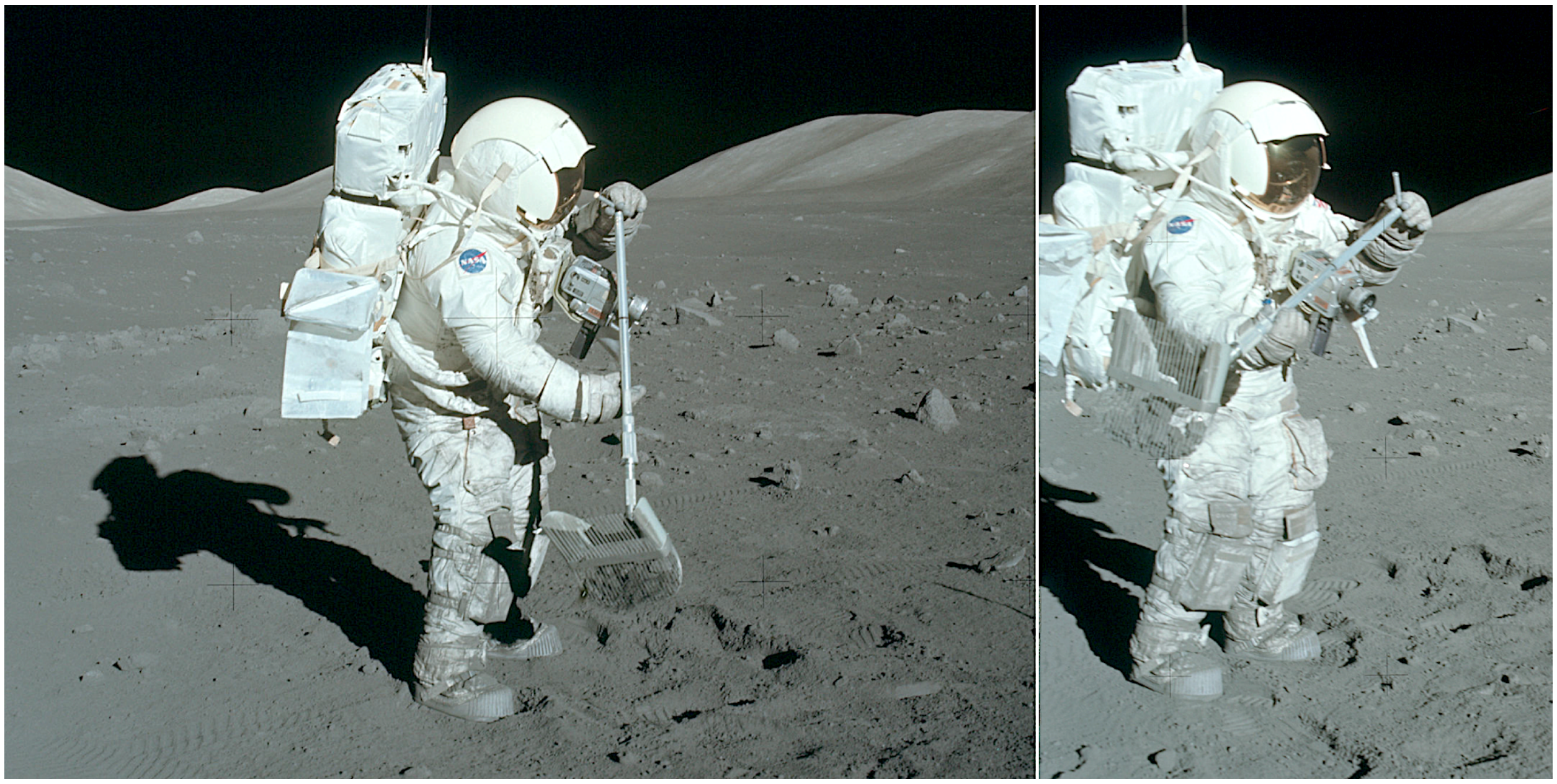

**Figure 5.16.** *Example of scooping, sieving, and raking lunar regolith from Apollo 17. Image credit: NASA (Left: AS17-134-20425; Right: AS17-134-20426).*

### 5.5.4 Local Mobility, Robotic Arms, and Sample Acquisition

In determining further aspects of the sample acquisition system, it is important to determine whether significant mobility is required by the landed asset for acquiring samples, and if so, what degree of mobility is required. The SPARX Sample Requirements (Section 4.3.6, FO Sample Requirements), define specific sizes, numbers, and lithologies of samples that must be collected for each SPARX Experiment. The SPARX surface platform must therefore be capable of acquiring and verifying samples meeting these requirements. **As described above, the sample acquisition requirements for SPARX can be met with relatively simple surface interactions, including scooping and sieving of lunar regolith. The recommendation of this sample acquisition method naturally leads to the recommendation of a robotic arm to enable collection of samples in the vicinity of the SPARX surface asset.**

Robotic arms have been flown on spacecraft for decades—going all the way back to the first Surveyor landers that scouted the surface of the Moon (Figure 5.17). A wide range of robotic arms have been flown on many missions to the Moon and Mars, and their capabilities have ranged from very simple robotic arms that scooped the lunar surface (e.g., Surveyor) to very complicated, high-degree-of-freedom arms with tools and *in situ* instruments (e.g., Perseverance).

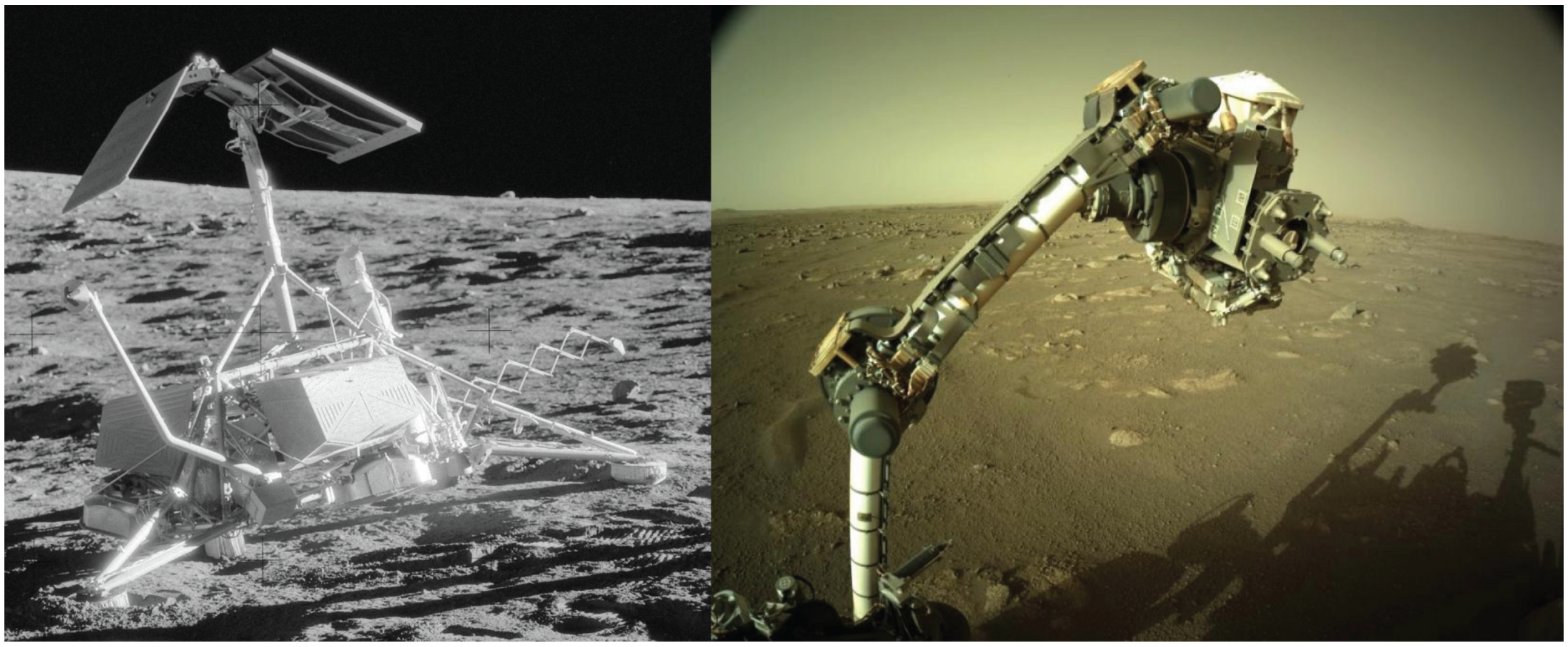

**Figure 5.17.** *Two planetary robotic arms, separated by over 50 years. Left: the Surveyor 3 spacecraft, imaged by the crew of Apollo 12. A simple robotic arm can be seen sticking off the righthand side of the spacecraft. Right: the robotic arm on the Mars2020 Perseverance rover, as imaged by the rover. The Perseverance arm is much more sophisticated, including more degrees-of-freedom and arm mounted tools, drills, and scientific instruments. Image credits: NASA / JPL-Caltech.*

Robotic arms are useful for interacting with material within ~meters of a spacecraft. Thus, in the context of SPARX, robotic arms can be essential for the collection of samples within a given site but clearly do not provide sufficient mobility to address any of the requirements for local-scale mobility at individual sites (10s to 100s of meters), let alone regional-scale mobility for accessing the geologic diversity of SPA (100s to 1,000s of kilometers).

There are many important considerations for designing a robotic arm for sampling. In general, as more capability is added, the cost and complexity increase, as illustrated in Figure 5.16. Key considerations for designing a robotic arm include: (1) the required reach and degrees-of-freedom of the arm; (2) required payloads or tools on the end of the arm; (3) Placement, pointing, and positioning accuracy of the robotic arm and end-effector; and (4) required interactions with the planetary surface, including any requirements for placing the arm within close-proximity, contact, or physically interacting with the surface (e.g., scooping). While it is possible to design extremely capable robotic arms, they can quickly become very high-cost spacecraft components.

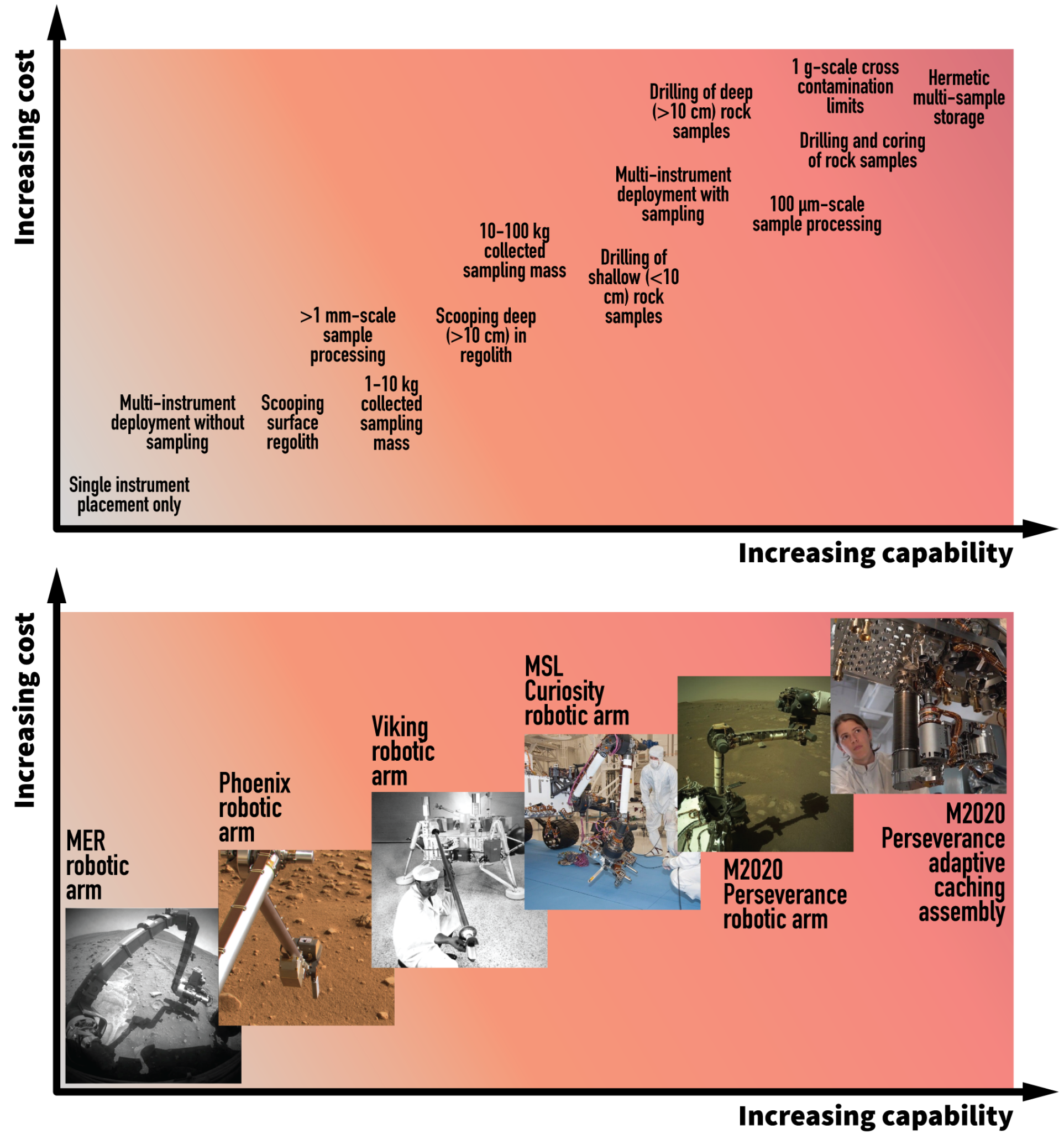

**Figure 5.18.** *Schematic illustration of the cost vs. capability for different robotic arm capabilities (top) and flown examples (bottom). Image credits: NASA / JPL-Caltech / University of Arizona / Texas A&M University.*

Based on the SPARX Science Goals and the sample requirements, the SPARX SDT made two determinations for possible SPARX robotic arms:

**Finding: The SPARX mission can achieve its science requirements with relatively simple interactions with the lunar regolith, such as scooping and sieving. More complicated interactions with lunar regolith, such as drilling, coring, or boulder chipping, are not necessary to address the SPARX science requirements.**

**Finding: The SPARX mission does not require contact or close-proximity measurements and therefore should not assume any arm-mounted science instruments. Excluding arm-mounted science instruments dramatically reduces the requirements on the robotic arm and thus reduce mission cost and complexity.**

## 5.6 MODES OF SURFACE MOBILITY

**The SPARX Science Requirements require surface mobility at both the regional and local scale.** The Science Requirements derived in Sections 4 and summarized in the STM (Section 3.6) and FO Site and Traverse Requirements (Section 5.8.3) define requirements for exploring a number of distinct geologic settings within SPA and also define local-scale requirements for sampling within these defined exploration regions. In this section we describe the different modes of surface mobility investigated by the SDT. For matters of time and practicality, the SDT did not investigate all potential variations of each mobility class. For completeness, several of these mobility variants are discussed in APPENDIX: Architecture and will be noted in the text.

### 5.6.1 Hoppers

A concept that is often proposed for regional-scale exploration is the use of "hoppers." A hopper concept envisions a lander with a significant propulsive reserve (i.e., extra fuel) that all allow it to launch off the surface and travel ballistically from one site to another (Figure 5.19). This concept was first demonstrated on the lunar surface in 1967 by the Surveyor 6 lander, which performed a short hop (~2.5 meters horizontally) as part of an investigation of surface mechanics and plume/surface interactions. While subsequent missions have also demonstrated hops (e.g., the Vikram lander of the ISRO Chandrayaan 3 mission), Surveyor 6 still maintains the record for furthest hop on the Moon. High-performance landers could conceivably hop great distances between two or more landing sites.

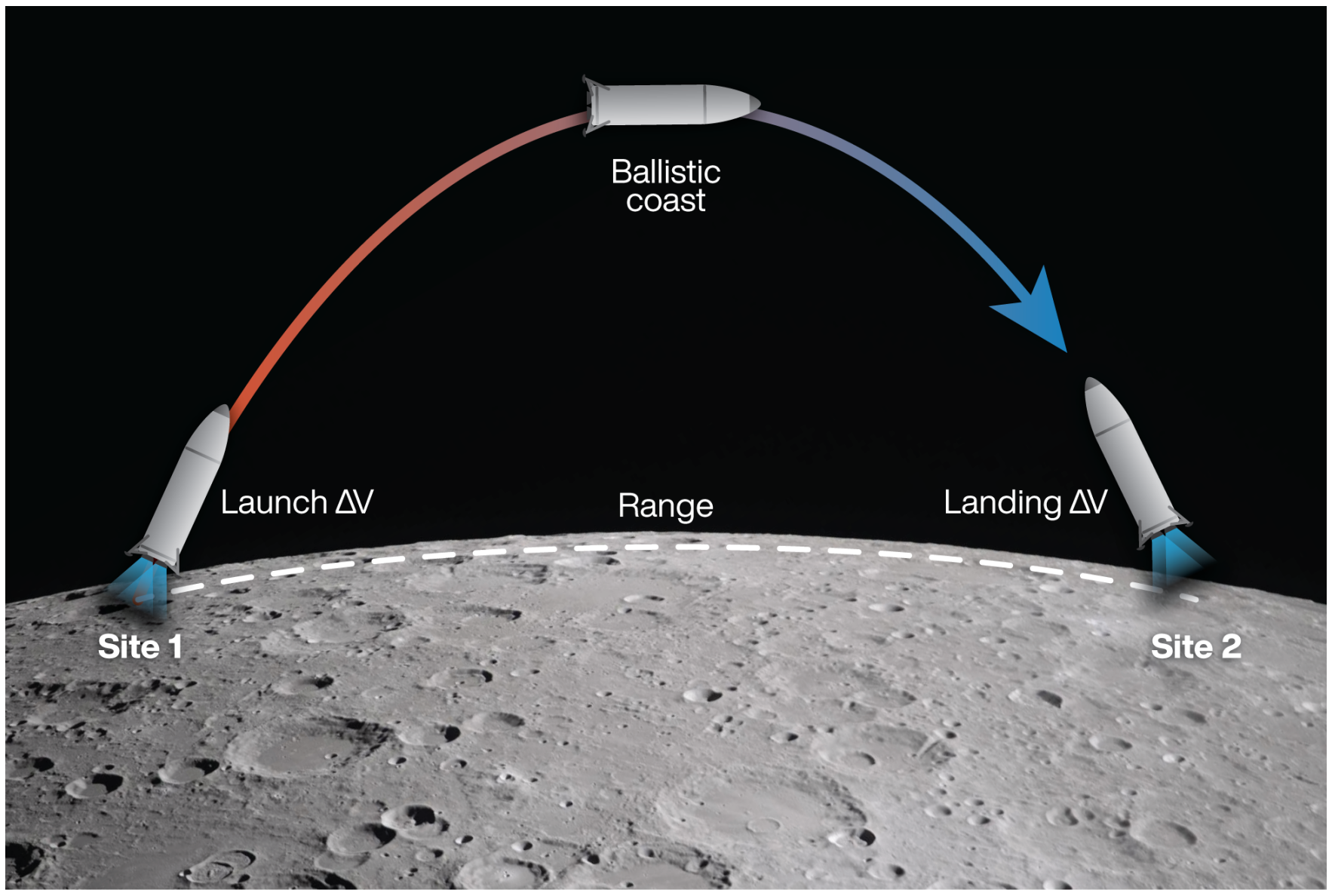

**Figure 5.19.** *Cartoon illustration of a lunar hopper. A ballistic hop consists of an initial propulsive boost, followed by a ballistic coast, and a propulsive landing. Image of the Moon taken by the crew of Artemis II on 6 April 2026 (NASA).*

For a single hop, the range is associated with the required velocity change ("delta V" or ΔV), both for the initial propulsive launch and the final propulsive landing (Figure 5.20). The ΔV for hops like this can be calculated analytically, using equations of motion for ballistic flight. Figure 5.20 shows the estimated range as a function of ΔV, which his independent of the details of the hopper (e.g., mass).

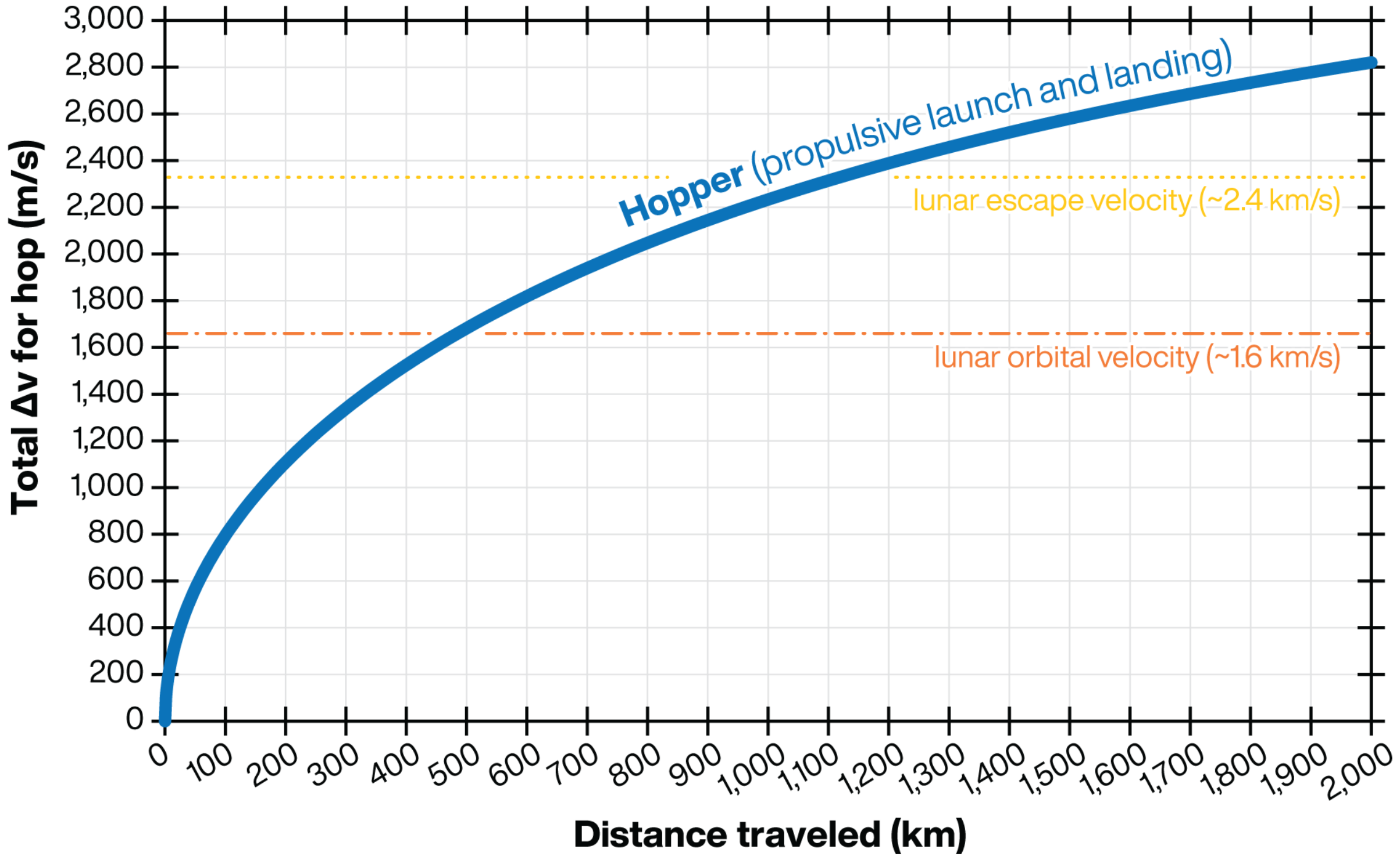

**Figure 5.20.** *The velocity change, ΔV, required for a lunar hopper to hop a given distance—assuming that the hopper propulsively launches and lands (as shown in Figure 5.19). This calculation assumes impulsive thrusts, and neglects higher-order effects.*

As a proof of concept, the technical team evaluated the feasibility of using a high-performance lander—the SpaceX Starship cargo lander—for accomplishing the SPARX mission. While a Starship has not yet landed on the surface of the Moon, and many aspects of its design are uncertain, it represents the largest delivery system planned for the near future and thus is a bounding case in terms of hopping capabilities. Table 5.8 states the assumptions for Starship performance utilized here. Effectively, the way to think about the Starship hopper concept is that it is a cargo lander where all of the landed payload capacity is allocated to additional fuel for hopping. In this scenario, the mass of the samples and sample collection system are negligible compared to the mass of fuel.

| | |
|---|---|
| Dry mass: | 110,000 kg |
| Raptor engine specific impulse, $I_{SP}$: | 380 s |
| Fuel capacity available for hops after first landing: | 200,000 kg |

**Table 5.8.** *SpaceX Starship cargo lander performance assumptions. The fuel capacity after landing assumes a maximum propellant load of 1,500,000 kg.*

Figure 5.21 shows the range as a function of the number of hops for this proof of concept based on SpaceX Starship cargo lander. While this hopper could make one or two long-range hops (~1,400 km for 2 hops), as the number of hops increase, the total range covered (the sum of each hop) rapidly decreases. As described in Section 4 and FO: Site and Traverse Requirements (Section 5.8.3), the threshold requirement for SPARX is 10 distinct sample sites. A notional single hopper architecture would require 9 hops to reach 10 sites (9 sites plus the 1 landing site). From Figure 5.21, 10 hops with a high-performance vehicle like Starship can only achieve a total range of ≤300 km—meaning that each hop is only ~30 km. This range limit is far below the threshold distance requirements for SPARX which exceed ≥1,000 km (Section 5.3.1). In this example, once the Starship completes its 9$^{th}$ hop, it is also completely out of fuel—meaning that a secondary sample return mechanism (either commercial or crewed) would need to rendezvous with the hopper, retrieve the samples, and return them to Earth.

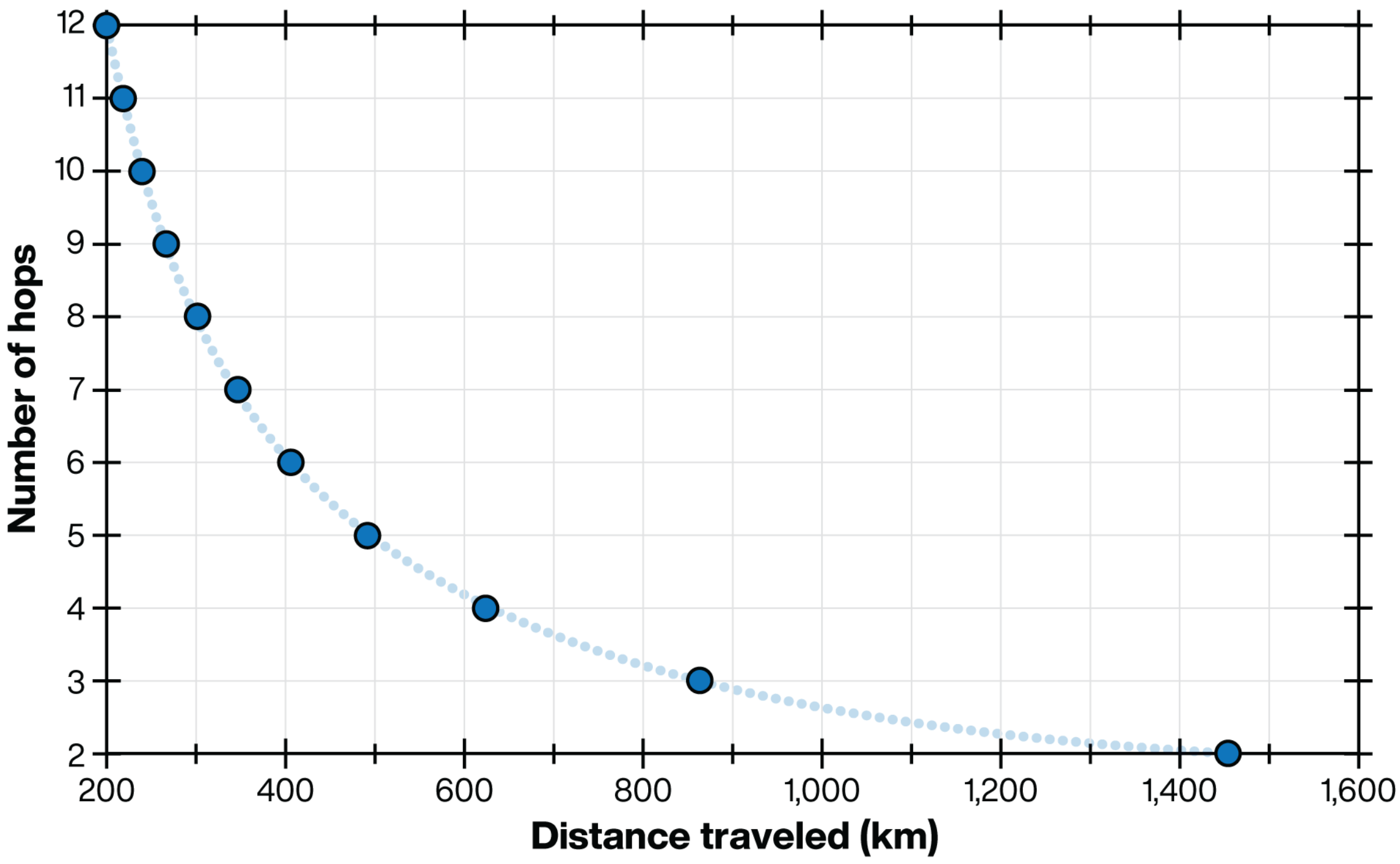

**Figure 5.21.** *The total range a hopper can travel exponentially decreases as the total number of hops increase. This calculation shows an example for a high-performance Starship-like hopper with characteristics defined in Table 5.8.*

The SDT investigated how much of the nominal SPARX traverse could be accomplished by a single hopper. Table 5.9 shows the baseline SPARX traverse, which is described further in Section 6.7, and the distance between each site along geodesics (i.e., "as the crow flies"), and the ΔV required to hop between successive sites. This analysis shows that the hopper could reach 3 of the 21 baseline science sites before it runs out of fuel. In that case, a second vehicle would still need to rendezvous with the hopper to retrieve the samples and return them to Earth. If the same hopper must launch back into lunar orbit for a rendezvous, then the hopper could only make 1 hop before returning to orbit. Given the gravity of the Moon and the required ranges, hopping is not an efficient mode of long-range transportation on the Moon. For a single hopper to accomplish the SPARX mission, it would require frequently refueling—almost after each hop. The fundamental challenge with hopping architectures arises from the rocket equation: For the spacecraft to hop further, it must carry both the fuel for each hop and the fuel necessary to loft the exponentially increasing fuel mass. There are at least two potential approaches to mitigate this: (1) refuellable landers with on-orbit refueling, or (2) using multiple hoppers in a "daisy chain" configuration, both of which are discussed in APPENDIX: Architectures.

| Site ID | Informal Name | Latitude (°N) | Longitude (°E) | Cumulative Distance to the next Site | Distance to the Next Site | Hop Total ΔV | Cumulative Fuel Mass Required | Could the hopper make this hop? | Could the Hopper get back to lunar orbit after the hop? |
|---|---|---|---|---|---|---|---|---|---|
| 1 | Mickey Mare (and Landing Site) | -52.16 | 193.23 | 129 km | 129 km | 898 m/s | 44,998 kg | Yes | Yes |
| 2 | Central SPA light plains | -53.54 | 199.88 | 146 km | 17 km | 331 m/s | 59,027 kg | Yes | No |
| 3 | Central SPA plains fresh crater | -53.54 | 200.82 | 276 km | 130 km | 903 m/s | 91,540 kg | Yes | - |
| 4 | Bhabha | -56.66 | 195.65 | 397 km | 120 km | 870 m/s | Insufficient fuel | No | - |
| 5 | Bose | -55.20 | 189.05 | 695 km | 298 km | 1,335 m/s | - | - | - |
| 6 | Abbe M northwestern rim | -61.31 | 174.31 | 728 km | 33 km | 464 m/s | - | - | - |
| 7 | Abbe M southwestern rim | -62.39 | 173.82 | 907 km | 179 km | 1,051 m/s | - | - | - |
| 8 | Mare in Poincaré E | -56.66 | 171.11 | 1,100 km | 194 km | 1,091 m/s | - | - | - |
| 9 | Poincaré X high albedo material | -54.19 | 160.72 | 1,116 km | 15 km | 316 m/s | - | - | - |
| 10 | Poincaré northern peak ring near Poincaré X | -54.57 | 160.13 | 1,254 km | 138 km | 928 m/s | - | - | - |
| 11 | Poincaré southern peak ring northwest of Poincaré Q | -59.10 | 160.77 | 1,271 km | 17 km | 336 m/s | - | - | - |
| 12 | Mare northwest of Poincaré Q | -59.16 | 161.89 | 1,297 km | 26 km | 407 m/s | - | - | - |
| 13 | Poincaré southern peak ring | -59.59 | 163.32 | 1,455 km | 158 km | 991 m/s | - | - | - |
| 14 | Lyman | -64.28 | 158.43 | 1,780 km | 324 km | 1,387 m/s | - | - | - |
| 15 | Schrödinger northern peak ring #2 | -73.16 | 141.60 | 1,787 km | 8 km | 224 m/s | - | - | - |
| 16 | Schrödinger northern peak ring #1 | -73.27 | 140.80 | 1,854 km | 67 km | 652 m/s | - | - | - |
| 17 | Schrödinger pyroclastics | -75.26 | 137.31 | 1,923 km | 69 km | 664 m/s | - | - | - |
| 18 | Schrödinger western peak ring | -74.45 | 129.15 | 2,010 km | 87 km | 742 m/s | - | - | - |
| 19 | Schrödinger southern peak ring massif | -77.16 | 132.88 | 2,077 km | 67 km | 654 m/s | - | - | - |
| 20 | Schrödinger southern rim | -79.00 | 138.86 | 2,242 km | 165 km | 1,010 m/s | - | - | - |
| 21 | South Pole-Aitken southern rim massif | -82.83 | 164.00 | 2,427 km | 186 km | 1,069 m/s | - | - | - |
| - | Notional Artemis rendezvous (site 007) | -88.70 | 131.17 | - | - | - | - | - | - |

**Total ΔV required to perform all hops: 16,324 m/s**

**Table 5.9.** *A single high-performance, Starship-like, hopper cannot accomplish the nominal SPARX traverse (described further in Section 6.7) where sample sites are spaced by many 100s to 1,000s of kilometers. Distances between sites are calculated here along great circles connecting sites, which is appropriate for a hopping architecture, but underestimate the total distances for rover architectures.*

Although the above discussion focused solely on the mechanical capabilities of a hopper system for achieving SPARX science, the SDT also notes that hopper mobility systems would be unable to achieve the SPARX geophysical experiments (E11, E12), which require a much closer spatial cadence of gravity measurements (Section 4.1.11–4.2.12). Additionally, a hopper system would face significant challenges in developing a sample acquisition system that meets the sample contamination requirements described in Section 5.3.2.3. Hoppers would also present challenges with sample contamination from exhaust, which may substantially perturb material in the vicinity of the lander (e.g., Clegg-Watkins et al., 2016). Mitigating sample contamination concerns may require additional assets, such as small sample collection rovers.

**Finding: The distance and range requirements of SPARX are incompatible with current hopping technology, even with a high-performance vehicle like Starship. Any hopping architectures would require the use of multiple hopping and sample return elements and face significant challenges in meeting SPARX science requirements E11 and E12, as well as SPARX sample contamination requirements.**

### 5.6.2 Rovers

Historically, the primary mode for surface mobility on other planetary bodies has been with rovers (Figure 5.22). Since the earliest Lunokhod rovers, decades of robotic rovers on the Moon and Mars have proven the efficacy of wheeled vehicles for characterizing local geology, conducting *in situ* analyses, and for collecting and caching samples over long traverses. For these reasons, it is natural to consider a robotic rover for the SPARX mission.

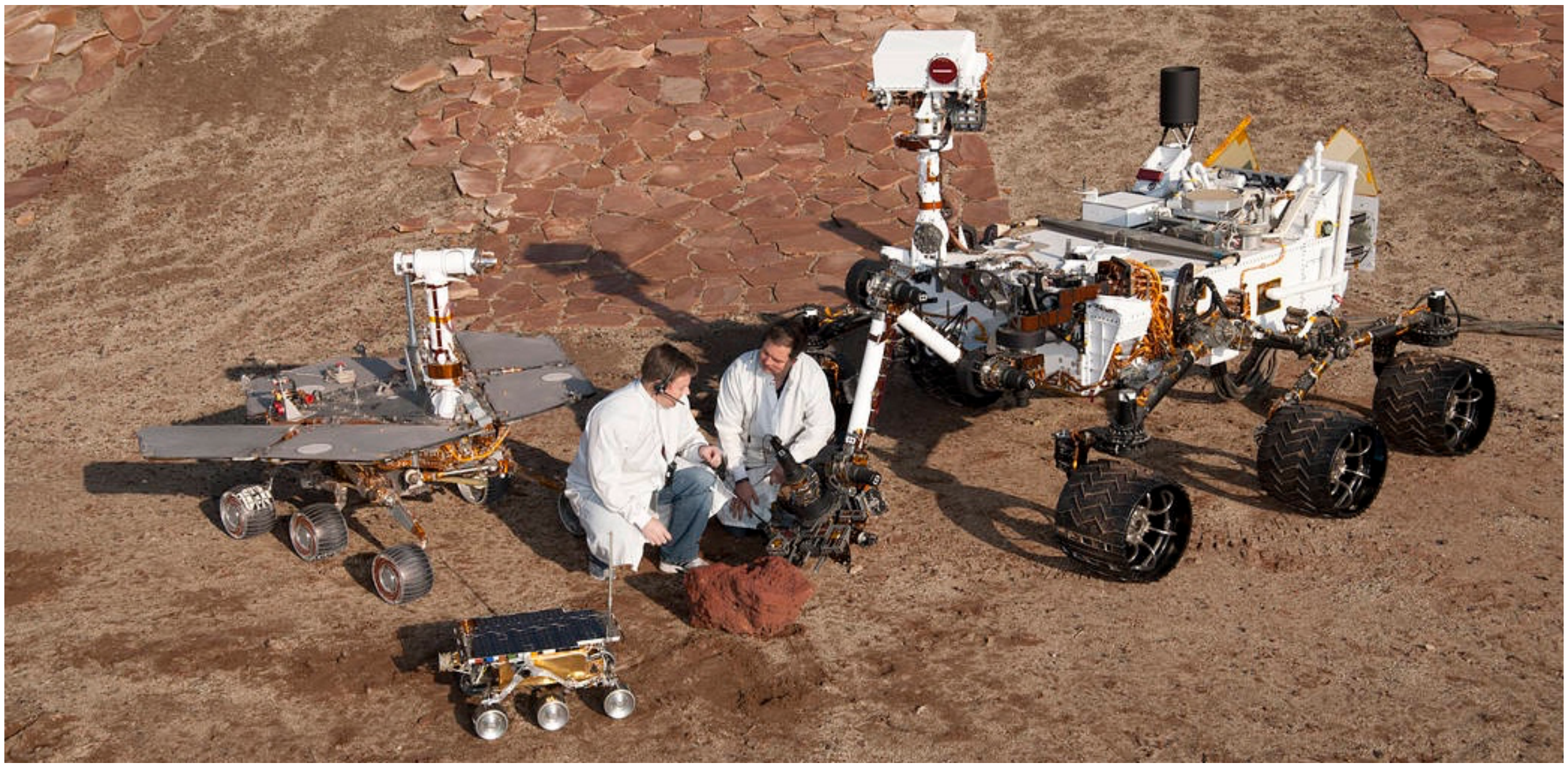

**Figure 5.22.** *Rovers have changed the paradigm for how we explore the surfaces of planetary bodies. This image shows three generations of Mars rover testbeds in the JPL Mars Yard testing area, including the testbeds for Mars Pathfinder (lower left), the Mars Exploration Rovers, Spirit and Opportunity (left), and the Mars Science Laboratory rover, Curiosity (right). Two JPL engineers (Matt Robinson and Wesley Kuykendall) are shown for scale. Image credit: NASA / JPL-Caltech.*

The central challenge for rover-based architectures is the scale of the traverse required for the SPARX mission dwarfs all past rover missions. Figure 5.24 shows the traverse distance of all past, current, and future planetary rovers. The current record holder for a planetary rover traverse is Opportunity (Mars Exploration Rover B), which traversed ~45 km over its 14-year mission on the surface. The Perseverance rover is rapidly approaching this record (in about 1/3rd the time). In contrast, SPARX has a threshold traverse requirement of ≥650 km (driven by Science Experiment E11), and most notional traverses with real science destinations exceed ≥1,100 km. The baseline traverse (described in more detail in Section 6.7) is ≥2,650 km long. **Despite the unprecedented scale, rovers remain one of the most viable architectures for accomplishing SPARX's Science Goals.** Unlike alternative approaches—like hoppers, static landers, or individual crewed Artemis sorties—rovers are uniquely equipped to perform quasi-continuous near-surface measurements along long transects (10s to 1,000s of kilometers). Rover-based mobility is enabling for exploring sites of interest and selecting priority samples (and not just being restricted to whatever is in reach of a static lander or hopper).

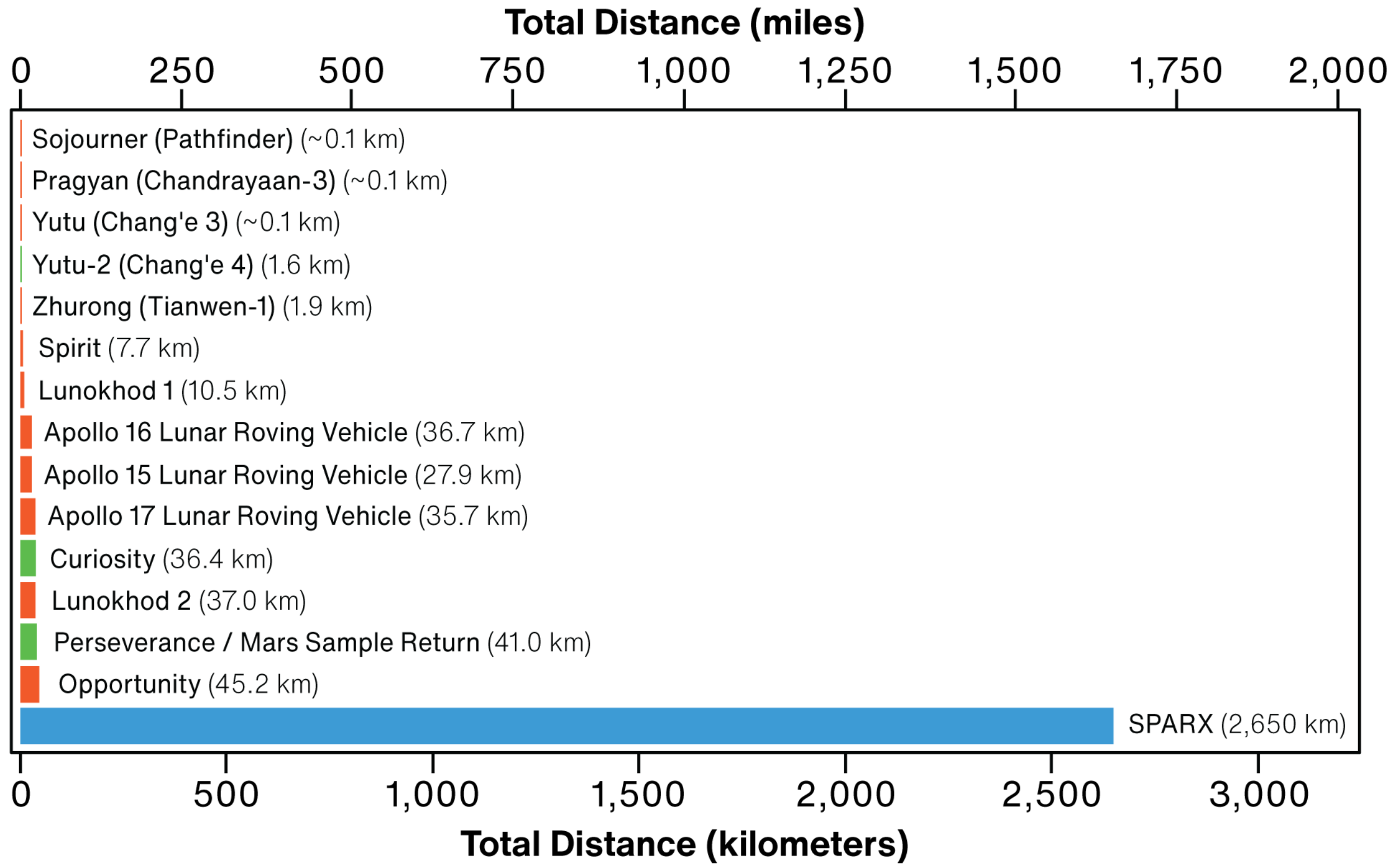

**Figure 5.23.** *Traverse distances of past, current, and future planetary rovers. This list excludes some small rovers and hoppers that traveled less than 100 meters. Past missions are shown in orange; missions that are still actively traversing are shown in green; future missions, including SPARX, are shown in blue.*

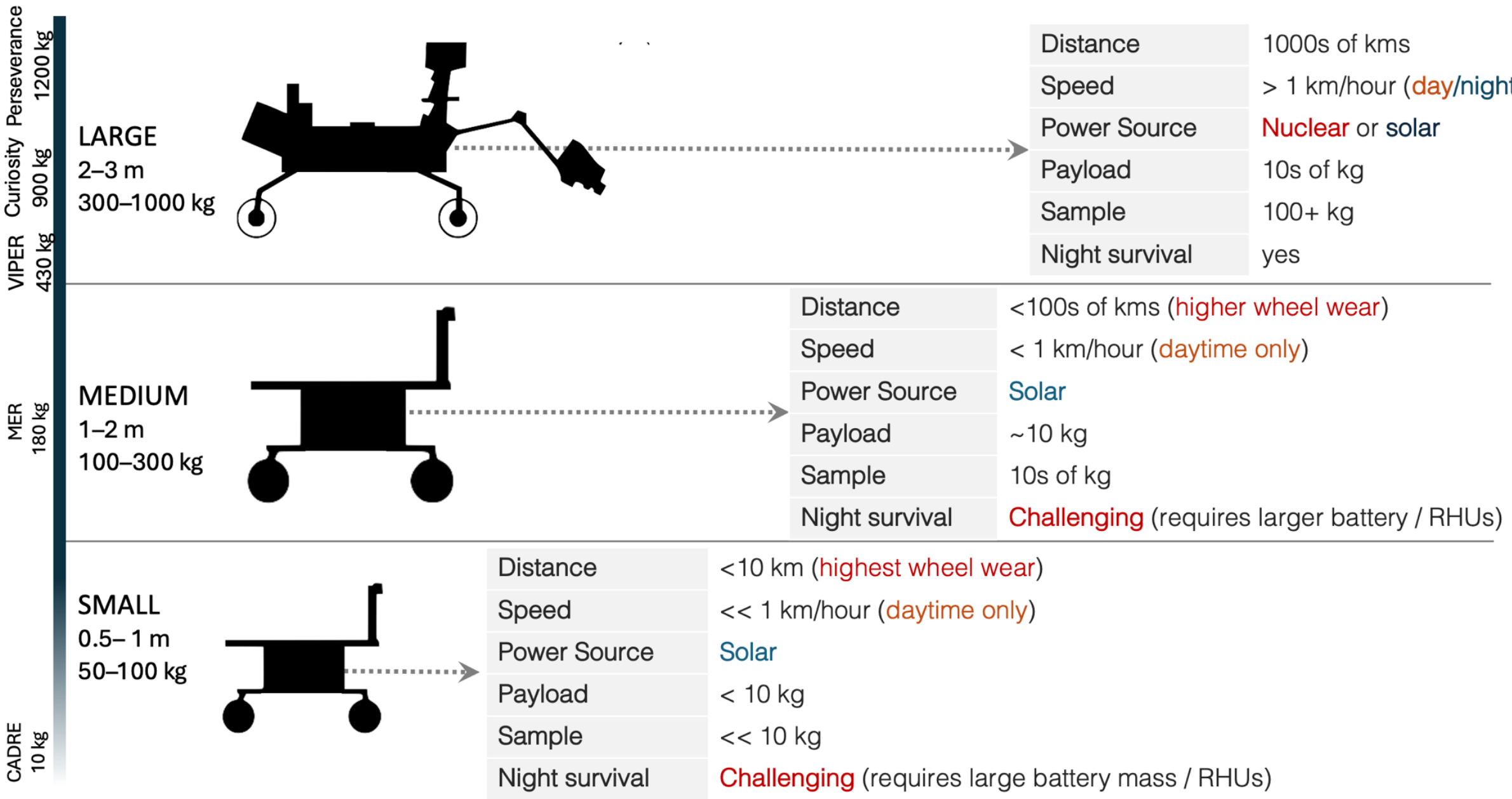

**Figure 5.24.** *Schematic illustration comparing the capabilities for three generic classes of rover sizes.*

While rovers are attractive, questions remain about the scale and cost of a rover system capable of accomplishing the SPARX science goals. Analysis of cost requires a relatively mature concept and will therefore be directed to the discussion of the baseline design reference mission (Section 6). This section will instead focus on the various size "classes" of rovers investigated by the SDT, and their ability to meet the SPARX science requirements.

At a high-level, it is helpful to categorize rover architectures by size. Figure 5.24 shows a schematic illustration of three rover size classes, ranging from small rovers (like Sojourner) to large rovers (like Perseverance). Rover size is a very rough proxy for cost, although there are exceptions; for example, it is possible to design a very large but simple rover with few capabilities.

### 5.6.2.1 "One Day Wonder" Rover

At present, one of the most important thresholds in lunar exploration is surviving the (14 earth day) lunar night. Night survivability is a significant design driver, and results in far more sophisticated (and thus costly) rover power and thermal systems. Therefore, the SDT evaluated what could be done with a rover that did not survive the lunar night, and instead only operated during a single lunar day. This notional architecture is referred to as the "One Day Wonder" rover.

The JPL Technical Team performed an analysis of the traverse distance possible for a "One Day Wonder" as a function of a variety of factors including the number and duration of sampling sites, traverse science along the route, the rate at which the rover encounters faults, and the rover's drive rate. Rover speeds and each of these factors are described in more detail in Section 6; here we focus on the overarching outcomes of the analysis. Figures 5.25–5.27 show three example models spanning from conservative approaches (Figure 5.25) to very aggressive rover assumptions (Figure 5.26). Meanwhile, Figure 5.28 shows an example of how these distances compare to exploration targets in Poincare basin, one of the SPA Regions of Interest (Section 5.2). In each case, the total mission duration is fixed to a single lunar daytime period (354 hours), and a set number of science stops (5). For the "conservative" rover, the rover speed is set to 1 kph, science interval stops are executed every 2 kilometers (Section 4.2), and sample acquisition stops include more ground-in-the-loop cycles. In this case, the rover only travels a distance of 3.4 kilometers away from the lander. Most of the time for the "conservative" rover is spent in sampling activities. In contrast, for the "aggressive" rover, the rover speed is set to 15 kph, with no science stops along the traverse, and dramatically reduced timeline for sampling and other margin. In this case, the rover can travel ~100 km out/back (~200 km total distance). One of the key factors that affects the traverse distance regardless of rover risk posture is the fault rate. Faults act to reduce the effective speed ("speed made good") of the rover. Even in the "aggressive" rover model where the rover has a top speed of 15 kph, faults act to reduce the effective speed down to ~1 kph. In all cases, faults rates are assumed to be substantially better than actual fault rates on Perseverance.

**Traverse Distance Calculator for SPARX**

Created July 8, 2025
Last update: July 22, 2025
**Version 1.6**
Created by Issa Nesnas, Edited by James Tuttle Keane

| Assumptions | Rationale |
|---|---|
| Solar rover | Provides sufficient power for higher speed navigation (up to 15 km/h) |
| Autonomous navigation | Ground-in-the-loop driving would be significantly slower due to operator decision making time, comm bandwidth, comm latency, orbiter availability, and ground station comm availability |
| Daytime only navigation | No night driving due to insufficient power to keep rover warm through the 354 lunar night hours |

| Reference Values | Duration |
|---|---|
| Lunar full day | 708 hrs |
| Lunar daytime | 354 hrs |
| Lunar nighttime | 354 hrs |

**OUTPUTS**

| | | |
|---|---|---|
| **Total traverse distance** | **7.7 km** | |
| **Total linear traverse distance accounting for tortuosity** | **6.7 km** | Assumes an average traverse tortuosity of 1.15 (see tortuosity tab) |
| **Total number of interval stops** | **3** | |
| **Total distance (if out/back)** | **3.4 km** | |
| SUCCESS | | |

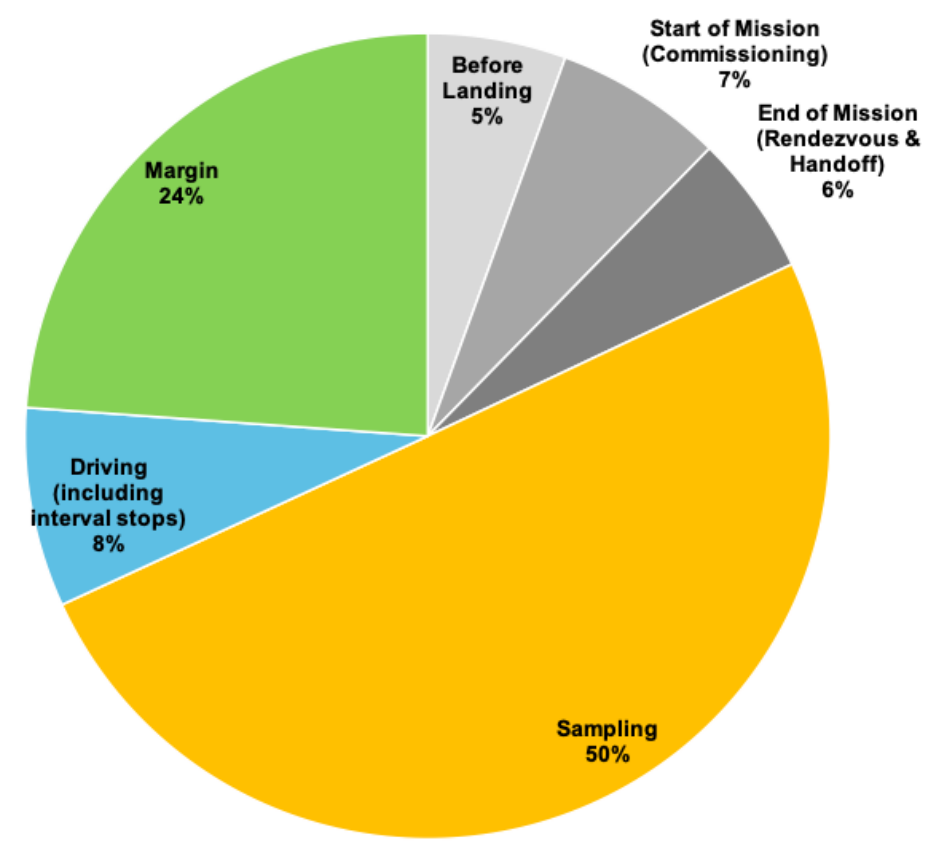

**INPUTS**

Purple cells are input parameters
Blue cells are calculated (do not change)
Orange cells relate to risk posture

**HIGH-LEVEL ASSUMPTIONS**

| Activity | Value | Comments |
|---|---|---|
| Sun elevation at landing | 10 deg | This determines when in the lunar day the mission can start |
| Number of sampling sites | 5 | Endurance baseline: 12 |
| Maximum drive speed | 1.0 km/hr | Endurance baseline maximum: 1 km/hr; LTV maximum: 15 km/hr |
| Risk posture | 1 | 1 = low risk; 2 = medium risk; 3 = high risk |

**SAMPLING ASSUMPTIONS**

| Activity | # | Every | Each Duration | Total Duration | Comments |
|---|---|---|---|---|---|
| # of sample recon stops per sample site | 5 | 0.03 km | 0.5 hrs | 2.5 hrs | |
| # of scoops per sample | 6 | N/A | 0.5 hrs | 3.0 hrs | Includes sieving and caching |
| Ground-in-the-loop (GITL) analysis & confirmation | N/A | N/A | 5.0 hrs | 30.0 hrs | If low risk, then use GITL to confirm each scoop (baseline is 5 hours per GITL); if medium/high risk, no GITL confirmation |
| Total duration per sampling site (unmargined) | | | | **35.5 hrs** | |

**TRAVERSE SCIENCE ("INTERVAL STOPS") ASSUMPTIONS**

| Activity | Duration for Science | Every | Each Duration | Total Duration | Comments |
|---|---|---|---|---|---|
| Interval stops | 1.0 hrs | 2 km | 1.00 hrs | **38.9 hrs** | Interval stops are also used for comms, which require a minimum of 15 minutes; note time for these stops is tracked by its effect on traverse rate |

**FAULT STATISTICS**

| Activity | Every | Average Recovery Duration | Comments |
|---|---|---|---|
| Minor faults | 6 km | 3 hrs | These fault rates are based on Endurance requirements (and represent a ~30x improvement over M2020's actual fault rates) |
| Major faults | 16 km | 7 hrs | |
| Average communications alert delay | 3 hrs | N/A | Time between fault happening and the ground knowing (constant ground-in-the-loop communications would be = 0) |

**TRAVERSE RATES**

| Traverse rates | Value | Comments |
|---|---|---|
| Average traverse rate (baseline) | 0.65 km/hr | Based on ERNEST field test data (small obstacles result in slight speed reduction) |
| Average traverse rate (w/ interval stops) | 0.49 km/hr | |
| Average traverse rate (w/ interval stops + faults) | 0.27 km/hr | See derivation tab |

**TOTALS**

| Description | | Duration | Comments |
|---|---|---|---|
| Time since lunar sunrise | | 19.7 hrs | |
| Commissioning | | 24.0 hrs | Scales with risk posture (24 hours for low risk; Endurance allocated 336 hours) |
| Sample collection | | 177.5 hrs | |
| Rendezvous and handoff with sample return vehicle/crew | | 20.0 hrs | Scales with risk posture (20 hours for low risk) |
| Subtotal | | **241.2 hrs** | |
| Non-traverse Margin | 30% | 72.4 hrs | |
| Subtotal (w/ Margin) | | **313.5 hrs** | |
| Remaining hours for traverse (w/o margin) | | 40.5 hrs | |
| Traverse Margin | 30% | 12.1 hrs | |
| Remaining hours for traverse (margined) | | 28.3 hrs | Includes science stops |
| Total | | **354.0 hrs** | Must equal 354 hours (1 lunar daytime period) |

**SUMMARY FOR PIE CHART**

| | | |
|---|---|---|
| Before Landing | 20 | hours |
| Start of Mission (Commissioning) | 24 | hours |
| End of Mission (Rendezvous & Handoff) | 20 | hours |
| Sampling | 178 | hours |
| Driving (including interval stops) | 28 | hours |
| Margin | 84 | hours |
| Total | 354 | hours |

**Figure 5.25.** *Example of the capabilities of a "One Day Wonder" rover assuming a "conservative" design with a slow speed, low risk tolerance, large margins, and ample time for science. In this mode, sampling (and any in situ science) easily dominate the rover timeline. Faults effectively reduce the drive speed.*

**Traverse Distance Calculator for SPARX**

Created July 8, 2025
Last update: July 22, 2025
**Version 1.6**
Created by Issa Nesnas, Edited by James Tuttle Keane

| Assumptions | Rationale |
|---|---|
| Solar rover | Provides sufficient power for higher speed navigation (up to 15 km/h) |
| Autonomous navigation | Ground-in-the-loop driving would be significantly slower due to operator decision making time, comm bandwidth, comm latency, orbiter availability, and ground station comm availability |
| Daytime only navigation | No night driving due to insufficient power to keep rover warm through the 354 lunar night hours |

| Reference Values | Duration |
|---|---|
| Lunar full day | 708 hrs |
| Lunar daytime | 354 hrs |
| Lunar nighttime | 354 hrs |

**OUTPUTS**

| | | |
|---|---|---|
| **Total traverse distance** | **229.3 km** | |
| Total linear traverse distance accounting for tortuosity | 199.4 km | Assumes an average traverse tortuosity of 1.15 (see tortuosity tab) |
| Total number of interval stops | 100 | |
| **Total distance (if out/back)** | **99.7 km** | |

SUCCESS

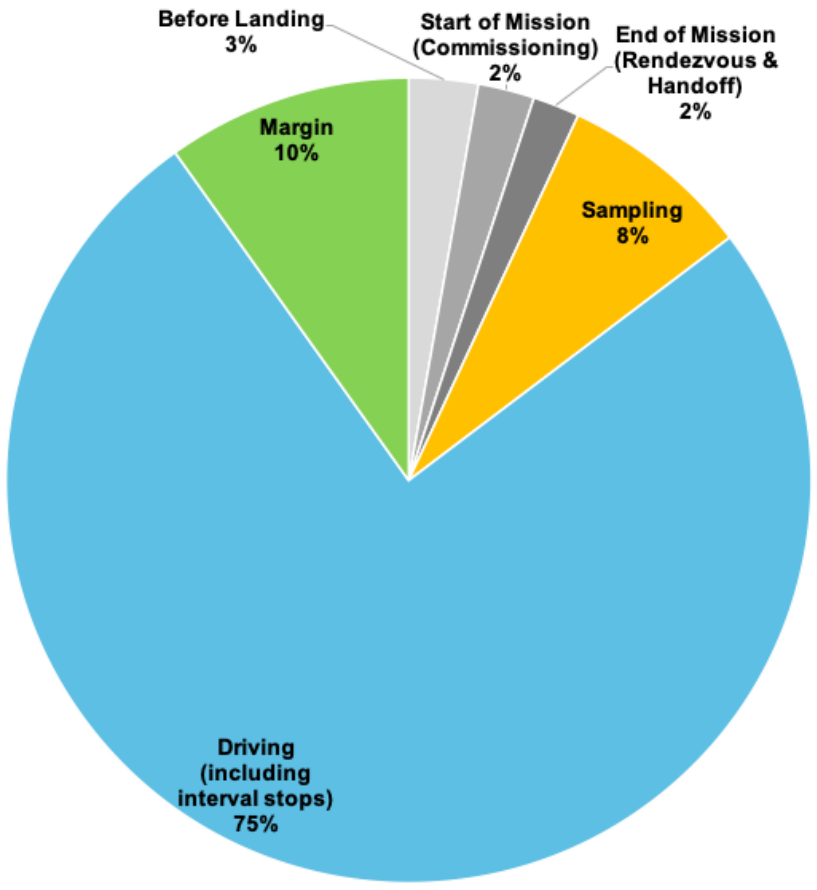

**INPUTS**

Purple cells are input parameters
Blue cells are calculated (do not change)
Orange cells relate to risk posture

**HIGH-LEVEL ASSUMPTIONS**

| Activity | Value | Comments |
|---|---|---|
| Sun elevation at landing | 5 deg | This determines when in the lunar day the mission can start |
| Number of sampling sites | 5 | Endurance baseline: 12 |
| Maximum drive speed | 15.0 km/hr | Endurance baseline maximum: 1 km/hr; LTV maximum: 15 km/hr |
| Risk posture | 3 | 1 = low risk; 2 = medium risk; 3 = high risk |

**SAMPLING ASSUMPTIONS**

| Activity | # | Every | Each Duration | Total Duration | Comments |
|---|---|---|---|---|---|
| # of sample recon stops per sample site | 5 | 0.03 km | 0.5 hrs | 2.5 hrs | |
| # of scoops per sample | 6 | N/A | 0.5 hrs | 3.0 hrs | Includes sieving and caching |
| Ground-in-the-loop (GITL) analysis & confirmation | N/A | N/A | 0.0 hrs | 0.0 hrs | If low risk, then use GITL to confirm each scoop (baseline is 5 hours per GITL); if medium/high risk, no GITL confirmation |
| Total duration per sampling site (unmargined) | | | | **5.5 hrs** | |

**TRAVERSE SCIENCE ("INTERVAL STOPS") ASSUMPTIONS**

| Activity | Duration for Science | Every | Each Duration | Total Duration | Comments |
|---|---|---|---|---|---|
| Interval stops | 0.0 hrs | 2 km | 0.25 hrs | **30.4 hrs** | Interval stops are also used for comms, which require a minimum of 15 minutes; note time for these stops is tracked by its effect on traverse rate |

**FAULT STATISTICS**

| Activity | Every | Average Recovery Duration | Comments |
|---|---|---|---|
| Minor faults | 6 km | 3 hrs | These fault rates are based on Endurance requirements (and represent a ~30x improvement over M2020's actual fault rates) |
| Major faults | 16 km | 7 hrs | |
| Average communications alert delay | 0 hrs | N/A | Time between fault happening and the ground knowing (constant ground-in-the-loop communications would be = 0) |

**TRAVERSE RATES**

| Traverse rates | Value | Comments |
|---|---|---|
| Average traverse rate (baseline) | 9.75 km/hr | Based on ERNEST field test data (small obstacles result in slight speed reduction) |
| Average traverse rate (w/ interval stops) | 4.39 km/hr | |
| Average traverse rate (w/ interval stops + faults) | 0.86 km/hr | See derivation tab |

**TOTALS**

| Description | | Duration | Comments |
|---|---|---|---|
| Time since lunar sunrise | | 9.8 hrs | |
| Commissioning | | 8.0 hrs | Scales with risk posture (24 hours for low risk; Endurance allocated 336 hours) |
| Sample collection | | 27.5 hrs | |
| Rendezvous and handoff with sample return vehicle/crew | | 6.7 hrs | Scales with risk posture (20 hours for low risk) |
| Subtotal | | **52.0 hrs** | |
| Non-traverse Margin | 10% | 5.2 hrs | |
| Subtotal (w/ Margin) | | **57.2 hrs** | |
| Remaining hours for traverse (w/o margin) | | 296.8 hrs | |
| Traverse Margin | 10% | 29.7 hrs | |
| Remaining hours for traverse (margined) | | 267.1 hrs | Includes science stops |
| Total | | **354.0 hrs** | Must equal 354 hours (1 lunar daytime period) |

**SUMMARY FOR PIE CHART**

| | | |
|---|---|---|
| Before Landing | 10 | hours |
| Start of Mission (Commissioning) | 8 | hours |
| End of Mission (Rendezvous & Handoff) | 7 | hours |
| Sampling | 28 | hours |
| Driving (including interval stops) | 267 | hours |
| Margin | 35 | hours |
| Total | 354 | hours |

**Figure 5.26.** *Example of the capabilities of a "One Day Wonder" rover assuming an "aggressive" risk posture design for high-speed, high-risk tolerance, and little time allocated for science. In this mode, the rover's effective speed is dominated by faults and engineering stops for communication/localization.*

**Traverse Distance Calculator for SPARX**

Created July 8, 2025
Last update: July 22, 2025
**Version 1.6**
Created by Issa Nesnas, Edited by James Tuttle Keane

| Assumptions | Rationale |
|---|---|
| Solar rover | Provides sufficient power for higher speed navigation (up to 15 km/h) |
| Autonomous navigation | Ground-in-the-loop driving would be significantly slower due to operator decision making time, comm bandwidth, comm latency, orbiter availability, and ground station comm availability |
| Daytime only navigation | No night driving due to insufficient power to keep rover warm through the 354 lunar night hours |

| Reference Values | Duration |
|---|---|
| Lunar full day | 708 hrs |
| Lunar daytime | 354 hrs |
| Lunar nighttime | 354 hrs |

**OUTPUTS**

| | | |
|---|---|---|
| **Total traverse distance** | **32.0 km** | |
| **Total linear traverse distance accounting for tortuosity** | **27.8 km** | Assumes an average traverse tortuosity of 1.15 (see tortuosity tab) |
| **Total number of interval stops** | **14** | |
| **Total distance (if out/back)** | **13.9 km** | |
| SUCCESS | | |

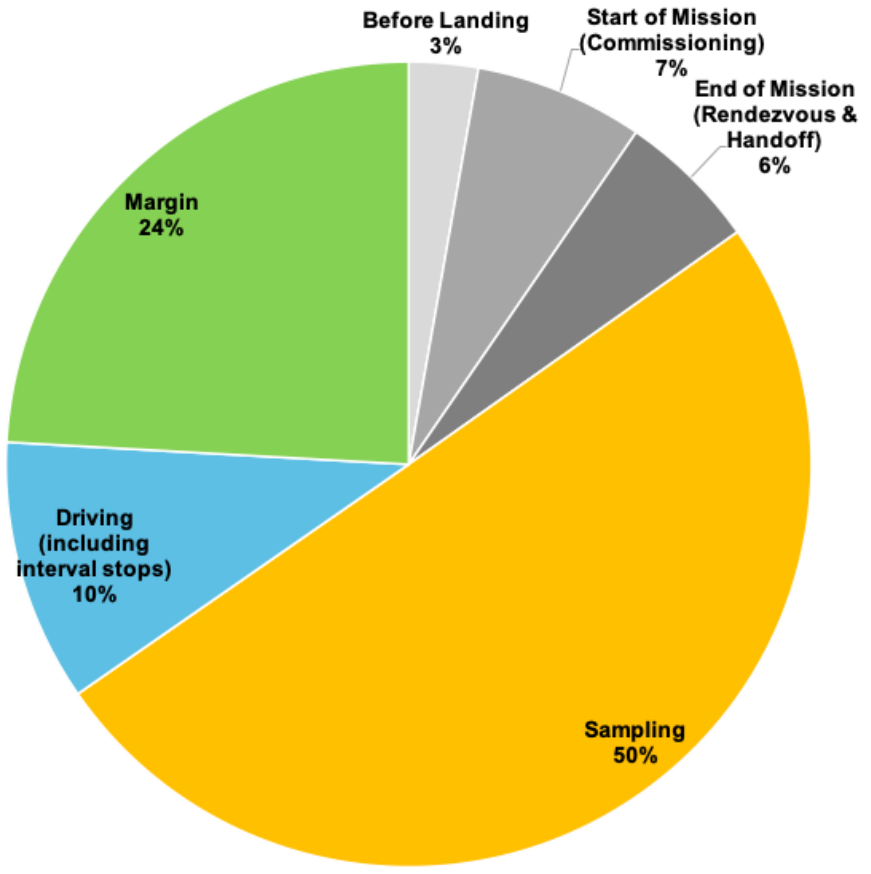

**INPUTS**

Purple cells are input parameters
Blue cells are calculated (do not change)
Orange cells relate to risk posture

**HIGH-LEVEL ASSUMPTIONS**

| Activity | Value | Comments |
|---|---|---|
| Sun elevation at landing | 5 deg | This determines when in the lunar day the mission can start |
| Number of sampling sites | 5 | Endurance baseline: 12 |
| Maximum drive speed | 15.0 km/hr | Endurance baseline maximum: 1 km/hr; LTV maximum: 15 km/hr |
| Risk posture | 1 | 1 = low risk; 2 = medium risk; 3 = high risk |

**SAMPLING ASSUMPTIONS**

| Activity | # | Every | Each Duration | Total Duration | Comments |
|---|---|---|---|---|---|
| # of sample recon stops per sample site | 5 | 0.03 km | 0.5 hrs | 2.5 hrs | |
| # of scoops per sample | 6 | N/A | 0.5 hrs | 3.0 hrs | Includes sieving and caching |
| Ground-in-the-loop (GITL) analysis & confirmation | N/A | N/A | 5.0 hrs | 30.0 hrs | If low risk, then use GITL to confirm each scoop (baseline is 5 hours per GITL); if medium/high risk, no GITL confirmation |
| **Total duration per sampling site** (unmargined) | | | | **35.5 hrs** | |

**TRAVERSE SCIENCE ("INTERVAL STOPS") ASSUMPTIONS**

| Activity | Duration for Science | Every | Each Duration | Total Duration | Comments |
|---|---|---|---|---|---|
| Interval stops | 0.0 hrs | 2 km | 0.25 hrs | **39.0 hrs** | Interval stops are also used for comms, which require a minimum of 15 minutes; note time for these stops is tracked by its effect on traverse rate |

**FAULT STATISTICS**

| Activity | Every | Average Recovery Duration | Comments |
|---|---|---|---|
| Minor faults | 6 km | 3 hrs | These fault rates are based on Endurance requirements (and represent a ~30x improvement over M2020's actual fault rates) |
| Major faults | 16 km | 7 hrs | |
| Average communications alert delay | 0 hrs | N/A | Time between fault happening and the ground knowing (constant ground-in-the-loop communications would be = 0) |

**TRAVERSE RATES**

| Traverse rates | Value | Comments |
|---|---|---|
| Average traverse rate (baseline) | 9.75 km/hr | Based on ERNEST field test data (small obstacles result in slight speed reduction) |
| Average traverse rate (w/ interval stops) | 4.39 km/hr | |
| Average traverse rate (w/ interval stops + faults) | 0.86 km/hr | See derivation tab |

**TOTALS**

| Description | | Duration | Comments |
|---|---|---|---|
| Time since lunar sunrise | | 9.8 hrs | |
| Commissioning | | 24.0 hrs | Scales with risk posture (24 hours for low risk; Endurance allocated 336 hours) |
| Sample collection | | 177.5 hrs | |
| Rendezvous and handoff with sample return vehicle/crew | | 20.0 hrs | Scales with risk posture (20 hours for low risk) |
| **Subtotal** | | **231.3 hrs** | |
| Non-traverse Margin | 30% | 69.4 hrs | |
| **Subtotal (w/ Margin)** | | **300.7 hrs** | |
| Remaining hours for traverse (w/o margin) | | 53.3 hrs | |
| Traverse Margin | 30% | 16.0 hrs | |
| Remaining hours for traverse (margined) | | 37.3 hrs | Includes science stops |
| **Total** | | **354.0 hrs** | Must equal 354 hours (1 lunar daytime period) |

**SUMMARY FOR PIE CHART**

| | | |
|---|---|---|
| Before Landing | 10 | hours |
| Start of Mission (Commissioning) | 24 | hours |
| End of Mission (Rendezvous & Handoff) | 20 | hours |
| Sampling | 178 | hours |
| Driving (including interval stops) | 37 | hours |
| Margin | 85 | hours |
| **Total** | **354** | **hours** |

**Figure 5.27.** *Example of the capabilities of a "One Day Wonder" rover assuming an "intermediate" risk posture design between the "conservative" case (Figure 5.25) and the "aggressive" case (Figure 5.26). In this case, all input parameters were selected to be identical to the "aggressive" case, but with a lower risk tolerance. This highlights how risk posture and margins are a major driver for traverse rate and total distance. Increasing drive speed does not always increase total distance.*

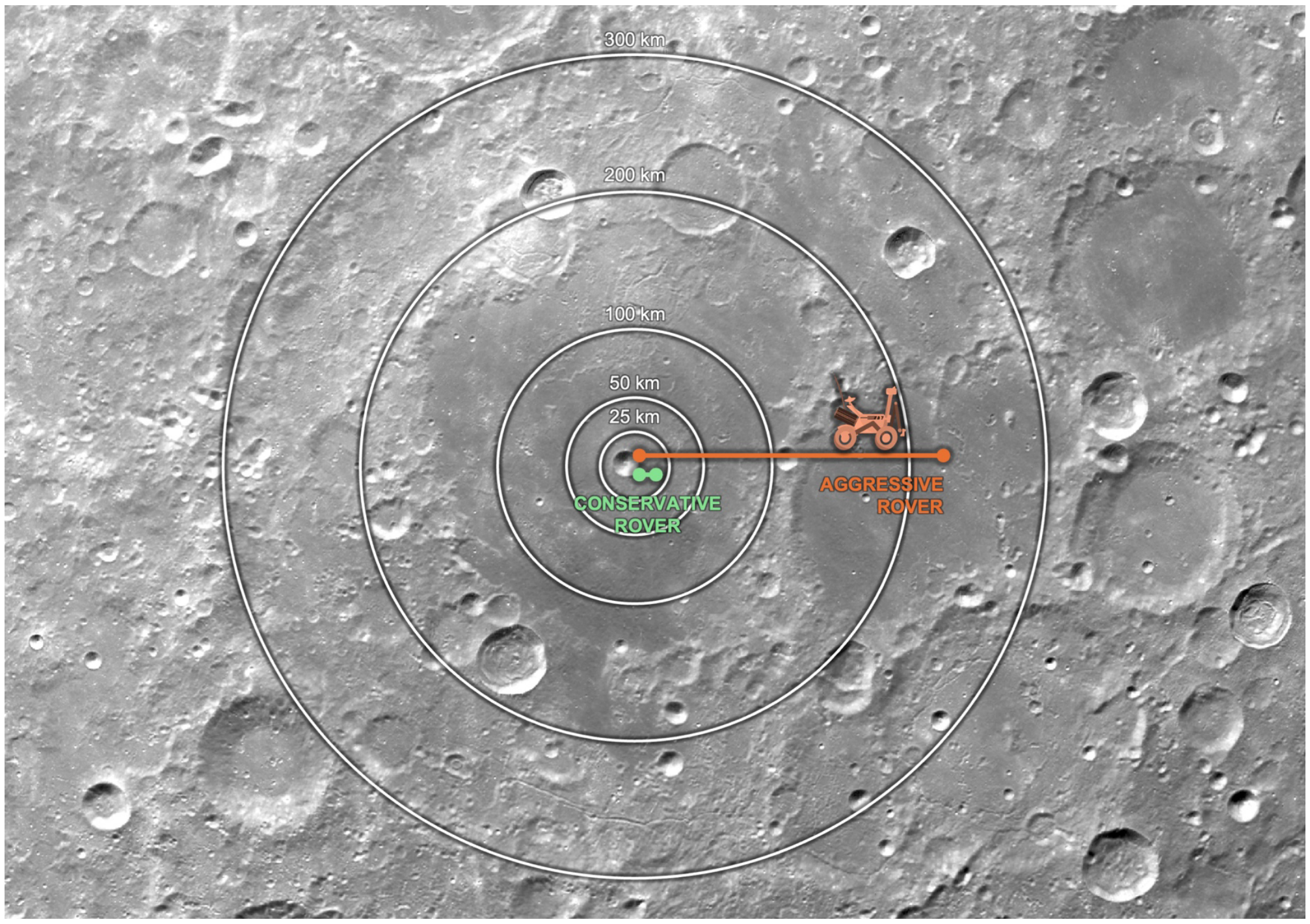

**Figure 5.28.** *Ranges of the "One Day Wonder" rovers described here compared to the scale of a single SPA region of interest: Poincaré (discussed further in Section 5.2.2).*

The main SDT conclusions from the "one day wonder" analysis are:

(1) A "one day wonder" rover would not have the range capability to either fully explore a SPARX region of interest or achieve the SPARX highest priority Science Goals (Section 3.4). In the most aggressive case, a "One Day Wonder" rover that only drives during lunar day could plausibly travel up to 200 km per day, while more conservative rover architectures may only traverse ~10 km per day (Figure 5.28). Even an "aggressive" rover of this type would not be able to completely explore a single SPARX region of interest (several 100s of kilometers in scale), suggesting that **single, "short-range" rovers are inadequate for achieving SPARX's Science Goals**.

(2) The mechanical speed of a rover is not always the primary driver of the rover's effective speed. Many other factors act to slow down the rover, including science activities (sampling and traverse science) and faults. Keeping science activities efficient and minimizing interruptions due to faults are keys to enabling faster rover operations. This is described further in Section 6.

**Finding: The SPARX science requirements cannot be achieved by a small- to medium-size rover designed to traverse over a single lunar day. The "one day wonder" architecture fails to meet the minimum required geologic sampling regions and number of feasible sampling stops.**

### 5.6.2.2 Long Distance Rover

To bridge the severe range and sampling limitations of static landers, hoppers (Section 5.6.1), and short-range or short-lived rovers (Section 5.6.2.1), a long-range robotic rover emerges as the only viable mobility platform capable of achieving the ambitious multi-region science objectives of SPARX. Because of this, Section 6 is dedicated to examining a long-range rover design reference mission.

### 5.6.2.3 Lunar Terrain Vehicle

The Lunar Terrain Vehicle (LTV) is an unpressurized surface mobility platform being conceptualized as part of the broader Artemis infrastructure to extend the exploration range of astronauts. When operated by a crew on the lunar surface, unpressurized rovers like the LTV are generally assumed to provide a maximum exploration range of approximately 10 kilometers from a safe haven, such as the lunar lander or Moon Base. During uncrewed phases, the LTV is envisioned to be capable of robotic ground-in-the-loop operations. At the time of the SPARX SDT, three companies were competing for the LTV service contract (Figure 5.29); however, since that time, selections have been made to further investigate designs from two of those companies. The SDT received briefings from all three companies in 2025 about their LTV plans, and the capabilities of each planned system for a potential SPARX mission scope.

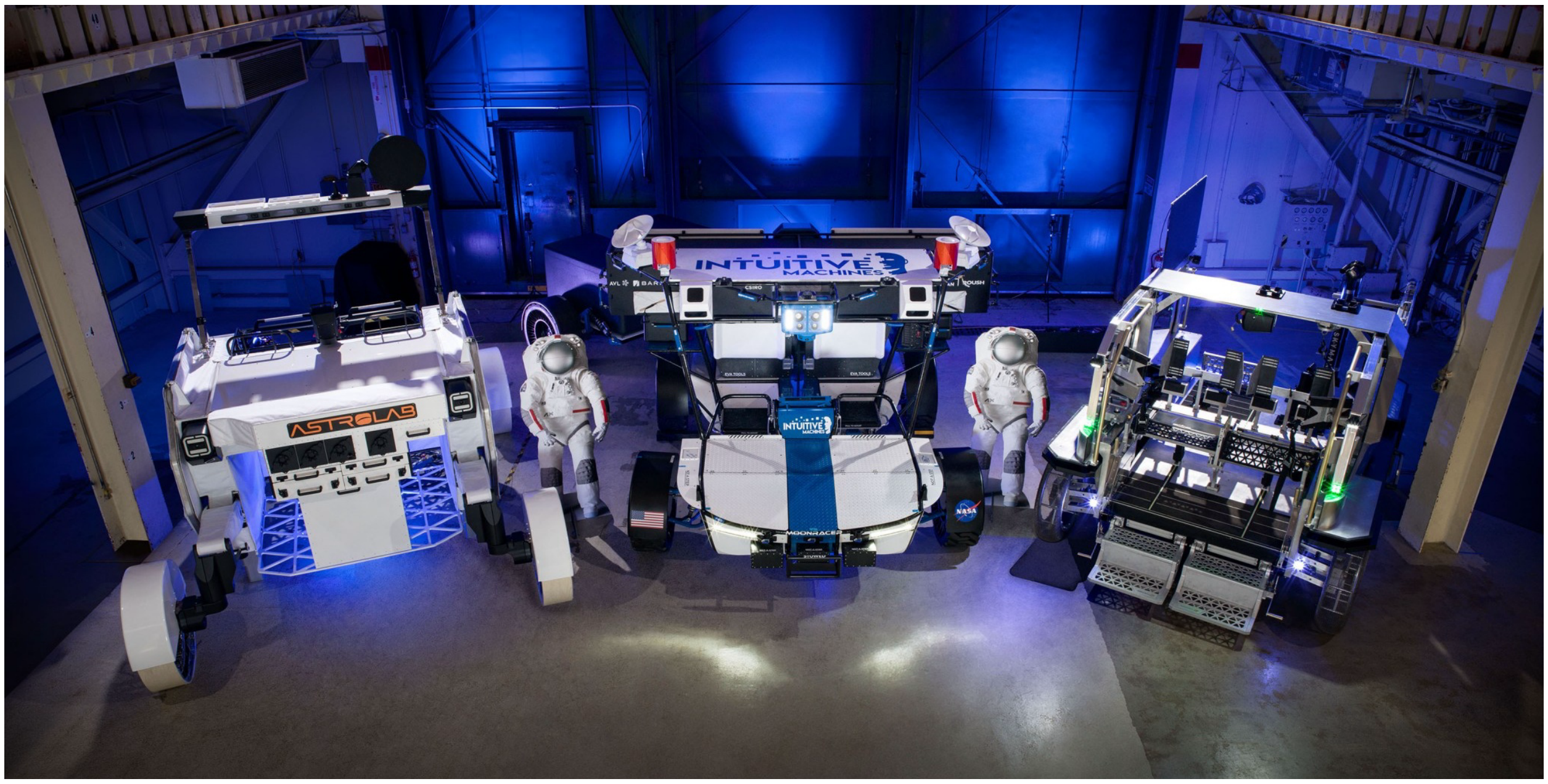

**Figure 5.29.** *Three Lunar Terrain Vehicle (LTV) prototypes from (left to right) Venturi Astrolab, Intuitive Machines, and Lunar Outpost. Image credit: NASA / Bill Stafford.*

The NASA-defined requirements for LTV (circa 2022) are similar to, but distinct from the SPARX's mission requirements (Table 5.10). This table only states the previous NASA-defined requirements, and in many cases LTV providers may exceed these requirements. In some cases, the LTV has more challenging requirements (e.g., payload mass, top speed, mission lifetime), while in others SPARX has more challenging requirements (e.g., autonomous driving, operations at mid-latitudes).

| | Lunar Terrain Vehicle (LTV) Requirements: | SPARX SDT Requirements: |
|---|---|---|
| Crew: | 2 suited astronauts | No crew |
| Science instruments: | ~5 (2 on a mast, 3 on the deck) | ~5 (2 on a mast, 3 in the body) |
| Payload mass: | 800 kg with standard performance<br>1,600 kg with reduced performance | 120 kg (100 kg for samples + 20 kg for science instruments) |
| Distance per year: | 1,300 km | ~500 km |
| Top speed when crewed: | 15 kph | N/A |
| Top speed without crew: | 0.25 km/hr (speed made good when teleoperated) | 1 km/hr (top speed with autonomous driving) |
| Autonomy: | Full driving autonomy is not required | Full driving autonomy is required |
| Slope capabilities: | <20° | <20° |
| Vehicle Lifetime: | 10 years | 5 years |
| Permanently Shadowed Regions (PSR) operations: | Required | Not required |
| Night survivability: | Not required (must survive >150 hours of shadow; 1 lunar night is 354 hour) | Required (must drive and operate at night) |
| Operating latitudes: | Poleward of 84°S (Artemis Exploration Zone) | Poleward of 52°S |
| Robotic manipulator: | Required | Required |
| Launch: | 2028 (for Artemis V) | 2031 |

**Table 5.10.** *Comparison of requirements for LTV (defined in LTV industry day documentation dated 31 August 2022) and SPARX. Yellow cells indicate which architecture has the more challenging requirement. Note that the LTV program is currently undergoing an evolution post the NASA Ignition event in early 2026, and the requirements are likely to change.*

Because the final technical capabilities, programmatic risk posture, and schedule availability of the LTV remain highly uncertain at the time of this report, it is difficult to reliably assess its viability for executing the extreme long-range (≥1,000 km) traverses and dedicated sampling campaigns required by SPARX. The recent NASA Ignition event further complicated this analysis, as NASA recently revectored the LTV program towards simpler, less-capable vehicles that could be flown sooner. Consequently, the SDT refrains from distinguishing between LTV architectures and other large, long-range rovers at this time.

### 5.6.3 Crewed Missions

While robotic missions have laid the essential groundwork for lunar exploration, their capabilities often pale in comparison to what astronaut crews can accomplish. For example:

- Human explorers can use their training and experience to synthesize geological context in real-time. Astronauts are adaptable and independent; they can think on their own to adjust plans and collect serendipitous samples and observations.
- Astronauts bring physical adaptability and fine motor dexterity beyond what is possible with robotic explorers (at least without great cost). Astronauts can dynamically interact with rocks while using a wide array of tools (e.g., hands, rock hammers, drive tubes, and drills). They can adjust their techniques based on tactile responses.

- Finally, astronauts can quickly solve problems on the spot, dramatically increasing overall mission efficiency and the probability of success.

In terms of mobility, astronauts can traverse much more varied terrains more quickly than most robotic approaches. Whether walking or driving vehicles, astronauts are adaptable and can replan based on the realities of the lunar environment. Table 5.11 states common assumptions about the range of crew in a single extravehicular excursion (EVA). Typical EVAs on the surface of the Moon may be able to traverse kilometers within a single ~8-hour long EVA. In most cases, mobility is limited by astronaut safety and the requirement to be able to walk back to a safe haven, such as the lunar lander. The single longest traverse performed by crew on a single EVA was on Apollo 17, where on EVA 2 the crew traversed ~20.4 kilometers using the Apollo Lunar Roving Vehicle (with a maximum range from the lander of ~7.6 kilometers) in ~7.5 hours (Figure 5.30).

| | |
|---|---|
| **Maximum range for a walking EVA:** | ~2 km |
| **Maximum range for crewed exploration with an unpressurized rover (e.g., LTV):** | ~10 km |
| **Maximum range for crewed exploration with a pressurized rover:** | ~12 km |
| **Maximum range for crewed exploration with dual pressurized rovers:** | ~100 km |

**Table 5.11.** *Artemis astronaut mobility assumptions, based on Coan 2020.*

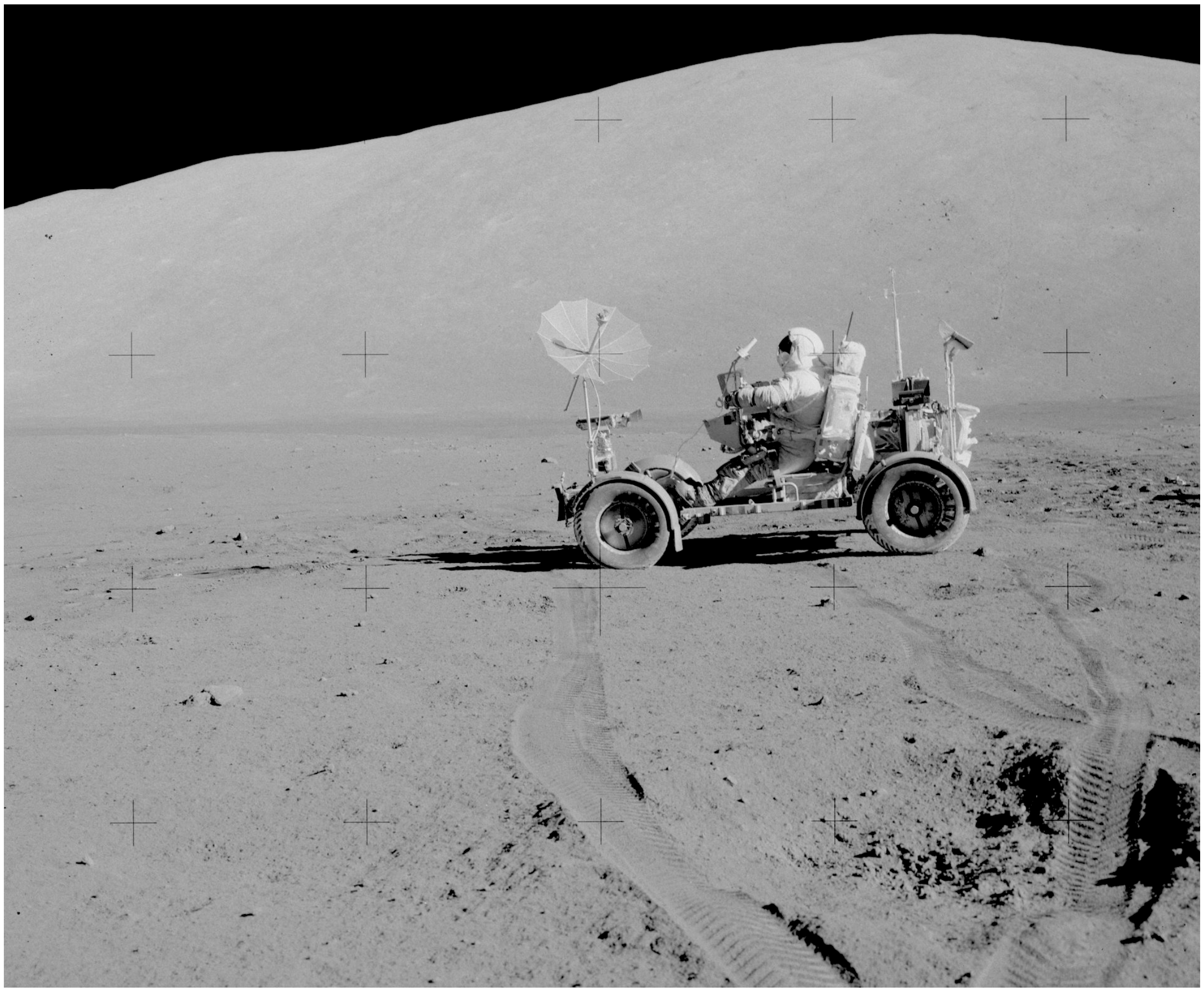

**Figure 5.30.** *Photograph taken on Apollo 17 EVA-2, which holds the record for the longest-range surface EVA: ~20.4 kilometers. Image credit: NASA (AS17-135-20544).*

While astronauts are the ultimate explorers, they do have limitations relevant to accomplishing the SPARX Science Goals. For example, as discussed in Sections 4 and 5.2–5.3, SPARX's Science Goals require extreme long-range mobility of many 100s of kilometers. **The required traverse ranges for SPARX Science Goals are well beyond near-term Artemis capabilities, even when accounting for unpressurized rovers like the Lunar Terrain Vehicle (LTV) or the pressurized rover.** This limitation results from walk-back requirements for astronaut safety.

At present, all Artemis missions are targeting the lunar South Pole. The Artemis exploration zone is defined as poleward of 84°S. **Artemis's focus on the South Pole significantly inhibits the ability of Artemis to address SPARX's Science Goals.** This can be mitigated either by using hybrid approaches were a robotic asset caches samples and delivers them to Artemis astronauts at the South Pole, and/or having non-polar Artemis missions. These combinations of options are discussed further in the following section and in Section 7.

## 5.7 MISSION ARCHITECTURES CONSIDERED

As described in the previous sections, it is possible to envision a large range of distinct mission architectures relevant to SPARX. Given the number of permutations, it can be challenging to provide a reasonable scope for the SPARX mission implementation. In this section, we describe and evaluate the mission architectures discussed in the most detail by the SDT. **These mission architectures do not represent detailed concept studies (although in many cases past concept studies are used to inform the overall evaluation), rather these schematics represent a distillation of the high-level SDT evaluation of the architecture against the SPARX objectives.** The SDT notes that a number of other architectures could be proposed, some are beyond the scope of the TOR (e.g., non-sample return SPA missions), and some are not envisioned to be technically feasible in the near term (e.g., refuellable hopper with on-orbit refueling and sample return). These concepts received only cursory discussion from the SDT but are nevertheless presented for completeness in APPENDIX: Architecture.

For each architecture, a schematic illustration is provided to quickly describe the architecture—an example of which Figure 5.31. The illustrations are not meant to depict an actual spacecraft or mission design but rather symbolize the type of architecture. Finally, for each architecture, an evaluation of the relative strengths and weakness is provided against the SPARX figures of merit (Section 5.3), and a single overall qualitative adjectival rating similar to NASA proposals (Table 5.12). Bold strengths and weaknesses indicate major strengths, weaknesses, or fatal flaws.

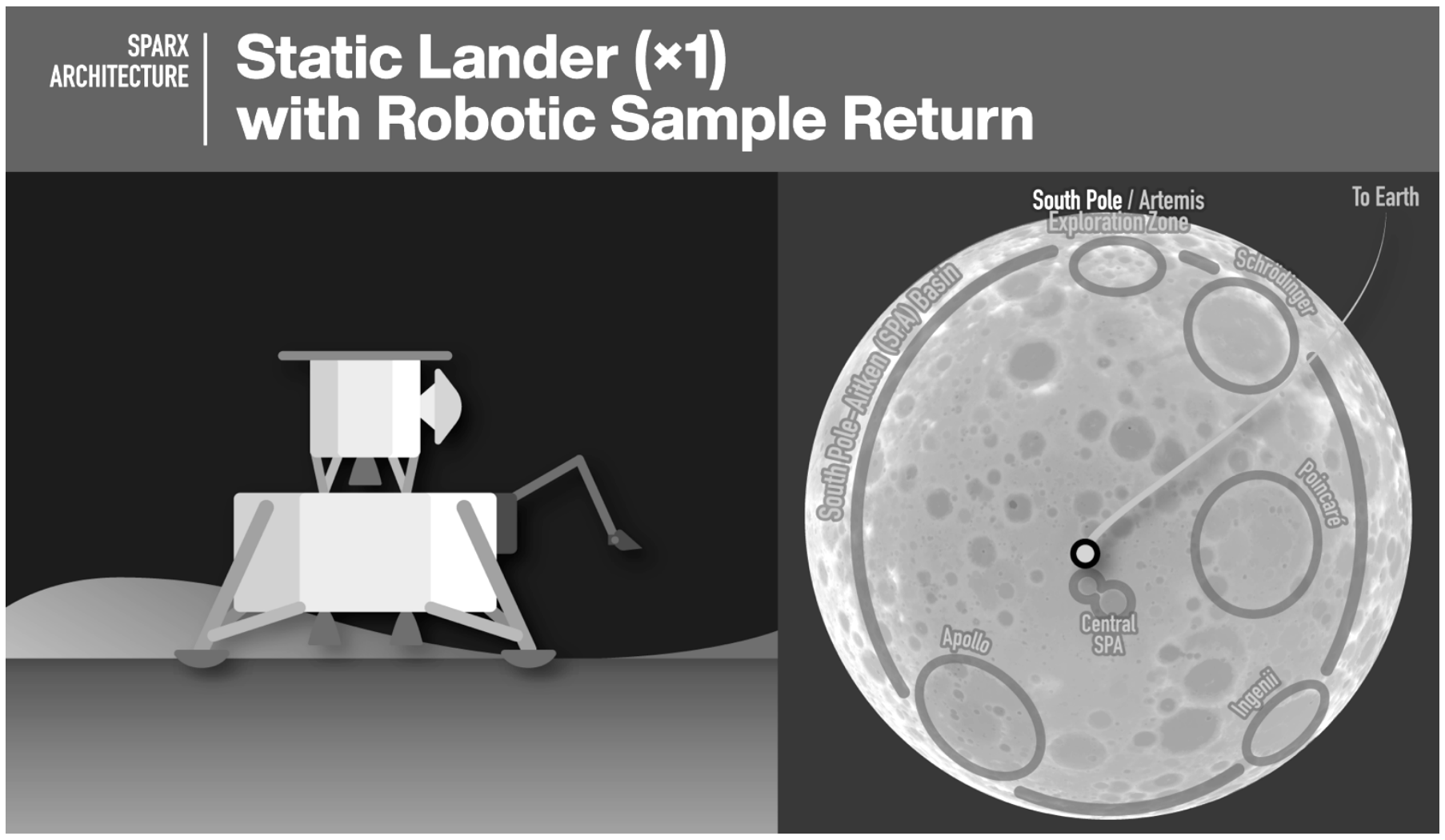

**Figure 5.31.** *An example for a notional SPARX mission architecture. Left: cartoon of the architecture. Right: notional map for how that architecture may be deployed across the surface. These diagrams are for illustrative purposes only, and do not represent real point-designs nor analyses of landing sites, traverses, etc.*

| | |
|---|---|
| EXCELLENT | Strengths substantially exceed weaknesses |
| VERY GOOD | Strengths outweigh weaknesses |
| GOOD | Strengths and weaknesses balance |
| FAIR | Weaknesses outweigh strengths |
| POOR | Weaknesses (potentially fatal flaws) substantially exceeds strength |

**Table 5.12.** *Qualitative adjectival rankings used to evaluate alternative architectures.*

### 5.7.1 Static Lander with Robotic Sample Return

SPARX ARCHITECTURE

## Static Lander with Robotic Sample Return

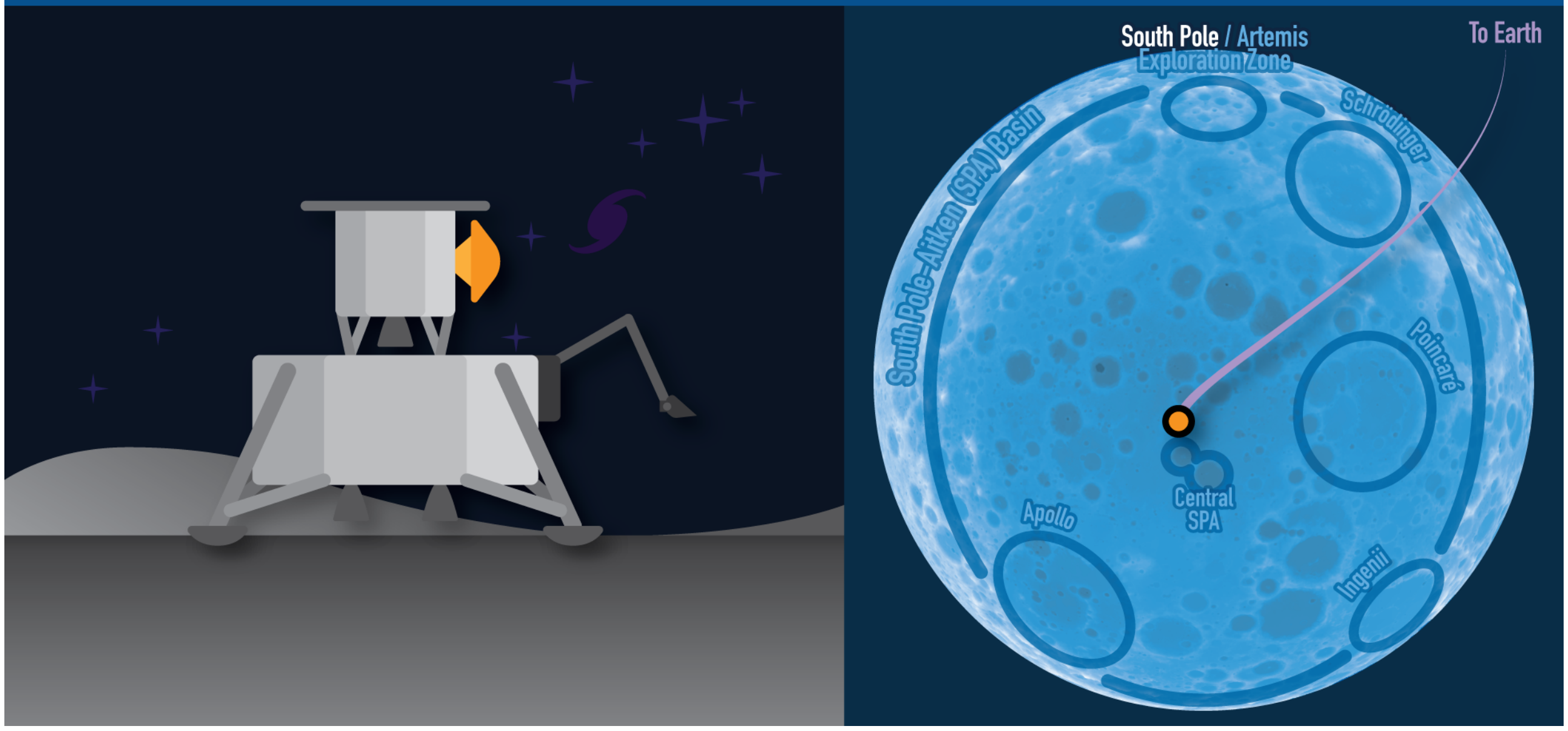

| SUMMARY OF THIS ARCHITECTURE: | |
|---|---|
| **DESCRIPTION:** | A single static landers lands in SPA, collects samples and *in situ* data, and returns the samples robotically. **Chang'e 6** and the **MoonRise** concept are examples of this type of architecture. |
| **NUMBER OF ARCHITECTURAL ELEMENTS:** | ~3 (1 lander, 1 earth return vehicle, 1 sample return capsule) |
| **MODE OF SAMPLE RETURN:** | Robotic sample return |
| **MODE OF SURFACE MOBILITY:** | Limited (robotic arm or similar) |

**EVALUATION OF THIS ARCHITECTURE AGAINST THE SPARX FIGURES OF MERIT:**

**STRENGTHS:**

– Architecture can meet SPARX sample return mass requirements for a single site (~2 kg)

**WEAKNESSES:**

– **Architecture does not meet SPARX Mission Requirements for sampling the geologic diversity of SPA (≥10 sites, ≥1,000 km range)**
– **Architecture does not meet the total SPARX sample mass requirements (≥10 kg from ≥10 sites)** owing to the low sample mass capability of robotic sample return
– **Architecture would not provide sufficient mobility to accomplish SPARX's traverse-centric Science Experiments (quasi-continuous near-surface measurements ≥650 km range)**
– **Architecture does not provide sufficient local-scale mobility for determining the geologic context for the samples (~10–100 m)**
– Samples would be collected from close to the lander, increasing risk of not meeting sample lithology requirements and increasing odds of contamination from the lander
– **Architecture is high cost:** Estimated cost for a single robotic sample return mission is ~$700M

**OVERALL EVALUATION:**

**POOR.** This architecture could not accomplish the SPARX science goals, primarily because of the lack of mobility necessary to collect diverse samples from across SPA. Sample return from a single site could provide incremental advancement to the SPARX Science Goals. Robotic sample return approaches may also be high cost, and the science return per cost is low.

### 5.7.2 Static Lander Campaign with Robotic Sample Return

SPARX ARCHITECTURE

## Static Lander Campaign with Robotic Sample Return

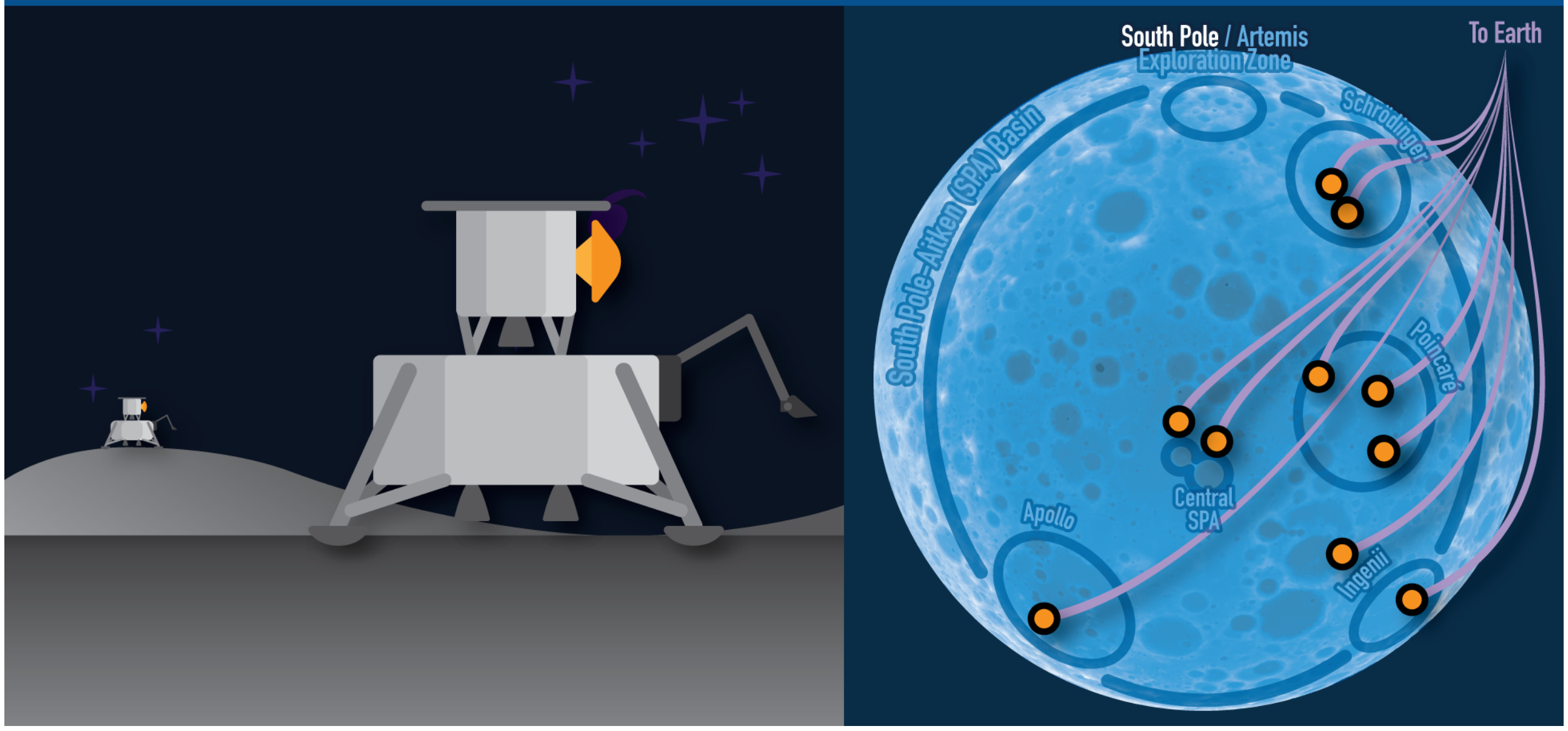

| SUMMARY OF THIS ARCHITECTURE: | |
|---|---|
| DESCRIPTION: | Multiple static landers lands across SPA, collects samples and *in situ* data, and returns the samples robotically. |
| NUMBER OF ARCHITECTURAL ELEMENTS: | ~30 to 60 (10–20 landers, each with 1 lander, 1 earth return vehicle, 1 sample return capsule) |
| MODE OF SAMPLE RETURN: | Robotic sample return (~20 to 40 kg total across all landers) |
| MODE OF SURFACE MOBILITY: | Limited (robotic arm or similar) |

**EVALUATION OF THIS ARCHITECTURE AGAINST THE SPARX FIGURES OF MERIT:**

**STRENGTHS:**

- **Architecture can meet the total SPARX sample mass requirements (≥10 kg)**
- **Architecture can meet the SPARX Mission Requirements for sampling the geologic diversity of SPA (≥10 sites, ≥1,000 km range)**
- Missions could be deployed across SPA with no constraints driven by sample return architecture or surface mobility
- The campaign approach would be more resilient to failure of individual missions (i.e., if one mission fails the SPARX Science Goals my still be achievable)

**WEAKNESSES:**

- **Architecture would not provide sufficient mobility to accomplish SPARX's traverse-centric Science Experiments (quasi-continuous near-surface measurements ≥650 km range)**
- **Architecture does not provide sufficient local-scale mobility for determining the geologic context for the samples (~10–100 m)**
- **Samples would be collected from close to the lander, increasing risk of not meeting sample lithology requirements and increasing odds of contamination from the lander**
- **Architecture is very high cost:** Estimated cost for the total campaign of 10–20 missions would be ~$7B–$14B (~$700M for each robotic sample return mission); an order-of-magnitude improvement in cost would be necessary for this architecture to be viable

**OVERALL EVALUATION:**

**FAIR.** While this architecture could address most of the sample return requirements, this approach would still provide inadequate *in situ* data and geologic context due to the lack of mobility. While this architecture could make substantial progress to addressing SPARX's Science Goals, the extremely high project cost (≥$7B) combined with the significant scientific risks make it untenable.

### 5.7.3 Hopper with Robotic Sample Return

SPARX ARCHITECTURE

## Hopper with Robotic Sample Return

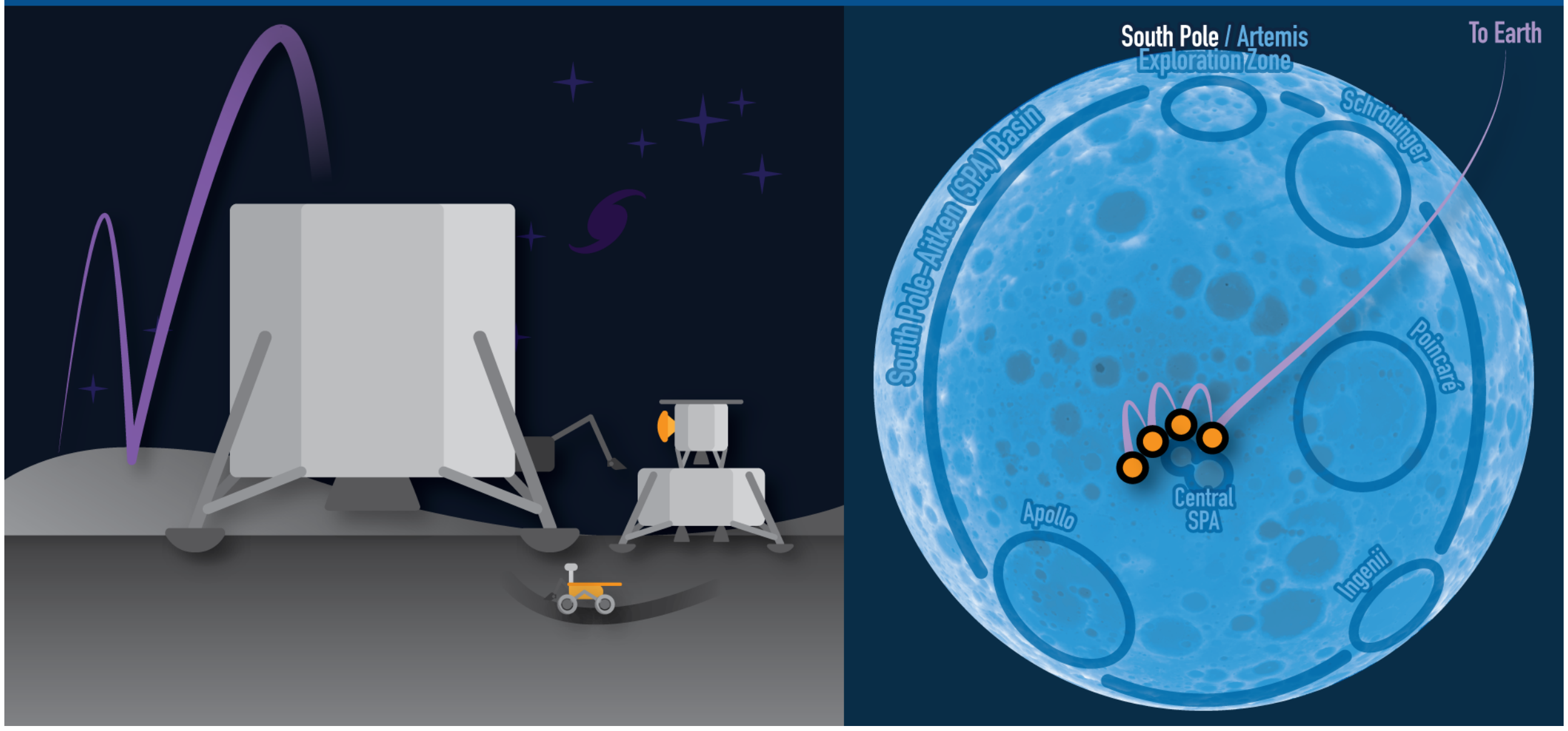

**SUMMARY OF THIS ARCHITECTURE:**

| | |
|---|---|
| **DESCRIPTION:** | A single hopper lands in SPA, collects samples and performs *in situ* analyses only, and hops to 3~4 sites before running out of fuel. |
| **NUMBER OF ARCHITECTURAL ELEMENTS:** | ~5 (1 hopper, 1 sample return lander, 1 earth return vehicle, 1 sample return capsule, 1 sample transfer rover) |
| **MODE OF SAMPLE RETURN:** | Robotic sample return |
| **MODE OF SURFACE MOBILITY:** | Hopper (~3 to 4 hops with a total range of 10s to 100s of km for Starship-class hoppers) with limited local-scale mobility (robotic arm or similar) |

**EVALUATION OF THIS ARCHITECTURE AGAINST THE SPARX FIGURES OF MERIT:**

**STRENGTHS:**

– None noted

**WEAKNESSES:**

– **Architecture does not meet SPARX Mission Requirements for sampling the geologic diversity of SPA (≥10 sites, ≥1,000 km range)**

– **Architecture does not meet the total SPARX sample mass requirements (≥10 kg from ≥10 sites) and may not meet the per-site sample mass requirements (~2 kg)** due to the low sample mass capability of robotic sample return

– **Architecture would not provide sufficient mobility to accomplish SPARX's traverse-centric Science Experiments (quasi-continuous near-surface measurements ≥650 km range)**: optimal hopping would take the hopper to great altitude, limiting effectiveness of remote sensing instruments

– **Architecture does not provide sufficient local-scale mobility for determining the geologic context for the samples (~10–100 m)**

– Samples would be collected from close to the lander, increasing risk of not meeting sample lithology requirements and increasing odds of contamination from the lander

– **Sample transfer between two surface assets (hopper and sample return vehicle) is a major technology development and may require the addition of a sample transfer rover**

– **Architecture is high cost:** Hopper cost is unknown, but the estimated cost for a single robotic sample return mission is ~$700M; the cost for sample transfer capability is likely large

**OVERALL EVALUATION:**

**POOR.** This architecture could not accomplish the SPARX science goals, primarily because of the lack of mobility necessary to collect diverse samples from across large regions of SPA. This architecture could not accomplish the SPARX Science Goals. A single mission of this type would only provide incremental advances to the SPARX Science Goals. Robotic sample return approaches may also be high cost, and the science return per cost is low.

### 5.7.4 Hopper Rendezvous with Polar Artemis Mission and/or Moon Base

SPARX ARCHITECTURE

## Hopper Rendezvous with Polar Artemis Mission and/or Moon Base

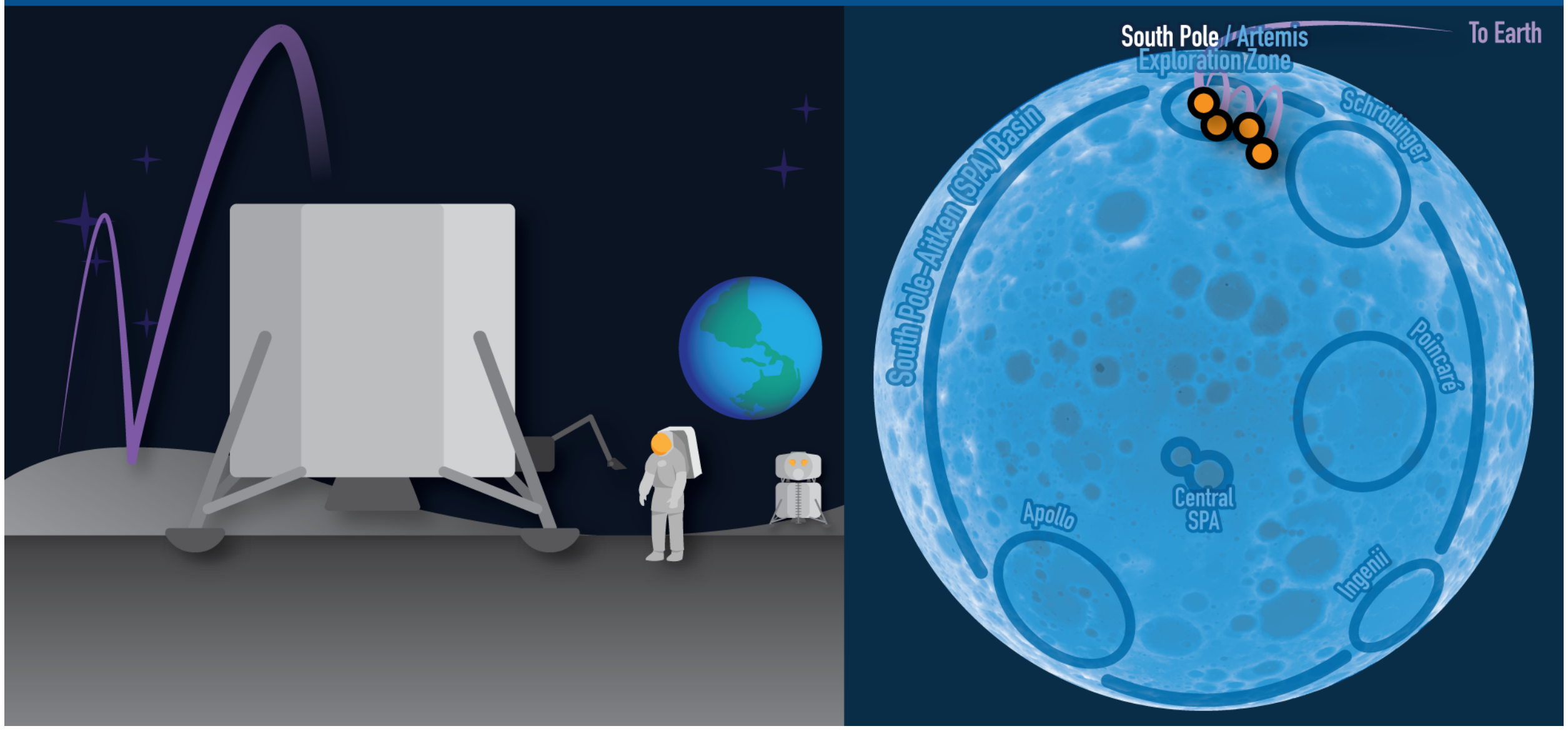

| SUMMARY OF THIS ARCHITECTURE: | |
|---|---|
| **DESCRIPTION:** | A single hopper lands in SPA, collects samples and performs *in situ* analyses, and hops to 3~4 sites before running out of fuel in the vicinity of a polar Artemis mission |
| **NUMBER OF ARCHITECTURAL ELEMENTS:** | 1 hopper plus 1 South Pole Artemis mission |
| **MODE OF SAMPLE RETURN:** | Artemis astronaut sample return (up to ~100 kg) |
| **MODE OF SURFACE MOBILITY:** | Hopper (~3 to 4 hops with a total range of 10s to 100s of km for Starship-class hoppers) with limited local-scale mobility (robotic arm or similar) |

**EVALUATION OF THIS ARCHITECTURE AGAINST THE SPARX FIGURES OF MERIT:**

**STRENGTHS:**

– None noted

**WEAKNESSES:**

– **Architecture does not meet SPARX Mission Requirements for sampling the geologic diversity of SPA (≥10 sites, ≥1,000 km range)**
– **Architecture does not meet the total SPARX sample mass requirements (≥10 kg from ≥10 sites)**, and architecture may not meet the per-site SPARX sample mass requirements (≥2 kg) due to limitations with robotic sample return
– **Architecture cannot access high-priority SPARX regions of interest beyond the South Pole**
– **Architecture would not provide sufficient mobility to accomplish SPARX's traverse-centric Science Experiments (quasi-continuous near-surface measurements ≥650 km range)**: optimal hopping would take the hopper to great altitude, limiting effectiveness of remote sensing instruments
– **Architecture does not provide sufficient local-scale mobility for determining the geologic context for the samples (~10–100 m)**
– Samples would be collected from close to the lander, increasing risk of not meeting sample lithology requirements and increasing odds of contamination from the lander
– The traverse and mission schedule are constrained by the requirement to rendezvous with Artemis astronauts at the South Pole

**OVERALL EVALUATION:**

**POOR.** This architecture could not accomplish the SPARX science goals, primarily because of the lack of mobility necessary to collect diverse samples from across large regions of SPA. Hopping to the South Pole limits the total range to regions proximal to the South Pole, which are not high value for SPARX. A single mission of this type would only provide incremental advances to the SPARX Science Goals.

### 5.7.5 Short-Range Rover with Robotic Sample Return

SPARX ARCHITECTURE

## Short-Range Rover with Robotic Sample Return

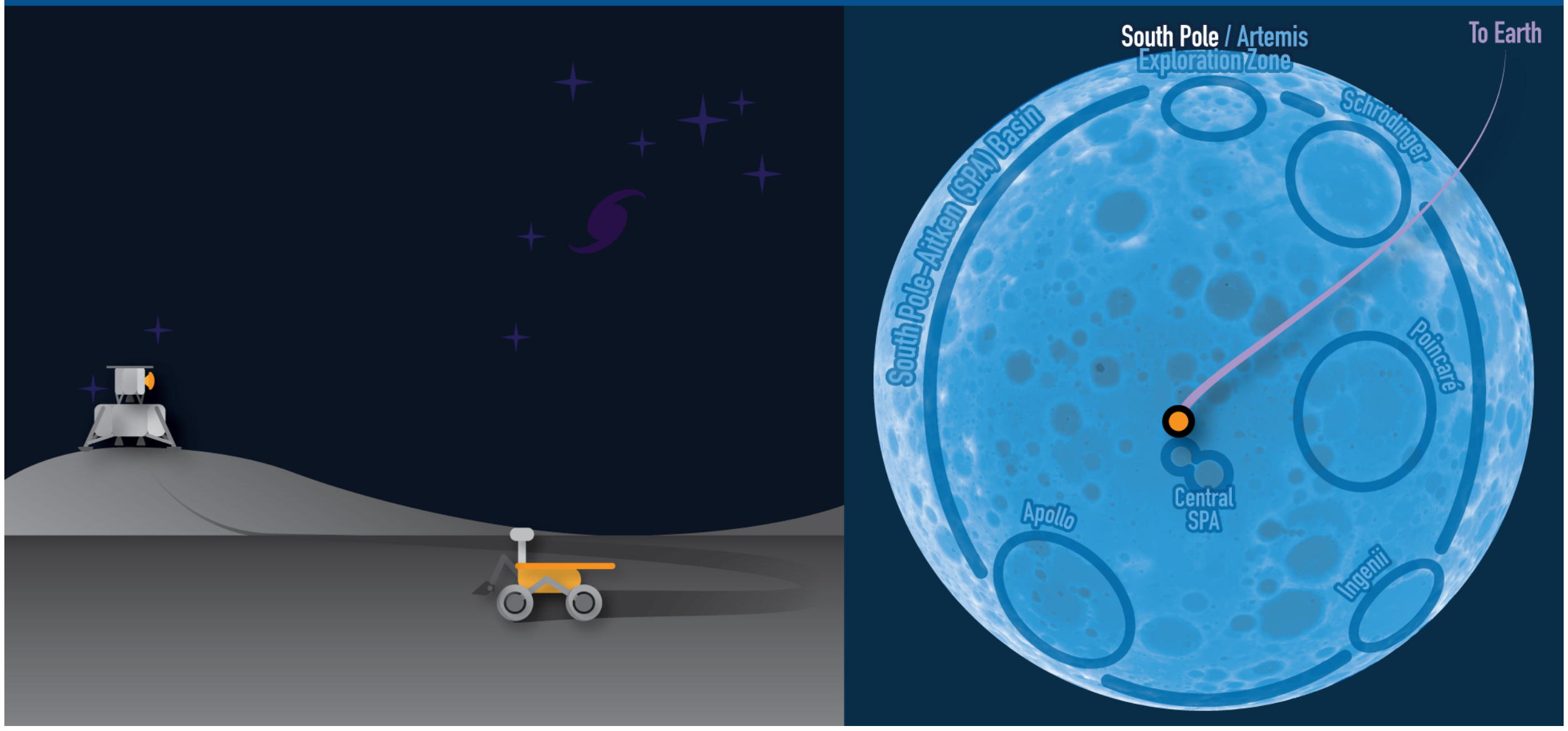

| SUMMARY OF THIS ARCHITECTURE: | |
|---|---|
| **DESCRIPTION:** | A single static landers lands in SPA, and deploys a small (low mass, short-lived) rover that collects samples and *in situ* data that it returns to Earth robotically. |
| **NUMBER OF ARCHITECTURAL ELEMENTS:** | ~4 (1 small rover, 1 lander, 1 earth return vehicle, 1 sample return capsule) |
| **MODE OF SAMPLE RETURN:** | Robotic sample return |
| **MODE OF SURFACE MOBILITY:** | Small rover with ~10–100 km range |

**EVALUATION OF THIS ARCHITECTURE AGAINST THE SPARX FIGURES OF MERIT:**

**STRENGTHS:**

- Architecture can meet SPARX sample return mass requirements for a single site (~2 kg)
- Samples could be collected far from the lander, decreasing the risk of contamination from the lander and increasing the probability of meeting sample lithology requirements

**WEAKNESSES:**

- **Architecture does not meet SPARX Mission Requirements for sampling the geologic diversity of SPA (≥10 sites, ≥1,000 km range)**
- **Architecture does not meet the total SPARX sample mass requirements (≥10 kg from ≥10 sites)** owing to the low sample mass capability of robotic sample return
- **Architecture would not provide sufficient mobility to accomplish SPARX's traverse-centric Science Experiments (quasi-continuous near-surface measurements ≥650 km range)**; a small rover may contribute to these Experiments but cannot accomplish them
- **Architecture is high cost:** Estimated cost for a single robotic sample return mission is ~$700M; the cost for a small rover depends on the capabilities of the rover and could plausibly cost ~$500M or more based on analogies to other Moon/Mars rovers (e.g., VIPER).

**OVERALL EVALUATION:**

**POOR.** This architecture could not accomplish the SPARX Science Goals. A single mission of this type would only provide incremental advances to the SPARX Science Goals. The science return per cost is low.

### 5.7.6 Long-Range Rover with Robotic Sample Return

SPARX ARCHITECTURE

## Long-Range Rover with Robotic Sample Return

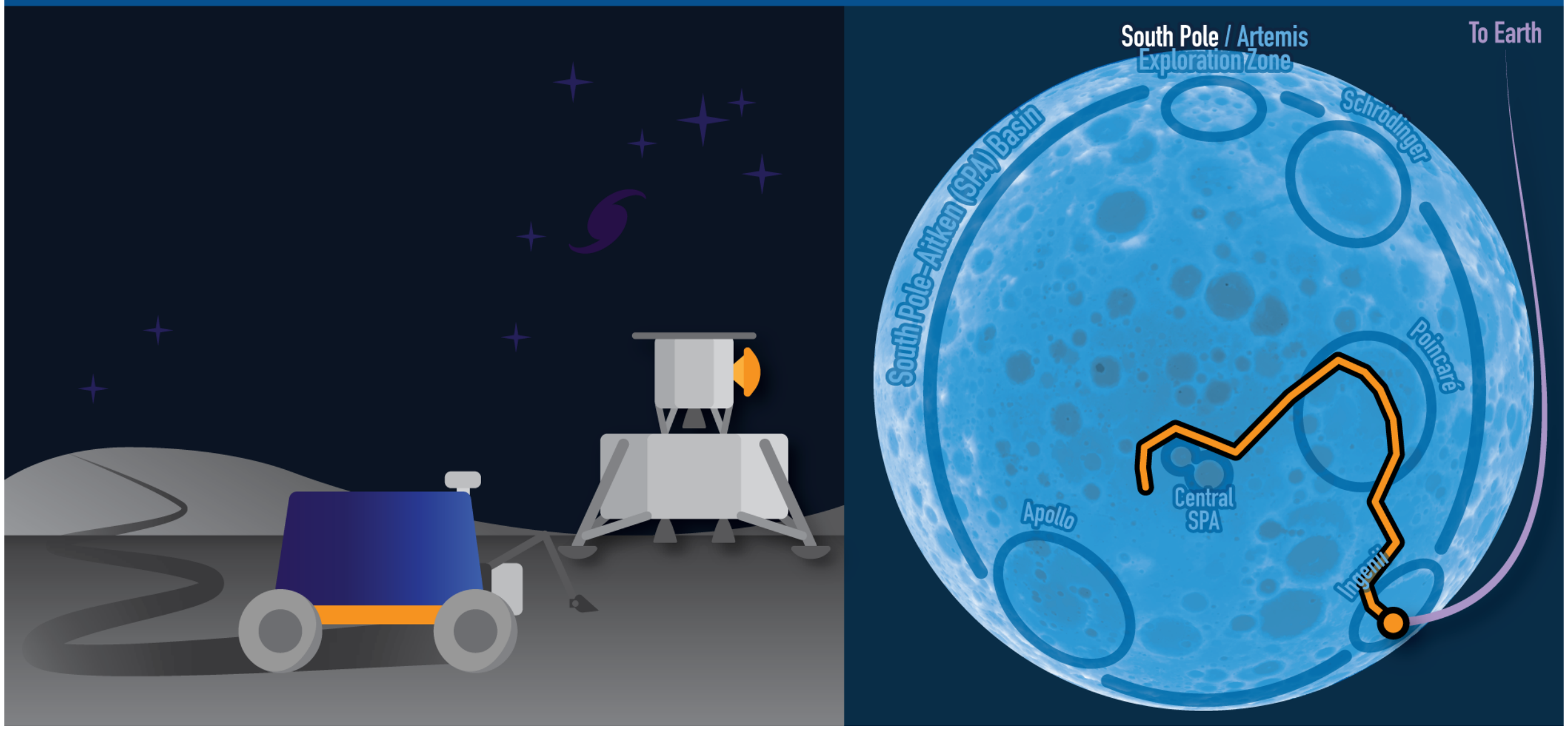

| SUMMARY OF THIS ARCHITECTURE: | |
|---|---|
| **DESCRIPTION:** | A large rover lands in SPA, traverses the basin collecting samples and *in situ* data, and ultimately rendezvous with an independently landed robotic sample return vehicle elsewhere in SPA. This architecture is consistent with the **Endurance-R** concept study. |
| **NUMBER OF ARCHITECTURAL ELEMENTS:** | ~5 (1 large rover delivered by 1 lander, plus 1 sample return mission consisting of 1 lander, 1 earth return vehicle, 1 sample return capsule) |
| **MODE OF SAMPLE RETURN:** | Robotic sample return |
| **MODE OF SURFACE MOBILITY:** | Large rover with >2,000 km range |

**EVALUATION OF THIS ARCHITECTURE AGAINST THE SPARX FIGURES OF MERIT:**

**STRENGTHS:**

- **Architecture can meet the SPARX Mission Requirements for sampling the geologic diversity of SPA (≥10 sites, ≥1,000 km range)**
- **Architecture can provide sufficient mobility to accomplish SPARX's traverse-centric Science Experiments (quasi-continuous near-surface measurements ≥650 km range)**
- **Architecture can provide sufficient local-scale mobility for determining the geologic context for the samples (~10–100 m)**
- Samples could be collected far from the lander, decreasing the risk of contamination from the lander and increasing the probability of meeting sample lithology requirements
- The traverse can be entirely driven by science requirements, with no constrains determined by the sample return architecture (e.g., needing to rendezvous with Artemis)

**WEAKNESSES:**

- **Architecture might not meet the total SPARX sample mass requirements (≥10 kg from ≥10 sites)** owing to the low sample mass capability of robotic sample return. While the TOR prescribed sample mass limit is consistent with the SPARX threshold sample return mass, the lack of clarification about additional mass margin raises concerns.

**OVERALL EVALUATION:**

**GOOD.** This architecture could make significant advances to the SPARX Science Goals. The primary limitation is the total sample return mass. If robotic sample return capabilities were improved to > 10 kg plus additional mass for sample containers and margin (e.g., 20 kg total return mass), the architecture would rate "Excellent."

### 5.7.7 Long-Range Rover Rendezvous with Polar Artemis Mission and/or Moon Base

SPARX ARCHITECTURE | **Long-Range Rover Rendezvous with Polar Artemis Mission and/or Moon Base**

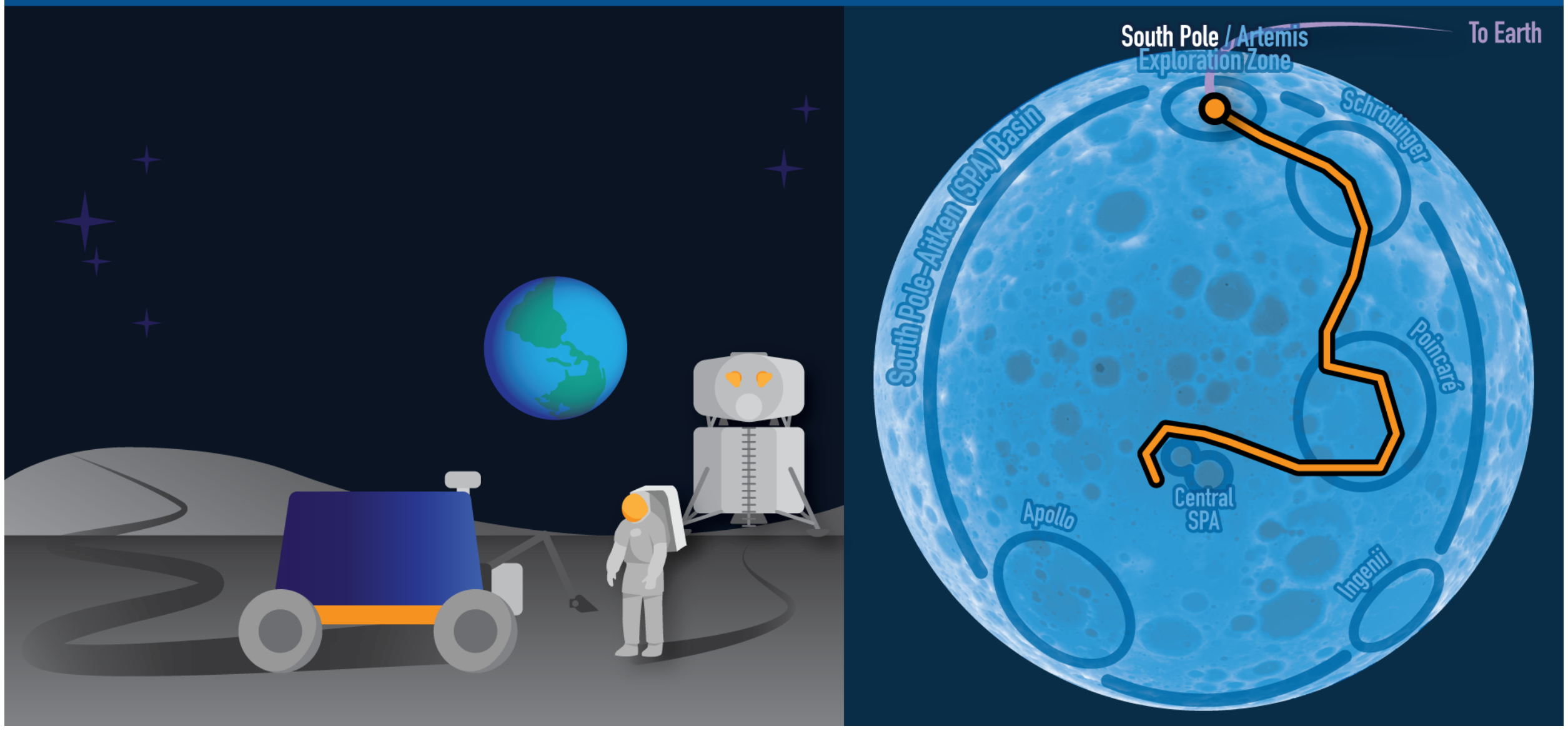

| SUMMARY OF THIS ARCHITECTURE | |
|---|---|
| **DESCRIPTION:** | A large rover lands in SPA, traverses the basin collecting samples and *in situ* data, and ultimately rendezvous with Artemis at the lunar South Pole. This architecture is consistent with the **Endurance-A** concept study. |
| **NUMBER OF ARCHITECTURAL ELEMENTS:** | ~2 (1 large rover delivered by 1 lander) plus 1 South Pole Artemis mission |
| **MODE OF SAMPLE RETURN:** | Artemis astronaut sample return (up to ~100 kg) |
| **MODE OF SURFACE MOBILITY:** | Large rover with >2,000 km range |

**EVALUATION OF THIS ARCHITECTURE AGAINST THE SPARX FIGURES OF MERIT:**

**STRENGTHS:**

- **Architecture can meet the total SPARX sample mass requirements (≥10 kg from ≥10 sites)**
- **Architecture can meet the SPARX Mission Requirements for sampling the geologic diversity of SPA (≥10 sites, ≥1,000 km range)**
- **Architecture can provide sufficient mobility to accomplish SPARX's traverse-centric Science Experiments (quasi-continuous near-surface measurements ≥650 km range)**
- **Architecture can provide sufficient local-scale mobility for determining the geologic context for the samples (~10–100 m)**
- Samples could be collected far from the lander, decreasing the risk of contamination from the lander and increasing the probability of meeting sample lithology requirements
- **Architecture has high science value per dollar:** This architecture completely addresses all of the SPARX Science Goals and does so with one of the lowest project costs: ~$840M for the rover

**WEAKNESSES:**

- The traverse and mission schedule are constrained by the requirement to rendezvous with Artemis astronauts at the South Pole

**OVERALL EVALUATION:**

**EXCELLENT** This architecture would completely address the SPARX Science Goals. This architecture has extremely high science value per dollar.

### 5.7.8 Long-Range Rover Rendezvous with Non-Polar Artemis Sortie Mission

SPARX ARCHITECTURE

## Long-Range Rover Rendezvous with Non-Polar Artemis Sortie Mission

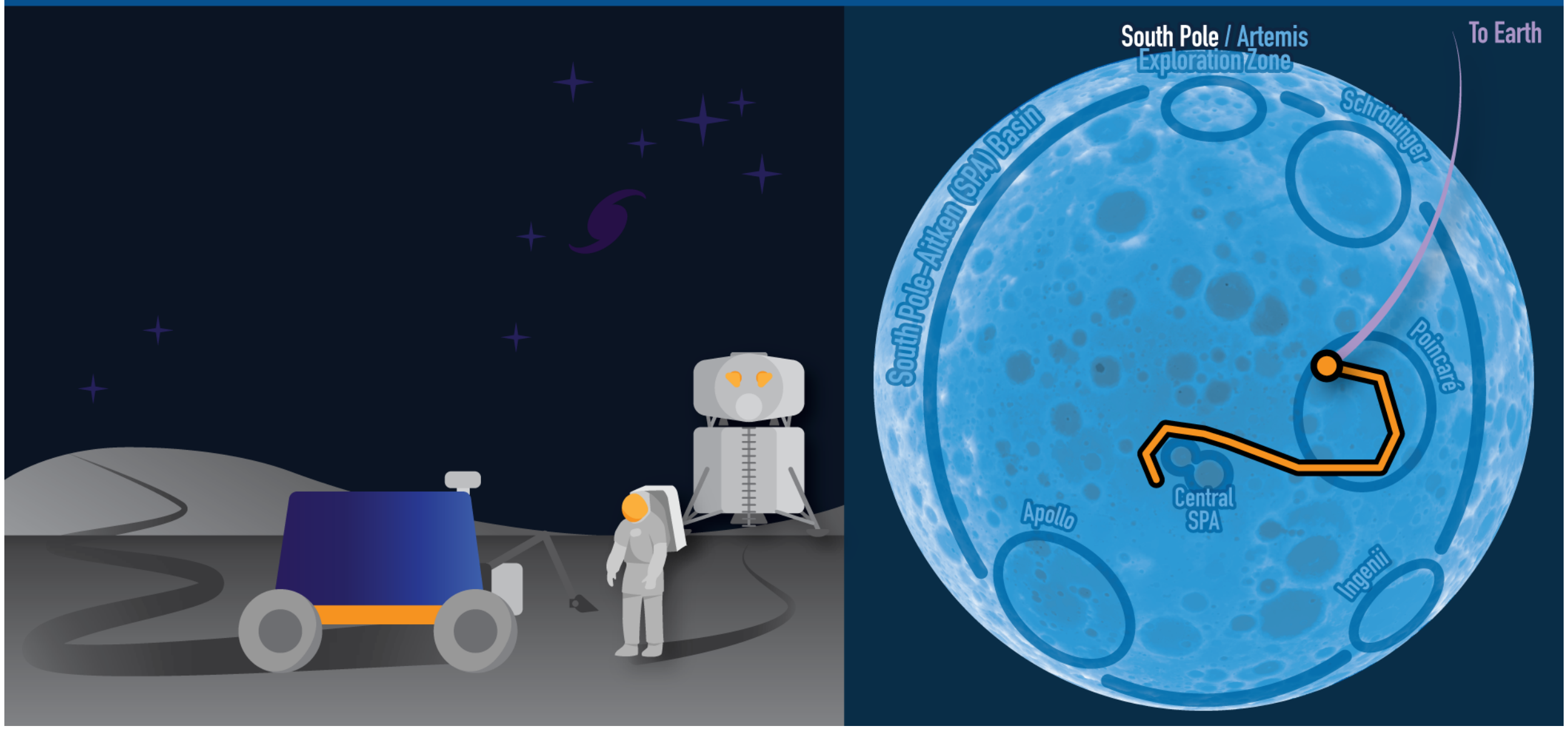

**SUMMARY OF THIS ARCHITECTURE:**

| | |
|---|---|
| **DESCRIPTION:** | A large rover lands in SPA, traverses the basin collecting samples and *in situ* data, and ultimately rendezvous with Artemis at a non-polar landing site. |
| **NUMBER OF ARCHITECTURAL ELEMENTS:** | ~2 (1 large rover delivered by 1 lander) plus 1 non-polar Artemis sortie mission |
| **MODE OF SAMPLE RETURN:** | Artemis astronaut sample return (up to ~100 kg) |
| **MODE OF SURFACE MOBILITY:** | Large rover with >2,000 km range, plus crewed explorers (range ~2–10 km) |

**EVALUATION OF THIS ARCHITECTURE AGAINST THE SPARX FIGURES OF MERIT:**

**STRENGTHS:**

- **Architecture can meet the total SPARX sample mass requirements (≥10 kg from ≥10 sites)**
- **Architecture can meet the SPARX Mission Requirements for sampling the geologic diversity of SPA (≥10 sites, ≥1,000 km range)**
- **Architecture can provide sufficient mobility to accomplish SPARX's traverse-centric Science Experiments (quasi-continuous near-surface measurements ≥650 km range)**
- **Architecture can provide sufficient local-scale mobility for determining the geologic context for the samples (~10–100 m)**
- Samples could be collected far from the lander, decreasing the risk of contamination from the lander and increasing the probability of meeting sample lithology requirements
- **Architecture provides a compelling combination of robotic and crewed explorers**, as the crewed mission can directly contribute to addressing SPARX Science Goals by exploring a SPARX site
- Of the large rover options, this approach maximizes science with the shortest traverse
- **Architecture has high science value per dollar:** This architecture completely addresses all of the SPARX Science Goals and does so with one of the lowest project costs: ~$840M for the rover

**WEAKNESSES:**

- This architecture requires a non-polar Artemis mission which must also sync up with the delivery of the rover or a rover sample cache.

**OVERALL EVALUATION:**

**EXCELLENT** This architecture would completely address the SPARX Science Goals. This architecture fully leverages the capabilities of well-trained Artemis astronauts by incorporating field geology by a non-polar Artemis mission. This architecture has extremely high science value per dollar.

### 5.7.9 Polar Artemis Sortie Mission

SPARX ARCHITECTURE

## Polar Artemis Sortie Mission

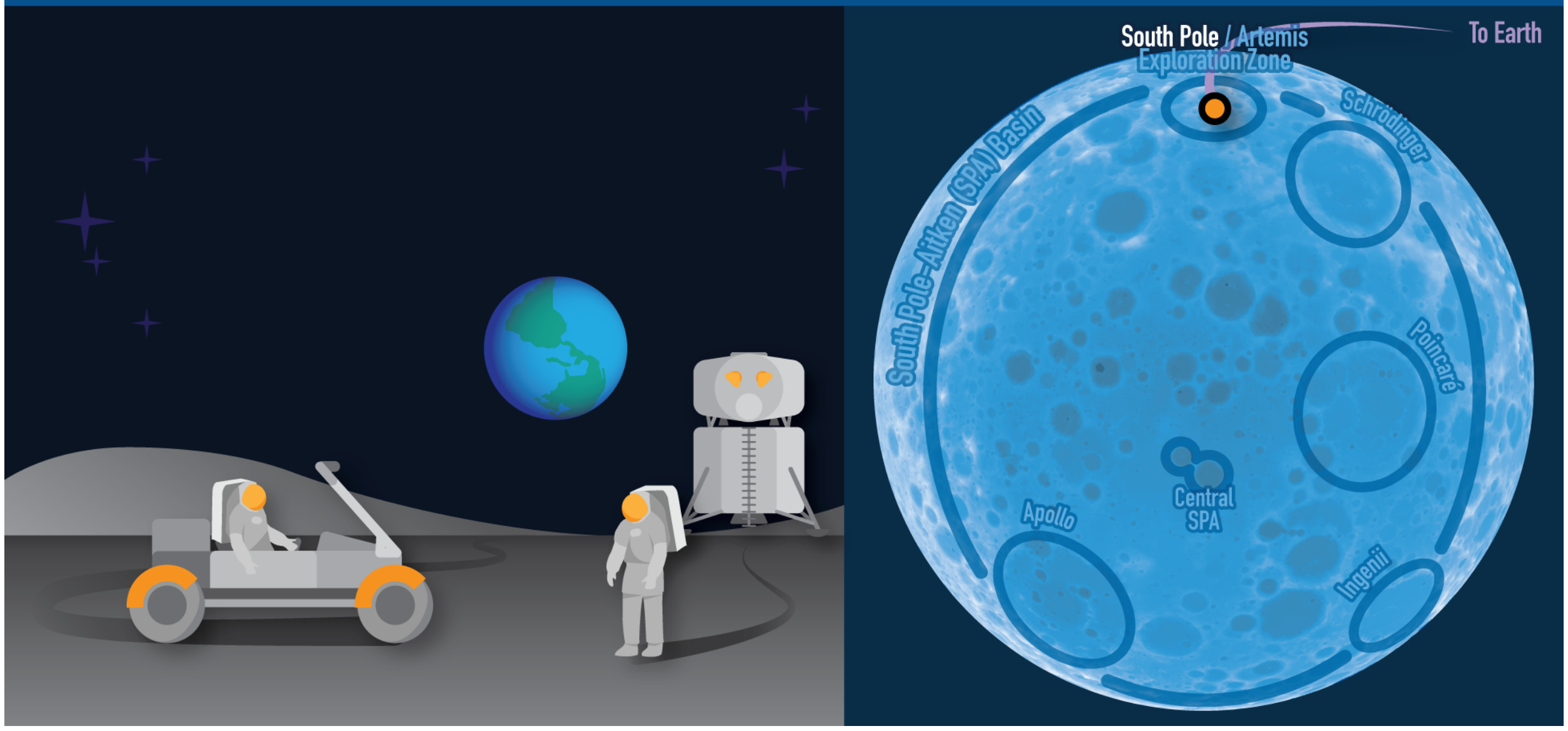

| SUMMARY OF THIS ARCHITECTURE: | |
|---|---|
| **DESCRIPTION:** | A crewed Artemis mission lands near the lunar South Pole in the Artemis exploration zone (poleward of 84°S), performs EVAs, collects samples and *in situ* measurements, and ultimately returns to Earth |
| **NUMBER OF ARCHITECTURAL ELEMENTS:** | ~1 polar Artemis mission |
| **MODE OF SAMPLE RETURN:** | Artemis astronaut sample return (up to ~100 kg) |
| **MODE OF SURFACE MOBILITY:** | Astronauts on walking EVAs (≤2 km range) or potentially with unpressurized rovers (≤10 km range) |

**EVALUATION OF THIS ARCHITECTURE AGAINST THE SPARX FIGURES OF MERIT:**

| STRENGTHS: | WEAKNESSES: |
|---|---|
| – **Architecture leverages well-trained astronaut crew who are adaptable and efficient, maximizing science return**<br>– Samples could be collected far from the lander, decreasing the risk of contamination from the lander and increasing the probability of meeting sample lithology requirements | – **Architecture does not meet SPARX Mission Requirements for sampling the geologic diversity of SPA (≥10 sites, ≥1,000 km range)**<br>– **Architecture does not meet the total SPARX sample mass requirements (≥10 kg from ≥10 sites)**<br>– **Architecture cannot access high-priority SPARX regions of interest beyond the South Pole**<br>– **Architecture would not provide sufficient mobility to accomplish SPARX's traverse-centric Science Experiments (quasi-continuous near-surface measurements ≥650 km range)** |

**OVERALL EVALUATION:**

**POOR** This architecture could not accomplish the SPARX Science Goals. A single mission of this type would only provide incremental advances to the SPARX Science Goals, primarily because the Moon's south polar region does not include any of the priority science targets. Furthermore, the range of astronaut explorers—even with crewed rovers—is insufficient for SPARX's "traverse science" requirements.

### 5.7.10 Architecture Summary

Table 5.13 summarizes all of the mission architectures considered in the previous section, plus the additional architectures shown in Appendix 10.4, and their overall evaluation. This section summarizes several major SDT findings from the overall mission architecture investigation.

| Architecture: | | Evaluation: | |
|---|---|---|---|
| Static Landers | Static Lander with *In situ* Science Only | POOR | |
| | Static Lander Campaign with *In situ* Science Only | POOR | |
| | Static Lander with Robotic Sample Return | POOR | |
| | Static Lander Campaign with Robotic Sample Return | FAIR | |
| Hoppers | Hopper with *In situ* Science Only | POOR | |
| | Hopper with Robotic Sample Return | POOR | |
| | Hopper Rendezvous with Polar Artemis Mission and/or Moon Base | POOR | |
| | Hopper with On-Orbit Sample Return | POOR | |
| | Refuellable Hopper with On-Orbit Refueling and Sample Return | FAIR | |
| | "Daisy Chain" Hopper Campaign | FAIR | |
| Rovers | Short-Range Rover with Robotic Sample Return | POOR | |
| | Long-Range Rover with Robotic Sample Return | GOOD | |
| | Long-Range Rover Rendezvous with Polar Artemis Mission and/or Moon Base | EXCELLENT | ←SPARX Baseline |
| | Long-Range Rover Rendezvous with Non-Polar Artemis Sortie Mission | EXCELLENT | |
| Crewed Missions Only | Polar Artemis Sortie Mission | POOR | |
| | Polar Artemis Campaign and/or Moon Base | POOR | |
| | Non-Polar Artemis Sortie Mission | GOOD | |
| | Non-Polar Artemis Campaign | VERY GOOD | |

**Table 5.13.** *Summary of analysis of mission architectures for SPARX.*

The major SDT findings from the overall mission architecture investigation are listed below:

1. **Large, long-range rovers that rendezvous with Artemis astronauts fully accomplish the SPARX science objectives and provide the highest science value per dollar.** Of all of the architectures considered, only two achieved an "excellent" ranking: Large rovers that rendezvous with Artemis astronauts either at the South Pole Moon Base, or at a non-polar SPARX traverse location. These architectures meet, or exceed, all of the SPARX science requirements and also provide an opportunity for human-robot teaming. Of these two options, only the "South Pole/Moon Base rendezvous" option is consistent with the guidance of the SPARX TOR.
2. **Static landers, hoppers, and short-range rovers are insufficient for addressing the SPARX Science Goals—even if deployed en masse.** These architectures are unable to meet the SPARX requirements for sampling specific geologic terrains across SPA which are often separated by 100s of kilometers. Even hypothetical Starship-class landers can only hop a fraction of the required distance (Sections 5.6). These architectures also suffer from the inability to accomplish SPARX's required traverse-centric Science Experiments.
3. **Astronaut-facilitated sample return is strongly enabling for accomplishing SPARX's Science Goals.** The SPARX total sample mass requirements (baseline: ≥20.5 kg; threshold: ≥9.8 kg, unmargined) may not be achievable with robotic sample return capabilities as defined by the SPARX TOR (capability: ≤10 kg). Thus, the only remaining avenue for returning large masses of samples is astronaut-facilitated sample return (capability: ~100 kg). Astronaut-facilitated sample return also provides an opportunity for astronauts to work synergistically with surface assets, increasing the overall science return.

4. **Commercial robotic sample return mass of ≤10 kg is within the scope of the proposed SPARX threshold sample mass of ≥9.8 kg.** Further development of robotic sample return mass capabilities to return this mass with sufficient margin for additional required systems (sample transfer, sample containers) would be significantly enabling for SPARX architecture options that do not require astronauts. Furthermore, the development of large sample return mass capabilities would also add resiliency to a mission architecture that originally baselines Artemis sample return, should that capability not be sustained. Neither NASA nor industry have accomplished robotic lunar sample return, and so developing this capability may be a substantial (and costly) technological development.
5. **Even with multiple missions and a future Moon Base, crewed exploration of the Artemis South Pole exploration zone (poleward of 84°S) alone is insufficient for addressing SPARX's Science Goals.** While the lunar South Pole is scientifically interesting for a variety of investigations (e.g., understanding polar volatiles), the South Pole is not a priority target for addressing SPARX's Science Experiments. None of the geologic settings required for SPARX investigations are known to be accessible in the Artemis South Pole exploration zone, and the nearest sites of interest are several 100s of kilometers away, and far beyond the reach of crewed expeditions from Moon Base. While Artemis missions to the South Pole have the potential to make serendipitous discoveries related to SPARX (e.g., by analyzing impact breccias mixed within the regolith), fundamental questions will remain until samples are collected directly from the required terrains within SPA.
6. **A non-polar Artemis rendezvous with SPARX could substantially enhance the science return of SPARX, while also requiring the shortest traverse.** While non-polar Artemis sorties were not within the scope of the original SPARX TOR (Appendix 10.1), the SDT did consider options for this architecture at the direction of NASA (Section 2). This architecture is particularly compelling if combined with a long-range rover, as such a scenario would provide an opportunity to blend the strengths of both robotic and crewed approaches, which is discussed in more detail in Section 7. The long-range rover could "cover the distance"—collecting samples and *in situ* data from long distances before the arrival of the astronauts. The astronauts could do a "deep dive" on a single location, ideally with the most complicated geology that would be challenging to interrogate robotically and meet the rover on its arrival. There may also be other science and exploration drivers for non-polar Artemis sortie missions in SPA. For example, SPA and the lunar farside may be an attractive destination for radio astronomy beyond the interference of Earth. In parallel to the SPARX SDT study, NASA requested a study from the National Academies of Science, Engineering, and Medicine (NASEM) about non-polar Artemis missions ("Key Non-Polar Destinations Across the Moon to Address Decadal-level Science Objectives with Human Explorers"). NASA may wish to investigate this hybrid approach in more detail pending the results of this NASEM study.

## 5.8 Baseline and Threshold Mission Concept

Having evaluated the geologic settings required to meet the SPARX sample requirements (Section 5.2), SPARX science requirement Figures of Merit (Section 5.3), the viability of various sample return architectures (Section 5.4), sample acquisition systems (Section 5.5), mobility systems (Section 5.6), and overall architecture profiles (Section 5.7), we describe here the final SDT recommendation for a SPARX baseline and threshold mission concept.

### 5.8.1 SPARX Baseline Mission Concept

**Finding: The recommendation of the SDT is that a SPARX baseline mission be implemented as a long-distance rover that acquires samples using a robotic arm to scoop and sieve lunar regolith while traversing from the central SPA region through Poincare basin, and Schrödinger basin before rendezvousing with Artemis astronauts at the lunar South Pole.**

The baseline mission is designed to address all of the Science Objectives listed in the SPARX STM (Section 3.6). The SPARX baseline mission includes 21 science traverse stops, with samples being acquired from 20 locations. The details of the geologic settings and proposed locations for each of these sampling and science activities are shown in Section 5.8.3 and Foldout 5.1. The SDT notes that while the proposed baseline traverse meets all of the SPARX requirements, this solution is non-unique, and a future SPARX Science Team may decide to select alternative locations based on new interpretations or acquisition of orbital remote sensing data, mobility asset restrictions, or other programmatic requirements arising.

The total traverse length of the baseline mission is comprised of a ~2,230 km science traverse from central SPA through Poincare and Schrödinger basin, plus an additional 450 km traverse to the Artemis exploration zone and Moon Base, for a total traverse length of ~2,680 km (Figure 5.32).

The SPARX baseline mission sample requirements are described in Section 4 and shown in Foldout 4.2 (Section 4.3.6). The SPARX baseline mission requires ≥20.5 kg of material representing both sieved rocks and rocklets of specified diameters, and also un-sieved bulk regolith samples. The sample mass provided above is considered an unmargined lower limit, and does not consider the mass of sample containers, or additional sample mass that might be acquired pursuant of meeting the specified sample requirements.

The SPARX baseline mission requires 7 classes of *in situ* measurements: context imaging, context mineralogy, fine-scale mineralogy, bulk elemental chemistry, gravity field, fine-scale imaging, and magnetic field (further description in Section 4.2). The SPARX strawman payload consists of 5 instruments, with the requirements for both context imaging and fine-scale imaging being met by the same instrument, and the requirements for both context mineralogy and fine-scale mineralogy being met by the same instrument. The SPARX baseline mission does not require any contact measurements, and consequently none of the instruments need to be accommodated on the robotic sampling arm.

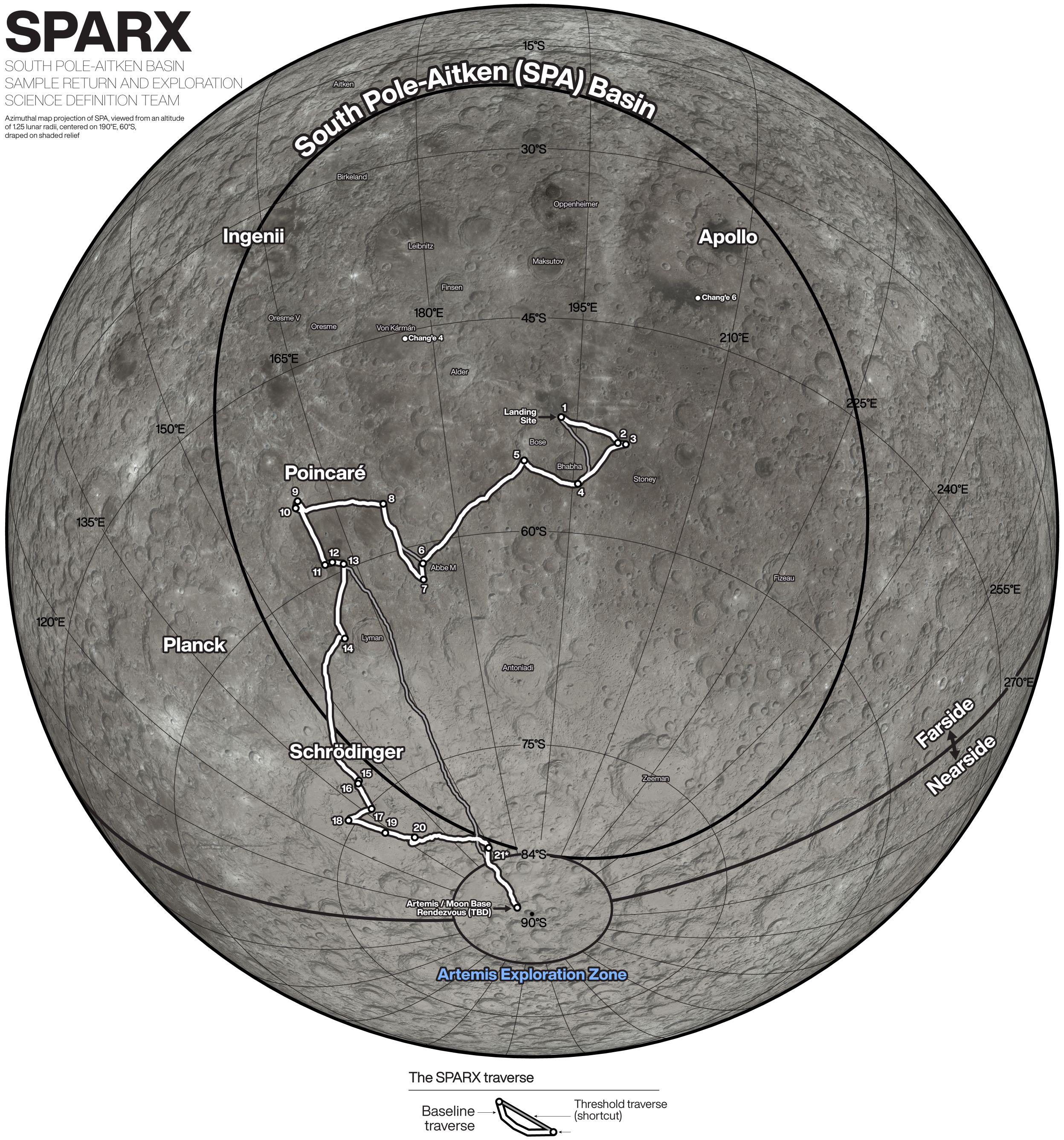

**Figure 5.32.** *Recommended baseline (white) and threshold (grey) traverse paths for the SDT recommended SPARX mission. The science sampling and traverse locations are numbers 1 to 21 for the baseline traverse. Both traverse paths are designed to rendezvous with Moon Base in the south polar Artemis Exploration Zone after completing their science activities. More details of this traverse and additional maps can be found in Section 6.7.*

### 5.8.2 SPARX Threshold Mission Concept

The SPARX threshold mission concept is designed to address the three highest priority SPARX objectives (Section 3.4) and make progress on five additional objectives (see SPARX STM, Section 3.6) (Figure 5.32). A comparison of the major characteristics of the threshold mission vs. the baseline mission is shown in Table 5.14. The SPARX threshold mission design prioritizes accomplishing the highest priority objectives within an acceptable level of scientific risk (Table 5.15). Using the traverse space defined by the sampling locations for the three highest priority objectives, the SDT identified additional SPARX Experiments and Objectives that could be addressed through either nearby or collocated sample acquisitions, maximizing the science return within the threshold envelope. The SDT notes that while the additional objectives cannot be addressed with the scientific rigor of the baseline mission, progress can still be made.

| MISSION CHARACTERISTICS | THRESHOLD | BASELINE |
|---|---|---|
| SPARX Objective Completion: | 75% | 100% |
| Total Traverse Stops: | 10 | 21 |
| Total Number of Samples: | 10 | 20 |
| Total Sample Mass: | ≥9.8 kg | ≥20.5 kg |
| Total Science Traverse: | 1,178 km | 2,227 km |
| Total Traverse with South Pole rendezvous: | 2,134 km | 2,672 km |
| Sample Acquisition Mode: | Scooping and sieving | Scooping and sieving |
| Surface Mobility: | Long-distance rover* | Long-distance rover* |
| Sample Return Mode(s): | Commercially derived robotic sample return†, or return from Moon Base or a non-polar Artemis sortie | Return from Moon Base or a non-polar Artemis sortie |
| Mission Duration‡: | ~2.7 years | ~3.7 years |

**Table 5.14.** *Comparison of major mission characteristics for the SPARX baseline and threshold mission concepts. *Long-distance rover includes all classes of custom and LTV-derived rovers that meet the minimum mission design requirements. †See discussion in Section 5.4 for caveats. ‡Average of mission duration simulations assuming an Artemis / Moon Base rendezvous scenario as described in Section 6.8.*

The SPARX threshold traverse (Figure 5.32) acquires samples from 10 sampling locations and does not include any non-sampling science traverse locations. The threshold traverse (described further in Section 6) starts in the central SPA region, and the primary science phases of the traverse end in Poincare basin. The total science traverse length of the threshold mission is ~1,180 km. The notional threshold traverse then includes an ~960 km traverse to the Artemis exploration zone for astronaut rendezvous and sample return. However, the SDT notes that commercially derived robotic sample return is potentially within scope, given the threshold sample mass requirements.

The SPARX threshold sample mass is ≥9.8 kg, including sieved rocks and rocklets of specified diameters, as well as un-sieved bulk regolith (Section 4.3.6, FO: Sample Requirements). As with the baseline mission, the sample mass requirements derived here do not include additional mass considerations, such as sample containers or mass of additional sample mass acquired while meeting the specified sample requirements.

| GOALS | OBJECTIVES | EXPERIMENTS | THRESHOLD | BASELINE |
|---|---|---|---|---|
| (G1) What was the chronology of key processes and events during the first half billion years of solar system history? | (O1.1) Quantify the age of the Moon to distinguish between early- and late-Moon formation models | (E1) Determine the composition and age of the anorthositic farside highlands crust | – | ✓✓✓ |
| | | (E2) Determine the age of SPA impact event | ✓✓ | ✓✓✓ |
| | | (E3) Determine the model age of the mantle underlying SPA | ✓✓ | ✓✓✓ |
| | (O1.2) Distinguish between fast- and slow-cooling models for lunar magma ocean crystallization | (E1) Determine the composition and age of the anorthositic farside highlands crust | – | ✓✓✓ |
| | | (E2) Determine the age of SPA impact event | ✓✓ | ✓✓✓ |
| | | (E3) Determine the model age of the mantle underlying SPA | ✓✓ | ✓✓✓ |
| | (O1.3) Determine if and when a giant planet instability drove a sharp change in the inner Solar System impactor flux | (E2) Determine the age of SPA impact event | ✓✓ | ✓✓✓ |
| | | (E4) Determine the age of ≥1 farside Imbrian-aged impact crater (D≥80 km) or basin (D≥300 km) with surface dateable by crater counts | X | ✓✓✓ |
| | | (E5) Determine the age of ≥1 non-SPA farside pre-Imbrian-aged impact basin (D≥300 km) | ✓✓✓ | ✓✓✓ |
| | (O1.4) Constrain the evolution of the lunar dynamo in the pre-4.0 Ga epoch of lunar history | (E6) Determine the age and remnant magnetization signatures from ≥3 returned samples with inferred ages ≥4.0 Ga | X | ✓✓✓ |
| (G2) Does the lunar farside preserve evidence of asymmetric magma ocean evolution and planetary crustal formation processes? | (O2.1) Determine if the farside highlands crust is similar to nearside crust in age and/or composition | (E1) Determine the composition and age of the anorthositic farside highlands crust | – | ✓✓✓ |
| | (O2.2) Determine if there was a farside KREEP reservoir, or other late-stage magma ocean products, prior to the SPA impact | (E7) Determine the composition and age of farside Th-rich materials | ✓✓✓ | ✓✓✓ |
| | (O2.3) Determine if and how the temporal and compositional evolution of the farside mantle differs from the nearside | (E8) Determine the composition and age of farside basaltic units | ✓✓ | ✓✓✓ |
| | | (E9) Sample hypothesized excavated lunar mantle material | X | ✓✓✓ |
| | | (E10) Determine the volatile content of the farside mantle as recorded by pyroclastic and basaltic materials | ✓ | ✓✓✓ |
| (G3) What are the effects of mega-basin impacts on terrestrial bodies? | (O3.1) Determine whether there are discontinuities in the structure (crustal thickness, density, magnetic field) across SPA to constrain models for the formation of a mega-basin | (E11) Determine the crustal structure (e.g., thickness, density, remnant magnetic field strength) across the SPA basin (from the center to the putative rim) to constrain mega-basin formation and evolution models | ✓✓✓ | ✓✓✓ |
| | | (E12) Determine if the massifs in the southwestern portion of the basin constitute the diffuse rim of SPA | X | ✓✓✓ |
| | (O3.2) Distinguish the relative contributions of impact melt differentiation, crustal backfilling, and volcanic infilling to the interior composition and stratigraphy of SPA | (E13) Determine the stratigraphy and composition of the unique deposits in central SPA | ✓ | ✓✓✓ |
| | | (E14) Determine the origin and extent of the low-calcium pyroxene annulus enclosing central SPA | X | ✓✓✓ |

**Table 5.15.** *Simplified version of the SPARX Science Traceability Matrix (STM), comparing the relative completion of SPARX Goals, Science Objectives, and Experiments between the Baseline and Threshold mission concepts. Dark green with three check marks (✓✓✓) indicates full sample and traverse requirements are met, lighter shades of green and fewer check marks indicate majority of science and traverse requirements are met with some science risk. Yellow with a dash (–) indicates partial completion with significant science risk. Red with an "X" indicates experiment unable to be met. More detailed STMs are in Foldouts 3.2, 4.1, 4.2, and 5.1.*

The SPARX threshold mission requires 5 classes of *in situ* measurements: context imaging, context mineralogy, fine-scale mineralogy, bulk elemental chemistry, and gravity field. The threshold strawman payload includes 4 instruments, and, as with the baseline mission, the threshold strawman instrument payload imaging and mineralogy elements are able to meet the requirements for both context and fine-scale measurements. Nevertheless, the SDT notes that fine-scale imaging in not a required measurement class for the threshold mission, and neither are magnetic field measurements.

### 5.8.3 SPARX Site and Traverse Requirements

The site and traverse requirements are summarized in Foldout 5.1.

| EXPERIMENTS | | MISSION REQUIREMENTS - SITES | DESIGN REFERENCE MISSION: SITES | | | | | | SITE PRECISION | | SAMPLE MASS (SEE SAMPLE REQUIREMENTS TABLE) |
|---|---|---|---|---|---|---|---|---|---|---|---|
| | | | SITE LABEL | INFORMAL NAME | REGION | JUSTIFICATION | LATITUDE | LONGITUDE | APPROXIMATE SCALE | JUSTIFICATION | |
| (E1) | Determine the composition and age of the anorthositic farside highlands crust | Sample ≥3 (≥[1]) geologically distinct exposures of farside anorthisitic highlands material | 9 | Poincaré X high albedo material | Poincaré | The ejecta blanket of Poincaré X is anomalously high albedo, consistent with exhumation of anorthositic material. | -54.190°N | 160.716°E | ~10.0 km | Scale of high albedo feature consistent with anorthosite-rich material | ≥2,244.6 g |
| | | | 18 | Schrödinger western peak ring | Schrödinger | M3 observations suggest the presence of anorthositic material in the peak ring of Schrödinger (Kramer et al. 2013). This specific 5-km diameter crater has been observed to have anorthositic (and noritic) material from M3 (Czaplinski et al. 2021). LROC NAC shows the presence of large, accessible boulders. Note, some of the rim massifs further west may also be adequate. | -74.450°N | 129.145°E | ~1.0 km | Scale of the crater rim and crater ejecta. | ≥2,244.6 g |
| | | | 16 | Schrödinger northern peak ring #1 | Schrödinger | M3 observations suggest the presence of anorthositic material in the peak ring of Schrödinger (Kramer et al. 2013). This specific site is selected at the base of the northern peak rim, and LROC NAC shows the presence of many large, accessible boulders at the base of the massif. | -73.270°N | 140.796°E | ~3.0 km | Scale of massif exposing and shedding targeted lithology. Note it is desirable to get to the base of the massif (which is a relatively sharp contact, but extends many kilometers laterally). | ≥2,244.6 g |
| (E2) | Determine the age of SPA impact event | Sample ≥[3] geologically distinct units hypothesized to be excavated or deposited SPA impact melt | 4 | Bhaba | Central SPA | Bhaba is a 64km diameter impact crater that impacted into the center of the SPA impact melt sheet (e.g., Moriarty & Pieters 2018). This specific location was selected based on M3 observations as the best plausible location for sampling noritic impact melt (from depths of >5 km). | -56.660°N | 195.650°E | ~0.1 km | Scale of M3 pixel size associated hypothesized to trace noritic impact melt | ≥970.7 g |
| | | | 5 | Bose | Central SPA | Bose is a 91 km diameter impact crater that impacted into the center of the SPA impact melt sheet (e.g., Moriarty & Pieters 2018). This specific location was selected based on M3 observations as the best plausible location for sampling noritic impact melt (from depths of >5 km). | -55.200°N | 189.050°E | ~0.1 km | Scale of M3 pixel size associated hypothesized to trace noritic impact melt | ≥970.7 g |
| | | | 20 | Schrödinger southern rim | Schrödinger | The southern rim of Schrödinger has been hypothesized as a good candidate for sampling SPA impact melt material (Hurwitz & Kring 2015) due to the presence of elevated FeO material and orthopyroxene/norite material as identified by M3. | -79.004°N | 138.861°E | ~0.1 × 10 km | It is preferential to sample near as close to the base of the rim slope as possible (~0.1 km), although the rim extends laterally (~10 km) | ≥970.7 g |
| (E3) | Determine the model age of the mantle underlying SPA | Sample ≥3 (≥[2]) spatially and/or temporally distinct mare units as mapped from orbital datasets (e.g., Pasckert et al. 2018) | 1 | "Mickey Mare" | Central SPA | "Mickey Mare" is a collection of mare basalt units in central SPA. This specific site was mapped as "CP1" by Pasckert et al. 2018. | -52.160°N | 193.226°E | ~20.0 km | Scale of mare unit | ≥484.0 g |
| | | | 8 | Mare in Poincaré E | Poincaré | Mare basalt filling the floor of Poincaré E. This specific site was mapped as "Po3" by Pasckert et al. 2018. | -56.656°N | 171.105°E | ~20.0 km | Scale of mare unit | ≥484.0 g |
| | | | 12 | Mare northwest of Poincaré Q | Poincaré | These mare basalts fill the central floor of Poincaré. | -59.156°N | 161.893°E | ~5.0 km | Scale of contact between Poincaré peak ring and mare basalt | ≥484.0 g |
| (E4) | Determine the age of ≥1 farside Imbrian-aged impact crater (D≥80 km) or basin (D≥300 km) | Sample ≥1 farside imbrian-aged impact crater or basin with a dateable surface (e.g., peak ring or ejecta blanket) | 15 | Schrödinger northern peak ring #2 | Schrödinger | Small (3 km diameter) impact crater in the northern peak ring of Schrödinger that is associated with orthopyrxone/norite material as identified by Kramer et al. 2013. | -73.161°N | 141.597°E | ~1.0 km | Approximate scale of the crater rim to be sampled | ≥970.7 g |
| (E5) | Determine the age of ≥[1] non-SPA farside pre-Imbrian-aged impact basin (D≥300 km) | For ≥[1] non-SPA farside pre-Imbrian impact basins (either Poincaré or Ingenii), sample ≥[3] geologically distinct units associated with the crater likely to retain a basin age (e.g., peak ring, rim) | 13 | Poincaré southern peak ring | Poincaré | Located on the southern peak ring of Poincaré. Peak rings are desirable targets for collecting impact melt, as they may exhume material from depth. The specific site was selected to be proximal to a small (~100-meter), fresh, rocky crater that plausibly excavated into the peak ring. | -59.594°N | 163.320°E | ~5 km | Approximate scale of the peak ring at this location. | ≥970.7 g |
| | | | 11** | Poincaré southern peak ring northwest of Poincaré Q | Poincaré | Located on the southern peak ring of Poincaré, northwest of Poincaré Q. Peak rings are desirable targets for collecting impact melt, as they may exhume material from depth. The specific site was selected to be proximal to a small (~100-meter), fresh, rocky crater that plausibly excavated into the peak ring. | -59.103°N | 160.773°E | ~5 km | Approximate scale of the peak ring at this location. | ≥970.7 g |
| | | | 10 | Poincaré northern peak ring near Poincaré X | Poincaré | Located on the southern peak ring of Poincaré, southwest of Poincaré X. Peak rings are desirable targets for collecting impact melt, as they may exhume material from depth. The specific site was selected to be proximal to a small (~50-meter), fresh, rocky crater that plausibly excavated into the peak ring. | -54.571°N | 160.134°E | ~0.1 × 5 km | It is preferential to sample near as close to the base of the peak ring slope as possible (~0.1 km), although the peak ring extends laterally (~5 km). | ≥970.7 g |
| (E6) | Determine the age and remnant magnetization signatures from ≥3 returned samples with inferred ages ≥4.0 Ga | (E6) does not drive sample specific sample site selections; this experiment is completed as part of the coordinated analysis of samples collected for other experiments | This experiment does not drive specific site or traverse requirements; it is accomplished by the collection of all other samples | | | | | | | | Addressed by coordinated analysis of all returned samples |
| (E7) | Determine the composition and age of farside Th-rich materials | Sample ≥2 (≥[1]) geographically distinct Th-rich regions identified by preexisting orbital datasets to have Th-abundances ≥3 ppm | 6 | Abbe M northwestern rim | Central SPA | Northwestern ejecta of **Abbe M** | -61.310°N | 174.313°E | ~20.0 km | Scale of Abbe M | ≥780.4 g |
| | | | 7 | Abbe M southwestern rim | Central SPA | Southwestern ejecta of **Abbe M** | -62.389°N | 173.821°E | ~20.0 km | Scale of Abbe M | ≥780.4 g |
| (E8) | Determine the composition and age of farside basaltic units | Same as (E3); no additional sites are required | **Fully adddressed by mare basalt samples acquired for (E3).** No additional sample mass or sites are required. | | | | | | | | Fully adddressed by mare basalt samples acquired for (E3). No additional sample mass or sites are required. |
| (E9) | Sample hypothesized excavated lunar mantle material | Sample ≥2 geologic unit that is hypothesized to be excavated mantle, including hypothesized low-calcium pyroxene and olivine exposures proposed to trace mantle ejecta | 19 | Schrödinger southern peak ring massif | Schrödinger | Massif selected based on M3 observations suggestive of olivine exposures (Kramer et al. 2013). This specific site was selected based on the presence of boulders, with clear boulder tracks up the massif linking them to exposures from the southern peak ring of Schrödinger. | -77.164°N | 132.884°E | ~0.1 km | Scale of the boulder field and approximate scale of M3 pixels used to identify possible olivine exposures. | ≥799.8 g |
| | | | 14 | Lyman | Poincaré | Location selected to be in the vicinity of Lyman crater, in the low-calcium pyroxene annulus. This location is also proximal to gravity anomalies identified by Gowman et al. as potential mantle material (although this is not the primary driver for this location). | -64.280°N | 158.430°E | ~20.0 km | This objective does not drive very precise sampling locations; it's largely intended to sample "average" material in this terrain type. | ≥780.4 g |
| (E10) | Determine the volatile content of the farside mantle as recorded by pyroclastic and basaltic materials | Basaltic materials are collected as part of (E3); no additional sites are required | **Fully adddressed by mare basalt samples acquired for (E3).** No additional sample mass or sites are required. | | | | | | | | Fully adddressed by mare basalt samples acquired for (E3). No additional sample mass or sites are required. |
| | | Sample ≥1 pyroclastic deposit | 17 | Schrödinger pyroclastics | Schrödinger | Location selected on the mapped pyroclastic deposit surrounding Schrödinger G. The location was selected to be on the flanks of the volcanic mound, not very close to the central vent, out of concerns for rover mobility. | -75.259°N | 137.314°E | ~25.0 km | Scale of mapped pyroclastic deposit | ≥400.0 g |
| (E11) | Determine the crustal structure (thickness, density, remnant magnetic field strength) across the SPA basin (from the center to the putative rim) to constrain mega-basin formation and evolution models | Perform a traverse of SPA that spans from central SPA out radially ≥[650] km towards the hypothesized rim region | This experiment does not drive specific site requirements, but it does drive a traverse requirement that is fully met by traversing from Central SPA to at least Poincaré. | | | | | | | | None; this experiment does not require returned samples |
| (E12) | Determine if the massifs in the southwestern portion of the basin constitute the diffuse rim of SPA | Traverse ≥[1] km in elevation up a hypothesized SPA rim massif | 21 | South Pole-Aitken southern rim massif | South Pole | Massif north of **Wiechert** | -82.830°N | 164.000°E | ~5.0 km | Scale of the plateau on top of the massif | None; this experiment does not require returned samples |
| (E13) | Determine the stratigraphy and composition of the unique deposits in central SPA | Sample the geologic diversity of central SPA including: ≥[1] sample of low-calcium pyroxene material excavated by Bose crate; ≥[1] sample of high-calcium pyroxene material excavated by Bhaba crater; ≥1 sample from the ejecta of a morphologically fresh ≥10-km diameter crater with a well-defined ejecta blanket (plausibly excavating through mare basalts); ≥1 sample of light plains material in central SPA | **Fully addressed by samples collected at Bhaba (E2).** No additional sample mass or sites are required. | | | | | | | | Addressed by samples collected at Bhaba for (E2) |
| | | | **Fully addressed by samples collected at Bose for (E2).** No additional sample mass or sites are required. | | | | | | | | Addressed by samples collected at Bose for (E2) |
| | | | 2 | Central SPA plains fresh crater | Central SPA | ≥10-km diameter young crater northeast of **Bhaba** | -53.540°N | 199.880°E | ~1.0 km | The approximate scale of the rim of the targeted young crater. | ≥970.7 g |
| | | | 3† | Central SPA light plains | Central SPA | Light plains northeast of **Bhaba** | -53.540°N | 200.820°E | ~40.0 km | The approximate scale of the light plains material. This objective does not drive very precise sampling locations; it's largely intended to sample "average" material in this terrain type. | ≥970.7 g |
| (E14) | Determine the origin and extent of the low-calcium pyroxene annulus enclosing central SPA | Acquire ≥1 sample from the low-calcium pyroxene annulus enclosing central SPA, but excluding mare basalts and non-SPA impact basin material | **Fully addressed by samples collected at Lyman (E9).** No additional sample mass or sites are required. | | | | | | | | Addressed by samples collected at Lyman for (E9) |

| TOTAL NUMBER OF DISTINCT SITES | |
|---|---|
| BASELINE | THRESHOLD |
| 21 | 10 |

| TOTAL SAMPLE MASS | |
|---|---|
| BASELINE | THRESHOLD |
| ≥20.5 kg | ≥ 9.8 kg |

**Foldout 5.1.** SPARX site and traverse requirements. Experiments shown in grey boxes are considered baseline only (not threshold). **Second sampling site for E2 (SPA impact melt) for the threshold mission profile. †Colloquially referred to as the "MoonRise site" as this location is proximal to the location targeted by the MoonRise mission concept.

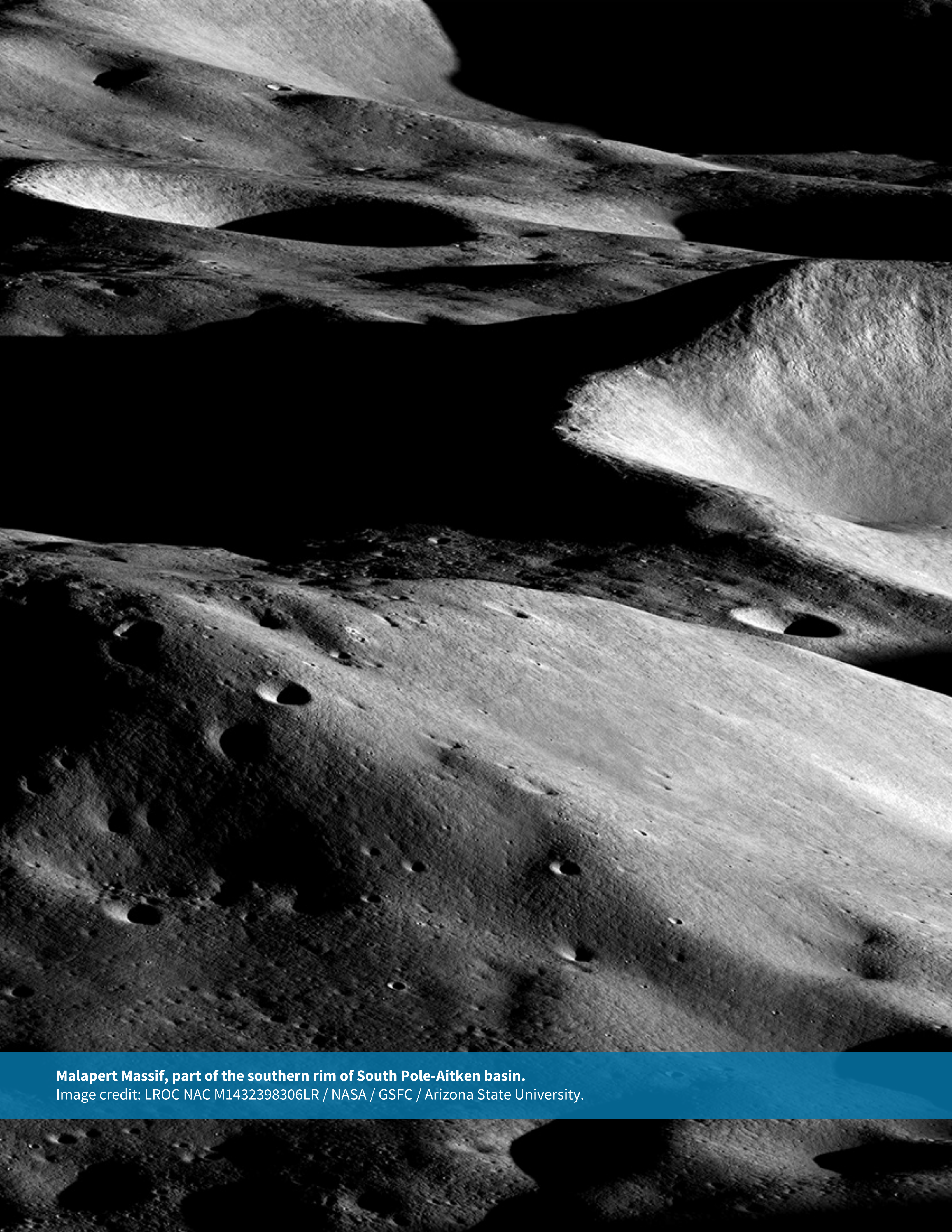

**Malapert Massif, part of the southern rim of South Pole-Aitken basin.**
Image credit: LROC NAC M1432398306LR / NASA / GSFC / Arizona State University.

# 6 DESIGN REFERENCE MISSION

## 6.1 INTRODUCTION

In this section, we present a design reference mission for the SPARX baseline mission concept defined in Section 5.8. Deriving from the rationale presented in Section 5, this design features a long-range lunar rover capable of meeting the science and exploration requirements outlined by the SDT. Figure 6.1 shows an artist concept of this design, which was developed in partnership with the JPL technical support team. The rover presented here represents a proof-of-concept for a technical solution that would accomplish the SPARX science requirements within the constraints defined by NASA. While this work was performed by the JPL and is framed around a custom-built rover, the challenges described here would be relevant to any lunar rover concept, including a rover derived from the Lunar Terrain Vehicle (LTV). This structure of this section is summarized in Table 6.1.

The SPARX rover design presented here shares technical heritage with the Endurance-A mission concept rover developed in the Planetary Science and Astrobiology Decadal Survey (Keane et al. 2021), although substantial changes have been implemented. Henceforth, we use “Rover” (with a capital “R”) to indicate the specific proof-of-concept rover developed here.

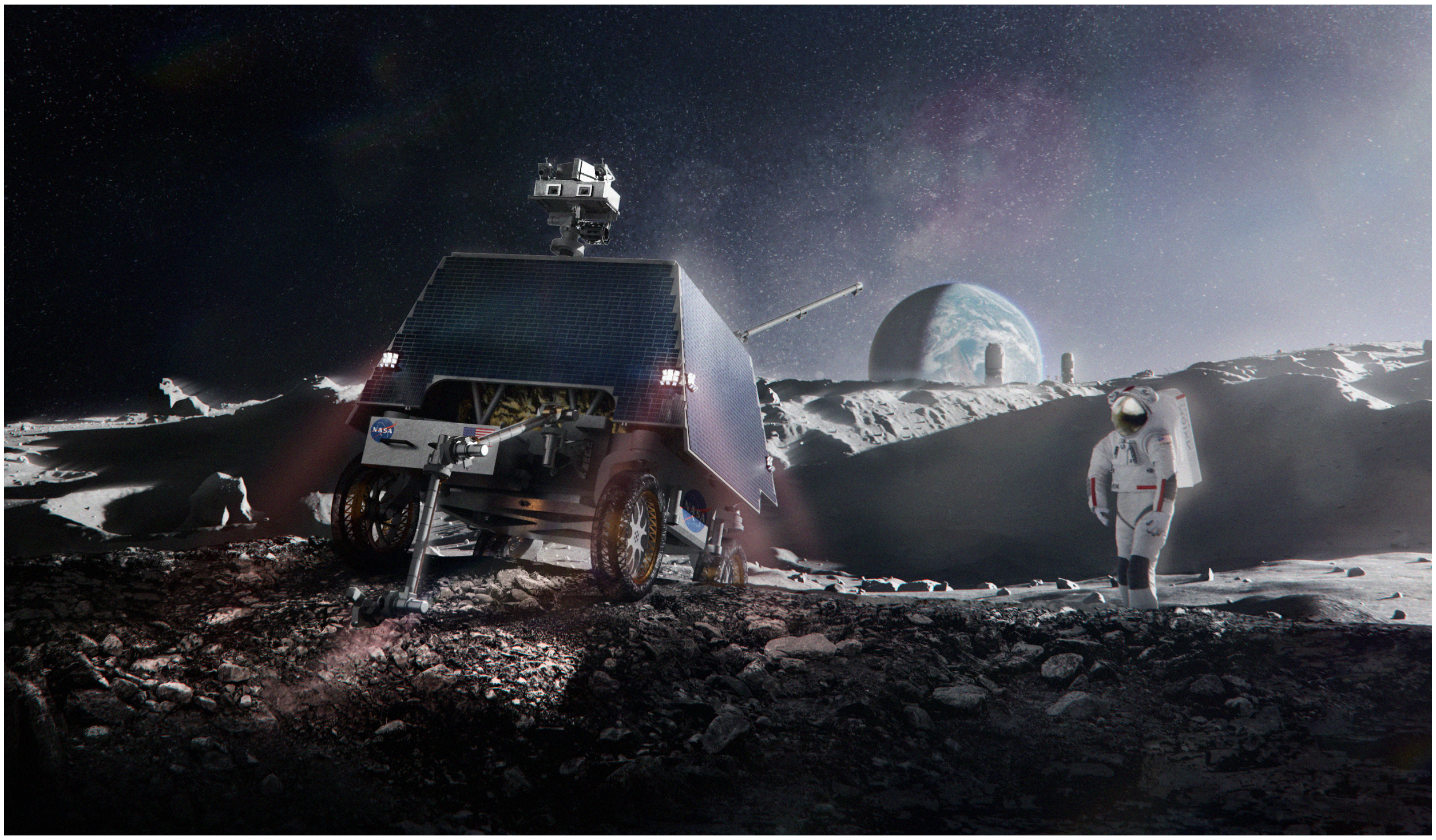

**Figure 6.1.** *Artist concept for the SPARX design reference mission rover. Credit: NASA / JPL-Caltech / Daniel Demirdjian.*

**Table 6.1.** *Table of contents for this section of the report focused on the SPARX Rover design reference mission.*

| Section | Description |
|---|---|
| Section 6.2 | The architectural principles guiding the technical design |
| Section 6.3 | The basics of operating a long-range rover and why fully autonomous driving is baselined |
| Section 6.4 | The strawman science payload, including a suite of 5 heritage science instruments designed for in situ measurements along the traverse and identifying/characterizing samples |
| Section 6.5 | The rover flight system including a description of major subsystems and key trades |
| Section 6.6 | The concept of operations (ConOps) for both the traverse and sample operations |
| Section 6.7 | The traverse design |
| Section 6.8 | The mission duration |
| Section 6.9 | The mission risks |
| Section 6.10 | The cost and schedule for the mission |
| Section 6.11 | Discussion of technologies that are either enabling or enhancing for the Rover |

## 6.2 ARCHITECTURAL PRINCIPLES

To design a rover that can confidently meet SPARX's ambitious science requirements (Sections 3–4), while also fitting within the cost, schedule, risk, and programmatic constraints defined by NASA (APPENDIX: TOR) the technical team established a set of architectural principles that guided the development of the Rover. These architectural principles are drawn from JPL's experience in designing and operating Mars rovers and informed by the recognized priority of the SPARX mission by the Decadal Survey, this SDT, and NASA. In this section, we summarize these architectural principles and provide examples of how they were applied throughout the Rover design process.

**(1) Robust environmental operability: The mission design shall allow operation at any time of the lunar day, and in any terrain along the pre-planned traverse (within reasonable limits).** The SPARX Rover must operate across a wide range of latitudes (52°S to 90°S), over multiple lunar days, and across a wide range of terrain types. The technical team designed towards maximizing operational flexibility and vehicle safety.

Early revisions to the SPARX TOR limited the design space to solar-power solutions. This presented an immediate change from the Endurance-A concept, which utilized RTG power. From an operational stand-point, this change made lunar daylight hours a limited resource. The SDT considered the impacts on mission ConOps, duration, and cost for solutions that operated in more conservative regimes (e.g. with hibernation modes during peak lunar temperatures); however, the SDT determined that such accommodations resulted in a cascading series of challenges for mission operations, duration, and ultimate cost. Ultimately, the SDT determined that all designs for a SPARX rover should be able to operate at any time of the lunar day, along any part of the SPARX traverse.

- Example Implementation: The Rover's power (Section 6.5.6) and thermal system (Section 6.5.5) are designed so that the Rover can operate at any time of the lunar day across the full latitude range (52°S to 90°S), without requiring a recharge or hibernation mode. This contrasts with many other planetary rovers (e.g., Yutu-2), which often cease operations and go into hibernation during midday to survive high temperatures. By designing towards operations with no dependence of time-of-day (at least during daytime), the Rover can maximize the amount of time available for science operations and traversing, while also simplifying operations planning and staffing. The Rover was designed to maintain payloads in their operational temperature ranges during the lunar day. The body-fixed, wrap-around solar arrays also mean that the rover can fault in any orientation and remain power positive.
- Example Implementation: In the final leg of the baseline traverse, the Rover must navigate through deeply shadowed terrains near the lunar south pole on its way to rendezvous with Artemis astronauts and/or Moon Base. The Rover's power system (Section 6.5.6) and traverse (Section 6.7) are designed such that: (1) The Rover always traverses when daylight is available (i.e., the Rover does not "sprint" across large shadows); and (2) the Rover will always drive through safe corridors, where if the Rover faulted, it would be able to survive through the long, polar lunar night. Both of these proposed implementations are designed to minimize mission risk in the challenging circum-south polar environment.

**(2) Expeditious mission timeline: The mission design shall prioritize efficiently accomplishing the stated mission science goals and include generous margins for fault recovery.** Accomplishing a ≥2,650 km long traverse in five years, or less, requires design that prioritizes progress towards stated mission science goals and efficient operations, and should de-emphasize flexibility in service of unknown or undefined future goals (e.g., opportunistic science). Furthermore, the Rover must include generous margins for fault recovery.

The long-distance of the SPARX baseline traverse and the large number of planned science interval stops and traverse stops, coupled with the restrictions of a solar-powered architecture mean that all unplanned or inefficient operations have the potential to cause cascading delays in the mission timeline. The SDT recognized this balance and prioritized a set of in-situ measurement types that can gather the required data in an efficient, and repeatable manner (Sections 4.2 and 6.4). The SDT strongly recognized the potential benefit to unplanned "science of opportunity" actions; however, determined that these could best be accomplished during pre-planned traverse stops, where regional reconnaissance is already accounted for, as opposed to, during the mid-traverse science interval stops, which focus on acquiring contextual geologic data (Section 4.2).

- Example Implementation: To maximize progress, the mission design should baseline fully autonomous navigation that can navigate lunar terrains without ground intervention (Sections 6.3 and 6.6). The Rover can traverse for several earth days at a time, without ground-in-the-loop planning cycles.
- Example Implementation: The SPARX baseline in situ measurements and strawman instrument suite were specifically selected to minimize complexity and maximize the rate of progress (Section 6.4). Instruments are accommodated only on the mast or body of the Rover, with no science instruments on the arm; therefore, reducing the requirements levied on the arm, and removing the need for frequent and time-consuming arm placements. Instruments were preferentially selected to minimize the amount of time required while stopped (≤30 minutes), or to allow measurements while driving. Instrument designs were assumed to allow operation at any time of lunar day.

**(3) The mission design shall be designed to make efficient use of future lunar communications infrastructure.** By utilizing on-board autonomy for traverse activities, which make up ~50% of daytime operations, the Rover reduces the need for continuous "ground-in-the-loop" operations. This minimizes relay requirements, reducing infrastructure costs and preserving bandwidth for other Artemis or Moon Base assets (Section 6.5.7).

**(4) The mission shall be designed to a NASA Class C+ mission standard.** Following direction from NASA, the project is governed by a Class C+ risk classification, representing a moderate-risk exploratory mission with tailored mission assurance practices and robust redundancy strategy to ensure system recoverability (Section 6.9).

## 6.3 LONG-DISTANCE FARSIDE TRAVERSE PRINCIPLES

The central challenge for the SPARX mission design is the requirement to collect a diverse suite of samples across an extremely long distance (≥2,650 km). This is well over an order of magnitude farther than any previous crewed or robotic rover (Figure 5.19). Accomplishing a traverse of this scale drives the entire design and cost of the Rover—and thus warrants special consideration. In this section, we summarize the trade space for operating a long-range rover on the farside of the Moon (Section 6.3.1), and ultimately, why this mission requires a highly autonomous system (Section 6.3.2). The subsequent sections (Section 6.4–6.9) describe the technical details of the Rover in more detail, including aspects that must be adapted for long-range and long-life on the lunar surface.

### 6.3.1 Operational Modes for Planetary Rovers

Operating a rover on another planetary body is a complex technical challenge, and multiple different operational modes have been developed to address these challenges. One of the fundamental challenges involves deciding on the level of onboard autonomy vs. human-assisted operation to employ (Figure 6.2). In this section, we will summarize the trade between these different operational modes, concluding in why a fully autonomous onboard navigation system is baselined for the SPARX Rover traverse.

#### 6.3.1.1 Rover Operational Mode 1: Human Control (i.e., "Joysticking")

***Human operators on Earth receive telemetry from the rover, make judgments, and send actions to the rover. The Soviet Lunokhod rovers operated in this mode.***

The drive rate for a rover operating under human control is set by: (1) the communications delay between the rover and the Earth (plausibly ~5 seconds one-way); (2) the perception horizon of the rover; (3) how quickly the rover can downlink navigational imagery to rover operators (dependent on rover perception system and data-rate relay); and (4) the response time of the human operators. For a rover operating under human control to make continuous progress, 24/7 operations (i.e., three 8-hour shifts) and near constant communication with the rover would be required during lunar daytime periods (14 Earth days), with traverse progress halted during lunar night.

The requirement for near-constant, low-latency communications is a particularly challenging requirement. To make continuous progress, the rover would need 24/7 availability of communications satellites, ground stations, and operators during lunar daytime. Any gaps in communications availability result in traverse delays and increases to the total mission duration. The requirement for near-constant communications raises two main concerns in the current operational environment: Infrastructure availability and operational design. At the time of this report writing, the farside communications infrastructure around the Moon is in development. While the mission design requires at least some form of farside data relay, there is significant risk in assuming continuous, high-bandwidth farside communications will be available for SPARX. Indeed, with the majority of planned communications assets focused on the south polar Artemis Exploration Zone and Moon Base, it is anticipated that the SPARX mission telecommunication requirements will need to be balanced with other exploration assets. Additionally, full human-control operations levy significant requirements on the rover telecommunications system for antenna pointing and operations that include frequent stops (every ~10-15 m) to allow for navigational images to be downlinked and drive commands to be uplinked. Preliminary analyses, assuming reasonable communications infrastructure, result in mission durations in excess of 10 years.

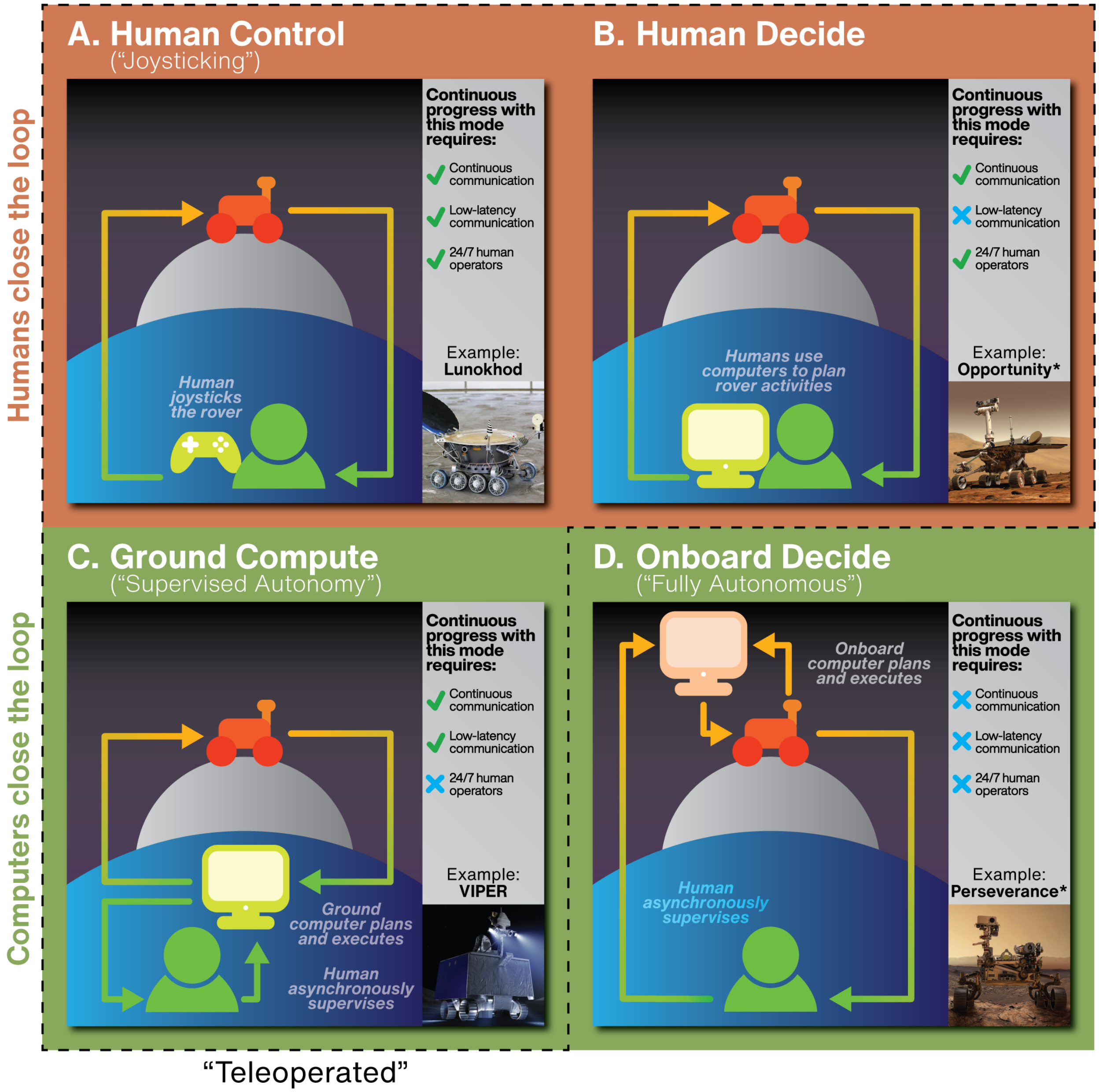

**Figure 6.2.** *A schematic illustration comparing operational modes for planetary rovers. For each mode, requirements for continuous progress are identified (green check indicates: "yes" that function is required, whereas a blue "X" indicates: "no" that function is not required). The examples show previous missions that have operated primarily with the associated operational mode; in reality, missions may use multiple operational modes. For example, Opportunity operated under the "human decide" mode for ~75% of the total distance, and under the "onboard decide" mode for ~25% of its total distance. In contrast, Perseverance operates ~75% under the "onboard decide" mode, but still occasionally uses "human decide."*

An additional challenge for "joystick" teleoperations is the challenges and risks associated with human operators. Human operators will take risks inconsistently based on personal experience and judgment, leading to more dangerous situations for the rover. A NASA study in 2023 investigated the performance of human teleoperated driving of lunar vehicles under various communication delays between 0 to 8 seconds (Litaker et al., 2024a-b). While this study achieved impressive effective drive rates (even 1.76 km/hr with an 8 second communications latency), operators only drove 1.5 km (less than 3 hours) in each scenario. Longer driving shifts per operator would be required to traverse the ≥2,650 km required by SPARX, which would lead to operator fatigue and a large number of operators required to staff round-the-clock overtime shifts for multiple years. This is reinforced by anecdotal reports of the high stress of teleoperating the Lunokhod rovers in this way. Furthermore, the 2023 study report concludes that, "Remote driving on the lunar surface can be challenging with no communication latency; however, with a time-delay, on the order of 6 to 8-seconds, can result in a dangerous state where the vehicle is either damage by contacting a large rock or loss due to falling into a deep crater compromising the future missions." These catastrophic mission-ending failures actually occurred during the relatively short drives performed in this study. A further look into the rock contact data from this study (see Section 7.5.1 of Litaker et al., 2024b) indicated that even in the 0 second latency regime, operators were unable to prevent the rover from contacting 49 "red" rocks over the 31.3 km of total driving performed by all operators. "Red" rocks were categorized as rocks greater than 0.38 m in height that "could cause severe instantaneous vehicle damage." This data suggests that over a ≥2,650 km route the chances of collision with a mission-ending rock would be very high if human operators were solely responsible for the safety of the vehicle.

From this combination of factors, plus the increased costs required to qualify a rover system for such an extreme operational lifetime, the SDT and technical team determined that a human-control "joysticking" operational mode is infeasible.

#### 6.3.1.2 Rover Operational Mode 2: Human Decide

*Human operators on Earth receive telemetry from the rover and utilize computers to assess the situation and generate action plans for the rover. The Mars Exploration Rovers (MER: Spirit and Opportunity) and the Mars Science Laboratory (MSL: Curiosity) operated predominantly in this mode.*

In this mode, while humans do not directly "joystick" the rover, progress is still paced by human operations which are required to approve every step. The advantage of this mode over the human control mode is that this mode reduces some of the burden on the operations team, as they are no longer "joysticking" the rover, but rather approving computer-defined plans. However, this mode shares almost all the other constraints as the human control mode. For a rover operating in the human decide mode to make continuous progress, the mission requires 24/7 availability of communications satellites, ground stations, and operators during lunar daytime. Any gaps in either the operations or the communications availability means the rover is not driving; thus, reducing mission progress and increasing the total mission duration. For many of the same reasons as "joysticking," operating in this mode would impose substantial telecommunications infrastructure requirements and high operations costs (as described in Section 6.3.1.1), although the actual operations may be less stressful and more sustainable.

#### 6.3.1.3 Rover Operational Mode 3: Ground Compute (i.e., "Supervised Autonomy")

*Computers on Earth receive near-real-time telemetry from the rover and generate actions autonomously that are directly uplinked to the rover. Human operators provide limited, asynchronous oversight. The VIPER rover is planning to execute drives in this mode.*

In this mode, rover pacing is determined by: (1) the ground computing capabilities (which are effectively infinite by planetary science mission standards); (2) the communications infrastructure; and (3) the risk posture for operating in this mode. Like the human control and human decide modes, this mode would require near constant communication with the rover during lunar daytime periods (14 days), thus incurring similar stringent (and costly) requirements on the communications infrastructure (as described in Section 6.3.1.1). The key difference is that this would not necessarily require continuous 24/7 operations (i.e., three 8-hour shifts), depending ultimately on how much supervision is deemed necessary for the various operational modes (e.g. traversing vs. reconnaissance vs. sampling).

#### 6.3.1.4 Rover Operational Mode 4: Onboard Decide (i.e., “Fully Autonomous”)

*Computers on the rover assess, generate, and execute actions autonomously. Human operators provide limited, asynchronous oversight. The Mars 2020 Perseverance rover has been operating in this mode for most of its long-range driving.*

In this mode the rover pacing is determined solely by the rover’s own perception, compute, and mobility capabilities. As a result, the onboard decide mode enables the fastest traverse speed, while also placing lower demands on communication infrastructure, and reducing operations costs. Compared to other modes, this mode requires the least amount of ground support for the rover to make continuous progress during lunar daytime. Ground support would only be required for check-ins and dealing with faults that the rover cannot address on its own. The downside to this operations mode is that this level of onboard autonomy does incur a higher development and maturation cost.

### 6.3.2 Trade Results for the SPARX Rover Operational Mode

Building on the technical work performed for the Intrepid (Robinson et al., 2020) and Endurance (Keane et al., 2022) concept study reports, the design team revisited trade between rover operational modes for the SPARX rover, accounting for updates to the mission’s science requirements. A summary table comparing these operational modes and relevant figures of merit is shown in Table 6.2.

**Table 6.2.** *Operational modes and figures of merit relevant to the SPARX mission. Asterisks indicate that durations are only relevant during the lunar day (14 earth days). Green, yellow, and red coloration indicates a qualitative evaluation of whether that factor is beneficial (green) or detrimental (red) for accomplishing the SPARX mission concept within the NASA-defined constraints (e.g., cost, mission duration, risk posture).*

<table>
<tr><th></th><th></th><th></th><th>Communications requirements</th><th>Ground staffing requirements</th><th>Achievable rover speed</th><th>Operations cost</th><th>Development cost</th><th></th></tr>
<tr><td>1. Human control (“joysticking”)</td><td rowspan="3">“Tele-operated”</td><td rowspan="2">Human closes the loop</td><td>High (24/7*)</td><td>High (3 shifts*)</td><td>Slow (~0.1 kph)</td><td>High</td><td>Low</td><td></td></tr>
<tr><td>2. Human decide</td><td>High (24/7*)</td><td>High (3 shifts*)</td><td>Slow (~0.1 kph)</td><td>High</td><td>Medium</td><td></td></tr>
<tr><td>3. Ground compute (“supervised autonomy”)</td><td rowspan="2">Computer closes the loop</td><td>High (24/7*)</td><td>Medium (1~3 shifts* depending on risk posture)</td><td>Intermediate (0.1~1 kph)</td><td>Medium</td><td>High</td><td></td></tr>
<tr><td colspan="2">4. Onboard decide (“fully autonomous”)</td><td>Low (1~2 communication windows per day*)</td><td>Low (1 shift*)</td><td>Fast (>1 kph)</td><td>Low</td><td>High</td><td>←Baseline for SPARX rover</td></tr>
</table>

The design team analysis concluded that the onboard decide (i.e., "fully autonomous") operations mode was the only viable approach for accomplishing a traverse of this magnitude. There are several factors for this conclusion:

(1) Fully autonomous driving allows for significantly faster rover speeds than teleoperated and supervised autonomy driving modes, thus reducing overall mission duration.

(2) Fully autonomous driving significantly reduces the operations cost and complexity for rover operations, and reduces requirements on ground staff availability and scheduling. The comparison in operations and telecommunications requirements for "teleoperated" vs. "fully autonomous" is illustrated schematically in Figure 6.3. The reduction in required ground staffing coverage for autonomous driving activities has the further benefit of preserving staff availability and attention for phases of the mission that require active, iterative operations, like site reconnaissance and sampling activities.

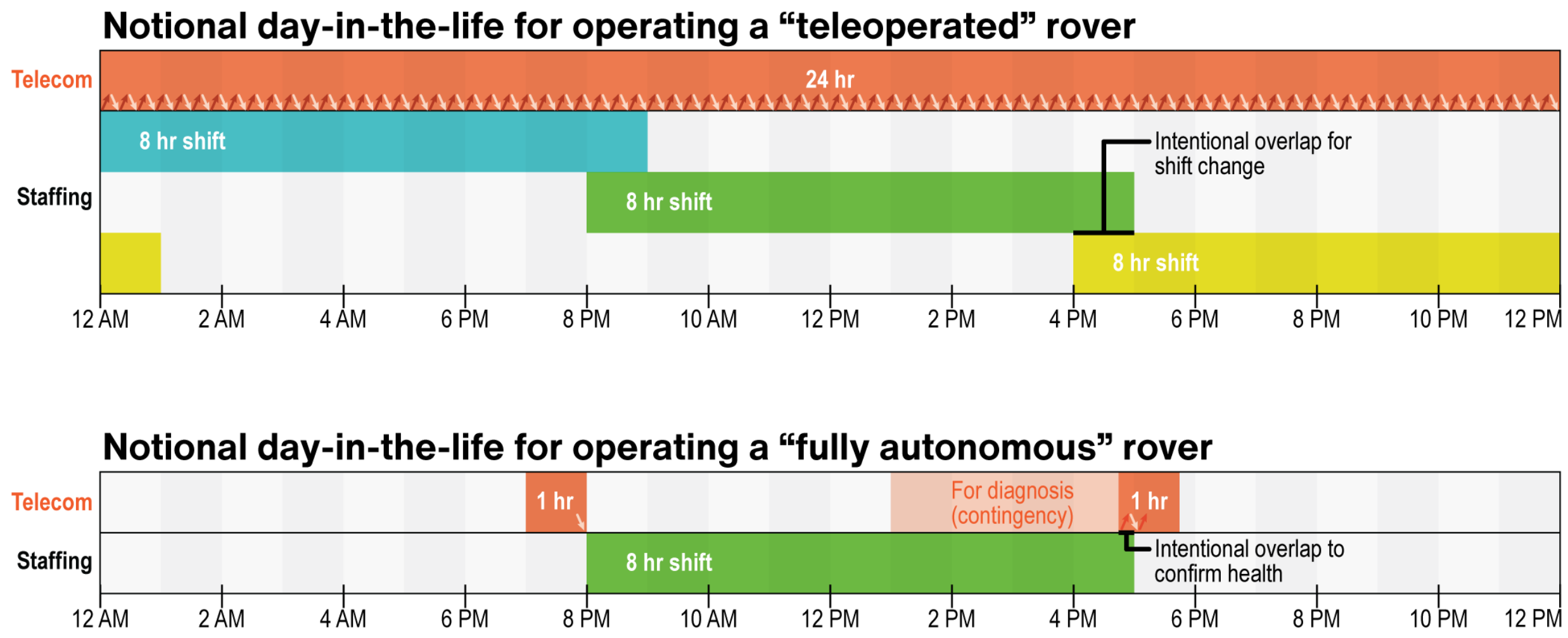

**Figure 6.3.** *Schematic illustration of a notional day-in-the-life for operating a rover, either in a teleoperated mode (top) or a "fully autonomous" mode (bottom). A rover in a teleoperated mode would require continuous telecommunications (with constant uplinks and downlinks, visualized as small arrows), and at least three shifts of continuously staffed operations teams on the ground. In contrast, a fully autonomous rover could operate with a single shift and a small number of communications windows.*

(3) Fully autonomous driving reduces the reliance on lunar communications infrastructure and reduces the required cost for communications time. For a teleoperated rover to have continuous progress, it requires near continuous (~24/7) communication with operators on the Earth. A fully autonomous rover affords a faster and cheaper mode for traversing, while also being more resilient to uncertainty in network capabilities—ensuring that even a minimally viable network is sufficient to complete the mission.

**Finding: Based on analysis of operational modes by the technical team, the SDT finds that the SPARX rover should be able to operate in a "fully autonomous" mode for the required long-distance driving modes. The SDT emphasizes that full autonomy is only desired for driving modes, and that SPARX science requires more iterative ground-in-the-loop operations during other mission activity phases, such as sampling and reconnaissance.**

### Why is onboard autonomous navigation necessary for the SPARX Rover?

There are many reasons why the SPARX Rover baselines autonomous navigation instead of relying on teleoperated "human-in-the-loop" driving. **A fully autonomous approach is faster, safer, simpler and cheaper to operate, and reduces burden on lunar telecommunications infrastructure.**

1. A fully autonomous rover can drive an order-of-magnitude faster than a teleoperated rover: ≥1 kph (fully autonomous) vs. ~0.1 kph (teleoperated). Driving faster reduces the overall mission duration and total operations cost (at least to a point, as described in Section 6.8.3). A faster rover also enables a higher cadence of science activities.
2. A fully autonomous rover significantly reduces requirements on the ground operations team compared to a teleoperated rover. A fully autonomous rover would have an order-of-magnitude fewer operational shifts, lower stress operations, and lower operations cost compared to a teleoperated rover. For a teleoperated rover, these effects would be negatively compounded by the longer mission duration.
3. Fully autonomous driving is potentially safer than teleoperated driving. Experience from the Lunokhod rovers and NASA studies for the LTV (Litaker et al., 2024a-b) have shown that rover operators take more risk, ultimately increasing the likelihood of mission failure.
4. Fully autonomous driving reduces the reliance on future lunar communications infrastructure by at least an order-of-magnitude compared to teleoperated rovers which require 24/7 communications to make progress. 24/7 communications may not be plausible depending on a variety of factors and would come at high cost.

**When taking all these factors into account, a rover capable of fully autonomous driving is most consistent with being able to accomplish the baselined SPARX traverse (~2,650 km) within the cost and schedule constraints defined by NASA.**

## 6.4 PAYLOAD

The SPARX baseline science requires the use of in situ instruments (Section 4.2) that must be accommodated by the rover. For the baseline SPARX Rover, the SDT worked with the technical team to identify a minimal subset of science instruments that have either flown, or are proposed for flight, that could meet the seven types of required in situ measurements (Section 4.2). The end-result was a list of five "strawman" instruments. Table 6.3 shows the in-situ measurement types, and how they trace to these five instruments. It is important to emphasize that these instruments are notional, and do not represent the only instrument types capable of meeting the SPARX requirements.

Based on discussions with the SDT, the technical team developed the following set of guiding principles for identifying potential strawman instruments and sketching out their operations:

- Accommodation: Instruments are only permitted to be mounted on/in the Rover body, or on the remote sensing mast. No instruments shall be located on the arm or otherwise require contact or close proximity to the lunar surface. No instruments shall require delivery of samples to the instrument. The guiding rationale for this requirement is that no SPARX in situ measurements have spatial or operational requirements that would necessitate contact/close proximity with the lunar surface, and therefore accommodation designs should avoid adding that complexity. Further scientific rationale for the balance of scope of in situ characterization can be found in the Section 4.2.2 call out box, and Section 7.
- Time to acquire observations: Instruments shall not require operations that significantly impact mission progress. While it is preferable to have instruments that can operate while the Rover is moving, this is not often practical. Therefore, preference is given to instruments that can acquire the required in situ measurements as efficiently as possible while the Rover is stopped. The current baseline ConOps assumes stops science interval stops (Section 4.2.1) that last ≤30 minutes.
- Time of day sensitivities: Like the Rover, instruments shall be operable at any time during the lunar daytime period. If instruments require additional illumination or significant thermal management (e.g., cryocooling), it is assumed that the instrument will provide it. The current Rover ConOps does not include science operations during lunar night
- Minimize impact to rover design: In some cases, instruments like gamma-ray and neutron spectrometers and magnetometers may levy requirements on the spacecraft in terms of allowable materials, magnetic cleanliness, etc. These requirements should be minimized to the greatest extent possible.
- Data volume: Communications relay to Earth from the farside of the Moon may be challenging and will be dependent on the future available infrastructure (Section 6.5.7). Instruments must consider pathways to reducing data volume.
- Operations: The traversing ConOps are designed to largely use fully autonomous operations (Section 6.3.2), potentially going several earth days without a ground-in-the-loop cycle while traversing. Instrument must be designed to operate in this mode as well, and to be able to meet the required geological context measurements of the Science Interval Stops (Section 4.2.1) without ground-in-the-loop operations. The SDT notes that operations during sample campaigns will include more ground-in-the-loop activities.

While these principles guided the selection of the strawman payload, and development of the baseline reference mission, the SDT recognizes that they are not immutable, and final implementation will rest with NASA, the ultimately selected implementation architecture, and the requirements of the selected payload elements.

As shown in Table 6.3, and described in Section 4.2, the requirements for Context Imaging and Fine-Scale Imaging can be met by the same instrument, as can the requirements for Context Mineralogy and Fine-Scale Mineralogy. The baseline design reference mission leverages these dual capabilities to decrease accommodation complexity (i.e., fewer payload elements that need to be integrated and accommodated) and also decrease overall mission cost.

Below we will briefly describe each of the selected strawman payload elements, including a discussion of the instrument capabilities, and why it was selected as a strawman instrument. It is worth noting that multiple existing or proposed instruments could meet the requirements for SPARX, and these are detailed in each of the measurement subsection descriptions in Section 4.2.

**Table 6.3.** *Traceability between the 7 classes of in situ measurements, and the 5 strawman science instruments included here for the baseline Rover design. The requirements for these measurements can be found in Section 4.2.9, Table: In Situ Measurement Requirements. Dark green (with two checkmarks) indicates that the strawman science instrument is fully responsive to the in-situ measurement requirements; light green (with one checkmark) indicates that the strawman science instrument is relevant but does not fully address the in-situ measurement requirements, and white indicates that the instrument does not meet the requirements.*

| | | Strawman Science Instrument Type (and Analog) | | | | | |
|---|---|---|---|---|---|---|---|
| | | **Visible Color Cameras** | **Imaging Spectrometer** | **Gamma-Ray and Neutron Spectrometer (GRNS)** | **Magnetometer** | **Gravimeter** | |
| | | (Mars 2020 MastcamZ) | (UCIS) | (LunarVISE GRNS) | (Insight IFG) | (DTU IMU) | |
| In Situ Measurement Requirements | (M1) Context Imaging | ✓✓ | | | | | |
| | (M2) Context Mineralogy | ✓ | ✓✓ | | | | |
| | (M3) Fine-scale Mineralogy | ✓ | ✓✓ | | | | |
| | (M4) Bulk Elemental Chemistry | | | ✓✓ | | | |
| | (M5) Gravity Field | | | | | ✓✓ | ↑ Threshold |
| | (M6) Fine-Scale Imaging | ✓✓ | | | | | |
| | (M7) Magnetic Field | | | | ✓✓ | | ↑ Baseline |

### 6.4.1 Stereo Visible Color Cameras: Mastcam-Z

Mastcam-Z rover is pair of focusable, 4:1 zoomable CCD cameras mounted on the Perseverance rover's remote sensing mast (Figure 6.4). Table 6.4 summarizes the basic capabilities of the instrument, which are detailed further in Bell et al. (2021). Mastcam-Z's color capabilities come from Bayer pattern filters mounted to the CCD detector, and they provide detailed morphology, topography, and geologic context along the rover's traverse. Each camera also has an 8-position filter wheel with UV, visible, and near-IR filters that provide information into about surface mineralogy.  Data from this camera are also used by the Perseverance team to support sample selection, drive planning, and other instrument operational decisions. At their shortest focal length (26 mm), the cameras provide context images that have fields of view (FOV) of 25.6° × 19.2° with instantaneous field of view (IFOV) of 283 µrad/pixel.  The cameras capture detailed textural information and imaging of distant features at their longest focal length (110 mm), which provides a FOV of 6.2° × 4.6° and IFOV of 67.4 µrad/pixel. The two cameras are separated by 24.4 cm on the instrument mast, allowing for stereo imaging. The SDT notes that while Mastcam-Z is a stereo camera system, the stereo capability is not required for the SPARX baseline measurements, although proposers may identify additional science that could be done using this capability. MastCam-Z was selected as a strawman payload element because of the color imaging capability, and the zoomable nature of the cameras, which allow the system to meet both the Context and Fine-Scale imaging requirements of SPARX (Section 4.2.9).

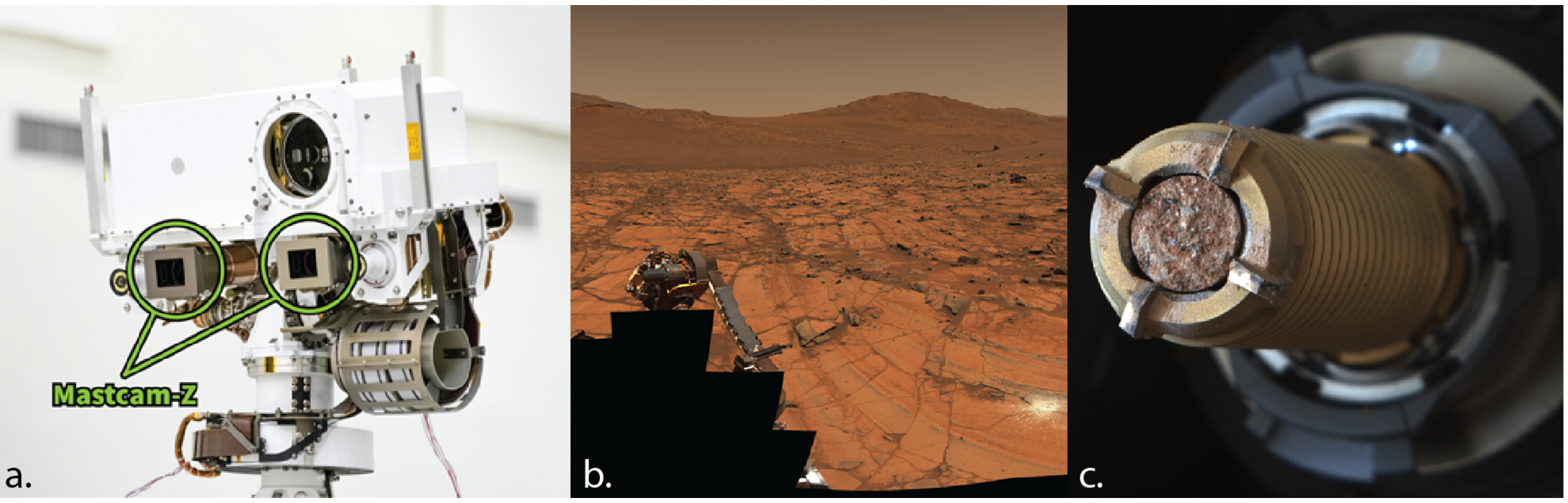

**Figure 6.4.** *(a)Location of the MastCam-Z camera suite on Mars 2020 (Perseverance) rover's remote sensing mast. (b) Cropped example of a full 360° context mosaic using left Mastcam-Z 34 mm focal length. Partial or full-panoramic imaging is required for SPARX geologic context measurements. (c) Example of a close-up image of a Perseverance drill sample (approximately 0.5 inches or 13 millimeters in diameter) from Swift Rock, collected on sol 490. The image was collected by the right Mastcam-Z at 110 mm focal length. This type of fine-scale imaging can provide important insight to samples collected by the SPARX Rover, as well as characterizing unsampleable material (e.g., boulders). Image credit: Mastcam-Z / NASA / JPL-Caltech / ASU / MSSS.*

**Table 6.4.** *Summary of Mastcam-Z characteristics.*

| Metric: | Estimated Performance: |
|---|---|
| Number of Cameras | 2 |
| Mass | Each camera (2×): 1.38 kg<br>Digital Electronics Assembly: 1.47 kg<br>Calibration targets: 0.12 kg |
| Volume | Each camera (2×): 11 × 12 × 26 cm<br>Digital Electronics Assembly: 22 × 12 × 5 cm<br>Calibration targets: 10 × 10 × 7 cm |
| Data | Variable depending on compression. For M2020, images were digitized as 11 bits/pixel and could be compressed 1.7:1. M2020 returns ~175 Mbit per sol, and ~120 GB raw for the prime mission. |
| Power | Standby: 7.5 W<br>Imaging per camera (2×): 11.8 W |
| Focal Length | 26 mm – 110 mm |
| FOV | 25.5° × 19.1° – 6.2° × 4.6° |
| IFOV | 280 µrad – 67.4 µrad |
| Image Size | Maximum: 1600 × 1200 pixels |
| Resolution | Variable: 0.7 mm at 2 meters range, to 3.3 cm at 100 meters range |
| Color | Broadband RGB (Bayer pattern on CCD)<br>Filter wheel: 400 – 1,000 nm |

### 6.4.2 Imaging Spectrometer: UCIS

UCIS is a push broom imaging spectrometer with an Offner design that measures surface reflectance from 500–2,500 nm at 10 nm spectral sampling (Blaney et al. 2014, Van Gorp et al. 2014), and was matured through the NASA MATISSE and DALI programs. As an imaging spectrometer, UCIS builds up a hyperspectral image of a full scene (Figure 6.5), as required for the SPARX Context Mineralogy measurement (Section 4.2.4). UCIS can complete a 360˚ scan in minutes to 10s of minutes depending in required integration times. Table 6.5 summarizes the basic characteristics of UCIS. Additionally, many common lunar bearing minerals including pyroxene, Fe-bearing plagioclase, and olivine have diagnostic absorptions in the UCIS wavelength range. Finally, the UCIS optical design allows it to meet both the Context and Fine-Scale Mineralogy requirements for SPARX (Section 4.2.9).

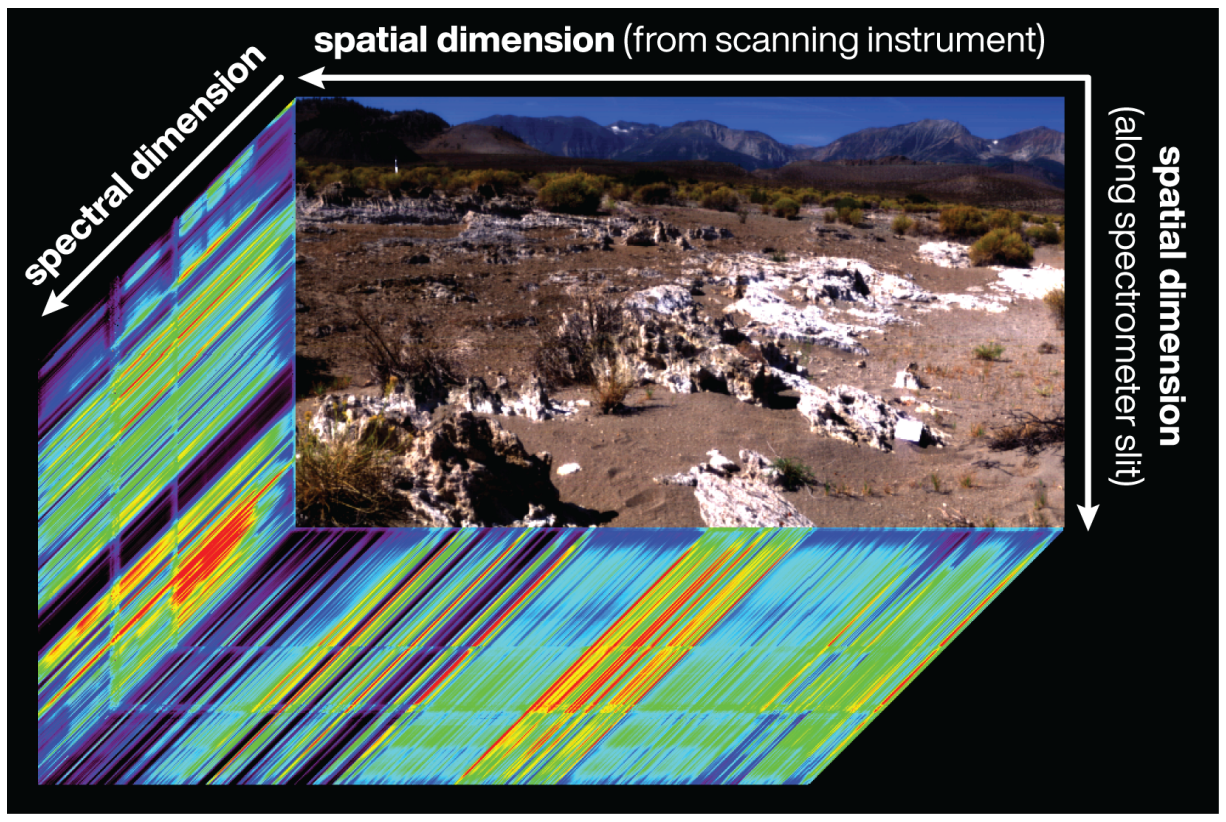

**Figure 6.5.** *Terrestrial example for how UCIS would construct a hyperspectral image.*

**Table 6.5.** *Summary of UCIS characteristics.*

| Metric: | Estimated Performance: |
|---|---|
| Mass | 6 kg (including electronics) |
| Volume | 12 ×18 × 24 cm |
| Data | Variable depending on compression and parameterization. A single panorama may be ~1.5 GB. |
| Spectral Range | 500 – 2,500 nm |
| Spectral Sampling | 10 nm |
| FOV | 30° |
| IFOV | 1.35 mrad |
| Spectral cross track uniformity | >95% |

### 6.4.3 Gamma-Ray and Neutron Spectrometer: LV-GRNS

The Lunar Vulkan Imaging and Spectroscopy Explorer (Lunar-VISE) gamma-ray and neutron spectrometer (LV-GRNS) is composed of two Li-6 enriched Cs26 LiYCl6 (CLYC) scintillator sensor heads, which is sensitive to both gamma rays and neutrons. The larger diameter CLYC crystal is encased in a gadolinium shell to veto thermal neutrons and to provide a large volume for stopping high energy gamma rays up to 9 MeV. The smaller CLYC crystal is not shielded and is used to measure neutrons from thermal to epithermal energies (<100 keV). The sensors are electrically connected to single-channel readout-control electronics (CE) that are designed to simultaneously identify which sensor (NS or GRS) produced a signal, whether the signal is from a neutron or gamma ray using pulse shape analysis, and to provide gamma-ray spectroscopy. Table 6.6 describes the basic characteristics of the LV-GRNS, and Figure 6.6 shows an example laboratory data.

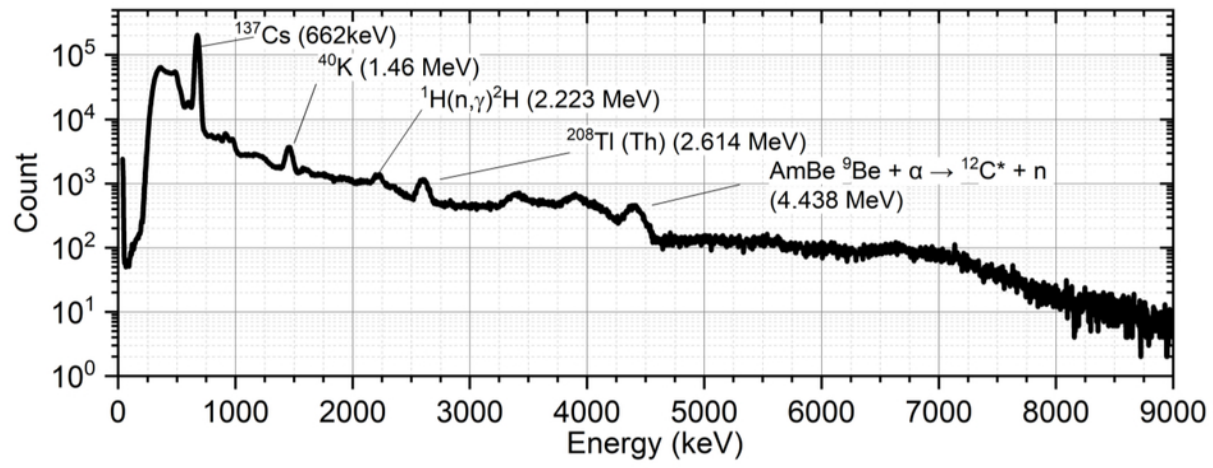

**Figure 6.6.** *Gamma-ray spectrum from laboratory test data demonstrating detection capabilities and characteristic elemental peaks. Notable peaks include $^{137}$Cs (662 keV), $^{40}$K (1.46 MeV), neutron capture on hydrogen (2.223 MeV), $^{208}$Tl (2.614 MeV), and the characteristic 4.438 MeV gamma-ray from Am-Be neutron source reactions. Data confirm system sensitivity to key gamma-ray emissions for elemental identification.*

**Table 6.6.** *Summary of LV-GRNS characteristics.*

| Metric: | Estimated Performance: |
|---|---|
| Mass | 4.75 kg |
| Volume | 47 × 25 ×16 cm |
| Power | Standby: <3 W<br>Data acquisition: <6 W |
| Data | ~25 MB/hour of science data for typical data types and a nominal event rate; SOH at 824 bps |
| Energy Resolution | 5% FWHM at 662 keV |
| Energy Efficiency | Intrinsic γ-rays (at 662 keV) 25.8 ± 0.2 %; Intrinsic neutrons (AmBe) 35.5 ± 0.4 % (both 2" CLYC) |
| Crystal Sizes | GRS: 2.5”Ø × 2” Gd-wrapped CLYC GRS and epithermal neutron detector<br>NS: 2”Ø × 1” bare CLYC thermal + epithermal neutron detector |
| Typical Elemental Targets | Fe, Si, Mg, Ca, K, Th, H |
| Thermal Requirements | 0 to +40°C<br>(+20C operational) |

## 6.4.4 Magnetometer: IFG

The InSight Fluxgate Magnetometer (IFG) on the InSight Mars lander is a triaxial, high-sensitivity instrument designed to measure the magnetic field at the Martian surface (Banfield et al. 2019). The IFG is one component of the broader InSight Auxiliary Payload Sensor Suite (APSS), which included a variety of sensors used to calibrate other instruments on InSight (e.g., the seismometer). The IFG consists of consists of a single triaxial fluxgate magnetometer with a range of ±20,000 nT and a sample rate up to 20 Hz, which meets the SPARX magnetometer requirements (Section 4.2.9). The data can be low pass filtered to reduce data volumes as necessary. Electronics are housed in the Payload Auxiliary Electronics (PAE) unit. The magnetometer transmits three 24-bit magnetic field components (X, Y, Z) together with housekeeping information at a rate of 20 vectors per second to the processing and telemetry system, which can average the data and perform statistical analysis on the data prior to transmission. The basic characteristics of the IFG are described in Table 6.7.

**Table 6.7.** *Summary IFG characteristics.*

| Metric: | Estimated Performance: |
|---|---|
| Mass | Sensor: 171 g<br>Electronics unit: 101 g<br>Cabling: 286 g |
| Volume | Sensor: 7 × 7 × 5 cm (excluding boom)<br>Electronics unit: 14 × 10 × 1.4 cm |
| Power | ~1 W |
| Data | 60 kBytes per hour |
| Dynamic range | ±20,000 nT |
| Sample rate | 20 Hz |
| Thermal qualifications | −105°C to +25°C |

## What About Magnetic Cleanliness?

Measuring the magnetic field on the surface of a planetary surface can be challenging especially when the spacecraft making the measurement is generating its own magnetic field. Electronics, motors, instruments, and a variety of other components generate confounding signals that can obscure the ambient magnetic field. For many missions, this is addressed by levying magnetic cleanliness requirements on the spacecraft, which can be costly to implement. However, this need not be the case if the experiment is well-designed.

For the SPARX Rover there may be multiple ways to accommodate magnetic cleanliness requirements without levying overly stringent or complex magnetic cleanliness requirements on the vehicle:

- Utilizing a boom: Placing the magnetometer on a boom away from the vehicle helps isolate the magnetic field contribution from the rover. The baseline SPARX Rover design utilizes this approach with a 1.5-meter boom. The length and orientation of the boom can be optimized to minimize the contribution from individual noise sources. A longer boom is beneficial, although there may be trades associated with magnetometer sensitivity, the rover design, and/or the use of "keep out zones" for certain components or materials.
- Utilizing more than one magnetometer: Having at least two well-placed magnetometers can allow for more accurate modeling and removal of the magnetic field signature from the rover via gradiometry. The Lunar Vertex mission (which will explore the Reiner Gamma magnetic anomaly on the nearside of the Moon) will use this approach, utilizing sets of four magnetometers in a tetrahedral configuration to minimize sources of noise (Waller et al. 2025).
- Leveraging "quiet time": One approach is to provide dedicated time for magnetic field measurements while the rover is otherwise "quiet", i.e., not moving or using other components that generate magnetic noise. The current Baseline ConOps allocates approximately 10 minutes at each Interval Stop (every 2 kilometer) for "quiet time" that serves both the magnetometer and gravimeter. Another potential period of "quiet time" could be during lunar twilight or night, if power permits operations of the magnetometer.
- Leveraging the rover's mobility to calibrate the instrument: Three-axis stabilized spacecraft, such as Europa Clipper, utilize spacecraft rolls to calibrate magnetometers and better determine the ambient field. From the perspective of a magnetometer on a rotating spacecraft, the external, ambient magnetic field (the "true" science signal) appears to rotate, whereas the spacecraft field will appear constant as it rotates with the magnetometer. For SPARX, calibration rolls could be accomplished by driving the rover in a circle (i.e., doing "donuts") or rotating in place. There may be opportunities where these calibrations are synergistic with other operations—like spinning the rover to acquire panoramas.
- Providing housekeeping data: Timed housekeeping data (e.g., records of when certain components are on/off) can be helpful for identifying, modeling, and mitigating rover noise sources. Temperature measurements can also inform interpretations of magnetometer data.

### 6.4.5 Gravimeter: DTU IMU

The SPARX SDT determined that the gravity measurement requirements (Section 4.2.9) could be met by the Inertial Measurement Unit (IMU) onboard the SPARX rover. The baseline SPARX Rover utilizes the Denmark Technical University (DTU) IMU, which is an integrated IMU and SRU (Stellar Reference Unit). This instrument is described further in Section 6.5.8.

# 6.5 FLIGHT SYSTEM

## 6.5.1 Overview

The baseline Rover design presented here was developed to address the science requirements defined by the SPARX SDT.

The baseline Rover design presented here builds on previous robotic lunar rover concept studies, including both Intrepid (a pre-Decadal Survey mission concept study for a long-range rover that would explore the diverse magmatic terrains on the nearside of the Moon; Robinson et al. 2020), and Endurance (a Decadal Survey mission concept for a long-range rover to explore SPA and deliver samples to astronauts or a robotic sample return vehicle; Keane et al. 2022). The design was further refined to address the science and mission requirements specifically defined by the SPARX SDT.

Design requirements include the ability to sample rock and regolith (≥ 20.5 kg, unmargined, Section 4.4) from predetermined traverse sites spread over a > 2,650 km traverse across SPA. The traverse further requires a rover design that can operate on the farside of the Moon, across a wide range of latitudes, and can survive the lunar night. Additional design principles were discussed in Section 6.2. It is also noted that for purposes of rover design, the JPL design team used the maximum sample mass permitted by the SPARX TOR, 100 kg. Although well in excess of the baseline required SPARX sample mass, this decision provides a maximum design envelop. In the following subsections, we will provide a high-level overview of each of the flight subsystems, and, as relevant, will identify any places where design choices are mission drivers (i.e., “tall tentpoles”). The baseline design Rover configuration is shown in Figure 6.7.

System mass and power modes are shown in Table 6.8 and Table 6.9, respectively. The Rover landed mass maximum expected value (MEV) is 1,133 kg. Carrying the JPL design principle 23% system margin on top of this value, the requirement on a delivery vehicle would be a capability of 1,352 kg. This mass is significantly larger than the mass of the previous Intrepid and Endurance concepts owing to a change in the power system design from Radioisotope Thermoelectric Generator (RTG) to solar power, due to RTG budget reductions in the President's FY26 budget request. NASA updated the SDT guidance to assess solar powered rovers as the primary option for SPARX (Section 2). The Rover now baselines solar power for daytime operations and incorporates a large battery to store energy for overnight survival. The wrap-around solar array provides ample power for all operations during the lunar day, independent of rover azimuth and solar elevation, allowing continuous traverse without interruptions or power-related operational constraints for the duration of the lunar day (Section 6.2). A highly efficient thermal design is incorporated in the Warm Electronics Box (WEB) to minimize overnight heating requirements while maintaining the ability to passively dissipate heat during the hottest periods of the lunar day. A summary of Rover characteristics is detailed in Table 6.10.

Figure 6.8 shows the accommodation of the strawman payload elements, described further in Section 6.4.

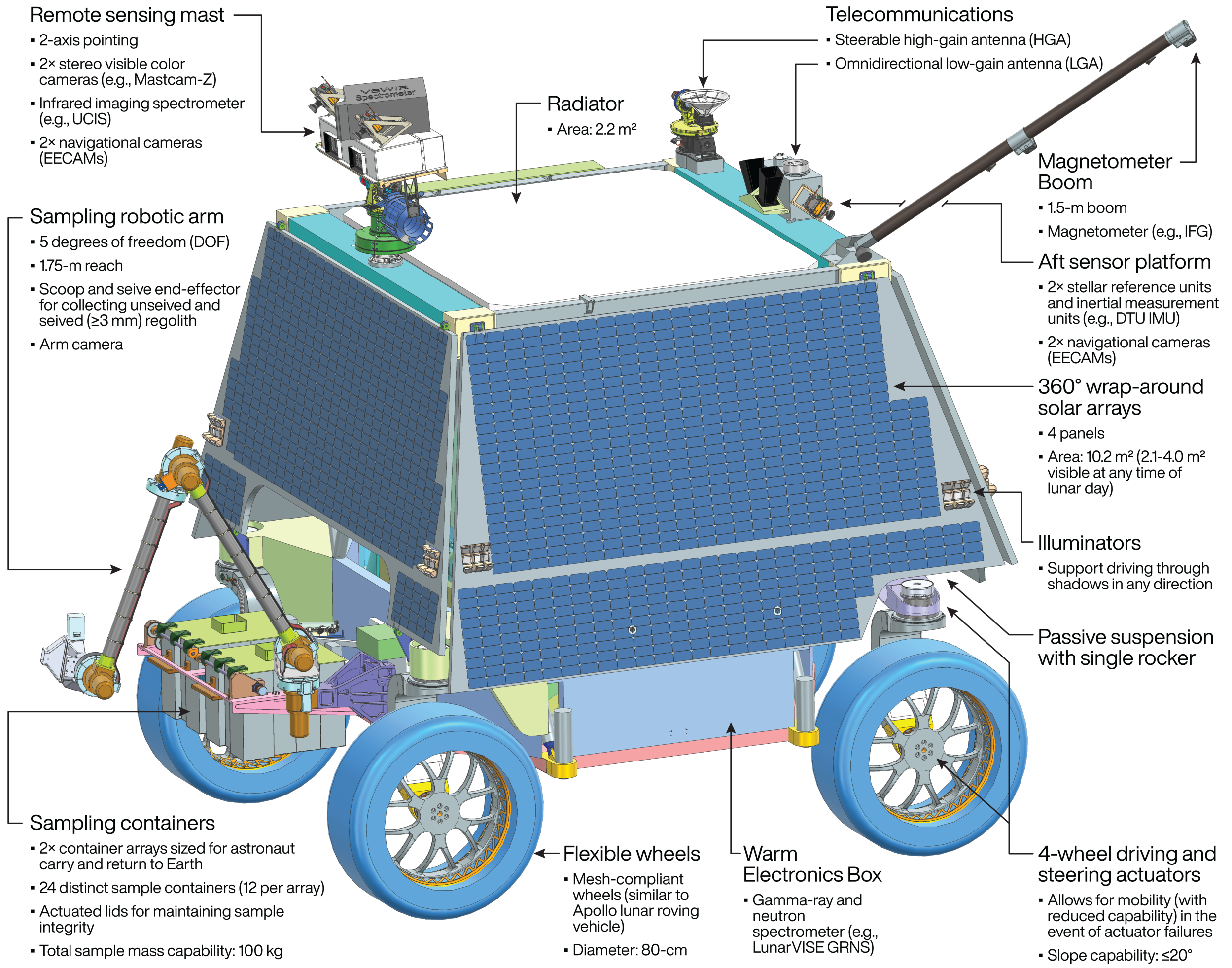

**Figure 6.7.** *Rover configuration and design highlights.*

**Table 6.8.** *Summary of the Master Equipment List (MEL). The mass growth allowance is determined by subject matter evaluation of the design maturity and likelihood for growth.*

| | Mass (kg) | | |
|---|---|---|---|
| | Current Best Estimate (CBE) | Mass Growth Allowance | Maximum Expected Value (MEV) |
| Instruments | 16.2 | 17% | 19.1 |
| Sampling | 38.6 | 33% | 51.34 |
| C&DH | 16.4 | 23% | 20.5 |
| Telecom | 17.4 | 29% | 22.4 |
| GNC | 4.9 | 23% | 6.0 |
| Power | 373.4 | 12% | 419.3 |
| Thermal | 29.3 | 34% | 39.3 |
| Structures | 292.0 | 20% | 350.9 |
| Mobility | 110.7 | 22% | 135.3 |
| Harness | 52.8 | 30% | 68.6 |
| Rover Total | 952.0 | 19% | 1,132.8 |
| Lander Allocation (MPV) | 1,352 | | |
| System Margin (MPV–CBE) / CBE | 23% | | |

**Table 6.9.** *Summary of Rover power modes.*

| | Major Power Modes (W) | | | | |
|---|---|---|---|---|---|
| | **Daytime** Traverse (autonomous driving) | **Interval Science Stop** | **Telecom Stop** | **Charging** (instruments idle, and spectrometer cryocooler operating) | **Night Hibernation** (overnight survival mode with rover in minimal power, thermally safe state) |
| Science Instruments | 43.6 | 49.2 | 36.6 | 35.6 | 0 |
| C&DH | 72.7 | 44.3 | 44.3 | 0 | 0 |
| Telecom | 10.1 | 10.1 | 75.4 | 10.1 | 0 |
| GNC | 11.5 | 0.5 | 0 | 0 | 0 |
| Power | 30.1 | 30.1 | 30.1 | 6.4 | 6.4 |
| Thermal | 0 | 0 | 0 | 0 | 41.6 |
| Mobility | 168.5 | 0 | 0 | 0 | 0 |
| Mast | 1.0 | 5.0 | 2.5 | 0 | 0 |
| Rover Total | 337.2 | 139.1 | 188.8 | 52.0 | 48 |
| Contingency | 43% | 43% | 43% | 43% | 43% |
| Maximum Expected Value (MEV) | 482.2 | 198.9 | 270.0 | 74.4 | 68.6 |
| Minimum Power Production | 399 | 399 | 399 | 399 | Battery sized for 550 hours of night survival |

**Table 6.10.** *Rover implementation highlights.*

| | | |
|---|---|---|
| **General** | Design life | 72 months (6 years) |
| **Structure** | Mass | 1,352 kg (fully margined) |
| | Size | 2.98 m (length) × 2.45 m (width) × 3.00 m (height) (with the boom and arm stowed) |
| | Material | Aluminum and composite |
| | Deployed structures | 1 (magnetometer boom) |
| **Mobility & Articulation** | Control method | 4-wheeled rover, 4-wheel steering |
| | Control reference | Stellar, terrain recognition |
| | Maximum slope capability | 20° |
| | Maximum rover speed on flat terrain | 1.0 kilometer per hour |
| | Ground clearance | 60 centimeters |
| | Robotic arm | 1.75-meter length, 5 degrees of freedom, scoop/sieve end-effector |
| | Mast | 2 degrees of freedom |
| **Telecom** | High-gain antenna (HGA) | Steerable (2 degrees of freedom) 15-cm diameter antenna, Ka-band |
| | Low-gain antenna (LGA) | Omnidirectional, S-band |
| | Radio | UST-lite |
| | Relay | NASA Lunar Communications Relay and Navigation Systems (LCRNS) constellation |
| **Thermal Control** | Type of thermal control used | Passive, heat pipes, radiators, electric heaters, thermal switches |
| **Power** | Type of solar array | Fixed, body mounted, 360° wrap-around arrays |
| | Array Size | 10.2 $m^2$ total (2.1–4.0 $m^2$ illuminated at any given time, depending on rover latitude and geometry) |
| | Solar Cell Type | Multi-junction |
| | Minimum and maximum power generation | Beginning of mission: 423 – 817 W<br>End of mission: 361 – 690 W |
| | Average power consumption | 382 W (daytime traverse mode) |
| | Battery type | Li-ion |
| | Battery storage capacity | 65.8 kWh at beginning of mission |

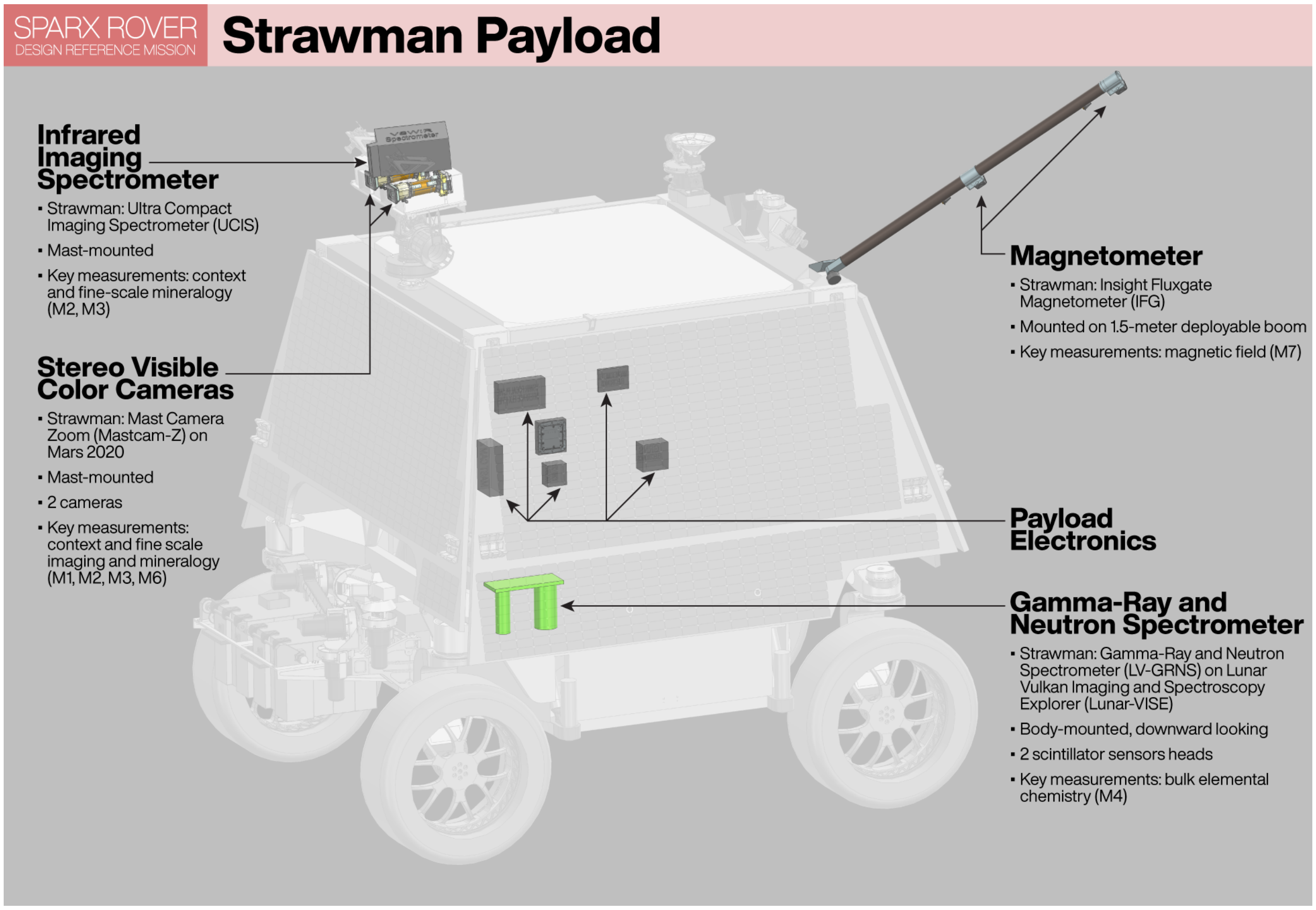

**Figure 6.8.** *Accommodation of the strawman payload elements on the Rover. The payload elements are described further in Section 6.4.*

### 6.5.2 Mechanical Configuration

The mechanical configuration of the Rover was designed to meet driving thermal, mobility, and structural requirements while enabling simplified integration and testing (I&T) and verification and validation (V&V). Figures 6.9–6.10 show orthographic and perspective views of the Rover, respectively.

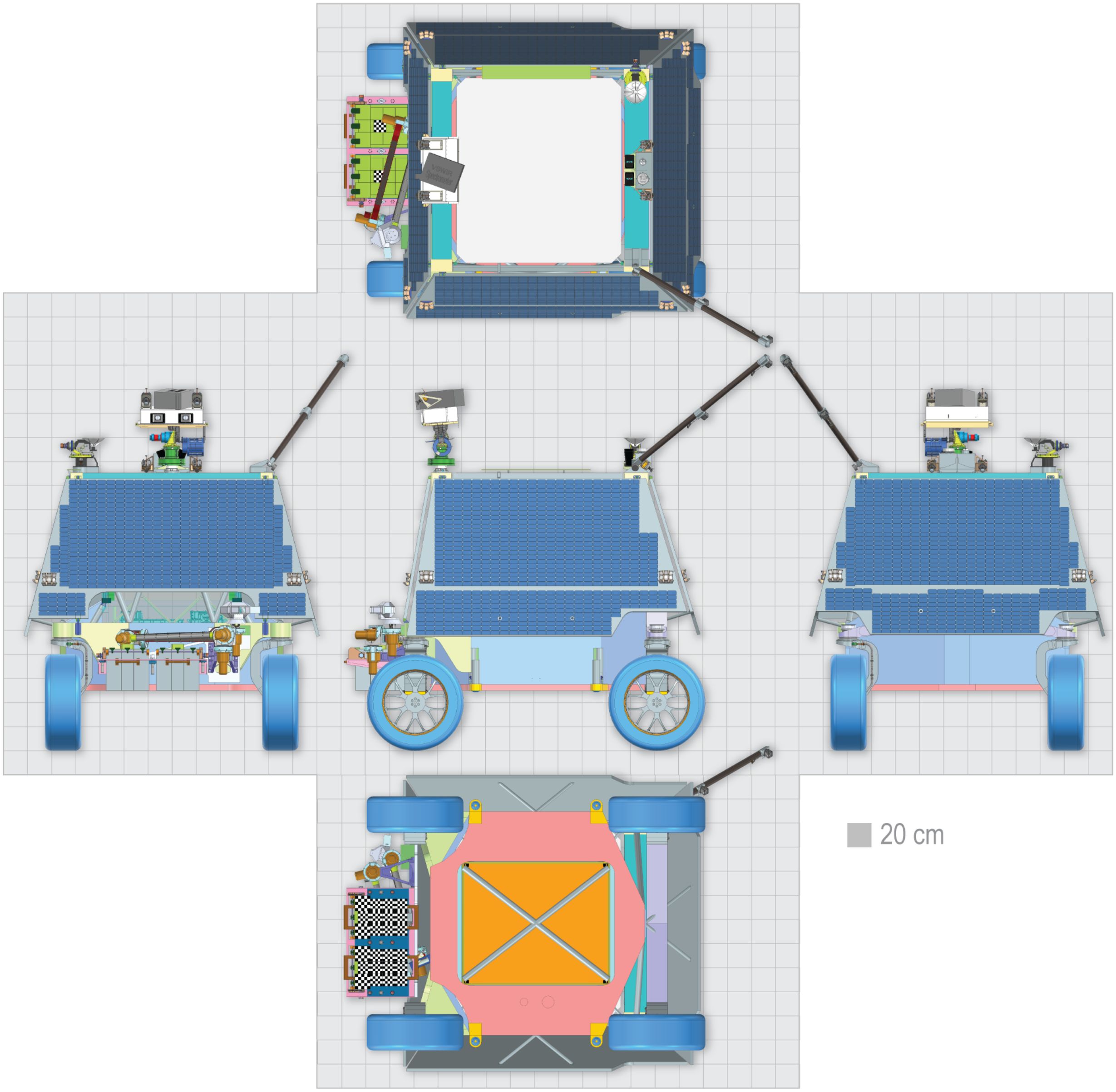

**Figure 6.9.** *Orthographic views of the Rover.*

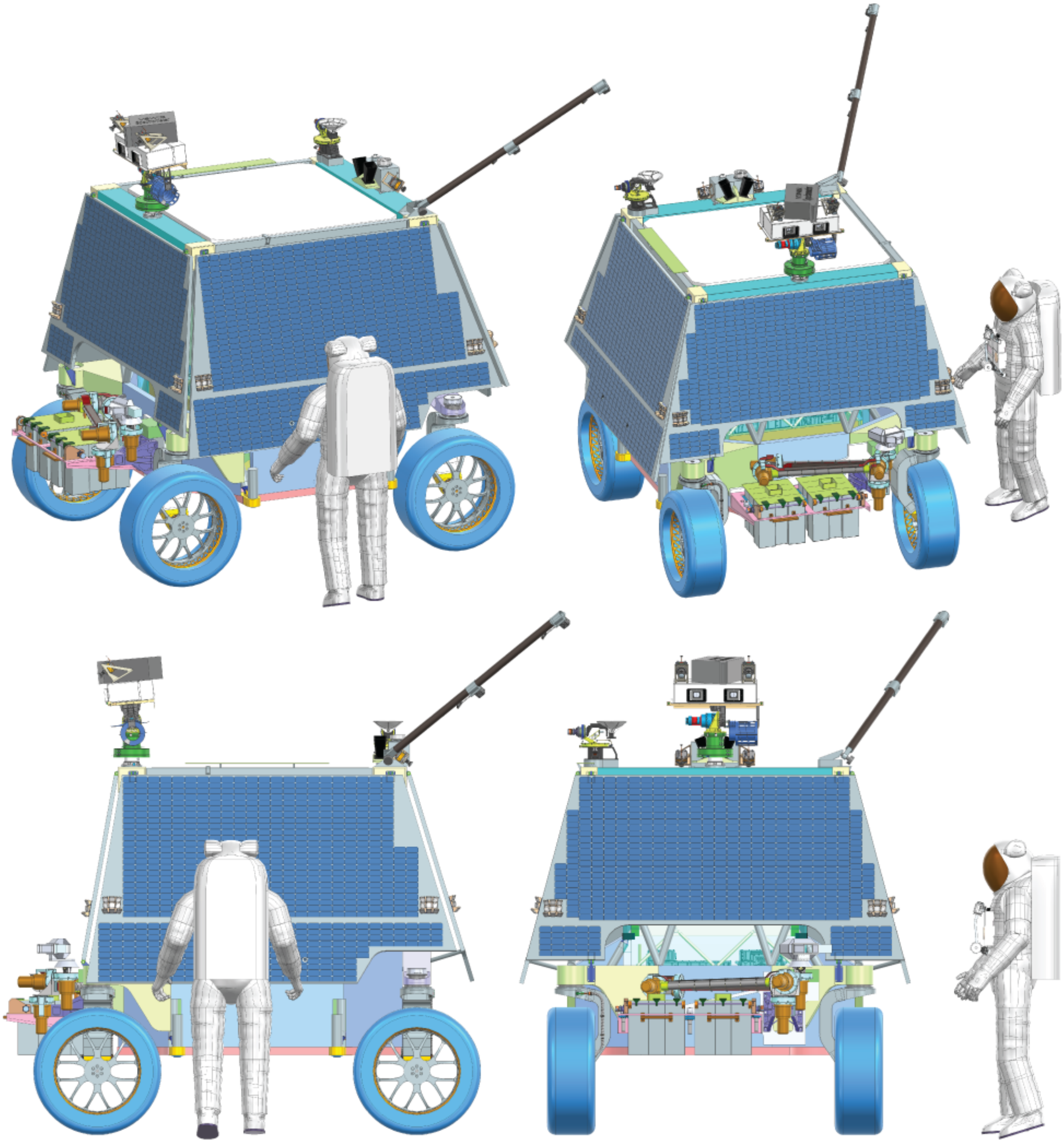

**Figure 6.10.** *Perspective views of the Rover, with an astronaut for scale.*

The mechanical system is primarily built using lightweight aluminum or aluminum honeycomb panels. This reduces thermal stress by having interfaces with common materials and reduces cost and schedule. The rover comprises of several key subsystems: The Warm Electronics Box (WEB), the mobility system, the upper deck assembly, the solar array assembly, and the sampling assembly.

The WEB (Figure 6.11) consists of the enclosure for the electronics and battery as well as the spacerless multilayer insulation (MLI), radiator, WEB struts, and WEB interface panel. This enables full integration and qualification testing of the WEB at this level of subassembly including thermal performance tests in thermal vacuum (TVAC) and vibration qualification testing. The WEB is kinematically mounted to the rover to ensure decoupled loads between it and the rest of the system. Long titanium struts for the WEB provide good thermal isolation to minimize heat loss. The WEB has two removable walls that allow access to the interior electronics from the sides for initial assembly as well as potential rework in later stages of I&T. The batteries are accessible and removable through two panels at the bottom of the rover.

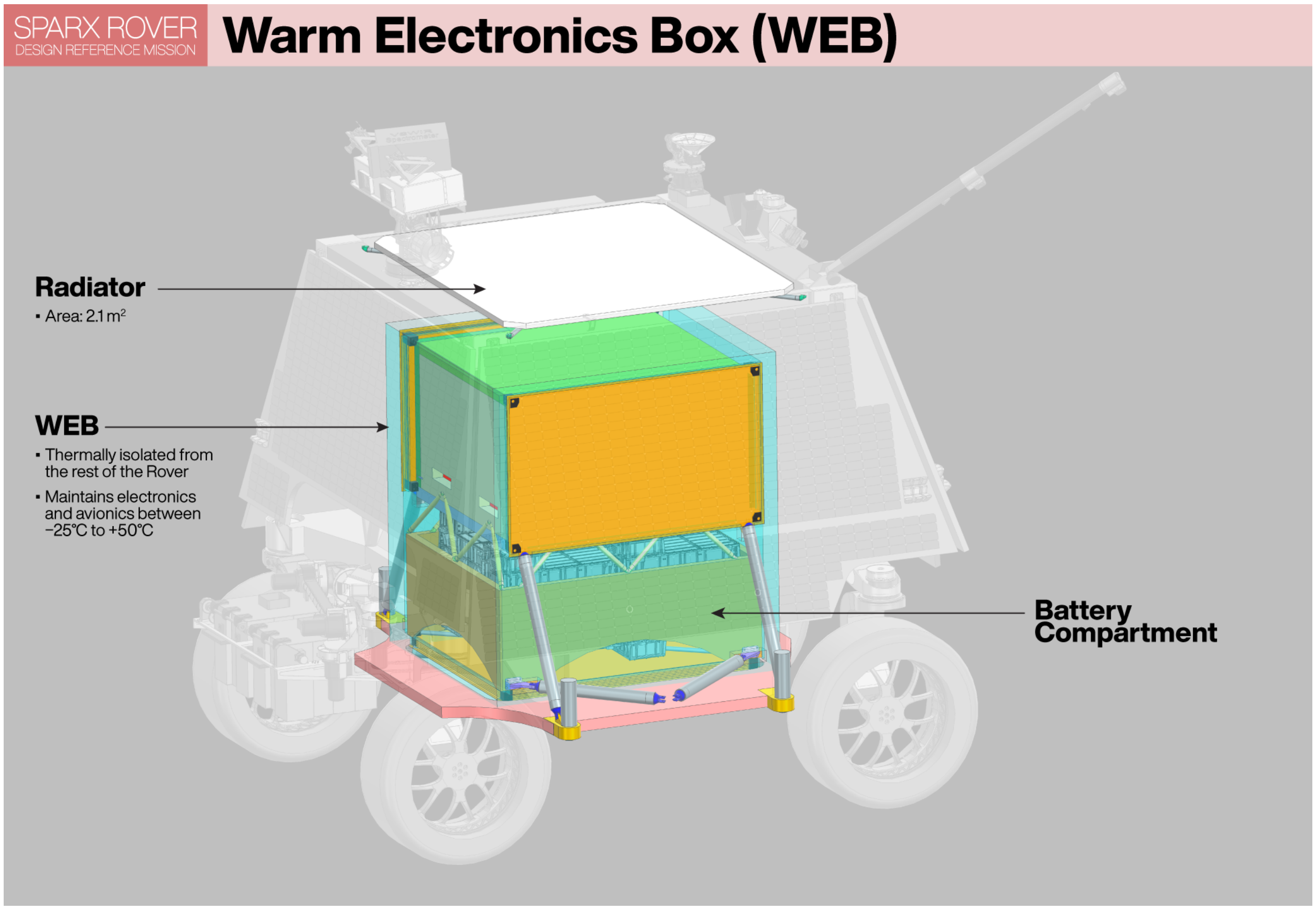

**Figure 6.11.** *Cross-section of the Rover showing the Warm Electronics Box (WEB).*

The mobility system (described further in Section 6.5.3) has been designed to be able to be fully tested and qualified independently from other subsystems. It consists of the wheels, actuators, rocker arm and pivot, as well as the lower half of the mechanical chassis. Volume in the mechanical configuration has been reserved to allow full rotation of each wheel individually and ±20° of rocker rotation.

The upper deck of the rover provides a single reference plane to install and align the various GNC, telecom, and science instruments mounted to it. These components can be integrated and co-aligned to one another at the lower level of assembly, simplifying system assembly by requiring only the integration of the upper deck to the mobility assembly via a system of struts.

The solar arrays are mounted independently to the upper deck and the mobility chassis to minimize loads going into them and not react global system loads. They can be installed late in I&T and can be individually removed to enable either rework of the arrays or allow access into the WEB.

The sampling assembly (described further in Section 6.5.4) consists of the robotic arm and the sample containers. These co-mount to a common plate and then this common plate mounts to the mobility chassis. This enables these items to be aligned and undergo V&V as a subsystem without requiring re-alignment and more functional checking after installation at system I&T.

### 6.5.3 Mobility Subsystem

Key drivers for the design of the mobility system are the mission traverse distance (≥2,650 km), mass of samples to be collected (up to 100 kg, per the SPARX TOR; APPENDIX: TOR), and the environmental conditions of the Moon along the traverse, including topography, terramechanics of lunar regolith, and the distribution of hazards such as rocks and craters. The mobility subsystem components of the Rover design are shown in Figure 6.12.

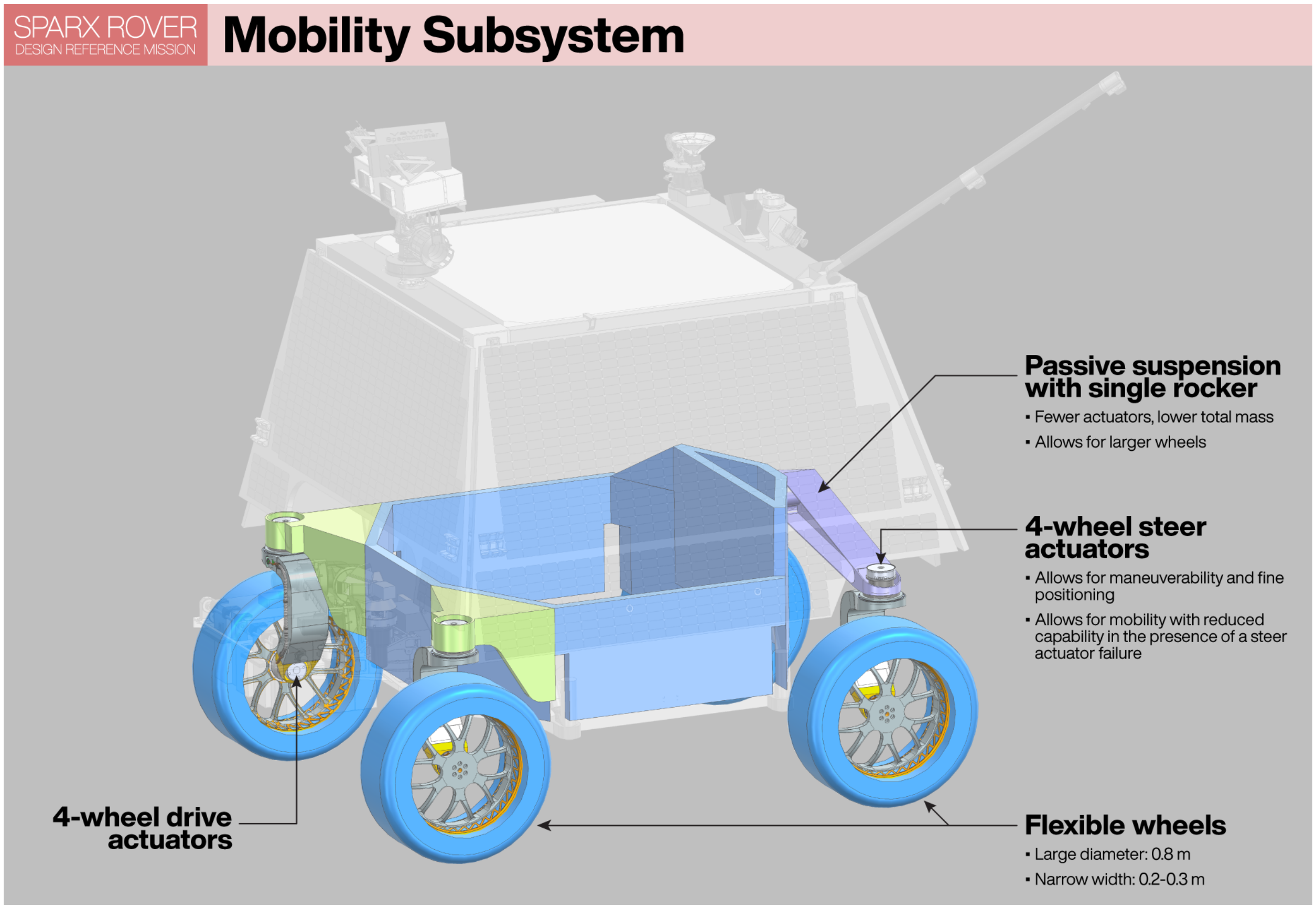

**Figure 6.12.** *Mobility subsystem components of the Rover design.*

The Rover has four steered and driven wheels that enable omni-directional maneuverability on lunar regolith slopes up to 20°. Analysis of the Rover has shown that it is statically stable on even steeper slopes, up to 30°, with the presence of large (30 centimeter) rocks and craters. This configuration maintains non-skid-steered (i.e., Ackermann steered) driving even with the loss of a steer actuator. A four-wheel configuration was chosen over the six-wheel rocker-bogie configuration used on NASA's previous Mars rovers because of the need to accommodate larger wheels (enabling better traction and energy efficiency) and to reduce mass and cost. A more thorough analysis of mobility trades can be found in the Intrepid and Endurance concept study reports (Robinson et al. 2020, Keane et al. 2022). To minimize power usage, the Rover could be nominally driven with the rear-wheel steering active and with the front-wheel steering for contingencies. If needed, the Rover can respond to unexpected terrain or actuator failure by engaging or changing the active steering actuators. A single-joint passive suspension mechanism rotating about the longitudinal axis of the Rover keeps all wheels in contact with the ground for maximum traction. The single-joint suspension has a lower mass compared to the rocker-bogie suspension used on Perseverance and Curiosity. A passive suspension has fewer joints and less mass, takes up a smaller swept volume, and automatically complies with terrain height changes enabling the rover to traverse at higher speeds without a sophisticated suspension control system. Active suspension, which has the advantage of tolerating more rugged terrains and steeper slopes, was considered; however, it was not baselined because the anticipated lunar terrain along the traverse is relatively benign and route planning combined with on-board obstacle detection and navigation will avoid insurmountable hazards. The range-of-motion of the suspension system allows the Rover to drive over rocks up to 30 centimeters tall (the "worst case" rocks), while conforming to natural undulations induced by small craters and surface roughness. To accommodate the sampling system at the front of the rover, the suspension mechanism is placed at the rear and articulates the back wheels to comply with the terrain.

To maximize efficiency and provide long life, the drive actuators use a 3-stage planetary gearbox with a modest gear ratio, in the range of 64:1. The gearboxes are driven by torque-dense, outrunner motors, with holding torque provided by magnetic detents. The detents are chosen in favor of solenoid brakes to maximize mechanical robustness (no moving parts) and minimize power required while driving. The motor windings and its Hall sensors have redundant current carrying paths to increase robustness. Testing has been performed on a representative gearbox and output seal, successfully achieving a simulated traverse of >7,500 km, in representative conditions.

JPL developed ERNEST (Exploration Rover for Navigating Extreme Terrain) to further refine mobility systems for the Moon. ERNEST represents a half-scale prototype vehicle of the Endurance-A rover design (Figure 6.13), and the ERNEST testbed has the kinematics and appropriately scaled mass of the SPARX Rover. ERNEST has enabled the technical team to investigate a variety of technical trades, including four-wheel drive, four-wheel steering, a single-side rocker, suspension location, and flexible wheel designs. The spring wheels on ERNEST, based on the design from NASA Glenn Research Center (GRC), stand in for the flexible wheels that would be leveraged from the LTV program for the SPARX Rover.

One of the major trades for the mobility system is the design of the Rover's wheels. Key drivers for wheel design for this mission include rover mass, speed, traverse distance, terramechanics, topography, power budget, and the environmental conditions on the Moon along the traverse. Additional considered factors included reliability, mass, steerability, stability, and development risk. A benefit of the four-wheel configuration is the larger volume it allows for wheels. Large wheels have reduced rolling resistance, improved rock traversal properties, better energy efficiency, and improved traction. For a long traverse, a larger wheel undergoes fewer actuator rotations and fewer cycles of contact with the ground thereby reducing wear and extending wheel lifetime. The trade analysis evaluating wheel options also considered rigid and several alternative flexible wheels. The advantage of a rigid wheel is that it is easier to design to meet performance specifications. While more challenging to design, a flexible wheel has better traction in soft ground, is more energy efficient, and complies when encountering embedded rocks thereby reducing impact loads on the rover structure. The large flexible wheel design baselined for the rover has the added benefit of leveraging industry-developed wheels being matured by the Lunar Terrain Vehicle (LTV) contracts and other efforts. By partnering with vendors who are currently developing wheels for lunar vehicles like Michelin, Goodyear,

Bridgestone, or Venturi Astrolab (Figure 6.14) SPARX can capitalize on their experience and expertise. A rigid wheel remains a descope option if the flexible wheel designs do not meet requirements or are too expensive.

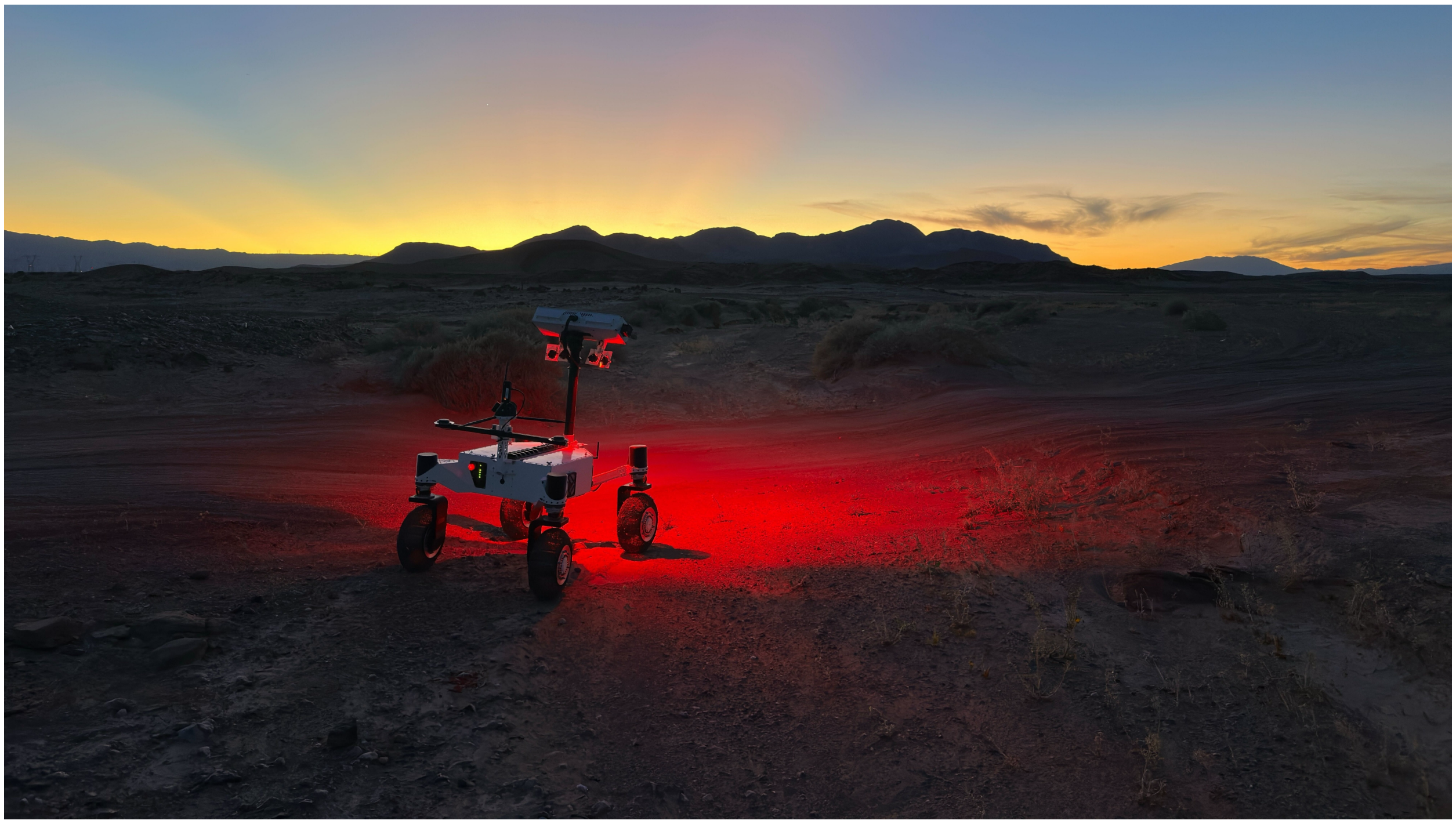

**Figure 6.13.** *Rover prototype ERNEST (Exploration Rover for Navigating Extreme Sloped Terrain) demonstrates some of its advanced mobility and autonomy capabilities in a desert field test. ERNEST is a half-scale prototype, derived from the Endurance concept. Image credit: JPL-Caltech.*

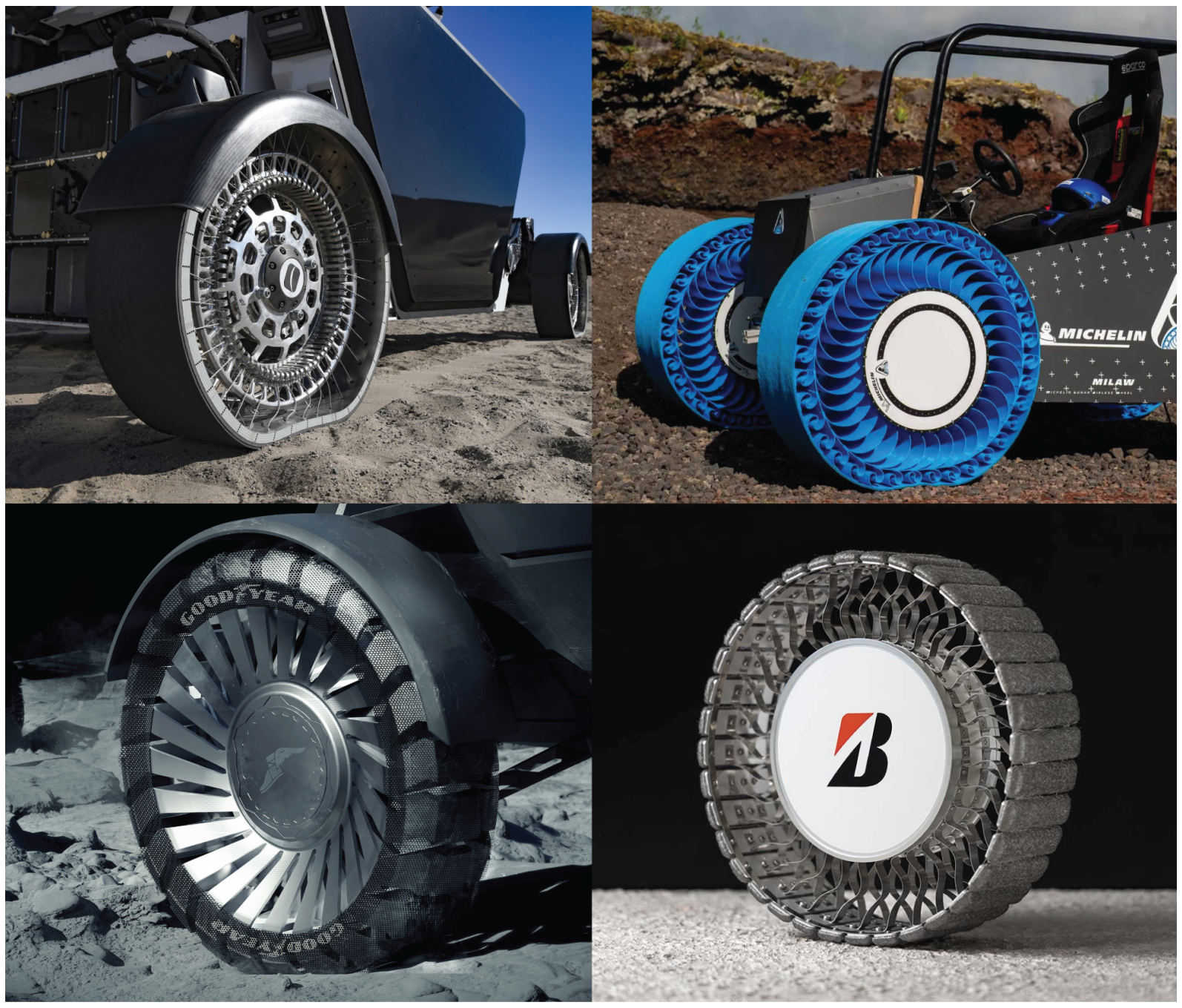

**Figure 6.14.** *Prototype lunar rover wheels developed by several potential vendors for lunar rover vehicles, including: Astrolab Venturi (top left), Michelin (top right), Goodyear (bottom left), and Bridgestone (bottom right).*

### 6.5.4 Sampling Subsystem

As a sample return mission, the sampling subsystem is one of the most critical for mission success. The following sections describe the driving requirements, key challenges and trades, and baseline sampling subsystem design. The concept of operations for sampling is discussed further in Section 6.6.

#### 6.5.4.1 Driving Requirements for the Sampling Subsystem

The sample requirements for SPARX are described in the Science Traceability Matrix (Section 3.6), the Science Experiment descriptions of Section 4.1, and summarized in the Sample Requirements Table in Section 4.3.6. Here, we distill these requirements as they relate to the design of the sampling subsystem.

The primary driving science requirements for the sampling subsystem are the number of sample sites and, for each site, the number (or mass) and size of rocks and regolith to be collected. The baseline number of sample sites required is 20, and the sample system design includes 20 site-specific containers, and 3 opportunistic or contingency sampling containers. The required rock sizes vary, from as small as ≥3 mm, to as large as ≥40 mm, and also include unsieved bulk regolith; meanwhile, the total mass requirements vary by site, ranging from 480 g to 2,244 g. The required characteristics for all 20 required sample sites can be distilled into 6 distinct "classes" (A–F) summarized in Table 6.11. **An additional SDT requirement for the sampling subsystem is the ability to provide documentation imaging throughout the sample handling chain to confirm that acquisition and storage activities met the science requirements.**

**Table 6.11.** *Simplified table of SPARX sample requirements. For more detail, refer to the Science Traceability Matrix (Section 3.6), and Sample Requirements table in Section 4.3.6. These values already account for factors associated with curation and sample quality.*

| | | Class A | Class B | Class C | Class D | Class E | Class F |
|---|---|---|---|---|---|---|---|
| | Number of samples sites of this class (baseline) | 3 | 2 | 9 | 2 | 3 | 1 |
| | Science experiments | E1 | E9 | E2, E4, E5, E13 | E7, E14 | E3 | E10 |
| | Driving lithology | Anorthosite | Low-Ca Pyroxene, Olivine | Impact Melt | (varies) | Basalt | Pyroclastics |
| | Sample collection modality | Targeted scooping and sieving | | Untargeted scooping and sieving | | | |
| Requirements per site | Number of ≥40 mm rocks | ≥12 | — | — | — | — | — |
| | Number of ≥20 mm rocks | — | ≥12 | — | — | — | — |
| | Number of ≥8 mm rocks | ≥32 | — | ≥96 | ≥64 | — | — |
| | Number of ≥5 mm rocks | — | — | — | — | ≥100 | — |
| | Number of ≥3 mm rocks | — | ≥800 | ≥1,200 | ≥800 | ≥500 | — |
| | Bulk regolith (including fines) | ≥400 g | ≥400 g | ≥400 g | ≥400 g | ≥400 g | ≥400 g |
| | Total mass | ≥2,245 g | ≥800 g | ≥971 g | ≥780 g | ≥484 g | ≥400 g |

### 6.5.4.2 Key Sampling Subsystem Challenges

Key challenges for design of the sampling subsystem are: (1) the quantity, mass, and size-frequency distribution of rocks required by science objectives, and (2) minimizing and simplifying the sample transfer interfaces between elements of the design, and (3) designing a system that accomplishes the necessary sample collection requirements as efficiently as possible.

As described in the Science Traceability Matrix (Section 3.6) and summarized in Table 4.3.6, accomplishing the science objectives of SPARX requires collecting rocks and regolith from the lunar regolith. Lunar regolith is the ubiquitous unconsolidated blanket of fractured, broken, and thoroughly mixed rock that has formed over billions of years of impact bombardment. Unlike Earth or Mars, there is very little exposed and accessible bedrock on the Moon. The size of rocks in lunar regolith is very heavily skewed towards the microscopic finest fraction, with a size-frequency distribution of rock sizes following a negative power-law distribution (Figure 6.15). This statistical behavior dictates that there are exponentially more smaller fragments than there is larger rock fragments contained in lunar regolith. To quantify these surface distributions for sampling system design, we rely on a combination of historical sample data, established geotechnical models, and recent robotic surveys. These foundational sources include digitized clast surface size frequency surveys from Apollo 14 landing site (Swann et al. 1977), log-normal mass distribution models of returned Apollo lunar soil samples derived by Carrier (Carrier III 2003), recent macro-scale rock size-frequency distributions measured at the Chang'e 3 landing site (Di et al. 2016), and updated log-normal mass distribution models of returned Apollo lunar soil samples including coarse particles (Kovtun et al. 2024).

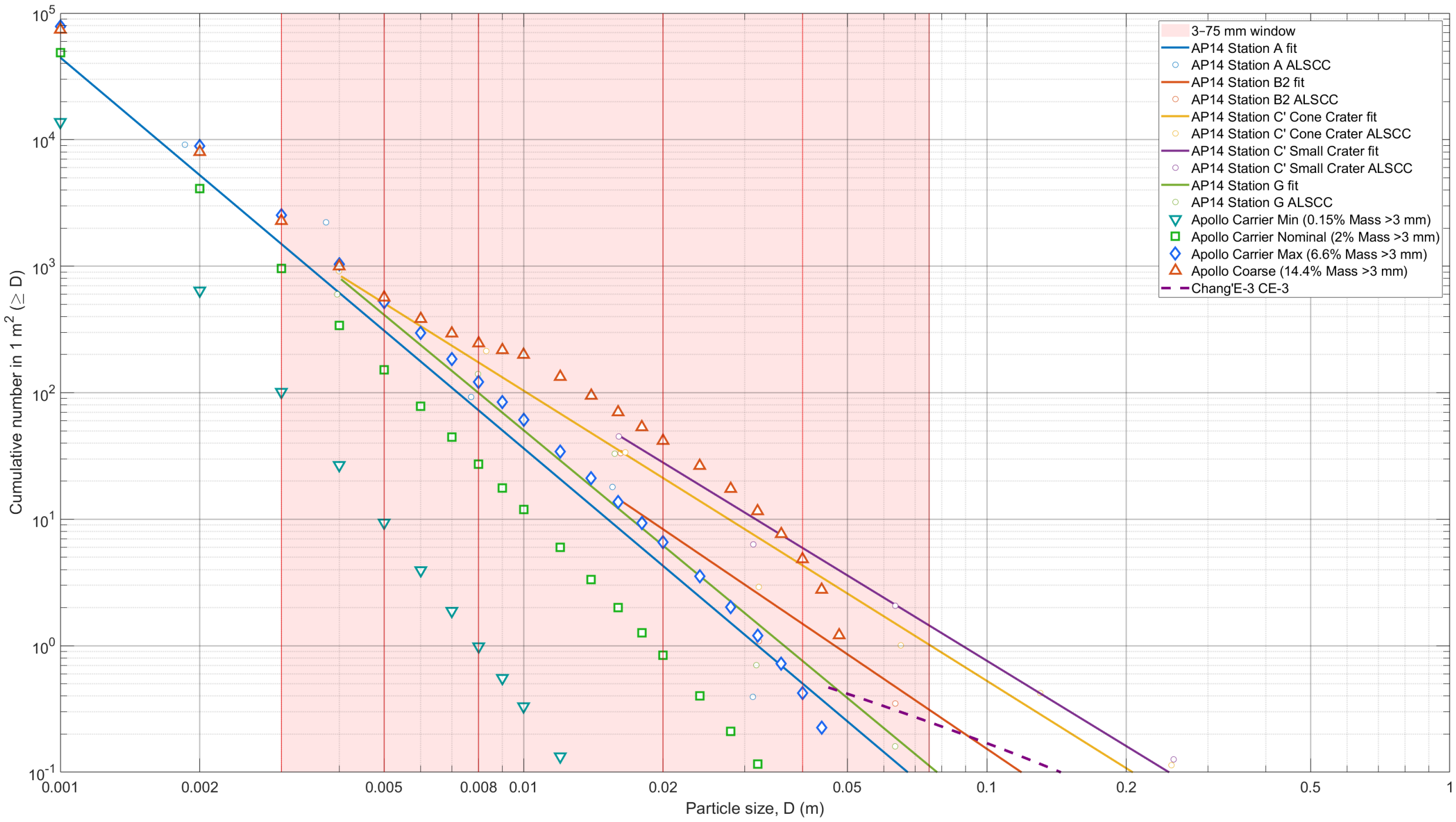

**Figure 6.15.** *The size-frequency distribution of rocks in lunar regolith follows a steeply negative power law. Rock abundance decreases exponentially as rock size increases—which directly informs the sampling requirements for SPARX. This plot summarizes a range of models, including digitized clast surveys from Apollo missions (Swann et al. 1977), log-normal mass distribution models of Apollo lunar soil samples (Carrier III 2003), recent macro-scale rock size-frequency distributions*

*measured at the Chang'e 3 landing site (Di et al. 2016), and updated log-normal mass distribution models of Apollo lunar soil samples including coarse particles (Kovtun et al., 2024)*

Table 6.12 details the expected cumulative rock counts for critical sampling thresholds ranging from 3 mm to 40 mm, corresponding to the range of sample thresholds required by SPARX. To capture the uncertainty and variability across the lunar surface, these expected counts are bracketed a range of abundances represented by the Carrier (Carrier III 2003) and Apollo coarse models (Kovtun et al. 2024). The Carrier model represents the standard particle size distribution interpolated from Apollo regolith samples; however, it only includes data from the < 1 cm size fraction, with interpolation used for the larger, SPARX-critical fraction. The newer Apollo coarse models (Kovtun et al., 2024) are a composite distribution combining the previous fine-fraction data with the records for the sieved > 1 cm clasts. Including both models allows our analysis to more confidently bracket the expected probability of critical SPARX grain size fractions. The data highlights a critical sampling challenge: even in a 1 $m^2$ sampling workspace, the probability of encountering rocks larger than 8 mm drops off significantly. When translated to the volume of a 500 g scoop, the system is statistically unlikely to capture a single 8 mm rock in a single pass under low or medium abundance models, necessitating a bulk processing and sieving strategy. **The low frequency across all models of 20 mm and 40 mm sized rocks necessitates a targeted sampling approach for those size scales.**

**Table 6.12.** *Values represent the predicted cumulative count of rocks larger than the specified target threshold across a range of rock abundance models for either a 1 m2 sampling workspace (top half) or a 500-gram capacity scoop (bottom half). Colors range from ≤1 (red) to ≥500 (green). Percentages in parentheses denote the model's total mass abundance of particles >3 mm under log-normal mass distribution models derived by Carrier (Carrier III 2003) and the Apollo coarse log-normal mass distribution model (Kovtun et al. 2024).*

| | | Rock Abundance Models | | | |
|---|---|---|---|---|---|
| **Operational Scale** | **Target Threshold** | **Apollo: Carrier Minimum** (0.15% mass >3 mm) | **Apollo: Carrier Nominal** (2% mass >3 mm) | **Apollo: Carrier Maximum** (6.6% mass >3 mm) | **Apollo: Coarse** (14.4% mass >3 mm) |
| 1 square meter workspace | >3 mm | 100 | 50 | 2520 | 2280 |
| | >5 mm | 9 | 152 | 521 | 566 |
| | >8 mm | 1 | 27 | 122 | 246 |
| | >20 mm | <1 | 1 | 7 | 42 |
| | >40 mm | <1 | <1 | <1 | 5 |
| 500 g scoop | >3 mm | 10 | 100 | 255 | 222 |
| | >5 mm | <1 | 11 | 37 | 35 |
| | >8 mm | <1 | 1 | 6 | 10 |
| | >20 mm | <1 | <1 | <1 | 1 |
| | >40 mm | <1 | <1 | <1 | <1 |

The diminishing abundance of larger rocks on the lunar surface presents a challenge to meeting site-specific mass requirements for coarse rock size bins. **To mitigate this scarcity, the sampling strategy relies on a robotic arm with an expansive workspace and a sieving-based concentration method.** Through sieving, the rover can process high volumes of unfractionated regolith to selectively retain the limited population of rocks larger than 3 mm while confirming their capture through imaging of the sieved coarse material. Furthermore, **the** **extensive targeting workspace of the robotic arm allows the sampling subsystem to perform precise, targeted capture of stochastically distributed, scarce large rocks previously identified in site survey imaging.**

Development costs and operational risks of planetary surface robotic sampling systems are primarily driven by the number of robotic and sample transfer interfaces across system elements. Interfaces such as robotic docking, the mechanical handling of sample containers, or the dynamic selection of sample flow paths require significant development and verification testing to address the uncertainty introduced by the presence of natural materials, specifically lunar regolith. These interfaces typically require specialized sensors and additional actuation elements, which serve as primary cost drivers for space robotic systems. In developing the Rover sampling subsystem baseline design, the goal was to minimize these interfaces while offering a robust capability to meet science requirements. By utilizing a free space transfer method to an open container array, the design eliminates the need for internal flow-path switching and docking maneuvers. This approach reduces the reliance on specialty hardware and extensive testing cycles. By eliminating circuitous and unobservable sample handling pathways, the baseline design becomes highly inspectable, while providing sufficient flexibility through the 5 Degrees-of-Freedom (DOF) robotic arm to adapt to the uncertain behavior of the lunar regolith.

#### 6.5.4.3 Sampling Subsystem Baseline Design

To develop the Rover sampling subsystem, the technical team executed several key trades, which resulted in a system with characteristics described in Table 6.13.

**Table 6.13.** *Summary of key trades and the resulting characteristics of the Rover sampling subsystem.*

| | |
|---|---|
| Sampling Architecture | Serial robotic arm and scoop for free space sample transfer to a non-docking container array. |
| Class of Scoop | Spade-like sieving scoop featuring a single sieve plate for rock concentration and a narrow chute for fines transfer. |
| Robotic Arm Degree of Freedom and Kinematic Arrangement | 5-DOF serial rotary arrangement utilizing a Yaw-Pitch-Pitch-Roll-Pitch configuration. |
| Sample Container Placement | Multi-container array mounted on the rover's front chassis, adjacent to the robotic arm base. |
| Sampling Confirmation Strategy | Imaging acquired by the rover remote sensing mast and an end effector camera on the robotic arm. |
| Sample Container Lid Strategy | Single lid over an array of containers to mitigate container contamination during rover landing and traverse. |

The sampling subsystem components of the Rover design are shown in Figure 6.16. The subsystem was designed to acquire, process, and store up to 100 kg of lunar regolith and rock samples across 24 dedicated sample containers. The technical team opted to design towards a 100 kg capability to ensure ample margin over the baseline science requirement (≥ 20.5 kg total). Similarly, the 24 sample containers provide additional margin to the 20 required sample sites, with margin provided to allow for redundancy for critical samples, unanticipated challenges in sampling, and unanticipated samples of opportunity. The design team notes that 100 kg may stress the near-term capabilities of Artemis architecture elements, such as the Human Landing System (HLS) and Orion; therefore, **early iteration between the sample return system and** **the design team is strongly recommended to characterize the true sample return margin and ensure consistency between system elements.**

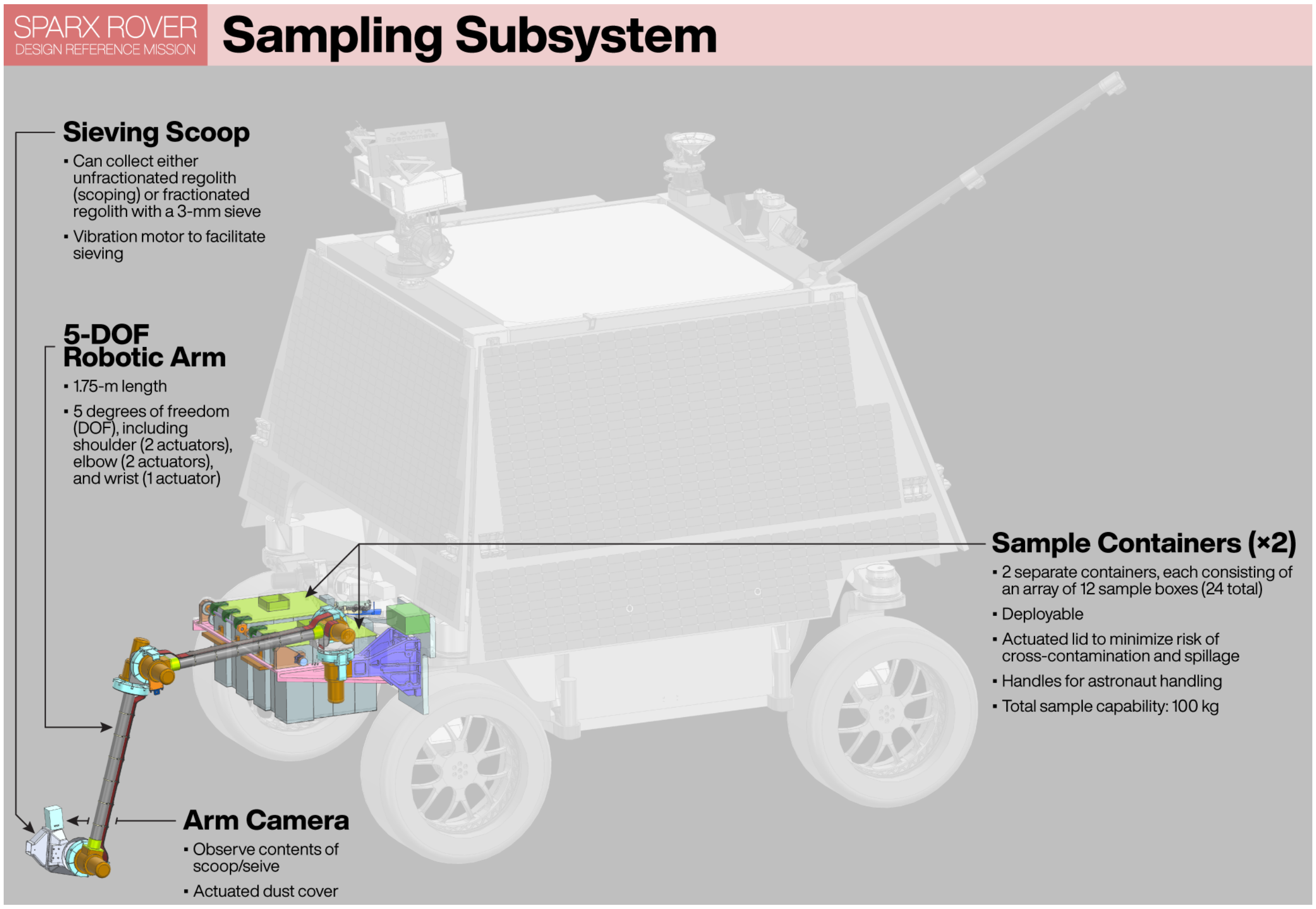

**Figure 6.16.** *Sample subsystem components of the Rover design.*

Sample collection and transfer is performed by a 1.75-meter length robotic arm with five rotary actuators, and an end effector with both scoop and sieving functionalities. In a single scoop pass, the scoop can collect ~500 g of lunar regolith, which it can fractionate through a 3-mm sieve to concentrate rocks ≥3 mm in size, while allowing finer material to exit via a narrow chute. Depending on the sample requirements for the specific site, either the concentrated rocks, sieved fines, or unsieved material can be directed into a sample container or returned to the lunar terrain. A vibration motor integrated into the scoop facilitates sample flow for processing and transfer of material. Vibration also is used to reduce the mechanical loads applied to the robotic arm during sample acquisition. The subsystem's sensor suite includes joint encoders and a wrist-mounted force-torque sensor to manage mechanical loads and facilitate autonomous scoop passes, while an integrated sample camera provides high-resolution inspection of the hardware and sample status. Mast-mounted instruments may also be able to inspect samples in the scoop and/or sample containers. Operational longevity and hardware integrity are maintained through integrated dust seals, protective covers, and a container lid designed to prevent contamination.

The robotic arm actuators are derived from those used on Mars 2020's sampling system. Each consists of a brushless DC motor, solenoid actuated brake, planetary gearbox, and strain wave gear output stage. The motor, brake, and planetary gearboxes are largely inherited as-is, while the output stage is resized for the lower torques required for this Rover. The total revolutions are like what was required for Mars 2020, giving high confidence in the actuator's ability to meet this Rover's operational life requirements. While the Mars 2020 actuators were qualified to narrower survival temperature range, they were qualified to many more cycles than is required for SPARX and there is no known low temperature survival limit (although a thermal cycling life test should be conducted early in development to confirm).

A set of 24 sample containers supporting up to 100kg of total sample mass is arrayed on the front of the rover for direct access by the robotic arm. The set of containers features a single, actuated lid to avoid contamination during landing and traverse activities, as well as protect sample integrity during the Artemis sample handling mission phase. One or more arrays may be used. Container arrays feature separation devices that enable the release of containers for handoff to Artemis.

**Finding: The SDT finds that the sampling subsystem requires documentation imaging throughout the sample acquisition and storage process to verify successful acquisition of the appropriate sample material.**

**Finding: The SDT finds that as a result of the expected distribution of rock sizes in the lunar regolith, the sampling subsystem arm should allow access to a large workspace volume, in order to support the required sieving-based strategy, and should also support more targeted acquisition modes for the larger clasts.**

**Finding: The SDT recommends a 3 mm sieve size for collecting and concentrating the required rock fractions and also requires the ability to collect unsieved regolith.**

### 6.5.5 Thermal Control Subsystem

Operating on the surface of the Moon is inherently challenging, as it requires needing to survive, operate, and thrive during the hot 14-day long lunar day, while also enduring the extremely cold temperatures of the equally long lunar night. Close to the lunar pole, the definition of day and night becomes complicated, as the low obliquity of the Moon results in extremely long shadows, periods of prolonged shadow, and even Permanently Shadowed Regions (PSRs). **The thermal control subsystem for the SPARX Rover must be capable of operating across a wide range of latitudes—both in the warm mid-latitudes (up to 52°S) and around the PSRs near the south pole (~90°S) for the final rendezvous with Artemis.** Figure 6.17 shows the surface temperature for this range of latitude determined from LRO Diviner thermal measurements.

To balance Rover capability, cost, reliability, and other factors, **the key design philosophies of the thermal control subsystem should be to minimize night survival power and allow for Rover operations at any time of lunar day regardless of latitude, orientation, terrain, or other environmental factors** (consistent with the overarching architecture guidelines described in Section 6.2). The thermal control subsystem design is intimately linked to other engineering design elements, particularly the mechanical subsystem (Section 6.5.2), and must be carefully considered and accounted for as a primary driver of overall system design. Thermal analysis is also coupled with assumptions about the terrain and environment, and must account for thermal hazards arising from reflected or re-radiated thermal energy, view of the sky, tilts due to slopes and rocks, etc.

The thermal control subsystem must maintain all hardware on the Rover within Allowable Flight Temperature (AFT) limits for all mission phases. Table 6.14 shows examples of common AFT limits for different types of components in the Rover. The thermal control subsystem components of the rover design are shown in Figure 6.18. As shown in Figure 6.17, worst-case hot conditions are sensitive to latitude, and there is more than 200 °C difference in the maximum ground temperature from the mid-latitudes to the pole. Worst-case cold conditions are relatively insensitive to latitude; as soon as the sun sets, the lunar environment is very cold no matter the latitude. For the baseline Rover described here, the northernmost latitude was assumed to be 52° S (necessary for collecting samples in the central SPA and Poincaré regions).

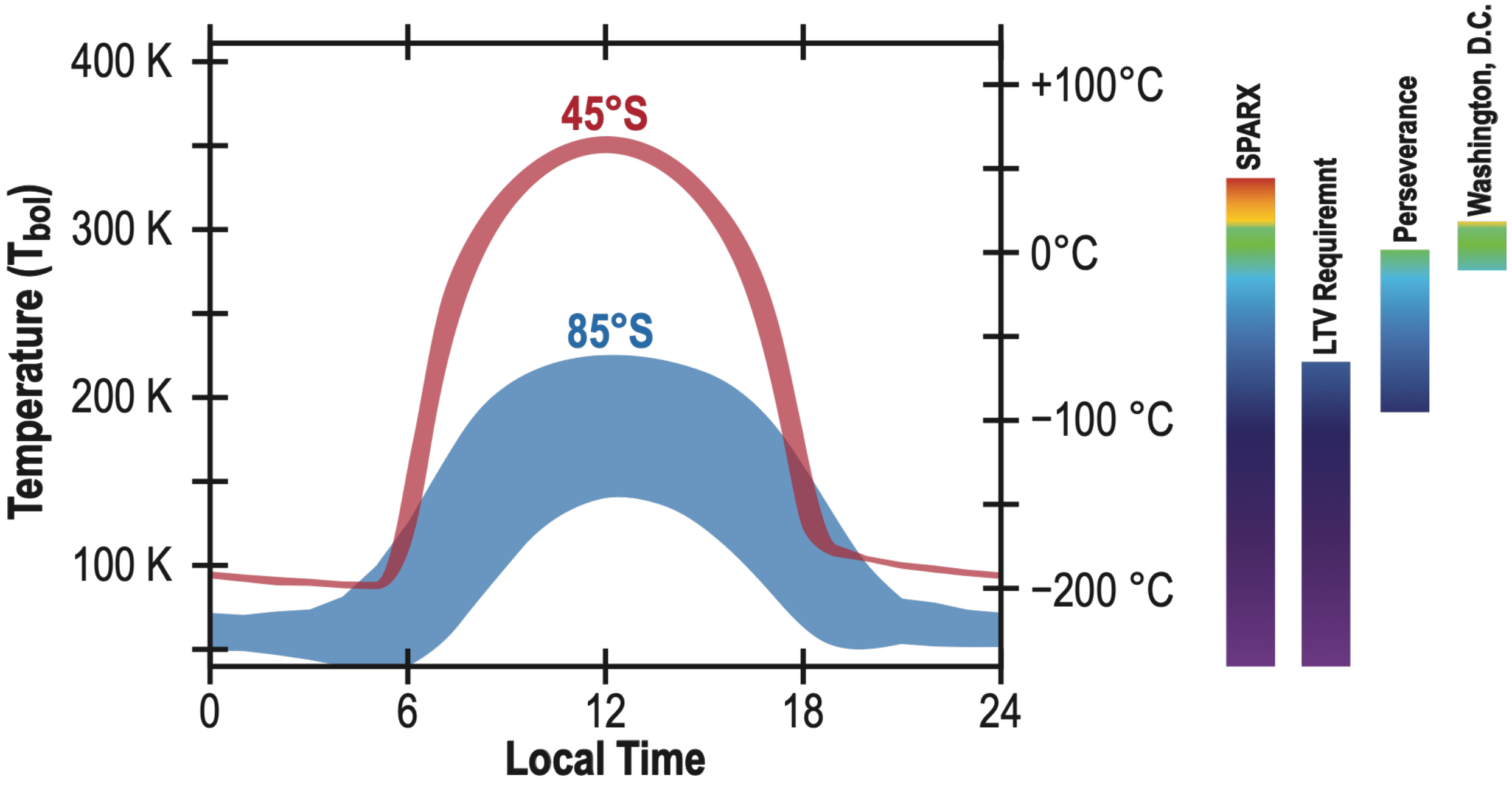

**Figure 6.17.** *Surface temperatures on the surface of the Moon for the relevant latitude ranges required by SPARX. Temperatures are from Williams et al. (2017) from the LRO Diviner thermal instrument.*

**One of the biggest drivers for the whole Rover system is the need to survive the extreme cold of the lunar night.** For the solar powered Rover to survive the cold lunar night, it is necessary to reduce the required amount of survival power, as this directly impacts battery

**Table 6.14.** *The Allowable Flight Temperature (AFT) limits for different components in the Rover.*

| | Non-Operational | | Operational | |
|---|---|---|---|---|
| | Minimum AFT (°C) | Maximum AFT (°C) | Minimum AFT (°C) | Maximum AFT (°C) |
| Actuators | −200 | +100 | −55 | +100 |
| Antennas | −150 | +130 | −150 | +130 |
| Battery | 0 | +30 | N/A | N/A |
| Electronics | −40 | +50 | −40 | +50 |
| Engineering Cameras | −130 | +50 | −55 | +50 |
| Science Cameras | −40 | +50 | −40 | +50 |
| Solar Array | −230 | +140 | −230 | +140 |

size, and thus rover size and mass. **Using governing design metrics, every 1 W of survival power requires ~7 kg of battery mass.** Most components are put into hibernation mode at night, and most of the power draw during hibernation is for survival heating to maintain hardware above non-operational AFT limits. Electronics and avionics are housed in the central Warm Electronic Box (WEB), which is thermally isolated in the interior of the Rover and maintained within the AFT range of −25 °C to +50 °C. High-performance thermal technologies exist (such as thermal switches, loop heat pipes, spacerless multi-layer insulation, etc.) to isolate the WEB and maintain survival temperatures overnight with ~15 W of electrical heating during the lunar day at the lowest proposed mission latitudes and highest relative sun elevation angles, the local regolith temperature can exceed +70 °C. The thermal design is cold biased, meaning that all components remain below the maximum AFT without the need for active coolers, even in the worst-case hot scenario. Because of this, some components may be too cold in the worst-case cold case without active thermal control (i.e., heaters). Cold biased thermal designs are the most common thermal designs for spacecraft. Potential exceptions to the cold biased thermal design are any science instruments that may have more stringent thermal requirements. For example, infrared spectrometers that may require their own cryocoolers to maintain optics at tight performance temperature ranges during data collection. A zenith-facing thermal radiator rejects heat from the Rover electronics in the hot case and is sized by considering factors such as maximum power dissipation, unfavorable Rover orientation, local terrain features, and optical property degradation due to dust coverage. The thermal connection from the WEB to the radiator is via Loop Heat Pipes (LHPs) with integrated Thermal Control Valves (TCVs) that passively control the flow of heat with a large turn-down ratio to bypass the radiator in the cold case and conserve heat within the WEB during night survival.

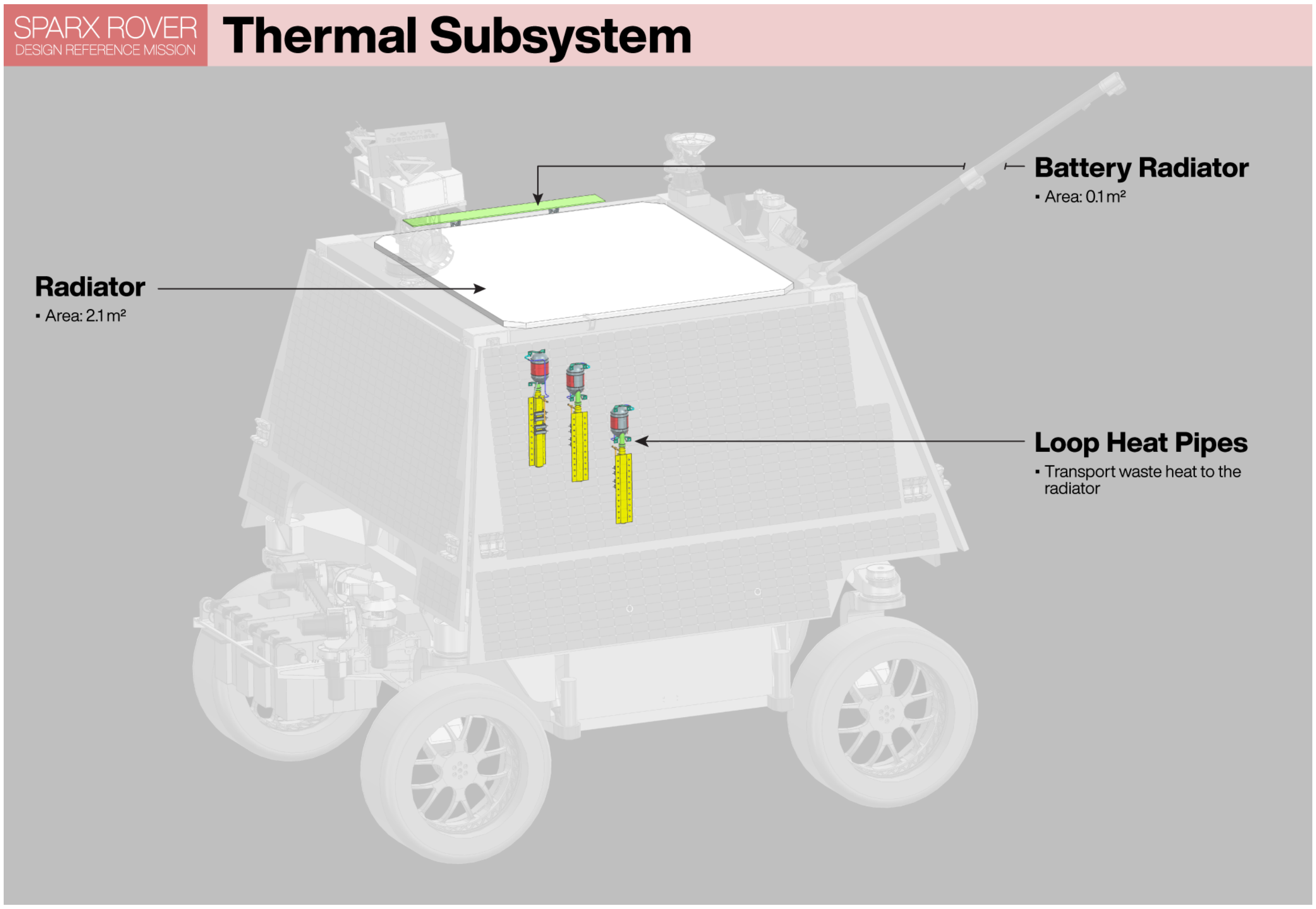

**Figure 6.18.** *Thermal Control subsystem components of the Rover design.*

Components on the exterior of the Rover are exposed to the extreme temperatures of the ambient environment and are therefore expected to be qualified to the widest temperature ranges possible to minimize the amount of survival heating needed in the cold case and heat rejection provisions such as local radiators in the hot case. For example, cold actuator technology developed for surface applications can reduce overall survival power by ~30%, significantly reducing Rover mass.

The Rover must provide and maintain defined thermal interfaces to accommodate science instruments. Understanding potential requirements for science instruments is critical for design of the TCS, particularly to estimate the allocation of survival heating needed and the impact to the overall system power budget. Instruments located on the exterior of the Rover will be subjected to extreme hot and cold environments and may be required to provide local thermal control, while any components of instruments located within the WEB (such as electronics) are expected to comply with the −25 °C to +50 °C WEB AFT range.

**Finding: The SPARX traverse requires thermal accommodation for latitudes as high at 52° S down to the lunar South Pole (90° S).**

**Finding: The thermal subsystem must be designed to allow the rover to operate at any part of the lunar day, regardless of latitude, orientation, terrain, or other environmental factors. The SDT strongly recommends against thermal subsystem designs that do not meet these criteria (e.g., requiring a pause in operations during the hotter parts of the lunar day).**

### 6.5.6 Power Subsystem

**For a solar powered rover architecture, the driving requirements are the need to provide sufficient power during the lunar day to support traverse and sampling operations in a stressing hot environment, and to provide enough stored energy to**

**sustain the rover survival throughout the lunar night.** Night survival also drives the need for electronics capable of operating at very low power in a hibernation mode. To meet these mission needs, the power subsystem comprises solar panels for power generation, a rechargeable battery for energy storage, and power electronics for power bus regulation and distribution. The power subsystem components of the rover design are shown in Figure 6.19.

The Rover is surrounded by 360° of wrap-around solar arrays, consisting of four panels, with a total solar array area of about 10.2 $m^2$. This configuration is designed to maximize sunlight exposure across all sun elevation and azimuth angles expected while the Rover is driving throughout the lunar day for the full range of latitudes covered by the baseline traverse (52°S to 90°S), including when the rover is not on flat terrain. The available solar power varies across a range of 360–690 W, with an average of about 550 W (Figure 6.20). While the panels leverage design heritage from the Farside Seismic Suite (FSS) and Cooperative Autonomous Distributed Robotic Exploration (CADRE) lunar projects, the higher predicted max temperatures and larger number of thermal cycles over the SPARX mission duration would necessitate extended testing, and potential materials modifications.

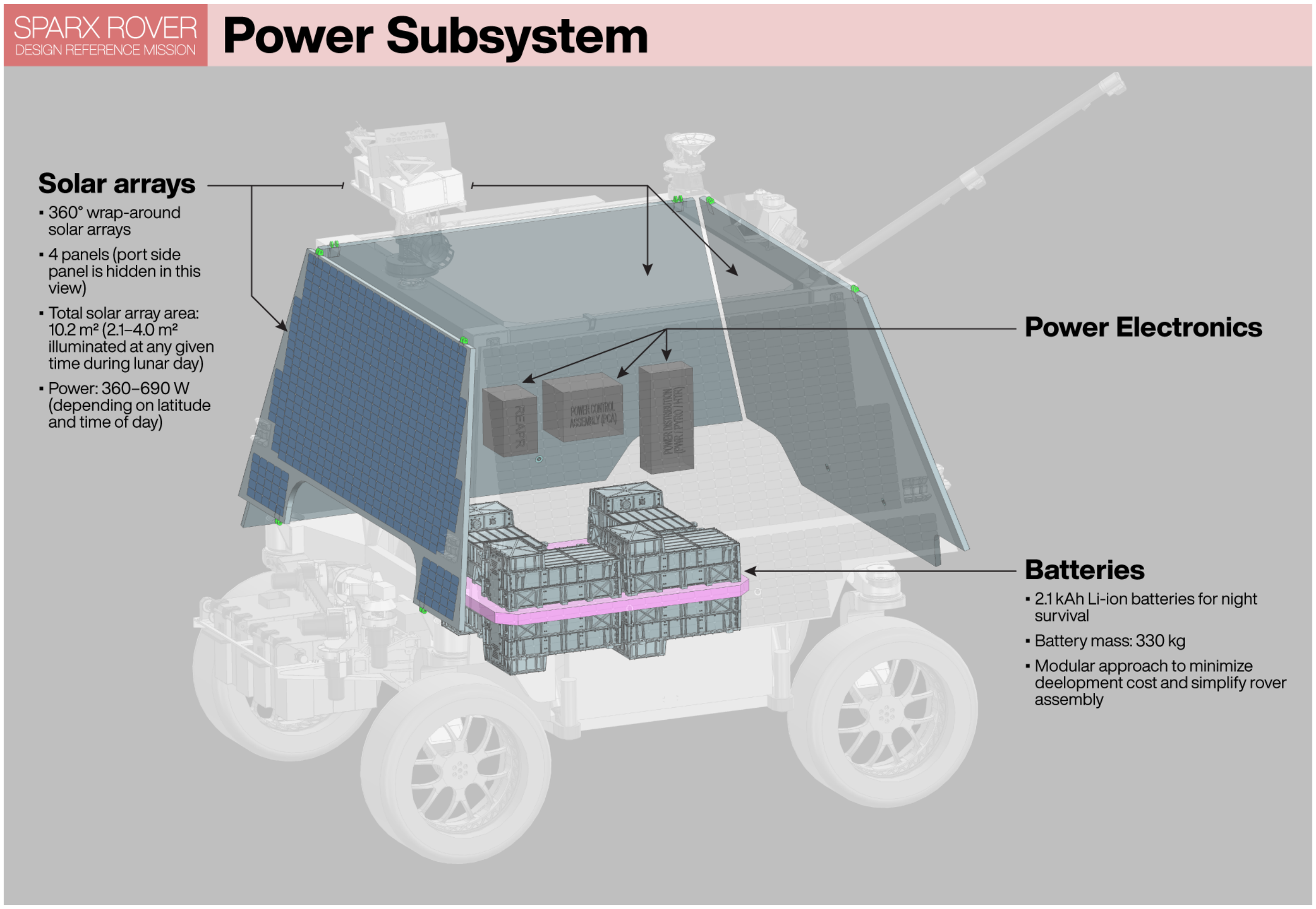

**Figure 6.19.** *Power subsystem components of the Rover design.*

During the night, the rover is powered by a large 2.1 kAh Li-ion secondary battery, with a mass of about 330 kg. Given its size and tighter thermal constraints relative to the rover electronics assemblies, the battery is accommodated in a dedicated warm enclosure. While commercial off-the-shelf (COTS) 18650 form factor battery cells are used for this design, battery packaging remains a challenge given the significant size. For this Rover design, a modular approach would be recommended to reduce development costs and simplify assembly.

The power electronics provide maximum power point tracking (MPPT) for solar panel regulation to extract the peak possible power from the panels; each panel has a dedicated MPPT converter to account for its unique operation conditions (insolation, temperature, etc.) Power is then distributed to electronic loads, heaters, and release devices via power switching cards controlled through a communications interface with the C&DH compute element. The electronics designs leverage past mission heritage, but they would be evolved to target low power consumption to minimize night hibernation power and reduce the demand on the battery (e.g., the use of microcontrollers).

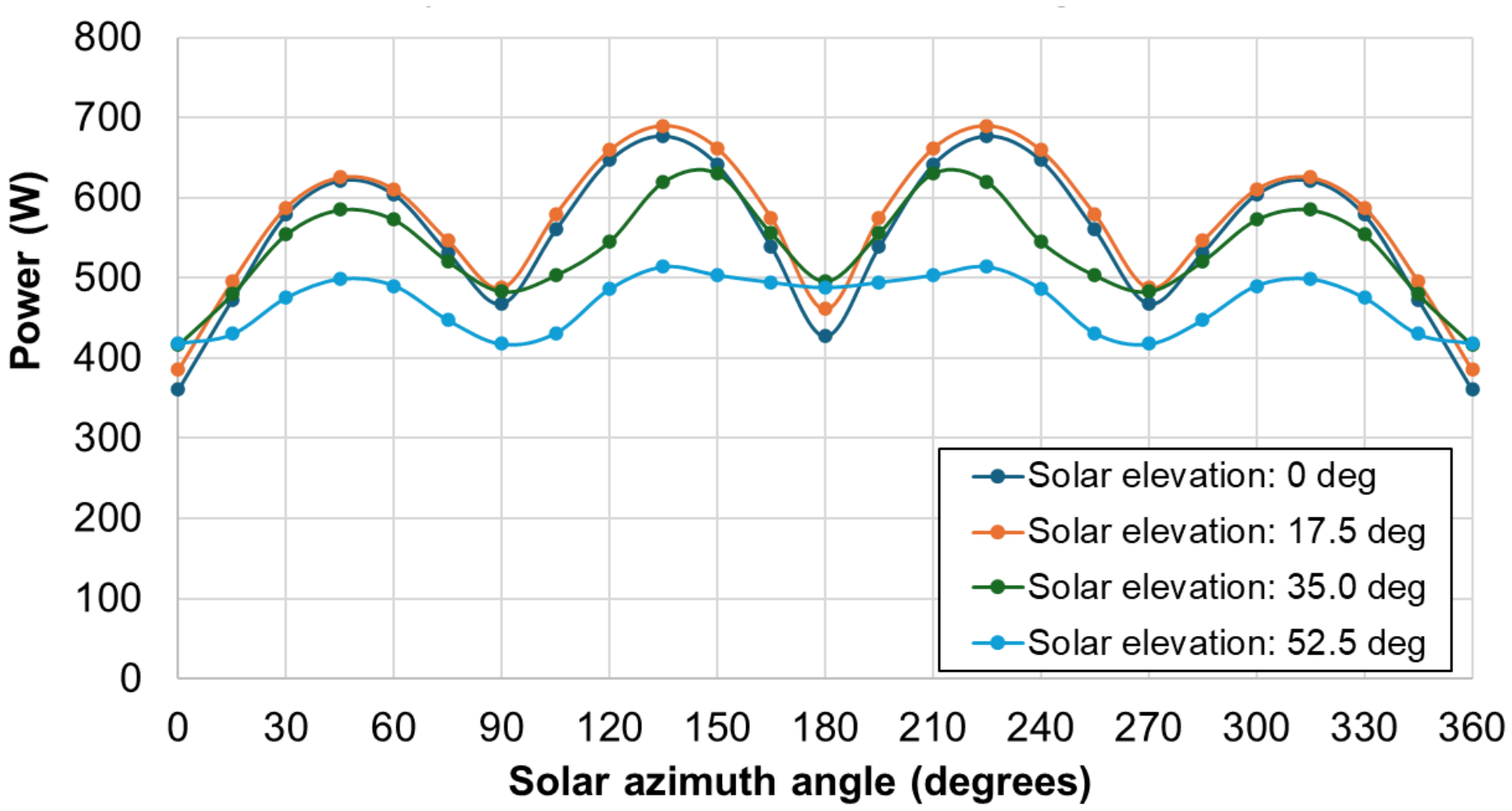

**Figure 6.20.** *Available solar power as a function of solar azimuth angle and elevation in a rover-fixed reference frame. These calculations assume a power subsystem loss factor of 0.95, and an additional loss from dust on solar arrays and shadowing from terrain of 0.90.*

### 6.5.7 Telecommunications Subsystem

#### 6.5.7.1 Telecommunications Relay

For most of the mission, the Rover will be on the farside of the Moon, and out of direct line-of-sight with Earth; therefore, the Rover telecommunications subsystem must utilize orbital communications relay satellites. Several lunar communications relay constellations are in development and are anticipated for deployment by the time this mission would be operating. We baseline the NASA Lunar Communications Relay and Navigation Systems (LCRNS) constellation as the relay service for the Rover. LCRNS is designed to fulfill two major service functions: data services, and Position, Navigation, and Timing (PNT) services. Therefore, the telecom subsystem architecture for the Rover combines these two functionalities in one unit to optimize the mass, power, and cost of the system. At the time of the planned Rover operations (c. 2032), LCRNS will consist of 5 satellites in Elliptical Lunar Frozen Orbits (ELFO), designed to provide sustained, high-elevation communication and navigation services, particularly to support Artemis at the lunar south pole (Figure 6.21). These orbits are "frozen," meaning the gravitational perturbations from the Earth and the Moon's non-spherical shape are balanced to prevent the orbit from drifting, thereby minimizing station-keeping maneuvers. Current plans for this orbital infrastructure involve placing the relay satellites on three orbital planes to ensure continuous, global coverage. Analysis of the planned lunar communications relay architecture suggests that the Rover will often have multiple satellites visible at any given time (Figure 6.22).

Terrain effects (e.g., multipath) can be important for communications especially if and when the relay is low on the horizon. These effects were modeled and considered when evaluating the telecommunications link budget.

While the Rover baselines LCRNS for communications relay, it is expected that there will be several other commercial lunar orbiters in service in the early 2030s that may be leveraged by the SPARX rover. For example, ESA is working with Surrey Satellite Technology Limited (SSTL) limited company to build and deploy an orbiter, called Lunar Pathfinder (LPF), in cooperation with NASA Commercial Lunar Payload Services (CLPS) program. It is expected that Lunar Pathfinder will be launched in early 2027 with the Firefly Aerospace Blue Ghost Mission 2. Firefly Aerospace is expected to deploy a commercial lunar orbiter called Elytra on the same CLPS mission as LPF. Both LPF and Elytra are planning to use the S-band version of the Proximity-1 relay protocol (described below).

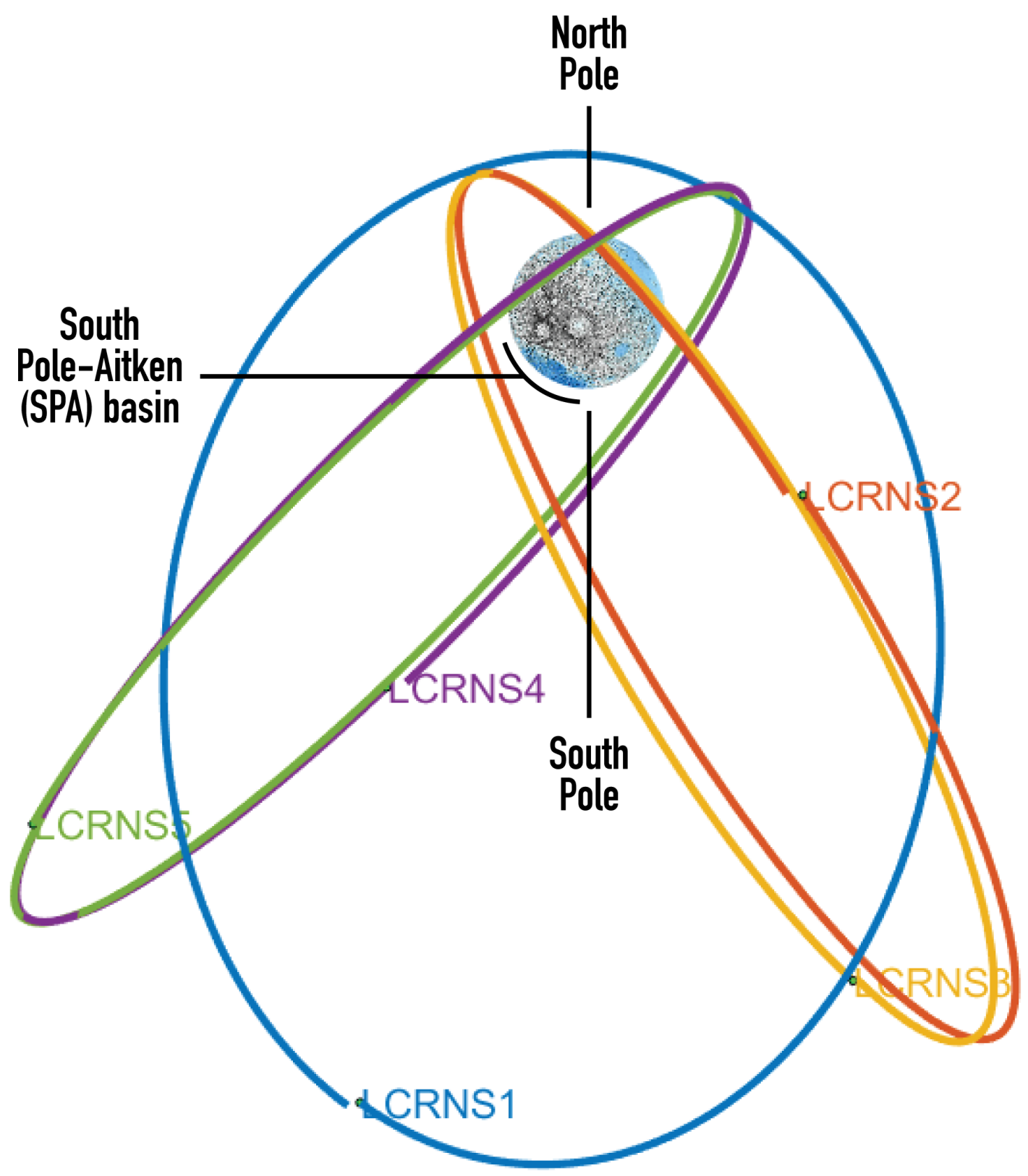

**Figure 6.21.** *The SPARX Rover baseline telecommunications plan would involve utilizing NASA's LCRNS communications infrastructure which will consist of at least five satellites. These orbits have an orbital period of ~32.8 hours, a high inclination of ~56° (designed to provide persistent coverage over the lunar south pole), and a large eccentricity of ~0.7*

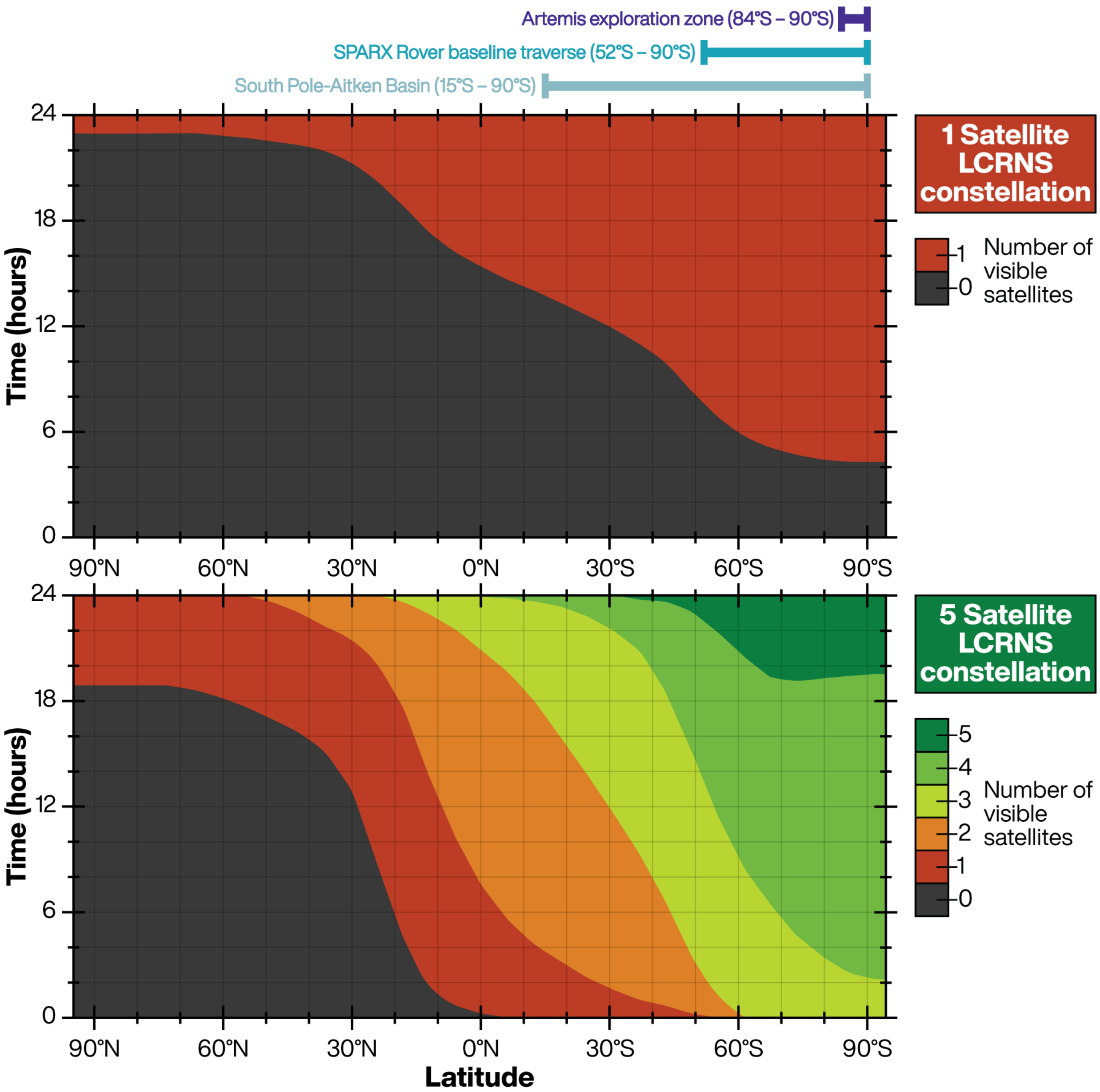

**Figure 6.22.** *Predicted daily contacts between the Rover and relay satellites, assuming a 1 satellite LCRNS configuration (top) and a 5 satellite LCRNS configuration (bottom). These calculations assume a 10° elevation mask (i.e., they assume that satellites near the horizon may not make for suitable relays).*

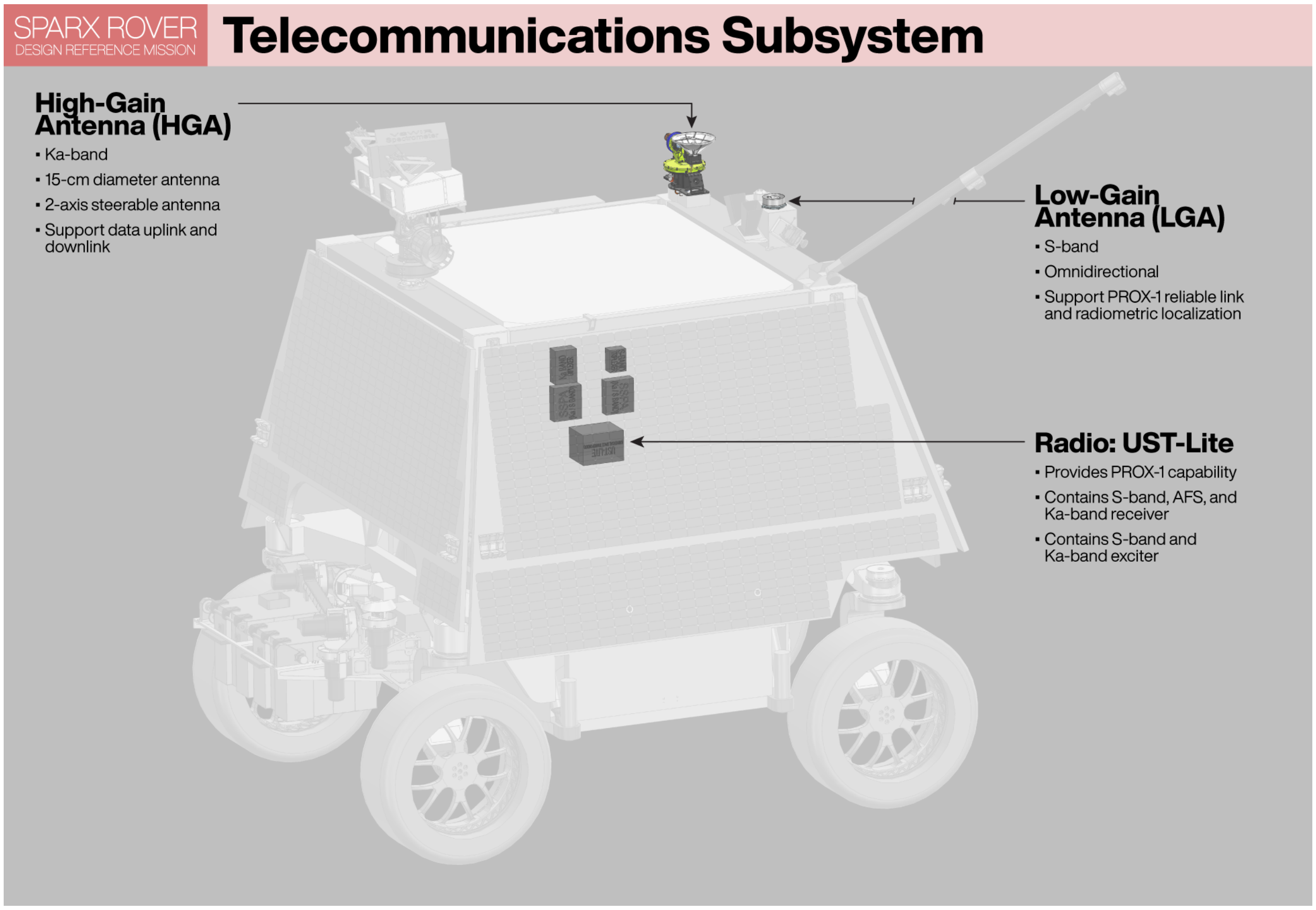

**Figure 6.23.** *Telecom subsystem components of the Rover design.*

### 6.5.7.2 Rover Telecommunications Subsystem

The telecom subsystem components of the rover design are shown in Figure 6.23. The Rover telecommunications subsystem supports all mission uplink and downlink requirements using S-band and Ka-band frequencies and components.

#### High-Gain Antenna (HGA)

A 15-cm directional Ka-band high gain antenna (HGA) with 30 dBi gain combined with a 5 W Ka-band power amplifier supports a maximum of 32 Mbps transmit data rate and a maximum of 64 Mbps receive data rate with the LCRNS constellation with 3 dB margin. The HGA is sized to maximize the data rate, while minimizing the pointing requirements as well as the multipath losses. The Ka-band HGA would be utilized while the rover is stationary.

#### Low-Gain Antenna (LGA)

The Rover also carries an S-band low gain antenna (LGA), which supports link establishment, navigation needs, a backup to the HGA and supports the Rover during safe mode. The LGA is boresight gain of 8.6 dBi. The LGA is fixed at zenith. The LGA combined with a 5 W S-band power amplifier supports a maximum of 128 kbps transmit data rate and a maximum of 512 kbps receive data rate with the LCRNS constellation with 3 dB margin. The S-band LGA is needed for PNT and for back-up/safe mode. The S-band forward link needs to be closed for SPARX to hail, track, and point its K-band antenna towards LCRNS. The S-band LGA can be operated while the rover is either stationary or driving.

#### Radio

The Rover utilizes a capable radio, UST-Lite, to support the required data rates. This radio, currently under development at JPL, is a lighter variant of the JPL-built Universal Space Transponder (UST). The radio is designed to incorporate a receiver for the LCRNS S-band PNT component called Augmented Forward Signal (AFS). Combining the PNT as well as the telecom functionalities into one radio simplifies the operation of the rover and reduces the mass and power consumption. Based on analysis of the technical team, there is no other radio option with adequate flight heritage. The Rover baselined a single UST-Lite. While this conceivably represents a single point of failure, the pre-project decided that a single string radio was an acceptable risk based on the mission classification of SPARX (Class C+ as described in Section 6.9) and given the high reliability of radios that UST-Lite derived heritage from. For example, MRO, MSL, and Mars2020 all have dual string radios and are still operating on radio #1 despite many years (or decades) of operations.

The radio uses Proximity-1 (Prox-1) protocol for communicating with the relay constellation. For the rover concept of operation, we will operate the Prox-1 on either S-band or K-band depending on the need of the data volume transfer but not on both. The telecom will be at most operating in Full Duplex mode on either K-band or S-band channel and in Receive mode on S-band for the AFS localization functionality.

**Finding: The SDT finds that the rover telecommunications subsystem should include, at a minimum, a Ka-band high gain antenna (HGA) to support data transfer; as well as, an S-band low gain antenna (LGA) to support pointing, navigation, and timing (PNT) data, and back-up/safe mode.**

### 6.5.8 Guidance and Control Subsystem

The primary functions for the Rover's Guidance and Control (G&C) subsystem are to provide the necessary sensors, algorithms, and capabilities to enable to navigate on the surface of the Moon. The Rover guidance and control subsystem components of the rover design are shown in Figure 6.24.

#### 6.5.8.1 Sensors

For autonomous driving, Navigation Cameras (NavCams) are required for path planning, hazard avoidance, and visual odometry. A total of four NavCams are baselined for the Rover: one forward-facing pair is mounted on a pan-tilt mast head, while a second non-articulated camera pair is mounted on the rear of the Rover. The rear-mounted NavCams are required both for backing up and providing redundant perception sensors. To aid navigation, LED-based illuminators are mounted on the Rover body. Illuminators are necessary for two reasons: (1) Illumination significantly aids navigation through shadowed terrains, that otherwise may require long integration times to properly see. (2) Illumination with the appropriate geometry can mitigate the opposition effect that occurs in scenarios where the Sun may be directly behind the rover, which can wash out surface textures and terrain, obscuring hazards. The Rover's LED-illuminators are mounted to the Rover body in optimized locations to maximize the camera/light separation to minimize self-illumination opposition effects that reduce perceived texture. The camera-based perception system is designed for a wide dynamic range to handle sunlit and shadowed regions and with short exposures to allow imaging-while-driving at the targeted speeds. The mast-mounted cameras may also provide incidental science context imaging and support sample collection operations.

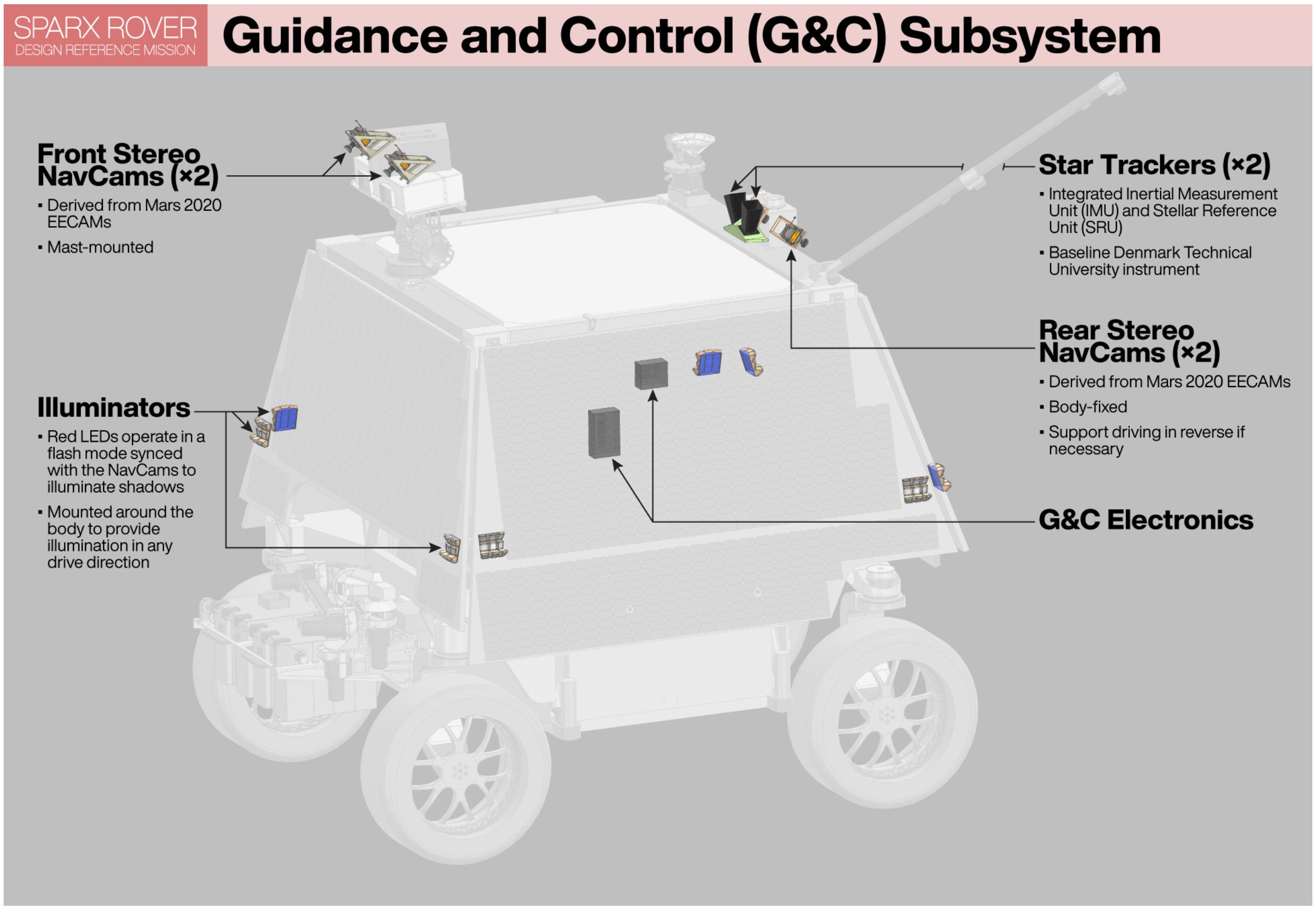

**Figure 6.24.** *Guidance and Control (G&C) subsystem components of the Rover design.*

The baseline Rover NavCam is the Enhanced Engineering Camera (EECAM) which has significant flight heritage through multiple Mars rover missions (Figure 6.25, Table 6.15). The power and thermal constraints established by the lunar night survival requirements drive sensor procurement requirements. Sensor procurement is biased towards hardware that can survive below -200° C. JPL testing has shown the EECAM electronics can survive non-operational temperatures down to −200 °C, which far exceeds the listed limits of current commercial camera options.

**Table 6.15.** *Specifications of the EECAM, based on Mars 2020 Perseverance's NavCams.*

| NavCam (EECAM) Capabilities | |
|---|---|
| Detector type | 20 Megapixel CMOS |
| Optical format | 35 mm |
| Array size | 5120 × 3840 pixels |
| Pixel Size and Pitch | 6.4 microns |
| Full well charge | 15,000 e- |
| Pixel dark noise | 8 e- RMS |
| Pixel dark current | 125 e-/s @ 25 °C |
| Windowing | Region of Interest (ROI) |
| Shutter | Global |
| Pixel quantization | 12 bits |

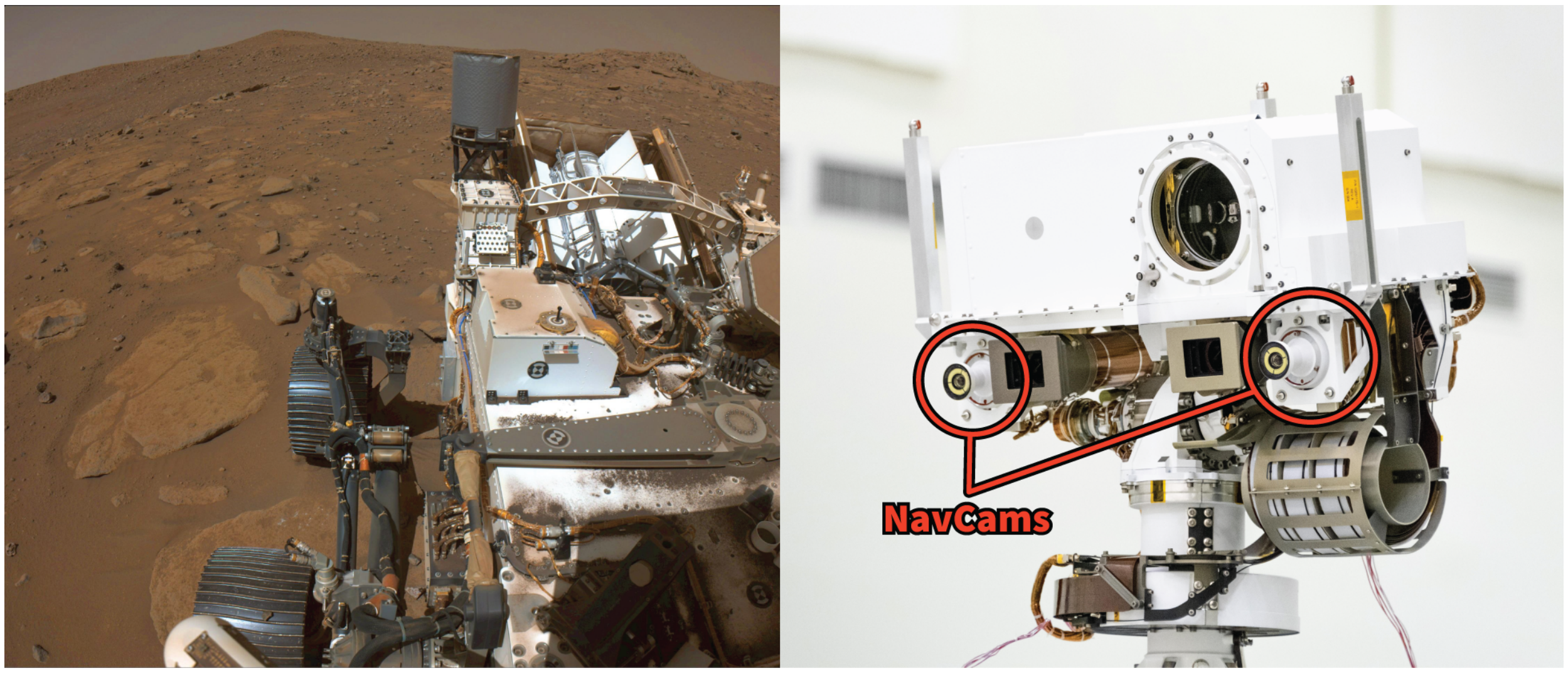

**Figure 6.25.** *Left: An example image of the Perseverance rover taken by the left NavCam (1 November 2023). Right: Close-up image of the head of the Perseverance rover's remote sensing mast, taken while the rover was being assembled on Earth. Image credit: JPL-Caltech.*

An IMU (Inertial Measurement Unit) and SRU (Stellar Reference Unit) are also required to provide odometry measurements, and more accurate heading (yaw) measurements. **The Rover IMU also has the added benefit of providing sufficiently accurate measurements of the lunar gravity field to meet the science requirements for the gravimeter in the strawman science payload** (Sections 4.2.8, 6.4.5). The current baseline for the IMU and SRU is one integrated instrument provided by the Denmark Technical University (DTU). This integrated sensor provides mass savings by minimizing the need for a separate IMU, as each SRU head contains four MEMS IMUs. The output of these IMUs can be fused to achieve similar performance metrics to an LN-200 IMU, which have flown on multiple missions (e.g., Perseverance). This is especially important as the IMU accelerometer measurements can be used for science gravimeter measurements as well. Additionally, the SRU heads have been qualified to survive down to −120 °C, whereas most other available star trackers have only been qualified to −40°C. While this may still require survival heating, this lower qualification temperature represents significant power savings. Figure 6.26 shows an example of the DTU SRU head, while specifications are provided in Table 6.16.

**Table 6.16.** *Specifications of the DTU IMU and SRU.*

| IMU Capabilities | | |
|---|---|---|
| Dynamic range | Accelerometers | ±2.0 to ± 16.0 g |
| | Rate gyroscope | ± 125 °/s |
| 3 db Bandwidth | | up to 3.3 kHz |
| Sample/output rate | | 6.7 kHz / 100 Hz |
| Resolution | Accelerometers | 61 µg |
| | Rate Gyroscope | $4 \cdot 10^{-3}$ °/s |
| Random walk | Accelerometers | < 0.015 m/s/√hr |
| | Rate gyroscope | < 0.13 °/√hr |
| Bias stability | Accelerometers | < 0.014 mg |
| | Rate gyroscope | < 0.7 °/hr |
| **SRU Capabilities** | | |
| Attitude Accuracy | | 0.02 ° (3σ) |
| Field of View | | 13 x 18 ° |
| Output Rate | | 10 Hz |
| Slew Rate | Cross-boresight | up to 8 °/s |
| | Roll | up to 30 RPM |

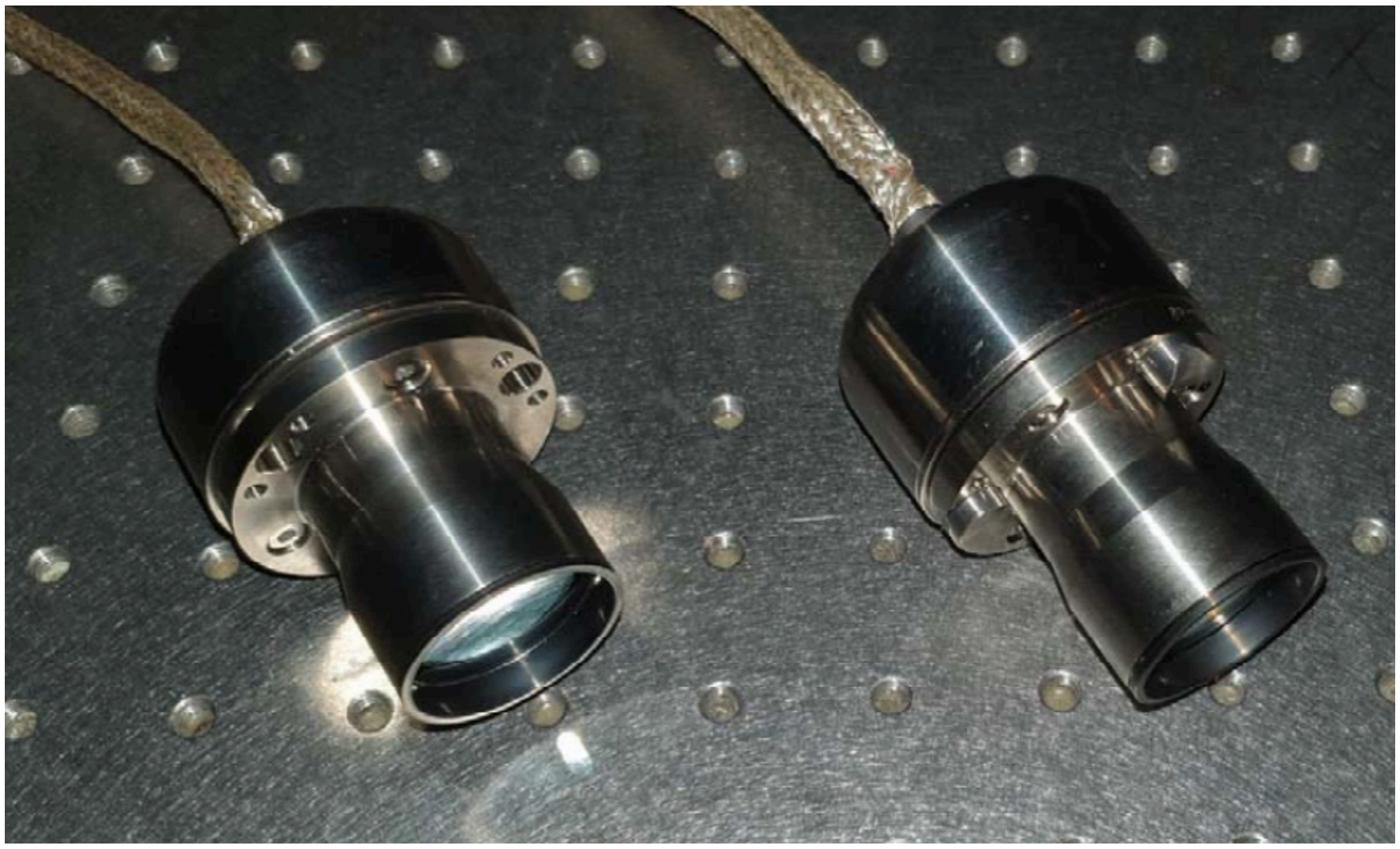

**Figure 6.26.** *Two example DTU SRU heads, with (left) and without (right) the integrated IMU. Image credit: DTU.*

In addition to procurement of G&C sensors, the G&C subsystem also provides the algorithms to point the high-gain antenna (HGA) at the lunar relay constellation for uplink and downlink communications. There are two notable differences for this rover's pointing requirements compared to the Mars 2020 Perseverance rover. First, the Rover has an improved attitude estimation accuracy due to the inclusion of an SRU (that Mars 2020 lacks). Second, the Rover has a reduced mechanical alignment knowledge accuracy due to the higher temperature cycling on the lunar surface compared to Mars. Taking these into consideration, an end-to-end HGA pointing error budget was developed, and is expected to be better than 3° (3σ).

#### 6.5.8.2 Global Localization

Global localization determines the vehicle's position and heading relative to a lunar-centric coordinate frame or orbital maps—which is essential for both navigating across the surface of the Moon, and for pointing the high-gain antenna for communications relay. Global localization corrects for drift that accumulates from relative pose estimation and can be executed on the ground or onboard. Most past rover missions have relied on ground operators and tools to correct for rover positional drift, while using sun sensing techniques to correct for global heading onboard. **For long-range roving at higher speeds, onboard global localization becomes a necessity to meet the target traverse rate and reduce the requirements for communication and operations shifts.**

Key trades for onboard global localization include: (1) radiometric techniques from the lunar relay orbiter network; (2) optical techniques; (3) hybrid optical and radiometric techniques; and (4) celestial techniques linking to the lunar gravity field. Many of these capabilities leverage heritage navigation and global localization approaches used by the Mars rovers. With radiometric techniques, one-way ranging and Doppler tracking can be done from the LGA while driving and provide an accuracy of 100–200 m, assuming two to five satellites, consistent with plans for NASA's LCRNS (Section 6.5.7.1). This mode is similar to terrestrial Global Position Systems (GPS), but with orders-of-magnitude fewer satellites. Higher accuracies of 10–80 m can be achieved using two-way ranging and Doppler using the HGA, although this would require stopping the Rover to point the HGA. The accuracy of such radiometric techniques depends on the quality of the onboard rover clock (e.g. atomic clocks produce higher accuracies) as well as the relay network's orbital configuration. Optical techniques include matching images, digital elevation models, horizon features, and semantic features (craters and rocks) observed from the surface to corresponding orbital data. Among the multiple techniques, an image-based technique, originally developed for the Perseverance rover, produces extremely high reliability global localization with ≤3 m accuracy when tested on a broad range of lunar surface and corresponding orbital images (using high-fidelity rendered lunar images under varying lighting conditions). Celestial techniques would require extremely high-resolution surface gravity maps to produce similar quality results, but this capability does not currently exist. **The current Rover design baselines optical global localization techniques for its high reliability, highest accuracy, and its ability to localize relative to the orbital maps that are used for route planning. In the baseline, a radiometric technique provides an additional safety constraint.**

Global localization operations can have a significant impact on rover operations. For example, optical techniques may require stopping the rover and slewing the mast to acquire a panorama, and two-way radiometric approaches require a full 30-minute stop for ranging to a relay satellite. The baseline design mitigates this by planning towards a cadence for global localization, at 1–2 km intervals, which could coincide with the science requirements for interval stops every ~2 km. It may also be possible to implement optical and radiometric techniques while driving (i.e., without stopping), although further development would be needed to assess the cost and benefit of doing so. Heading corrections from star trackers do not require the rover to stop and hence can occur at much higher cadence while driving. In addition to operational importance, global localization can also be important for certain science investigations. For SPARX, gravity science investigations (Section 4.1.11, 4.2.8) drive the global localization requirement, as described in Table 6.17. Unlike global localization for navigation, these science-driven requirements can be achieved after the completion of the mission.

**Table 6.17.** *Notional science requirements for global localization, post mission operations.*

| Requirement | Baseline | Threshold |
|---|---|---|
| Horizontal | Within ±26 m | Within ±130 m |
| Vertical | Within ±5 m | Within ±27 m |

### 6.5.8.3 Surface Navigation and Hazard Avoidance

Surface navigation comprises perception, pose estimation (dead-reckoning), 3D terrain mapping, hazard assessment, motion planning, hazard avoidance, and motion control for continuous steering while driving.

For the Moon, hazards are primarily dominated by terrain topography, craters, and rocks. Currently available orbital datasets are limited, with topographic data at 10 ~100 meters per pixel (depending on latitude), and imagery at 0.5 ~ 2 meters per pixel. While larger hazards (~10 meters in scale) may be detectable in these datasets, hazards of the scale of the rover (~2 meters) would not be detectable. Thus, it is necessary for the Rover to have the ability to detect and navigate around smaller hazards.

Many of the largest hazards—particularly rocks and steep craters larger than several meters—can be detected from orbital datasets (e.g., LROC-NAC) and avoided with thoughtful route planning (Section 6.6.1). Smaller scale hazards must be detected by the rover's perception system. Some notable types of hazards include:

- Rocks: Rock hazards pose a risk of collision with the rover body as well as a risk of high centering, where the rover's wheels would not all be in contact with the terrain.
- Craters: Craters can represent both topographic obstacles and may have different regolith characteristics. Young craters are steeper, and some crater walls may exhibit higher slippage compared to other terrains. When taken together, some craters may become slip hazards and pose a risk of entrapment.
- Steep slopes: Traversing steeper slopes pose a risk as they increase rover slip. The higher the slopes, the higher the slip. Sinkage is tied to slip; higher slip results in higher sinkage and eventual entrapment.
- Sinkage hazards: Sinkage hazards are expected to be scarce. Sinkage is correlated to slip and occurs on high slopes, and plausibly in regions with extensive pyroclastic deposits. The exact terramechanical properties of large lunar pyroclastic deposits are uncertain. However, Apollo 17 reported no difficulties in traversing through the regional Taurus-Littrow pyroclastic deposit.
- Thermal hazards: The Moon presents a challenging thermal environment (Section 6.5.5), and it is plausible that there may be localized thermal hazards, such as large boulders, that reflect/emit thermal radiation on the rover.

The baseline Rover design uses the aforementioned stereo navigational cameras with an integrated IMU and star tracker sensor from DTU (Section 6.5.8.1) to estimate the rover's relative pose. Visual features are tracked in consecutive images and fused with gyroscopic and accelerometer measurements in a Kalman Filter. Absolute heading measurements are acquired from the star tracker while driving and are used to correct heading for the pose estimates. Terrestrial experiments with the DTU star tracker onboard a prototype rover (ERNEST, Section 6.5.3) have shown promising results under much worse light pollution conditions than would be encountered on the Moon. Global positional corrections would execute at a much lower cadence to correct from positional drift of the dead reckoning pose estimator.

A mapping function co-registers and fuses 3D point clouds produced from consecutive stereo navigation impacts, which can be coupled with the pose estimator to build situational awareness and assess hazards within the rover's perception horizon. Hazards are assessed based on the terrain topography and the rover's mechanical abilities like traversable rock height or crater slope, ground clearance, and slip characteristics. Perception, pose estimation, and mapping would be updated at a cadence of 0.5–1 Hz. An on-board path planner produces a route around hazards and is combined with a global cost map generated by ground operators (Section 6.7) to guide the rover to the next traverse location. Given the homogeneity of the lunar topography, route guidance from the ground would be primarily captured through a global cost map that defines a re-planned direction to guide the rover toward the next sampling region. Keep-in- and keep-out-zones would be embedded in the cost and would occur at a scale that would allow the rover to go multiple kilometers before needing positional corrections.

### 6.5.9 Avionics

The avionics for the rover must satisfy several stringent requirements: low power consumption, robustness to the lunar radiation and thermal environment, and sufficient processing capability to support high-rate onboard vision processing. Low-power avionics are critical for a mission in which solar energy is limited and available for only approximately two weeks out of every four weeks. The avionics need to survive a Total Ionizing Dose (TID) of 15 krad and survive high energy particles of at least 75 MeV·cm2/mg against Single Event Latchup (SEL). In addition, the avionics must operate across a wide temperature range, from approximately −40°C to +50°C. Finally, to achieve an average traverse speed of 1 kph, the Rover must execute vision processing algorithms at substantially higher rates than those used on prior planetary missions.

The baseline rover avionics subsystem consists of three parts: a Compute Element (CE), a Remote Engineering Unit (REU), and a set of motor controllers. The baseline reference design CE utilizes a new JPL design that allows for significantly more computing power than prior missions. Prototypes versions of the CE currently exist. Similarly, the reference REU leverages a previous JPL design. Commercially available, space-qualified motor controllers are available, and would be recommended for procurement. The Rover avionics subsystem components of the rover design are shown in Figure 6.27.

The compute element utilizes a single board computer (SBC) based on the NASA High Performance Spaceflight Computer (HPSC). The SBC and supporting hardware will have the performance required to handle the complex processing needed for the rover to drive autonomously, and the mission environmental requirements. The compute element design also includes a PCU (Power Conditioning Unit) based on gallium nitride (GaN) technology. Lastly, it has a custom interface card that allows it to connect to all the various interfaces the mission might have, including SpaceWire, Ethernet, and UART.

The REUs serve three main functions. The first is to maintain the mission clock since the Compute Element will get powered off during lunar night. It also contains the oversight functionality over the CE and acts like a “lizard brain” to monitor and react to faults where the CE becomes non-operational. Lastly, it is responsible for waking the CE up when lunar day starts or when there is a fault at night that requires the CE to react to it.

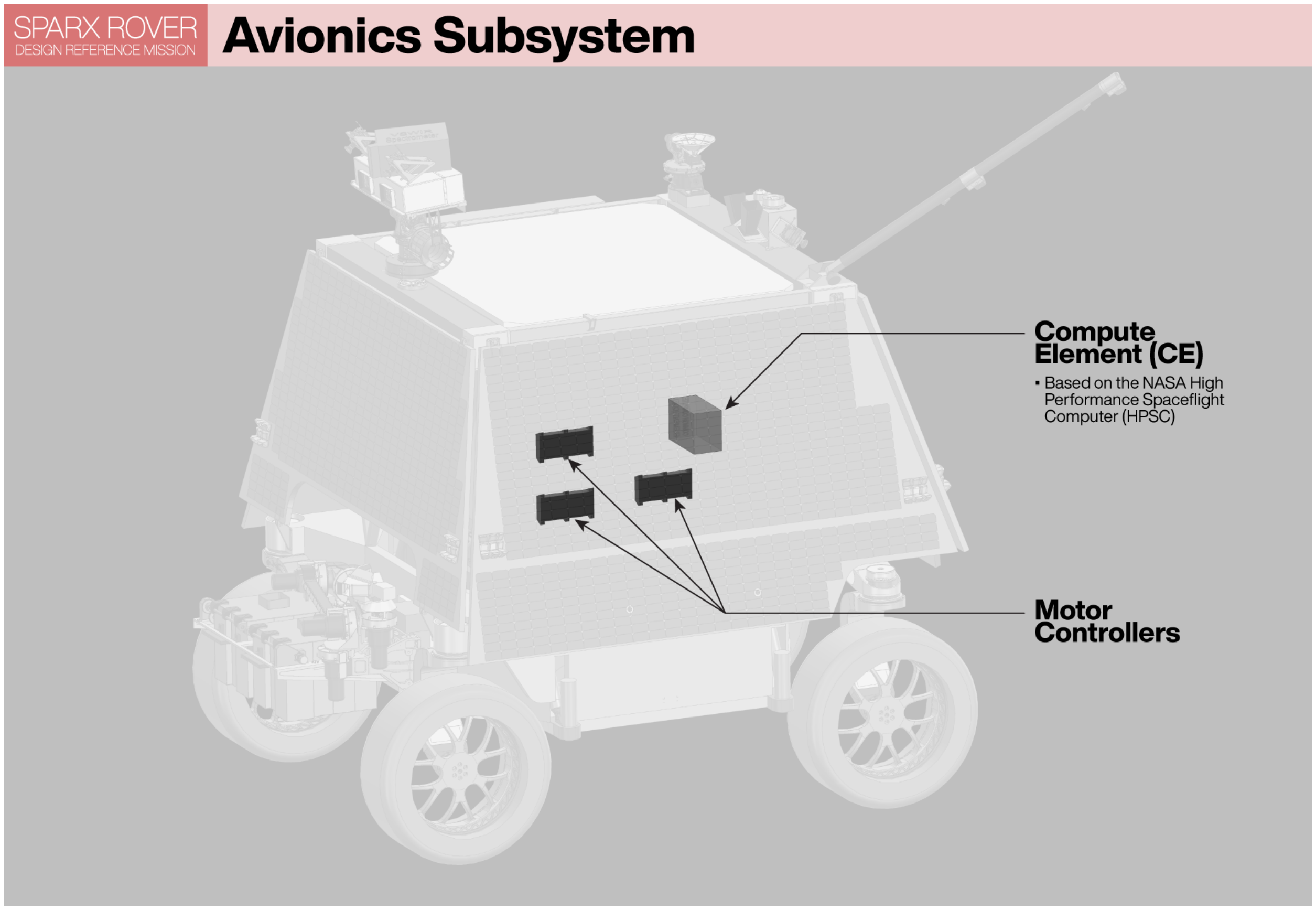

**Figure 6.27.** *Avionics subsystem components of the Rover design*

### 6.5.10 Activity, Resource, and Fault Management

#### 6.5.10.1 System-Level Activity and Resource Management

To minimize reliance on ground commanding, a long-range lunar rover must autonomously manage its activities, resources, and health to handle environmental variability and overcome the difficulty in predicting resource usage. The reference Rover design leverages the onboard autonomous scheduler "Onboard Planner" used by Perseverance (e.g., Verma et al., 2023).

To facilitate autonomous system-level activities and resource management, activities are planned based on intent from the ground: route to traverse, stops to make for communication, and the type and cadence of science measurements along the route. The Rover schedules the planned activities based on the Rover's state and pre-defined constraints to maintain efficient resource utilization and operations. As activities are executed, the onboard autonomy reacts to evolving conditions and continuously updates the schedule to eliminate inefficiencies and potential faults. This strategy allows the Rover to deal with environmental conditions that are impossible to predict on the ground.

#### 6.5.10.2 Fault Protection

Similar to past missions, the on-board system-health manager (fault protection) detects and identifies all faults and failure and also responds to many of these faults in ways that allow the Rover to continue making progress without intervention from the ground.  For off-nominal situations that cannot be handled onboard, operations fall back to ground-in-the-loop control (i.e., “human decide” operations mode). Such residual faults, ones that cannot be recovered onboard, would have to be diagnosed and dispositioned by ground operators to resume driving. Emphasis is placed on safely resuming driving even if the root cause may require a longer duration to diagnose. Such residual faults must be driven to a minimum to allow the rover to continue to make progress, with faults requiring ground engagement kept on the order of no more than a couple per week on average during the traverse phases. During sampling operations, continuous communication and ground operations allow for greater ground engagement in decision making and fault diagnosis part of the planned operations.

## 6.6 CONCEPT OF OPERATIONS

### 6.6.1 Traverse Concept of Operations

Traversing >2,650 km across the lunar farside drives a novel traverse concept of operations (ConOps) that utilizes onboard autonomy for navigation and high-level system resource management for traversing. Figure

The details of the actual traverse (i.e., the path across the lunar surface) are described further in Section 6.7. Figure 6.28 shows a summary of the SPARX Rover mission ConOps from launch to sample delivery. Subsequent sections examine important mission phases in more detail.

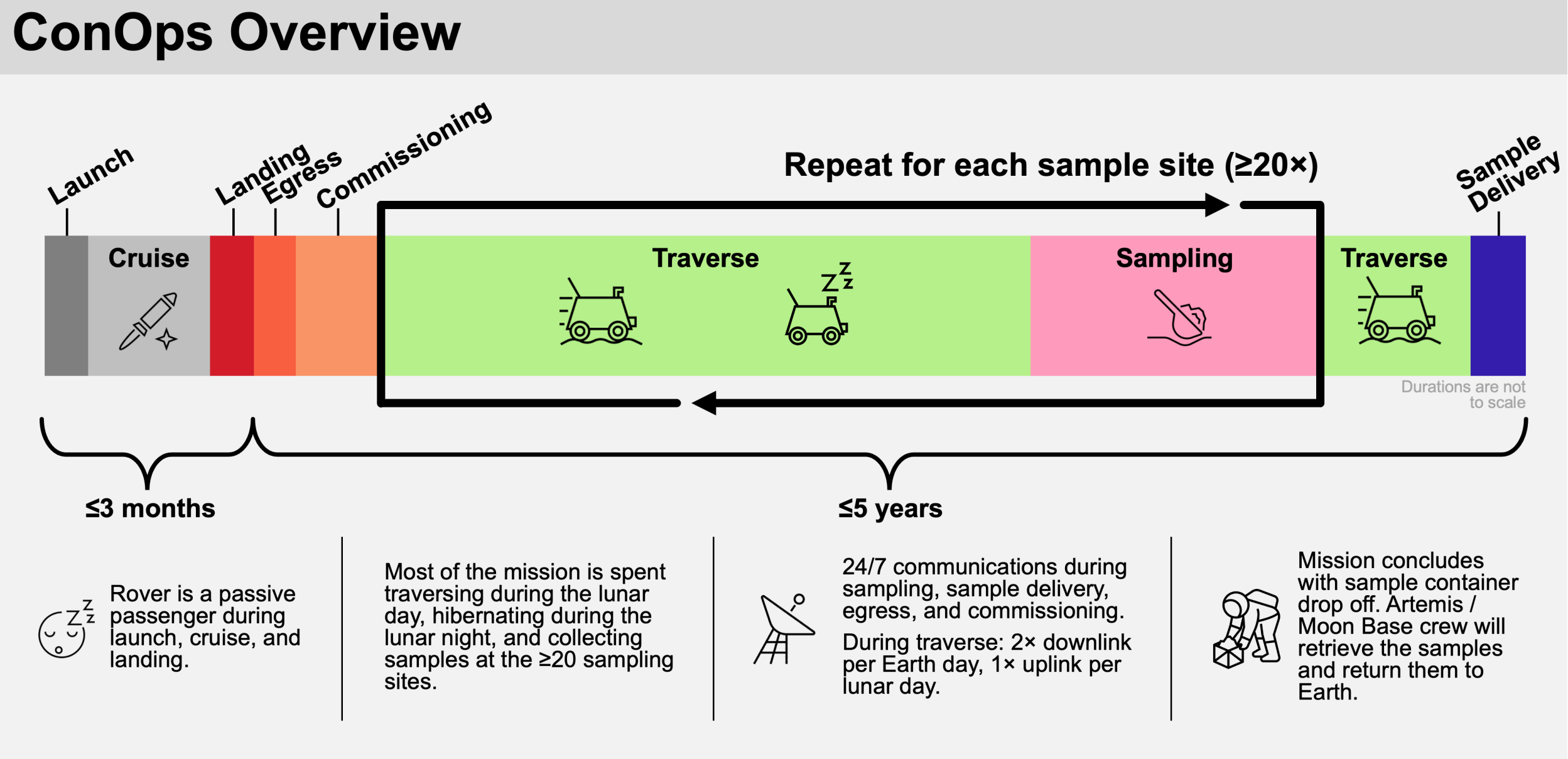

**Figure 6.28.** *Summary of the SPARX Rover mission ConOps.*

#### 6.6.1.1 Traverse ConOps

The SPARX Rover mission is a mission dominated by the need to drive long distances. Figure 6.29 summarizes the ConOps during for a typical lunar day which may be dominated by driving. Science aspects of the Traverse ConOps are described in more detail in the Section 6.6.1.2.

The Rover navigation system (Section 6.5.8) is designed to traverse between 21 science sites (20 sampling sites and 1 non-sampling science site) and reach the Artemis / Moon Base rendezvous point autonomously, with minimal reliance on human-directed driving and ground-derived products. Human-directed driving should be an available capability to support rare cases where anomalies or exceptionally challenging terrain prevent the onboard autonomy from making progress. These situations are expected to be less than 0.5% of the total traverse distance based on analysis of the slope and hazards along the route. The vast majority of the Baseline traverse is benign (Section 6.7), with 95.5% of the terrain having slopes less than 12.5°, which is well below the Rover's maximum slope capability of 20°. Hazards on the lunar terrain are primarily geometric in nature (rocks, craters, etc.) which means they can be recognized and avoided by an onboard autonomous navigation system. Ground operators will likely only need to manually direct rover driving during short periods of fault recovery (10s of meters) or crossing difficult boundaries like particularly steep crater rims (100s of meters).

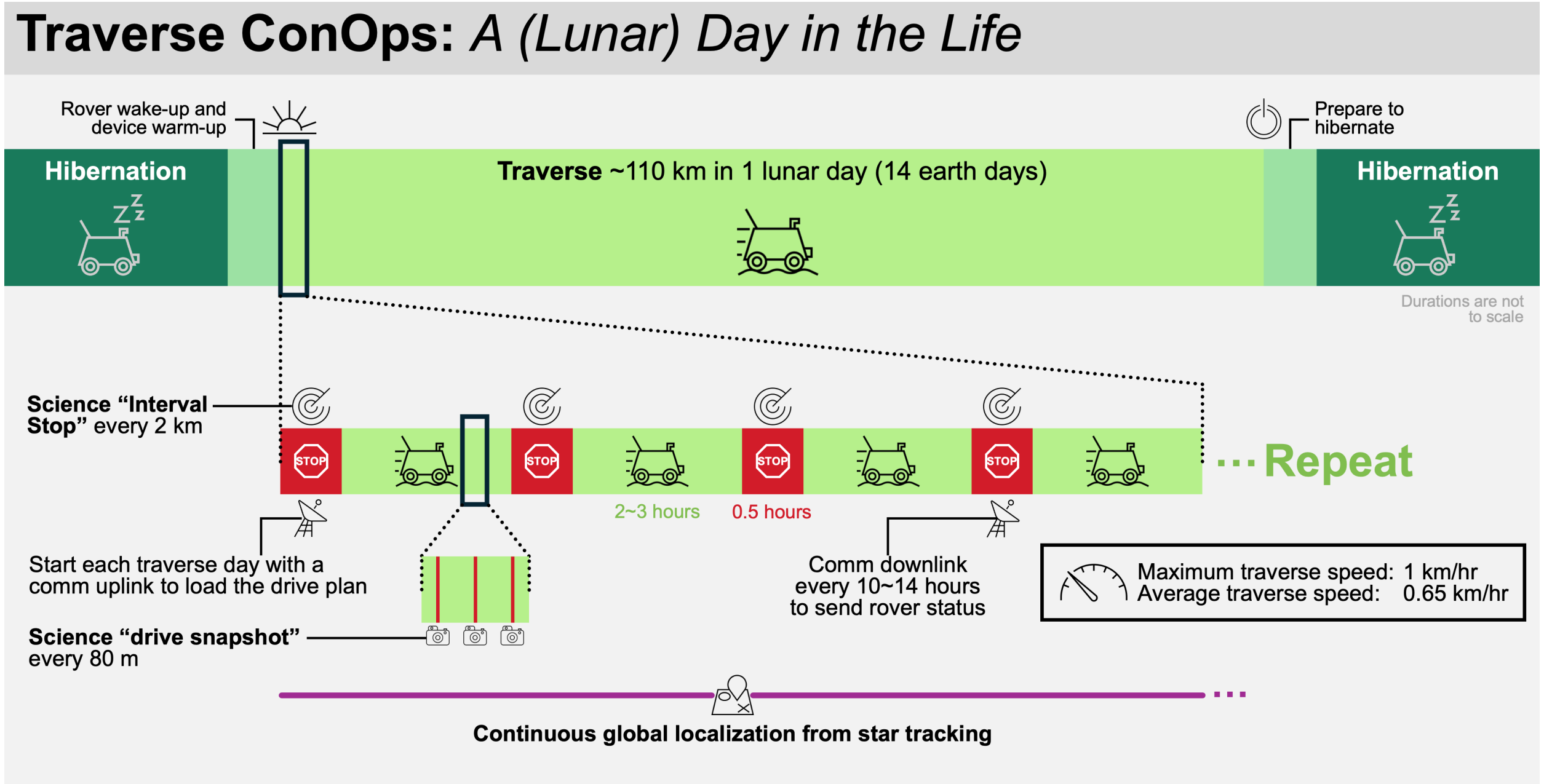

**Figure 6.29.** *Rover ConOps for a typical lunar day of traversing.*

The ground-derived products that the Rover needs on-board to navigate consist of (1) route waypoints, and (2) a traversability cost map covering the route. Waypoints are used to specify the 21 science sites and any intermediate locations to use as engineering guidance. The cost map is a map that encodes predicted traversability of each map cell, which can be related to surface slopes, illumination conditions, rocks or other hazards. Untraversable regions (i.e. keep-out zones) are defined as infinite cost. Cost maps could also encode science value as "beneficial" costs to maximize science return from instruments operating along the traverse (for example, preferentially stopping near craters or boulders). Using the cost map, the onboard navigation system can merge local hazard information with global costs to calculate the optimal route to each waypoint and enable flexible re-routing and backtracking around difficult areas without additional ground direction. The route connecting the science sites is assumed to be defined entirely by engineering constraints, i.e., there are no additional science constraints or requirements that affect the route trajectory. Cost and hazard maps are further described in Section 6.7.2.2.

The baseline Rover also includes the use of onboard autonomy for system health management, activity scheduling, and resource management during traverse. The design revolves around an onboard scheduler that predicts resource usage according to models in addition to reacting to resource usage in real-time (Section 6.5.10.1). This strategy allows the Rover to quickly overcome significant environmental uncertainties without relying on ground assessment or commanding. This is particularly important when dealing with the extreme solar illumination and thermal environment as differential heating of rover components will require a responsive system to manage temperatures at a component level while driving through shadows and at variable azimuth angles relative to the sun.

During the traverse, ground operations personnel will be assigned to on-call shifts to respond to mobility faults and anomalies as needed, with one on-call shift scheduled per day (Figure 6.3). The personnel schedule for these on-call shifts may be released if the rover is healthy and making progress. This strategy increases ops efficiency as personnel can work other tasks while the rover is being productive.

#### 6.6.1.2 Traverse Science ConOps

Along the traverse, the Rover will collect a wide range of in situ science measurements which are required to meet the SPARX science objectives (e.g., Section 4.2). The measurements are necessary to address many SPARX science experiments (Section 4.1), provide essential geologic context for the samples, and provide opportunities for unexpected discoveries. However, for the mission to accomplish its long traverse in a reasonable amount of time, traverse science must be constrained; there is no time for the rover to explore every terrain along the way or get distracted by every interesting feature. Thus, the traverse science ConOps is designed to be as efficient as possible. The following reference mission traverse ConOps derive from the Intrepid mission concept study report (Robinson et al. 2020); although, all aspects of the ConOps were discussed and either affirmed or modified by the SDT, as necessary. Different traverse science ConOps modes are shown schematically in Figure 6.29, Table 6.18, and described below.

**Table 6.18.** *Summary of traverse science ConOps modes. Green indicates the instrument is active and red indicates the instrument is inactive.*

| | **Interval Stop** (30-minute science stop every 2 km) | **Drive Snapshot** (Quick science observation every 80 m while driving) | **While Driving** (Continuous science measurements while the rover is in motion) | **Twilight Science** (Opportunistic science making use of the rover being stopped while charging at dawn and dusk) | **Night** (Rover survival is prioritized over nighttime science measurements) |
|---|---|---|---|---|---|
| Context Imaging (MastcamZ) | 1 full color panorama | 1 forward-looking full color image | Not acquiring science data | 1 full color panorama, if possible | Not acquiring science data |
| Context Mineralogy (UCIS) | 1 hyperspectral panorama | Not acquiring science data | Not acquiring science data | 1 hyperspectral panorama, if possible | Not acquiring science data |
| Bulk Elemental Chemistry (LunarVISE GRNS) | Continuously acquiring data | Continuously acquiring data to build GRNS “noodles” | Continuously acquiring data to build GRNS “noodles” | Acquiring data for as long as possible | Not acquiring science data |
| Magnetic Field (Insight IFG) | Acquiring data w/ ≥10 minutes of “quiet time” | Continuously acquiring data to build magnetic field “noodles” | Continuously acquiring data to build magnetic field “noodles” | Acquiring data for as long as possible | Not acquiring science data |
| Gravity Field (DTU IMU) | Acquiring data w/ ≥2 minutes of “quiet time” | Not acquiring science data | Not acquiring science data | Acquiring data for as long as possible | Not acquiring science data |

- **<u>Science Interval stop:</u> Every ~2 km the Rover will stop to perform both science and engineering tasks.** The choice of 2 kilometers is driven by a few factors, including that this is the approximate distance to the lunar horizon from the perspective of the Rover (so stops every ~2 km allows for building up a quasi-contiguous in situ dataset), and that many terrains are mapped on the surface of the Moon at the scale of kilometers. The specific interval stop locations are not at preplanned locations; rather the Rover plans for autonomous stops using a combination of the odometer reading and the previously uploaded cost-map (Section 6.6.1.1., 6.7.2.2). This weighting allows the Rover flexibility to stop near scientifically valuable terrains, while simultaneously not deviating too strongly from the nominal traverse path or stop cadence.

**During the interval stops, the rover will employ its full suite of remote sensing instruments, including taking panoramas with the color cameras and imaging spectrometer, having longer integrations with the gamma ray and neutron spectrometer, and having mechanical quite for magnetometer and gravimeter measurements.** For both the magnetometer and gravimeter, it is preferable for the Rover to be stationary for 2~10 minutes with no movement to minimize noise sources for those measurements. The driver for total duration of the interval stop in the baseline reference mission is the imaging spectrometer, which may take up to ~30 minutes to complete the panoramic imaging of the scene. **In addition to science measurements, interval stops are the primary time for the rover to accomplish other stationary engineering tasks, including pointing the high gain antenna at the relay satellite for uplink and downlink of data, and accurately localizing its position on the surface of the Moon.**

- **Drive snapshot:** Every 80 meters along the drive, the Rover will take a "snapshot" from the mast-mounted stereo visible cameras. The current baseline assumes the Rover stops to take these snapshots to minimize motion blur and to allow time for auto-exposure strategies to adapt to the natural lighting—although a future trade may revisit this and instead opt for taking the snapshot while moving to maximize the Rover's "speed made good". The science motivation for the drive snapshot is that, in aggregate, they allow for the rover to build up long high-resolution (~cm scale) "noodles" of images along the traverse. These images can provide important geologic context for the terrains the rover is traversing, and how terrains may change between interval stops or sample sites. The drive snapshot mode is considered baseline only, and is descopable should constraints arise in mission duration, data storage, or other operational drivers.
- **Driving:** While the Rover drives, the passive instruments (gamma ray and neutron spectrometer and magnetometer) continue to take measurements. Both instruments can continue to integrate and build up long "noodles" of data along the traverse. The quality of the magnetometer data may be particularly impacted by the electromagnetic noise arising from the Rover (e.g., motors); however, it may be possible to reconstruct some science measurements in post. Depending on the details of the selected Context Mineralogy instrument, the instrument may need to continue to operate a cryocooler while driving to remain ready for operations at interval stops.
- **Twilight science:** As the sun gets low on the sky, the Rover will eventually stop traverse operations to charge for night (Section 6.6.1.3). During this period, science measurements can be collected on a "best effort" basis. While the priority for these periods is charging the battery, long stops (many hours) provide an excellent opportunity for many instruments—such as the gamma ray and neutron spectrometer, magnetometer, and gravimeter—to acquire long integrations, improving sensitivity and driving down noise.
- **Night:** For a solar-powered Rover, the priority at night is survival (Section 6.6.1.3); therefore, no science operations are baselined during lunar night. However, if excess energy is available, there may be opportunities for opportunistic science similar to the twilight science mode.

The exact details of these traverse science ConOps modes will invariably depend on the final instrument choices for the rover.

### 6.6.1.3 Hibernation ConOps

**As a solar-powered rover operating at the mid-latitudes, about 50% of the SPARX mission duration is spent in night hibernation.** The Rover's solar powered architecture (Section 6.5.6) has the limitation that during lunar night there is no power production. To simplify the engineering design and reduce rover mass and size, the Rover goes into hibernation mode during all lunar nights instead of carrying significant extra battery mass to provide power beyond survival needs overnight. Figure 6.30 shows the Hibernation ConOps.

Hibernation mode is utilized during both traverse and sampling operations. As the Sun starts to set towards the end of the lunar day, the Rover will stop, autonomously charge to full battery state of charge, and transition to hibernation mode before night. In the morning, the Rover will autonomously wake up when the solar panels are illuminated, and power production is sufficient to begin wakeup activities such as heating components. Once the battery state of charge has recovered and required components have reached operational temperature, the Rover can continue nominal operations. Telecom windows during the lunar morning wake-up period will be used to uplink instructions to the Rover, which could include route updates if the rover is about to resume the traverse, or new sampling instructions if hibernation occurred during sampling operations.

Excess power may be available for science observations around lunar sunset if the Rover is fully charged and still receiving enough solar power to remain power positive. These "twilight" periods may be particularly useful for acquiring long-integrations with some science instruments, such as the gamma-ray and neutron spectrometer, gravimeter, or magnetometer. However, **in night hibernation, the Rover minimizes survival power—so no overnight science or telecommunications is baselined.**

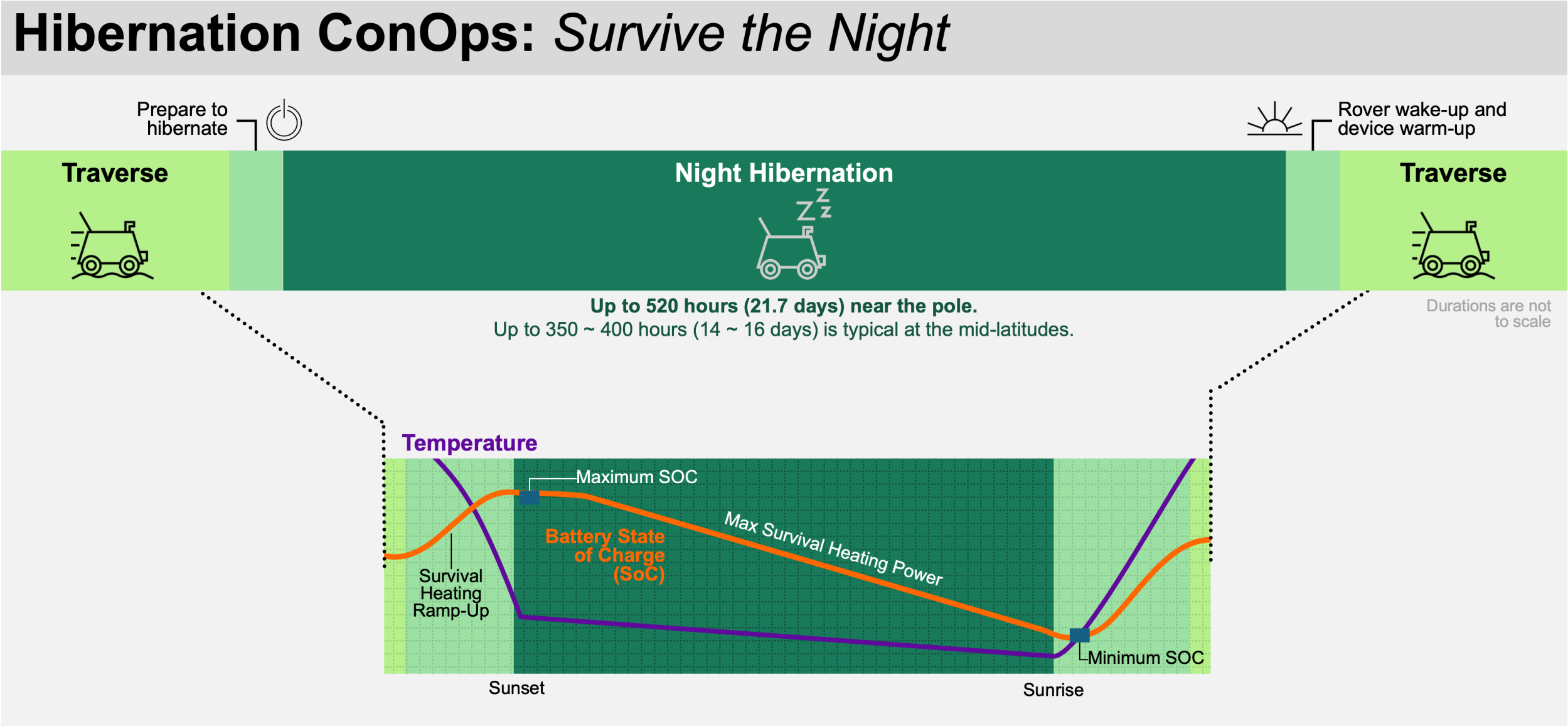

**Figure 6.30.** *Hibernation ConOps.*

### 6.6.2 Sampling Concept of Operations

The sampling ConOps are designed to meet science and engineering requirements for reconnaissance, and the identification, acquisition, processing, transfer, and stowage of samples. Each of these elements involve some level of ground-in-the-loop (GITL) operations to direct the Rover's activities and make decisions based on downlinked science and engineering data products. The following sections describe the sampling ConOps in more detail, which are summarized in Figure 6.31.

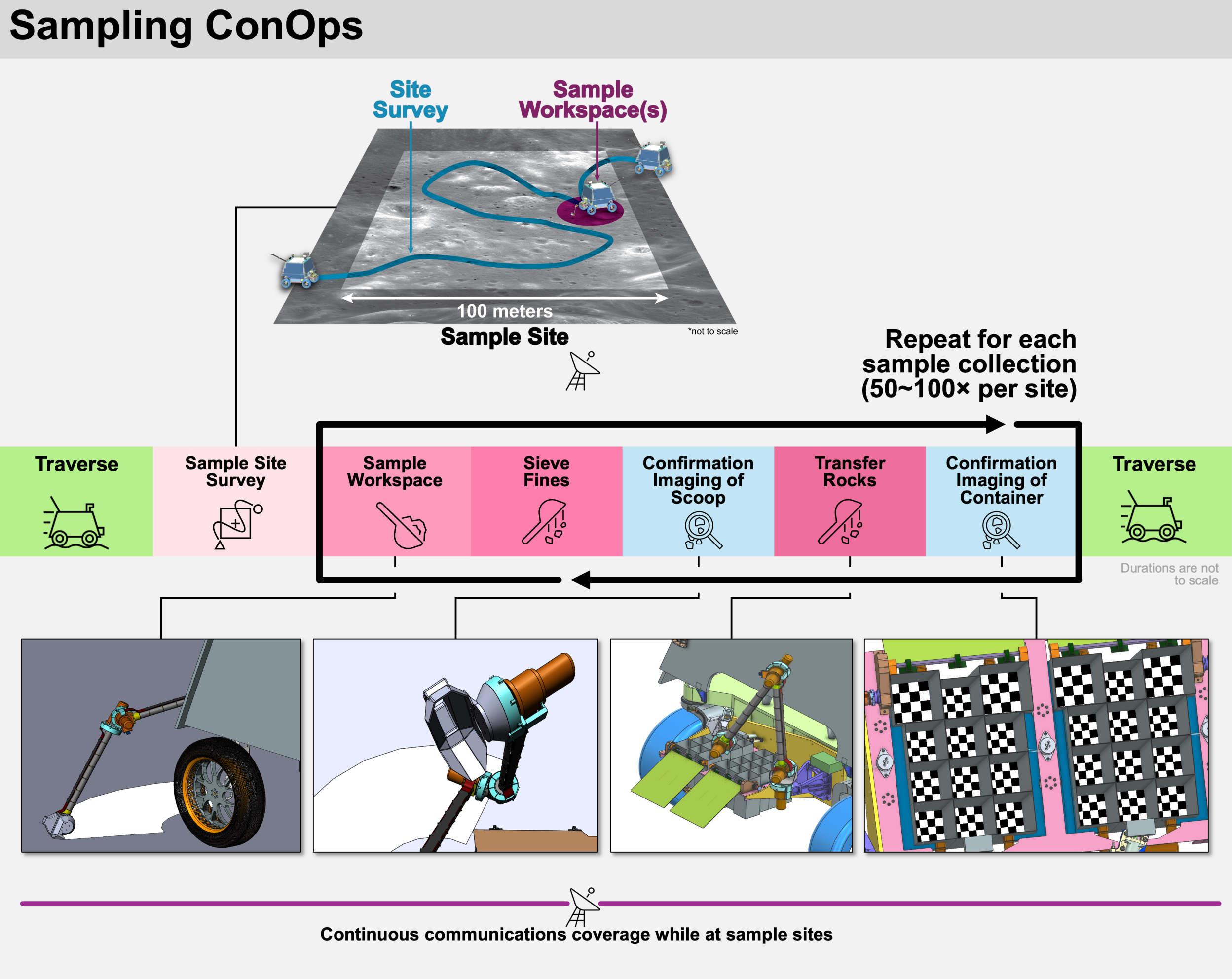

**Figure 6.31.** *Sampling ConOps.*

#### 6.6.2.1 Sample Site Reconnaissance

Individual sample and traverse sites are planned by the science team before the mission launches based on available orbital remote sensing datasets, using characteristics defined in Section 4.1. While there are some exceptions, most sites are defined as 100 m × 100 m quasi-rectangular quadrangles on the surface of the Moon. Once the Rover reaches the site, the first action is to perform a **Sample Site Reconnaissance Survey**. The goal of this reconnaissance is to characterize the site, verify the site meets the required science characteristics, collect the data necessary for scientists on the ground to pick the ultimate sampling location(s), and collect the in-situ measurements that provide essential geologic context. A schematic illustration of Sample Site Reconnaissance Survey is shown in Figure 6.31.

The Rover drives to ≥5 parking locations spaced-out across the site and acquires remote sensing data from its science and engineering instruments to characterize the site and allow scientists on the ground to ultimately pick a final sampling location. Engineering data collected during the survey is used to identify regions that are compatible with sample acquisition requirements for vehicle parking (i.e., parking tilt) and robotic arm surface workspace topography (i.e., local slopes in the workspace). When necessary, data is also collected to geolocate larger rocks (≥20 mm in diameter) to support targeted rock acquisition modes (described below).

#### 6.6.2.2 Sample Acquisition

After the Sample Site Reconnaissance Survey has been completed, the science team selects an initial workspace and directs the rover to that location. Upon arrival at the workspace, the rover will acquire science and engineering imaging (and other remote sensing data as necessary) for terrain reconstruction, target selection, and scoop pass planning.

The sampling ConOps is designed around two general sampling modalities:

- **Untargeted scooping and sieving:** Most SPARX Experiments (E2, E3, E4, E5, E7, E10, E13, E14—as shown in Foldout 4.2 and Section 4.3.6) can be accomplished by untargeted, or minimally targeted, scooping and sieving regolith to collect either unsieved regolith or small rocks (≤8 mm in diameter). While this mode is defined as "untargeted", the remote sensing measurements and sample site reconnaissance are still required to verify that the site meets the pre-defined sample site requirements (Section 4.1); however, once this has been done, there is not a requirement to further target specific rocks for collection. The expected size-frequency distribution of these rocks (Section 6.5.4) is such that rocks of this size occur with sufficient abundance that repeated scooping and sieving of typical lunar regolith should be able to acquire the required number and mass of samples within a reasonable amount of time, and without targeting.
- **Targeted scooping and sieving:** Some SPARX Experiments (E1 and E9—as shown in Foldout 4.2 and Section 4.3.6) require the collection of larger rocks (≥20 mm in diameter). Given the steep power-law nature of the size-frequency distribution of rocks in lunar regolith (Section 6.5.4), untargeted scooping and sieving is an inefficient approach to collecting these rocks. For these Experiments, a targeted approach is necessary. In these scenarios, the Rover would use its science and engineering instruments to surveil the scene, scientists and engineers would identify rocks in the scene that meet the sampling requirements, the team would then geolocate these larger samples and then utilize the arm and scoop to collect them.

For all sampling modalities, the initial scooping, sieving, and sample transfer will involve a full "ground-in-the-loop" assessment to triage performance and adjust subsequent activities. After this initial pass, the Rover will then autonomously execute a full sample collection workflow without ground-in-the-loop gateways. During these pre-planned operations, engineering data and confirmation imaging are downlinked continuously for parallel assessment (i.e., the science team can asynchronously monitor as the Rover autonomously scoops, sieves, and stows samples). If the ground team identifies a need to adjust activities, a "stop" command can be issued, prompting the flight system to enter a pause state while awaiting the next command uplink.

At each robotic arm workspace, the Rover can execute ≥13 scoop passes, excavating the top ~22 millimeters of lunar regolith without repeating scoop passes. Projections based on size frequency distributions and numbers of rocks required, indicate that rock collection requirements for most sites will necessitate 50–150 total scoop passes; consequently, multiple robotic arm workspaces are required per site. After a workspace is exhausted, the Rover will perform a short (<1 meter) “bump” drive or rotation to provide a new arm workspace.

### Why do SPARX’s sampling campaigns require “human-in-the-loop” operations?

It may seem contradictory that the SPARX Rover requires fully autonomous operations for traversing (Section 6.3), but “human-in-the-loop” operations for sampling campaigns. However, this is a consequence of the sampling requirements and state of autonomous sampling technology. **“Human-in-the-loop” operations are required to satisfy science requirements for sampling while also providing operational flexibility and limiting the dependence on new technologies.**

1. **The science objectives are sufficiently complex that it is required that the ground-based science team select individual sample sites (i.e., rover workspaces), and in some cases even target individual rocks for collection.** Orbital datasets do not have sufficient resolution to pre-select sampling workspaces, so the Rover’s science instruments (which afford ~cm-scale resolution) are required. Human operators will coordinate and react to the in-situ measurements, taking many variables into consideration that onboard autonomy would not be able to handle.
2. **Sampling interactions with the terrain have significant uncertainty that is difficult to predict or plan for ahead of time, especially if relying on autonomous behaviors alone.** Human operators provide flexibility and can react to unexpected situations and expedite fault recovery without the overhead and limitations of an autonomous system.
3. **Developing and testing a fully autonomous system to characterize sample quality and make critical sampling decisions would be a challenge.** This level of autonomy is beyond current capabilities, and also highly specialized such that it would need to be developed from the ground-up for the SPARX Rover, resulting in significant increases to development cost. Because the technology would be new for the lunar environment, there would be significant risk that it would fail to meet performance requirements, even after extensive and costly testing. With only 20 sampling sites in the Baseline mission—many of which with bespoke sampling requirements (**Section 4.1**)—the benefit of developing such a specialized autonomous system is minimal.

While fully autonomous sampling is not required for the SPARX rover, sampling can be enhanced by combining both direct commanding and supervised autonomy behaviors with limited scope—allowing ground operators to monitor and react via low-latency communications links.

### 6.6.2.3 Sample Processing, Transfer, and Confirmation

Once the arm acquires samples in the scoop, it can be transferred to a sample container. The mode of transfer depends on whether the samples are sieved or unsieved.

- Sieved samples are delivered to the rover by first sieving the samples while the arm is extended far from the Rover. Fines are discarded onto the lunar surface, while >3 mm rocks are retained in the sieve. Sieving while the arm is far from the Rover minimizes the risk that the sieving action lofts material that can land on other sensitive Rover components (e.g., radiators, optics). The remaining rock fragments are poured into the sample container.
- Unsieved samples are delivered to containers in two steps. First fines are delivered by sieving the sample over the site's dedicated container. The remining >3 mm rock fragments can then be poured out of the scoop into the container. Vibration is used to motivate fines through the sieve and rock fragments out of the scoop.

After each step of sample acquisition, processing, and sample transfer, the arm-mounted sample acquisition camera collects documentation imaging that can aid in characterization and confirmation of sample delivery by the ground team. The mast-mounted science and engineering cameras also provide context. After all the sample material has cleared from the scoop, the arm end effector camera captures an image of the interior of the sample bin using LEDs integrated into the image for lighting.

### 6.6.2.4 Sample transfer to Artemis

After the last samples have been collected, the container lid is driven to a latching position that will maintain sample integrity through the Artemis handling activities.

At the conclusion of the prime mission, the Rover will arrive at a pre-defined rendezvous point with Moon Base or a crewed Artemis mission where the Rover will deploy the sample container to the surface of the Moon (Figure 6.32). The final drop-off site will be selected to provide a quasi-level, stable surface for the sample container, and to minimize obstacles for astronauts on EVA to reach and lift the container (potentially with the aid of other vehicles, such as the Lunar Terrain Vehicle). Artemis astronauts may perform simple mechanical activities to secure the sample container for return to Earth.

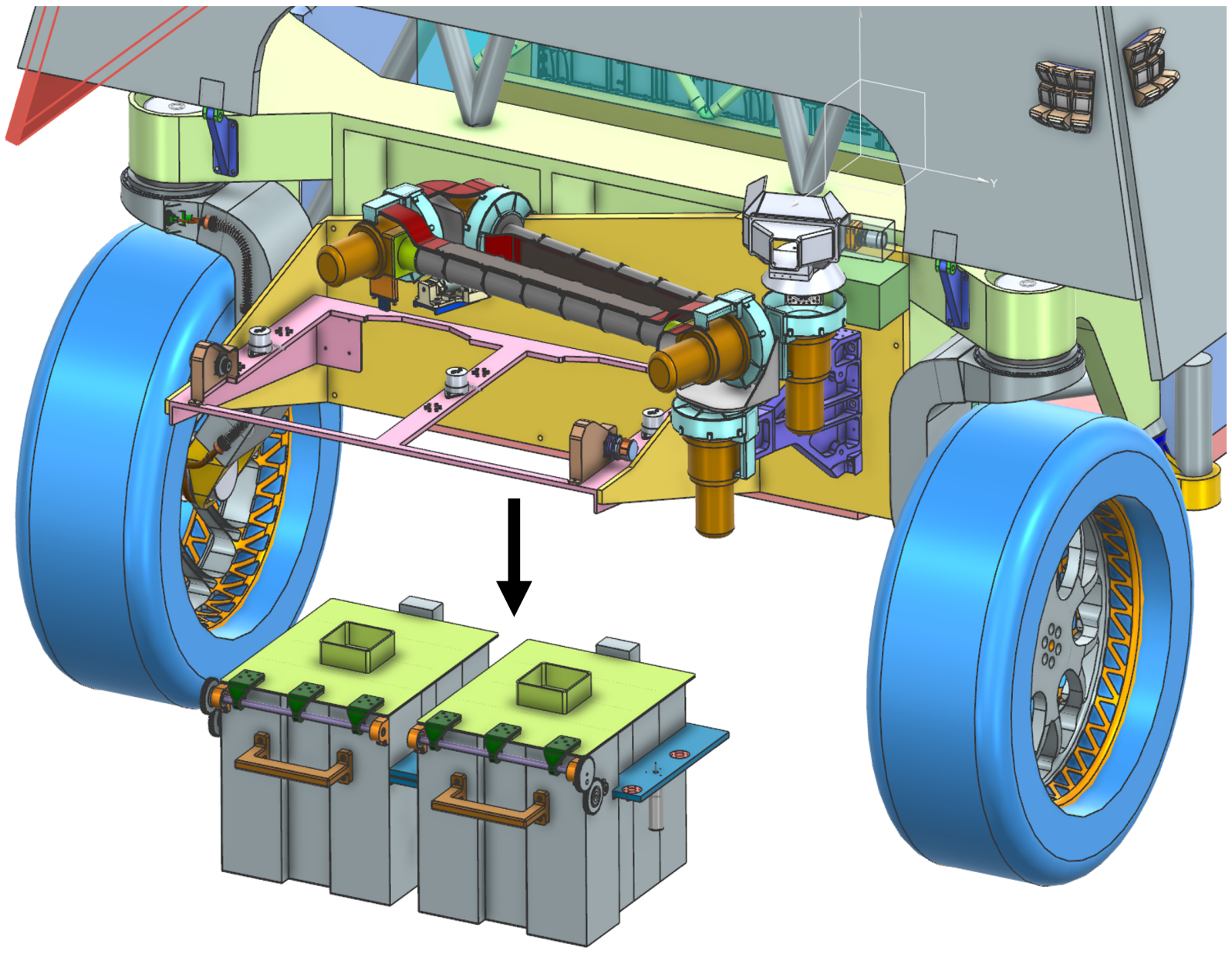

**Figure 6.32.** *At the end of the mission, the Roer will deploy the sample containers on the lunar surface for collection by astronauts.*

After the Rover deploys the sample container, the Rover will back away to provide a clear workspace for Artemis astronauts. Dropping off the sample container on the surface of the Moon is advantageous over other sample transfer approaches, as it removes the need for astronauts to interact with the Rover, and the additional astronaut safety requirements that would result from that integration. It also frees the rover to perform other scientific and exploration related activities in the area—especially if Artemis astronauts do not arrive exactly when the Rover reaches rendezvous point.

**Finding: The SPARX science objectives require reconnaissance of the pre-defined traverse sample location to acquire geologic context, perform in situ measurements, and identify the locations within the local (100 m × 100 m) region most conducive for meeting the sampling requirements.**

**Finding: The SPARX science objectives require ground-in-the-loop verification of sample properties during the initial phases of scooping. Verification is determined using the fine-scale mineralogy instrument to confirm that the sampling area meets the specified lithology requirements. Once the science team verifies sample properties, autonomous sampling can begin.**

**Finding: The SDT finds that the Sampling ConOps described above should be treated as a universal process, regardless of the final rover design. Due to the importance of sample selection and collection for the mission Science Objectives, the SDT has worked iteratively with the technical team to define a set of ConOps procedures that should be utilized by any final mission design.**

## 6.7 TRAVERSE DESIGN

Traverse planning establishes the methodology for providing navigational guidance to the Rover using a priori maps. While conceptually similar to drawing a "line on a map," the process may utilize complex representations—such as multi-resolution, uncertainty-aware cost-to-go maps—to guide the Rover while maintaining sufficient flexibility for onboard autonomy to respond to immediate ground conditions.

Fundamentally, the driving requirement for traverse planning is to provide sufficient guidance to the rover for safe and reliable navigation to the science sites, which are typically separated by 10s to 100s of kilometers. The Rover's onboard autonomy (Section 6.3.2) is responsible for all local hazard avoidance, but it has a very limited visibility horizon of 10–30 meters. Thus, the primary role of traverse planning is to "steer" the rover around larger hazards that can be identified from orbit before they are encountered, and to otherwise bias the traverse towards regions that are more probably benign. While the most common hazards are those for mobility (like crater, slopes, and rocks) operational hazards for other subsystems such as power, thermal, and communications must also be accounted for.

Planning for the SPARX Rover mission presents two distinct challenges compared to prior rover missions:

- Scale: The distance of the traverse is roughly 100× greater than previous traverse on the Moon or Mars.
- Resolution mismatch: While Rover hazards can be as small as 10s of centimeters, the currently available topography datasets (primarily derived from LRO LOLA) are limited to 10–100 meters in resolution.

However, two key assets render this problem tractable:

- High-resolution imagery: There is broad availability of high-resolution imagery (specifically LRO-NAC). While not yet in a calibrated ortho-mosaic form readily available for planning, the raw data exists for the entirety of the traverse. Although only a small fraction of the traverse has "stereo" coverage amenable for traditional photogrammetric DTM generation (i.e., multiple images with appropriate viewing and illumination geometry that can be analyzed via SOCET SET or Ames Stereo Pipeline), other methods can be deployed on single images (e.g. "point-photoclinometry" and deep-learning-based methods), or a collection of images with varying illumination (e.g. shape-from-shading).
- Geologic uniformity: The relative geologic uniformity of the lunar surface—particularly below ~100 meters where the surface is generally in cratering equilibrium—allows for high confidence in the statistical models used to predict small-scale hazard abundance. Unlike Mars, which features a surface shaped by a diverse range of geologic processes, lunar geology is dominated by impact cratering and gardening. Lunar topography and cratering mechanics have been studied for decades, and due to the global similarity of the cratering process, data collected from the many missions that have been to the surface can be used to inform statistical models we use to infer/extrapolate the prevalence/abundance of rover-scale hazards from orbital data.

### 6.7.1 Key Hazards and Characteristics of the Moon Relevant to Traverse Planning

From the perspective of a Rover driving on the surface, there at least three common hazards that must be considered: craters, rocks, and terramechanical properties.

#### 6.7.1.1 Craters

Crater-induced slopes (at all scales) are by far the most abundant mobility hazard for the Rover. The size-frequency distribution of craters is well studied and generally falls within one of two families of power-law cumulative number densities. The end result is only a few percent of "geometric saturation," meaning the rover can always find a path between craters. Moreover, not all craters are hazardous. In fact, most small craters are heavily degraded with benign wall slopes. Less than about 20% of craters are likely to be steep enough to pose a hazard for a rover.

In general, the mean free path between hazardous rover-scale craters is 10s of meters, requiring frequent hazard avoidance.

#### 6.7.1.2 Rocks

Rocks are sparse on the moon with the global average area coverage <0.5%, but they are concentrated near fresh crater rims. While most individual wheel-hazard-scale rocks may not be detectable from orbital images, small rocks generally come from the breakdown of larger rocks and are therefore highly spatially correlated. In other words, where there are big rocks, there are likely small rocks, and vice versa. Thus, we can safely avoid high-rock density areas by identifying (1) regions with above-average prevalence of large boulders (big enough to be seen in LRO-NAC imagery), and/or (2) rim regions around craters that are likely to be rock-bearing.

Typically, the mean free path between hazardous rocks in inter-crater regions will be 10s to 100s of meters.

#### 6.7.1.3 Terramechanical Properties

The terramechanical properties of the lunar regolith have been characterized on prior missions and tend to vary within a limited range (generally based on proximity to crater rims), with an angle of repose generally >35°. Terramechanical hazards can be mitigated by varying the allowable slope base on local uncertainties (e.g., the presence of pyroclastic deposits, crater ejecta blankets).

### 6.7.2 Traverse Planning Methodology

The fundamental challenge in traverse planning lies in balancing the effort to produce "exquisite" or "perfect" maps against reliance on statistical models and safety margins. While maximally exploiting orbital data reduces risk, maps will inherently contain uncertainties and fail to capture all rover-scale hazards. Regardless of map quality, fully autonomous navigation and hazard avoidance are necessary to address small scale hazards. Thus, the fundamental trade is how and where to exercise higher fidelity mapping and route refinement and how to best utilize existing orbital data.

The baseline traverse planning methodology for the Rover follows a hierarchical approach, summarized in Figure 6.33, and described in the following sections.

#### 6.7.2.1 Data and Map Curation

The traverse design utilizes an "Improved LOLA" south polar DEM (Barker et al. 2023), ranging in nominal resolution from ~60 meters/pixel at mid-latitudes to ~10 meters/pixel near the south pole. An accompanying map layer quantifies local true resolution based on the location of LOLA ground tracks. Where available, auxiliary layers such as Diviner rock abundance models can inform the local hazard environment.

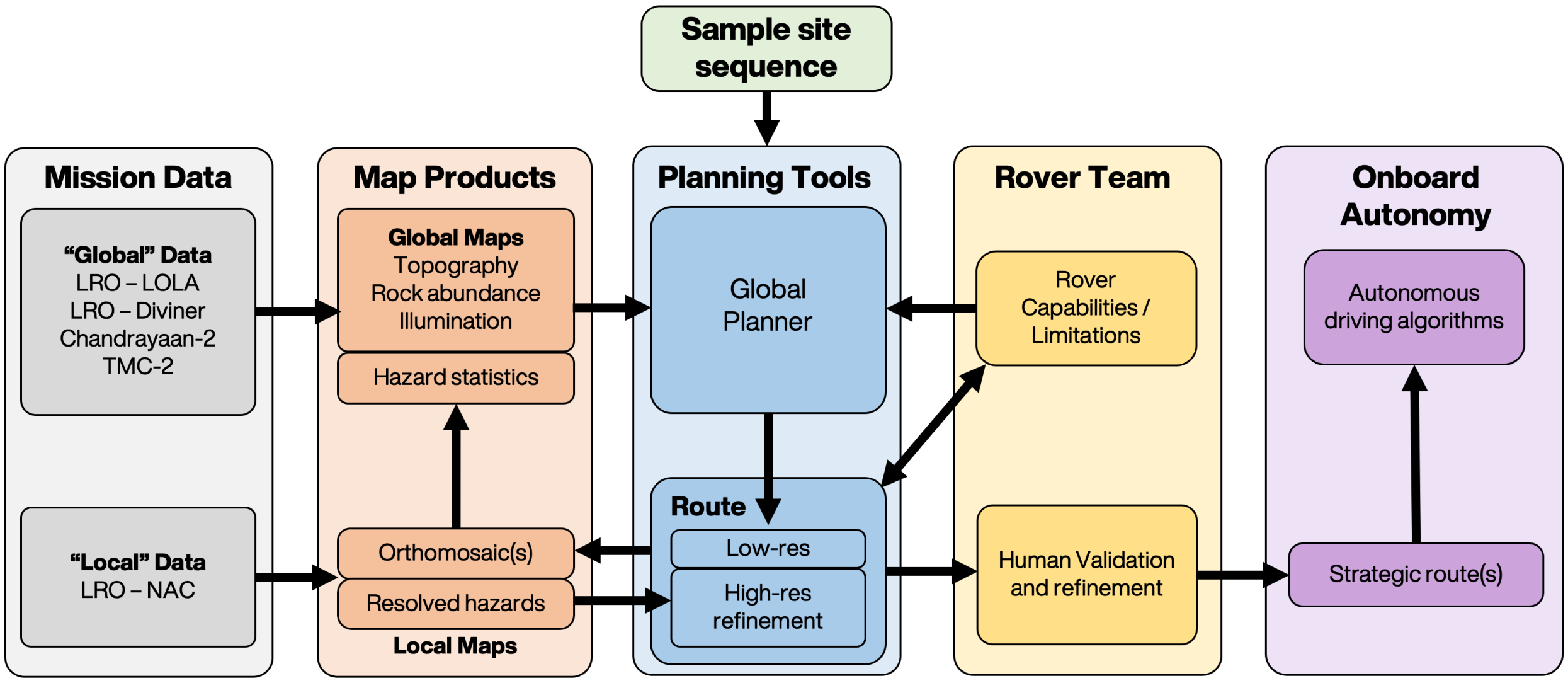

**Figure 6.33.** *Developing a >2,650-kilometer-long traverse across the farside of the Moon requires integrating a combination of datasets, map products, statistical analyses, and planning. This chart summarizes the traverse planning methodology and key components.*

### 6.7.2.2 Cost and Hazard Maps

A map is computed based on a margined model of the Rover's mobility capabilities, local terrain slopes, and statistics. This map consists of two components: a cost map and a hazard map.

The role of the hazard map is to prevent the rover from driving through areas that are, with some confidence, untraversable or unsafe. The primary hazards for this traverse are slope hazards, which are currently identified from the "Improved LOLA" slope map, inflated by 2× the local root-mean-square slope error, as defined by Barker et al. (2023). The slopes threshold is set at 15°, which, when coupled with the Rover's 20° slope capability, leaves an additional 5° of slope margin for smaller scale topography that is unresolved in the LOLA slope map. The choice of 5° slope margin is informed by studies of the topographic roughness spectra with higher resolution datasets.

A second notable class of hazards for the hazard map are illumination hazards. As the solar-powered Rover drives to the south pole for rendezvous with Artemis, seasonal and permanent shadows can represent obstacles to be avoided to maintain solar charging. As described in Section 6.9, the Rover design team assumed an operational safety policy whereby the Rover would never drive through a location where, if it faulted, could result in loss of mission. Thus, permanent shadows are deemed hazards. Transient or seasonal shadows are deemed hazards if the length of "night" in that shadow is longer than 520 hours during equinox. (This duration is notably longer than the length of lunar night at mid-latitudes, which is approximately 354 hours.) The choice of 520 hours flows from a trade analysis for minimizing battery mass, while also still affording traversable paths to potential Artemis landing sites and Moon Base locations.

The role of the cost map is to reflect the expected time to traverse a given map pixel. The cost map is currently computed as a function of the absolute slope of a given pixel of the "Improved LOLA" DEM, with an expression that approximates the increased effects of wheel slip, power consumption, and fault likelihood associated with steeper terrain: $C = (s_{max} / (s_{max} - s))^k$, where $s_{max}$ is the maximum slope threshold (set at 15° here), s is the local slope, and k is a tuning parameter to adjust the bias against intermediate slopes (currently set at 0.5). Finally, this cost is scaled by a “solar factor” heuristic: F = 354 / L, where L is the total number of hours of daylight at that location. The solar factor (Section 6.7.2.5) is intended to capture the operational impact of shorter drive times in regions with earlier sunsets and/or later sunrises. While this approach is sufficient for early planning, ultimately the final traverse planning would need to be done by a diurnal planner with an explicit time-state.

These formulations for cost and hazard maps represent a present-day snapshot of the traverse planning process for the purposes of the baseline reference mission design. Like all planetary traverses, the fidelity of these planning tools is expected to improve through formulation and development. An important future update could involve using high-resolution LRO-NAC orthoimages to further refine statistics, including the path tortuosity, based on local rock, crater, and slope hazards.

#### 6.7.2.3 Route Planning

A route planning algorithm determined the optimal sequence of SPARX science sites and planned the route between them. Specifically, the technical team used a fast-marching method (FMM) to propagate a cost-to-go wavefront from each science site to every pixel in the map, whereby the optimal path from any point was simply the gradient descent through the cost-to-go map. Importantly, this approach yielded an "any-angle" solution comparable to algorithms like Theta* or Field-D*. Crucially, it avoided the sub-optimal "Manhattan" (grid-constrained) paths that are intrinsic to conventional discrete planners such as Dijkstra or A*, ensuring a more efficient and realistic traverse path. A fully connected graph was then constructed representing optimal paths between all science sites, and a traveling salesman solver was applied to find the optimal sequencing of sites to produce the shortest overall path. Ongoing improvements to the algorithm include the integration of anisotropic cost functions (e.g. representing the different costs of going “uphill” vs. “downhill”).

#### 6.7.2.4 Route Refinement

The preliminary route developed by the process described here establishes a “corridor” for high-resolution map curation. The plan includes developing a NAC ortho-mosaic for the entire traverse to aggregate regional hazard statistics. In areas of higher risk or uncertainty, high-quality local digital terrain models (e.g., shape-from-shading) will be generated for safety validation and meter-scale refinement.

#### 6.7.2.5 The “Solar Factor”

As the Rover baselines solar power, surviving the lunar night is a core safety requirement. While mid-latitudes offer consistent day-night cycles with ~354 hours of daylight followed by ~354 hours of night, the high-latitude portion of the traverse presents extreme sunlight variations due to local topography. The technical team assumed an operational requirement that the Rover shall never drive in a location in which it would be unsafe if it were to fault or stop for one entire day/night cycle (Section 6.2). While there are many transiently illuminated areas that the rover could drive through by timing its transit, these would be a potential trap that would fully drain the battery if the rover were to fault within them. Therefore, the design choice was taken to never have the Rover “race” across permanent or transient shadows. This is a notable architectural difference from other lunar rovers, such as VIPER which must race in and out of PSRs or chase “islands of light.” The Rover's night-survival capabilities impose strict constraints on feasible driving region:

- <u>Battery capacity:</u> While a ~400-hour night survival capability suffices for low-mid latitudes (where the night is a relatively consistent ~354 hours long), the Rover must be capable of driving through the south polar region, which has many locations with long transient shadows and even permanent shadow. The Rover requires a 500–550-hour capability to provide safety and flexibility for the terminal rendezvous at Moon Base or a polar Artemis mission.

- Seasonal constraints: Even with increased battery capacity, seasonal constraints dictate when the rover can safely execute the final ~30 kilometers through heavily shadowed regions. A 520-hour battery affords a 6-month window for this final approach to maintain fault-safe operations. If the Rover arrives at the south polar region in an unfavorable season (i.e., south polar winter), a safe staging point is identified if an extended wait is needed.

To visualize the effect of seasonality, Figure 6.34 shows maps of the south polar illumination during summer solstice (left) and winter solstice (right). While the ecliptic obliquity of the Moon (angle of Moon's axis with respect to the ecliptic plane) is only about 1.5°, there are still substantial variations in regional illumination throughout the year. Many of the candidate Artemis landing sites (including the current baseline: 007) are surrounded by heavily shadowed regions in the winter months, making them inaccessible for part of the year. Figure 6.35 shows a zoom-in of the final leg of a possible SPARX Rover traverse, which highlights how the final leg of the traverse may become seasonally dependent—with safe (and expeditious) corridors only opening up outside of south pole winter.

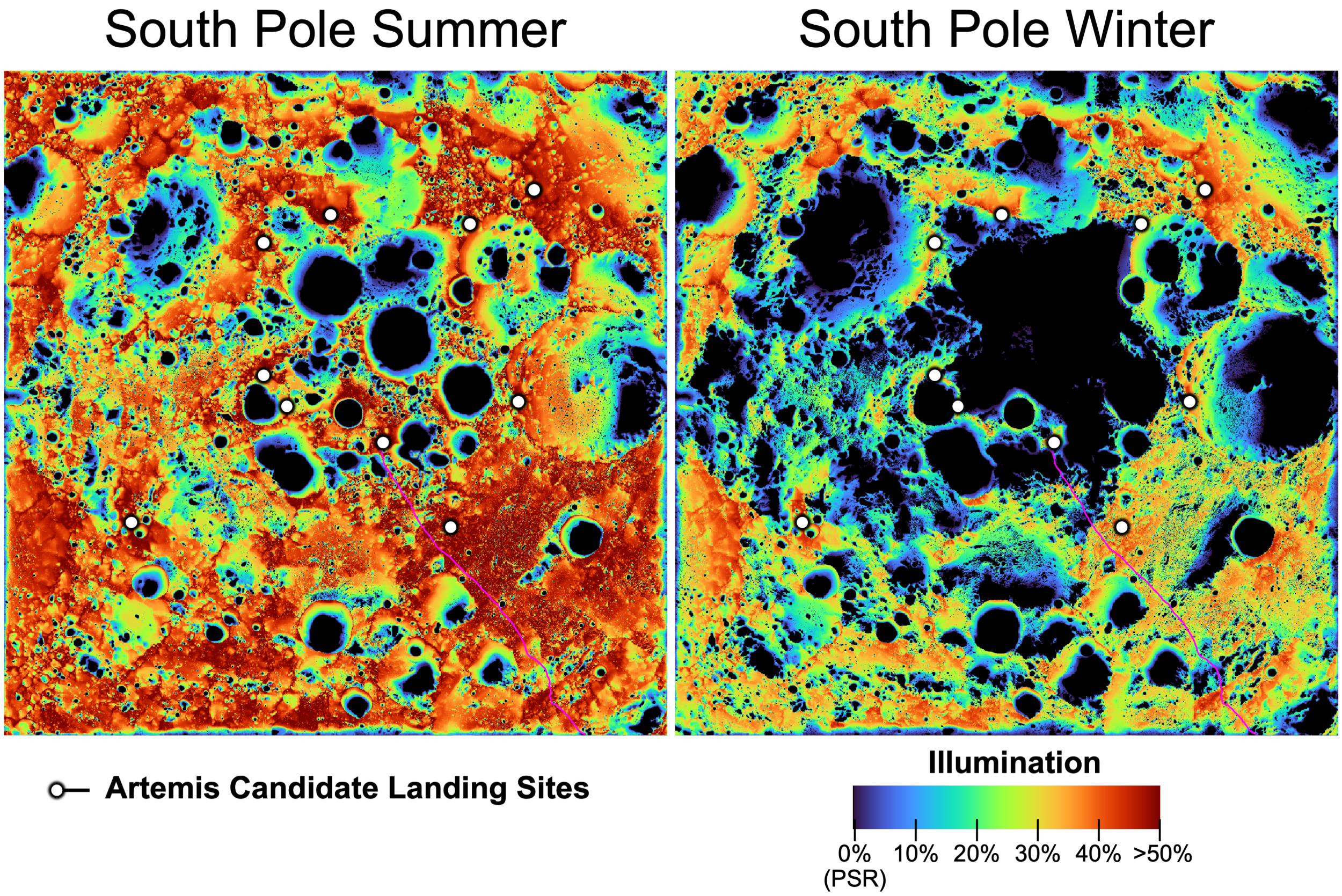

**Figure 6.34.** *Seasonal illumination (fractional daylight) of the south polar region can be a major driver for traverse planning for a solar-powered rover. The map is derived from the 80°S south pole stereographic Improved LOLA DEM (Barker et al. 2023). The pink line shows a notional SPARX rover traverse.*

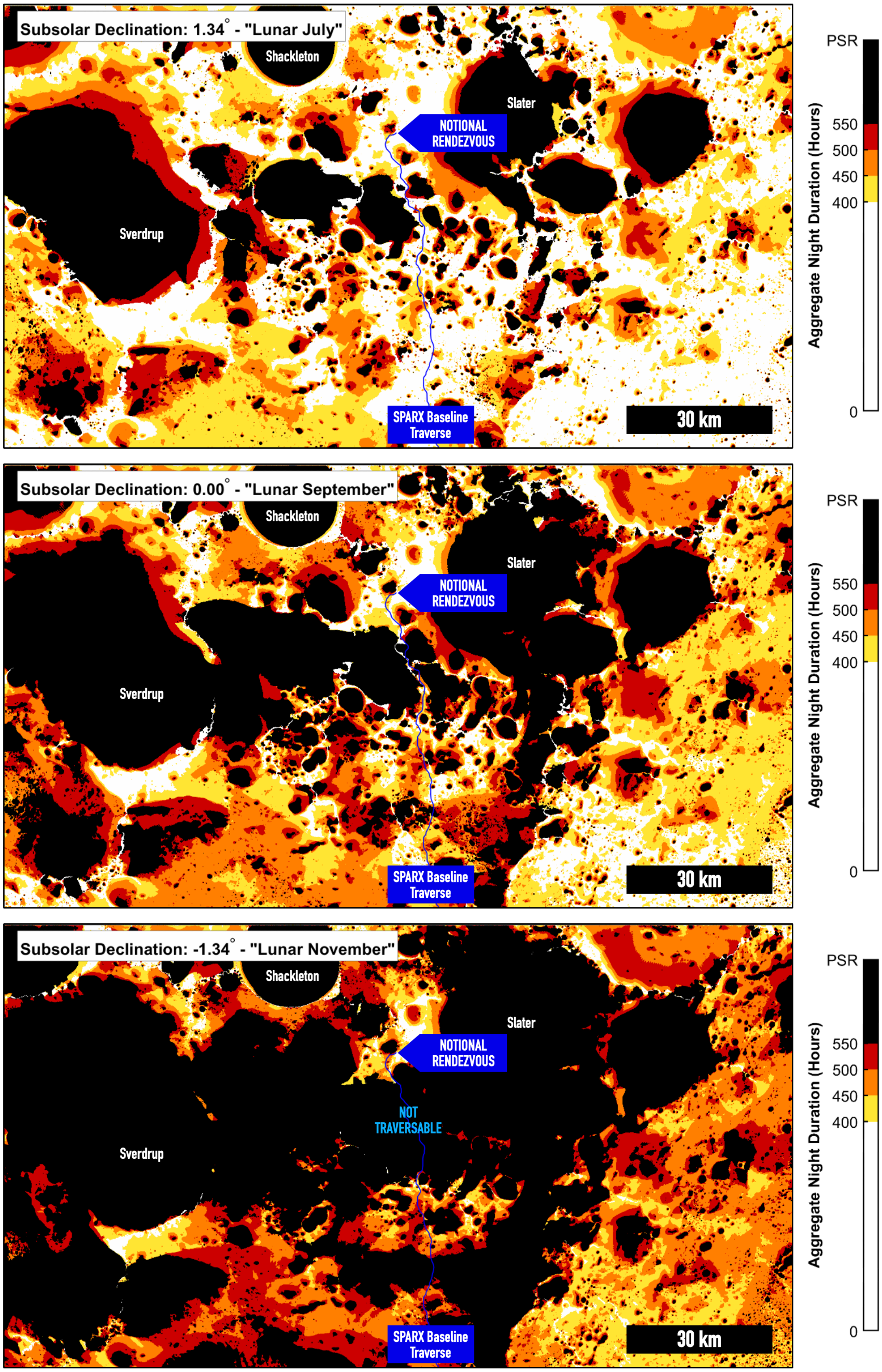

**Figure 6.35.** *The final leg of the SPARX traverse to a notional Moon Base rendezvous is challenging for a solar-powered rover, requiring a seasonal final approach.*

### 6.7.3 Baseline Sample Sites and Traverse Waypoint

The SPARX science requirements flow to a set of site and sample characteristics, described in Section 5.8. Foldout 5.1 shows the Sample Site and Traverse Requirements Traceability Matrix. Table 6.19 provides a list of the sample sites and traverse waypoint.

**Table 6.19.** *SPARX baseline sample sites (Sites 1 to 20) and traverse waypoint (Site 21). *SPARX Science Experiments E6 and E11 are addressed with samples and measurements from all sites and are omitted from this table.*

| ID | Site Name | Latitude | Longitude | Cumulative Distance | Fractional Distance | Distance to the Next Site | SPARX Science Experiments* |
|---|---|---|---|---|---|---|---|
| 1 | Mickey Mare (and landing site) | -52.160 °N | 193.226 °E | 0 km | 0.0% | 134 km | E3, E8, E10 |
| 2 | Central SPA light plains | -53.540 °N | 199.880 °E | 134 km | 5.0% | 17 km | E13 |
| 3 | Central SPA plains fresh crater | -53.540 °N | 200.820 °E | 151 km | 5.7% | 135 km | E13 |
| 4 | Bhabha | -56.660 °N | 195.650 °E | 286 km | 10.7% | 127 km | E2, E13 |
| 5 | Bose | -55.200 °N | 189.050 °E | 413 km | 15.5% | 312 km | E2, E13 |
| 6 | Abbe M northwestern rim | -61.310 °N | 174.313 °E | 725 km | 27.1% | 35 km | E7 |
| 7 | Abbe M southwestern rim | -62.389 °N | 173.821 °E | 760 km | 28.4% | 188 km | E7 |
| 8 | Mare in Poincaré E | -56.656 °N | 171.105 °E | 948 km | 35.5% | 209 km | E3, E8, E10 |
| 9 | Poincaré X high albedo material | -54.190 °N | 160.716 °E | 1,157 km | 43.3% | 15 km | E1 |
| 10 | Poincaré northern peak ring near Poincaré X | -54.571 °N | 160.134 °E | 1,172 km | 43.9% | 143 km | E5 |
| 11 | Poincaré southern peak ring northwest of Poincaré Q | -59.103 °N | 160.773 °E | 1,315 km | 49.2% | 18 km | E5 |
| 12 | Mare northwest of Poincaré Q | -59.156 °N | 161.893 °E | 1,333 km | 49.9% | 26 km | E3, E8, E10 |
| 13 | Poincaré southern peak ring | -59.594 °N | 163.320 °E | 1,359 km | 50.9% | 170 km | E5 |
| 14 | Lyman | -64.280 °N | 158.430 °E | 1,529 km | 57.2% | 383 km | E9, E14 |
| 15 | Schrödinger northern peak ring #2 | -73.161 °N | 141.597 °E | 1,912 km | 71.6% | 8 km | E4 |
| 16 | Schrödinger northern peak ring #1 | -73.270 °N | 140.796 °E | 1,920 km | 71.9% | 73 km | E1 |
| 17 | Schrödinger pyroclastics | -75.259 °N | 137.314 °E | 1,993 km | 74.6% | 71 km | E10 |
| 18 | Schrödinger western peak ring | -74.450 °N | 129.145 °E | 2,064 km | 77.2% | 90 km | E1 |
| 19 | Schrödinger southern peak ring massif | -77.164 °N | 132.884 °E | 2,154 km | 80.6% | 73 km | E9, E14 |
| 20 | Schrödinger southern rim | -79.004 °N | 138.861 °E | 2,227 km | 83.3% | 229 km | E2 |
| 21 | South Pole-Aitken southern rim massif | -82.830 °N | 164.000 °E | 2,456 km | 91.9% | 216 km | E12 |
| - | Notional Artemis Moon Base rendezvous (site 007) | -88.700 °N | 131.170 °E | 2,672 km | 100.0% | - | - |

### 6.7.4 Baseline Traverse

Utilizing the traverse planning methodology described above, and the science sites defined in Section 5, the technical team developed a new baseline traverse for the Rover, shown in Figures 6.36–6.40.

The total length of the traverse is an important metric for the Rover design. The total length of the baseline traverse is 2,672 km when planned at the LOLA map resolution (ranging from ~10 to ~100 meters per pixel). Based on a variety of studies of traverse planning with higher resolution datasets, we expect an additional “tortuosity” factor to increase the actual drive distance by about 5–6% to drive around smaller features (e.g., unresolved craters and rocks), for a total of 2,800 km. An additional 5% margin to account for unplanned diverts and higher-than average tortuosity takes the final margined drive distance to 2,950 km. These total distances are summarized in Table 6.19.

**Table 6.20.** *The total driving distance for the SPARX Rover.*

| Distance as planned from LOLA | Distance when accounting for traverse tortuosity around hazards that are unresolved in LOLA | Margined distance for designing the Rover |
|---|---|---|
| 2,672 km | 2,800 km | 2,950 km |

Figure 6.41 shows the distribution of several key environmental statistics along the traverse, including the from the “Improved LOLA” south polar DEM. The maximum slopes do not exceed 14° (providing ample margin to the rover’s 20° capability), and 97% of the traverse is less than 10°.

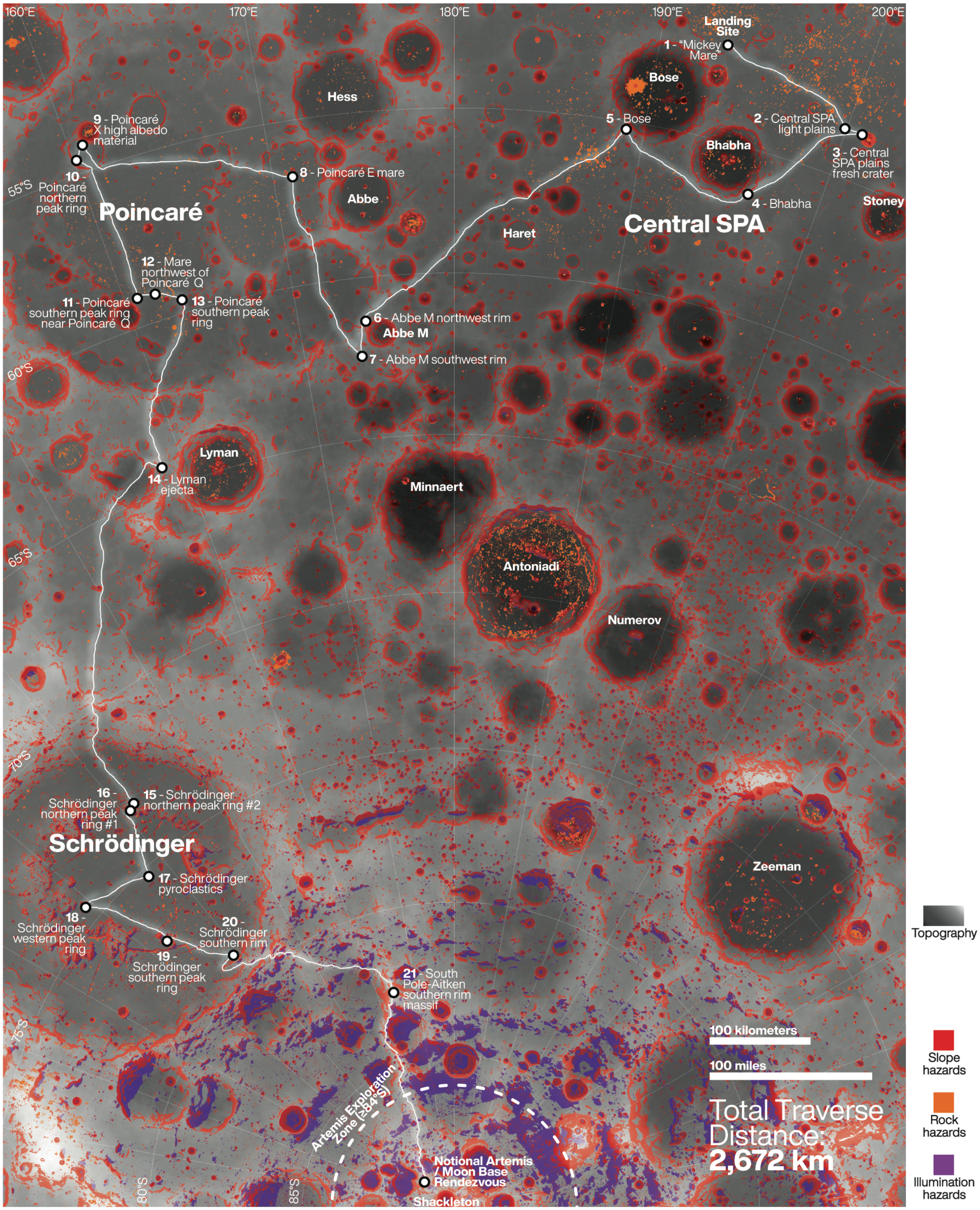

**Figure 6.36.** *The SPARX baseline traverse traverses ≥2,650 kilometers of the South Pole-Aitken basin, visiting 20 sample sites and 1 traverse waypoint before rendezvous at Moon Base. This map shows topography (grayscale), and then key hazards, including slopes, rocks, and poor illumination.*

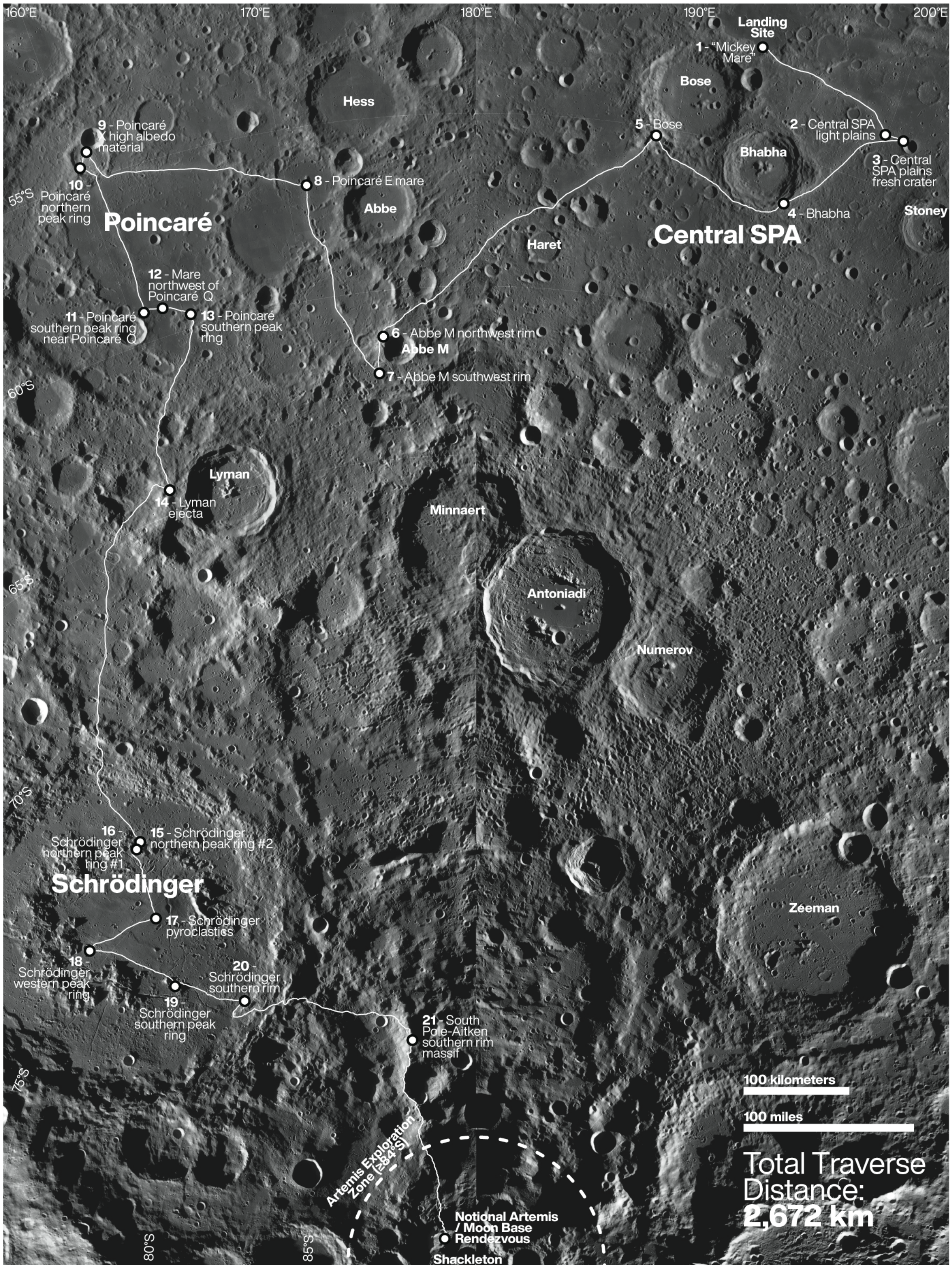

**Figure 6.37.** *The SPARX baseline traverse traverses ≥2,650 kilometers of the South Pole-Aitken basin, visiting 20 sample sites and 1 traverse waypoint before rendezvous at Moon Base. This map shows the traverse draped across an LRO image mosaic of the farside of the Moon.*

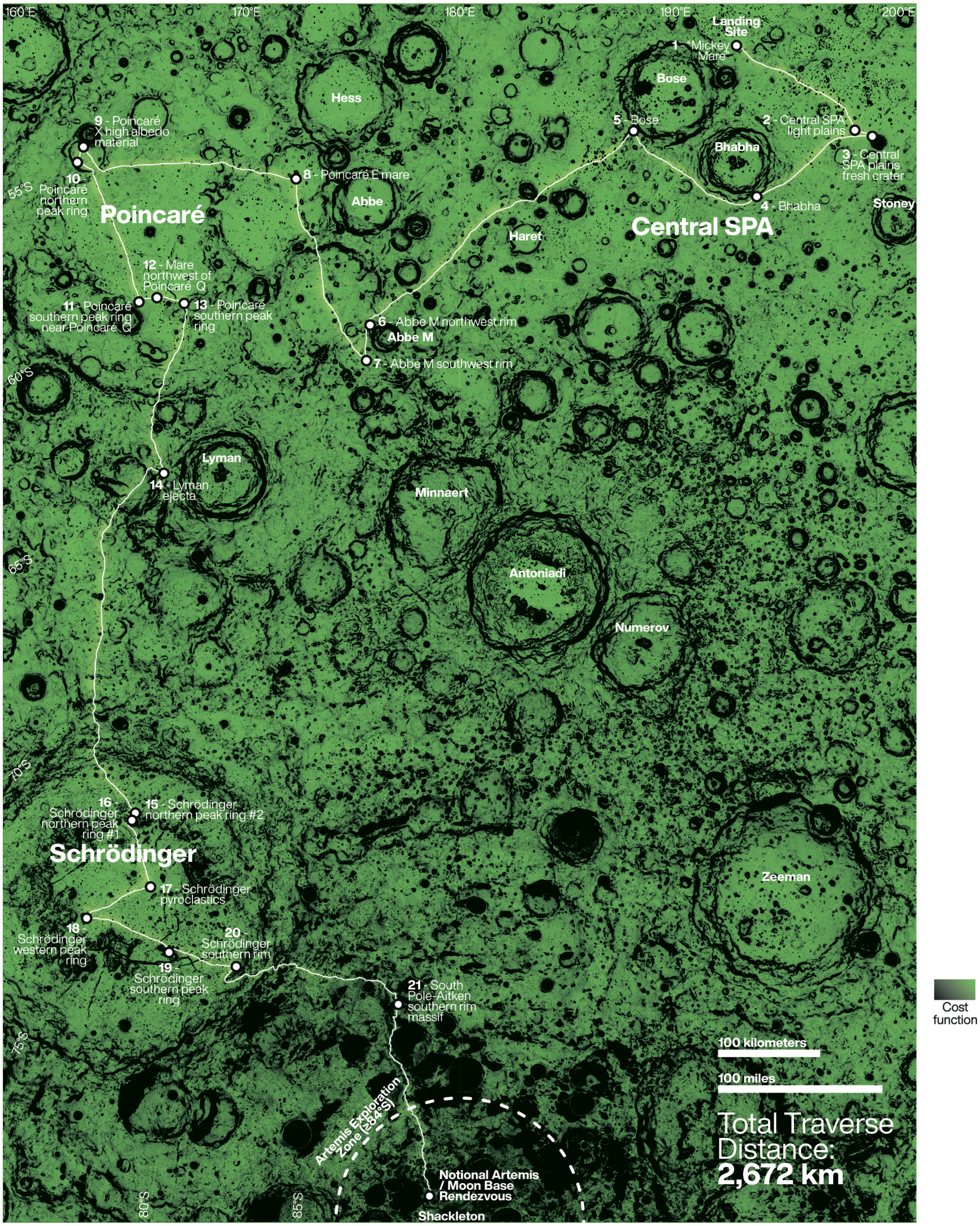

**Figure 6.38.** *Developing the Endurance traverse required developing a cost function that includes key hazards (slopes, rocks, illumination, etc.) across much of South Pole-Aitken basin. Black indicates high (or infinite) cost; green indicates low cost.*

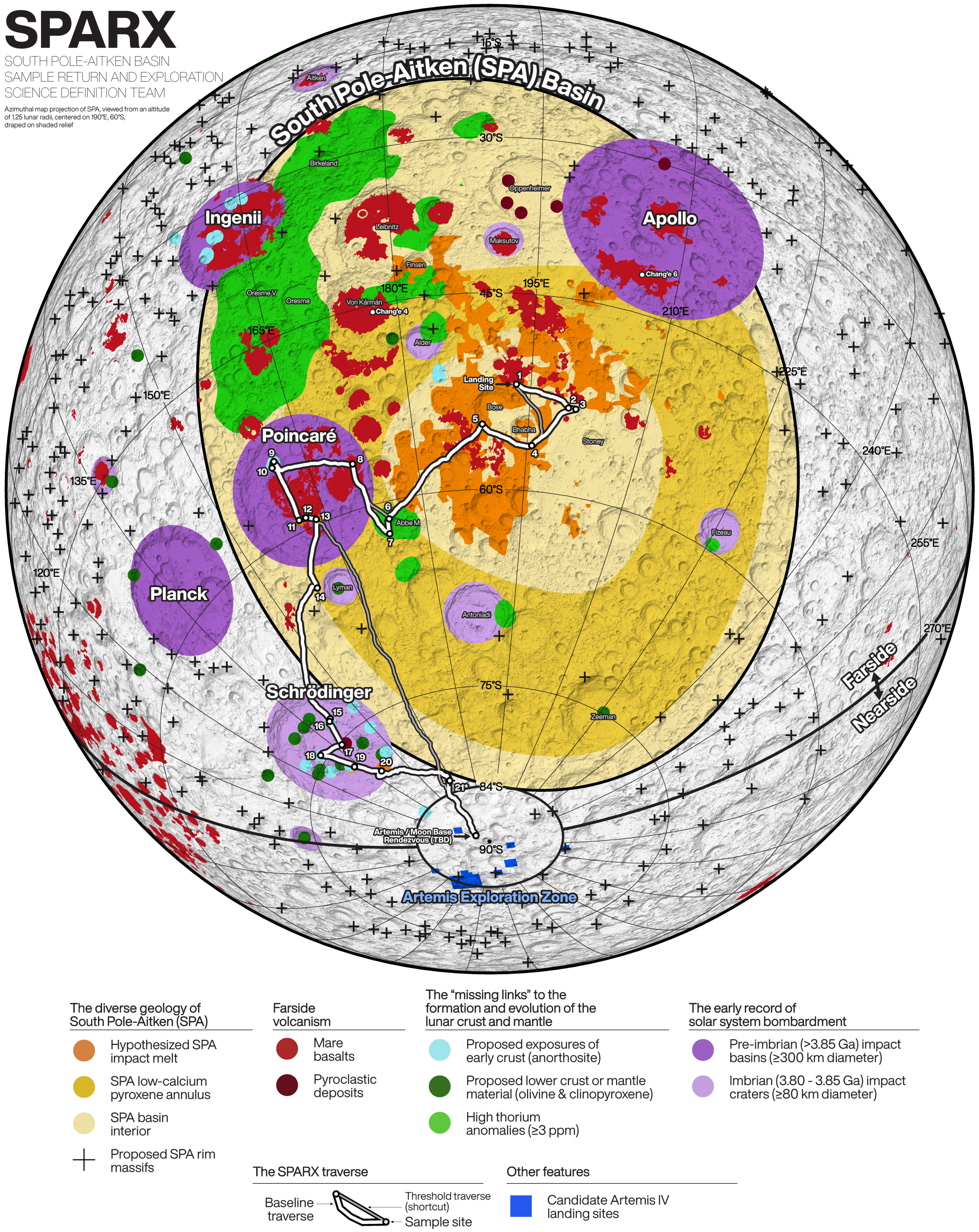

**Figure 6.39.** *The SPARX baseline (and threshold) traverse on a map of the entire South Pole-Aitken (SPA) basin, highlighting several of the key terrains that drove site and sample requirements (described further in Section 4).*

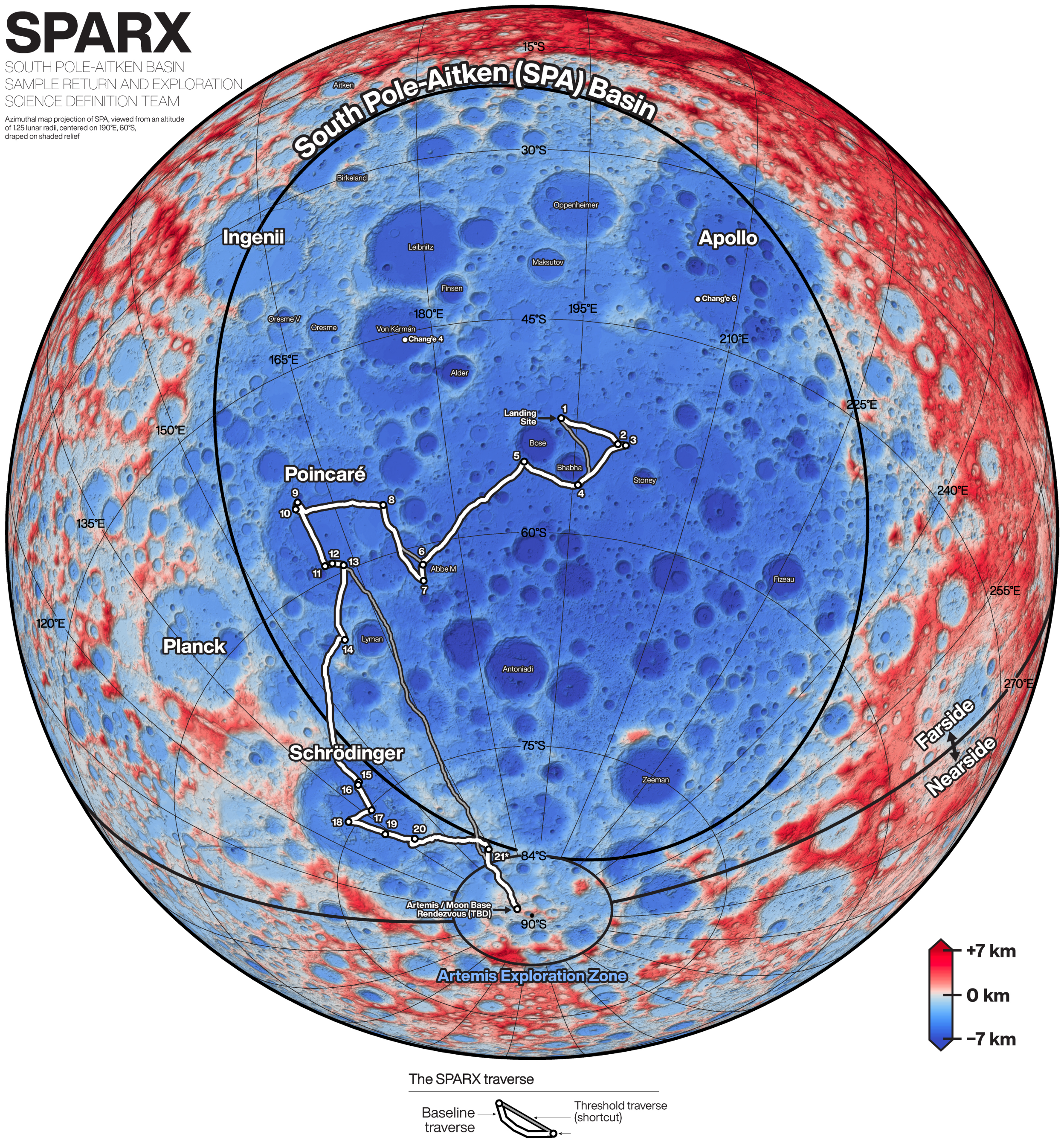

**Figure 6.40.** *The SPARX baseline (and threshold) traverse on a map of the entire South Pole-Aitken (SPA) basin, draped over an LRO/LOLA topographic map.*

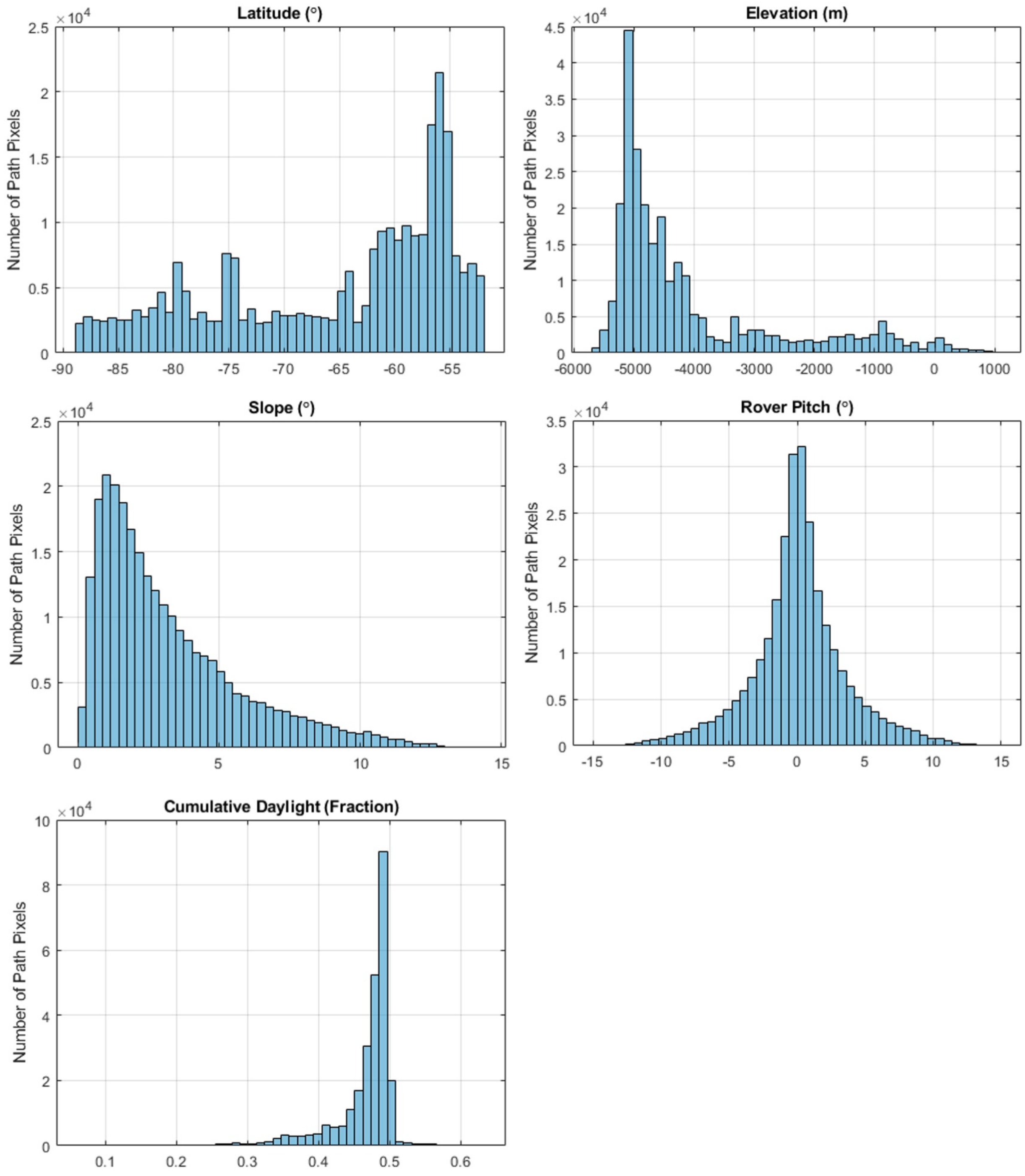

**Figure 6.41.** *Statistical distribution of key environmental parameters along the baseline SPARX traverse, including latitude (top left), elevation (top right), absolute value of slope (middle left), rover pitch (middle right), and cumulative daylight (bottom left).*

# 6.8 MISSION DURATION

A major technical and programmatic factor for the design of this concept is the mission duration. The mission duration determines the number of thermal cycles the Rover will experience, how long parts must be qualified, and ultimately the cost for operations. In the following sections we describe the model, MissionSim, that was developed to estimate mission duration (Section 6.8.1) and predictions for the mission duration (Section 6.8.2).

## 6.8.1 MissionSim

To estimate mission duration, the JPL technical team built a mission simulator to model and validate the baseline Rover design and ConOps. This simulator, known as MissionSim, was developed to model key mission variables, resources, and nominal rover behaviors with high accuracy, enabling a thorough simulation of nominal scenarios. In addition, MissionSim can model anomalies and off-nominal situations, and the variability arising from these conditions. Finally, simulation results are margined to account for delays and issues that are currently unknown and not modeled. In this fashion, MissionSim can calculate mission performance metrics like mission duration while taking nominal, off-nominal, and margin factors into account. In the following sections, we describe the key sub-models and assumptions of MissionSim.

### 6.8.1.1 Mission Route and Autonomous Driving Model

The complete route and all sample site destinations are modeled in MissionSim. To simplify and expedite simulation over the entire route, terrain interactions are modeled behaviors, not simulated interactions. For example, the simulation includes a path inefficiency factor estimated from orbital data to capture the effect of autonomously navigating around small hazards – it does not simulate onboard hazard detection or avoidance maneuvers. The route (described in more detail in Section 6.7.4) is derived from orbital products (e.g., LOLA digital elevation models, LROC images) and includes slope measurements that are used to estimate skid/slip ratios while modeling driving. A mobility power model estimates how much power the Rover would use to traverse each segment of the route, based on the slope as well.

### 6.8.1.2 Solar Illumination and Orbital Geometries

MissionSim uses NASA SPICE models to compute geometries within the Earth-Moon-Sun system. These geometries are used for evaluating solar illumination and communication network configurations.

### 6.8.1.3 Communication Relay Network

In the baseline mission design, NASA's Lunar Communications Relay and Navigation Systems (LCRNS) network is the primary communication network supporting the Rover (Section 6.5.7.1). MissionSim simulates notional orbital trajectories of the LCRNS satellites in order to compute view periods and communication opportunities. Communications windows are modeled as a fixed schedule of contacts that would be negotiated with the LCRNS provider during the mission. The visibility of satellites is geometrically computed based on the actual position of the Rover along the route at each simulation timestep.

The telecom data transfer rates used in the simulator are computed by a suite of JPL link budget analysis tools that model the communication paths between the LCRNS satellites and the Rover's telecom system. This analysis includes the effects of multipath interference on the antenna-specific transmit/receive patterns.

#### 6.8.1.4 Ground Operations Model

MissionSim includes a schedule of ground operation shifts that allows it to directly simulate interactions between the Rover and the ground operations team. This enables studies that help understand how sensitive the mission is to ground support, especially as it pertains to fault response.

#### 6.8.1.5 Rover Modes and State Machine

Operational modes are used in the simulation to track resource utilization and mission activities onboard the rover. For example, there is an Autonomous Drive mode to model navigating the route, a Sample Collection mode to model sampling activities, a Telecom Stop mode to model a communication window, a Hibernation mode to model night survival activities, and so on.

A state machine is used to transition the Rover between these different operational modes in the simulation. Each mode has entrance and exit conditions in the state machine that encode the engineering and science ConOps through all mission phases. While an operational mode is active, the simulation evaluates progress towards that mode's objectives (driving, sampling, data collection, etc.) and updates the Rover's resources to reflect time spent in that mode.

#### 6.8.1.6 Sampling Campaign Model

Sampling campaigns are modeled using detailed timelines of the science and engineering activities at each site based on the site-specific sampling requirements (described in Section 4). Those timelines produce estimates of the total campaign duration for each site and include duration held for anomaly response during the campaign. MissionSim takes these duration estimates and combines them with real environmental and operational constraints to evaluate campaign progress during the simulation.

#### 6.8.1.7 Thermal Management Model

During a simulation, the energy utilization of the thermal management system is approximated without including a complex thermal dynamics model. Temperature of each component is tracked and the rate of temperature change for each component is predicted based on environmental factors and reference data from the thermal subsystem design team analysis. Based on the Rover mode in simulation, the model will look up the required temperature for each component and simulate which heaters should be powered on.

During night hibernation, the simulation assumes that all survival heaters are required for the entire night duration. This allows the simulation to model a constant power draw as a conservative prediction of night hibernation energy utilization.

#### 6.8.1.8 Power Production and Use Model

The simulation tracks power production from the solar array, power usage by the current mode, and power usage by the heaters. A geometric model of the Rover attitude relative to the Sun is simulated, such that the total area of illuminated solar panels and corresponding power production can be calculated.

If the battery nears the minimum state of charge, the state machine automatically schedules and executes a charge cycle. Similarly, in the mornings after night hibernation, batteries are charged to a nominal state of charge before starting productive activities.

#### 6.8.1.9 Instrument Suite and Science ConOps Model

The science instruments in the baseline reference design are modeled in simulation by tracking their power usage and data production, as well as using the Rover mode state machine to model the instrument operation schedule.

#### 6.8.1.10 Data Production, Storage, and Downlink Model

All modes have data production estimates based on the most relevant information. For Rover engineering subsystems, data production estimates are based on actual Perseverance Rover data production rates and predictions using the baseline reference design models. For the science payload, simulated data production is derived from the strawman instruments (Section 6.4).

Engineering and science data are each modeled in three different priority "bins" in the simulation (for a total of six data bins):

- Decisional data: Required to be downlinked at next available opportunity, for example to support anomaly response.
- Routine data: Standard engineering trending and science observation data return from the mission that does not need to be acted upon by the ground operations.
- Optional data: Additional data that is collected but not queued for downlink unless there is a need to do so. Typically, just additional context engineering data, and not science data.

The simulation accumulates data in each of the six separate bins and models downlinking the data during comm passes. During communications windows, the Rover model is only required to downlink decisional and routine data. Once this data is sent to the ground, the communications window can be autonomously ended so the rover can continue making progress.

#### 6.8.1.11 Seasonal Traverse Constraint

Navigating a solar-powered rover to sites at the south pole is a challenge, due to the extreme seasonal variation in lighting conditions within the polar region. The baseline reference Rover is capable of traversing the south pole region during favorable seasons, but not the full year (see Section 6.7). Due to the route length and the uncertainty in traverse progress, the mission start time may be constrained to ensure the rover arrives at the south pole in a favorable season. To evaluate this uncertainty, MissionSim starts the surface mission at a random date during the year and enforces a "Seasonal Standby" period if the season is unfavorable when the rover reaches far southern latitudes (typically around 85˚S). In actual operations, this standby period could be used for productive activities, but in the simulation, it is modeled as simply waiting. When the season becomes favorable, the rover is able to traverse the south pole region and complete the mission.

#### 6.8.1.12 Modeling Faults and Off-Nominal Situations

Perhaps the most underappreciated contribution to mission duration is faults and off-nominal situations. Neglecting off-nominal scenarios (which occur on *every* planetary science mission) can result in extremely optimistic, but entirely implausible mission durations.

To more accurately capture the effects of faults and off-nominal situations, the MissionSim utilizes a Monte Carlo framework to run hundreds of simulations with randomly injected faults. The results of this Monte Carlo study helped the team understand the sensitivity to off-nominal variables, better predict mission performance metrics, and motivate improvements in design areas that have the largest effects.

Through the simulation studies, the team found that the rate of traverse faults and the corresponding ground response strategy have significant effect on mission duration. Using the baseline Rover design and supporting analysis work, the team was able to formulate a mobility and navigation system that is anticipated to achieve a mean distance between faults of 20 km. Combining this with the communications and ground operations model in MissionSim, the study was able to directly simulate the impact of traverse fault rate on mission duration and other mission performance metrics.

#### 6.8.1.13 System Margin

Even with the models included in this mission simulator, margin is needed to represent unmodeled operational delays and anomalies (the “unknown unknowns”). Although rare, anomalies that cause moderate to significant downtime are possible and have been experienced on previous rover missions. The mission duration results reported here in this report include a standard 33% margin used in mission formulation work on most daytime activities. The only exception is that, to prevent double margining, traverse fault recovery receives no additional system margin, as fault recovery is already a directly modeled margin activity.

### 6.8.2 Mission Duration Estimates

Figure 6.42 shows a probability distribution for the estimated Rover duration based on a Monte Carlo simulation of 300 distinct MissionSim runs, and an average breakdown of mission phase duration. This estimate includes a 14 Earth-day commissioning phase, a 14 Earth-day sample drop-off phase, and system margin as described above. The shape of the distribution is a result of the randomly distributed traverse faults combined with the seasonal traverse constraint on the end of the route. **The range of mission durations, between 3.5 years and 4.3 years, is believed to be an accurate prediction based on the reference baseline design.**

With the modeling parameters described above, on average, 23% of the mission is spent on traverse activities (including fault response), and 10% of the mission is spent on sampling activities. **For a solar powered rover, approximately half of the mission duration (55%, or approximately two years) is spent hibernating at night.** The 9% system margin portion is a result of the 33% margining strategy for all daytime activities except traverse fault recovery, which is already a directly modeled margin.

### 6.8.3 The Role of Rover Drive Speed in Mission Duration

Since the total mission duration is an important programmatic factor for the SPARX Rover, the technical team investigated the effect of rover drive speed on mission duration. While the SPARX Rover has a top autonomous drive speed of 1 kph (kilometer per hour), consistent with the constraints of the TOR (Appendix 10.1) capabilities exist or are being developed for traverse speeds as high as 10 kph.

These speeds are fast compared to past Mars rovers such as Perseverance, which has a top autonomous drive speed of ~0.1 kph, but slower are than human-operated lunar rovers like the Apollo Lunar Roving Vehicle which had a top speed of ~18 kph. However, the mechanical drive speed of a rover is only one factor of a rover’s “speed-made-good” (i.e., the effective speed of the vehicle between two waypoints). The SPARX baseline mission design includes several key factors that affect—and typically reduce—the speed-made-good:

- <u>Traverse science stops</u>: The SPARX mission has requirements for in-situ science along the traverse. The primary mode of traverse science is the interval stops (Sections 4.2 and 6.6), which require science stops every ~2 kilometers. During these stops, the Rover uses remote sensing instruments to collect essential geologic context measurements while stopped.

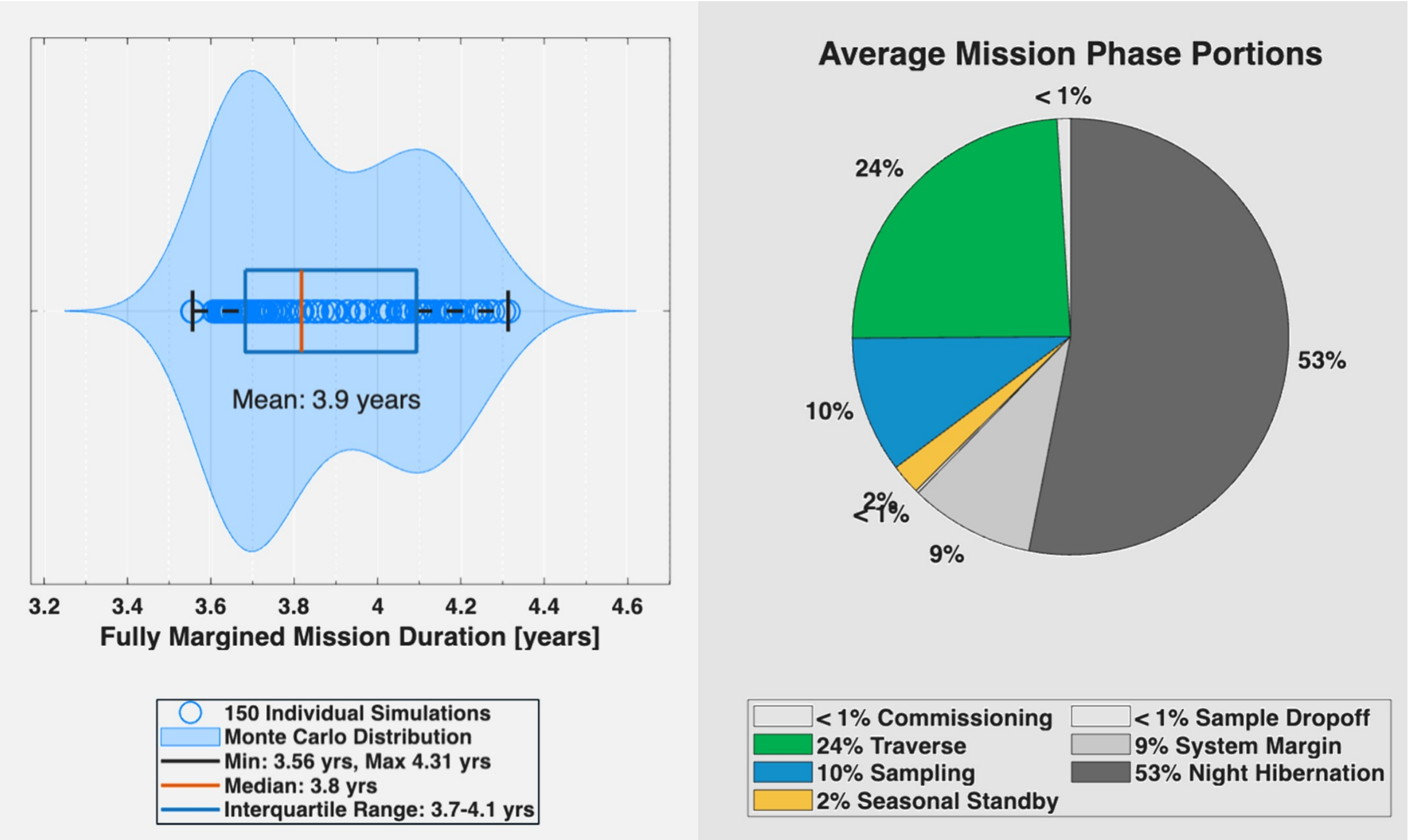

**Figure 6.42.** *The estimated SPARX Rover mission duration is well within the NASA-defined requirement of five years. Left: Probability distribution for fully margined mission duration for the Rover described here. Individual blue circles represent distinct MissionSim runs, each with their own randomized distribution of faults and off-nominal situations. Right: Average breakdown of the mission duration by major activities, including traversing (green), sampling (blue), night hibernation (dark gray), and system margin (light gray).*

- Sampling stops: The SPARX mission is a sample return mission with 20 sample collection campaigns in the baseline mission. Each of these sample campaigns requires a substantial amount of time (currently baselining approximately 14 earth days, equivalent to one lunar daytime period, per sample campaign, or ~16% of the entire mission duration as shown in Figure 6.42). Driving faster does not directly reduce the required time for sampling, as the duration for sampling campaigns is largely driven by the decision-making process of the ground-based science team.
- Lunar night: For a solar-powered rover, the Rover will spend approximately half of its mission duration hibernating to survive. Driving faster does not reduce the fraction of the mission spent in lunar night.
- Faults: Every planetary mission experience faults (i.e., off-nominal scenarios) that impact the rate of progress. Driving faster does not directly reduce the frequency of faults.

Figure 6.43 shows a proof-of-concept for how the Rover's drive speed translates to "speed-made-good" and how it depends on two of the most impactful factors: the duration of science stops and the mean distance between faults. The most significant result is that there is a rapid drop-off in speed-made-good with increasing drive speed. As drive speed increases, the time it takes to cover a given distance asymptotically approaches the amount of time spent doing science and recovering from faults. Between these two factors, the distance between faults is by far the most important driving factor. Removing science stops entirely (light blue curve) only results in incremental gains in speed-made-good.

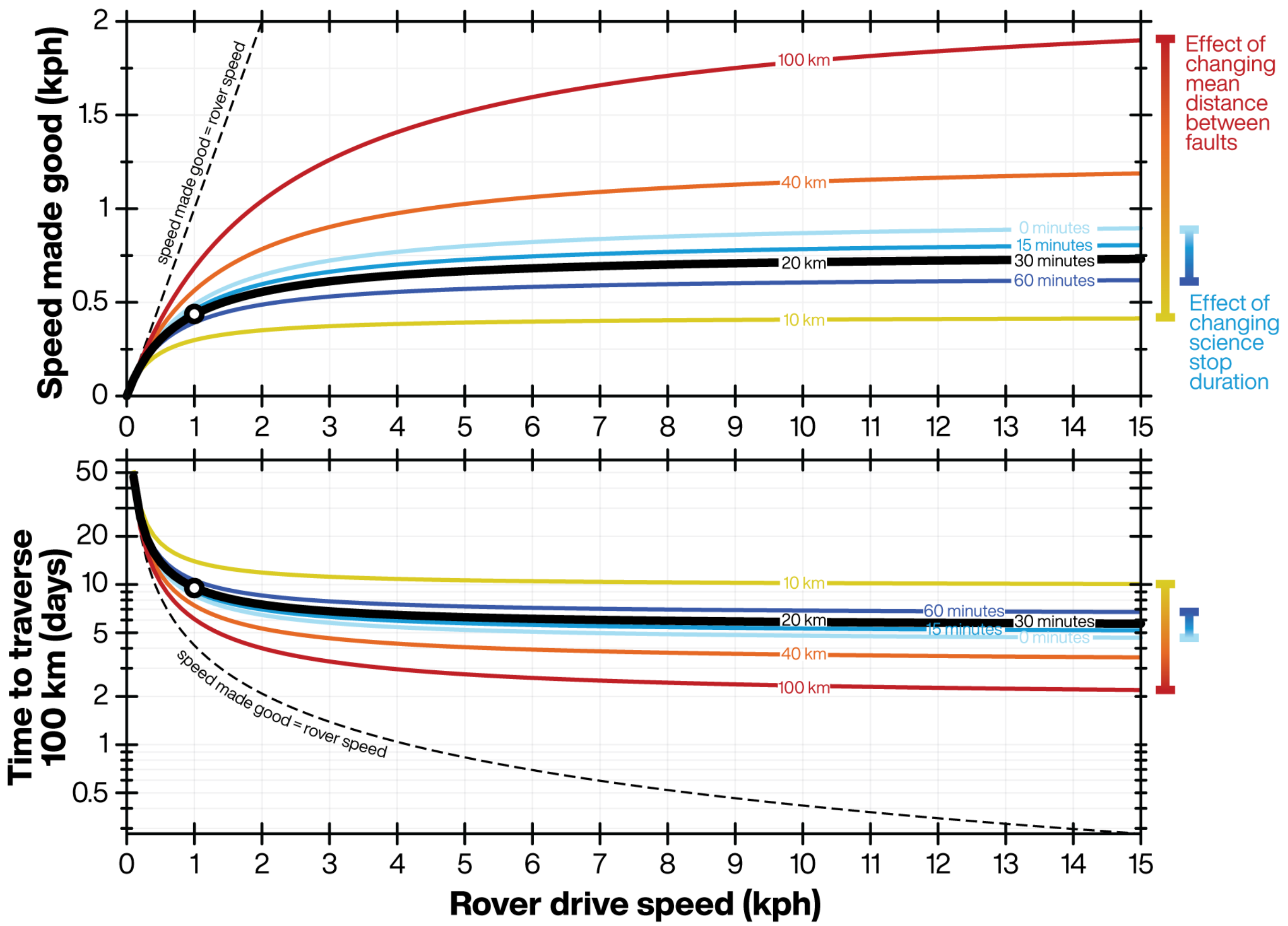

**Figure 6.43.** *Proof-of-concept of how drive speed translates to "speed-made-good" (top panel) and how long the Rover traveling at that speed would take to traverse a fixed distance of 100 km (bottom panel). The black line shows a nominal scenario with 30-minute-long science stops every 2 km, a mean distance between faults of 20 km, and a mean fault recovery time of 21 hours. The black/white dot indicates where the baseline Rover is on this spectrum with a top speed of 1 kph. Blue curves show the effect of changing the duration of science stops: no science stops (light blue), shortened science stops (15 minutes; medium blue), and lengthened science stops (60 minutes; dark blue). Red/orange curves show the effect of changing the mean distance between faults: much shorter (10 km; yellow), further (40 km; orange), and much longer (100 km; red). This analysis shows that there is diminishing returns for increasing rover speed, and that faults are the primary driver of speed-made-good, and thus mission duration.*

We use MissionSim (Section 6.8.1) to investigate the effect of drive rate on mission duration (Figure 6.44). As drive speed drops below 1 kph, the mission duration rapidly grows. Above 1 kph, the mission duration is only weakly dependent on drive speed.

Increasing the speed also has major consequences for rover design that can impact the overall size, mass, cost, and risk posture, summarized graphically in Figure 6.45. Significant downstream effects include increased onboard computation requirements to perceive and process the autonomous navigation algorithms at an increased rate, while increased computing requirements drive increased power consumption, and resultingly larger solar panels. In this manner, increases to rover drive speed propagate throughout the entire system. The baseline Rover design presented here optimizes rover design constraints, science constraints, and mission duration with the TOR-compliant traverse speed. Other rover or mobility platform designs (e.g., LTV) should carefully weigh the design effects and science requirements in planning for traverse speed.

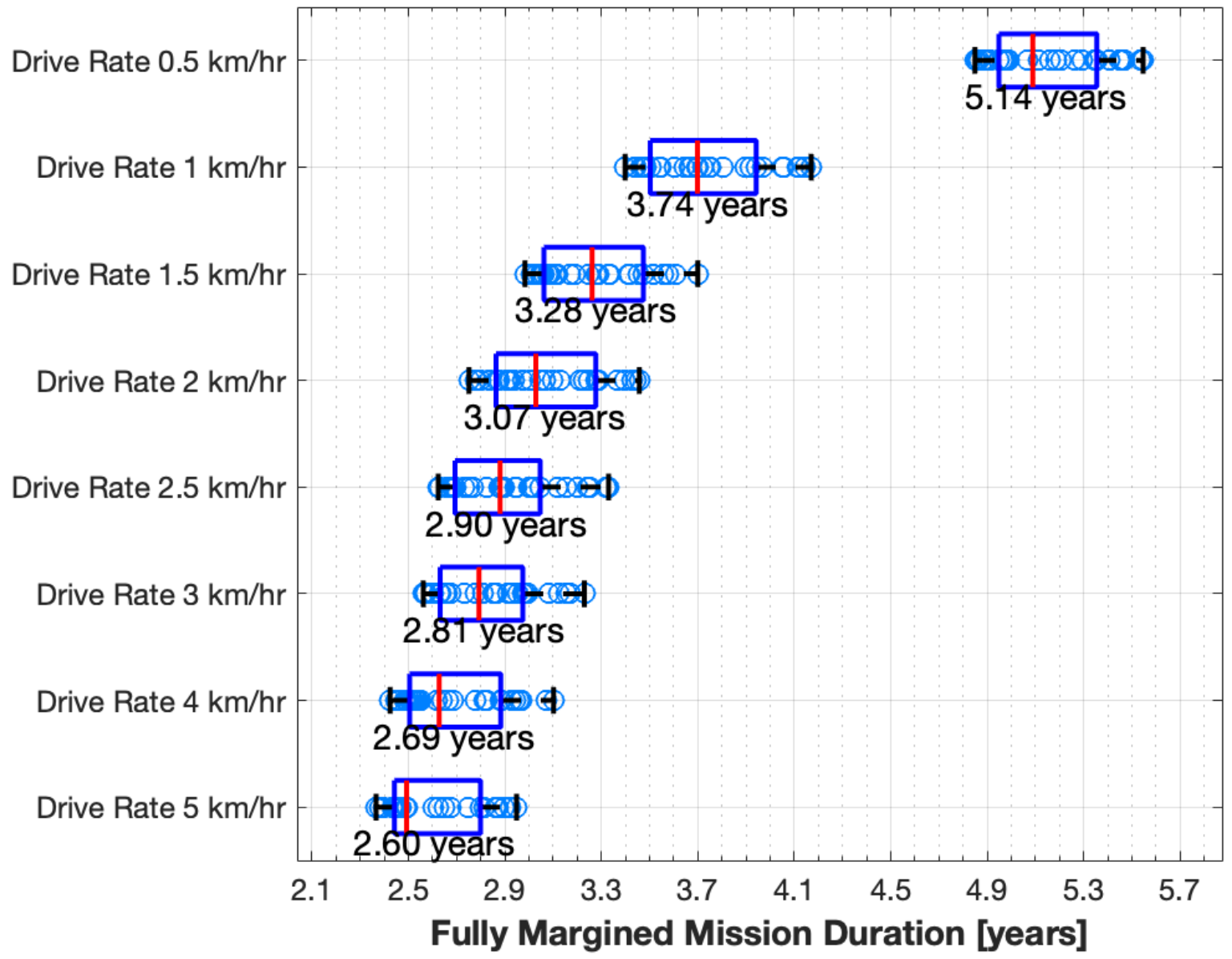

**Figure 6.44.** *MissionSim simulations for the fully margined mission duration of the SPARX Rover for different assumptions of the Rover's drive speed.*

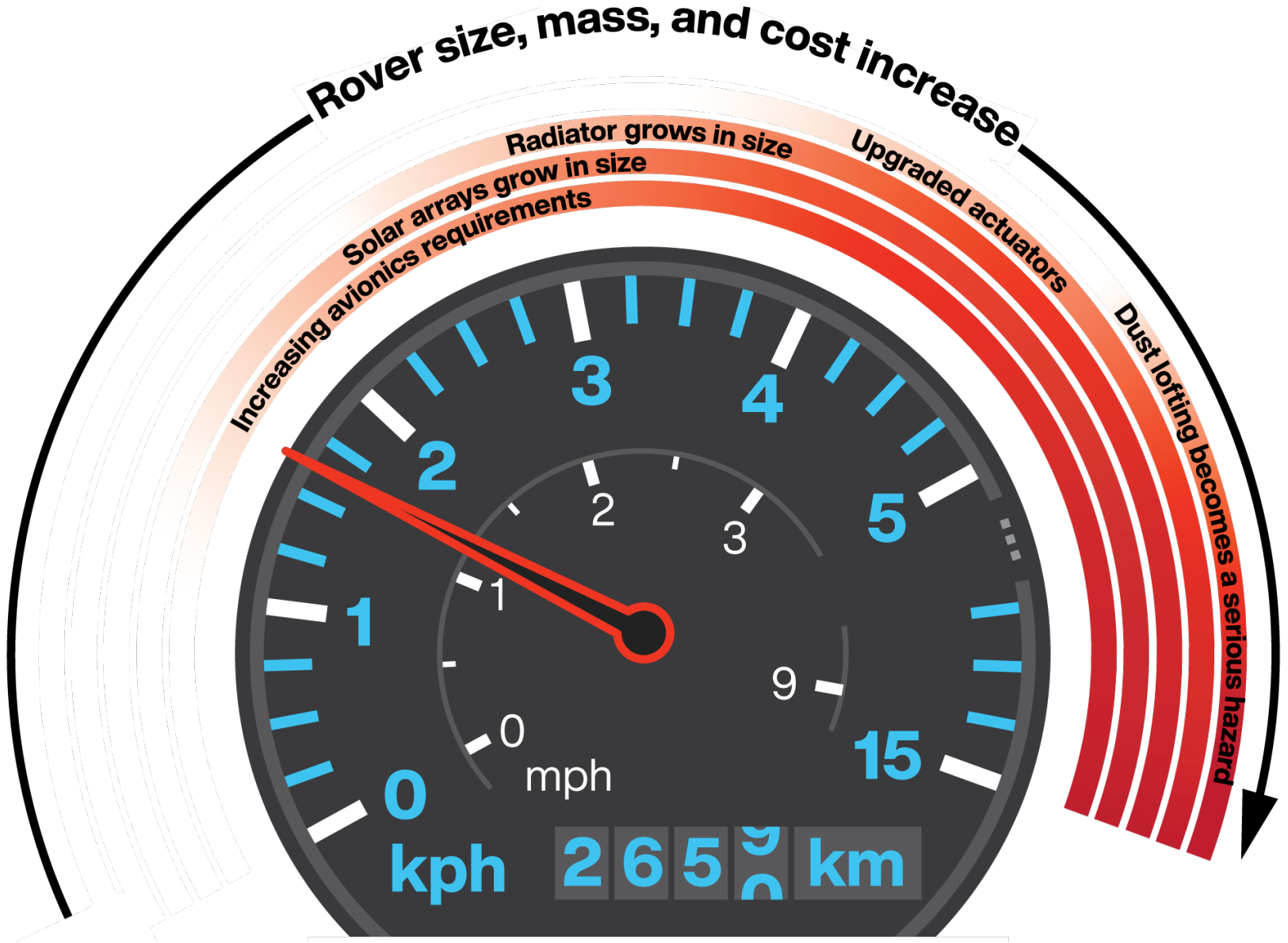

**Figure 6.45.** *Cartoon illustration of how increasing rover speed affects the rover design. For a long-range, autonomous rover it will be necessary to find the "sweet spot" in rover speed that balances the desire for shorter mission duration without dramatically increasing rover complexity (and cost).*

# 6.9 RISKS

This section addresses the risks associated with the SPARX Rover concept. While the baseline Rover concept provides a framework for interrogating risk, this section is intended to be independent of implementation choices.

## 6.9.1 Mission Risk Classification

NASA classifies missions by risk classification (see NASA Procedural Requirements NPR 8705.A, “Risk Classification for NASA Payloads”), summarized in Table 6.21. The mission risk classification sets the standards for designing a spacecraft, and in particular the approach to architectural redundancy and component selection. Class A and B missions often involve multiple-string designs and minimize the probability of a single-point failure. Class C and D missions often allow for more “tailored” redundancy approaches, utilizing commercial-off-the-shelf (COTS) hardware, and streamlining development. **Per NASA revised guidance to the SPARX SDT, the SPARX mission is categorized as Risk Class C (NPR 8705.4A), augmented with targeted resiliency and robustness enhancements—a posture referred to hereafter as “Class C+.”**

**Table 6.21.** *Summary comparing selected aspects of NASA risk classifications, based on NPR 8705.4A.*

| | Agency Priority | Cost | Risk Tolerance | Examples |
|---|---|---|---|---|
| Class A | Very High | High | Lowest | Flagship class missions<br>(e.g., Cassini, Europa Clipper) |
| Class B | High | High to Medium | Low | New Frontiers and Discovery class missions<br>(e.g., Juno, OSIRIS-REx) |
| Class C | Medium | Medium to Low | Moderate | Selected directed missions<br>(e.g., Lunar Reconnaissance Orbiter) |
| Class D | Low | Low | High | SIMPLEx missions<br>(e.g., Lunar Trailblazer) |

## 6.9.2 Programmatic Risks

The following section discusses a variety of programmatic risks for the SPARX baseline reference mission, and the mechanisms by which the technical team addressed these risks.

### 6.9.2.1 Mission Duration

Based on Monte Carlo simulations, the current best-estimate for the SPARX mission duration is less than 5 years (3.39 – 4.18 years, Section 6.8). The mission must find a balance between the deterministic risks of hardware longevity, the complex risks associated with validating reliable high-speed and autonomous performance, and programmatic and cost risks of significantly increased surface operations. The current design balances these risks to the best of the currently known ability. The technical team has identified several additional mitigations to ensure the reliability of the autonomy flight software, including: (1) conducting extensive field-testing complemented with a validated high-fidelity simulation using digital twins for terrain, lighting, hazard avoidance, localization errors and traverse rate performance; (2) including worst-case terrain types in rover mobility design and test plans; (3) increasing ground engagement in autonomous operations that have the least reliable performance; (4) include adequate margin and flexibility in mission plan to allow for anomaly resolution in operations and alternate route planning.

### Why is the SPARX baseline reference mission Risk Class C+ ?

**Class C+ is strategically aligned with NASA's current priorities.** A traditional Class A or B designation would necessitate exhaustive qualification and broad-spectrum redundancy, which may not align with NASA's current "Agile Exploration" model.

**Class C+ provides the right balance between risk, robustness and science return.** The Class C+ model allows for a tailored approach, striking the right balance between high-priority science and the fiscal realities of a multi-year robotic mission. By prioritizing redundancy and robustness options in targeted areas with the highest benefit for our specific vehicle design and objectives, the mission maximizes science return per dollar.

The Rover architecture establishes a moderate-risk baseline and strategically scales up robustness and resiliency where the unique lunar environment demands it. This bottom-up approach serves as a critical mitigation against cost and mass growth by prioritizing mission-specific requirements, leveraging know-how and experience while avoiding heritage traps.

The JPL technical team took this approach for the baseline concept to analyze every line item in every subsystem and add the right amount of block redundancy, functional redundancy and graceful degradation, as well as taking a tailored approach to parts qualification, qualification testing and appropriate margins. This addresses reliability, robustness and resilience as defined below:

- Reliability: consistent performance in nominal conditions.
- Robustness: capability of performing in variety of stressing scenarios.
- Resilience: ability to recover and adapt after failure.

**The Class C+ posture is supported by significant flight heritage.** Perhaps the most relevant examples are the Mars Exploration Rovers (MER), Spirit and Opportunity, which demonstrated that thoughtfully designed "single string" systems can achieve extraordinary longevity, far exceeding their primary mission requirements through operational flexibility and mechanical robustness.

**Robust, high-confidence missions have historically succeeded by prioritizing system-level resiliency over component-level redundancy.** Even within Class A flagship missions, such as Curiosity and Perseverance, mission designers have strategically utilized single-string motor controller systems to manage complexity, and vehicle mass and volume, while maintaining high reliability.

**These precedents confirm that a Class C+ architecture is viable and appropriate for this mission and can be leveraged to design a mission with high confidence for delivering samples to the Artemis Base Camp while remaining within the NASA resource constraints.**

Other programmatic factors outside the scope of the TOR that could substantially reduce the mission duration and associated risks include:

- Availability of an RPS: While the SPARX TOR initially included Radioisotopes Power Systems (RPS), such as Radioisotope Thermoelectric Generators (RTGs), NASA ultimately updated the guidance for SPARX to focus on solar-powered implementation options after funding was cut for future RTG generation in the FY26 President's budget. While an RTG power source has many pros and cons compared to solar power, one definite advantage is that an RTG power source can enable the Rover to operate and drive during the lunar night. An RTG-powered Rover would also be able to freely navigate through the deeply shadowed south polar region of the Moon and not require seasonal staging for the final rendezvous (Section 6.7). As shown in Section 6.8.2, and Figure 6.42 in particular, the solar-powered Rover spends half of the total mission duration in hibernation during lunar night. While the solar-powered Rover has a mission duration of ≤5 years, preliminary analyses suggest that an RTG-powered Rover could have a mission duration closer to ≤2 years.
- Availability of a non-polar Artemis rendezvous: The science return SPARX is maximized through partnership with Artemis; however, at this time, all planned crewed Artemis missions are targeting the lunar South Pole. If this constraint was removed, and the SPARX Rover could rendezvous with a crewed Artemis mission outside of the Artemis exploration zone, it could substantially reduce the total traverse length and mission duration. Avoiding the South Pole would also remove the need for a seasonally staging the final leg of the traverse and would simplify thermal design and power requirements. The science implications of this architecture are discussed further in Section 7.1.

#### 6.9.2.2 Sample Handover and Schedule Integration with Artemis

In the baseline mission design, sample return is carried out by Artemis astronauts utilizing Artemis infrastructure, e.g., the Human Landing System (HLS). In order to simplify risks associated with asynchronous mission timing and crew safety concerns, the technical team prepared a baseline reference architecture that delivers the sealed sample canisters to a designed “drop off” location, and the Rover then egresses from the immediate vicinity, allowing the crew to retrieve the samples at any time during their surface operations. The drop off could plausibly happen days or years before astronauts retrieve the samples. This approach also has the added benefits of freeing up the Rover to do other science and exploration tasks before astronauts arrive, and also removing any risks associated with astronauts directly interacting with the Rover.

#### 6.9.2.3 Telecommunications Relay Availability and Continuity

Since the SPARX Rover operates on the lunar farside with no direct communication with Earth, it is entirely dependent on lunar communications relay infrastructure. A single point failure in the relay link could result in a total loss of mission. The preferred mitigation for this is to assume and leverage the commercially provided Lunar Communications Relay and Navigation Systems (LCRNS) constellation of communication satellites. LCRNS is part of NASA’s Moon-to-Mars architecture, and current plans suggest that a 5-satellite network will be operational by the start of the SPARX Rover mission in the early 2030s. A 5-satellite network greatly exceeds the actual telecommunications requirements of the Rover, and as few as 1~3 satellites may be sufficient (as described in Section 6.5.7) although it may come at the cost of mission duration.

### 6.9.3 Environmental Risks

#### 6.9.3.1 Thermal Cycling and Night Survival

One of the largest environmental challenges for the Rover is the need to survive both the cold lunar nights, and hot lunar days, over a large latitude range (Section 6.5.5). The Rover must be operable both in the mid-latitudes with regular 14-day long days and 14-day long nights, and near the south pole where the definition of day and night become blurred by long shadows and seasonal affects.

The technical team adopted a "safety-by-design" principle, which ensures that the Rover can thermally survive faulting at any point during the traverse (Section 6.2). This is accomplished through battery size, and season-dependent traverse planning for the south polar region. Thus, while the thermal subsystem is a primary design driver, it is not a mission-limiting factor.

#### 6.9.3.2 Dust

Lunar dust is a well-known hazard for lunar missions. With the absence of terrestrial weathering processes, lunar dust can be extremely angular and abrasive. Dust may be lofted naturally, or by the actions of the Rover, other spacecraft, or astronauts. The unprecedented traverse distance exposes all external mechanisms to lunar dust that poses a mission longevity threat due to cumulative wear on joints, seals, and optical surfaces.

To mitigate the dust hazard, the Rover employs a defense-in-depth strategy against dust from landing, mobility, and the natural environment. Key baselined mitigations include:

- <u>Active systems:</u> Integration of Electrodynamic Dust Shields (EDS) and specialized anti-static optical coatings (described further in Section 6.11.1.2). EDS was recently demonstrated on Blue Ghost Mission 1.
- <u>Mechanical barriers:</u> Use of three-stage seals, leveraging Mars rover heritage and labyrinthine paths to eliminate direct line-of-sight dust ingress to sensitive internal volumes.
- <u>Performance margins:</u> System performance analysis incorporates a conservative degradation factor for solar arrays and radiators, ensuring mission success even with significant dust accumulation.
- <u>V&V:</u> The mobility system will undergo rigorous life cycle testing in high-fidelity regolith simulants to validate performance requirements.

#### 6.9.3.3 Power Resource Management

Unlike RTG-powered missions with a constant power floor, the solar-powered Rover faces variable energy availability due to seasonal sun angles and topographic shadowing. Unexpected terrain obstacles or "stalls" could lead to a critical battery depletion if the rover cannot reach an illuminated area before a lunar night or seasonal eclipse.

To mitigate the hazards associated with power resource management for a solar-powered rover, the architecture utilizes a Look-Ahead Energy Strategy. Using high-resolution Digital Elevation Models (DEMs) from the Lunar Reconnaissance Orbiter (LRO), the rover's mission planning software predicts the "energy cost" of the terrain in advance (described further in Section 6.7). The mission is baselined with a seasonally aware traverse plan, prioritizing high-speed travel during peak solar summers and utilizing a "corridor of sunlight" closer to the poles. Two options are considered in the architecture to balance drive speed and power generation needs: one is a design philosophy to always ensure rover is power positive while driving; the second is to energy balance over a 2 km driving period without the need for a charge mode. The rover autonomously manages its activities throughout the lunar day, considering sun elevation and latitude, such that it guarantees 100% state of charge by sunset to survive the lunar night. The system is designed to minimize survival power during the lunar night and is designed to survive up to 520 hours of lunar night. This ensures that even if the rover were to stall at the pole due to an unforeseen navigational challenge, it can survive the night with enough charge to resume operations in the lunar day.

## 6.10 COST & SCHEDULE

The SPARX TOR provided a notional Phase A-E cost cap of $1.5B (FY25), exclusive of lunar delivery service and sample return costs, as a guideline for evaluating the feasibility of the recommended baseline SPARX concept. The technical team produced the cost and schedule estimate described in this section. This estimate leverages extensive experience in past robotic surface exploration systems and sample collection/caching technologies, including prior mission concepts (such as Endurance-A).

The cost estimate here is focused on the cost of the custom-built SPARX Rover, including its development (design, build, assembly, test), integration of the science payload, and operations from launch through the sample drop off at the Artemis rendezvous point. In concordance with NASA, several components were deemed outside of scope and not included in the cost estimate: the launch vehicle, lunar delivery and landing system (whether accomplished by CLPS or otherwise), the cost associated with sample retrieval by Artemis astronauts and return to Earth, or the cost associated with Earth-based curation and analyses of samples. Furthermore, per the TOR, payload elements were assigned a pass-through cost of $25M (FY25) for each element, and individual design, development, or procurement costs were not considered beyond this. The cost estimate also includes Phase A-E support for a science team consisting of ~25 scientists (Section 7), but not inclusive of payload-specific science team members.

#### 6.10.1.1 Development

Figure 6.46 shows the full Phase A-D development lifecycle, spanning a total of 65 months (~5.4 years) from the start of Phase A through the transition to Phase E. Assuming a Phase A start in October 2026, the Rover could be ready for launch in 2031, contingent on the availability of the launch vehicle and lunar delivery system. Opportunities exist to compress the development schedule through targeted implementation strategies or adjustments to resource levels and cost profile, which could be evaluated in future studies. No schedule drivers or long-lead procurements have been identified that extend beyond the proposed development timeline.

#### 6.10.1.2 Operations

The Operations, Phase E, timeline lasts approximately 63 months, consisting of:

- Launch, cruise, and landing: ~2 months. The exact duration between launch and landing will depend on the chosen launch vehicle, design and operations of the lunar delivery system, and the design of the trans-lunar trajectory—all of which is beyond the scope of this report. ~2 months is a bounding case for a high-efficiency lunar transfer.
- Commissioning: ~1 month. Commissioning will occur after the Rover is placed on the surface and likely occur in the near vicinity of the lander.
- Primary mission: ≤60 months (5 years). The total duration of 60 months is used as a bounding number and covers the time from the start of the science traverse up until the Rover drops the sample containers at the Artemis rendezvous point. This estimate is based on numerical simulation (Section 6.8) and includes all associated operational margins and an unencumbered operational system margin of 12 months.

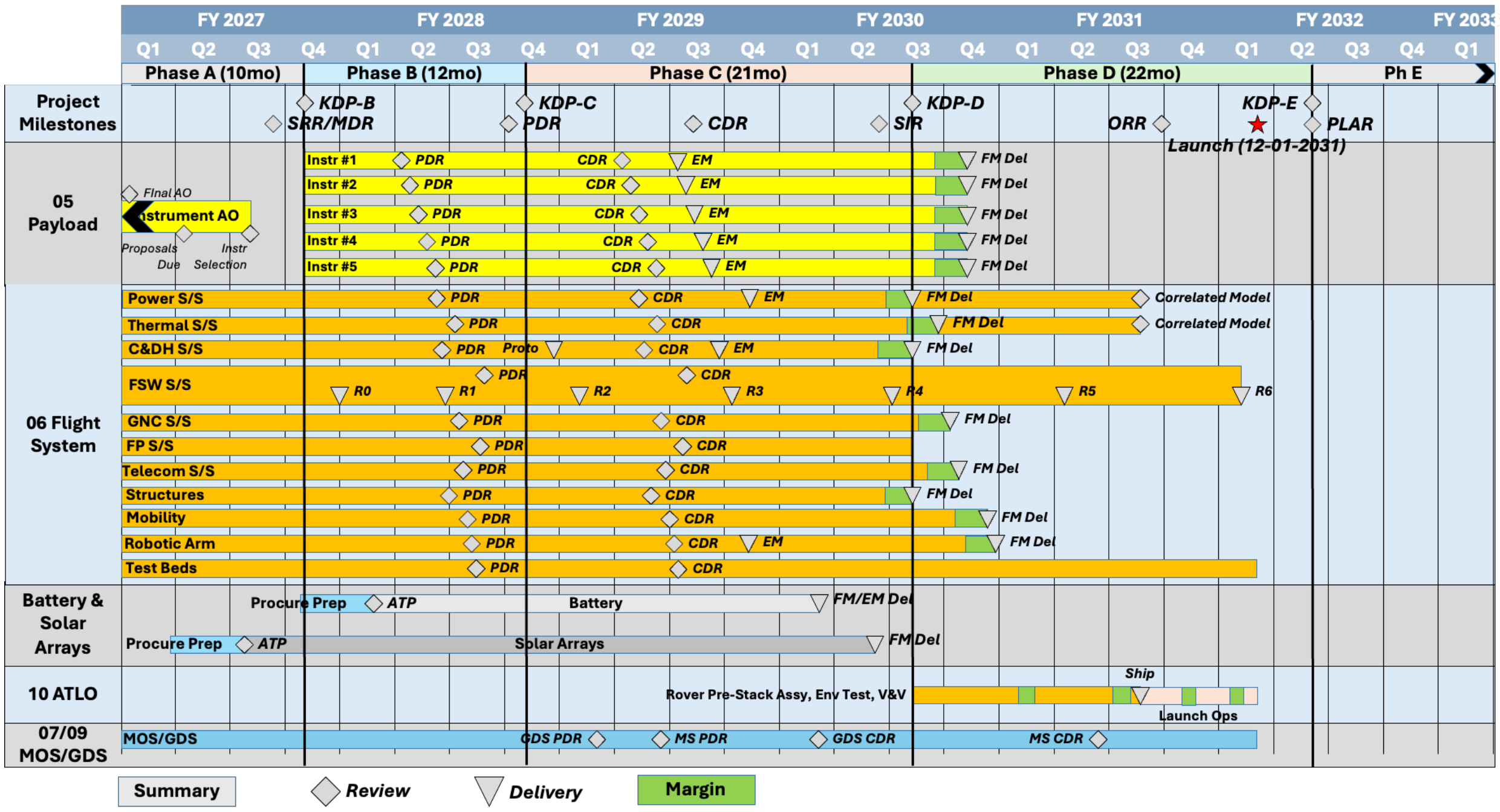

**Figure 6.46.** *High-level mission schedule for the development portion of the SPARX Rover (Phase A-D), assuming a start in October 2026 and no programmatic constraints (e.g., funding profiles, launch windows, launch/lander availability). This cost and schedule information is of a budgetary and planning nature and is intended for informational purposes only; it does not constitute a commitment on the part of JPL and/or Caltech.*

## What is the SPARX Rover basis of cost estimate in a nutshell?

**The SPARX Rover cost estimate uses a hybrid methodology, applying the most appropriate estimating approach to each component and phase of the mission.** Hardware subsystems are estimated using mass-based cost factors derived from cost actuals from analogous rover missions, such as Curiosity and Perseverance. Non hardware components are estimated using wrap factors based on the same analogous missions. Management, systems engineering, and other level-of-effort activities are estimated using a schedule-aligned workforce plan informed by rover development and operations experience tailored to the SPARX Rover concept. Some components that are commercially available, such as batteries, tires, and wheels, are estimated using discrete engineering vendor cost estimates.

**Why is the SPARX Rover cost estimate credible?**

**The SPARX Rover lifecycle cost estimate is based on a fully traceable, high fidelity, analogy-driven model in which each cost element is linked to a defined estimation method and explicitly parameterized inputs.**

The SPARX Rover estimate draws directly on development-cycle cost actuals from the Curiosity (Mars Science Laboratory) and Perseverance (Mars 2020) rover missions, which represent the closest analogs for robotic surface exploration systems. Subject matter experts familiar with Curiosity, Perseverance, and the SPARX Rover design assessed the relative complexity of each subsystem to select the most appropriate cost analog.

It SPARX Rover cost estimate represents a full lifecycle estimate, including reserves applied in accordance with standard flight project practices. The estimate accounts for mission risk classification, advances in enabling technologies, and efficiencies enabled by modern development practices.

Together, these elements provide a transparent and empirically grounded basis of estimate, anchored in demonstrated rover mission development and operations experience, while accounting for the architectural and operational characteristics unique to the SPARX Rover concept. This provides a stronger basis of estimate than typically available at this stage of mission formulation.

### 6.10.2 Cost Estimate

The cost estimate (in FY25$M) by FY for the development and operation of the SPARX Rover is $840M shown in Table 6.22. The lifecycle cost estimate by fiscal year (FY) is shown in Figure 6.46. The estimate includes reserves appropriate for this stage of formulation.

**Table 6.22.** *Summary of the cost estimate of the SPARX Rover development and operations. This cost information is of a budgetary and planning nature and is intended for informational purposes only; it does not constitute a commitment on the part of JPL and/or Caltech.*

| | Cost (FY25 $M) |
|---|---|
| **Rover** | $543M |
| **Instruments** | $137M |
| **Operations** | $160M |
| **Total** | $840M |

This cost estimate does not account for potential contributions from international partners that may reduce the overall cost to NASA. There are many possible avenues for contributions, including mission elements, subsystems, or instruments. Preliminary discussion suggests that compelling Rover contributions may include the robotic arm, telecommunications subsystem (Ka-band antenna, gimbals, radios), power electronics, guidance and control hardware (star scanners, IMUs), and Rover wheels. Rough estimates suggest that leveraging contributions could reduce the total cost to NASA by ~$220M. Additional contributions, including instruments, elements of the command and data handling subsystem, motor controllers, actuators, batteries, and solar arrays could further reduce the cost. The mission architecture incorporates design features—including well-defined subsystem interfaces and functional modularity—to facilitate the integration of externally provided elements should such partnerships be pursued.

**SPARX Rover Cost Estimate**

| | FY25 (FY25$M) | FY26 (FY25$M) | FY27 (FY25$M) | FY28 (FY25$M) | FY29 (FY25$M) | FY30 (FY25$M) | FY31 (FY25$M) | FY32 (FY25$M) | FY33 (FY25$M) | FY34 (FY25$M) | FY35 (FY25$M) | FY36 (FY25$M) | FY37 (FY25$M) | Grand Total (FY25$M) |
|---|---|---|---|---|---|---|---|---|---|---|---|---|---|---|
| Rover - Phase A->Launch | - | - | 28.9 | 57.2 | 100.3 | 124.5 | 106.2 | 11.6 | - | - | - | - | - | 428.8 |
| Reserves | - | - | 3.7 | 15.2 | 27.5 | 34.4 | 30.5 | 3.3 | - | - | - | - | - | 114.7 |
| Subtotal | - | - | 32.6 | 72.5 | 127.9 | 158.9 | 136.7 | 14.9 | - | - | - | - | - | 543.4 |
| Science Instruments - Phase A->Launch | - | - | 4.9 | 19.4 | 27.9 | 30.1 | 25.8 | 3.4 | - | - | - | - | - | 111.6 |
| Reserves (Project held) | - | - | 1.1 | 4.4 | 6.3 | 6.7 | 5.8 | 0.8 | - | - | - | - | - | 25.0 |
| Subtotal | - | - | 6.0 | 23.8 | 34.2 | 36.9 | 31.6 | 4.2 | - | - | - | - | - | 136.6 |
| Operations - Launch->EOM | - | - | - | - | - | - | - | 32.0 | 24.9 | 24.4 | 24.2 | 24.0 | 10.0 | 139.6 |
| Reserves | - | - | - | - | - | - | - | 4.7 | 3.6 | 3.5 | 3.5 | 3.5 | 1.5 | 20.3 |
| Subtotal | - | - | - | - | - | - | - | 36.7 | 28.5 | 28.0 | 27.8 | 27.5 | 11.4 | 159.9 |
| GRAND TOTALS | | | | | | | | | | | | | | |
| GRAND TOTAL (wo/Reserves) | - | - | 33.8 | 76.7 | 128.3 | 154.6 | 132.0 | 47.1 | 24.9 | 24.4 | 24.2 | 24.0 | 10.0 | 680.0 |
| RESERVES | - | - | 4.8 | 19.6 | 33.8 | 41.2 | 36.3 | 8.7 | 3.6 | 3.5 | 3.5 | 3.5 | 1.5 | 159.9 |
| GRAND TOTAL | - | - | 38.6 | 96.3 | 162.1 | 195.8 | 168.3 | 55.8 | 28.5 | 28.0 | 27.8 | 27.5 | 11.4 | 839.9 |

**Figure 6.47.** *Total cost for development and operations of the SPARX Rover by mission phase. This cost information is of a budgetary and planning nature and is intended for informational purposes only; it does not constitute a commitment on the part of JPL and/or Caltech.*

## 6.11 TECHNOLOGY DEVELOPMENT

**Although ambitious, the SPARX baseline mission, benefits from decades of developments in planetary exploration, and does not require the use or development of any currently non-existent technologies.** As is the case with any new mission concept, the SPARX rover will apply specific engineering developments and the application of key technologies currently in development.

### 6.11.1 Enabling Technologies

The following listed technologies that are enabling for the SPARX Rover mission, meaning that they are necessary to accomplish the mission. While these technologies are currently at a range of maturity levels, there is substantial development (and even flight demonstrations) ongoing today, that provides confidence that these technologies will be ready for the SPARX mission.

#### 6.11.1.1 Long-Distance Lunar Traverse Capabilities

The SPARX Rover mission requires a traverse capability beyond what has been demonstrated by previous planetary missions. While components of this technology have been developed for previous rovers and terrestrial applications, there is still work needed to mature and adapt these technologies to the lunar environment, especially for a long distance, long duration mission.

The most recent Mars rover, Perseverance, has been operating for over 5 years and has completed ~45 km to date without any mobility anomalies affecting long-term drive performance; however, the life expectancy of the Perseverance rover hardware is estimated to only be around 100 km. Therefore, hardware design and reliability testing for the ~2,800 km SPARX baseline traverse (Sections 5.8 and 6.7) is a significant driver of mission technology.

Additional technologies that are existing but require further advancement prior to utilization in a SPARX mission are algorithms for onboard perception enabling autonomous driving, and avionics component selection and thermal design.

Perhaps the most critical capability to advance prior to a SPARX mission is the reduction in projected rover fault rate. The technical team identified modest technology developments in mobility hardware, avionics, global navigation, and system health management, and implemented them as part of the SPARX baseline reference mission design. Developing a sufficiently robust system further requires extensive testing in a relevant environment, and the SPARX project will need to leverage high fidelity simulations combined with physical testing of flight relevant rover hardware to validate the fault rate requirement.

#### 6.11.1.2 Electrodynamic Dust Shield

Given the pervasiveness of lunar dust and its propensity for covering, adhering, and damaging surfaces, the study team recommends utilizing electrodynamic dust shield (EDS) technology for sensitive surfaces. Thermal radiators are of particular concern, as dust coverage can negatively impact their heat rejection ability. EDS uses an electrode wire mesh below the radiator paint layer to generate an electric field that drives accumulated dust off the radiator surface. EDS has undergone extensive research at NASA Kennedy Space Center, and a small-scale demonstration was recently successfully demonstrated on the Moon by Firefly Aerospace's Blue Ghost Mission 1 (Figure 6.48). However, there is still development work to go to implement the dust mitigation technology for long-term operation in the lunar environment.

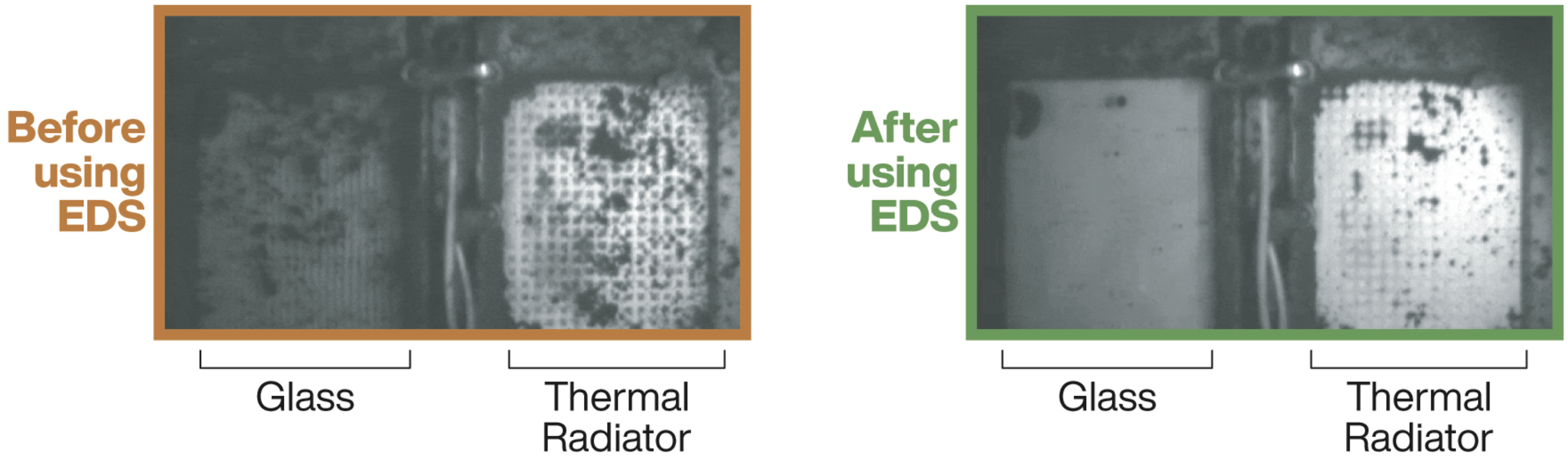

**Figure 6.48.** *Preliminary results from the Electrodynamic Dust Shield (EDS) demonstration on Firefly Aerospace Blue Ghost Mission 1. Top image: lunar dust is seen covering test surfaces made of glass and thermal radiator material in the EDS experiment, before EDS is turned on. Bottom image: after operating the EDS, much of the dust has been cleared from these sensitive surfaces. Image credit: NASA.*

## 6.11.2 Potentially Enhancing Technologies

The following listed technologies represent a non-exhaustive list of technologies that could enhance the SPARX mission but are not required for accomplishing the science objectives or meeting the baseline reference mission design.

### 6.11.2.1 Batteries

One of the largest design drivers for a large solar powered rover is the size and mass of the battery. The current baseline over design includes a ~330 kg battery. While the baseline Rover battery uses commercial off-the-shelf 18650 Li-ion cells, alternative battery solutions could be considered for use. Pouch cells (Figure 6.49) offer high specific energy due to their low mass and would reduce the overall battery mass by about 30%, i.e., near 100 kg. Decreasing the rover mass reduces the overall scale of the rover, requirements on the mobility system, etc. Additionally, pouch cells can operate at a lower temperature than the commercial Li-ion chemistry, simplifying the battery's thermal accommodation and associated required electrical heating.

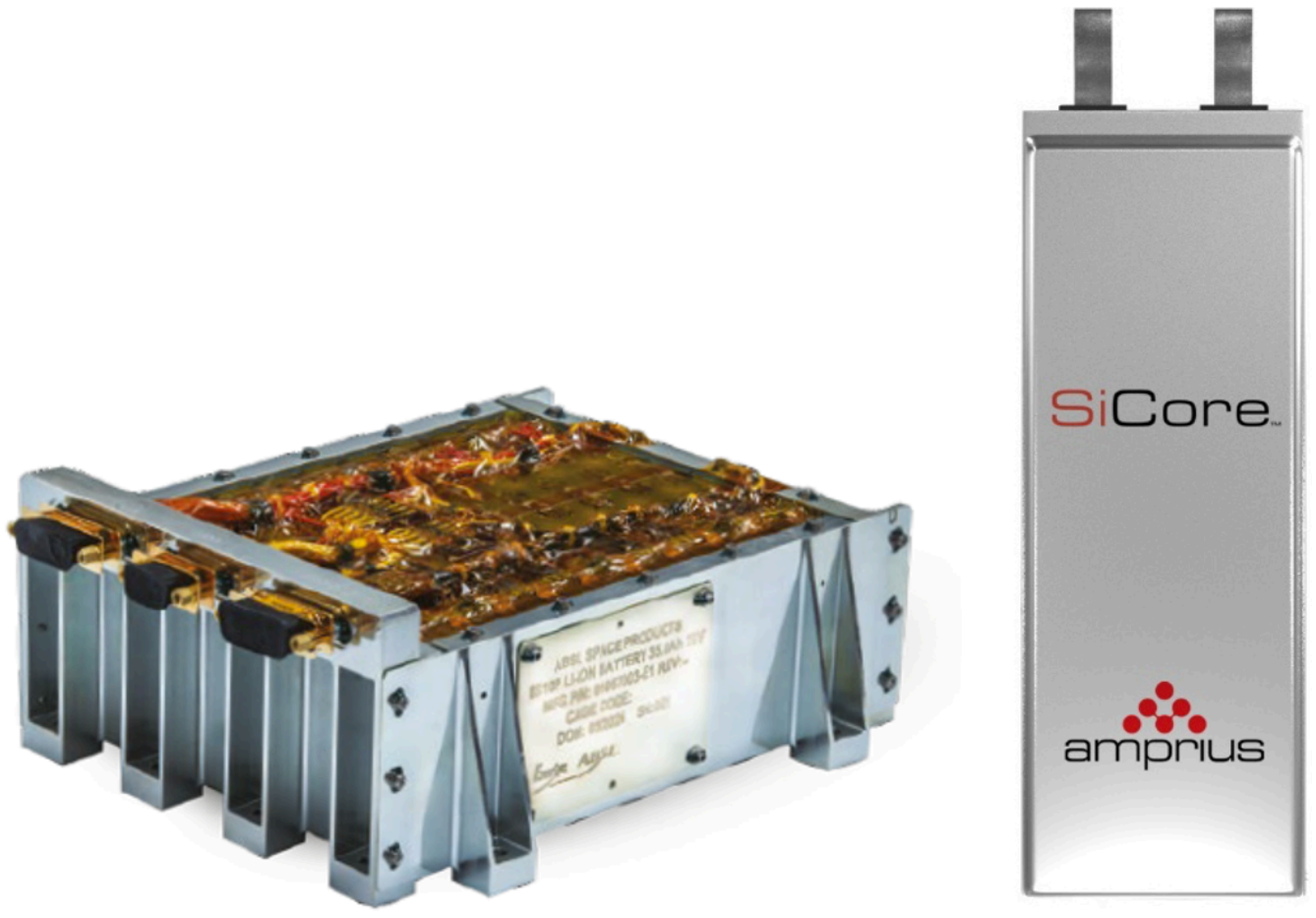

**Figure 6.49.** *18650 cylindrical cell based ABSL battery (from Enersys) compared to pouch cell battery (from Amprius).*

### 6.11.2.2 LIDAR

At present, the Rover's perception system is done via stereo vision (Section 6.5.8). LIDARs provide an attractive alternative, as they are less prone to variations in lunar lighting conditions, and other perception issues (such as the "opposition effect" whereby surface features and hazards can be hidden when the Sun is directly behind the camera). However, state-of-the-art

LIDARs has not achieved a level of maturity necessary for baselining them in cost-constrained lunar surface missions. The most relevant LIDARs may have large power draw, limited vertical resolution, and performance limitations or variations under thermal vacuum conditions. The longevity of the laser would have to be demonstrated for the mission duration.

An alternative approach could be solid-state Frequency Modulation Continuous Wave (FMCW) LIDARs, although these are years away. FMCW LIDARS use frequency modulation to measure range of individual points in 3D space, rather than time-of-flight responses. FMCW allows for longer ranges (several kilometers), use less power, and provide direct velocity measurements (e.g., rover speed relative to terrain) in addition to range and reflection intensity. However, FMCW requires more computation power than time-of-flight LIDARs, and more importantly, FMCW is a newer and less mature technology that would require technology development.

LIDARs with moving parts would have to also be tested to ensure that their components could handle launch vibration loads.

#### 6.11.2.3 Science Autonomy

Onboard science autonomy describes the ability of a spacecraft to autonomously analyze raw scientific data, make real-time decisions based on those analyses, and summarize or prioritize data for transmission back to Earth. Science autonomy has been infused in active missions to varying degrees. For example, the Perseverance rover on Mars utilizes the AEGIS (Autonomous Exploration for Gathering Increased Science) system to analyze rover camera images, identify and prioritize targets based on science criteria (e.g., rock size, shape, brightness), and then acquire follow-up, high-resolution data using other instruments including LIBS and other spectroscopic data (Verma et al. 2023). The Decadal Survey identified science autonomy as a key technology area needing development in the coming decade and beyond.

In this section, we describe several potential use cases for science autonomy. Some of these could leverage existing technologies (like AEGIS), whereas others may require new developments.

##### Science Autonomy Use Case: Autonomous Detection of Unknown, Anomalous Material and Enabling Opportunistic Sampling

Although the SPARX baseline traverse has been planned with extensive use of remote sensing observations, decades of experience with planetary surface exploration suggests that there will be many scientific surprises. This is well demonstrated by Apollo 17, where astronauts stumbled upon orange-colored regolith excavated by their rover and while walking on a crater rim (Figure 6.50). The collection of these unexpected, opportunistic samples resulted in a number of important discoveries. The SDT recognizes the potential scientific importance of such “science of opportunity” activities but elected not to prescribe ConOps activities related to their operation. Increased science autonomy is one potential avenue for increasing opportunistic science options without significantly burdening the established ConOps.

How science autonomy could help: With the wealth of data provided by the frequent Science Interval Stops, tools onboard the rover could autonomously analyze that data looking for signatures of unexpected materials. These signatures could either be pre-defined (e.g., some catalog of anomalous material that might be found on the Moon), or by the rover building up its own catalog of what the Moon “typically looks like” and identifying anomalies. In an extreme case, this process could trigger collection of samples of opportunity. While opportunistic samples are not part of the current baseline, they may substantially increase the science value of the mission.

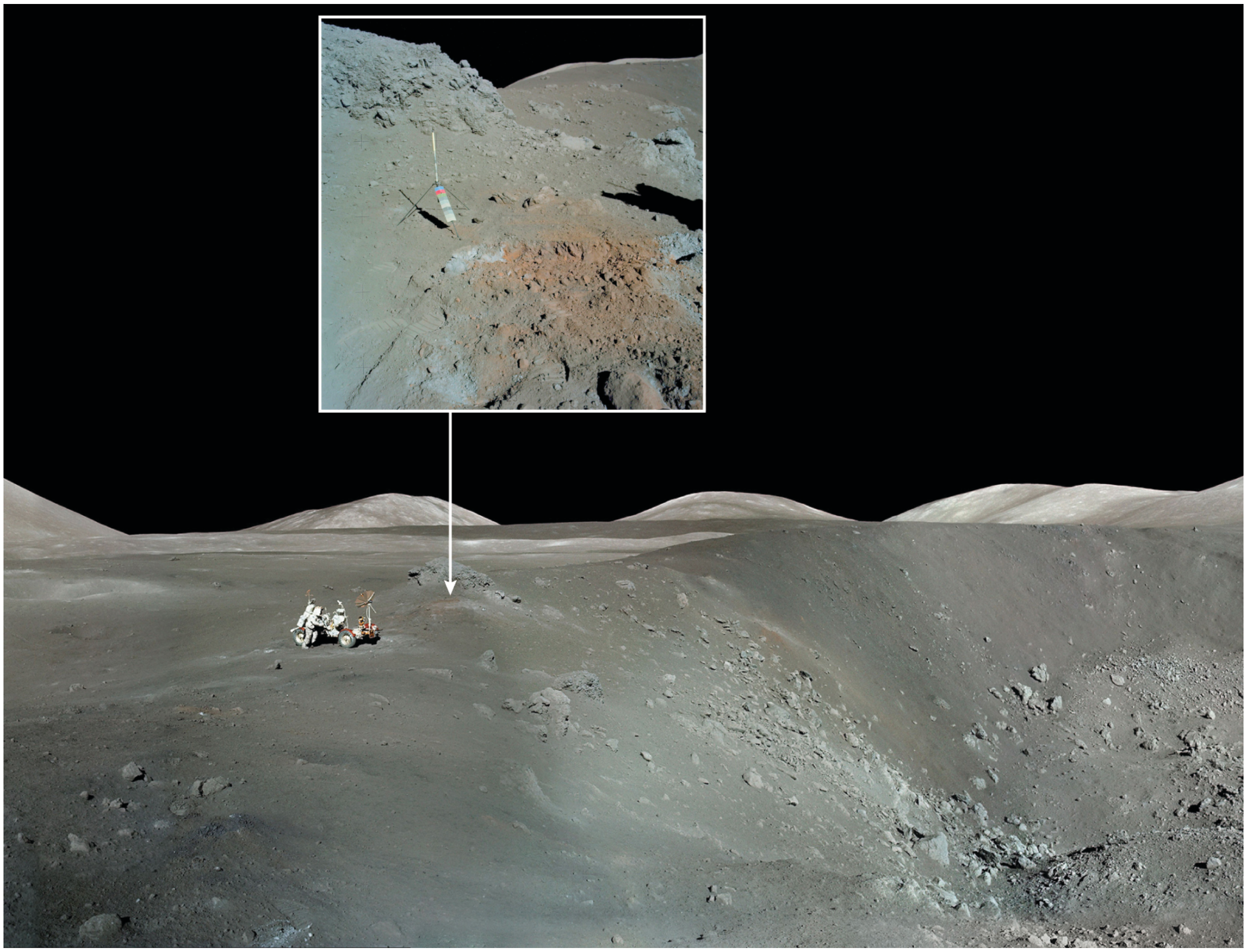

**Figure 6.50.** *The Apollo 17 orange soil is an excellent example of an unplanned scientific discovery made when exploring the Moon. Top image: panorama showing Harrison Schmitt near the Apollo Lunar Roving Vehicle at the rim of Shorty Crater on 12 December 1972. Around the crater rim, orange soils can be seen. A white box shows the approximate location of the close-up bottom image. Background image by Gene Cernan, edited by Tom Dahl. Inset image: color enhanced close-up image of the orange soil. Image AS17-137-20990*

The SDT recognizes this modality as one of the most powerful potential applications of increased science autonomy capabilities.

### Science Autonomy Use Case: Faster and Better Science Instrument Operations

Many science observations require careful preparation and iterative operations to achieve science requirements. For example, constructing an imaging mosaic requires some understanding of illumination conditions, rover orientations, distances to relative terrains. Some geochemical and geophysical measurements may require different integration times depending on the nature of the targeted material—which can only be determined after starting the observation. Traditionally, this is done either with ground-in-the-loop planning, or some time-intensive rover operations.

<u>How science autonomy could help:</u> If the Rover was able to analyze its own scientific data, or ingest other engineering data (e.g., rover orientation, navigation-derived data, sun-location, etc.) it may be able to expeditiously acquire scientific measurements. For example, consider a scenario where the Rover utilizes its GNC subsystem to accurately determine its orientation and direction of the Sun, and then data derived from the navigational camera—such as a DEM of the surrounding environment, or preliminary map of brightness in different directions—to preplan a time-efficient image mosaic sequence where it already has good estimates for integration times and focus parameters.

### Science Autonomy Use Case: Optimizing the Location of Science Interval Stops

In the current baseline, the Rover stops every 2 km for an interval science stop where it utilizes its full set of in situ instruments (Sections 4.2 and 6.6). For the purposes of defining a baseline reference mission, this stop happens when the rover odometer hits 2 km, somewhat independent of where the rover is located/positioned. However, the SDT recognizes that there is significant benefit to allowing flexibility within this stop cadence to maximize stopping at scientifically valuable locations.

<u>How science autonomy could help:</u> The 2 km requirement is flexible; therefore, the rover could utilize its engineering instruments (e.g., NavCams) or science instruments (e.g., cameras and spectrometer) to preferentially stop closer to features of interest (e.g., a boulder, crater, hill-top, etc.). Since some of these features are hazards, the rover may already be identifying them. The rover could also make small detours from the “straight line path” that it otherwise drives to optimize science. Any path deviations would, of course, be limited by flight rules regarding mission duration and expediency.

### Science Autonomy Use Case: Reduce Data Volumes while still Maximizing Science Return to Earth

Many of the current strawman instruments, like the cameras and spectrometers, have the potential to produce very large datasets. Transmitting these datasets in their entirety back to Earth can require substantial time, which may be limited due to the availability of relay satellites.

<u>How science autonomy could help:</u> It may be possible for the Rover to perform some initial data analysis onboard to reduce large data volume products (e.g., hyperspectral data cubes) into smaller data volume products (e.g., a map of mineralogy). Similarly, the Rover could identify the most interesting aspects of a large dataset and prioritize return of that element. Relatedly, depending on the communications infrastructure of the Moon, it may be advantageous for the Rover to have some sort of autonomous communication scheduling capability to return as much data as possible through the various satellite services.

### Science Autonomy Use Case: Getting the Most out of Challenging Phases of the Mission, including Dusk, Dawn, and Night

In the current baseline mission, there are time periods where rover operations are substantially curtailed. At dawn, the Rover requires some finite amount of time (~24 hours) to warm up before it can drive. At dusk, the illumination will eventually become sufficiently poor that the Rover will stop driving forward. At night, the Rover will have extremely limited power, currently all reserved for survival.

<u>How science autonomy could help:</u> In time periods where Rover operations are limited, an adaptive Rover planning system that balances Rover operations with power and other constraints, could also find time for inserting opportunistic science observations. Examples of such opportunistic science observations could include: (1) observations of dust near the horizon at dusk/dawn; (2) observations looking for surface frosts at dawn before sunlight hits the surface; (3) observations at dusk/dawn/night monitoring for impact flashes; (4) continuous measurements from the GRNS, magnetometer, and IMUs for longer integrations and improved signal to noise for geochemical, magnetic field, and gravity field determination; (5) monitoring of the IMU accelerometers for potential seismic signatures; (6) observations of the solar corona while the Sun is occulted by topography.

### Additional Considerations for the use of Science Autonomy

While science autonomy is undeniably exciting, there are at least three cross-cutting factors that must also be considered:

- Science autonomy is not a substitute for a well-integrated science team that has ground-in-the-loop interactions with the rover for key activities (e.g., sample collection).
- While autonomy technologies can be developed on the Earth, a sufficiently rigorous validation approach must be undertaken to show that any autonomous system deployed on the Moon is capable of achieving the motivating science requirements reliably and repeatedly. One potential approach may be to begin the mission with little science autonomy and gradually let more science autonomy be incorporated as those systems are proven out on the surface of the Moon. Such an approach is nicely aligned with some of the constraints of the current baseline mission, where the mission becomes more operationally and environmentally challenging to execute towards the end of the mission (due to power, thermal, and illumination conditions near the pole).
- The inclusion of science autonomy may be a significant driver for the computation requirements of the rover, and designs should balance requirements for the primary flight computer system with any additional requirements, or requirements from including an additional secondary computer.

### 6.11.2.4 Radioisotope Heating Systems

Radioisotope Heating Units (RHUs) are a common approach to providing thermal heat for spacecraft missions. Given the thermal challenges associated with the SPARX Rover mission (Section 6.5.5), RHUs warrant special consideration. Historically, DOE has produced Radioisotope Power Systems (RPS) and RHUs but in recent years there has been a pivot towards commercial RPS and RHU production and acquisition models. In this section we describe the current state-of-the-art for both government-furnished and commercial RHUs. While the former are developed (and thus are not a new technology), the latter are still in development.

A fundamental difference between government-furnished and commercially developed RHUs (and RPS) is the fuel source. Government-furnished RHUs have traditionally utilized plutonium ($^{238}$Pu), while commercial RHU providers propose to utilize americium ($^{241}$Am). Table 6.22 and Figure 6.51 shows some of the key differences between $^{238}$Pu and $^{241}$Am systems.

There are two key differences between plutonium and americium fueled RHU and RPS systems: (1) Americium fuels have a lower heat density than plutonium fuels, limiting the thermal conversion efficiency. This results in larger and heavier heating and power systems that produce less heat and power. (2) Americium fuels have a longer half-life, which enables long duration storage with minimal performance degradation compared to Plutonium fuel sources.

**Table 6.23.** *Key differences between plutonium and americium fuel sources for radioisotope power and heating units.*

| | Half-life | Decay mode | Heat density |
|---|---|---|---|
| Plutonium ($^{238}$Pu) | 87.7 years | α-decay | 220 cm$^3$ / kW |
| Americium ($^{241}$Am) | 432.2 years | α-decay | 865 cm$^3$ / kW |

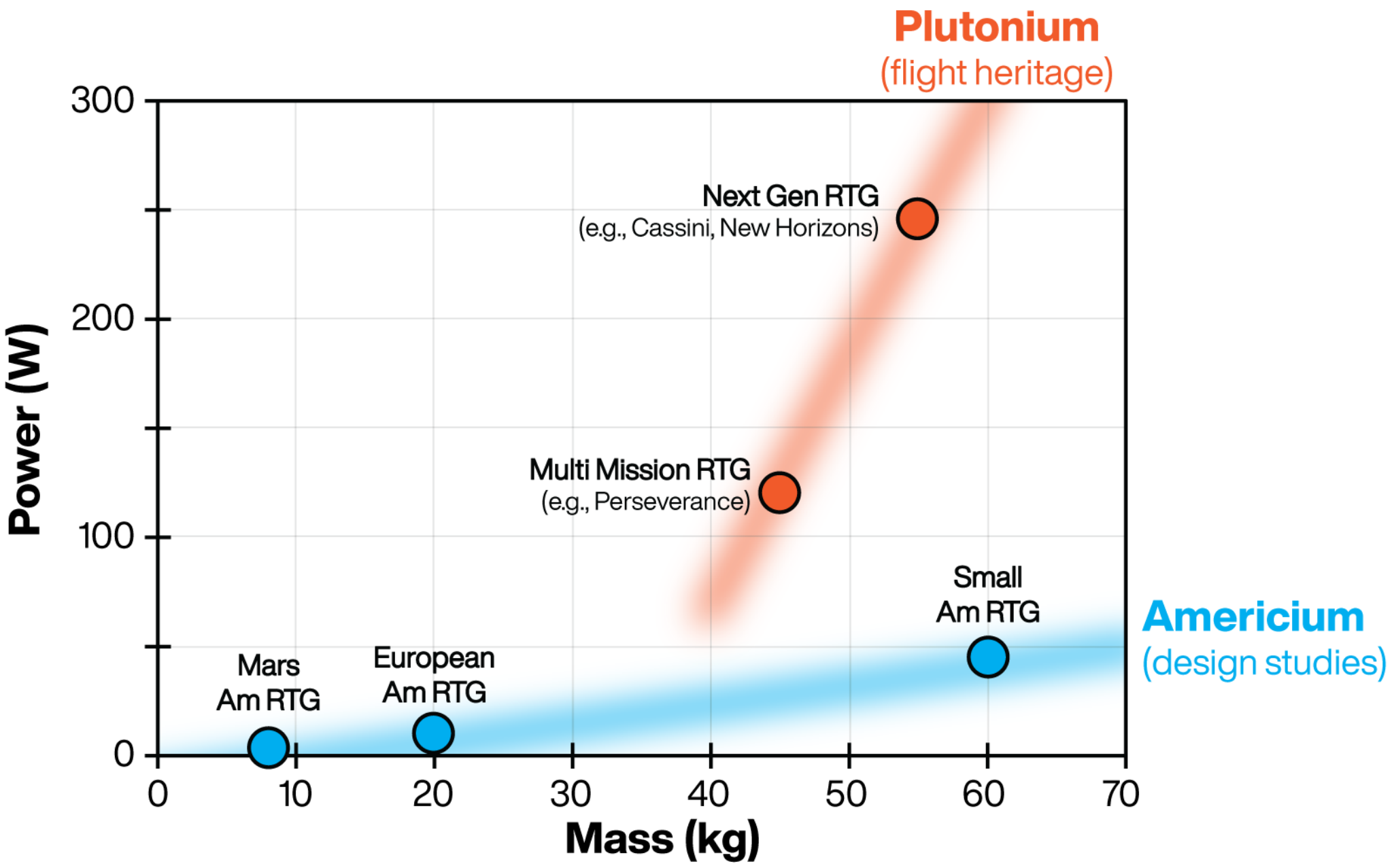

**Figure 6.51.** *Comparison of the power density of plutonium and americium fuel sources.*

### Plutonium RHUs

For decades the US Department of Energy (DOE) has provided plutonium-based RHUs to NASA to provide heat to spacecraft and their payloads, ensuring proper functioning in cold environments. Plutonium RHUs are not new technology; but we describe them here for context. Plutonium RHUs were first used on the Apollo 11 and later on Pioneer and Voyager missions. A redesigned, lightweight version has been in use since the Galileo and Cassini missions. Three Mars rovers: Sojourner, Spirit, and Opportunity use LWRHUs. Dragonfly will be the next NASA mission to use DOE-provided plutonium-based RHUs.

The current standard DOE provided RHU is the Light-Weight Radioisotope Heater Unit (LWRHU). The LWRHU generates heat through a fuel pellet containing 2.66 g of $^{238}PuO_2$, the same isotope used in radioisotope power sources like the Multi-Mission Radioisotope Thermoelectric Generator (MMRTG) used on missions like Perseverance and Dragonfly. A single LWRHU has a mass of 42 grams and a nominal thermal output of 1.1 W at the beginning of the mission (although NASA missions have frequently used LWRHUs with lower heat outputs). The device is designed to survive reentry through the atmosphere and withstand maximum loads of 425 G. Its mass is 42 grams.

Missions are responsible for developing structures that enable integration of LWRHUs at the Kennedy Space Center (KSC). To facilitate this, the US DOE and Idaho National Lab ship LWRHUs to KSC 4 to 6 months before launch. The Payload Hazardous Servicing Facility (PHSF) is used for spacecraft integration. To ensure mission success and safety, DOE and Flight System Engineers jointly develop hardware and integration procedures.

With their radioisotope components, missions using any RHU are classified as nuclear missions. Therefore, they go through the same Launch Safety Analysis and Launch Authorization process as missions operating radioisotope power generation systems. Due to funding for radioisotope power systems being removed from the President's budget during the time of this SDT and the decision for DOE to cease plutonium fuel production in the near future, NASA directed the SPARX team to assess solar powered options.

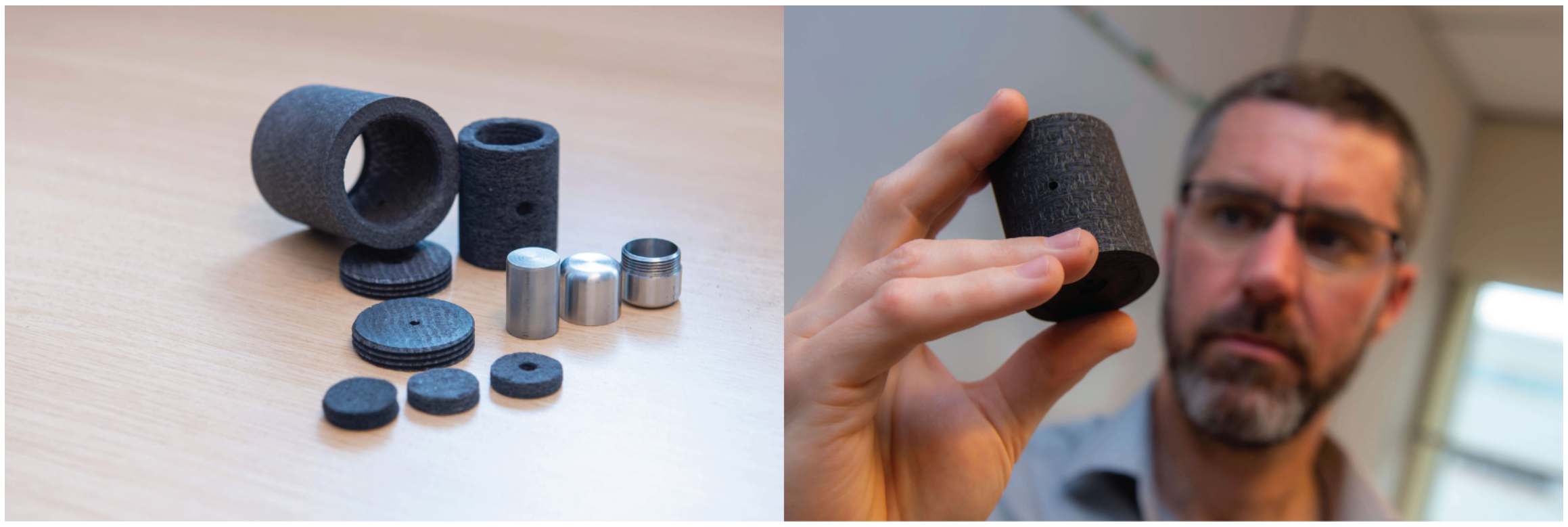

**Figure 6.52.** *Left: Disassembled Perpetual Atomics americium-RHU. Right: Keith Stephenson, nuclear and power systems engineer, holding the RHU prototype. Image credits: ESA / G. Porter.*

### Americium RHUs

Americium ($^{241}$Am) is emerging as a leading candidate to supplement the supply of plutonium-based radioisotope systems. The $^{241}$Am radioisotope is available for extraction from used nuclear fuels. Successful commercialization of this process would establish a supply of fuel for both radioisotope power and heat sources. Currently, two companies have announced the future availability of $^{241}$Am-based radioisotope heater units:

- Perpetual Atomics: Perpetual Atomics is a radioisotope technology spin off from the University of Leicester in the United Kingdom. With funding from the UK Space Agency and the European Space Agency, they have developed several americium-based heat and power sources. The design of this RHU (Figure 6.43) follows the same concept as the DOE LWRHU, with the basic building blocks including a heat source, clad, graphite impact shell, and aeroshell. This RHU is slightly larger than the DOE-developed LWRHU, weighing ~200 grams and providing ~3 W thermal heat. The first unit may be available for integration with the Rosalind Franklin rover.
- Zeno Power: Zeno Power is a US-based company developing strontium and americium products for terrestrial and space applications. Currently, they are maturing several products, including $^{241}$Am radioisotope heater units. Based on publicly released information, the first unit may be available in the next few years and produce <10 W of thermal power.  Unit mass and dimensions are not publicly available, but they will likely have a similar mass to other RHUs discussed here.

If these commercial RHUs are developed and enter production in the near future, they may worth considering in future thermal design trades.

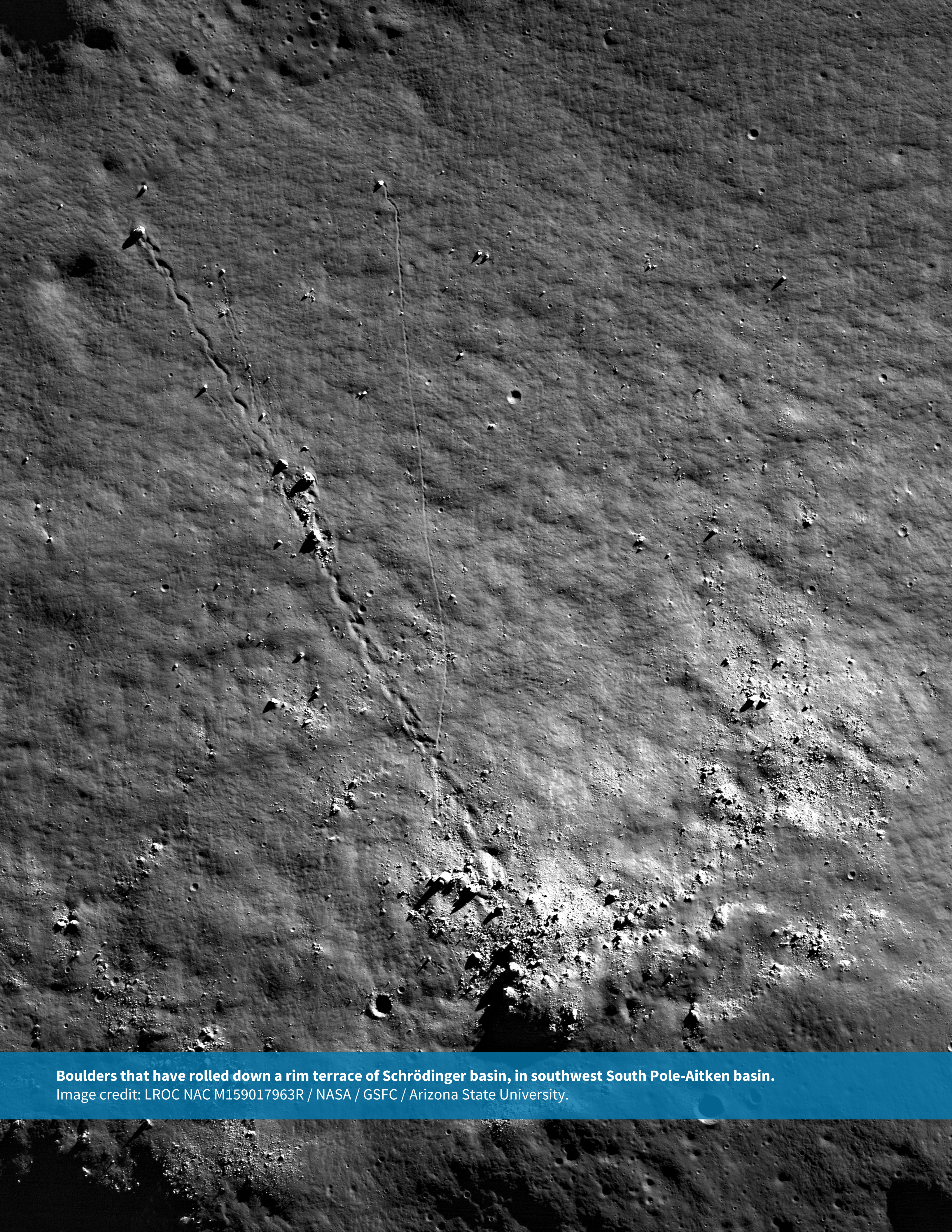

**Boulders that have rolled down a rim terrace of Schrödinger basin, in southwest South Pole-Aitken basin.**
Image credit: LROC NAC M159017963R / NASA / GSFC / Arizona State University.

# 7 PROGRAMMATIC FACTORS

The chartering of the SPARX SDT and its study and report come at a particularly dynamic time for NASA and the broader space exploration community. Here we discuss where the SPARX mission has programmatic touchpoints with broader NASA plans and the larger space environment. We also discuss how decisions about how to pursue a future SPARX mission design could drive these interactions. In this section, we will also reiterate and discuss the rationale behind critical SPARX infrastructure choices, and ways in which the future scope of a SPARX mission could be implemented. Because of the broad scope of topics in this section, abbreviated high-level findings and recommendations will be provided at the end of every subsection.

## 7.1 ASTRONAUTS

As a new era of human exploration on the Moon dawns, it is natural to consider the potential role of astronauts in achieving SPARX mission objectives. The OWL decadal mission concept study Endurance included a mission profile, Endurance-A, that utilized interactions with astronauts at a south polar lunar base station to return rover-collected samples to Earth. The SPARX TOR, therefore, asked the SDT to consider what the role for astronauts could be in a future SPARX mission. To facilitate this discussion, the SPARX Chair assembled a list of questions from the SDT, and with the assistance of the SPARX HQ Liaison, engaged in a discussion with several members of NASA ESDMD (now HSMD).

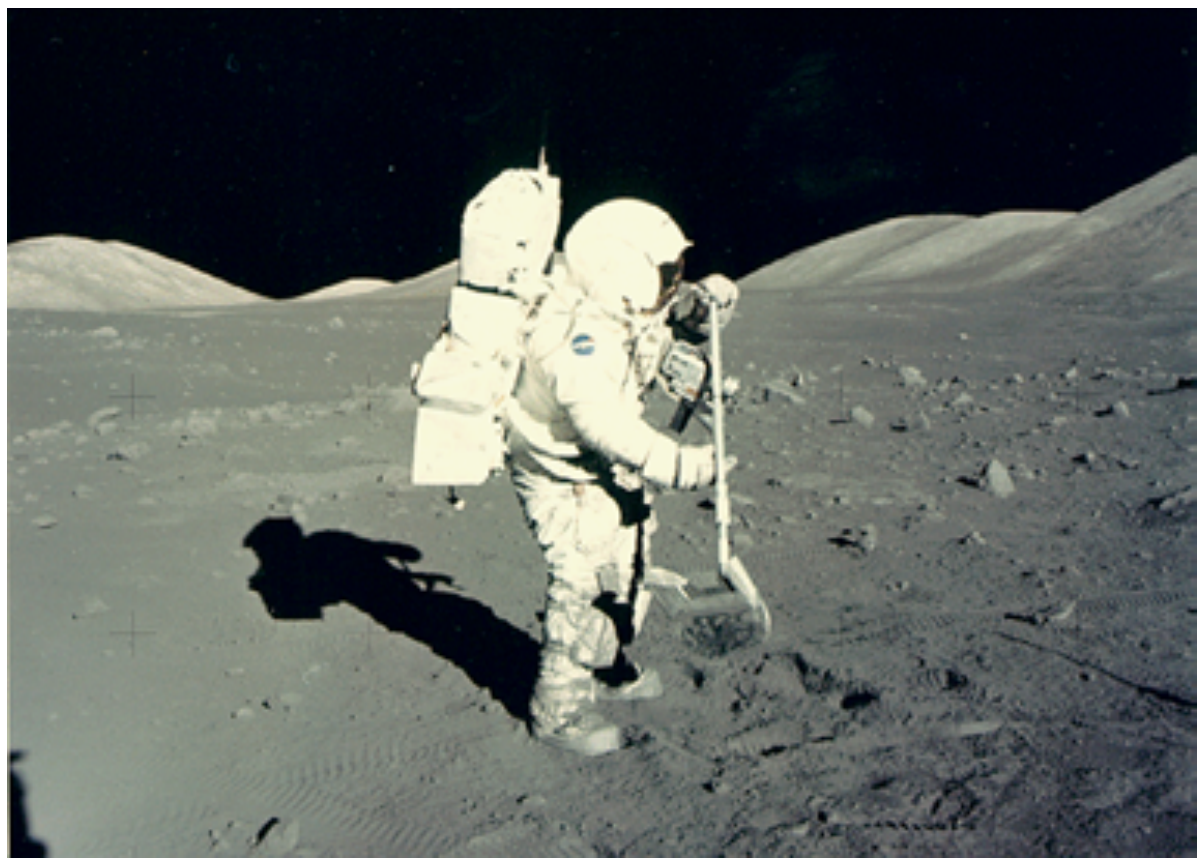

**Figure 7.1.** *Apollo 17 image AS17-134-20425, showing Jack Schmitt raking the lunar surface for rocklets.*

A central conclusion of the SPARX SDT was that astronauts are best deployed in situations that maximize human decision-making, critical thinking, training and experience, and situational awareness (e.g., Figure 7.1). The limiting factor for astronauts in contributing to SPARX objectives is their anticipated limited surface exploration range (and EVA time) compared to the proposed scope required to achieve most SPARX objectives. Even when crew has access to future mobility assets, such as the unpressurized LTV or a pressurized lunar rover, astronauts will be restricted to excursions that extend no farther than one lunar day from the main crew habitation zone. As detailed in the "one-day-wonder" medium-class mobility section of the Mission Architectures section (Section 5.6.2.1), this spatial restriction limits the likelihood of crew-centric missions achieving the SPARX threshold science requirements (Section 3.6, 5.8) or any of the highest priority science objectives (Section 3.4) with a high degree of confidence. Therefore, without the ability to bring the astronauts to all of the SPARX sampling locations, the SDT then discussed options for how astronauts could best support the mission at its end point (i.e., as samples reached a south polar basecamp, Moon Base, or other non-polar endpoint).

We will first discuss options for interactions between astronauts and the SPARX rover at a notional south polar basecamp, in Section 7.6.2 we will discuss other architectures with mixed robotic and astronaut activities. Using the SPARX design reference baseline traverse (Section 5.8, 6.7), all of the SPARX samples will have been collected prior to entering the circum-south polar region, therefore any astronaut interactions are limited in scope to (1) potential sample triage activities, and/or (2) transferring and packaging samples for return to Earth. **The SDT considered a range of crew involvement in triage but ultimately recommends that astronauts do not triage material from within individual sample containers.** Interacting with sample containers prior to Earth return risks cross-contamination with material from the local South polar region, and without specialized laboratory analyses, the crew is not particularly well-situated to triage on the Moon. **Only in the event that the sample return-mass capability is lower than the collected mass of the total sample suite would it make sense for astronauts to triage and prioritize return of a subset of the containers containing samples from the locations that address the highest-priority SPARX objectives.** Regardless, having astronauts able to package samples for return is advantageous: although it is feasible to move and manipulate sample containers robotically, crew will be much faster and potentially more effective at the task.

The use of astronauts would more significantly assist achieving the SPARX mission objectives, however, if NASA undertakes a non-polar crewed mission and astronauts land at a strategically selected sample collection region of the SPARX traverse which is also selected to be the terminal location of the SPARX traverse. Astronauts could utilize a combination of geologic training, hand-held and deployed instruments, and their local field investigation to identify and collect important target lithologies such as anorthosite, hypothesized mantle lithologies, and large rocks in a highly complementary partnership with rover-based reconnaissance and collection strategies. Potential profiles for this exploration mode will be discussed in further detail in Section 7.6.2, below, and in Appendix 10.5.

**Finding: For mission profiles involving SPARX-Astronaut rendezvous at a south polar basecamp/Moon Base, astronaut activities are best utilized for sample packaging for Earth return, or to assist in sample container triage if Earth return mass is limited. The SPARX SDT finds that the preferred profile for astronaut utilization is rover rendezvous with non-polar crewed deployment to a strategic terminal traverse sampling location.**

## 7.2 COMMERCIAL SPACE COMPANIES

The growing capabilities of the commercial space enterprise was one of the critical enabling factors noted by the OWL Endurance mission concept for the mission profile and scope. When considering these capabilities for a SPARX mission profile, the SDT considered contributions in three broad categories: surface delivery, sample return, and surface mobility and operations.

### Delivery:

A growing number of commercial companies are developing technologies for delivery of payloads to the lunar surface. While the majority of currently allocated delivery task orders are for payloads with a smaller (mass and volume) scope than the baselined SPARX surface asset, the capabilities of commercial providers are rapidly growing. The VIPER rover is now scheduled to be delivered to the lunar surface by a commercial partner (Blue Origin), and the original Endurance mission design study baselined delivery of the surface asset via a commercial partner. **Given this, the SDT finds that if a commercial partner can demonstrate surface delivery capabilities for an asset commensurate with the scale recommended by SPARX, this is one of the stronger avenues for commercial involvement.** However, the SPARX SDT cautions that due to the cost and complexity of the surface asset, care should be taken to ensure that any potential delivery partner has sufficiently demonstrated landing success prior to a SPARX mission delivery. In other words, it would not make sense to have a radically different risk posture and risk acceptance for the delivery asset than for the rover itself.

### Sample Return:

Perhaps the most technologically challenging portion of a potential SPARX mission would be the robotic return of samples to Earth. Per the SPARX TOR, the SDT's consideration of commercial sample return was primarily concerned with the scope of returned sample mass. The proposed SPARX threshold design mission would collect approximately 9.8 kg of sample material for return (Section 4.4), which falls within the TOR guidelines of NTE 10 kg for commercial robotic sample return profiles. It should be noted that the threshold sample return mass does not include sample containers. While this supports the potential feasibility of commercial robotic sample return for a SPARX threshold mission profile, the SDT did not consider it as part of the design reference mission. As such, a number of systems and accommodations which would be required for the SPARX rover to robotically transfer samples are not defined in this report, although are addressed briefly in Section 5.4.1. Those looking for details on the scope of those system changes are directed to the OWL Endurance-R report.

### Surface Mobility and Operations:

At the time of the SPARX SDT report writing, the only viable commercial mobility options for a SPARX-like mission profile are the proposed Lunar Terrain Vehicle (LTV) designs (Figure 7.2). During the course of the study, the SPARX team met with representatives from all three companies to discuss their mobility architecture and how it might interface with a SPARX mission profile. **The SDT determined that based on the publicly available technical specifications of the proposed LTVs, any of the three options could potentially serve as a viable mobility asset. However, the substantially different environmental conditions (e.g., thermal, illumination) associated with the latitude range of the proposed SPARX exploration region compared to the south polar region the LTVs are designed for would require design changes that are not yet costed.** Whenever possible, the SDT report has noted the critical design constraints for each of the mobility subsystems that any credible mobility asset should meet to be compatible with the SPARX profile (Sections 5 and 6). The SDT also notes specifically that any design accommodations should pay careful attention to the proposed SPARX ConOps, and that deviations from the proposed ConOps (for example, to accommodate thermal blackouts around lunar noon or slower average drive speeds) would have cascading effects for mission duration and cost.

While the SPARX SDT ultimately acknowledges commercial LTV designs as described by the providers would be a viable mobility platform for achieving a SPARX mission profile, the decision was not unanimous. A number of panelists argued that complete mobility systems were not a realistic commercial contribution at this time given the challenging requirements for the SPARX traverse length within a relatively short mission timeline. The primary concerns raised were with regard to the maturity, lack of experience, adaptations required to SPARX needs, and costs of the currently unproven commercial mobility options, particularly when traded against the deep institutional design and operations knowledge, and proven performance of a NASA-provided rover.

The SPARX SDT also considered options for commercial involvement in other aspects of the surface mobility and operations, including contributions to the mobility subsystems such as motors and actuators, the instrument systems, the sampling system, and other aspects of the operations systems. Although the SDT did not receive specific briefings related to all of these items (the number of possible contributors was too extensive to make this option feasible), the SDT was nevertheless comfortable acknowledging the significant history of successful commercial partnerships on a range of subsystems and instrument developments across generations of NASA exploration assets. If NASA elects to pursue a custom design rover for a SPARX-like mission, **the SDT strongly recommends commercial partnerships on technologies and systems where commercial companies have demonstrated capabilities**.

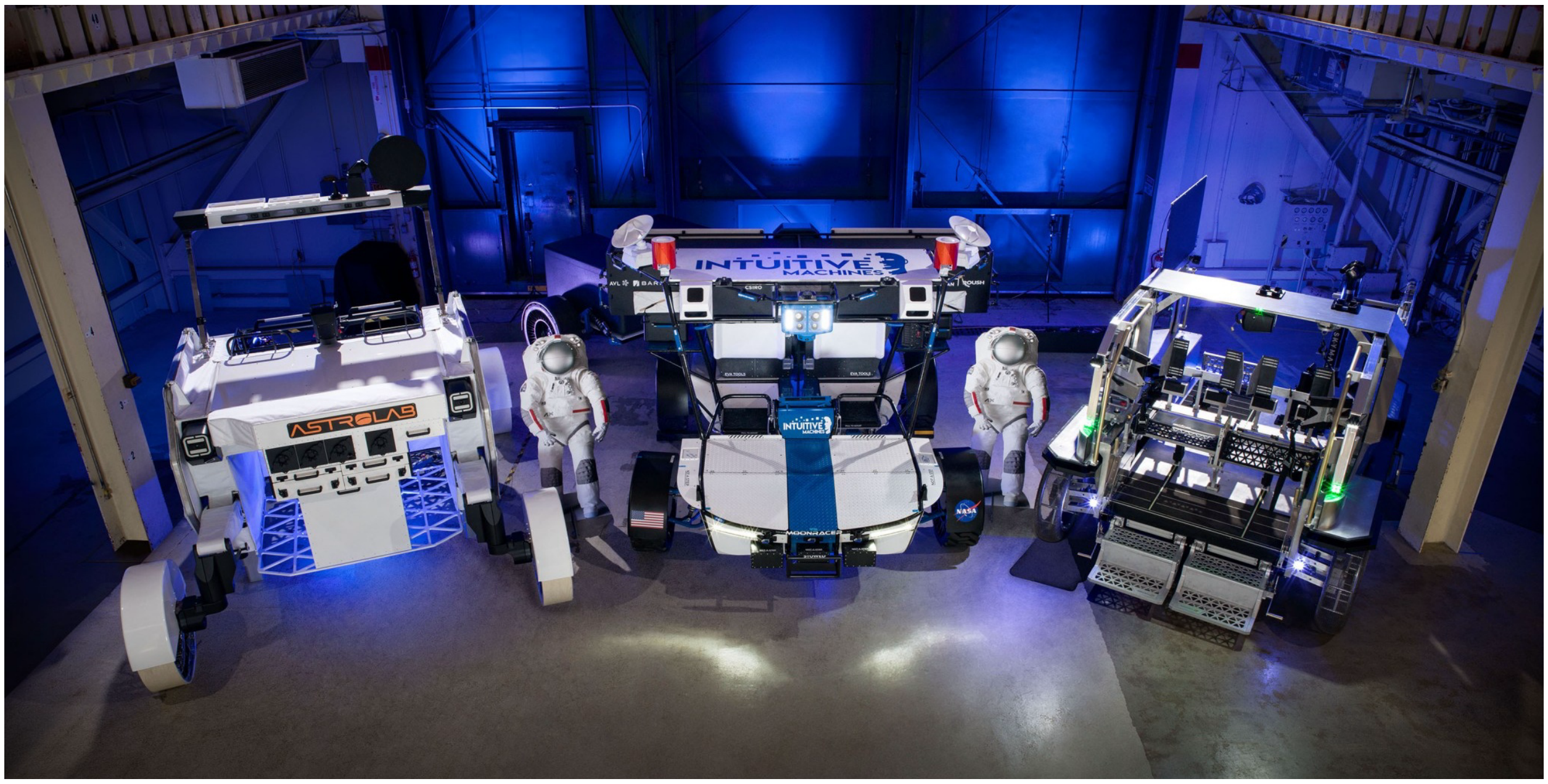

**Figure 7.2.** *Three Lunar Terrain Vehicle (LTV) prototypes from (left to right) Venturi Astrolab, Intuitive Machines, and Lunar Outpost. Image credit: NASA / Bill Stafford.*

**Finding: The SDT finds that commercial delivery of a SPARX-like mission to the lunar surface remains viable if large-delivery platforms are proven successful, for example by successfully delivering the VIPER mission to the lunar surface.**

**Finding: The SDT finds that while commercial robotic sample return could be viable for the reduced scope of the threshold mission, significant technical challenges remain in demonstrating this technology. The SDT recommends that decisions about commercial sample return be made as soon in the design process as possible, as this modality will drive multiple design points across the rover, sampling system, and eventual return asset.**

**Finding: The SDT finds non-unanimously that a commercial LTV platform could accomplish the proposed SPARX mission; however, the SDT strongly notes that any credible proposal for such must take into account the environmental and ConOps requirements for SPARX, which differ from the current LTV requirements. The SDT recommends that commercial partnerships on appropriate subsystems, components, and instrumentation continue to be pursued in the manner that is commensurate with historical NASA developments.**

## 7.3 INTERNATIONAL CONTRIBUTIONS

Although not charged with specifically investigating aspects of international contributions for a SPARX mission, the SDT did contain non-voting representative members from ESA, JAXA, and CSA. The SDT recognizes the contributions of these participants, and supports their contributions based on a strong history of many fruitful mission collaborations that have occurred between these agencies and NASA on past missions. **The SDT recommends that NASA continue to engage international agencies and other collaborators in the development of a SPARX mission**. Particular areas of collaboration could include contributed science instruments, subsystem and component developments (e.g., solar arrays, IMUs, sample acquisition arms). The SDT recognizes the value of international partners not only to mission development, but also to mission science operations teams and analysis of SPARX-returned samples. The SPARX mission represents a new-generation of planetary surface exploration, and participation in such a mission would bring the benefit of these developments to international partners as agencies continue to build on the Moon and look forward to future Mars human exploration.

**Finding: The SPARX SDT recommends NASA continue to pursue international collaboration on the development of a SPARX-like mission in ways that are commensurate with existing and past partnerships. The SDT notes that these partnerships could include a range of technical and team contributions and should also hold reciprocal benefits for partner institutions.**

## 7.4 BALANCE OF *IN SITU* MEASUREMENTS AND RETURNED SAMPLE MEASUREMENTS

The goals and objectives of the SPARX mission are derived from the highest priority science goals defined by the lunar science community and the planetary Decadal survey, and they drive the balance between the recommended requirements for *in situ* measurement vs. measurements on returned samples.

This approach is compliant with the terms of the TOR to "prioritize SPA Sample Return and Exploration goals and objectives" "within the resources and risk posture constraints provided by NASA". **The SDT developed the set of science objectives (Section 3) and experiments (Section 4), which require both in situ measurements and measurements of returned samples to holistically address the highest priority questions for lunar science, as exposed in the SPA basin**.

As discussed in Section 4.2, *in situ* measurements are required for three primary purposes in the SPARX mission: (1) directly addressing science objectives related to the mapping and characterization of the SPA impact structure, (2) placing samples and sampling locations in geologic context, and (3) ensuring the collected samples meet science requirements (Table 7.1).

The requirement most open to interpretation is (2) placing samples and sampling locations in geologic context. The SDT defined geologic context as "sufficient morphological and compositional data to contextualize the geologic setting of samples, the origin and modification process for local units and features, and the identification of compositional or morphologic outliers". The crux of this definition is the term "sufficient," which led the SDT to define a **minimum** set of measurements both in location and in type that would be needed to contextualize the lunar environment.

**Table 7.1.** *Traceability of in situ measurments to SPARX science experiements and objectives (described further in Section 4).*

| IN SITU MEASUREMENTS | | |
|---|---|---|
| MEASUREMENT | APPLICABLE EXPERIMENTS | APPLICABLE OBJECTIVES |
| M1 Context Imaging | E1 - E3, E4¢ - E7¢, E8 - E10, E13, E14 | 1.1, 1.2, 1.3, 1.4¢, 2.1, 2.2¢, 2.3, 3.2 |
| M2 Context Mineralogy | E1, E2¢, E4¢ - E7¢, E9¢, E10¢, E13, E14 | 1.1, 1.2, 1.3¢, 1.4¢, 2.1, 2.2¢, 2.3¢, 3.2 |
| M3 Fine-scale Mineralogy | E1, [E4], [E5], E6¢ | 1.1, 1.2, [1.3], 1.4¢, 2.1 |
| M4 Bulk Elemental Chemistry | E1¢, E2¢, E4¢ - E6¢, E7, E9¢, E10¢, E13, E14¢ | 1.1¢, 1.2¢, 1.3¢, 1.4¢, 2.1¢, 2.2, 2.3¢, 3.2 |
| M5 Gravity Field | E11, E12 | 3.1 |
| M6 Fine-scale Imaging | E1, [E4], [E5], E6¢ | 1.1, 1.2, [1.3], 1.4¢, 2.1 |
| M7 Magnetic Field | E6¢, E11 | 1.4¢, 3.1 |
| ¢ denotes applicable items associated with context measurements; bracketed items are associated with baseline measurements | | |

The SDT concurred that geologic context cannot be acquired from local measurements taken only at a sampling location (10s to 100s of meters), but must instead include the broader region to understand regional-scale context (10s to 100s of kilometers). In order to meet this requirement, the SDT recommends a set of geologic context measurements be taken approximately every 2 km along the traverse route, which define the ConOps mode "Science Interval Stops" described in Section 4.2, 6.6.1.3. These required stops not only provide required geologic context, but also provide *continuous* science during the mission that will allow future science teams to fully trace the geologic evolution of the basin throughout the mission, while also providing a rich opportunity for unexpected scientific discoveries. Although not within the purview of the SDT to prescribe, the team nevertheless notes that the acquisition of the described *in situ* measurement dataset would continue to serve large portions of the lunar science community both throughout the SPARX mission, and also well into the future.

When defining a strawman payload that would be able to collect "sufficient morphological and compositional data," the SDT carefully considered the trade space of instrument capability with science requirements and mission programmatic considerations. The *in situ* analysis capabilities of previous surface investigation missions can be viewed as a spectrum of capability, from remote sensing characterization that does not require contacting or sampling lunar material, all the way to highly capable on-board geochemistry "laboratories," such as modern versions of the Sample Analysis at Mars (SAM) and Chemistry and Mineralogy (CheMin) instruments on the Mars Science Laboratory (MSL) (Figure 7.3).

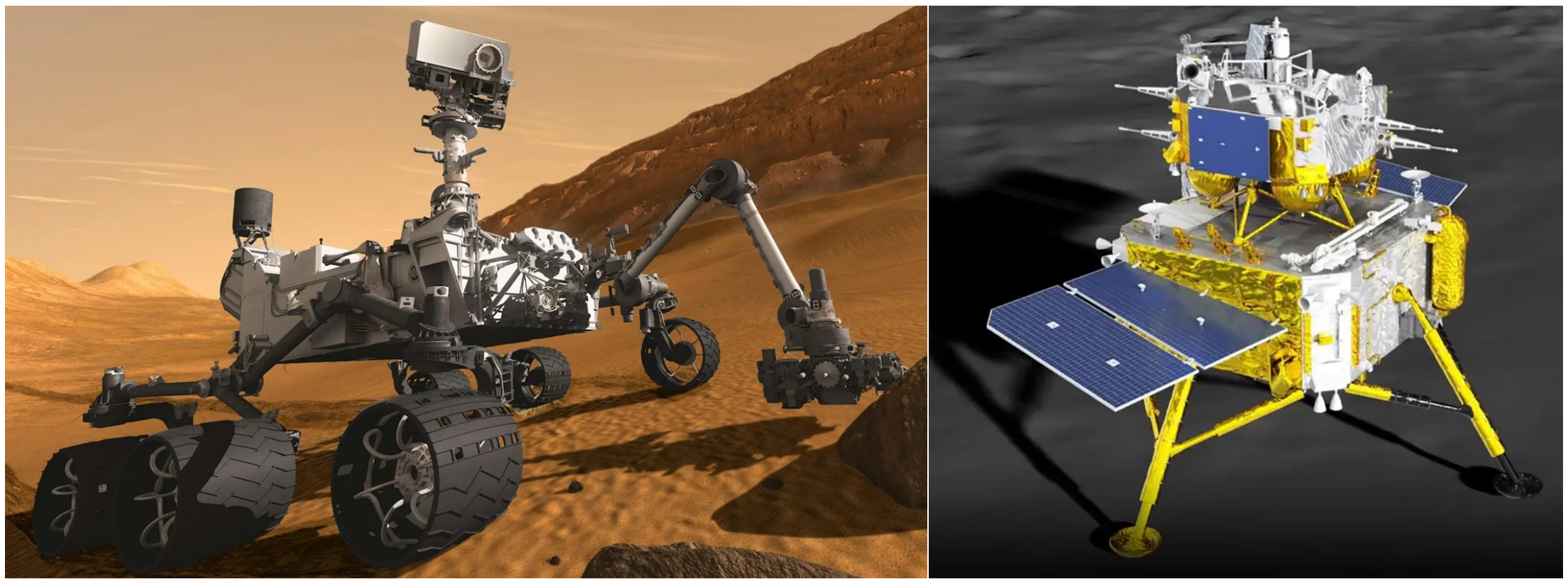

**Figure 7.3.** *In situ analysis capabilities of surface exploration assets exist along a spectrum and are tailored to the science and profile of the mission. The Chang'e 6 lander (right) focused on basic landing site characterization to support the returned samples, while the Mars Science Laboratory (MSL) Curiosity (left) rover includes multiple complex analysis instruments across the rover body and arm to support the fully in situ geochemical analyses.*

**The SDT-defined strawman payload suite was selected to fully address the requirements of the in situ measurements using the lowest complexity of payload elements, commensurate with the requirements of the SPARX TOR to keep the mission "within the resources and risk posture constraints provided by NASA".** The SDT argues that a sophisticated remote sensing suite that documents the mineralogy, lithologies, bulk elemental composition, and relationships of materials on the lunar surface is all that is *required* to establish geologic context and collect other *in situ* measurements needed for SPARX mission objectives. A drawback to this approach is that if the SPARX mission is not able to return its sample cache, the science return from the mission will be more limited than if it had a larger *in situ* measurement suite which collected more detailed chemical, mineralogic, and textural analyses, similar to an upgraded suite of instruments having high heritage from Curiosity (Cannon et al., 2023). However, the SDT determined that the contributions towards the SPARX science objectives by these additional analytical instruments did not justify the increase in resources required for the mission, both by demanding an increased complexity for the SPARX vehicle (e.g., potentially adding an arm with contact science instruments or additional sampling requirements), and adding mission time needed to collect additional data. This assessment does not preclude NASA from soliciting additional *in situ* instruments should programmatic factors encourage and/or allow it.

The SDT also considered whether there existed additional payload elements which, if added, could reduce or remove the need for terrestrial analysis of samples. While both MSL and Mars 2020 have generated incredible results from their onboard analysis capabilities without having returned samples, the SDT determined that the nature of the SPARX science objectives did not show the same potential for benefit. The crux of this decision lies in the fundamental difference in science goals between the Mars rover missions, which are most broadly interested in the *geochemistry* of Mars and its past potential for habitability, and the science goals of SPARX, which are most broadly focused on the *geochronology* of the Moon and the early Solar System. Inherent to the majority of SPARX objectives is the component of time, and unraveling when units formed, and how their composition changed as a function of time. *At present, there are no instruments capable of in situ radiometric dating with the precision or accuracy required for SPARX objectives.* Therefore, all SPARX objectives requiring sample age also, by association, require sample return. Many of these objectives also include geochemistry measurements, and the SDT weighed whether it was beneficial from either a science perspective or a risk posture perspective to include payload elements which could perform this portion of the analysis *in situ*. Here again, the SDT determined that for the defined science requirements, existing *in situ* capabilities were not capable of meeting SPARX requirements for spatial scale, precision, and in some cases are not capable of even detecting the necessary elements. Additional discussion of the above points can be found in a call-out box in Section 4.3.5. **Ultimately, the SDT determined that adding additional payload elements with the goal of reducing or supplanting the need for terrestrial sample analyses was not a viable strategy, and the marginal additional science gains did not sufficiently address SPARX science objectives to warrant the associated increases in mission cost, complexity, and operations.**

Finally, the SDT emphasizes that while the SPARX mission requires the return of samples to fully achieve the mission science objectives, this does not diminish the significant value of the *in situ* science that would be accomplished over the duration of the surface mission. The long-distance traverse of the SPARX mission would facilitate unprecedented exploration of the lunar surface through contiguous data acquisition, supporting SPARX goals and unexpected science of opportunity.

**Finding: The SPARX science objectives require a balance of robust in situ measurements to address and support the terrestrial sample analyses. The SDT further affirms that the highest-priority science objectives for an SPA mission cannot be met without the combination of in situ measurements and analyses of returned samples.**

**Findin: The SPARX strawman payload suite is tailored to meet the science measurement requirements with the greatest degree of flexibility and leveraging synergies with existing remote sensing datasets. The SDT finds that the inclusion of high-cost, high-complexity in situ analytical payload elements does not sufficiently address the SPARX measurement requirements to justify the resulting increases to mission cost, complexity, and duration.**

## 7.5 BASELINE VS. THRESHOLD MISSION SCOPE

Threshold missions are defined by a science floor which is used to identify technical or instrument-related descopes that can be applied to keep the cost of a mission in scope while remaining scientifically viable. The threshold mission prioritizes accomplishing the highest priority SPARX objectives with a moderate science risk posture, and preserves several SPARX objectives that can be accomplished through coincident sampling at critical locations in the proposed threshold traverse design. The SPARX threshold mission (Section 5.8) descopes include 1 science objective (SO1.4), 5 constituent Experiments (E4, E6, E9, E12, E14), and 10 sampling locations (Table 7.1 and Figure 7.4).

The threshold science mission requires mobility beyond that offered by a static lander or small to medium class rover and does not lead to major mission architecture descopes from the baseline mission. Indeed, the SDT found no viable threshold science mission that achieves a preponderance of the most important objectives from a static lander or from a local-to-regional rover traverse.

The payload to achieve threshold science goals requires an instrument suite that provides the minimum set of *in situ* measurements for identifying scientifically rewarding samples and providing geologic context. This includes color context and workspace imagers, a context mineral mapper, and a gamma-ray spectrometer. A magnetometer is not required for a threshold mission as the measurements form that instrument do not contribute to any of the identified threshold Experiments.

A major driver of SPARX mission architecture design is the mission terminus. The threshold SPARX reference mission terminates in the Poincare basin. Rendezvous with assets at the lunar South Pole requires >1,000 km additional traverse length, and the seasonality of South polar approach can also drive mission duration. As described in Sections 5.4.1 and 7.2, the SPARX threshold sample mass (excluding sample containers) fits within the TOR parameters for a notional commercial robotic sample return architecture. If selecting the threshold SPARX mission and non-polar human landings are not an option for sample return, NASA should define and evaluate the cost and risk tradeoffs between robotic sample return from a non-polar SPARX terminus vs continuing on to rendezvous with a crewed mission from sample return from the South Pole.

In summary, the proposed SPARX threshold mission criteria offer 1 instrument descope, 1 science objective and 5 Experiment descopes, 10 fewer traverse sampling stops, and up to 1,000 km shorter traverse requirements, pending decisions about the sample return architecture.

**Finding: The threshold level of SPARX science cannot be achieved from either a static or local (<100 km) scale investigation. The SPARX threshold investigation makes progress on eight of the nine SPARX objectives, although the reduced scope of sampling locations and returned sample mass decreases the degree to which these investigations can be addressed, and increases scientific risk. NASA should weigh cost and design trades when considering both the baseline and threshold mission profiles, particularly the costs associated with the sample return mechanism.**

| GOALS | OBJECTIVES | EXPERIMENTS | THRESHOLD | BASELINE |
|---|---|---|---|---|
| (G1) Geochronology | **(O1.1) Age of the Moon** | (E1) | – | ✓✓✓ |
| | | (E2) | ✓✓ | ✓✓✓ |
| | | (E3) | ✓✓ | ✓✓✓ |
| | (O1.2) Crystallization of the Moon | (E1) | – | ✓✓✓ |
| | | (E2) | ✓✓ | ✓✓✓ |
| | | (E3) | ✓✓ | ✓✓✓ |
| | **(O1.3) Shape of the Early Impact Flux** | (E2) | ✓✓ | ✓✓✓ |
| | | (E4) | X | ✓✓✓ |
| | | (E5) | ✓✓✓ | ✓✓✓ |
| | (O1.4) Ancient Lunar Dynamo | (E6) | X | ✓✓✓ |
| (G2) Planetary Formation | (O2.1) Farside Crust Formation | (E1) | – | ✓✓✓ |
| | **(O2.2) Existence of Farside KREEP** | (E7) | ✓✓✓ | ✓✓✓ |
| | (O2.3) Evolution of the Farside Mantle | (E8) | ✓✓ | ✓✓✓ |
| | | (E9) | X | ✓✓✓ |
| | | (E10) | ✓ | ✓✓✓ |
| (G3) Megabasin Impacts | (O3.1) Physical Structure of the Basin | (E11) | ✓✓✓ | ✓✓✓ |
| | | (E12) | X | ✓✓✓ |
| | (O3.2) Compositional Structure of the Basin | (E13) | ✓ | ✓✓✓ |
| | | (E14) | X | ✓✓✓ |

**Table 7.2.** *Very simplified version of the SPARX Science Traceability Matrix (STM) comparing the relative completion of SPARX Goals, Science Objectives, and Experiments between the Baseline and Threshold mission concepts (see also Table 5.15). Goals and objectives are abbreviated for brevity. Dark green with three check marks (✓✓✓) indicates full sample and traverse requirements are met, lighter shades of green and fewer check marks indicate majority of science and traverse requirements are met with some science risk. Yellow with a dash (–) indicates partial completion with significant science risk. Red with an "X" indicates experiment unable to be met. More detailed STMs are in Foldouts 3.2, 4.1, 4.2, and 5.1.*

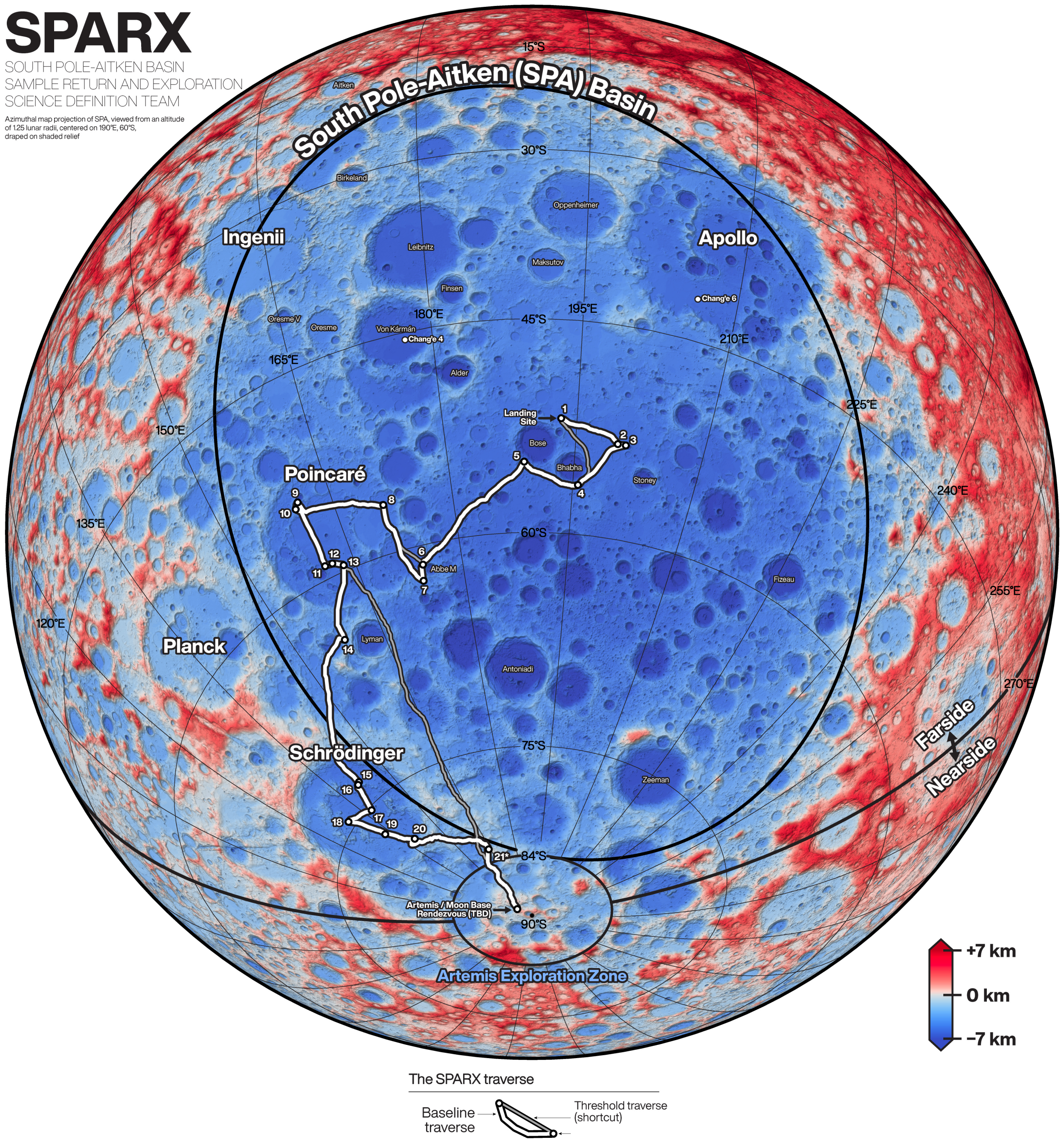

**Figure 7.4.** *Comparison of SPARX baseline and threshold traverses. The baseline traverse is designed to fully accomplish all of the SPARX objectives and constituent experiments. The threshold mission prioritizes the accomplishment of the three highest priority SPARX objectives (Section 3.4) and makes progress on a number of other objectives and experiments that can be accomplished within the scope of the highest priority objectives (e.g., through sampling and coincident locations). While both traverse profiles ultimately rendezvous at Moon Base in the Artemis Exploration Zone, the active science sampling stops for the traverse mission stop after leaving Poincare basin*

# 7.6 PROGRAMMATIC ASSESSMENT OF ARCHITECTURAL OPTION SPACE

## 7.6.1 SPARX as Campaign Science

As described in Section 5.7, the SPARX SDT considered multiple mission architectures, including static landers, hoppers, local- (<10 km) to regional-scale (<100 km) rovers, and long-distance rovers. **When compared against the threshold highest priority science objectives of SPARX, only the long-distance rover option proved capable of achieving the minimum science requirements. For this reason, the SPARX SDT settled on a nominal mission concept architecture using a long-distance rover**. However, given that the resulting SPARX baseline design reference mission traverse is comparable in total distance to a traverse spanning a path across all of the Apollo landing sites, it is worth considering if a campaign style architecture could feasibly achieve the SPARX science objectives. “Campaign” here is defined as a series of >1 landed assets whose science goals cumulatively address the SPARX science goals and objectives. The rationale for exploring this architecture option stems from the recent rise of the CLPS (Commercial Lunar Payload Services) program which aims to support a higher volume of small-scale surface landings, each at lower cost to NASA.

When considering the SPARX objectives and associated experiments, one of the most important items to note is the variety of targeted lithologies required to address the objectives. Identifying these different lithologies for sampling present a range of challenges, with some being able to be credibly identified by *in situ* methods prior to sampling (e.g., olivine, anorthosite), while others rely upon strategic choice of sampling location and a statistically viable returned sample population (e.g., SPA impact melt). This kind of mixed modality presents challenges for any static lander approaches— while material scooped from a single location could address some aspects of SPARX objectives, ensuring the desired, centimeter-sized target lithologies are within “arms reach” at a specific landing point is impossible to determine *a priori* from orbit. The samples returned by the Chang’E-6 rover are an excellent example of this principle. This mission landed on an orbitally identified mare deposit in the Apollo basin, and the samples it collected could in theory support, but not conclusively address two of the SPARX objectives that require analysis of mare deposits (SO 2.2 and 2.3). However, the limited spatial diversity of collected samples (i.e., falling within a workspace where the arm could reach) means that ages and analyses reported for non-mare investigations, including the age of SPA, ultimately suffer from geologic uncertainty and need to be viewed critically. The end result is such that a SPARX science static lander campaign would likely need to include at minimum a lander for every SPARX traverse location; but even doing so, there would be irreducible scientific risk that certain high-priority lithologies (including the highest priority anorthosite and Th-rich material) might not be retrieved in sufficient or analyzable quantities due to the difficulty in pin-pointing exact sampling locations for these materials with existing remote sensing datasets. This is to say nothing of the increased costs of allocating for >10 instrumented static landers, lander delivery costs, and lander sample return costs.

When considering either local-scale (<10 km) or regional scale (<100 km) medium class rovers, this same problem presents itself at different scales. Although a mobility system allows for local reconnaissance to enhance targeting of specific lithologies, the fundamental challenge remains the distance between viable locations for sampling the most important lithologies (e.g., Figure 7.5). Indeed, the SDT was challenged to identify a single 100 $km^2$ region where the highest-priority science objectives could be accomplished, and none were identified. Therefore, campaign science utilizing either of these exploration baselines would still require approximately one mobility asset per sample target location, and an estimated minimum of 7 landed assets to complete the SPARX threshold science.

While the above discussions focus only on the scientific viability of limited-mobility campaign science, the SDT also notes that the science goals for each of these profiles would require delivery and sample return services for each surface asset; although projecting those costs are beyond scope of this SDT, we encourage NASA to strongly weigh these costs, and those of any required supporting *in situ* instrumentation, as they factored heavily into the analysis of the SDT.

**Finding: The SPARX SDT does not consider “campaign science” consisting of multiple (likely 10s) of low-capability surface assets (static landers, small rovers) to be sufficient for accomplishing the science objectives of SPARX on a timescale that is commensurate with the proposed baseline mission, nor does the SDT consider this to be a cost-effective approach for NASA given the greatly increased number of deliveries, sample returns, and in situ instruments required to support each campaign science asset.**

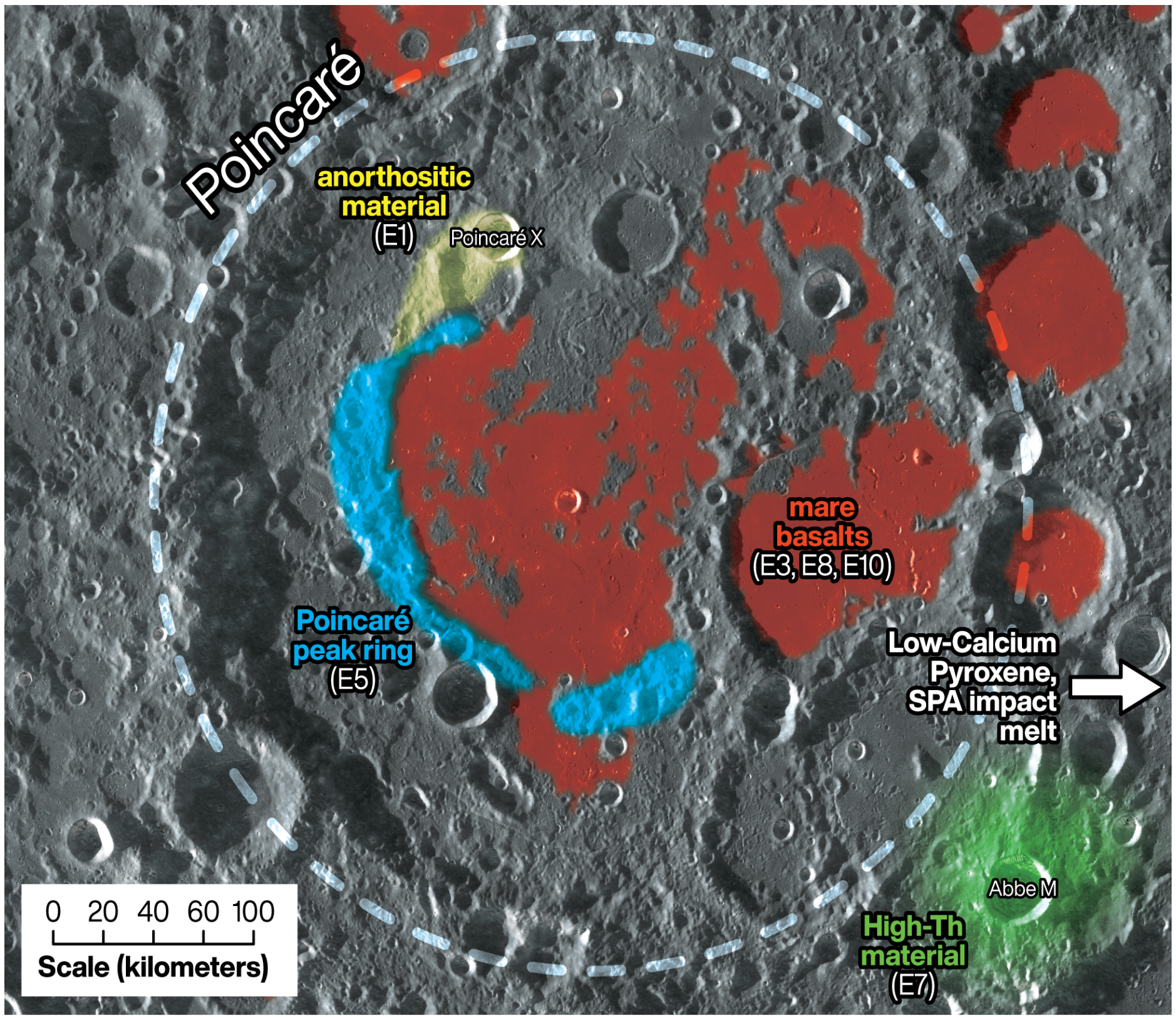

**Figure 7.5.** *Sketch geologic map of the Poincaré basin, showing that even within an individual SPA Region of Interest, long-range mobility is required to access the required number and diversity of terrain types. Figure generated with LRO Quickmap.*

### 7.6.2 SPARX Mixed Astronaut-Robotic Mission

As discussed in Section 7.1 and previously in Section 5.6.3, the greatest utility for astronaut crews is leveraging their ability to combine critical thinking and situational awareness for more efficient, and generally more successful collection of specific lithologies that can be adequately characterized *in situ*. In the case of the SPARX objectives, these include lithologies like anorthosite, hypothesized mantle lithologies, and Th-rich material. The largest hinderance to utilizing astronauts in the baseline SPARX mission concept remains that all of the required SPARX sampling and *in situ* measurements are completed before entering the circum-south polar region. Further, proposing that astronauts stationed at a south polar basecamp/ Moon Base traverse out to the final SPARX science locations for concurrent operations before traversing back to basecamp/Moon Base raises so many technical and logistical complications as to be infeasible for discussion at this time.

The SPARX SDT discussed a concurrent exploration mission architecture, i.e., one where both astronauts and an autonomous rover land together, and astronauts work in the near-lander environment while the rover completes farther distance sample fetching activities. As discussed previously (Section 5.6.2.1, 7.1, 7.6.1), this variation on the “one-day-wonder” design struggles to identify a location within SPA where the highest-priority science objectives can be accomplished in the limited spatial and temporal scope of such a mission.

**There is, however, a potential implementation approach utilizing both astronaut and robotic exploration that could significantly increase the scientific return of a SPARX-type mission—an architecture combining a non-polar astronaut landing at the terminal traverse location of a long-distance rover**. While the discussion on non-polar astronaut interactions was not part of the original SPARX SDT TOR scope, at the request of ESSIO, the SDT considered the options for this style of architecture during the final SDT meeting in December 2025.

Within this proposed implementation, the SPARX long-distance rover would be delivered to the start of the proposed traverse region. The rover would then carry out the SPARX mission traverse before progressing to the final traverse location. The astronaut crew would land at the final SPARX traverse location and use their human perception and tools, complemented by *in situ* measurements of the site collected by the SPARX rover, to find and collect the best samples for achieving SPARX science objectives at this location. They would also package all samples already collected by the SPARX rover for the return trip to Earth.

The proposed implementation has a number of advantages over other implementation architectures. First, this architecture leverages the technical abilities of astronauts for the acquisition of high-priority lithologies, decreasing the science risk and increasing overall science value. Secondly, by having the rover rendezvous with the astronauts, the mission preserves the ability to return a higher mass of samples, consistent with the baseline science requirements. Finally, by removing the need for a long traverse to the circum-south polar region, this approach decreases the engineering complexity needed for operations at very high latitudes while maintaining the benefits of astronaut involvement (reduced sample transfer interface complexity and ease of prioritizing what to return based on available return mass).

Although the SDT did not perform a full option space study of the non-polar astronaut rendezvous scenario, by comparing the implementation architecture with the existing SPARX STM (Section 3.6), recommendations were made for more ideal rendezvous locations. As stated previously, the desire to utilize astronaut capabilities for important targeted lithologies with unique acquisition ConOps means that the central SPA region (Bose crater, Bhabha crater, CSPA mare units) are the least ideal fit for this implementation. The science targets within these regions (mare deposit, SPA impact melt, and crater ejecta) do not have any unique acquisition modes that would strongly benefit from astronaut presence. In contrast, Poincare basin, Schrödinger basin, and Ingenii basin all have significant potential as viable end-traverse rendezvous locations. Poincare basin and Schrödinger basin, in particular, both have targeted regions where high-priority anorthosite samples are suggested to have shed from basin peak rings. Astronauts have the ability to assess the geology of these regions, and locate and identify targeted lithologies in a more efficient manner than the specialized ConOps developed by SPARX. Indeed, this hypothesized scenario echoes operations from Apollo 15, where astronauts specifically identified a large anorthositic rock by its visual properties (the so-called "Genesis rock" Figure 7.6).

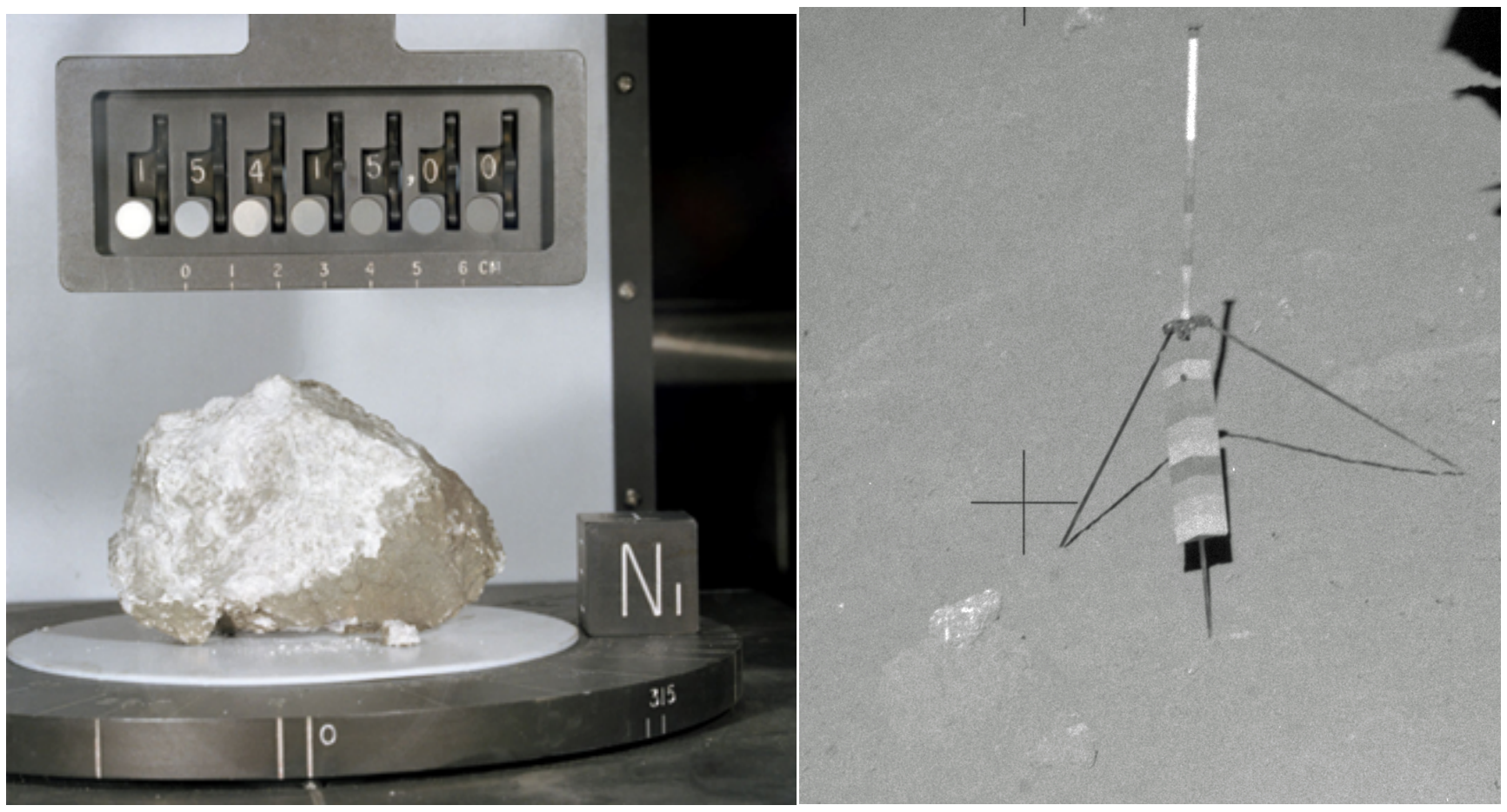

**Figure 7.6.** *Apollo 15 sample 15415, the so-called "Genesis rock" (left) is an anorthositic sample identified by Commander Dave Scott by its distinctive physical properties, particularly when observed on the surface next to the surrounding mare deposits (right) (Scott: "Guess what we just found! I think we found what we came for." ALSJ, 145:42.47). Image credit: NASA.*

The request of ESSIO, the SDT has compiled a further series of recommendations for proposed crew landing sites, science objectives, and required tools and measurements in Appendix 10.5.

**Finding: The SDT finds that the optimal implementation for mixed-astronaut and robotic SPARX mission is non-polar astronaut landing at the terminal location of a long-distance rover traverse. This implementation would optimize astronaut surface operations for targeting specific high-priority lithologies (anorthosite and hypothesized mantle rocks), while allowing the long-distance rover to collect samples and perform in situ traverse science from across the traverse route. This implementation brings the sample return benefits of an astronaut architecture without the added engineering and mission duration challenges of a south polar rendezvous.**

### 7.6.3 Overall Implementation Assessment

When assessing the overall range of possible implementation architectures for SPARX, presented both here, and in detail in Section 5.7, the SDT considered that, at the broadest level, a sample return mission consists of three key elements: delivery, surface operations, and return. Although costing for the delivery and return was outside of the purview of the SDT, historical understanding suggests that the cost of these elements is non-trivial, and can easily represent half, or more, of any implementation's potential cost profile. In making the final assessment of recommended implementation, the SDT sought to optimize science-per-dollar, though this is conceptually elusive; it is also possible to spend a substantial amount of money on notionally cheaper implementation architectures with a high risk of failing to achieve the threshold science.

**The SPARX recommended baseline mission concept is a long-distance large rover that makes a rendezvous with astronauts for sample return.** This has the highest probable science return within a reasonably scoped overall mission cost profile.

The SPARX SDT would like to emphasize that given the initial parameters of the TOR, astronaut-enabled return was recommended insofar as this option allowed for the highest possible return sample mass. Traversing within the south polar environment does not offer any particular scientific benefits to the SPARX mission; indeed, the south polar environment places additional engineering burden on the mobility asset. Additionally, to the extent that other polar Artemis missions make progress relevant to SPARX objectives, there may be complementarity and efficiency in *not* traversing to the circum-south polar region.

However, the SDT weighed the challenges associated with a south circum-polar rendezvous against a fully robotic return scenario, which has increased costs for both the rover and return system and decreases the science return by reducing the return mass. The SDT determined that astronaut rendezvous was the preferred optimization. **Although not initially part of the SPARX TOR, the potential availability of a non-polar astronaut rendezvous at the terminal SPARX traverse location (Poincare, Schrödinger, or Ingenii) would represent the most optimized implementation for a SPARX mission**.

Ultimately, the SPARX SDT recommends that NASA evaluate these potential implementation options and make a decision at the earliest possible opportunity. The return architecture is anticipated to be a significant driver of overall mission design and cost, and the SDT cautions that down-stream changes have the potential to grow mission cost and scope beyond reasonable bounds.

**Finding: A long-distance rover rendezvous with astronauts remains the optimal solution to maximizing SPARX science return while minimizing overall programmatic cost. Astronaut rendezvous at a nonpolar, terminal SPARX sampling location is potentially more optimal than a South Pole rendezvous.**

## 7.7 SCIENCE TEAM STRUCTURE

A core mission-level SPARX science team should be established immediately at the start of SPARX mission development, led by a Science Team Lead who is responsible for SPARX's scientific integrity and overall scientific success. The Science Team Lead should work closely with, and be a peer of the Project Scientist and Project Manager. The SPARX science team must be comprised of members with expertise that covers lunar geology, geochemistry, and geophysics, lunar petrology and sample analysis, and rover science operations, including instrument operations. The SPARX SDT contained members with this wide range of expertise, which the team found was necessary to address the range of TOR charges. This core science team should have a size that is large enough to accommodate the required expertise, estimated to be ~20-30 individuals (e.g., Figure 7.7).

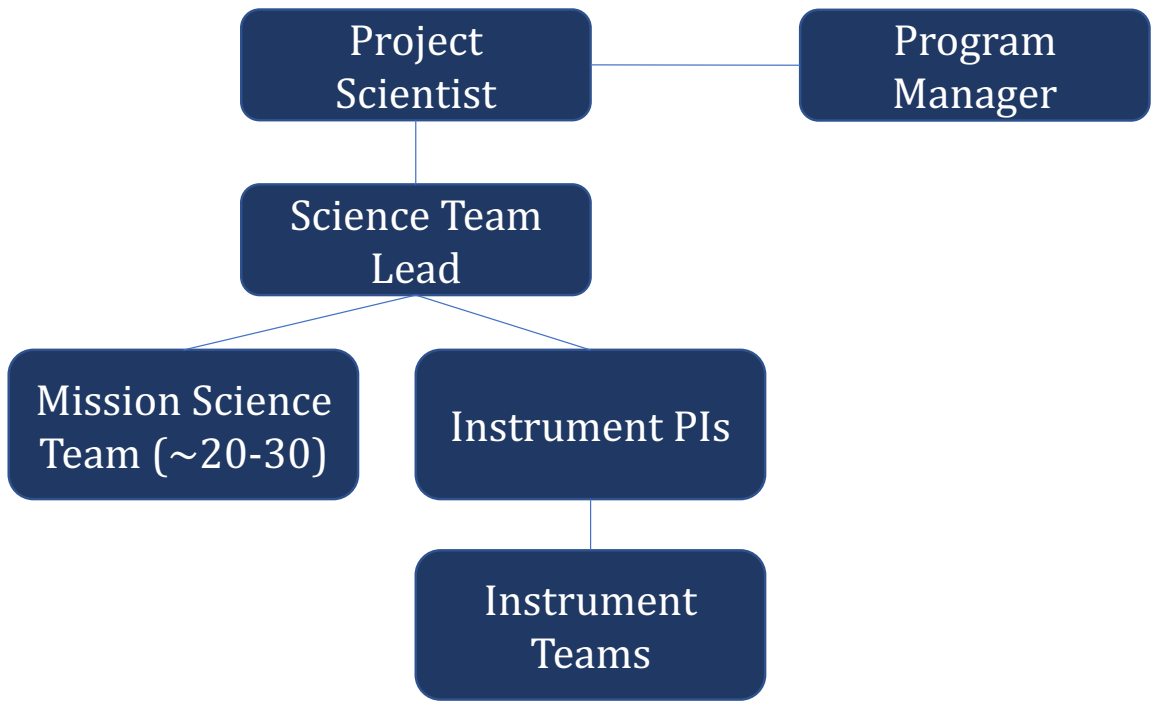

**Figure 7.7.** *Notional schematic of recommended SPARX science team structure. The SDT emphasizes the necessity of a core mission science team, in addition to, payload instrument teams, and recommends a Science Team Lead to coordinate efforts by all science team and payload team members.*

NASA should separately solicit instruments for the SPARX science payload, and a mission-level review of the top-rated instrument proposals should be provided by the core science team (e.g., which combinations of highly rated instruments work together best to achieve science goals), and considered as input by NASA selecting officials. Payload instruments will be led by science PIs with small science teams that have expertise in operating the instrument and interpreting data from it, as well as the lunar science and science analysis expertise described above. After payload instruments are selected, their associated instrument teams should become equal members with the core-science team, and instrument PIs be assigned a title of "Instrument Lead". Structurally speaking, the SDT recommends a science team structure that is closer to a PI-led mission structure of Discovery and New Frontiers scale missions versus the instrument PI-led councils of large Flagship missions in order to encourage scientific integration across instrument datasets. Regardless of actual science team structure, it will be imperative for teams to develop a "Rules of the Road" document that clearly defines science team member roles and responsibilities early in the process.

A key lesson learned from NASA's Mars rover experiences is that frequent interactions and close partnerships between science and engineering teams throughout a mission's lifetime, from earliest concept development and design through daily surface operations, is critical to maximize science return within available resources for mobile surface missions. The Project Scientist will be a peer of the Project Manager in all matters impacting science and in science-engineering design trades. SDT further emphasizes that SPARX science team integration should occur regardless of mission or payload operations center, i.e., if a commercial mobility provider is selected, contracts should include stipulations for the science team to work closely with said provider during the planning and execution of operations. To not include the science team during these operations risks not achieving the full scientific success of the mission.

**Finding: The SDT strongly recommends that NASA organize a SPARX mission-level science team with a range of expertise relevant to the design, formulation, and implementation of the mission, using the expertise solicited for the SPARX SDT as a suggested topical guide. This team should be intimately integrated with payload leads, engineering and operations leads, and other relevant personnel. The SDT further finds that this mission-level science team should be integrated into mission operations regardless of the nature of the operating partner.**

## 7.8 TECHNOLOGY MATURATION AND FEED-FORWARD CAPABILITIES

As discussed in Section 6 the SPARX mission presents an advancement of multiple surface exploration technologies that have been developed over many generations of successful NASA mission operations. **The overall assessment of the SDT is that no new technologies are required for development prior to initiating a SPARX mission program. This finding comes with the understanding that developments will continue into the critical mission-enabling capabilities of lunar night survival, long-distance autonomous driving, sample acquisitions subsystems, and farside lunar communications assets, as discussed in Section 6.11.**

The SDT specifically notes that one place where current technology capabilities do not meet the full profile of the SPARX mission is robotic sample return. As assessed above, the TOR provided guidelines for commercial robotic sample return masses NTE 10 kg. The SPARX threshold sample return mass is within these parameters (≥ 9.8 kg); however, the SDT urges NASA to consider encouraging commercial development of capabilities that can comfortably accommodate that SPARX threshold sample return mass and include additional margin for variations in sample container conformation and robotic sample transfer mechanisms that would increase the practical return mass above 10 kg, with suggestions for capabilities in the 20 kg total return mass range.

At a programmatic level, the SDT finds that SPARX would demonstrate a broad range of technologies with significant feed-forward capabilities for the next-generation of planetary exploration, including several Moon to Mars objectives. A detailed traceability for the SPARX objectives to Moon to Mars objectives can be found in Foldout 3.1 (Section 3.5). In brief, these technologies include long-distance autonomous rover navigation, farside human exploration (implementation dependent), robotic sample return (implementation dependent), human and robotic teaming (implementation dependent), long-term lunar night survival, and long-baseline lunar environmental monitoring.

Finally, although not a new technology, **the SDT would like to specifically state that the availability of radioisotope power systems (RPS) for a SPARX mission would be strongly enabling**. Although considered at the time of the initial SDT chartering, programmatic changes early in the SDT study directed all proposed implementation solutions towards solar-power only solutions. The baseline SPARX long-distance rover design presented in Section 6 complies with this request, and the system design shows the required system designs for a solar implementation. While the solar-powered rover option fits within the cost parameters given by the TOR, the mission duration exceeds the requested 2–4-year baseline duration. If nuclear powered options were to become available, the SPARX mission duration could be meaningfully decreased, with a comparative example being found in the Endurance-A notional mission duration timeline.

**Findings: The SDT finds that no new technologies are required for development prior to initiating a SPARX mission program. Within this finding, the SDT does recognize the importance of continued development on a number of enabling technologies, including technologies supporting extreme long-distance roving, electrodynamic dust shielding, and autonomous systems.**

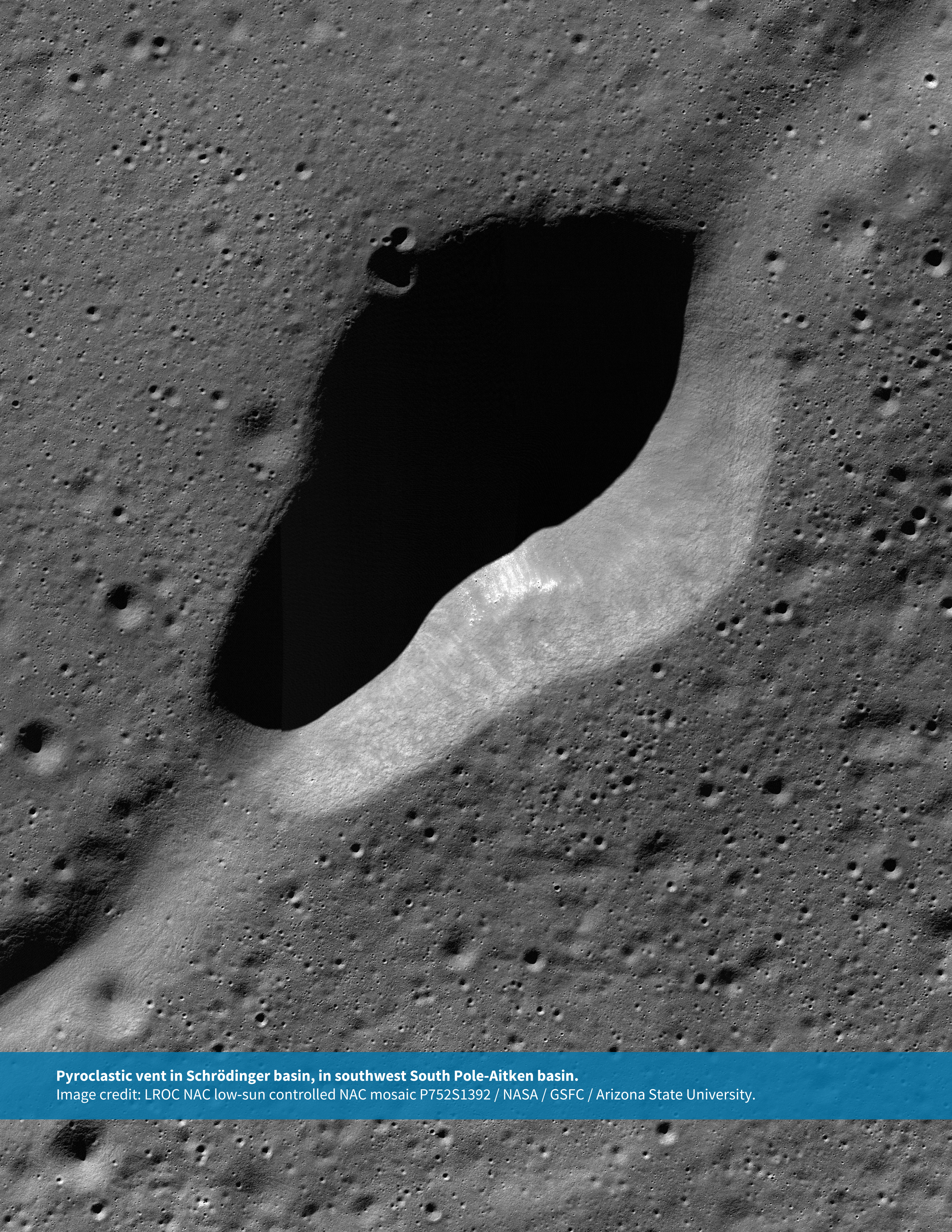

**Pyroclastic vent in Schrödinger basin, in southwest South Pole-Aitken basin.**
Image credit: LROC NAC low-sun controlled NAC mosaic P752S1392 / NASA / GSFC / Arizona State University.

# 8 HIGH-LEVEL CONCLUSIONS OF THE SPARX SDT

Over half a century after the Apollo missions first returned samples from the lunar surface, and bolstered by several generations of remote orbital observation of the Moon, we are now at the dawn of a new era of lunar exploration with sustained human presence as the goal. The SPARX mission will be a crowning scientific achievement of this new era, rewriting the lunar science textbooks and answering science questions that affect our understanding of the entire Solar System. Analyses of the samples returned by Apollo directly established foundational hypotheses about planetary dynamics during the end stages of Solar System formation, hypotheses for the formation, growth, and evolution of planetary crusts, the impact record tie-points that underpin the system for planetary surface dating used across the Solar System, and many others. However, through the intervening decades, the organic process of science continued as new remote satellite data were acquired which highlighted inconsistencies and insufficiencies in those original hypotheses. The lunar community has therefore advocated with increasing fervor for the importance of returning new samples from the Moon, this time from the unsampled lunar farside, in one of the only remaining places in the Solar System that taps the earliest periods of formation—the South Pole-Aitken basin (SPA). **The recent Decadal Survey on Planetary Science and Astrobiology: Origins, Worlds, and Life (2023) recommended a sample return mission from the Moon's South Pole-Aitken basin as the highest-priority mission for NASA's Lunar Discover and Exploration Program.**

This report defines science goals and objectives for a South Pole-Aitken Basin Sample Return and eXploration (SPARX) mission, the *in situ* and terrestrial laboratory measurements required for the completion of these objectives, and the lithologies, sample properties, and masses of samples that must be returned for the required measurements. This report also explores a range of potential mission implementation and programmatic architectures for achieving various levels of the SPARX science objectives, including a full range of mobility options, commercial capabilities, the role of astronauts, and the potential for multi-landing campaign-style science. From these investigations, the SPARX SDT has developed a baseline mission concept that includes a traverse design that can accomplish the SPARX science goals, optimizes the programmatic understanding of the implementation space, and fits within the cost guidance provided by NASA.

## 8.1 SCIENCE OBJECTIVES AND GOALS

The SPARX SDT defined three overarching science goals and nine constituent objectives, presented below:

**SG1: What was the chronology of key processes and events during the first half billion years of solar system history?**

- O1.1: Quantify the age of the Moon to distinguish between early- and late-Moon formation models
- O1.2: Distinguish between fast- and slow-cooling models for lunar magma ocean crystallization

- O1.3: Determine if and when a giant planet instability drove a sharp change in the inner Solar System impactor flux
- O1.4: Constrain the evolution of the lunar dynamo in the pre-4.0 Ga epoch of lunar history

**SG2: Does the lunar farside preserve evidence of asymmetric magma ocean evolution and planetary crustal formation processes?**

- O2.1: Determine if the farside highlands crust is similar to nearside crust in age and/or composition
- O2.2: Determine if there was a farside KREEP reservoir, or other late-stage magma ocean products, prior to the SPA impact
- O2.3: Determine if and how the temporal and compositional evolution of the farside mantle differs from the nearside

**SG3: What are the effects of mega-basin impacts on terrestrial bodies?**

- O3.1: Determine whether there are discontinuities in the structure (crustal thickness, density, magnetic field) across SPA to constrain models for the formation of a mega-basin
- O3.2: Distinguish the relative contributions of impact melt differentiation, crustal backfilling, and volcanic infilling to the interior composition and stratigraphy of SPA

The scientific background and motivation for each of these goals and objectives is presented in detail in Section 3.1-3.3. These goals and objectives represent the highest-priority lunar science questions that should be addressed by a mission to SPA, and they trace to multiple OWL Decadal science main question and sub-questions, as well as, numerous lunar community documents, and other priority NASA documents, such as the Moon to Mars objectives (Section 3.5). Notably, the SPARX mission achieves Decadal-level (not simply Decadal-relevant) science, which is defined at the "significant, unambiguous progress in addressing at least one of the survey's 12 priority science questions." (OWL, footnote on Summary page 5).

At the request of NASA, the SDT has prioritized both the goals and objectives for a SPARX mission (Section 3.4). T**he goals as presented are in priority order, with SG1 being the highest priority goal. The priority of this goal is in recognition of the broad implications of this goal for understanding the early history of the Earth-Moon system, and for establishing the chronology and impact crater flux that define the surface dating systems used for the rest of the Solar System.**

> **The prioritized objective order is as follows: O1.3, O1.1, O2.2, O2.1, O2.3, O3.2, O3.1, O1.2, O1.4. From this prioritized list, the SDT highlights objectives 1.3, 1.1, and 2.1 as being the highest-priority objectives that should form the science threshold for any mission to SPA.**

## 8.2 *IN SITU* MEASUREMENTS, LABORATORY MEASUREMENTS, AND SAMPLE MASSES

In constructing an STM for the SPARX mission (Section 3.6), the SDT defined a series of 14 experiments, through which the science objectives and goals would be accomplished (Section 4.1). These experiments (E1-E14) encompass the specific measurement objectives, sample lithology requirements, sample property and mass requirements, *in situ* measurement requirements, and guidance on traverse requirements for the overarching mission goals and objectives.

### 8.2.1 *In situ* Measurements

As the nature of the SPARX mission science focuses heavily on science gathered from returned samples, the *in situ* measurements defined by the SDT are primarily designed to support the sampling at a range of scales (Section 4.2,). At the widest scale, the *in situ* measurements are designed to capture required geologic context of the basin along the scale of the traverse, marking key progressions in composition that are challenging to resolve with orbital dataset resolution, and linking these observations with geologic landforms. In parallel, the *in situ* measurements are designed to capture the geophysical variation of the basin, forming a robust dataset for current and future investigations. The *in situ* measurements are also designed to assist in sampling ConOps, aiding both in the initial sampling region reconnaissance, and also in ground-in-the-loop verification of sample acquisition activities. The classes of *in situ* measurements defined for the SPARX mission are presented below, and additional information about required resolutions and the associated strawman payload are presented in Section 4.2 and Section 6.4.

#### In situ Measurement Requirements:

- M1: Context Imaging
- M2: Context Mineralogy
- M3: Fine-scale Mineralogy
- M4: Bulk Elemental Chemistry
- M5: Gravity Field
- M6: Fine-scale Imaging
- M7: Magnetic Field

The strawman payload the SPARX baseline mission concept uses to address these 7 measurement classes consists of 5 payload elements, with the context and fine-scale mineralogy requirements being met by a single instrument, and context and fine-scale imaging requirements similarly being met by a single instrument. Measurement classes M1-M5 are required to achieve the SPARX threshold science, with M6: Fine-Scale Imaging and M7: Magnetic Field measurements being considered baseline only (Section 4.2).

### 8.2.2 Terrestrial Laboratory Measurements

Commensurate with the returned sample nature of the mission, the majority of SPARX objectives require the future examination of SPARX samples using a wide range of laboratory techniques (Section 4.3). These techniques fall into five categories: (1) petrologic characterization, (2) bulk elemental chemistry, (3) geochronologic measurements, (4) volatile measurements, and (5) remanent magnetization measurements. Details about the recommended investigations and associated laboratory techniques can be found in Section 4.3 and APPENDIX: Samples, in addition to, an example investigation workflow.

The SPARX SDT weighed the nature of returned sample material, and determined that sampling systems should be designed to acquire small rocks (diameters ranging from 40mm to 3mm) and unsieved bulk regolith. This sampling approach is tailored to the target lithologies, and prioritizes acquiring a larger number of small samples, from which statistically robust analyses can be acquired.

Recent LEAG/ExMag Sample SAT findings (ASR-SAT, 2026) highlight an impending critical lack of both laboratory facilities and trained personnel in the United States to carry out many of the highest-priority SPARX investigations. The SDT strongly recommends NASA address the findings of this report, particularly for the critical laboratory facilities required for SPARX-related investigations.

### 8.2.3 Sample Masses

The SPARX SDT arrived at calculations of baseline and threshold sample masses by considering first the number, grain sizes, and masses of specific samples required to address the SPARX objectives. The team then increased this requirement to address the NASA sample curation requirement, which stipulates that 75% of sampled materials returned to Earth must be allocated for curation and future investigations. Finally, the SDT applied a number of scaling factors intended to capture both the fractional likelihood of acquiring the targeted lithology at a specific sampling location, and an additional scaling factor reflecting the known variability of the lunar regolith in terms of particle size frequency distribution and agglutinate fraction. These masses were then applied across the 20 sampling locations of the baseline mission concept (10 for threshold) to arrive at the projected return sample mass estimates. Both baseline and threshold estimates are provided without including additional allocation for sample containers.

- **Baseline Return Sample Mass: ≥20.5 kg**
- **Threshold Return Sample Mass: ≥9.8 kg**

The SPARX TOR requirements for return sample mass indicate a NTE (not to exceed) mass of 100 kg (inclusive of sample containers) for mission designs that incorporate astronaut return from the lunar South Pole, and NTE 10 kg mass for commercial robotic return. The SPARX baseline mission architecture profile will be discussed in further detail below, however, the projected baseline and threshold return sample mass allocations are within the scope provided by the TOR. Additional information about the sample mass calculations can be found in Section 4.4.

## 8.3 MISSION IMPLEMENTATION ARCHITECTURES

The SPARX TOR directed the SDT to investigate a range of possible mission implementation architectures which may accomplish some set of the SPARX science priorities. The SDT was further directed to offer commentary on the viability of these architectures for achieving the SPARX objectives, with emphasis on the SPARX highest-priority objectives, before selecting a single architecture for the SPARX baseline mission concept. In defining these implementation architectures, the SDT considered the axis of mobility, ranging from static lander to long-distance (>100 km) rover, and sample return architecture, encompassing both robotic return and astronaut-assisted return. The implementation architectures are discussed in full detail in Section 5 and Section 7.5.

Accomplishing either the baseline or threshold SPARX objectives requires the ability to traverse to multiple different regions within SPA, distances significantly greater than 100 km. **A long-distance rover is the only implementation architecture capable of meeting the highest-priority SPARX objectives.**

Given the return mass envelope presented in the TOR, robotic sample return is a viable implementation option for the threshold SPARX mission; however, the lack of currently existing capabilities for either NASA-developed or commercial robotic sample return options increases the risk of committing to this return architecture. **The SDT determined that astronaut-assisted sample return is the preferred implementation for SPARX baseline sample return.** Samples from the circum-south polar/Moon Base region do not address any SPARX science objectives, but requiring to traverse this region for astronaut rendezvous does provide multiple additional engineering constraints for the SPARX rover. **A non-polar astronaut landing rendezvous with the SPARX rover at the terminal traverse location is the ideal SPARX architecture; although, designs can also accommodate rendezvous at a south polar basecamp/Moon Base location as shown in the design reference mission.**

## 8.4 SPARX DESIGN REFERENCE MISSION

Using the requirements set by the SDT, the JPL technical team created a design reference mission for a long-distance rover. The purpose of this mission design exercise was to enable an initial feasibility and cost assessment for meeting SPARX science goals. Detailed discussion of reference mission subsystems, including the sample acquisition subsystem, can be found in Section 6.

The baseline SPARX traverse from central SPA to the lunar south polar basecamp region is ~2,800 km and includes 20 science sampling locations and 1 science non-sampling traverse location. The total science traverse of the SPARX baseline mission is ~2,500 km, excluding the final traverse to the south polar basecamp. The SPARX threshold traverse includes 10 science sampling locations from central SPA to Poincare basin, and is ~1,500 km to the terminal sampling location, although traverses requiring south polar astronaut-assisted return are ~2,600 km.

## 8.5 COST

The cost numbers presented here represent the Phases A-E costs for JPL design reference mission, and assuming a class C+ risk posture. Following TOR guidance, instrument costs are treated as pass-through, with a flat-rate cost of $25M for each of the five strawman payload elements. The costs do not include surface delivery costs, sample return costs, curation costs, or other Phase F costs. The SPARX baseline mission cost is $840M in FY25 dollars, inclusive of reserves, which is well below the TOR recommended cap of $1.5B. Additional details about the cost modeling approach can be found in Section 6.10.

Areas with opportunity for cost savings or high risk of cost growth without sufficient planning are:

1. The SPARX threshold mission concept includes a descope of 1 instrument and 11 science (10 sampling) traverse locations. Depending on the sample return architecture, the traverse is up to ~1,300 km shorter. NASA should, however, strongly weigh the cost savings of these science and instrumentation descopes against any required changes to the sample return architecture.
2. International partnerships on contributed payload elements could decrease overall mission cost, as could the selection of high-heritage, mature payload designs.
3. For all possible baseline or threshold implementations, the SDT strongly recommends NASA commit to a sample return architecture at the earliest possible stage, particularly as robotic sample return designs will strongly drive rover design and will incur cost increases from those projected by the baseline mission design.

4. Although costing for surface delivery and sample return were not part of the SDT purview, the SDT nevertheless notes that, given successful demonstration, commercial providers could provide a cost-effective surface delivery mechanism.
5. The SDT strongly recommends NASA consider the operation space for non-polar astronaut landings, as the recommended mixed long-distance SPARX rover rendezvous with non-polar astronaut mission could provide cost savings by decreasing mission traverse length, overall mission duration, and decreasing engineering burden on certain rover subsystems.

## 8.6 FINAL REPORT CONCLUSIONS

The SPARX mission concept presented in this report leverages and highlights the confluence of several factors that have made sample return from SPA programmatically viable in the current exploration climate. Technological advancements in remote surface exploration technology, the rise of a U.S. commercial space industry focused on the lunar surface market, and the programmatic direction of human exploration and sustained presence at the lunar South Pole all play a factor in enabling a new generation of robotic lunar surface exploration. The goals and objectives defined by this report to present a scientific outline for a mission that would transform lunar science, would address multiple high-level Decadal science questions at the only place in the Solar System where they can be credibly addressed, would pioneer a new paradigm for planetary surface exploration in the 21st century, would leverage the burgeoning space economy to pursue ground-breaking science on a more cost-efficient basis, and would unite robotic exploration with human exploration in a model for a future sustained lunar enterprise.

## Acknowledgements:

The SPARX SDT would like to acknowledge several groups of people who contributed to the analysis and preparation of this report:

The SDT wishes to acknowledge the scientists and engineers at JPL who provided engineering design and support for the SPARX design reference mission, contributed material to the report, and provided subject matter expert briefings, including:

- Katarina Aguayo
- James W. Alexander
- Steven J. April
- Avak G. Archanian
- Andrija Bajovic
- John D. Baker
- Saptarshi Bandyopadhyay
- Dr. Ansel Barchowsky
- Charles F. Bergh
- Sriramya Bhamidipati
- Dr. Shyamkumar Bhaskaran
- Jeffrey J. Biesiadecki
- Dr. Diana L. Blaney
- Jonathan B. Blossom
- Robert L. Bocchino
- Arthur Bouton
- Joseph J. Bowkett
- Charles B. Brinkley
- Dr. Travis L. Brown
- G. Mark Brown
- David C. Bugby
- Brandon J. Burgett
- Brandon Burns
- Paige Butler
- Bailey E. Cassler
- Dr. Nacer E. Chahat
- Dr. Yang Cheng
- Jacob B. Chesin
- Subha Comandur
- Donald W. Cook
- Michael Cox
- Dr. Alex Davis
- Stephen F. Dawson
- Tyler Del Sesto
- Jeff H. Delaune
- Daniel I Demirdjian
- Ann S. Devereaux
- Daniel Duarte
- John O. Elliott
- Asher Elmland
- Dr. Todd A. Ely
- Blair J. Emanuel
- Nickolas Emis
- Carlos Esproles
- Michael J. Fong
- Daniel T. Forgette
- Hardeep K. Fuller
- Natalie Gallegos
- John C. Gallon
- Juan Garcia Bonilla
- Matthew J. Garcia
- Nikola Z. Georgiev
- Matthew Gildner
- Dr. Ashish Goel
- Jonathan F. Grinblat
- David C. Gruel
- Jonathan C. Hamel
- Corey C. Harmon
- Cathleen M. Harris
- David A. Henriquez
- Diego Hernandez Maciel
- Dr. Benjamin J. Hockman
- Dr. William A. Hoey
- Louise Jandura
- Michael R. Johnson
- Dr. John-Paul Jones
- Vladimir Jovovic
- Dr. Anup B. Katake
- Dr. James T. Keane
- Andrew Kennett
- Dr. Laura H. Kerber
- Shehryar Masaud Khan Khattak
- John S. Koehler
- Dr. Michael Kokorowski
- Daniel C. Kolenz
- Dr. Richard P. Kornfeld
- Todd M. Krafchak
- Joshua D. Krohn
- Dr. Damon F. Landau
- Ryan A. Lannom
- Kameron Larsen
- Dania D. Lazo
- David K. Lee
- Dr. Charles H. Lee
- Mallory S. Lefland
- Heather A. Lethcoe
- Daniel Limonadi
- Dr. Yang Liu
- Marcel Llopis
- Kevin D. Lo
- Christopher A. Lopez
- Corey A. Mack
- Dr. Mark W. Maimone
- Shantanu Malhotra
- Robert M. Manning
- Graydon W. Manzke
- John R. Mccann
- Ryan L. Mccormick
- Michael B. Meacham
- Patrick L. Meras
- Samuel E. Molleur
- Dhack Muthulingam
- Jeremy Nash
- Eleanor Naudzius
- Dr. Hari D. Nayar
- Dr. Issa A. Nesnas
- David E. Newill-Smith
- Dr. David Y. Oh
- Lupe Pacheco
- Dr. Kamrooz K. Parchamazad
- Dr. Michael Paton
- Jorge E. Quintas Del Campo
- Richard A.Rainen
- Dr. William Reid
- Kimberley D. Rizzo
- Esteben Rodriguez
- Preston Rogers
- Brandon Rothrock
- Dr. Nicolas F. Rouquette
- Alejandro M. San Martin
- Dr. Joel A. Schwartz

- Dr. Mazen M. Shihabi
- Milo C. Silverman
- Charna S. Stein
- Dr. Henry W. Stone
- Eric T. Sunada
- Rowan G. Swain
- Robert M. Swan
- Wilson Tang
- Kyle Thornburg
- Dr. Slava G. Turyshev
- Jose J. Uribe
- Carolina Uscategui
- Marsette A. Vona
- Gregory M. Waigand
- Jeffrey D. Waldman
- Dr. Kristopher T. Wehage
- Lindsay K. White
- Dr. Reginald G. Willson
- Kerry A. Wisden
- Yao Xu
- Garson J. Yee
- Michelle Zhong

The SDT especially wants to thank the JPL science liaison team, including Dr. Laura Kerber, Dr. Diana Blaney, and Dr. Yang Liu for their contributions to SPARX meeting discussions, and helpful reviews in the preparation of this report. The SDT further emphasizes the contributions Matt Gildner, Ben Hockman, Tyler Del Sesto, John Elliott, Issa Nesnas, Corey Harmon, Milo Silverman, Corey Harmon, Henry Stone, John Baker, and Subha Comandur who provided substantial contributions to the SDT.

The SDT wishes to acknowledge and thank the following community members for their valuable insights and discussion provided during team meetings and through correspondence with the team: Dr. Brad Jolliff, Dr. Lars Borg, Dr. Sonia Tikoo, Dr. Jared Epsley, Dr. Jacob Richardson, and Dr. Benjamin Weiss.

The SDT wishes to thank the officials in NASA’s Exploration Science Strategy Integration Office (ESSIO) who supported and guided the project, including Dr. Joel Kearns, Dr. Brad Bailey, Dr. Ryan Watkins, and Dr. Debra Needham.

Finally, the SDT gratefully acknowledges the work and contributions from the team at NRESS who organized and supported the SPARX meetings, including Annie Rattanaphone and Nicy Ross, as well as the rest of their team members. This SDT would not have been possible without the efforts of Justin Mathurin and the NRESS team, who found a way to ensure the study could be completed under all of NASA's desired requirements.

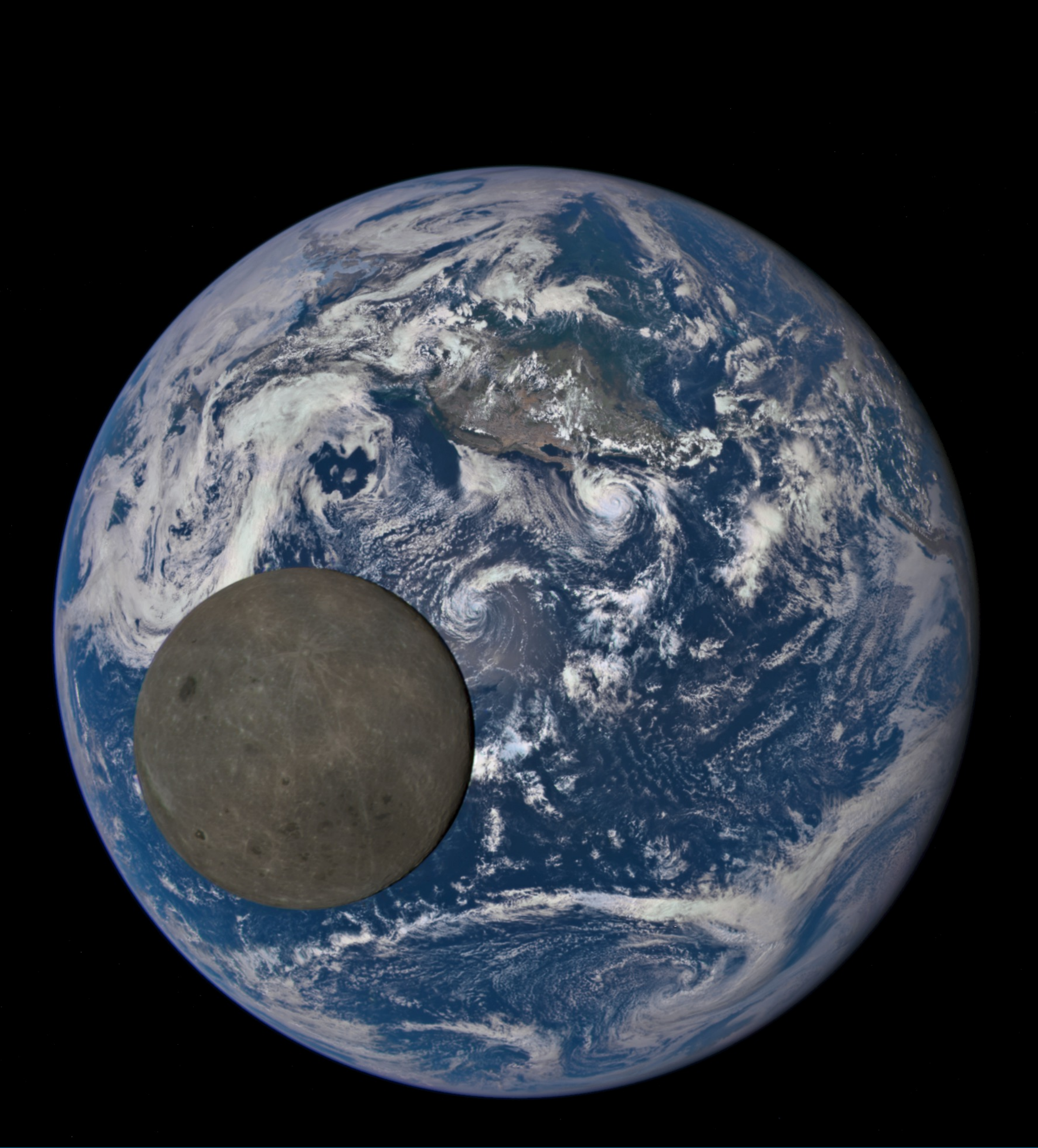

**The farside of the Moon, and the South Pole-Aitken basin, as viewed from the Deep Space Climate Observatory.**
Image credit: NASA / NOAA.

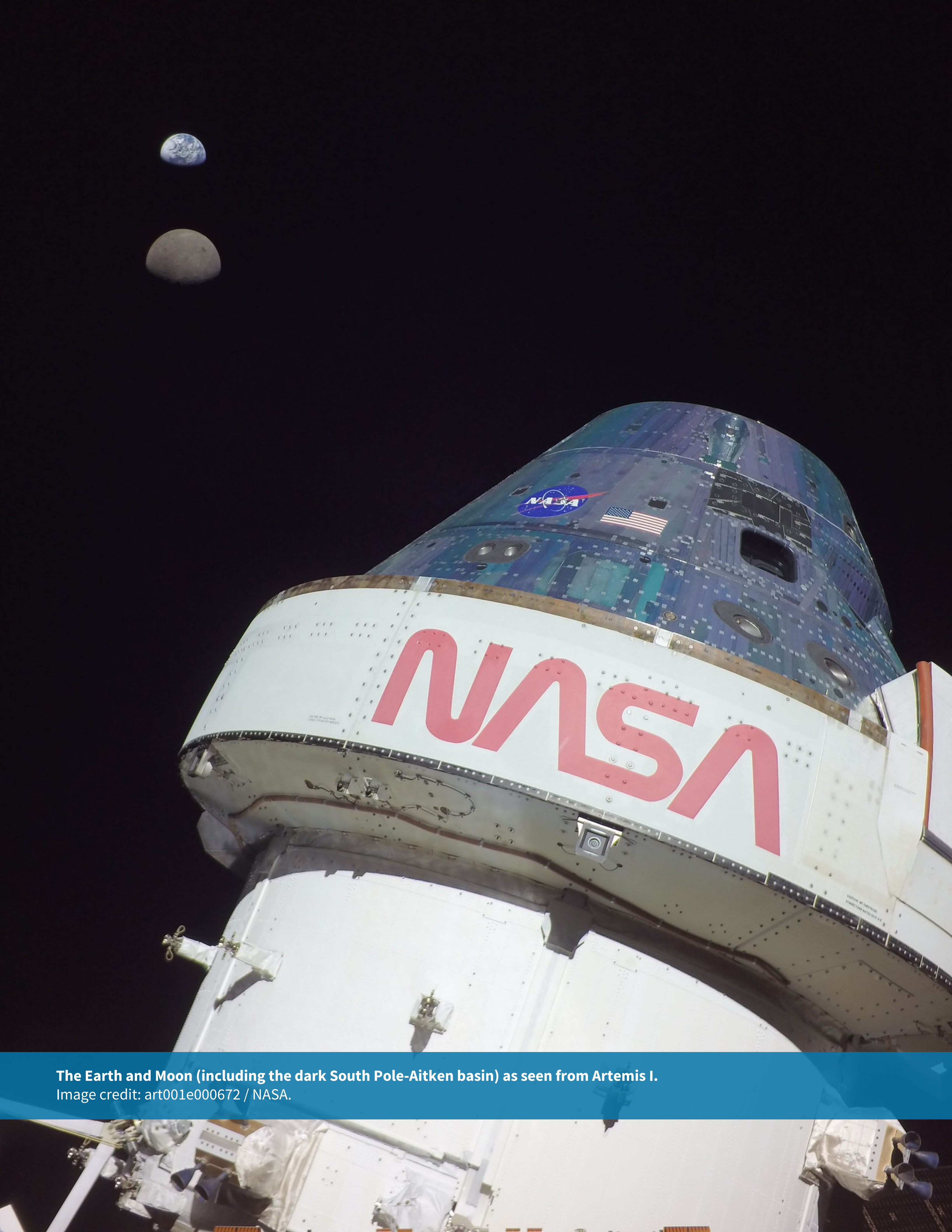

**The Earth and Moon (including the dark South Pole-Aitken basin) as seen from Artemis I.**
Image credit: art001e000672 / NASA.

# 10 APPENDICES

The following sections include supporting material for the SPARX SDT report, summarized in Table 10.1.

**Table 10.1.** *Order and contents of the SPARX appendices.*

| Section | Description |
|---|---|
| Section 10.1 | NASA's Terms of Reference (TOR) for the SPARX SDT |
| Section 10.2 | Discussion of science prioritization for goals |
| Section 10.3 | Discussion of sample analysis techniques |
| Section 10.4 | Additional mission architectures considered by the SPARX SDT |
| Section 10.5 | Discussion about Non-polar Artemis rendezvous architectures |
| Section 10.6 | Compilation of all SPARX findings |
| Section 10.7 | SPARX SDT membership, including CVs |

## 10.1 SPARX SDT TERMS OF REFERENCE

The following six pages include the Terms of Reference (TOR) for the SPARX SDT.

# **South Pole-Aitken Basin Sample Return and eXploration (SPARX) Science Definition Team (SDT) Terms of Reference**

## 1. Background and Scope

The recent planetary science decadal survey, *Origins, Worlds, and Life: A Decadal Strategy for Planetary Science and Astrobiology 2023-2032* (OWL), identified Endurance as the highest priority strategic mission for NASA's Lunar Discovery and Exploration Program. Endurance is a mission concept that would utilize Commercial Lunar Payload Services (CLPS) to explore and return samples from strategic locations across the Moon's South Pole–Aitken Basin (SPA), the largest and oldest impact basin on the lunar surface.

In response to the OWL recommendation, NASA is convening a Science Definition Team (SDT) to provide an analysis to the Deputy Associate Administrator for Exploration (DAAX) on prioritized science objectives and requirements and different implementation approaches for a South Pole-Aitken Basin Sample Return and Exploration mission. Specifically, the SDT shall:

- Define and prioritize science goals for a South Pole-Aitken Basin Sample Return and Exploration mission that is traceable to the science questions and objectives outlined in the OWL and other community documents such as, but not limited to, the Scientific Context for the Exploration of the Moon (SCEM) Report;
  - The committee shall consider and prioritize scientific objectives that are relevant to both sample return and in-situ measurements within SPA.
- Define a hierarchy of science objectives that are responsive to the prioritized goals above and identify the criteria that will enable these objectives to be met;
- Discuss the criteria used to prioritize the science goals and objectives;
- Determine example location(s) in SPA where the science goals can be met;
- Define the number and type of samples and in situ measurements that are needed to address the prioritized science objectives. The committee should specify key measurements, if any, that must be made in-situ or on the lunar surface as well as those that must be made in terrestrial labs in order to achieve the identified science objectives;
  - The committee shall also note the minimum expected number of kilograms of samples required to meet the baseline and threshold objectives.
- Define a set of strawman science instrument types that fully address the SDT-defined science objectives, as well as their required measurement capabilities;
  - At the proper time NASA will determine the final science payload suite for the mission.
- Develop a Science Traceability Matrix (STM) that flows from the top-level science goals through science objectives and derived measurements, culminating in functional requirements and capabilities for the strawman payload suite, mobility system (if applicable), and sample return system;
- Assess the role of astronauts in accomplishing these science goals and objectives, plausibly including triage of samples, in situ analyses, or other EVA activities.

- Determine what technology(ies) need to be matured before the project starts;
- Define and present implementation approaches to achieving the proposed science objectives, considering a variety of approaches that allow different levels of the hierarchy of science objectives to be met.
    - The committee should identify the key discriminators between the various implementation approaches and outline the rationale for why these approaches are best suited to meeting the science objectives.
- From the range of implementation approaches, the SDT shall develop one nominal mission concept that maximizes the overall science return and progress towards NASA's goals within the resources and risk posture constraints provided by NASA (see below).
    - For this mission concept, the SDT shall develop both a baseline and threshold mission.
    - For both the baseline and threshold mission, the SDT shall prioritize payload options and describe priorities for scaling the mission concept up or down (in cost and capability) and describe potential trades.

The inputs listed above shall be provided individually by members of the SDT and will be compiled into a single non-consensus summary report upon completion of the task.

The South Pole-Aitken Basin sample Return and eXploration (SPARX) SDT should not necessarily assume the Endurance implementation or science goals as described in the Decadal Survey concept study, but should instead prioritize SPA Sample Return and Exploration goals and objectives based on community documents and input, and then present a range of science implementation approaches based on how these objectives can best be achieved. The SDT will work concurrently with a technical team located at NASA's Jet Propulsion Laboratory who will assist with assessing implementation options.

The SDT's analyses will be used by NASA to inform decisions on the prioritized science goals and implementation of a near-term SPA Sample Return and Exploration mission, including through both human and/or robotic implementation paths. Any findings involving Artemis human exploration-related resources would be used by NASA to assess whether the identified science goals are best supported by Artemis.

## 2. Factors to Consider

A few factors and constraints the SDT committee should consider are listed below. The SDT shall assume:

- A cost cap of $1.5B (FY25), not including the lunar delivery service, for Phases A–E (including margin and reserves). Costs should be broken down by mission phase.
    - Curation and Earth-based sample analysis will occur during Phase F and is not included in the Phase A-E cost cap and associated schedule. However, the SDT shall document what analyses need to be done on the sample(s) in order to satisfy the STM requirements. The SDT will not be asked to generate a cost/schedule estimate for Phase F.

- A competed instrument suite with a nominal cost limit of $75M (FY25; including margin and reserves) and a notional mass limit of 20 kg. Note this cost limit is for instruments only; other components such as sample acquisition and retrieval mechanisms are not included in this cost limit.
- A launch nominally in 2031.
- A nominal mission duration of 2-4 years, depending on the scope of the science objectives.
- A maximum roving speed of 1 kilometer per hour, if roving is included as part of any implementation approaches.
- That any proposed sample return utilizes returning samples either to Artemis astronauts at the south pole, or via commercial-derived robotic sample return capabilities. The SDT should state and justify all sample needs.
    - Artemis sample return capability shall not exceed 100 kg, inclusive of sample containers.
    - CLPS-derived robotic sample return capability shall not exceed 10 kg.

Additionally,

- The SDT may consider the use of Radioisotope Power Systems (RPS).
  The SDT will work iteratively with JPL mission pre-project which will provide additional constraints and information on payload size, mass, power, data rates, configuration, and cost.

Detailed landing site selection and traverse planning is outside the scope of this SDT, but the SDT should identify key sites of interest for achieving the science objectives.

## 3. Final Deliverables

The SDT shall deliver a final non-consensus report to the DAAX that must contain prioritized science goals, a hierarchy of science objectives for a South Pole-Aitken Basin sample return and exploration mission, a Science Traceability Matrix (STM), an assessment of the various architectural options for executing the mission, and all supporting rationale. This final report must reflect all viewpoints of the committee members as individuals but will be compiled as a single summary non-consensus document.

The SDT shall report on its progress on a regular basis to NASA Headquarters management. Results from the SDT efforts will help NASA develop a strategy and implementation plan for South Pole-Aitken Basin Sample Return and Exploration. The SDT will produce a final presentation upon completion of work that will be verbally presented to SMD and other NASA stakeholders, followed by the provision of a final written report. This final report is due to NASA Headquarters no later than 6 months after the conclusion of the SDT unless otherwise negotiated before the end of the study, and will be made public upon completion.

## 4. Management

The SPARX SDT will deliver a final non-consensus report to the Deputy Associate Administrator for Exploration (DAAX), Exploration Science Strategy and Integration Office (ESSIO), Science Mission Directorate (SMD), NASA Headquarters. The NASA point of contact for this SDT is Dr. Ryan Watkins, ESSIO Program Scientist (ryan.n.watkins@nasa.gov) and the back-up point of contact is Dr. Debra Needham, ESSIO Program Scientist (debra.m.hurwitz@nasa.gov ).

SMD will assure the necessary financial and personnel support for the SDT. The SDT Chair and the ESSIO Program Scientist will run all activities of the SDT and ensure the quality of review deliverables.

## 5. Notional Timeline

The SPARX SDT will conduct their analyses over an approximately 12–18 month period from initial meeting to completion of the study, with the final non-consensus analysis report to be delivered no later than 6 months after the conclusion of the last meeting. The SDT shall meet approximately 4–6 times, or more as required by the DAAX, over the course of these 12–18 months. These meetings will be a mix of virtual and in-person/hybrid, likely at JPL or at a location TBD based on the geographic distribution of the members. Each meeting will last 2–3 days, or longer as determined by the SDT members and upon agreement by NASA. Meetings will not be open to the public, although the SDT may choose to solicit community input to help guide their discussions. SDT members who are eligible will receive compensation for their time and participation at each committee meeting.

Additional ad hoc virtual meetings (e.g., weekly telecons, subgroup meetings, etc.) may be required in order to complete the study in the anticipated timeline. The frequency of such ad hoc meetings will be decided upon by the Chair, with concurrence from the NASA POC(s), and will not be be included in the compensation rate.

| Milestone | Timing* | Format |
|---|---|---|
| Meeting 1 | ~January 2025 | Hybrid |
| Meeting 2 | ~March 2025 | Virtual |
| Meeting 3 | ~ May 2025 | Hybrid |
| Meeting 4 | ~ July 2025 | Virtual |
| Meeting 5 | ~ September 2025 | Hybrid |
| Meeting 6 | ~ November 2025 | Virtual |
| Draft Analysis Report to NASA | ~ January 2026 | |
| Incorporate NASA feedback into Report | Jan-April 2026 | |
| Analysis Report Due | NLT April 2026 | |
| SDT Briefing to NASA | April 2026 | |

*all timelines are tenatitve and subject to change at the discretion of NASA

## 6. Anticipated Membership Details

The SDT shall consist of 15 members to serve for the duration of the SDT, including one member who will serve as Chair and three who will represent international space agencies. NASA shall ensure a balanced representation in terms of the points of view represented and the functions to be performed, as well as diversity of team members (e.g., career stage, demographics, professional background, institutional representation, etc.). The SDT will also have up to two early career (<10 years post-terminal degree) and/or current graduate student

Documentarians who will assist with duties such as scheduling meetings, note-taking, and documentation during committee discussions. Documentarians are also encouraged to participate in discussions. NASA plans to include 3 international space agency representatives to be selected by their host agency. These international representatives will be knowledgeable about lunar science and able to speak both to SPA Basin Sample Return and Exploration objectives and on the behalf of their host agency.

Members of the SDT will be required to sign non-disclosure agreements and will not be conflicted from proposing to SPA sample return mission-related solicitations (e.g., instrument proposals) in the future. The timeline for soliciting instruments and/or mission team members has not yet been decided.

Subject matter expert (SME) areas to be covered with SDT membership:

- Geochronology
- Impact processes
- Geochemistry, mineralogy, and petrology
- Geology
- Geophysics
- Volcanism
- Sample Analysis
- Remote sensing
- Volatiles
- Instrumentation
- Rover science and operations

Recommended by:

Ryan Watkins Digitally signed by Ryan Watkins Date: 2024.10.07 11:56:14 -04'00'

Dr. Ryan Watkins
Program Scientist and SDT Point of Contact
Exploration Science Strategy and Integration Office
Science Mission Directorate

Concurrence:

Joel Kearns Digitally signed by Joel Kearns Date: 2024.10.07 12:01:24 -04'00'

Dr. Joel Kearns
Deputy Associate Administrator for Exploration
Exploration Science Strategy and Integration Office
Science Mission Directorate

Approval:

Nicola Fox Digitally signed by Nicola Fox
Date: 2024.10.14
06:12:45 -04'00'

Dr. Nicola Fox
Associate Administrator
Science Mission Directorate

## 10.2 SCIENCE PRIORITIZATION OF GOALS

The three goals were prioritized numerically, where Goal 1 represents the highest priority science to address with the future mission. The objectives within each goal framework were also prioritized numerically. Each goal and objective were given a score from 1 to 5 based on four categories: Traceable Science, Science Environment, Programmatic Environment, and Implementation Considerations.

Traceable Science corresponds to Explicit Decadal Level Science, as outlined in the OWL (National Academies of Science, 2023), and transformational science, corresponding to the degree to which the described science investigation can make significant progress on the proposed goal or objective. Science Environment incorporates three factors, namely science strongly enabled by visiting SPA, science unique to SPARX given the samples returned by the Chang'e-6 mission, and the ability to achieve the goal's science without returned samples. Programmatic Environment incorporates three factors: the applicability of the goal/objective supporting NASA's Moon to Mars Architecture (NASA, 2023), applicability to the Lunar Science Community Objectives (e.g., Scientific Context for the Exploration of the Moon (National Research Council, 2007)), and science unique to SPARX in the context of the Artemis surface science goals (NASA, 2023). Finally, Implementation Considerations incorporates three factors: confidence that the measurement and/or sample will address the desired science goal/objective, confidence in identifying the correct location to acquire the sample, and ease in acquiring the required sample and its proximity to other sampling locations.

The goal rankings were assigned based on the majority vote rather than an average of the votes. Goals were voted on by stating a numerical value representative of its ability to achieve the categories described above, where: 5 = strongly achieves the stated goal (excellent); 4 = achieves the goal well with some caveats (very good); 3 = archives the goal but caveats are present (good); 2 = concerns present to achieve the stated goal (fair); 1 = does not address the stated goal (poor). In contrast to the majority votes reported for goals, averages were reported for the objectives because the quantity of objectives far exceed the quantity of goals.

In the cases of non-consensus votes on prioritization scores, team members with different scores volunteered their rationale for their voting score during discussion, and these discussion points are summarized below. In some cases, a revote was undertaken after discussion. The scores post-re-vote are presented in this document and in Figure 3.17 and 3.18. The results of the majority votes ranked the Goals in order as Goal 1, Goal 2, and Goal 3. The results of the average votes ranked the Objectives in order of 1.3, 1.1, 2.2, 2.1, 2.3, 3.2, 3.1, 1.2, 1.4.

The final goal and objective rankings reflect the totality of discussions held by the SDT, guided by the voting matrix. The SDT recognizes that in several places the prioritization does not match the presented numerical values, in these cases the SDT recognized that specific aspects of the scientific importance of each goal or objective should be weighted more heavily. For example, the SDT determined that the scientific importance of Goal 1 outweighed the programmatic difficulty of acquiring the specific samples to address these objectives.

The following text summarizes the discussion and rationale pertaining to SDT member scoring of goals and objectives.

| | S.G.1 What was the chronology of key processes and events during the first half billion years of Solar System history? | S.G.2 Does the lunar farside preserve evidence of asymmetric magma ocean evolution and planetary crustal formation processes? | S.G.3 What are the effects of mega-basin impacts on terrestrial bodies? |
|---|---|---|---|
| Traceable Science: | | | |
| **Explicit Decadal Level Science** | U | U | U |
| **Transformational Science (degree to which described investigation can make significant progress on proposed objective)** | vote 9-4 | vote 10-2 | vote 8-4 |
| | | | |
| Science Environment: | | | |
| **Science strongly enabled by SPA** | U | U | U |
| Science unique to SPARX given the samples previosuly returned by Chang'E 6 | vote 9-5 | vote 8-7 | U |
| Ability to achieve science without returned sample | U | vote 8-5 | vote 8-4 |
| | | | |
| Programmatic Environment: | | | |
| Applicability to Moon to Mars Objectives | U | U | vote 10-2 |
| Lunar Community Science Objective (e.g. SCEM) | U | U | U |
| *Science unique to SPARX in the context of Artemis science goals* | U | vote 8-7 | vote 7-6 |
| | | | |
| Implementation Considerations: | | | |
| Confidence that the measurement/sample will address the desired science | vote 8-5 | vote 7-5 | vote 7-5 |
| Confidence in identifying the correct location to acquire the sample | vote 10-3 | vote 7-5 | vote 6-5 |
| Ease in acquiring required sample - proximity to other sampling locations | vote 7-5 (one 3) | vote 7-5 | vote 7-5 |

| Categorization Guide |
|---|
| 5 |
| 4 |
| 3 |
| 2 |
| 1 |

**Figure 10.1.** *Prioritization of Goals. The panel votes are recorded. U = unanimous vote.*

| | Obj. 1.1 Quantify the age of the Moon to within ± 30 Ma to distinguish between Early Moon and Late Moon formation models | Obj. 1.2 Distinguish between fast-cooling or slow-cooling models for lunar magma ocean crystallization | Obj. 1.3 Determine if and when a giant planet instability drove a sharp change in ISS impactor flux | Obj. 1.4 Constrain the evolution of the lunar dynamo in the pre-4.0 GA epoch | Obj. 2.1 Determine if the farside highlands crust is similar to nearside crust in age and/or composition | Obj. 2.2 Determine if there was a farside KREEP reservoir or other late stage magma ocean product prior to the SPA impact | Obj. 2.3 Determine if and how the temporal and compositional evolution of the farside mantle differs from the nearside | Obj. 3.1 Distinguish between models for the SPA forming impact based on asymmetries in basin structure and materials | Obj. 3.2 Distinguish the relative contributions of impact melt differentiation, crustal backfilling, or volcanic infilling to the interior composition of SPA |
|---|---|---|---|---|---|---|---|---|---|
| Traceable Science: | | | | | | | | | |
| **Explicit Decadal Level Science** | 5.00 | 4.92 | 5.00 | 4.92 | 4.92 | 4.92 | 4.92 | 4.85 | 4.54 |
| **Transformational Science (degree to which described investigation can make significant progress on proposed objective)** | 4.85 | 4.31 | 4.92 | 3.77 | 4.77 | 4.77 | 4.69 | 4.00 | 4.15 |
| | | | | | | | | | |
| Science Environment: | | | | | | | | | |
| **Science strongly enabled by SPA** | 4.69 | 4.31 | 4.69 | 3.31 | 3.85 | 4.85 | 4.54 | 4.85 | 4.92 |
| Science unique to SPARX given the samples previosuly returned by Chang'E 6 | 3.92 | 4.00 | 4.15 | 4.46 | 4.77 | 4.62 | 3.85 | 4.92 | 4.85 |
| Ability to achieve science without returned sample | 1.00 | 1.00 | 1.31 | 1.77 | 1.23 | 1.62 | 1.31 | 3.08 | 2.38 |
| | | | | | | | | | |
| Programmatic Environment: | | | | | | | | | |
| Applicability to Moon to Mars Objectives | 5.00 | 4.92 | 5.00 | 4.54 | 4.85 | 5.00 | 4.92 | 4.77 | 4.54 |
| Lunar Community Science Objective (e.g. SCEM) | 5.00 | 5.00 | 5.00 | 5.00 | 5.00 | 5.00 | 5.00 | 4.92 | 4.85 |
| *Science unique to SPARX in the context of Artemis science goals* | 4.08 | 3.85 | 4.15 | 4.00 | 3.69 | 4.54 | 4.54 | 4.23 | 4.85 |
| | | | | | | | | | |
| Implementation Considerations: | | | | | | | | | |
| measurement/sample will address the | 4.31 | 3.77 | 4.38 | 3.62 | 4.69 | 4.46 | 4.85 | 3.77 | 4.15 |
| Confidence in identifying the correct location to acquire the sample | 4.08 | 4.00 | 4.23 | 3.85 | 4.23 | 4.54 | 4.92 | 3.92 | 4.15 |
| Difficulty in acquiring required sample - proximity to other sampling locations | 4.69 | 4.54 | 4.62 | 3.69 | 4.08 | 3.46 | 4.77 | 2.85 | 3.62 |
| | | | | | | | | | |
| Averages: | **4.24** | 4.06 | **4.31** | 3.90 | 4.19 | **4.34** | 4.39 | 4.20 | 4.27 |
| Weighted Averages | **4.56** | 4.29 | **4.61** | 3.95 | 4.36 | **4.61** | 4.56 | 4.39 | 4.41 |

**Figure 10.2.** *Prioritization of Objectives. The average scores are presented in the table.*

### 10.2.1 Goals

#### 10.2.1.1 Goal 1: What was the chronology of key processes and events during the first half billion years of Solar System history?

Traceable Science:

Explicit decadal level science: unanimous.

Transformational science: nine voted 5, four voted 4. Some team members voted a score of 4 rather than 5 because the potential science results rely on statistical argument for ages of samples, LMO hypotheses, and ages of crystallization, and some degree of ambiguity in age or distinguishing between different models given the reliance on statistics. In addition, even with high-quality samples, there's a chance the events won't be resolvable from statistics or models, or events may be unresolvable.

Science Environment:

Science strongly enabled by SPA: unanimous.

Science unique to SPARX given Chang'e-6: eight voted 4, five voted 5. Most team members voted a score of 4, citing Chang'e 6 likely producing at least one possible SPA age, so the SPARX results would not be entirely unique. Some team members voted a 5 because Chang'e 6 landed in a young basalt and impact basin, so the SPA age obtained is still inconclusive, and because SPARX would fill the critical gap of having multiple ages from multiple locations to confidently assign an age to the SPA impact event.

Ability to achieve without returned samples: fourteen voted 1, one voted 0. One team member voted 0, of the opinion that the mission would not be worth flying if samples were not returned, given returned samples are critical for chronology of processes on the Moon early in solar system history.

Programmatic Environment:

Applicability to Moon to Mars objectives: unanimous.

Relevance to lunar community science objectives: unanimous.

Uniqueness of SPARX compared to Artemis surface science goals: initially seven voted 5, four voted four, and three voted 3. After revoting, unanimous vote of 4. Half voted either 3 or 4 citing optimism that there will be sufficient diversity of materials at the Artemis site and returned sample volume to obtain SPA impact material. Others voted 5 because they did not expect to find enough simple material from early bombardment locations other than SPA which are necessary to establish the timing of impact flux early in the Moon's history. After a revote, the consensus across the group was a vote of 4.

Implementation Considerations:

Confidence measurement/sample will address the desired science: nine voted 4, four voted 5, one voted 3. After revoting, eight voted 4, five voted 5. Those who voted four were not convinced it is possible to untangle the resetting story, stated concerns about collecting sufficient quantities of impact melt, and that the current Apollo and Chang'e samples already leave room for some interpretation. Those in favor of a higher score stated that regardless of the Moon's behavior, this sampling campaign will establish what to expect from the Moon when collecting this quantity of samples. A diversity of samples via a SPARX-like infrastructure is the best chance we have at achieving this goal. The revote yielded a non-consensus result but was reflective of the discussion ahead of the revote.

Confidence in identifying correct location for sample: ten voted 4, three voted 5. Those who voted 5 stated the locations where we'd sample floatation crust are clear and that going to multiple locations yields a high likelihood of sampling impact melt. Others stated it is not a certainty we will collect impact melt from multiple locations, which is needed to accomplish this goal with high certainty.

Ease in acquiring required sample / proximity to other locations: seven voted 5, five voted 5, and zone voted 3. Some members voted a score of 4 because collecting ferroan anorthosites of sufficient diversity and quality is difficult. To do so, the rover would need to travel to central SPA and a pre-Imbrian basin which is a long traverse; a short traverse would not accomplish this goal as stated. Those who voted a score of 5 acknowledged the difficulty of collection but were more optimistic given there are detections of ferroan anorthosites in a lot of places.

#### 10.2.1.2 Goal 2. Does the lunar farside preserve evidence of asymmetric magma ocean evolution and planetary crustal formation processes?

##### Traceable Science:

Explicit decadal level science: unanimous.

Transformational science: there were ten votes were for 5 and two votes for 4. One voter for 4 cited the decadal survey, where asymmetry appeared in Chapter 8, rather than Chapter 4, thus lowering their voting score.

##### Science Environment:

Science strongly enabled by SPA: unanimous.

Science unique to SPARX given Chang'e-6: was a split vote with eight votes for 4 and seven votes for 5. No discussion was noted.

Ability to achieve science without returned samples: vote was recorded as 8:5 in favor of 1 versus 2. One voter for 2 cited investigations listed in the STM where modal mineralogy factored into the measurements and instruments on the rover, which could potentially address the science without samples. Those in favor of a 1 felt the issue relies on isotopic and trace element analyses, and those measurements cannot be completed without returned samples.

##### Programmatic Element:

Applicability to Moon to Mars objectives: unanimous.

Relevance to lunar community science objectives: unanimous.

Uniqueness of SPARX compared to Artemis surface science goals: there were eight votes for 4 and seven votes for 5. One felt the south pole appears different from the farside in composition, so we need to visit SPA locations to address the farside component. One felt if basalts were not collected, questions regarding the Th anomaly cannot be addressed, therefore in favor of a vote for 5. One felt SPA ejecta is likely to be collected at the south pole, in favor of a vote for 4.

##### Implementation Considerations:

Confidence the measurement/sample will address desired science: there were seven votes in favor of 5 and five votes in favor of 4. One posed to the '4' voters, where else could we go that is better than where we are targeting? One suggested farther away from the center of the basin, and we don't know if the material is FANS or low iron-bearing material. One was unsure how much the SPA impact impact modified samples of the crust within the basin to the extent that they would be less useful than samples from outside the basin.

Confidence in identifying correct location for sample: there were seven votes for 5 and five votes for 4. One felt that it would not be trivial to locate Th rich material with the orbital data. Anorthosite locations cannot be identified easily. One felt that Th hotspots can be identified well, and with an onboard GRNS, we can refine locally where to sample. There is an issue that Th could be point sources. One felt that going to a Th hotspot guarantees Th rich material in the returned sample based on the Apollo 14 returned samples.

Ease in acquiring required sample / proximity to other locations: there were seven votes for 4 and five votes for 5. One voter felt some basalts are reasonably close but the Th spots are farther away and requires more travel for one aspect of the goal.

#### 10.2.1.3 Goal 3. What are the effects of mega-basin impacts on terrestrial bodies?

##### Traceable Science:

Explicit decadal level science: unanimous.

Transformational science: there were eight votes in favor of 5 and four votes in favor of 4. One felt that those outside lunar science may not feel strongly about the goal and its two objectives. One felt less interest in answering theories of an oblique impact. Others argued that this is the most accessible location to learn about megabasin formation and how they affect planets, and that it can translate well to understanding other solar system bodies.

##### Science Environment:

Science strongly enabled by SPA: unanimous.

Science unique to SPARX given Chang'e-6: unanimous.

Ability to achieve science without returned samples: there were 8 votes in favor of 3 and four votes in favor of 2. A few voters argued that surface compositional data would be useful to learn about what the interior SPA material is composed of.

##### Programmatic Environment:

Applicability to Moon to Mars objectives: there were ten votes in favor of 5 and two votes in favor of 4. One posed the question: Could this objective address habitability? The voter felt that could be the case, leading to a higher vote.

Relevance to lunar community science objectives: unanimous.

Uniqueness of SPARX compared to Artemis surface science goals: There were seven votes in favor of 5 and six votes in favor of 4. Artemis 3 will not provide information about the interior of the basin. SPARX will not go to the outside of the basin, but Artemis is. Compositional data and impact models suggest we should sample ejecta, but it may not be abundant.

##### Implementation Considerations:

Confidence the measurement/sample will address desired science: There were seven votes in favor of 4 and five votes in favor of 3 after a revote. There was disagreement in whether there would be a "slam dunk" answer to this goal. There is more ambiguity than in Goal 2. One felt that for the objective related to basin structure, they were not confident that significant headway would be made without a complete traverse in and out of the basin. Additionally, one felt that for 3.2, it has not been clearly shown that impact melt sheets strongly differentiate. Since this is not a geophysics mission, this was felt to be a weaker part of what the mission could accomplish. There was a revote with no further discussion.

Confidence in identifying correct location for sample: There were six votes in favor of 4 and five votes in favor of 5. One felt the gravity traverse would be uncertain for some massifs, specifically surrounding the ambiguity in relation to crater rims or other impact processes. Others felt that for Objective 3.2, the STM clearly defined where to get the materials (in ejecta blankets or resurfaced volcanic areas), and there likely is not a better spot to accomplish this goal.

Ease in acquiring required sample / proximity to other locations: There were seven votes in favor of 3 and five votes in favor of 4. One voter felt sampling would be accomplishable and it would be relatively easy to sample ejecta; however, geophysical measurements are more difficult. Any measurement requiring gravity would be difficult to accomplish in terms of climbing massifs.

#### 10.2.1.4 Summary

After ranking the goals based on the above factors, Goal 1 and Goal 2 were both ranked highly. Another vote was conducted, and Goal 1 was unanimously selected as the highest priority goal. The second-most important goal was unanimously decided as Goal 2. Therefore, the goals ranked by highest priority are ordered as Goal 1, Goal 2, and Goal 3.

### 10.2.2 Objectives

Each panel member recorded their scores for the Objectives within individual spreadsheets, and these votes were compiled to form an average score for each Objective. The Objectives were ranked according to the same factors applied to the Goals. Based on the averaged rankings for Goal 1, Objectives 1.3 and 1.1 were both selected as having high importance. One member argued that Objective 1.3 was higher priority than Objective 1.1 because 1.3 is most applicable to other solar system bodies.

Members discussed the ordering of Goal 2's Objectives. While Goal 2 is high priority, it does not have as high solar system science applications compared to Goal 1, so the Goal 2 Objectives are ranked below the highest priority Goal 1 objectives. Objective 2.2 was voted as most important. Objectives 2.1 and 2.3 were nearly equal based on the scores, and a vote was conducted. Seven votes were in favor of 2.1, and four votes were in favor of 2.3. After these discussions, Goal 2's Objectives were ranked in importance as 2.2, 2.1, and 2.3.

Members discussed the importance of Goal 3's Objectives with 3.2 ranking higher than 3.1 because 3.2 is a traverse-driver and most complimentary to Artemis. One felt the more interesting science questions pertain to 3.2, and that this objective could be easier to answer. After considering that 3.2 most closely aligns with lunar community priorities, it was ranked higher priority than 3.1.

The Objectives prioritization was finalized as 1.3, 1.1, 2.2, 2.1, 2.3, 3.2, 3.1, 1.2, 1.4.

## 10.3 SAMPLE ANALYSIS TECHNIQUES

The following appendix presents a list of the proposed sample analysis techniques listed by the SDT in Section 3.3. The technical aspects of this material are largely summarized in Tables 4.3–4.7, and this list presents a brief narrative description of the techniques, their uses, and (where applicable) notes about relevant sample masses and preparations.

### 10.3.1 Petrologic Sample Characterization Techniques

#### 10.3.1.1 Visible Light Microscopy

Visible light microscopy is one of the first techniques used to triage and document the state of returned samples. Samples are visually examined using either a binocular microscope for rough particles (centimeters to millimeter in size), or a petrographic microscope for polished sections. No sample preparation is required for studying samples using a binocular light microscope, making it a useful, non-destructive preliminary examination tool. To study samples using a petrographic microscope in transmitted and/or reflected light, samples must be flat. This sample preparation involves grinding and mechanical polishing. Petrography remains the fundamental tool for classifying lunar rock types—such as mare basalts or anorthosites—and interpreting their volcanic or impact histories. Features down to several micrometers in size can be imaged routinely using a petrographic microscope.

#### 10.3.1.2 Structured Light Scanning

Structured light scanning projects a known pattern of light onto a sample's surface and calculates 3D coordinates based on the deformation of that pattern as it hits the sample. This technique allows for the creation of high-fidelity digital photographs of lunar samples (e.g., Ryan et al., 2026). Structured light scanning also allows determination of sample volumes (typically 1% uncertainty in derived volumes) that combined with masses can lead to highly precise densities. Structured Light Scanning was used by the OSIRIS-REx team (Lauretta et a., 2024) and provided data on samples that were not scanned using Micro-XCT. This method is a simple, robust, and rapid technique that can provide basic information used to identify lithologies in the triage stage using density information.

#### 10.3.1.3 Micro-XCT (Micro-Computed Tomography)

Micro-XCT is a non-destructive imaging method that utilizes X-ray attenuation to create high-resolution, 3D internal density maps of a sample. Micro-XCT allows researchers to peer inside samples without physically cutting into them. As X-rays pass through the sample, they are absorbed differently by various minerals and voids. A high energy X-ray beam passes through the sample while the sample is rotating on the stage. The X-ray attenuation at a given position is recorded by a CCD detector as a 2D projection. These 2D projections are then reconstructed into a 3D volume through computational algorithm. Because X-ray attenuation is sensitive to both density and composition, sample textures and components (e.g., vesicles in basalts, dense mineral clasts in breccias) can be readily identified and further quantified. Micro-XCT provides a voxel precision typically ranging from 1 to 100 µm, depending on the size of the sample being scanned. Micro-XCT is also used to target areas of interest within a sample, like a coarse-grained clast in a breccia, for creating polished sections ("serial sectioning"), helping to minimize the loss of material (e.g., Eckley et al., 2025).

#### 10.3.1.4 Micro-XRF (X-ray Fluorescence)

Micro-XRF utilizes a primary X-ray beam to excite atoms in the sample, causing them to emit secondary fluorescent X-rays characteristic of specific elements. The intensities of these X-rays are measured to create elemental maps of large sample areas, helping researchers locate mineral phases or quantify bulk chemical trends. Micro-XRF is a non-destructive technique and can be conducted on both rough and polished surfaces (e.g., Cohen et al., 2025; Crow et al., 2024). It offers a spatial resolution on the order of 10s of µm and detection limits in the parts-per-million (ppm) range for some elements.

## 10.3.2 Additional Sample Characterization Techniques

### 10.3.2.1 Scanning Electron Microscopy (SEM)

SEM is a high-resolution imaging technique used to characterize the surface morphology and textural relationships of samples (e.g., Shearer and Borg, 2006). A focused beam of electrons scans the irregular or polished surface of a sample, interacting with atoms to produce signals like secondary electrons (useful for investigating topography) and backscatter electrons (useful for compositional contrast based on atomic number). SEM offers sub-micrometer imaging resolution.

### 10.3.2.2 Electron Microprobe Analysis (EMPA)

EMPA is a quantitative technique used to determine the major and minor element composition of minerals and glass within samples (e.g., Seddio et al., 2013; Shearer and Borg, 2006). EMPA uses an electron beam to excite electrons in the phase of interest, which then emit characteristic X-rays that are measured using Wavelength Dispersive Spectroscopy (WDS), which filters X-rays based on specific wavelengths. By comparing the X-ray intensities in the phase of interest to known mineral standards, the weight percentages of major and trace elements can be calculated. A chemical precision better than 0.1 wt.% is typical.

### 10.3.2.3 Raman Spectroscopy

Raman spectroscopy utilizes a laser beam to interact with the vibrational modes of a mineral's crystal lattice, resulting in a wavelength shift (Raman shift) that acts as a unique molecular fingerprint. In lunar research, it is essential for identifying specific mineral polymorphs and assessing the degree of shock caused by meteorite impacts (Ling et al., 2011; Cao et al., 2022). It provides a spatial precision on the order of 1 µm.

### 10.3.2.4 Electron Backscatter Diffraction (EBSD)

EBSD analyzes the diffraction patterns of electrons that backscatter off the crystal lattice of a sample within an SEM to determine crystallographic orientation. By mapping these orientations across a sample, researchers can quantify the internal strain and deformation history of lunar minerals, which serves as a record of the peak shock pressures the rock experienced (e.g., Cernok et al., 2019; Timms et al., 2012). It achieves a spatial resolution down to 50 nm.

### 10.3.2.5 Transmission Electron Microscopy (TEM)

TEM uses a high-energy electron beam to image extremely thin samples (typically < 100 nm thick) under high vacuum. The transmitted electrons provide detailed information on the internal crystal structure and chemical variations at the atomic scale, making it the primary method for studying space weathering effects and nanophase iron in lunar regolith (McFadden et al., 2024; Keller et al., 2021).

### 10.3.3 Whole Rock Composition Analysis Techniques

#### 10.3.3.1 Optical Emission Spectroscopy (OES)

OES is a common technique for analysis of major and trace elements in which the compositions can be determined using plasma and a spectrometer. Samples (in solution) are ionized which excites core electrons in the sample's atoms, and when they are de-excited, they return to their ground level energy and emit characteristic photons. Intensities of each element's wavelength are measured, compared to standards, and concentration determined.

#### 10.3.3.2 X-ray Fluorescence Spectroscopy (XRF)

X-ray Fluorescence spectroscopy is a non-destructive routine analytical technique used for the chemical analysis of rocks (e.g., Ryder and Schuraytz, 2001). The technique is based on excitation of electrons by X-ray radiation. Upon de-excitation, the elements emit fluorescent X-rays of characteristic energy and wavelength that can be used to quantify the concentrations of elements present in the sample. Wavelength dispersive spectroscopy (WDS) utilized the characteristic wavelengths, whereas energy dispersive spectroscopy (EDS) utilizes the characteristic energies. WDS offers higher precision due to more flexibility to correct for interfering lines than does EDS.

#### 10.3.3.3 Inductively Coupled Plasma Mass Spectroscopy (ICP-MS) (also ID-ICP-MS, MC-ICP-MS, and Quadrupole ICP-MS)

Inductively coupled plasma mass spectrometry (ICP-MS) is an analytical technique used for the quantification of bulk elemental abundances at concentrations as low as parts per billion or even trillion (e.g., Neuman et al., 2025; Liu et al., 2025). This sensitivity is achieved by ionizing the sample with an inductively coupled plasma and then using a mass spectrometer to separate and quantify those ions. For bulk elemental abundances, samples are introduced to the plasma as a solution prepared using acid digestion. Variations within this general approach include isotope dilution (ID-ICP-MS) in which samples are spiked with a known isotopic standard in advance of ionization which allows more precise measurements in the mass spectrometer. An addition, different types of mass spectrometers can be combined with an ICP source, such as quadrupole, sector field, and multi-collector (MC) ICP-MS instruments.

#### 10.3.3.4 Instrumental neutron activation analysis (INAA)

Historically, Apollo and other lunar samples have been analyzed for specific elements, including elements of interest to SPARX such as Th and Sm. Neutron activation analysis utilizes neutron capture and subsequent gamma ray (γ) emission to identify elements. Samples are placed in a nuclear reactor, allowing atoms in the sample to capture neutrons. Of the resulting atoms that are radioactive, they emit characteristic γ-rays upon decay, which can then be measured to determine abundances. Although this technique is limited to ~35 elements, it has historically provided highly precise measurements of Th, Sm and other elements that have helped to understand and define various lunar processes (Korotev, 2022; Korotev and Irving, 2021). Many active labs from the last 50 years have shut down or are no longer active, with few to no new labs being established. However, when SPARX samples are analyzed, new techniques will be utilized and compared to the numerous existing lunar sample INAA analyses. A thorough appreciation of INAA must exist for the optimal interpretation of analysis of new Th-rich materials, for example. INAA can be performed on a wide range of sample masses, including very small mg-sized fragments all the way up to small rocks.

### 10.3.4 Required Micro-Analytical Techniques

For samples with limited material availability, or brecciated samples with small clasts and mineral fragments, any of the following three techniques are well suited to achieve the relevant (requiring micro-analysis) objectives defined for Experiments E1-E10, and E13-E14. Some examples of recent or demonstrative papers for these techniques are given after each description.

#### 10.3.4.1 Micro–X-ray Fluorescence (µXRF)

Micro-XRF is a higher spatial resolution version of XRF (see description above) that allows nondestructive elemental analysis on the micron scale without special sample pretreatment (e.g., unlike XRF, which requires powdering of bulk sample and fusing with a binding material). Therefore, micro-XRF can be used to measure major and trace elements in small samples such as mounted clasts, mineral fragments, or glasses, or thin or thick sections of rocks.

#### 10.3.4.2 Secondary Ion Mass Spectrometry (SIMS) (also NanoSIMS and TOF-SIMS)

Secondary ion mass spectrometry (also called ion microprobe) is a common analytical technique used for isotopic and trace element analyses of solid samples. SIMS utilizes a primary ion beam (e.g., $O^-$ or $Cs^+$) directed onto a specimen, which produces secondary ions that are then separated according to their mass to charge (m/z) ratios. The detection limit is usually parts per billion level for areas of a few micrometers. Point, line, and image analysis can be obtained with a lateral resolution of sub-micrometer scale and depth resolution of 10 nm. SIMS allows analysis of elements and isotopes from H to U in a solid but is also limited by the useful yield (Hervig et al., 2006), and thus SIMS is optimal for some elements more than others.

Variations of the SIMS approach include SIMS and NanoSIMS. Although multipurpose SIMS instruments are able to measure the isotopic and trace element compositions of many lunar samples (e.g., Dottin et al., 2025; Hauri et al., 2015; Zhang et al., 2023), they have been limited by their small mass spectrometers and low mass-resolving powers. As a result, larger geometry ion probes (e.g., SHRIMP or CAMECA 1290HR) were developed for higher sensitivity at high mass-resolving power, especially useful for U–Pb dating (e.g., Arai et al., 2025). NanoSIMS techniques allow the primary ion beam to be focused to less than 100 nm and small sputtering depths, and can be used for trace elements, light elements, stable-isotope analysis, and radiochronology for smaller grains. Time-of-flight SIMS (ToF-SIMS) pulses ions from the ionizer to the detector. The time it takes for ions to pass down the flight tube indicates the m/z. ToF-SIMS can provide high to ultrahigh resolutions (5,000 to 50,000 m/m) with mass ranges from 100 to many thousands m/z, whereas masses below 100 have lower mass accuracy and sensitivity.

Geochronology can be determined using ion microprobe techniques. The target phases for these analyses would be zircon, baddeleyite, phosphates, mesostasis pockets, and others, and would be conducted on polished sections of samples via Secondary Ion Mass Spectrometry (SIMS), NanoSIMS, and potentially LA-ICP-MS (see next section). Polished sections prepared for petrologic characterization are sufficient for in situ dating provided datable phases are present.

#### 10.3.4.3 Laser-ablation Inductively Coupled Plasma Mass Spectroscopy (LA-ICP-MS)

ICP-MS was described above and is a routine and highly precise technique that is used to analyze a wide range of rocks (McIntosh et al., 2024; Nottingham et al., 2024). For spatially resolved analysis, laser ablation analysis (LA-ICP-MS) is used; material is ablated from the sample surface by a laser and is then introduced to the plasma. The spatial resolution varies depending on the mineral under study and the abundance of the target elements, but spot sizes of 5 to 200 microns have been used to analyze lunar materials. LA-ICP-MS analyses require a polished sample surface, and the sample should be at least 100 µm thick. Elements of interest for LA-ICP-MS include almost every element and thus lithophile, siderophile, and chalcophile elements can be analyzed in silicates, metals, and sulfides.

### 10.3.5 Geochronologic Measurement Techniques

#### 10.3.5.1 Thermal Ionization Mass Spectrometry (TIMS)

Thermal ionization mass spectrometry (TIMS) is a technique for high-precision measurement of isotope ratios. After purification of the element of interest, the sample is loaded onto a source filament, which is heated to ionize the sample. Ions are then accelerated (can be positive or negative ions) and focused by applying electrostatic potentials and sorted by mass using a magnet, before being directed into an array of multiple detectors for simultaneous detection of ions with different masses. TIMS is the optimal method for specific elements, including some chronometers (e.g., Sr, Nd, Pb). Typical mass requirements depend on sample lithology and mineralogy, and grain size, but can range from 100-700 mg for basaltic rocks (e.g., Torigoye-Kita et al. 1995; Gaffney et al., 2007; Rankenburg et al., 2007) and 300 mg to 5 g for anorthositic rocks (e.g., Nyquist et al., 2006; Borg et al., 2011), as explained in more detail below.

Mass requirements for dating anorthosites fall into three general categories. First, anorthosites ≥40 mm will provide the mass (up to 5 g), mineralogic variation, and sampling options that will be necessary to provide Sm-Nd and Lu-Hf ages of highest precision (e.g., Borg and Carlson, 2023; Borg et al., 2015). The large samples will be used for mineral separates, including pyroxene (e.g., Borg et al., 2011), that will provide ample spread of parent/daughter element ratios, which thus allows lowest uncertainty ages. Second, ≥8 mm rocklets will be used to determine crystallization and model ages using chronometers (Rb-Sr, Ar-Ar, and U-Pb) requiring less mass (100-700 mg) and also single mineral grains such as zircons (U-Pb, Lu-Hf) (e.g., Schneider and Kleine, 2025; Cohen and Coker, 2010). Coordinated measurements on 4 particles per site will yield ages using these complementary chronometers to those used on the larger 40 mm rocks. Because these samples will potentially contain a mixture of lithologies (and some non-anorthositic) mass estimates are based on our expectations and understanding of Apollo regolith that ~50% of the material sieved will be anorthositic (e.g., McKay et al., 1991; Carrier, 1973). Finally, bulk regolith will be scooped at these sites, and bulk regolith is expected to contain a fine-grained fraction that will include zircons and other mineral fragments (e.g., Nemchin et al., 2009; Meyer et al., 1996; Crow et al., 2017; Barboni et al., 2017; 2024; Zhang et al., 2021, Greer et al., 2023; Dauphas et al., 2025). These zircons will be identified and concentrated to allow chronologic studies and in particular to look for ages that might be associated with lunar formation and subsequent crustal differentiation. The number of zircons recovered in Apollo regolith samples, for example, has a very small mass (<< 100 mg) compared to the mass of the volume of material searched (~40 g) and thus most of the material remains unused (e.g., Taylor et al., 2009; Crow et al., 2017) and available for future scientific investigations.

For crystalline basalts, similar masses of ~100 – 700 mg (≥3 mm and portions or all of ≥5 mm rocklets, depending on elemental abundances) will be required for isochron dating methods. If the abundances of the elements of interest for dating are high enough, petrologic characterization and dating can be done on the same rocklets, conserving overall sample mass. In all cases where either Sm-Nd system and/or Lu-Hf are utilized, measurements are required to follow protocols for correcting Sm, Nd, and Hf isotopic compositions for thermal and epithermal neutron capture using established protocols (e.g., Borg et al., 2019; Gaffney and Borg, 2014;

Sprung et al., 2013). Basalts are also amenable to dating via Pb-Pb and potentially U-Pb using the same in situ approaches described above.

Isochron methods for impact melts will require typical masses of roughly 100 – 500 mg required (corresponding to ≥3 mm particles or portions of ≥8 mm particles), although this depends on the composition and mineralogy of the sample.

#### 10.3.5.2 Noble Gas Mass Spectrometry

Noble gas mass spectrometry (NGMS) is an analytical technique that is used for the elemental and isotopic measurements of the chemically inert noble gases. Because the abundances of these elements are low in lunar materials, noble gases trapped within solid materials are extracted from samples by heating, UV laser ablation, crushing, or etching. The $^{40}Ar/^{39}Ar$ chronometer is an improved derivative of the K/Ar method and is a very powerful tool to investigate the crystallization, igneous and/or cooling history of an extraterrestrial rock/particle (e.g., Bogard 2011; Cohen et al., 2025). Furthermore, Ar isotopes can provide a $^{38}Ar$ cosmic-ray exposure age which records the time span over which a given sample has been exposed unshielded to the bombardment by cosmic rays in the vacuum of space.

Depending on sample mineralogy and composition, it is possible to obtain Ar-Ar ages from samples as small as sub-micrograms to 10s of micrograms (e.g., Cohen et al., 2005). Fragments for Ar-Ar analyses can be subsamples of rocklets of any size as well as fragments collected from the Bulk Regolith sample. Even taking a more conservative approach where, to ensure samples contain enough K to enable a high precision Ar-Ar measurement, we assume required sample masses are on the order of 10s of milligrams, the overall sample mass requirements for E4 and E5 (and indeed all Experiments that involving sampling) will easily provide sufficient mass for many 10s of Ar-Ar measurements, allowing for a robust statistical approach to determining impact ages for these basins. **This statistical framework is particularly important for distinguishing primary basin-forming ages from later thermal overprinting events and for identifying potential age populations within sampled melt units.** Beyond the requirement of the Ar-Ar system, there is a strong possibility that impact melt samples collected for these experiments will be sufficiently crystalline to enable dating using isochron systems (e.g., $^{147}Sm$-$^{143}Nd$, Rb-Sr, U-Pb, Lu-Hf) and/or will contain mineral phases such as phosphates, baddeleyite, and zircon that are amenable to *in situ* U-Pb and Pb-Pb dating via SIMS or nanoSIMS, thus providing additional information on impact ages (e.g., Che et al., 2021; Crow et al., 2017; Grange et al., 2013, 2011; Liu et al., 2012; Merle et al., 2017; Nemchin et al., 2021, 2008; Snape et al., 2016). In some cases, U–Pb analyses can be conducted on the same mineral fragments used for Sm–Nd isochron dating, enabling complementary age constraints from a single sample (Borg et al., 2011).

Isochron methods such as Sm-Nd will require significantly more sample mass than Ar-Ar, with typical masses of roughly 100 – 500 mg required (corresponding to ≥3 mm particles or portions of ≥8 mm particles), reflecting lower parent element abundances and the need for multiple mineral or whole-rock fractions.

Noble gas cosmic ray exposure age (CRE) measurements can typically be conducted on small aliquots ranging from sub-milligram for regolith samples to a few milligrams for basaltic or impact melt lithologies, depending on cosmogenic nuclide concentrations and noble gas extraction sensitivity. Pyroclastic glass beads and fine-grained materials may require similarly small masses, though additional material may be allocated where solar wind and cosmogenic components must be resolved through stepwise heating or crushing techniques. These relatively low mass requirements make CRE analyses particularly compatible with precious returned samples and allow multiple aliquots to be analyzed to evaluate heterogeneity or complex exposure histories.

#### 10.3.5.3 MC-ICP-MS

MC-ICP-MS is also used for geochronologic measurements, as described previously.

### 10.3.6 Volatile Measurement Techniques

#### 10.3.6.1 Electron Microprobe Analysis (EMPA)

EMPA is also used for volatile measurements, as described previously.

#### 10.3.6.2 Secondary Ion Mass Spectrometry (SIMS) (also NanoSIMS and TOF-SIMS)

SIMS is also used for volatile measurements, as described previously.

#### 10.3.6.3 Fourier Transform Infrared Spectroscopy (also micro-FTIR)

FTIR (Fourier Transform Infrared) spectroscopy is a non-destructive analytical technique that identifies chemical bonds in materials by measuring infrared light absorption, creating a "molecular fingerprint." Micro-FTIR combines this with a microscope to analyze microscopic, heterogeneous, or trace samples (typically 5–20 microns), enabling pinpoint identification of various volatile species.

#### 10.3.6.4 X-ray Absorption Near Edge Structure (XANES) Spectroscopy

X-ray absorption near edge structure (XANES) spectroscopy utilizes monochromatized soft X-rays to provide information regarding the distribution of bonding environments for all elements in period 2 and greater. The XANES spectrum is produced by the absorption of monochromatized light that, when tuned to specific energies, excites core-level electrons to higher-energy electronic states. XANES spectra for C, N, and O organics are well calibrated to known standards, making it possible to acquire quantitative information about bonding environments at spatial resolutions down to 30 nm and detection limits approaching 1 ppm.

#### 10.3.6.5 Stepped Combustion

Rock samples decompose, oxidize, or degas volatiles over discrete and often distinct temperature ranges. Small bulk samples are heated in a stepwise manner from room temperature up to 1400C, and the abundances and isotopic compositions of carbon and nitrogen are analyzed at each temperature step. As different components combust or decompose at different temperatures, the stepped combustion combined with mass spectrometer can provide a measurement of the abundance and isotopic composition of the C and N associated with individual components within the bulk sample.

#### 10.3.6.6 Laser-ablation Inductively Coupled Plasma Mass Spectroscopy (LA-ICP-MS)

See 2.3.3 above

#### 10.3.6.7 Multicollector Inductively Coupled Plasma Mass Spectroscopy (MC-ICP-MS)

MC-ICP-MS is also used for volatile measurements, see described previously.

## 10.4 ADDITIONAL MISSION ARCHITETURES

The SPARX SDT discussed the following implementation architectures (Section 5.7) with various levels of fidelity relating to their level of technical maturity and ability to achieve the SPARX science objectives. The implementation architectures presented below represent additional mission architectures that were discussed, but considered less technically mature, or less compatible with SPARX science objectives. Each implementation architecture has a cartoon mockup of the architectural elements (e.g., lander(s), hopper(s), astronaut(s)), hypothetical areas of investigation within SPA, and a summary of the architecture as described in more detail in Section 5.7. Holistic evaluations based on the strengths and weaknesses of each architecture is also included. Although these did not rise to the level of being commensurate with the scope of the TOR (e.g., non-sample return architectures), or the currently understood technical landscape (on-orbit hopper refueling), the SDT presents them here for the sake of completeness and any future inquiries.

### 10.4.1 Static Lander with In Situ Science Only

SPARX ARCHITECTURE

## Static Lander with In Situ Science Only

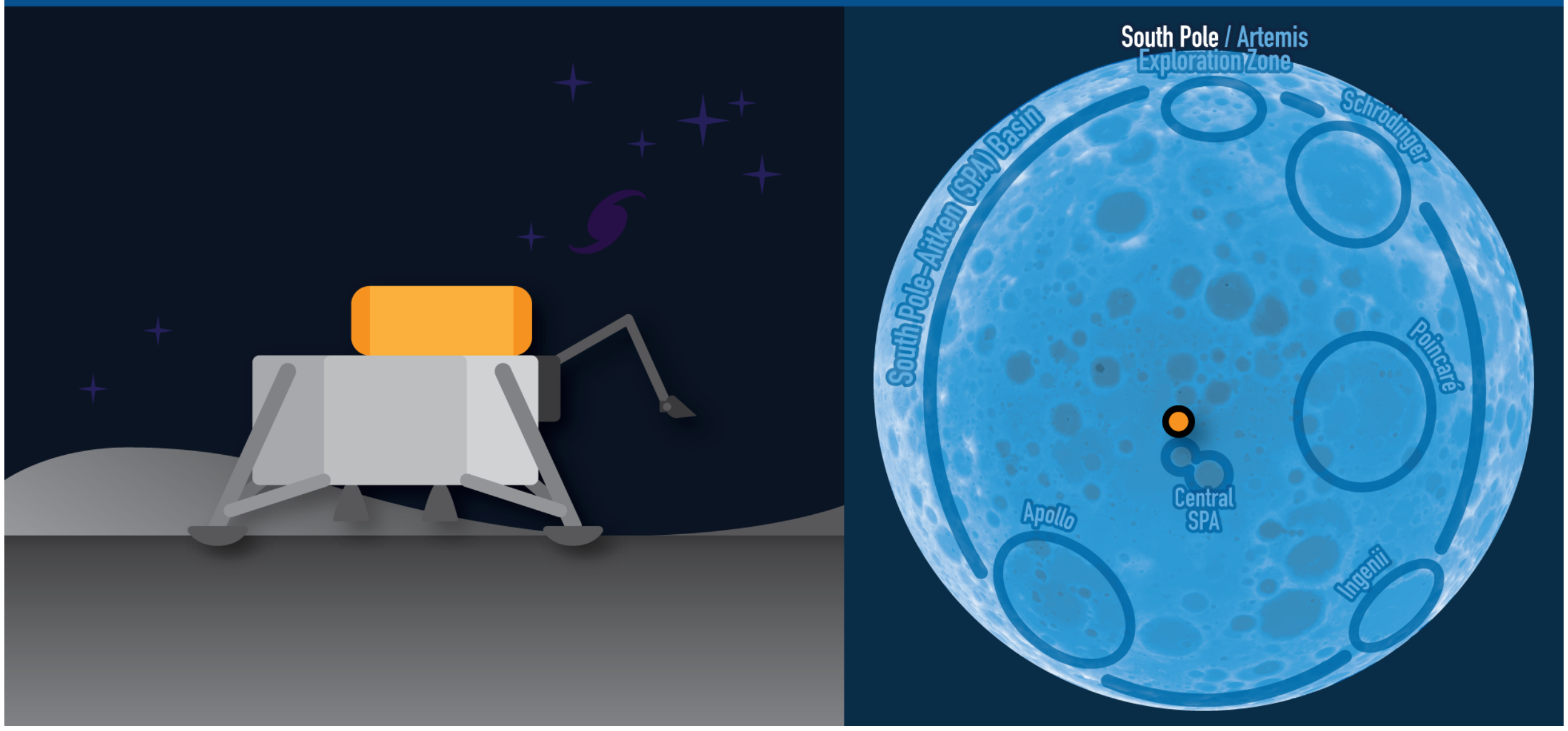

| SUMMARY OF THIS ARCHITECTURE: | |
|---|---|
| DESCRIPTION: | A single static lander lands in SPA and analyzes materials in its vicinity using in situ instruments only. No sample return is attempted. |
| NUMBER OF ARCHITECTURAL ELEMENTS: | 1 lander |
| MODE OF SAMPLE RETURN: | N/A |
| MODE OF SURFACE MOBILITY: | Limited (robotic arm or similar) |

**EVALUATION OF THIS ARCHITECTURE AGAINST THE SPARX FIGURES OF MERIT:**

| STRENGTHS: | WEAKNESSES: |
|---|---|
| – Architecture is low cost: Estimated cost of one CLPS lander is ~$200M | – **In-situ-only methods do not meet SPARX Science Objectives; sample return is required**<br>– **Architecture does not meet SPARX Experiment Requirements for sampling the geologic diversity of SPA (≥10 sites, ≥1,000 km range)**<br>– **Architecture would not provide sufficient mobility to accomplish SPARX's traverse-centric E11 (quasi-continuous near-surface measurements ≥650 km range)**<br>– **Architecture does not provide sufficient local-scale mobility for determining the geologic context for the samples (~10–100 m)**<br>– Samples would be collected from close to the lander, increasing risk of not meeting sample lithology requirements and increasing odds of contamination from the lander |

**OVERALL EVALUATION:**

**POOR.** This architecture could not accomplish the SPARX Science Goals, primarily because in-situ-only approaches are inadequate. The science return per cost is low.

### 10.4.2 Static Lander Campaign with In Situ Science Only

SPARX ARCHITECTURE

## Static Lander Campaign with In Situ Science Only

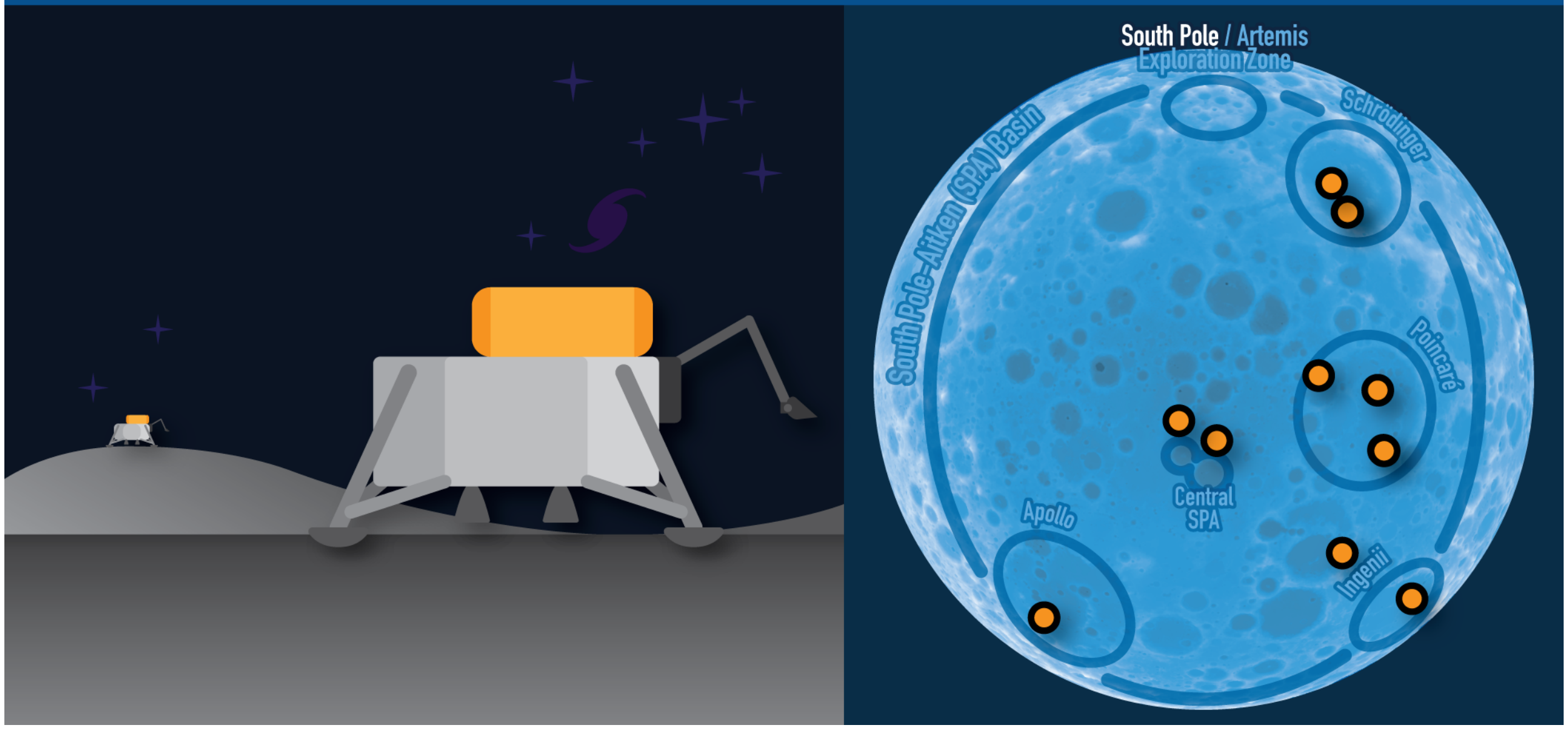

| SUMMARY OF THIS ARCHITECTURE: | |
|---|---|
| **DESCRIPTION:** | Multiple static landers are distributed across SPA and analyze materials in their vicinity utilizing in situ instruments only. No sample return is attempted. |
| **NUMBER OF ARCHITECTURAL ELEMENTS:** | 10–20 landers |
| **MODE OF SAMPLE RETURN:** | N/A |
| **MODE OF SURFACE MOBILITY:** | Limited (robotic arm or similar) |

**EVALUATION OF THIS ARCHITECTURE AGAINST THE SPARX FIGURES OF MERIT:**

**STRENGTHS:**

- Missions could be deployed across SPA with no constraints driven by sample return architecture or surface mobility

**WEAKNESSES:**

- **In-situ-only methods do not meet SPARX Sample Requirements; sample return is required**
- **Architecture would not provide sufficient mobility to accomplish SPARX's traverse-centric Science Experiments (quasi-continuous near-surface measurements ≥650 km range)**
- **Architecture does not provide sufficient local-scale mobility for determining the geologic context for the samples (~10–100 m)**
- Samples would be collected from close to the lander, increasing risk of not meeting sample lithology requirements and increasing odds of contamination from the lander
- **Architecture is high cost:** A full campaign of 10–20 landers would cost ~$2B–$4B based on current CLPS costs (~$200M per lander)

**OVERALL EVALUATION:**

**POOR.** This architecture could not accomplish the SPARX Science Goals, primarily because in-situ-only approaches are inadequate. Even a campaign with multiple missions of this type would only provide incremental advances to the SPARX Science Goals and at high cost. The science return per unit cost is low.

### 10.4.3 Hopper with In Situ Science Only

SPARX ARCHITECTURE

## Hopper with In Situ Science Only

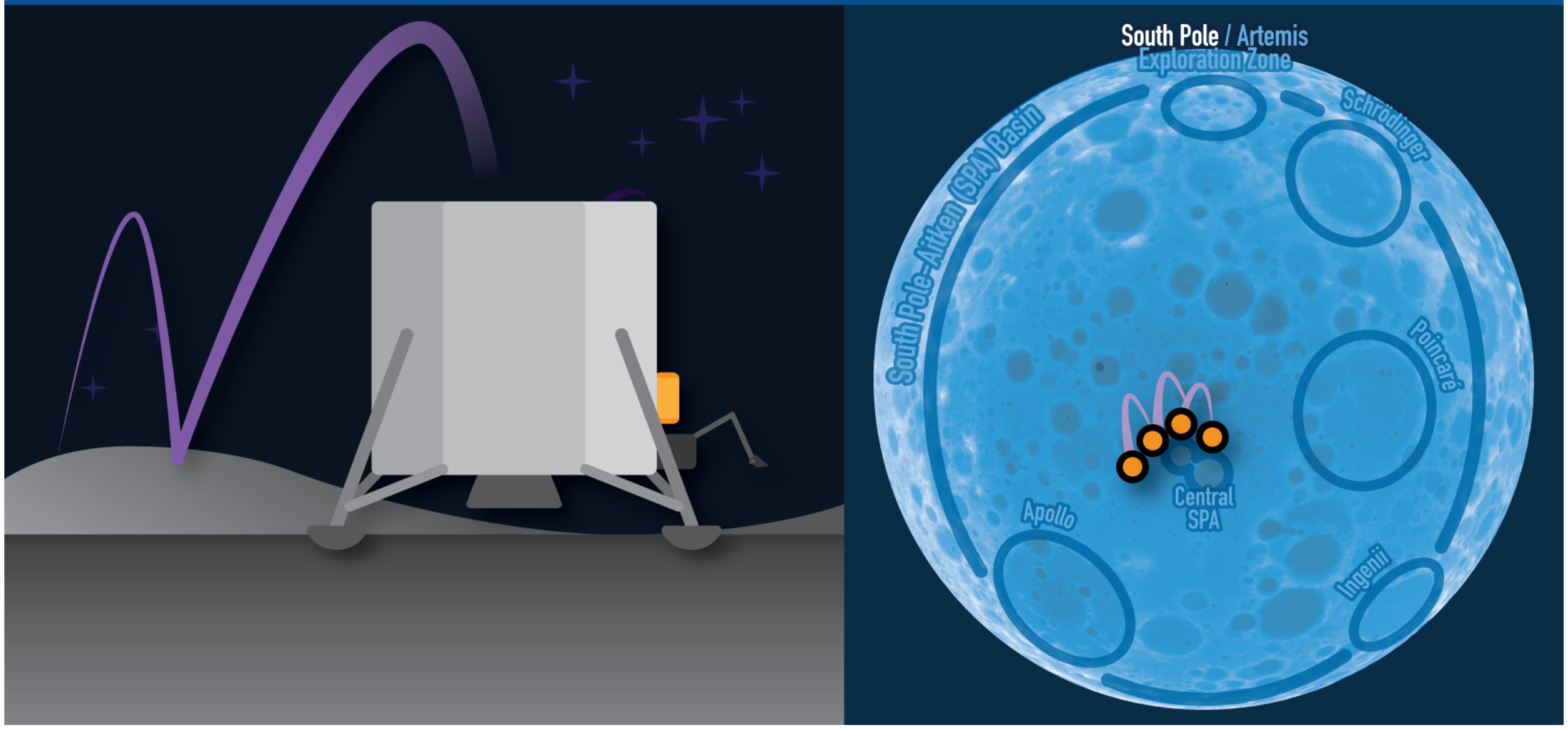

| SUMMARY OF THIS ARCHITECTURE: | |
|---|---|
| **DESCRIPTION:** | A single hopper lands in SPA, collects samples and performs in situ analyses only, and hops to 3~4 sites to perform repeat analyses before running out of fuel. |
| **NUMBER OF ARCHITECTURAL ELEMENTS:** | 1 hopper |
| **MODE OF SAMPLE RETURN:** | N/A |
| **MODE OF SURFACE MOBILITY:** | Hopper (~3 to 4 hops with a total range of 10s to 100s of km, depending on the architecture) with limited local-scale mobility (robotic arm or similar) |

**EVALUATION OF THIS ARCHITECTURE AGAINST THE SPARX FIGURES OF MERIT:**

| STRENGTHS: | WEAKNESSES: |
|---|---|
| – None noted | – **In-situ-only methods do not meet SPARX Sample Requirements; sample return is required**<br>– **Architecture does not meet SPARX Mission Requirements for sampling the geologic diversity of SPA (≥10 sites, ≥1,000 km range)**<br>– **Architecture would not provide sufficient mobility to accomplish SPARX's traverse-centric Science Experiments (quasi-continuous near-surface measurements ≥650 km range):** optimal hopping would take the hopper to great altitude, limiting effectiveness of remote sensing instruments<br>– **Architecture does not provide sufficient local-scale mobility for determining the geologic context for the samples (~10–100 m)**<br>– Samples would be collected from close to the lander, increasing risk of not meeting sample lithology requirements and increasing odds of contamination from the lander |

**OVERALL EVALUATION:**

**POOR.** This architecture could not accomplish the SPARX Science Goals primarily because in-situ-only approaches are inadequate. Hopper-based mobility is inadequate for addressing the SPARX Science Objectives that require mobility. A single mission of this type would only provide incremental advances to the SPARX Science Goals.

### 10.4.4 Hopper with On-Orbit Sample Return

SPARX ARCHITECTURE | **Hopper with On-Orbit Sample Return**

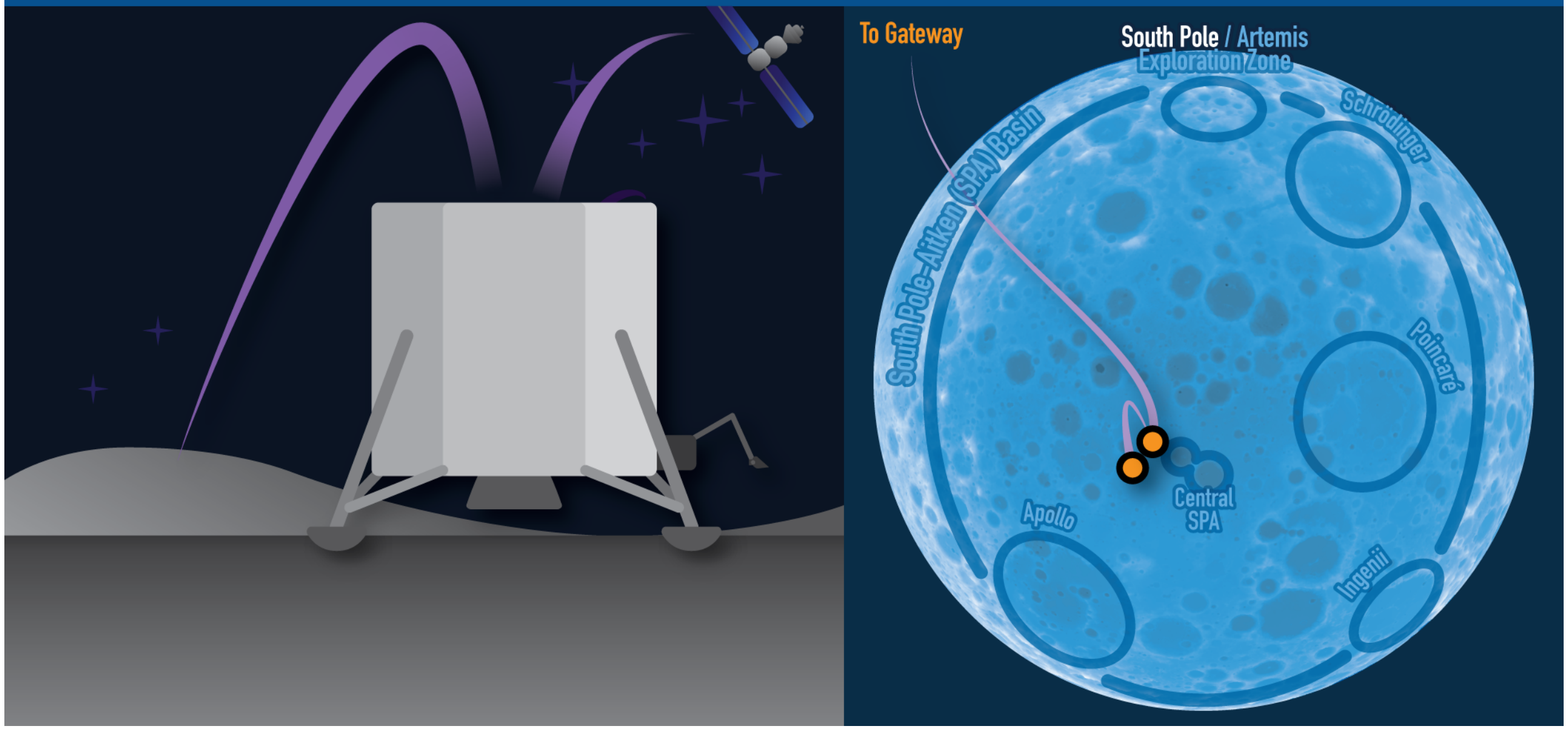

| SUMMARY OF THIS ARCHITECTURE: | |
|---|---|
| **DESCRIPTION:** | A single hopper lands in SPA, collects samples and in situ data, and hops to ~2 sites before launching back to rendezvous with an orbiting spacecraft which can receive the samples. |
| **NUMBER OF ARCHITECTURAL ELEMENTS:** | ~2 (1 hopper, 1 element in lunar orbit for retrieving and returning samples to Earth such as Orion) |
| **MODE OF SAMPLE RETURN:** | Sample return from lunar orbit; exact architecture is to be determined but could include Orion |
| **MODE OF SURFACE MOBILITY:** | Hopper (~2 hops with a total range of ~10–100 km for Starship-class hoppers) with limited local-scale mobility (robotic arm or similar) |

**EVALUATION OF THIS ARCHITECTURE AGAINST THE SPARX FIGURES OF MERIT:**

**STRENGTHS:**

- Samples returned from single investigation area, capable of addressing a part of some SPARX exeperiments.

**WEAKNESSES:**

- **Architecture does not meet SPARX Mission Requirements for sampling the geologic diversity of SPA (≥10 sites, ≥1,000 km range)**
- **Architecture does not meet the total SPARX sample mass requirements (≥10 kg from ≥10 sites)**
- **Architecture would not provide sufficient mobility to accomplish SPARX's traverse-centric Science Experiments (quasi-continuous near-surface measurements ≥650 km range)**: optimal hopping would take the hopper to great altitude, limiting effectiveness of remote sensing instruments
- **Architecture does not provide sufficient local-scale mobility for determining the geologic context for the samples (~10–100 m)**
- Samples would be collected from close to the lander, increasing risk of not meeting sample lithology requirements and increasing odds of contamination from the lander
- **Architecture relies on uncertain orbital Artemis infrastructure for rendezvous (e.g., Gateway) and return of samples to Earth (e.g., Orion)**

**OVERALL EVALUATION:**

**POOR.** This architecture could not accomplish the SPARX science goals, primarily because of the lack of mobility necessary to collect diverse samples from across large regions of SPA, and because the sample quantity would not be sufficient to accomplish the SPARX Science Goals. A single mission of this type would only provide incremental advances to the SPARX Science Goals.

### 10.4.5 Refuelable Hopper with On-Orbit Refueling and Sample Return

SPARX ARCHITECTURE

## Refuelable Hopper with On-Orbit Refueling & Sample Return

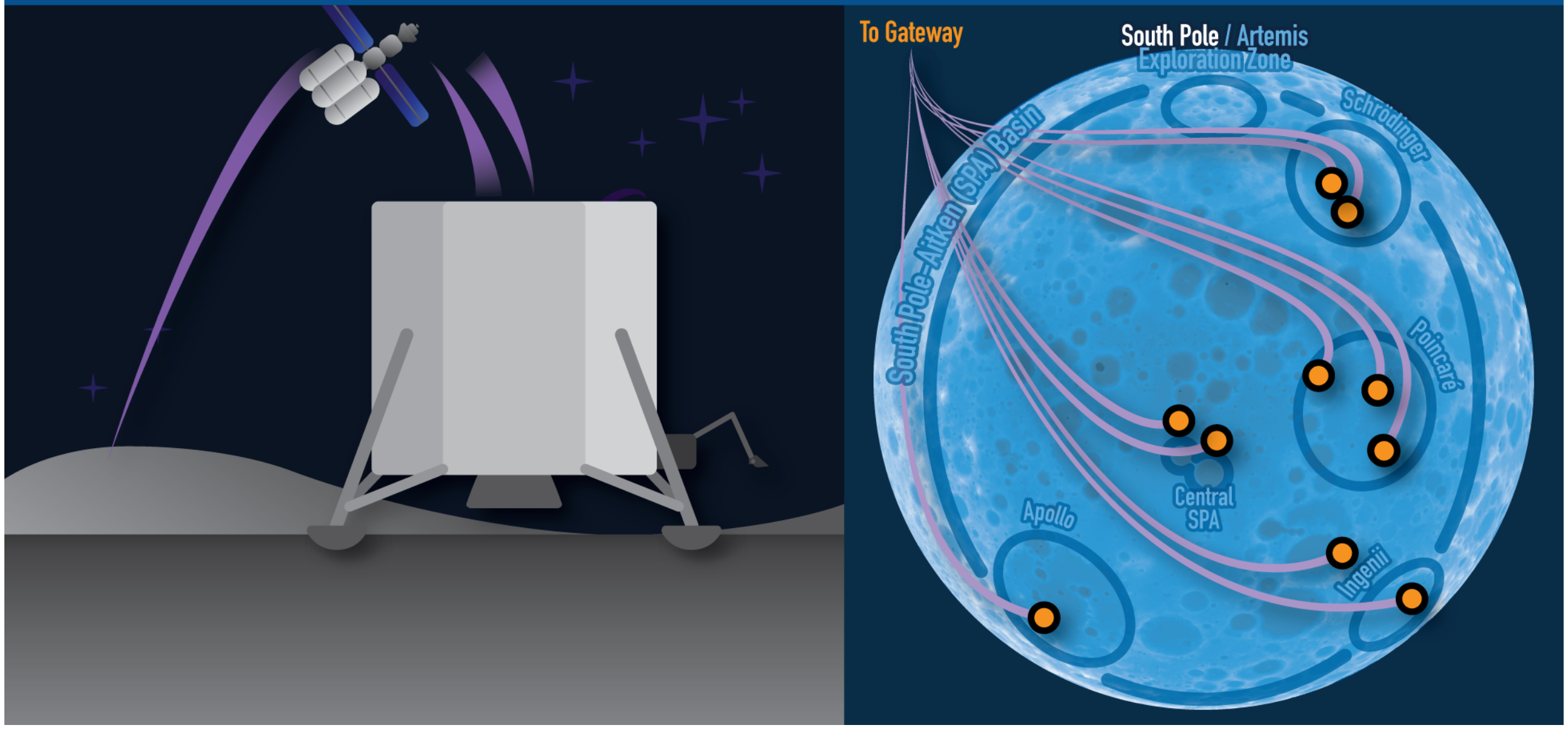

| SUMMARY OF THIS ARCHITECTURE: | |
|---|---|
| **DESCRIPTION:** | A single hopper lands in SPA, collects samples and *in situ* data, and hops back to orbit where it transfers samples and is refueled for 10–20 subsequent flights. |
| **NUMBER OF ARCHITECTURAL ELEMENTS:** | ~3 (1 hopper, 1 orbital rendezvous refueling depot, 1 element for returning samples to Earth such as Orion) |
| **MODE OF SAMPLE RETURN:** | Sample return from lunar orbit; exact architecture is to be determined but could include Orion |
| **MODE OF SURFACE MOBILITY:** | Orbital hopper with quasi global access with limited local-scale mobility (robotic arm or similar) |

**EVALUATION OF THIS ARCHITECTURE AGAINST THE SPARX FIGURES OF MERIT:**

**STRENGTHS:**

- **Architecture can meet the total SPARX sample mass requirements (≥10 kg)**
- **Architecture can meet the SPARX Mission Requirements for sampling the geologic diversity of SPA (≥10 sites, ≥1,000 km range)**
- Missions could be deployed across SPA with no constraints driven by sample return architecture or surface mobility, assuming that the hopper is designed for global access

**WEAKNESSES:**

- **Architecture would not provide sufficient mobility to accomplish SPARX's traverse-centric Science Experiments (quasi-continuous near-surface measurements ≥650 km range)**: optimal hopping would take the hopper to great altitude, limiting effectiveness of remote sensing instruments
- **Architecture does not provide sufficient local-scale mobility for determining the geologic context for the samples (~10–100 m)**
- Samples would be collected from close to the lander, increasing risk of not meeting sample lithology requirements and increasing odds of contamination from the lander
- **Architecture relies on the presence of orbital refueling depot in lunar orbit, which is not currently part of the Artemis infrastructure plans**
- **Architecture relies on orbital Artemis infrastructure for return of samples to Earth (e.g., Orion)**

**OVERALL EVALUATION:**

**FAIR.** While this architecture could address most of the SPARX sample return requirements, this approach would still provide inadequate in-situ data and geologic context due to the lack of local-scale mobility. Furthermore, this architecture relies on the development of a refuellable lunar lander and the presence of a refueling depot in lunar orbit, which is not part of the current Artemis infrastructure plans, and which has an unknown cost scope.

### 10.4.6 "Daisy Chain" Hopper Campaign

SPARX ARCHITECTURE | **"Daisy Chain" Hopper Campaign**

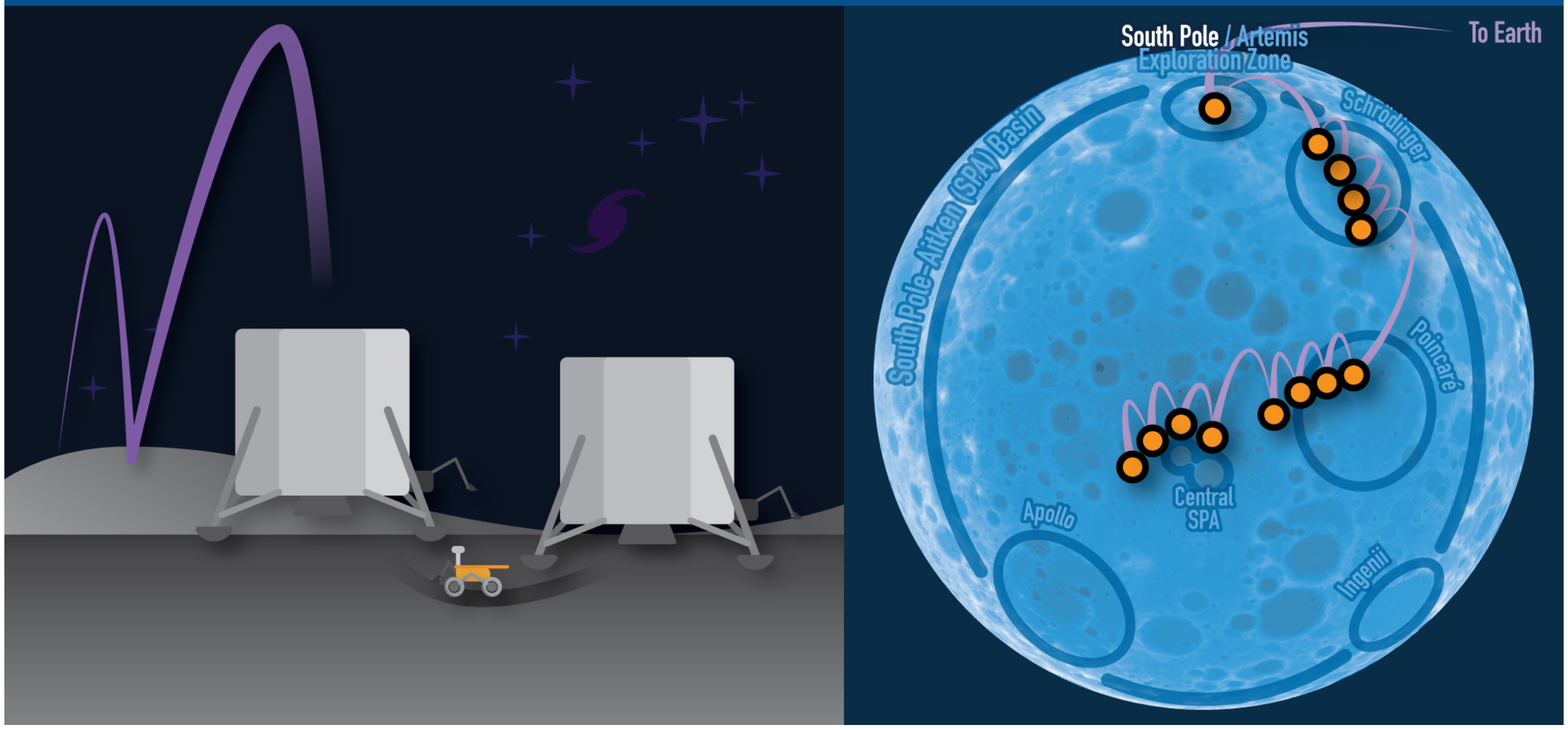

| SUMMARY OF THIS ARCHITECTURE: | |
|---|---|
| **DESCRIPTION:** | A hopper lands in SPA, collects samples and *in situ* data, performs a series of 3~4 hops before rendezvousing with a second fully fueled hopper. The samples are transferred, and the second hopper continues on. This cycle repeats for ~10 hoppers. |
| **NUMBER OF ARCHITECTURAL ELEMENTS:** | ~20 (~10 hoppers and ~10 sample transfer rovers) plus 1 south pole Artemis mission |
| **MODE OF SAMPLE RETURN:** | Landed Artemis mission (up to ~100 kg) |
| **MODE OF SURFACE MOBILITY:** | Multiple hoppers (~3 to 4 hops with a total range of 10s to 100s of km for Starship-class hoppers) with limited local-scale mobility (robotic arm, sample caching rover, or similar) |

**EVALUATION OF THIS ARCHITECTURE AGAINST THE SPARX FIGURES OF MERIT:**

| STRENGTHS: | WEAKNESSES: |
|---|---|
| – **Architecture can meet the total SPARX sample mass requirements (≥10 kg)**<br>– **Architecture can meet the SPARX Mission Requirements for sampling the geologic diversity of SPA (≥10 sites, ≥1,000 km range)** | – **Architecture would not provide sufficient mobility to accomplish SPARX's traverse-centric Science Experiments (quasi-continuous near-surface measurements ≥650 km range)**: optimal hopping would take the hopper to great altitude, limiting effectiveness of remote sensing instruments<br>– **Architecture does not provide sufficient local-scale mobility for determining the geologic context for the samples (~10–100 m), unless the scope of the sample transfer rover is expanded to include science**<br>– Samples would be collected from close to the lander, increasing risk of not meeting sample lithology requirements and increasing odds of contamination from the lander<br>– **Sample transfer between two surface assets (two hoppers) is a major technology development and may require the addition of a sample transfer rover**<br>– **Architecture is likely high cost:** While hopper cost is unknown, the number of required hoppers (~10 Starship-like landers) plus the likely need for sample transfer rover suggest that the cost for this architecture would be high |

**OVERALL EVALUATION:**

**FAIR.** While this architecture could address most of the sample return requirements, this approach would still provide inadequate *in-situ* data and geologic context due to the lack of local-scale mobility. The need for sample transfer between surface assets and the number of elements likely drives high cost, and the interfaces between multiple staged assets increases risk.

### 10.4.7 Polar Artemis Campaign and/or Moon Base

SPARX ARCHITECTURE

## Polar Artemis Campaign and Moon Base

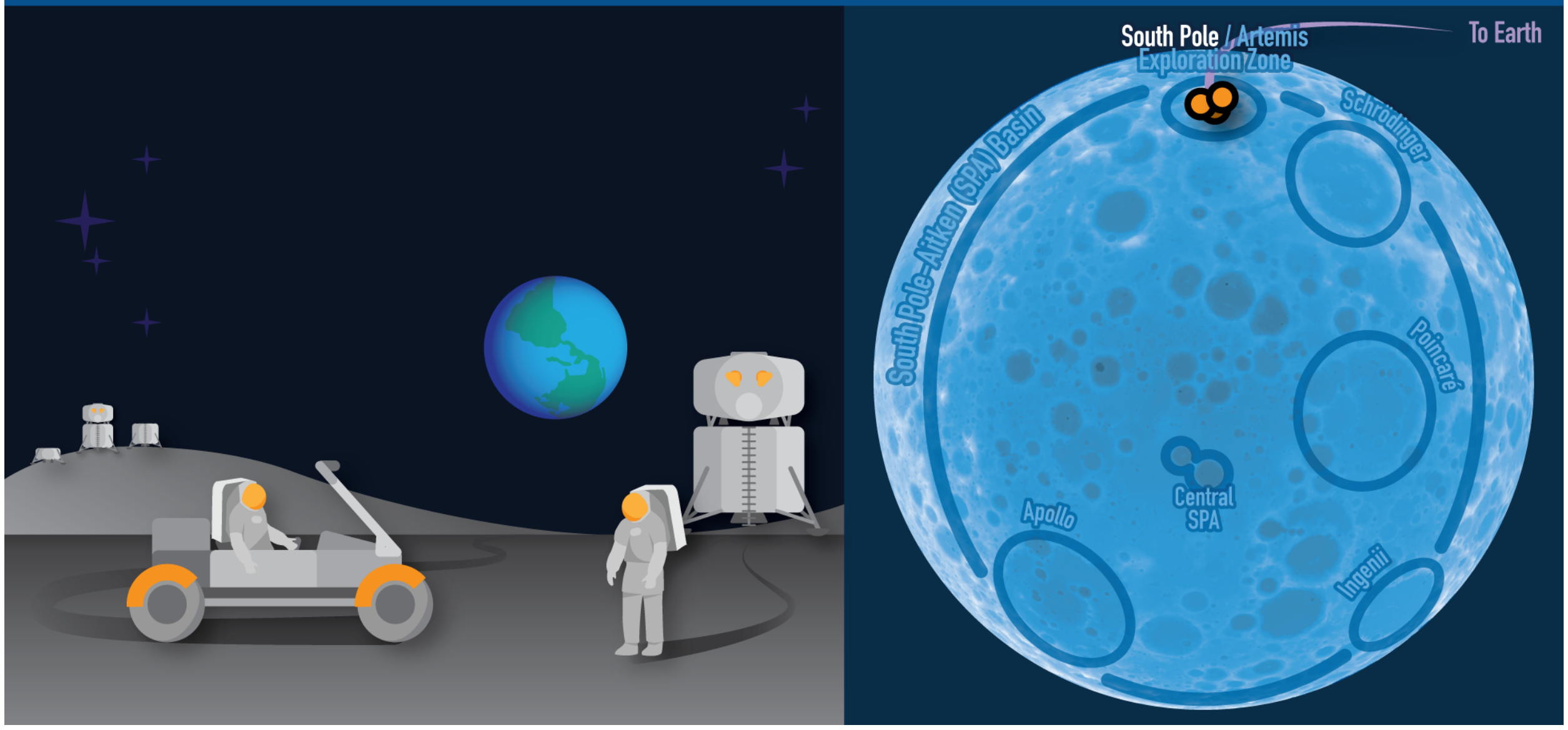

| SUMMARY OF THIS ARCHITECTURE: | |
|---|---|
| **DESCRIPTION:** | Multiple crewed Artemis mission lands near the lunar south pole in the Artemis exploration zone (poleward of 84°S), performing EVAs, collecting samples and *in-situ* measurements. These missions may also build up a persistent presence—Moon Base—which may support in-situ laboratory measurements. |
| **NUMBER OF ARCHITECTURAL ELEMENTS:** | Uncertain, but likely many polar Artemis missions, rovers, and landed assets |
| **MODE OF SAMPLE RETURN:** | Multiple Artemis astronaut sample return missions (plausibly many 100s of kg) |
| **MODE OF SURFACE MOBILITY:** | Astronauts on walking EVAs (≤2 km range), unpressurized rovers (≤10 km range), and potentially pressurized rovers (≤100 km range, depending on architecture) |

**EVALUATION OF THIS ARCHITECTURE AGAINST THE SPARX FIGURES OF MERIT:**

**STRENGTHS:**

- **Architecture leverages well-trained astronaut crew who are adaptable and efficient, maximizing science return**
- Samples could be collected far from the lander, decreasing the risk of contamination from the lander and increasing the probability of meeting sample lithology requirements
- The establishment of a well-instrumented Moon Base could enable more sophisticated in-situ measurements
- More frequent missions to/from to Moon Base may afford larger total returned sample masses

**WEAKNESSES:**

- **Architecture does not meet SPARX Mission Requirements for sampling the geologic diversity of SPA (≥10 sites, ≥1,000 km range)**
- **Architecture does not meet the total SPARX sample mass requirements (≥10 kg from ≥10 sites)**
- **Architecture cannot access high-priority SPARX regions of interest beyond the south pole**
- **Architecture would not provide sufficient mobility to accomplish SPARX's traverse-centric Science Experiments (quasi-continuous near-surface measurements ≥650 km range)**

**OVERALL EVALUATION:**

**POOR.** This architecture could not accomplish the SPARX Science Goals. Even with multiple missions, this architecture would only make incremental advances to the SPARX Science Goals because the Moon's south polar region does not include any of the priority science targets. Furthermore, the range of astronaut explorers—even with pressurized rovers—is insufficient for SPARX's "traverse science" requirements.

### 10.4.8 Non-Polar Artemis Sortie Mission

SPARX ARCHITECTURE

## Non-Polar Artemis Sortie Mission

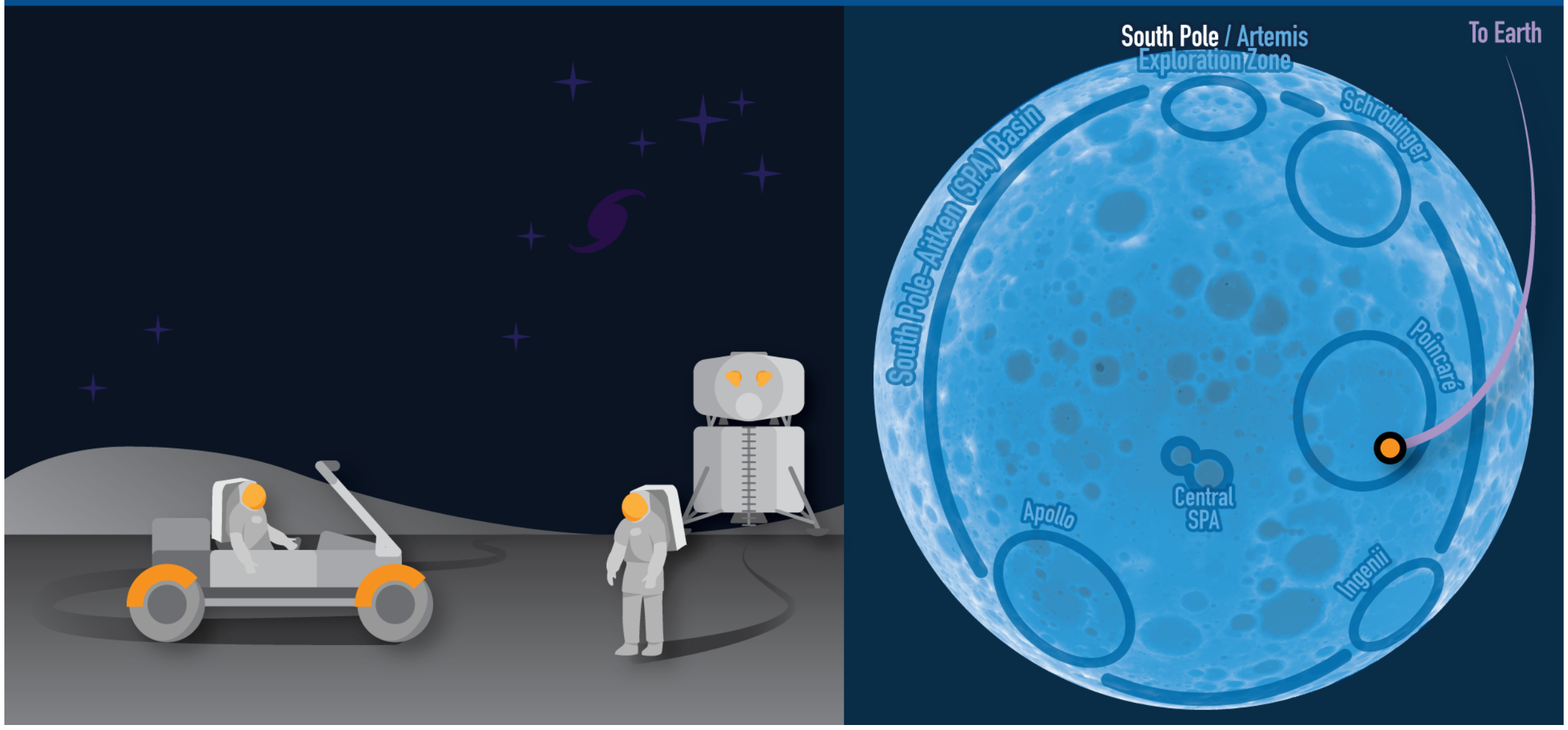

**SUMMARY OF THIS ARCHITECTURE:**

| | |
|---|---|
| **DESCRIPTION:** | A crewed Artemis mission lands at a location in SPA, performs EVAs, collects samples and *in-situ* measurements, and ultimately returns to Earth |
| **NUMBER OF ARCHITECTURAL ELEMENTS:** | 1 non-polar Artemis sortie mission, including an uncertain number of landers, crew, and mobility assets |
| **MODE OF SAMPLE RETURN:** | Artemis astronaut sample return (up to ~100 kg) |
| **MODE OF SURFACE MOBILITY:** | Astronauts on walking EVAs (≤2 km range) or potentially with unpressurized rovers (≤10 km range) |

**EVALUATION OF THIS ARCHITECTURE AGAINST THE SPARX FIGURES OF MERIT:**

**STRENGTHS:**

- **Architecture can provide sufficient local-scale mobility for determining the geologic context for the samples (~10–100 m)**
- **Architecture leverages well-trained astronaut crew who are adaptable and efficient, maximizing science return**
- Samples could be collected far from the lander, decreasing the risk of contamination from the lander and increasing the probability of meeting sample lithology requirements
- Architecture sample return capability (~100 kg) greatly exceeds the SPARX sample mass requirements per site (~2 kg)
- 

**WEAKNESSES:**

- **Architecture does not meet SPARX Mission Requirements for sampling the geologic diversity of SPA (≥10 sites, ≥1,000 km range)**, although a well-chosen landing site and the inclusion of mobility (e.g., an unpressurized rover) may allow the crew to access a ~2–3 such sites
- **Architecture does not meet the total SPARX sample mass requirements (≥10 kg from ≥10 sites)**
- **Architecture would not provide sufficient mobility to accomplish SPARX's traverse-centric Science Experiments (quasi-continuous near-surface measurements ≥650 km range)**; an unpressurized rover (range ≤10 km) may contribute to these Experiments but cannot accomplish them

**OVERALL EVALUATION:**

**GOOD/FAIR.** This architecture could accomplish a subset of the SPARX Science Goals, especially if the mission is well-equipped (e.g., with an unpressurized rover, *in situ* instruments), and goes to a well-chosen landing site. The fundamental limitations for this architecture are the limited range of crewed missions. This architecture is more responsive than either a static lander or a small rover, however, it still fails to meet the SPARX science threshold.

### 10.4.9 Non-Polar Artemis Campaign

SPARX ARCHITECTURE | **Non-Polar Artemis Campaign**

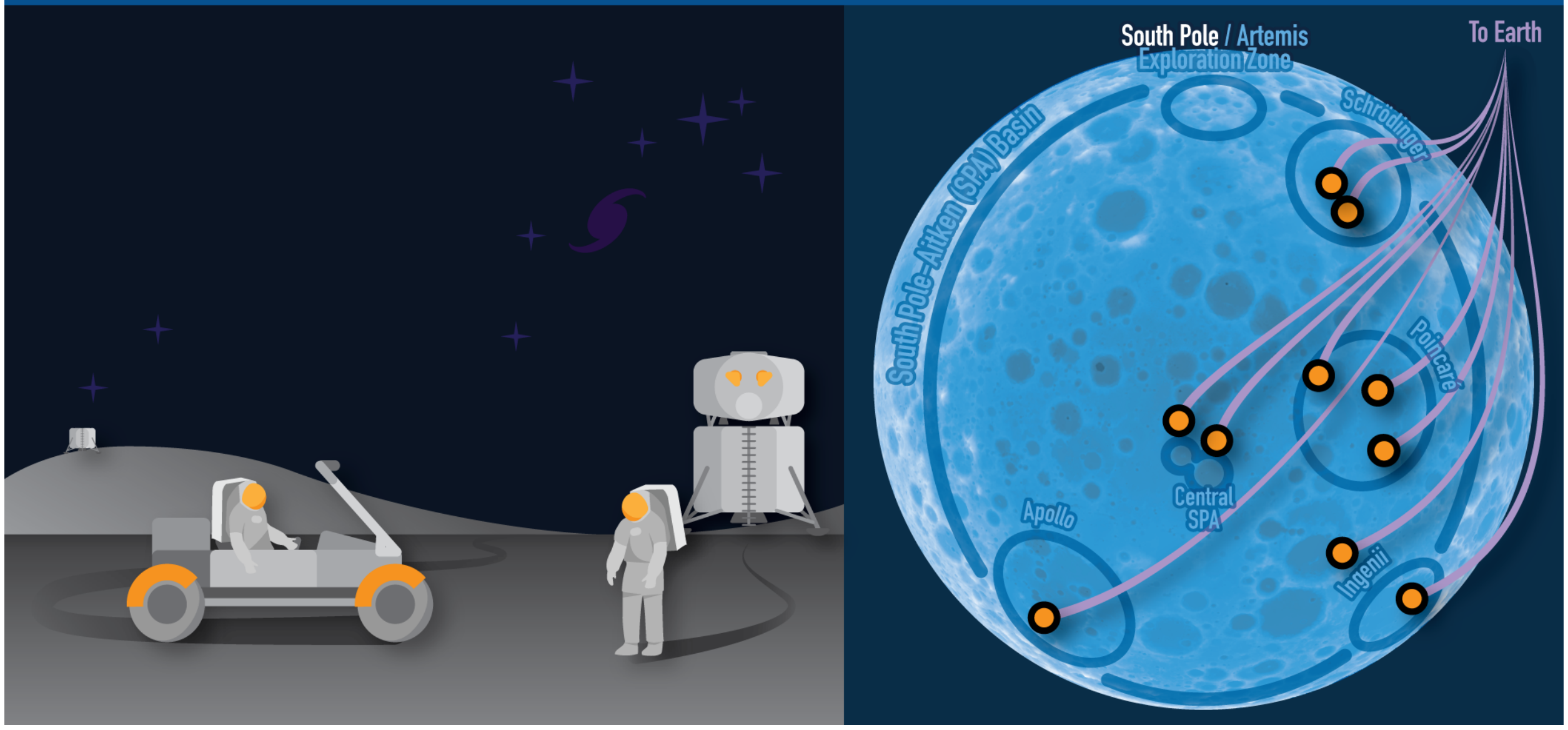

| SUMMARY OF THIS ARCHITECTURE: | |
|---|---|
| **DESCRIPTION:** | Multiple crewed Artemis mission land at distributed locations across SPA, performs EVAs, collects samples and in-situ measurements, and ultimately returns to Earth |
| **NUMBER OF ARCHITECTURAL ELEMENTS:** | ~10 non-polar Artemis sortie mission, including an uncertain number of landers, crew, and mobility assets |
| **MODE OF SAMPLE RETURN:** | Artemis astronaut sample return (up to ~100 kg per mission) |
| **MODE OF SURFACE MOBILITY:** | Astronauts on walking EVAs (≤2 km range) or potentially with unpressurized rovers (≤10 km range) |

**EVALUATION OF THIS ARCHITECTURE AGAINST THE SPARX FIGURES OF MERIT:**

| STRENGTHS: | WEAKNESSES: |
|---|---|
| – **Architecture can meet the total SPARX sample mass requirements (≥10 kg from ≥10 sites)**<br>– **Architecture can meet the SPARX Mission Requirements for sampling the geologic diversity of SPA (≥10 sites, ≥1,000 km range)**<br>– **Architecture can provide sufficient local-scale mobility for determining the geologic context for the samples (~10–100 m)**<br>– **Architecture leverages well-trained astronaut crew who are adaptable and efficient, maximizing science return**<br>– Samples could be collected far from the lander, decreasing the risk of contamination from the lander and increasing the probability of meeting sample lithology requirements | – **Architecture would not provide sufficient mobility to accomplish SPARX's traverse-centric Science Experiments (quasi-continuous near-surface measurements ≥650 km range)**; multiple unpressurized rovers (range ≤10 km) may contribute to these Experiments but cannot accomplish them<br>– **A campaign of this scale may not be programmatically favorable** as Artemis is currently prioritizing the lunar south pole and this would represent a radical change in Artemis plans |

**OVERALL EVALUATION:**

**VERY GOOD.** This architecture could accomplish a subset of the SPARX Science Goals, especially if the mission is well-equipped (e.g., with an unpressurized rover, in-situ instruments), and goes to well-chosen landing sites. The fundamental limitations for this architecture are the limited range of crewed missions, and the substantial cost of dedicating multiple Artemis landings to SPARX science destinations.

## 10.5 NON-POLAR ARTEMIS RENDEZVOUS IMPLEMENTATION

As noted in the SPARX SDT report, the commissioning and execution of the SPARX SDT study period came at a particularly dynamic time for NASA programmatic development, and included evolving architectures for the Artemis program, the development of the Moon Base architecture, and the initiation of the Ignition program. Reflecting these evolving programmatic settings, the SDT was additionally tasked with evaluating implementation architectures for SPARX utilizing Artemis astronauts in a non-polar exploration setting.

The main findings, described in detail in Sections 7.1 and 7.6.2 can be summarized as follows:

1. **The science of a SPARX mission is <u>significantly enhanced</u> by implementations utilizing crew rendezvous with the SPARX rover at a non-polar terminal exploration destination.** This strategy leverages crew science and exploration capabilities for high-priority targeted sample acquisition, while still allowing the full complement of SPARX science objectives to be met by in-situ measurements and sample acquisition along the planned traverse.
2. **A series of crewed, non-polar landings (i.e., a multi-landing campaign) is <u>not recommended</u> for accomplishing SPARX science goals.** This implementation would require a minimum of 5 landings to meet the SPARX science floor and does not meet the SPARX requirements for in-situ measurements.

Given the potential benefits to science for the non-polar astronaut rendezvous implementation, and the strength of the SDT recommendation for this architecture, the SDT presents here a detailed analysis of the proposed science and programmatic benefits of various non-polar rendezvous implementation scenarios (Section 1), a prioritized list of rendezvous locations and associated astronaut-assisted SPARX experiments (Section 2), and an assessment of the science risks and benefits to the proposed approaches (Section 3).

### 10.5.1 Benefits of Crewed Exploration to SPARX Science

A central conclusion of the SPARX SDT is that astronauts are best deployed in situations that take advantage of human decision-making, critical thinking, training and experience, and situational awareness. When placed within the context of the SPARX science objectives and mission concept of operations, the SDT determined that astronauts would best be implemented in identifying and collecting samples of specific lithologies where otherwise specialized ConOps acquisition modes have been defined (see Section 6.6, Table 6.11 for further detail). Two of the highest-priority SPARX lithologies (anorthosite and Th-rich material) require these specialized, highly targeted ConOps modes. As was observed in the Apollo missions, crews trained in the identification of key lunar lithologies were able to identify these samples on the surface (Figure 1), in both planned EVA locations, and along the exploration route. The collection of these samples, both as float rocks on the surface, and also chipped off of boulders, significantly increased the science return of these missions.

From the SPARX lithologies, the SDT identified anorthosite as the highest priority lithology class for targeted astronaut objectives. Anorthosite originating from the lunar farside highlands is critical to constraining the age and formation of the Moon and is a required lithology for two of the three SPARX science floor objectives (Objective 1.3 and 1.1, Section 3.6); therefore, the successful acquisition of high-quality anorthosite samples is critical to SPARX mission success. Anorthosite has a bright white color, and experience from Apollo astronauts suggests that large clasts of it are easy to identify on the Moon (e.g., Figure 10.3). While remote sensing datasets have been used to identify pixels with the highest fraction of anorthosite, geologic setting and experience from the Apollo missions suggests that anorthositic rocks will be discretely scattered throughout a sampling region. Therefore, in the case of robotic sample collection, a dedicated ground-in-the-loop "rock-hunting" ConOps mode is required to systematically search a target region with the goal of identifying the white anorthosite samples. In the case of astronaut assisted sample collection, the crew would deploy to a previously targeted sample region, and leveraging training and standard suit mobility, would be able to explore the region and identify promising anorthosite clasts with more efficient movement and ground-in-the-loop interactions.

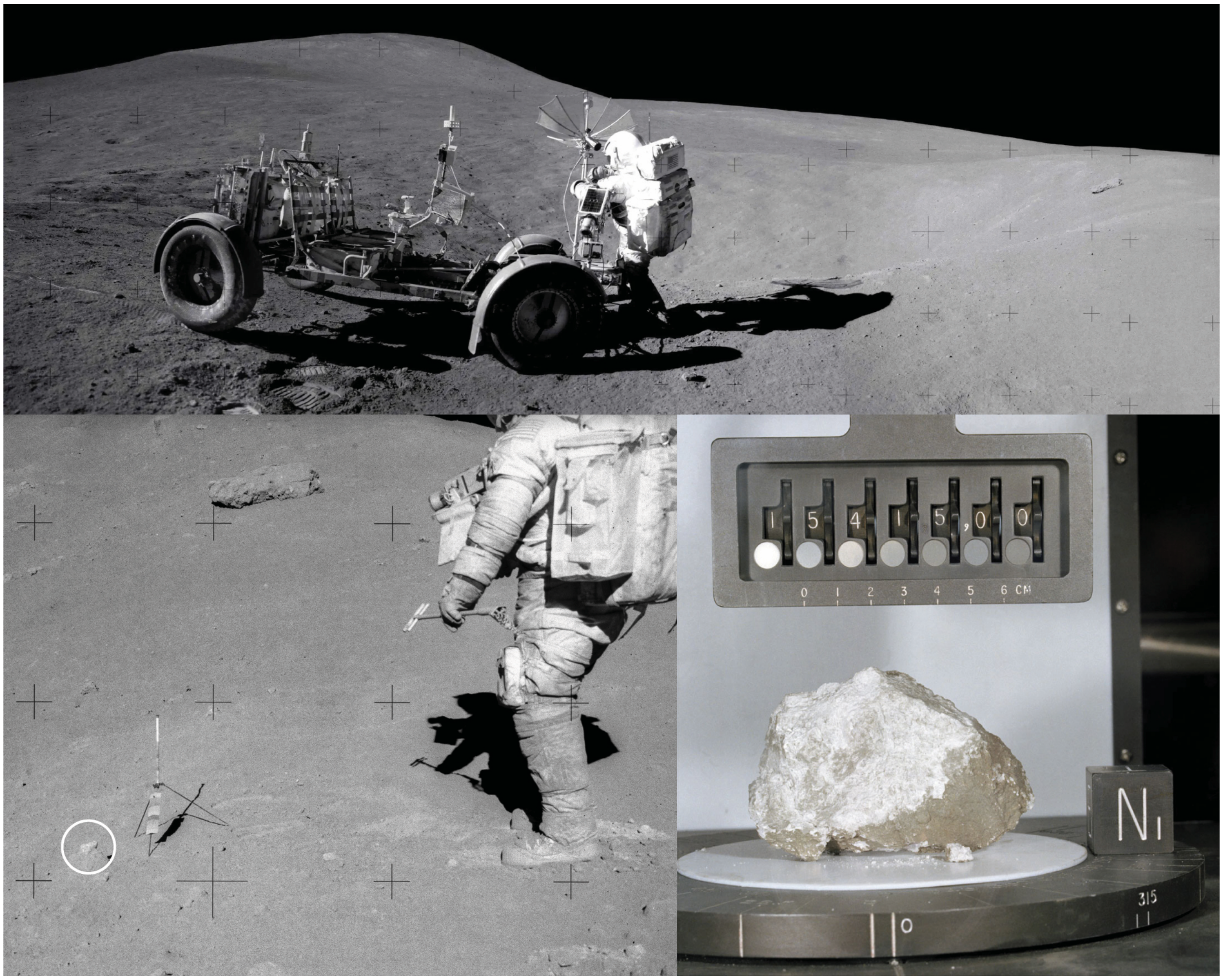

**Figure 10.3.** *Well-trained astronauts can identify and collect high-priority lunar samples. Top: panorama during Apollo 15 EVA 2, where the astronauts first spotted the "Genesis Rock." Bottom left: The "Genesis Rock" on the surface of the Moon, prior to collection. Astronauts were able to identify the rock based on its distinctive color and texture. Credit: NASA / Dave Scott / Irwin. Credit: NASA / AS15-90-12228. Bottom left: Photograph of Apollo 15 sample 15415, the "Genesis Rock" taken in curation. Credit: figure 6-8 of the Apollo 15 Preliminary Science Report (NASA SP-289, 1972).*

Crewed sampling of anorthosite has the additional benefits of sample size flexibility. While the majority of SPARX investigations preferentially utilize rocks and rocklets with diameters of a few millimeters, decades of laboratory experience studying lunar anorthosites suggests that these particular samples benefit from larger rock samples. For the SPARX baseline design, the SDT balanced this desire for larger samples with the operational realities of sieving activities, expected regolith particle size distributions, and overall sample acquisition and storage system complexity with a final requirement of sampling anorthosite rocks with diameters ≥40 mm. If present, astronauts could find and collect larger hand-clast sized samples that would be too large for the nominal SPARX sieve system, but which could be strongly beneficial to laboratory investigations. Astronauts also have the flexibility to acquire anorthosite samples from more complicated regions, such as chipped from a larger boulder, or from peak rings.

The science benefits described here for anorthosite (rapid identification, enhanced sampling capabilities) also extend to other lithologies that may be present in an exploration region. Trained crews would also be able to evaluate the geologic diversity of a given sample site and ensure that samples are acquired both of representative material, and also compositional or morphologic outliers. While the nominal SPARX sample acquisition emphasizes the collection of regionally representative material, exploration experience shows that investigations of outliers can be just as scientifically valuable. The use of astronauts for sample collection in strategic locations allows SPARX to broaden the potential science impact of the mission, without adding mechanical complexity or other operational constraints.

**Finding: The SDT finds that SPARX science is enabled by focusing crew efforts on acquiring SPARX-relevant lithologies with either “targeted” or “rock-hunting” ConOps modes. Of these, the SDT recommends anorthosite as the highest priority lithology for targeted astronaut acquisition.**

### 10.5.2 Prioritized Non-Polar Rendezvous Target Locations

Using the SPARX baseline traverse, and the scientific rationale for exploration defined above, the SDT identified and prioritized five potential astronaut rendezvous locations (yellow diamonds in Figure 10.4). In all cases, the Rover would start in central SPA before traversing outward towards one of these rendezvous locations. The locations are described below, along with a brief description of the scientific rationale, astronaut focus activities, and any tools or mobility that would be required for astronauts to complete the proposed science.

#### 10.5.2.1 Schrödinger southern peak ring massif, traverse site #19

The top recommended location for a non-polar Artemis rendezvous is at this Schrödinger peak-ring massif. This location hosts remotely identified exposures of both anorthosite and olivine, and the geologic setting provides opportunity for other unique lithologies to have been uplifted from depth. The location of this stop would allow the SPARX rover to complete all or nearly-all sampling activities prior to rendezvous. The SPARX Rover could then complete the final non-sampling traverse science objective (Site #21, E12, Section 4.1.12) after delivering the sample payload to the astronauts.

#### 10.5.2.2 Poincare southern peak ring, traverse site #11 or #13

The second recommended location for a non-polar Artemis rendezvous is at the southern peak ring of Poincare. As was noted above for the Schrödinger location, peak-rings are optimal sampling settings because they expose material that has been uplifted from depth, and whose regolith often contains coarser grain size fractions (i.e., more rocks) that have been eroded from the topographic peaks. The Poincare peak-ring region includes discrete contacts between basaltic mare units, and peak rings material that could be used to date the Poincare formation event. Astronauts could also identify other unique lithologies that are below the resolution of current remote sensing observations. The location of this stop would allow the SPARX rover to complete the threshold traverse mission before delivering the sample payload to the astronauts for return. If desired, and within Rover design capabilities, the SPARX Rover could then continue traversing, either along the remainder of the SPARX baseline traverse, or to other science destinations.

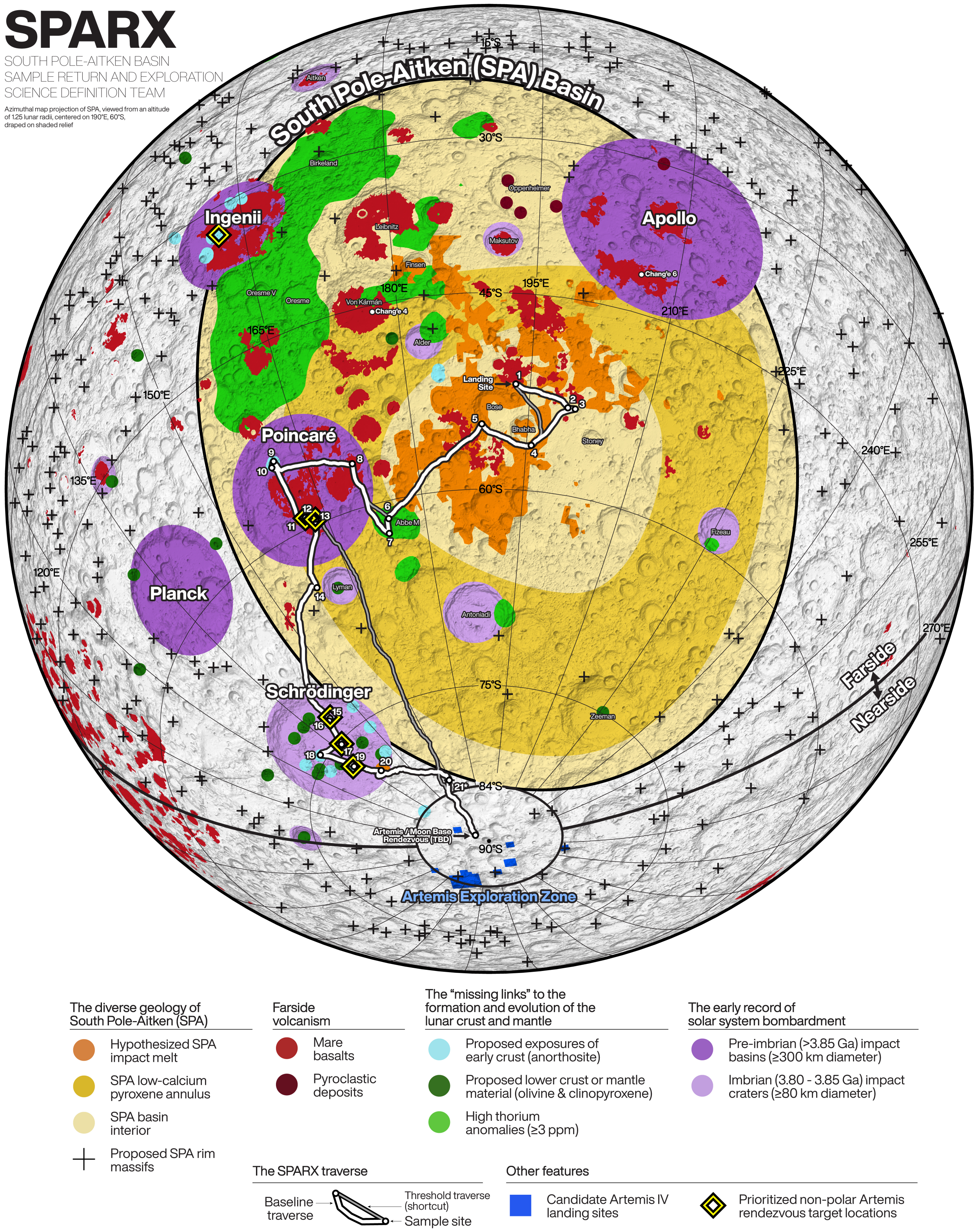

**Figure 10.4.** *The SPARX baseline (and threshold) traverse on a map of the entire South Pole-Aitken (SPA) basin (adapted from Figure 6.39), with prioritized non-polar rendezvous target locations highlighted in yellow diamonds.*

#### 10.5.2.3 Schrödinger northern peak ring, traverse site #16

Another proposed peak ring location within Schrödinger would be the northern peak ring region. This region similarly includes remote sensing observations of anorthosite and olivine in a geologic setting optimized for astronaut exploration. This location is considered less ideal than traverse site #19 because it would require the SPARX Rover to either not complete sampling at locations 17-19, or to backtrack to the rendezvous location after completing these locations.

#### 10.5.2.4 Schrödinger pyroclastic vent, traverse site #17

This location is considered a lower priority for SPARX science objectives, however, the location near a large pyroclastic vent and fracture system could provide a wide range of science opportunities for astronaut exploration objectives. This location is reminiscent of the Apollo 15 Hadley Rille exploration and would allow rare access to the source vent of an explosive pyroclastic deposit. Despite the scientific opportunities, this location is not anticipated to host any of the special targeted lithologies, and the location would likely require the SPARX Rover to backtrack to after completing the sampling at locations 18 and 19 to rendezvous with the crew.

#### 10.5.2.5 Ingenii kipuka (161.96° E, 33.22° S), not included in baseline traverse design

Although not included in either the baseline or threshold traverse, the SDT continues to recognize the scientific viability of the Ingenii basin for SPARX objectives (discussed further in Section 5). If this location fit within the design and safety parameters for astronaut landing, it could provide a strongly compelling target for a SPARX Rover-astronaut rendezvous. The location includes remote sensing observations of anorthosite, in addition to, contacts with several basaltic units of variable age and composition. This location is also geologically coincidence with a lunar swirl and would provide an excellent opportunity for measurements complementary to NASA's upcoming Lunar VERTEX PRISM payload. The challenges to this location are the significantly higher latitude, which would require specific design coordination for the SPARX rover, the astronaut support systems, and a redesign of the SPARX traverse.

**For all of these locations, the SDT would strongly recommend a detailed geologic study to identify complementary science goals and activities for the astronauts to complete, in addition to, the SPARX relevant activities.**

The SDT emphasizes that none of the locations identified above would require the astronauts to have additional mobility beyond the nominal EVA distance (~2 kilometers for a walking-only EVA). However, additional mobility (e.g., a crewed rover) could provide access to additional terrains of scientific interest. It is important to note that all of the proposed human landing systems are anticipated to disturb an extensive region of regolith, and therefore, efforts should be made to land as far away as possible from the target exploration location so as to provide minimal contamination. For similar reasons, the SDT would strongly recommend the stationing of the SPARX Rover well-outside the anticipated deformation region prior to landing, and then subsequent traversing into the region for characterization and sample transfer.

The locations and science targets identified above also would not require specific tools for astronaut use beyond basic surface tools similar to Apollo (e.g., rock hammer and rake). Because of the strongly diagnostic color characteristics of anorthosite, the crew do not require handheld geochemical analysis tools. Nevertheless, the SDT would recommend that the crew work synergistically with the SPARX Rover instrument suite to characterize the geologic setting of the sampling locations, and to verify target lithology prior to sample collection. It is noted that these measurements to not need to be made simultaneously. Therefore, if there are safety concerns for either the crew or the Rover, measurements and sample collection can be done asynchronously in the same region.

Other rendezvous sites were discussed (e.g., Bhabha crater) but were not prioritized. As discussed in Section 7.6.2, the SDT prioritized utilizing astronaut capabilities for lithologies and ConOps that strongly benefit from astronaut decision making and training. The lithologies present in central SPA (Bose, Bhabha, and central mare units), although scientifically critical to establishing the early SPA chronology, are also not readily identifiable on the surface using either *in situ* remote sensing instruments or astronaut geologic analysis (see Section 4.2.2 for discussion of impact melt identification). Instead, collection of these samples relies upon careful sample site selection and statistically robust sampling quantities. The SDT notes that there may be synergy with the forthcoming National Academies of Sciences, Engineering, and Medicine study “Key Non-Polar Destinations Across the Moon to Address Decadal-level Science Objectives with Human Explorers” which was conducted concurrently to the SDT.

**Finding: The SDT finds that the strongest science synergy for a SPARX-Non-Polar Artemis rendezvous is in Schrödinger basin, near central peak ring regions where remote sensing instruments have indicated a high confidence in the presence of desired target lithologies, and where the existing geologic setting (i.e., a peak ring kipuka, traverse stop #19) leverages crew awareness and mobility for efficiently finding target rocks that have been shed from steeper terrains, and/or enabling chipping of critical materials from boulders.**

### 10.5.3 Risks and Benefits of a Non-Polar Artemis Rendezvous

In addition to the science benefits provided by astronaut-assisted sample collection, the proposed non-polar Artemis rendezvous architecture could also provide engineering and programmatic benefits. The most straight-forward of these benefits are to decrease the engineering design constraints, and potentially decrease the mission duration, by no longer requiring traversing through the south polar region. Lessening these constraints can have direct implications for hardware; for example, not requiring operations in the lunar south pole can result in reduced battery sizing. As described throughout Section 6, the final SPARX traverse leg to Moon Base in the south pole Artemis exploration zone places multiple design constraints upon the rover that are not required to complete the SPARX science. By moving the rendezvous location to a non-polar location, these design constraints are removed, simplifying the rover design and decreasing mission risk.

The SDT also considered the mitigation to mission risk provided by delivering some or all of the SPARX samples to astronauts for return at the mid-point of the mission. If the astronaut rendezvous were to occur in one of the recommended Poincaré basin locations at the end of the SPARX threshold traverse, the astronauts could return this sample cache and allow earlier validation of SPARX science objectives. The SPARX Rover would then have the flexibility to continue the baseline traverse as designed or modify the traverse in response to the scientific results from the initial cache. For example, if the anorthosite samples from the threshold cache fully address the relevant objectives, the second half of the traverse could be modified to focus on other lithologies. Alternatively, if the threshold cache is found to sufficiently address the SPARX objectives, NASA could then direct the Rover towards completing other science investigations. The SDT does note, however, that these perceived programmatic benefits do have the science tradeoff of potentially not addressing all of the SPARX baseline objectives, and NASA should strongly weigh these programmatic benefits with the science tradeoff and a practical understanding of future rover utility.

**Finding: The SDT finds that the strongest benefits to a SPARX-Non-Polar Artemis rendezvous in Poincare basin are related to overall mission risk mitigation. A mid-traverse or terminal rendezvous in Poincare basin would coincide with the recommended SPARX threshold mission profile, allowing threshold science requirements to be met without the additional risks of requiring the post-sampling traverse to the lunar South Pole.**

## 10.6 COMPILED SPARX SDT FINDINGS

Throughout the report, the SPARX SDT has highlighted a series of major findings. These findings are compiled below for ease of reference.

### Science (Section 3)

- The SDT finds that the highest priority science goals of a South Pole-Aitken basin mission that includes sample return (e.g., SPARX) are to constrain the chronology of key processes in the earliest period of Solar System formation, to constrain and characterize the lunar crustal asymmetry as a mirror into planetary crustal formation processes, and to characterize the physical and compositional effects of mega-basin impacts as early drivers of planetary evolution.
- The SDT reaffirms the findings of the Origins, Worlds, and Life Planetary Science Decadal Survey in establishing sample return from the Moon's South Pole-Aitken basin as the highest priority lunar science mission resulting from the preponderance of Decadal Level Science Questions addressed by the mission, and the Solar System-wide implications of the proposed mission science goals.
- The SDT further emphasizes that the Moon's South Pole-Aitken basin is a geologically special region stemming from its location, size, and antiquity. As a result, the critical science questions identified here cannot be addressed by other missions either throughout the Solar System or in other non-SPA regions of the Moon.

### Science Requirements (Section 4)

- The SDT finds that a baseline set of 7 in situ measurement classes (context imaging, context mineralogy, fine-scale imagine, fine-scale mineralogy, bulk elemental chemistry, gravity field, and magnetic field) are required to achieve the SPARX objectives, and these measurements can be accomplished by a minimum payload of 5 instruments.
- The SDT finds that in situ measurements taken along the prescribed intervals of the SPARX mission, and in particular, in the sample acquisition regions are required to meet the science objectives. The SDT further notes that in situ measurements taken in a manner that are not coincident with paired sample activities (e.g., measurements taken at different locations on the Moon or as part of different missions) do not successfully meet the required criteria, and risk not achieving SPARX science objectives.
- The SDT finds that the measurements required to address the defined SPARX objectives cannot be achieved using available in situ analysis methods, and therefore the SPARX mission requires sample return to Earth to successfully address mission science objectives.
- The SDT finds that SPARX science objectives require the return of specifically targeted lithologies in a range of prescribed rock sizes (diameters ≥ 3mm, 5mm, 8mm, 20, and 40mm) and unsieved bulk regolith. The SDT notes that returning fewer, larger mass samples risks not meeting the defined measurement requirements and significantly increases scientific risk.
- The SDT finds that the results of the LEAG/ExMAG Artemis Sample Return Specific Action Team (ASR-SAT) report are highly applicable to future SPARX returned samples, and the SDT encourages NASA to implement key findings of this report in advance of SPARX sample return. In particular, the SDT highlights the ASR-SAT finding of an impending critical lack of both laboratory facilities and trained personnel in the United States to carry out many of the highest-priority SPARX investigations. The SDT strongly recommends NASA address the findings of this report, particularly for the critical laboratory facilities required for SPARX-related investigations

## Mission Architectures (Section 5)

- The SDT recommends baseline mission implementation architectures that can support exploration of the Central SPA region, Poincaré basin and surrounding region, and the Schrödinger basin and surrounding region. The SDT recommends a threshold mission implementation architecture that supports exploration of the Central SPA region through the Poincaré basin and surrounding region.
- The following findings represent Figures of Merit for evaluating SPARX implementation architectures:
- The SPARX baseline mission requires an implementation that can return ≥20.5 kg of lunar samples, and the threshold implementation must be able to return ≥9.8 kg of sample material.
- The SPARX sample acquisition and caching system must be able to: (1) Acquire samples specifically targeted from in-situ measurements, (2) acquire specified numbers of samples within certain size ranges, (3) acquire un-sieved regolith samples.
- SPARX requires that samples be acquired outside the contamination zone of landing rocket exhaust. Cross-contamination between acquired samples should be kept to ≤ 1% by volume, and all mission-level sample integrity and contamination requirements should be commensurate with NASA lunar sample curation requirements.
- SPARX implementations must acquire sufficient in-situ measurements at each site to adequately characterize the geologic environment.
- SPARX science objectives require the periodic acquisition of in-situ geologic context measurements in-between major regions of interest and designated sampling sites.
- The SPARX implementation architecture should optimize science return within a framework of minimized cost, complexity, and overall mission risk.
- The SDT finds that while the unmargined SPARX threshold sample return mass (9.8 kg) is technically below the TOR recommended commercially derived robotic NTE value of 10 kg, many practical challenges remain before recommending this implementation system. The SDT chiefly suggests that NASA consider the current and near-future capabilities, and capability demonstration of commercially derived sample return systems against the SPARX requirements.
- Sample return by Artemis astronauts is commensurate with the baseline and threshold sample mass required by SPARX. Astronaut architectures have the opportunity to provide additional benefits, however, are dependent on concordant schedules between SPARX and Artemis launches, and permissible astronaut exploration zones.
- The preponderance of science objectives and laboratory measurements required by SPARX necessitate implementation architectures with terrestrial sample return.
- The SPARX mission can achieve its science requirements with relatively simple interactions with the lunar regolith, such as scooping and sieving. More complicated interactions with lunar regolith, such as drilling, coring, or boulder chipping, are not necessary to address the SPARX science requirements.
- The SPARX mission does not require contact or close-proximity measurements and therefore should not assume any arm-mounted science instruments. Excluding arm-mounted science instruments dramatically reduces the requirements on the robotic arm and thus reduce mission cost and complexity.

- The distance and range requirements of SPARX are incompatible with current hopping technology, even with a high-performance vehicle like Starship. Any hopping architectures would require the use of multiple hopping and sample return elements and face significant challenges in meeting SPARX science requirements E11 and E12, as well as SPARX sample contamination requirements.
- The SPARX science requirements cannot be achieved by a small- to medium-size rover designed to traverse over a single lunar day. The "one day wonder" architecture fails to meet the minimum required geologic sampling regions and number of feasible sampling stops.
- Large, long-range rovers that rendezvous with Artemis astronauts fully accomplish the SPARX science objectives and provide the highest science value per dollar.
- The recommendation of the SDT is that a SPARX baseline mission be implemented as a long-distance rover that acquires samples using a robotic arm to scoop and sieve lunar regolith while traversing from the central SPA region through Poincare basin, and Schrödinger basin before rendezvousing with Artemis astronauts at the lunar South Pole.

### Design Reference Mission (Section 6)

- Based on analysis of operational modes by the technical team, the SDT finds that the SPARX rover should be able to operate in a "fully autonomous" mode for the required long-distance driving modes. The SDT emphasizes that full autonomy is only desired for driving modes, and that SPARX science requires more iterative ground-in-the-loop operations during other mission activity phases, such as sampling and reconnaissance.
- The SDT finds that the sampling subsystem requires documentation imaging throughout the sample acquisition and storage process to verify successful acquisition of the appropriate sample material.
- The SDT finds that as a result of the expected distribution of rock sizes in the lunar regolith, the sampling subsystem arm should allow access to a large workspace volume, in order to support the required sieving-based strategy and should also support more targeted acquisition modes for the larger clasts.
- The SDT recommends a 3 mm sieve size for collecting and concentrating the required rock fractions and also requires the ability to collect unsieved regolith.
- Finding: The SPARX traverse requires thermal accommodation for latitudes as high at 52° S down to the lunar South Pole (90° S).
- The thermal subsystem must be designed to allow the rover to operate at any part of the lunar day, regardless of latitude, orientation, terrain, or other environmental factors. The SDT strongly recommends against thermal subsystem designs that do not meet these criteria (e.g., requiring a pause in operations during the hotter parts of the lunar day).
- The SDT finds that the rover telecommunications subsystem should include, at a minimum, a Ka-band high gain antenna (HGA) to support data transfer; as well as an S-band low gain antenna (LGA) to support pointing, navigation, and timing (PNT) data, and back-up/safe mode.
- The SPARX science objectives require reconnaissance of the pre-defined traverse sample location to acquire geologic context, perform in situ measurements, and identify the locations within the local (100 m × 100 m) region most conducive for meeting the sampling requirements.
- The SPARX science objectives require ground-in-the-loop verification of sample properties during the initial phases of scooping. Verification is determined using the fine-scale mineralogy instrument to confirm that the sampling area meets the specified lithology requirements. Once the science team verifies sample properties, autonomous sampling can begin.

- The SDT finds that the Sampling ConOps described above should be treated as a universal process, regardless of the final rover design. Due to the importance of sample selection and collection for the mission Science Objectives, the SDT has worked iteratively with the technical team to define a set of ConOps procedures that should be utilized by any final mission design.

### Programmatic Factors (Section 7)

- For mission profiles involving SPARX-Astronaut rendezvous at a south polar basecamp/Moon Base, astronaut activities are best utilized for sample packaging for Earth return, or to assist in sample container triage if Earth return mass is limited. The SPARX SDT finds that the preferred profile for astronaut utilization is rover rendezvous with non-polar crewed deployment to a strategic terminal traverse sampling location.
- The SDT finds that commercial delivery of a SPARX-like mission to the lunar surface remains viable if large-delivery platforms are proven successful, for example by successfully delivering the VIPER mission to the lunar surface.
- The SDT finds that while commercial robotic sample return could be viable for the reduced scope of the threshold mission, significant technical challenges remain in demonstrating this technology. The SDT recommends that decisions about commercial sample return be made as soon in the design process as possible, as this modality will drive multiple design points across the rover, sampling system, and eventual return asset.
- The SDT finds non-unanimously that a commercial LTV platform could accomplish the proposed SPARX mission; however, the SDT strongly notes that any credible proposal for such must take into account the environmental and ConOps requirements for SPARX, which differ from the current LTV requirements. The SDT recommends that commercial partnerships on appropriate subsystems, components, and instrumentation continue to be pursued in the manner that is commensurate with historical NASA developments.
- The SPARX SDT recommends NASA continue to pursue international collaboration on the development of a SPARX-like mission in ways that are commensurate with existing and past partnerships. The SDT notes that these partnerships could include a range of technical and team contributions and should also hold reciprocal benefits for partner institutions.
- The SPARX science objectives require a balance of robust in-situ measurements to address and support the terrestrial sample analyses. The SDT further affirms that the highest-priority science objectives for an SPA mission cannot be met without the combination of in-situ measurements and analyses of returned samples.
- The SPARX strawman payload suite is tailored to meet the science measurement requirements with the greatest degree of flexibility and leveraging synergies with existing remote sensing datasets. The SDT finds that the inclusion of high-cost, high-complexity in-situ analytical payload elements does not sufficiently address the SPARX measurement requirements to justify the resulting increases to mission cost, complexity, and duration.
- The threshold level of SPARX science cannot be achieved from either a static or local (<100 km) scale investigation. The SPARX threshold investigation makes progress on eight of the nine SPARX objectives, although the reduced scope of sampling locations and returned sample mass decreases the degree to which these investigations can be addressed, and increases scientific risk. NASA should weigh cost and design trades when considering both the baseline and threshold mission profiles, particularly the costs associated with the sample return mechanism.

- The SPARX SDT does not consider “campaign science” consisting of multiple (likely 10s) of low-capability surface assets (static landers, small rovers) to be sufficient for accomplishing the science objectives of SPARX on a timescale that is commensurate with the proposed baseline mission, nor does the SDT consider this to be a cost-effective approach for NASA given the greatly increased number of deliveries, sample returns, and in-situ instruments required to support each campaign science asset.
- The SDT finds that the optimal implementation for mixed-astronaut and robotic SPARX mission is non-polar astronaut landing at the terminal location of a long-distance rover traverse. This implementation would optimize astronaut surface operations for targeting specific high-priority lithologies (anorthosite and hypothesized mantle rocks), while allowing the long-distance rover to collect samples and perform in-situ traverse science from across the traverse route. This implementation brings the sample return benefits of an astronaut architecture without the added engineering and mission duration challenges of a south polar rendezvous.
- A long-distance rover rendezvous with astronauts remains the optimal solution to maximizing SPARX science return while minimizing overall programmatic cost. Astronaut rendezvous at a nonpolar, terminal SPARX sampling location is potentially more optimal than a South Pole rendezvous.
- The SDT strongly recommends that NASA organize a SPARX mission-level science team with a range of expertise relevant to the design, formulation, and implementation of the mission, using the expertise solicited for the SPARX SDT as a suggested topical guide. This team should be intimately integrated with payload leads, engineering and operations leads, and other relevant personnel. The SDT further finds that this mission-level science team should be integrated into mission operations regardless of the nature of the operating partner.
- The SDT finds that no new technologies are required for development prior to initiating a SPARX mission program. Within this finding, the SDT does recognize the importance of continued development on a number of enabling technologies, including technologies supporting extreme long-distance roving, electrodynamic dust shielding, and autonomous systems.

### Non-Polar Artemis Rendezvous (Appendix 10.5)

- The SDT finds that SPARX science is enabled by focusing crew efforts on acquiring SPARX-relevant lithologies with either “targeted” or “rock-hunting” ConOps modes. Of these, the SDT recommends anorthosite as the highest priority lithology for targeted astronaut acquisition.
- The SDT finds that the strongest science synergy for a SPARX-Non-Polar Artemis rendezvous is in Schrödinger basin, near central peak ring regions where remote sensing instruments have indicated a high confidence in the presence of desired target lithologies, and where the existing geologic setting (i.e., a peak ring kipuka, traverse stop #19) leverages crew awareness and mobility for efficiently finding target rocks that have been shed from steeper terrains, and/or enabling chipping of critical materials from boulders.
- The SDT finds that the strongest benefits to a SPARX-Non-Polar Artemis rendezvous in Poincare basin are related to overall mission risk mitigation. A mid-traverse or terminal rendezvous in Poincare basin would coincide with the recommended SPARX threshold mission profile, allowing threshold science requirements to be met without the additional risks of requiring the post-sampling traverse to the lunar South Pole.

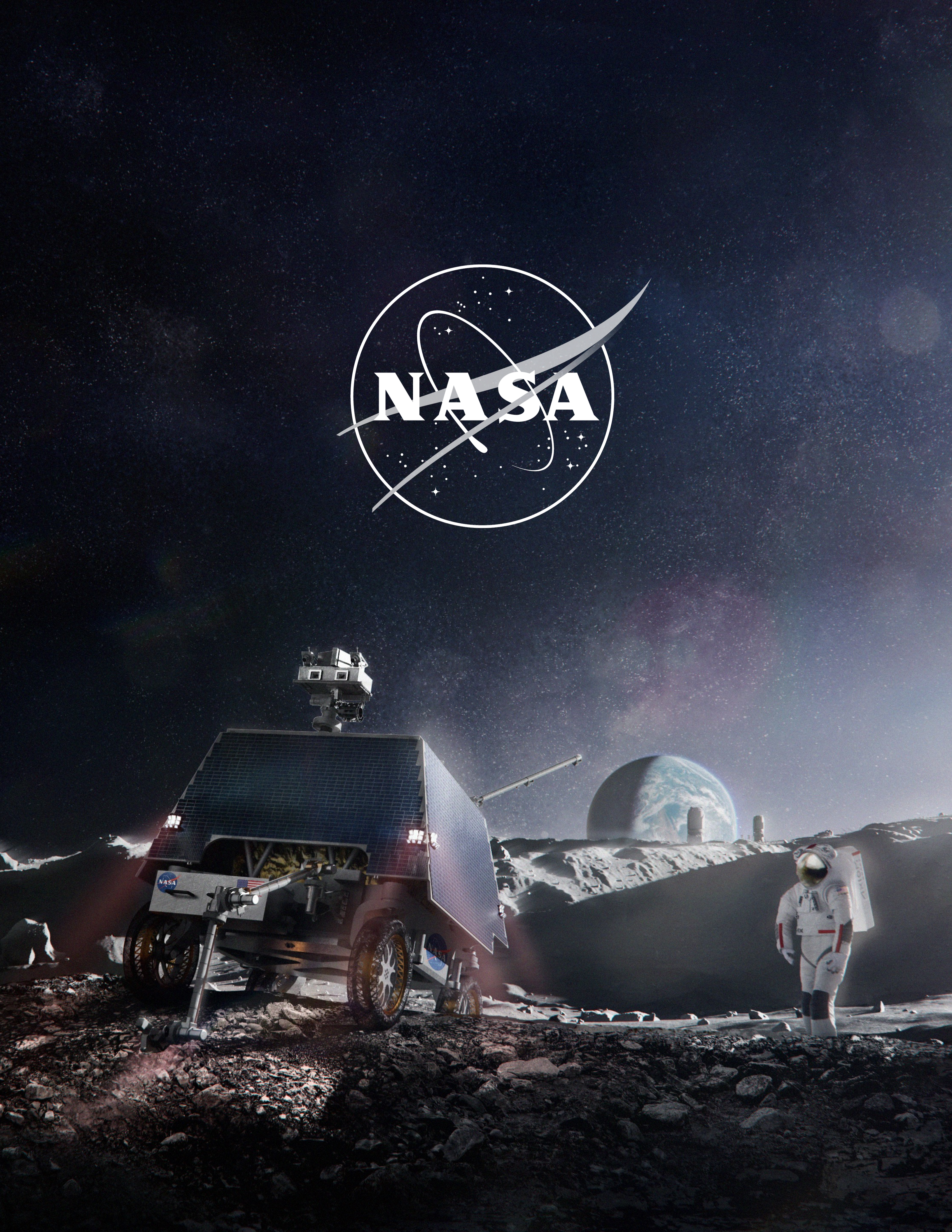
NASA